\documentclass[a4paper,12pt,twoside]{thesisbook}
\usepackage{amsmath} %mathematical symbols etc.
\usepackage{amssymb} %mathematical symbols etc.
\usepackage{amsthm}  %for theorems, propositions, proofs etc.

\usepackage{color}
\definecolor{azur}{rgb}{0.118,0.498,0.796}
\definecolor{green1}{rgb}{0.21,0.6,0.32}
\definecolor{indigo_elec}{rgb}{0.435,0,1}

\usepackage{latexsym}
\usepackage{amssymb}
\usepackage{bbm}
\usepackage{graphicx}
\usepackage{subfig}
\usepackage{wrapfig}
\usepackage{psfrag}
\usepackage{mathrsfs}
\usepackage{fancyhdr}
\usepackage{enumitem}
\usepackage[utf8]{inputenc}
\usepackage{setspace}
\usepackage{textcomp}
\usepackage[retainorgcmds]{IEEEtrantools}
\usepackage{url}
\usepackage{minitoc}
\usepackage{tikz}
\usepackage{multicol}
\usepackage[toc,page]{appendix}
\usepackage{feynmp}
\usepackage{marginnote}
\usepackage{cancel}
\usepackage{anyfontsize} % for chapter number
\usepackage{titlesec}
\usepackage{cancel}
\usepackage{empheq}  % eqs in boxes
\usepackage{etoolbox}
\makeatletter
\patchcmd{\@chap@pppage}{\thispagestyle{plain}}{\thispagestyle{empty}}{}{}
\makeatother

\usepackage[a4paper,twoside, hmarginratio={1:1}]{geometry}
\usepackage{geometry}
\usepackage[a4paper]{meta-donnees}
\ifx\pdfoutput\undefined
\usepackage[bookmarks]{hyperref}	% This is for arXiv.org
\else
\usepackage{hyperref}	% This is for pdftex
\fi
\hypersetup{colorlinks=true,urlcolor=indigo_elec,citecolor=azur,linkcolor=green1,linktocpage=true,
breaklinks=true, %permet le retour à la ligne dans les liens trop longs
pdftitle={Supersymmetric Dark Matter candidates in light of constraints from collider and astroparticle observables},
pdfauthor={Jonathan DA SILVA},
pdfsubject={PhD Thesis},
pdfkeywords ={Dark Matter, Supersymmetry, Collider and astroparticles constraints, Neutralino, Right-Handed sneutrino
, UMSSM}}

    \makeatletter
    \renewcommand\part{%
      \if@openright
        \cleardoublepage
      \else
        \clearpage
      \fi
      \thispagestyle{empty}%
      \if@twocolumn
        \onecolumn
        \@tempswatrue
      \else
        \@tempswafalse
      \fi
      \null\vfil
      \secdef\@part\@spart}
    \makeatother

\begin{document}
\dominitoc

\pagestyle{empty}
% la ligne ci-dessous est à insérer obligatoirement dans le préambule du document avant \begin{document}

%\usepackage[a4paper]{meta-donnees}

% les lignes en bas sont à insérer obligatoirement après \begin{document}

%%%%%%%%%%%%%%%%%%%%%%%%%%%%%%%%%%%%%%%%%%%%%%%%%%%%%%
%%             Commandes Meta-données               %%
%%   à renseigner par les auteurs pour générer      %%
%%     la couverture modèle Univ. Grenoble          %%
%%%%%%%%%%%%%%%%%%%%%%%%%%%%%%%%%%%%%%%%%%%%%%%%%%%%%%
%%      Fichier encodé au format ISO-8859-16        %%

%\Sethpageshift{???mm}   %%optionnel : à décommenter si besoin pour ajout d'espace afin de center la couverture horizontalement (valeur par défaut est -5.5mm)
%\Setvpageshift{???mm}   %%optionnel : à décommenter si besoin pour ajout d'espace afin de center la couverture verticalement (valeur par défaut est -15.5mm)

%\Universite{}    %%optionnel : à décommenter et à renseigenr si vous voulez changer le non d'université
%\Grade{}         %%optionnel : à décommenter et à renseigenr si vous voulez changer le grade
\Specialite{Physique Th\'eorique}
\Arrete{7 ao\^ut 2006}
\Auteur{Jonathan \textsc{Da} \textsc{Silva}}
\Directeur{Genevi\`eve \textsc{B\'elanger}}
%\CoDirecteur{}    %%optionnel : à décommenter et à renseigenr si présence d'un Co-directeur de thèse
\Laboratoire{Laboratoire d'Annecy-le-Vieux de Physique Th\'eorique (LAPTh)}
\EcoleDoctorale{l'\'Ecole Doctorale de Physique de Grenoble}         
\Titre{Supersymmetric Dark Matter candidates in light of constraints from collider and astroparticle observables}
%\Soustitre{}      %%optionnel : à décommenter et à renseigenr si présence d'un sous-titre de thèse
\Depot{3 juillet 2013}

% Commande pour création de nouvelles catégories dans le jury:

%\UGTNewJuryCategory{...NomDeLaCategorie...}{...Definition...}

% Exemple \UGTNewJuryCategory{UGTFamille}{Membre de la famille} que nous ajoutons dans la commande \Jury ci-dessous sous la forme \UGTFamille{Jean Rousseau}{(...titre_et_affiliation...s'il_y_en_a...)}

\Jury{
%\UGTPresident{...Civilit\'e, Pr\'enom-et-Nom...}{...titre-et-affiliation...}
\UGTPresidente{Dr. Rohini \textsc{Godbole}}{Professeur, CHEP Bangalore, Inde}

\UGTRapporteur{Dr. Farvah Nazila \textsc{Mahmoudi}}{Ma\^itre de Conf\'erences, LPC Clermont}      %% 1er rapporteur
\UGTRapporteur{Dr. Ulrich \textsc{Ellwanger}}{Professeur, LPT Orsay}      %% second rapporteur

\UGTExaminatrice{Dr. C\'eline \textsc{B\oe{}hm}}{Charg\'e de recherche, Durham University, Royaume-Uni}     %% second examinateur
\UGTExaminateur{Dr. Anupam \textsc{Mazumdar}}{Professeur, Lancaster University, Royaume-Uni}    %% 3ème examinateur

\UGTDirecteur{Dr. Genevi\`eve \textsc{B\'elanger}}{Directeur de Recherche, LAPTh}       %% Directeur de thèse
%\UGTCoDirecteur{...Civilit\'e, Pr\'enom-et-Nom...}{...titre-et-affiliation...}     %% Co-Directeur de thèse s'il y en a

%\UGTInvite{...Civilit\'e, Pr\'enom-et-Nom...}{...titre-et-affiliation...}
%\UGTInvitee{...Civilit\'e, Pr\'enom-et-Nom...}{...titre-et-affiliation...}
}

\MakeUGthesePDG    %% très important pour générer la couverture de thèse

\newpage\null
\newpage

\thispagestyle{empty}
\begin{flushright}\textit{A meus avós.}\end{flushright}

\newpage\null
\newpage
\pagestyle{plain}

\frontmatter
\tableofcontents
\newpage\null\newpage

\chapter*{Acknowledgements - Remerciements}
\addcontentsline{toc}{chapter}{Acknowledgements - Remerciements}

\textit{Caution, high level of mixing French-English in the following pages !! \\For a safe reading, \href{http://translate.google.com}{{\color{black}Google Translate}} may be your friend... or not !}\\

Avant de vous laisser seul devant le reste de ce manuscrit de th\`ese \`a conjecturer sur le domaine de validit\'e de l'\'equation~\ref{eq:7.c2b} ou sur l'intrigant graphique~\ref{fig:8.plus}a, voici les deux pages que vous devez absolument lire (sous peine de courroux tr\`es grand, oui tr\`es grand, attention j'aurais pr\'evenu) (\`a la limite laissez tomber le reste... non mais quand m\^eme ce graphique~\ref{fig:8.plus}a...).\\

Je tiens \`a remercier en premier lieu ma directrice de th\`ese Genevi\`eve. De mon stage de Master \`a ma soutenance de th\`ese de doctorat, j'ai pu grandement b\'enificier de ses connaissances, de ses pr\'ecieux conseils, de sa patience et de sa disponibilit\'e dans les diff\'erents aspects qui constituent ces trois ann\'ees d'initiation \`a la recherche en physique des particules. Cette exp\'erience enrichissante me servira quelles que soient mes activit\'es \`a venir.

I am very thankful to Rohini Godbole for being the president of my thesis jury, to my referees Nazila Mahmoudi and Ulrich Ellwanger and to C\'eline B\oe{}hm, Anupam Mazumdar and Genevi\`eve for being in my jury. Thanks for your valuable comments, suggestions and your careful reading of my PhD thesis.\\

Je remercie bien entendu l'ensemble des membres du LAPTh, laboratoire dans lequel j'ai effectu\'e la plus grande partie de ma th\`ese. La chaleureuse ambiance qui r\`egne dans ce laboratoire et son emplacement en font un excellent cadre pour la recherche scientifique. Je remercie son directeur Fawzi pour sa disponibilit\'e et ses r\'eponses \`a mes questions, l'\'equipe administrative, compos\'ee durant ces trois ann\'ees de Dominique, Nathalie, V\'eronique et Virginie, pour son aide pr\'ecieuse dans le quotidien d'un chercheur, et Mathieu pour son expertise en mati\`ere d'informatique. Je remercie \'egalement le CNRS et l'Universit\'e de Grenoble pour leur soutien dans cette \'etape de la recherche.

Je souhaite particuli\`erement remercier C\'eline de m'avoir permis d'effectuer un s\'ejour de six mois \`a Durham où les projets se sont multipli\'es et conclus tr\`es positivement. Je remercie d'ailleurs la r\'egion Rh\^one-Alpes pour le financement de ce s\'ejour \`a travers sa bourse CMIRA 2011 EXPLO’RA DOC. I would like to warmly thank the Institute for Particle Physics Phenomenology of the University of Durham for having received me. Special thanks to Linda, Trudy and Mike for their help on logistics and computing.\\

I also would like to thank my other collaborators during these three years : Sasha Pukhov who was a very precious help for understanding the {\tt CalcHEP\;}and {\tt micrOMEGAs\;}codes and for coding/debugging related topics, Daniel
Albornoz V\'asquez, Peter Richardson, Chris Wymant (merci pour ton aide concernant \textsf{Herwig++} !), Marco Cirelli, Anupam Mazumdar and Ernestas Pukartas (thank you for your updated plots !). I thank Andrei Semenov for his help on using the {\tt LanHEP\;}code.\\

I thank the 2011 International School on Astro-Particle Physics, the 2012 MCnet-LPCC Summer School and the Carg\`ese International School 2012 for the great quality of lectures I attended. J'en profite également pour remercier les personnes qui, par leurs enseignements, ont \'et\'e importantes durant mes \'etudes \`a l'Universit\'e Montpellier 2 : J\'er\^ome L\'eon, David Polarski, Gilbert Moultaka et Cyril Hugonie.\\

Je tiens \`a remercier ma tutrice de monitorat Isabelle pour ses conseils tout au long de ces trois ann\'ees d'enseignement \`a l'IUT d'Annecy. Je remercie Françoise avec qui j'ai pu d\'ebuter et apprivoiser mon activit\'e de moniteur. Enseigner en collaboration avec d'autres doctorants contractuels enseignants a aussi \'et\'e une exp\'erience enrichissante. Je tiens donc \`a remercier Laurent B., Florian et Timoth\'ee.\\

Je f\'elicite Guillaume DLR d'avoir surv\'ecu \`a mes blagues pendant pr\`es d'un an et demi (mais comment a-t-il fait ??). J'\'etais un peu plus sage avec Laurent G. (l'\'ecriture de cette th\`ese a-t-elle eu une quelconque influence ??). Je remercie l'ensemble de cette promotion 2010, en incluant les lappins : Andrey, Chi Linh, Sadek, le duo Guilhem - Laurent B., grands organisateurs de soir\'ees qui marqueront l'histoire du LAPP et du LAPTh pour les decennies \`a venir, au moins (et comme par hasard j'y \'etais pas !) et Maud. Je remercie B\'eranger et J\'er\'emie pour les discussions passionantes durant diverses \'ecoles d'\'et\'e ou conf\'erences. I thank Ilan, Alix, Jonathan, Daniel and the other PhD Students I met at Durham for the warm atmosphere. Pour finir ce paragraphe (parce que, au lieu d'oublier une ou deux personnes, mieux vaut en oublier plein !), je remercie Nicolas, mon successeur Vincent (l'informatique, c'est son dada) et tous les autres pour le cadeau offert lors de ma soutenance.\\

Mes derniers remerciements vont \`a ma famille proche : mes parents \`a qui je dois tout, mon fr\`ere Nuno qui par ces nuits \'etoil\'ees au-dessus de \textit{Coutada} a mis sournoisement dans la t\^ete de son fr\`ere la passion pour l'activit\'e qu'il pratique d\'esormais, ma belle-s\oe{}ur Carla et leur petit Mateo dont les vid\'eos m'ont \'egay\'e de nombreuses soir\'ees. Voil\`a.\\

Ah oui j'allais oublier, je remercie aussi le Centre Hospitalier de la R\'egion d'Annecy pour ces installations m'ayant permis de d\'emarrer \`a toute vitesse l'\'ecriture de cette th\`ese (quatre jours productifs tout de m\^eme !). Par la m\^eme occasion je remercie mon pancr\'eas de me faire vivre depuis maintenant plus de six ans des exp\'eriences... originales. 

\adjustmtc

\newpage\null\newpage\null\newpage

\phantomsection
\addcontentsline{toc}{chapter}{\listfigurename}
\listoffigures
\adjustmtc
\newpage\null\newpage

\cleardoublepage
\phantomsection
\addcontentsline{toc}{chapter}{\listtablename}
\listoftables
\adjustmtc
\newpage\null\newpage

\chapter*{List of Abbreviations}
\addcontentsline{toc}{chapter}{List of Abbreviations}

\begin{spacing}{1.5}
\textbf{$\mathbf{\Lambda}$CDM} : Lambda Cold Dark Matter\\
\textbf{AMSB} : Anomaly-Mediated Supersymmetry Breaking\\
\textbf{BAO} : Baryon Acoustic Oscillations\\
\textbf{BBN} : Big Bang Nucleosynthesis\\
\textbf{BNL} : Brookhaven National Laboratory\\
\textbf{BRST} : Becchi-Rouet-Stora-Tyutin\\
\textbf{BSM} : Beyond the Standard Model\\
\textbf{CDM} : Cold Dark Matter\\
\textbf{CKM} : Cabibbo-Kobayashi-Maskawa\\ 
\textbf{CMB} : Cosmic Microwave Background\\
\textbf{CMSSM} : Constrained MSSM\\ 
\textbf{DD} : Direct Detection\\ 
\textbf{DE} : Dark Energy\\ 
\textbf{DM} : Dark Matter\\
\textbf{dSph} : dwarf Spheroidal\\
\textbf{EW} : Electro-Weak\\ 
\textbf{EWSB} : Electro-Weak Symmetry Breaking\\ 
\textbf{FCNC} : Flavour-Changing Neutral Current\\ 
\textbf{FL} : Friedmann-Lema\^{i}tre\\ 
\textbf{GIM} : Glashow-Iliopoulos-Maiani\\ 
\textbf{GMSB} : Gauge-Mediated Supersymmetry Breaking\\ 
\textbf{GR} : General Relativity\\ 
\textbf{GUT} : Grand Unified Theories\\ 
\textbf{HDM} : Hot Dark Matter\\ 
\textbf{ID} : Indirect Detection\\ 
\textbf{KK} : Kaluza-Klein\\ 
\textbf{LBL} : Lawrence Berkeley Laboratory\\
\textbf{LH} : Left-Handed\\ 
\textbf{LHC} : Large Hadron Collider\\
\textbf{LO} : Leading Order\\ 
\textbf{LSP} : Lightest Supersymmtric Particle\\ 
\textbf{LSS} : Large Scale Structures\\ 
\textbf{MACHO} : MAssive Compact Halo Object\\ 
\textbf{MCMC} : Markov Chain Monte Carlo\\ 
\textbf{MFV} : Minimal Flavour Violation\\ 
\textbf{MOND} : MOdified Newtonian Dynamics\\
\textbf{MSSM} : Minimal Supersymmetric Standard Model\\ 
\textbf{mSUGRA} : minimal SUperGRAvity\\ 
\textbf{MW} : Milky Way\\ 
\textbf{NLO} : Next-to-LO\\ 
\textbf{NLSP} : Next-to-LSP\\ 
\textbf{NMSSM} : Next-to-MSSM\\
\textbf{NP} : New Physics\\ 
\textbf{NUHM} : Non-Universal Higgs Masses\\
\textbf{PMNS} : Pontecorvo-Maki-Nakagawa-Sakata\\ 
\textbf{pMSSM} : phenomenological MSSM\\ 
\textbf{PQ} : Peccei-Quinn\\ 
\textbf{QCD} : Quantum ChromoDynamics\\
\textbf{QED} : Quantum ElectroDynamics\\ 
\textbf{RGE} : Renormalization Group Equation\\ 
\textbf{RH} : Right-Handed\\ 
\textbf{RHSN} : RH sneutrino\\ 
\textbf{RW} : Robertson-Walker\\ 
\textbf{SD} : Spin-Dependent\\
\textbf{SDSS} : Sloan Digital Sky Survey\\
\textbf{SI} : Spin-Independent\\
\textbf{SLAC} : Stanford Linear Accelerator Center\\
\textbf{SM} : Standard Model\\
\textbf{SNIa} : Type Ia supernov\ae\\ 
\textbf{SR} : Special Relativity\\ 
\textbf{SSB} : Soft SUSY Breaking\\ 
\textbf{SUSY} : Supersymmetry\\
\textbf{TeVeS} : Tensor-Vector-Scalar gravity\\
\textbf{UED} : Universal Extra Dimensions\\
\textbf{VEV} : Vacuum Expectation Value\\
\textbf{WDM} : Warm Dark Matter\\ 
\textbf{WIMP} : Weakly Interacting Massive Particle\\
\end{spacing}

%------------------------------------------------------------------------
% new definitions, abreviations, etc
%------------------------------------------------------------------------

%codes
\def\NTools{{\tt NMSSMTools\;}}
\def\HB{{\tt HiggsBounds-3.8.0\;}}
\def\micro{{\tt micrOMEGAs\;}}
\def\chep{{\tt CalcHEP\;}}
\def\lhep{{\tt LanHEP\;}}
\def\DSUSY{{\tt DarkSUSY\;}}
\def\SisoR{{\tt SuperIso Relic\;}}
\def\Spect{{\tt SuSpect\;}}
\def\ISA{{\tt ISAJET\;}}
\def\SPno{{\tt SPheno\;}}
\def\SoSUSY{{\tt SOFTSUSY\;}}

%others
\def\ra{\rightarrow}
\def\RA{\Rightarrow}
\def\lsp{\tilde{\nu}_R}
\def\mlsp{m_{\tilde{\nu}_R}}
\def\snl{\tilde{\nu}_L}
\def\mneut{m_{\tilde{\chi}^0_1}}
\def\mchi{m_{\tilde{\chi}^0_i}}
\def\mneutt{m_{\tilde{\chi}^0_2}}
\def\mneuth{m_{\tilde{\chi}^0_3}}
\def\mneutf{m_{\tilde{\chi}^0_4}}
\def\mchar{m_{\tilde{\chi}^+_1}}
\def\mchart{m_{\tilde{\chi}^+_2}}
\def\msel{m_{\tilde{e}_L}}
\def\mser{m_{\tilde{e}_R}}
\def\mslo{m_{\tilde{\tau}_1}}
\def\mslt{m_{\tilde{\tau}_2}}
\def\msul{m_{\tilde{u}_L}}
\def\msur{m_{\tilde{u}_R}}
\def\msdl{m_{\tilde{d}_L}}
\def\msdr{m_{\tilde{d}_R}}
\def\msto{m_{\tilde{t}_1}}
\def\mstt{m_{\tilde{t}_2}}
\def\msbo{m_{\tilde{b}_1}}
\def\msbt{m_{\tilde{b}_2}}
\def\sw{s_W}
\def\cw{c_W}
\def\tb{\tan\beta}
\def\bsg{\mathscr{B}(\bar{B}^0 \to X_s\gamma)}
\def\bXsl{\mathscr{B}(\bar{B}^0 \to X_s\ell^+ \ell^-)}
\def\bXsmu{\mathscr{B}(\bar{B}^0 \to X_s\mu^+ \mu^-)}
\def\bsmu{\mathscr{B}(B^0_s \to \mu^+ \mu^-)}
\def\btau{\mathscr{B}(B^\pm \to \tau^\pm \nu_\tau)}
\def\Omg{\Omega h^2}
\def\sip{\sigma^{SI}_{\chi p}}
\def\amu{\delta a_\mu}
\def\neut1{\chi^0_1}
\def\neuti{\chi^0_i}
\def\neutj{\chi^0_j}
\def\neut2{\chi^0_2}
\def\neut3{\chi^0_3}
\def\neut4{\chi^0_4}
\def\chargi{\chi^\pm_i}
\def\charg1{\chi^\pm_1}
\def\charg2{\chi^\pm_2}
\def\gluino{\tilde{g}}
\def\ul{\tilde{u}_L}
\def\ur{\tilde{u}_R}
\def\stau{\tilde{\tau}}
\def\sl{\tilde{l}}
\def\sq{\tilde{q}}
\def\sneutrino{\tilde\nu}
\def\msnu{m_{\tilde\nu_R}}
\def\anu{A_{\tilde\nu}}
\def\mzp{M_{Z_2}}
\def\azz{\alpha_{Z}}
\def\te6{\theta_{E_6}}
\def\gp2{g'^{\,2}_1}
\def\beq{\begin{equation}}
\def\eeq{\end{equation}}
\def\wino{\tilde{W}}
\def\bino{\tilde{B}}
\def\binop{\tilde{B'}}
\def\cw{c_W}
\def\sw{s_W}
\def\cb{c_\b}
\def\sb{s_\b}
\def\tb{t_\b}
\def\Vub{\left|V_{ub}\right|}
\def\Vtbts{\left|V_{ts}^*V_{tb}\right|}
\def\Vtbtd{\left|V_{td}^*V_{tb}\right|}
\def\VtsbVcb{\left|\frac{V_{ts}^*V_{tb}}{V_{cb}}\right|}
\def\a{\alpha}
\def\b{\beta}
\def\g{\gamma}
\def\G{\Gamma}
\def\d{\delta}
\def\e{\epsilon}
\def\l{\lambda}
\def\s{\sigma}
\def\t{\theta}
\def\Om{\Omega}
\def\p{\partial}
\def\Qa{Q_{\a}}
\def\Qda{\bar{Q}^{\dot{\a}}}
\def\Qb{Q_{\b}}
\def\Qdb{\bar{Q}^{\dot{\b}}}
\def\ie{\;\textit{i.e.}\;}
\def\ca{\;\textit{ca.}\;}
\def\vs{\;\textit{vs.}\;}
\def\eg{\;\textit{e.g.}\;}

%labels in figures
\newcommand{\ablabels}[3]{
  \begin{picture}(100,0)\setlength{\unitlength}{1mm}
    \put(#1,#3){\bf (a)}
    \put(#2,#3){\bf (b)}
  \end{picture}\\[-8mm]
} 

\def\Fermi{{\sc Fermi}} %{{\sc Fermi-LAT}}
\def\PAMELA{{\sc Pamela}}

%comments
\newcommand{\com}[1]{{\bf\color{red} #1}\marginpar{$\bigstar$ $\bigstar$ $\bigstar$}}
\newcommand{\comm}[1]{{\bf\color{red} #1}\marginnote{$\bigstar$ $\bigstar$ $\bigstar$}}

\adjustmtc

\newpage\null\newpage

\chapter*{List of Publications}
\addcontentsline{toc}{chapter}{List of Publications}

Here are listed the papers written in the context of this PhD thesis.

\subsubsection*{Published articles}

\begin{itemize}
\item G. Bélanger, J. Da Silva and A. Pukhov. \textit{The Right-handed sneutrino as thermal dark matter in U(1) extensions of the MSSM}. \href{http://iopscience.iop.org/1475-7516/2011/12/014}{{\texttt JCAP \textbf{1112} (2011) 014.}} \href{http://arxiv.org/abs/1110.2414}{{[\texttt arXiv:1110.2414].}}
\item D. A. Vasquez, G. Bélanger, C. B\oe{}hm, J. Da Silva, P. Richardson and C. Wymant. \textit{The 125 GeV Higgs in the NMSSM in light of LHC results and astrophysics constraints}. \href{http://prd.aps.org/abstract/PRD/v86/i3/e035023}{{\texttt Phys. Rev. \textbf{D86} (2012) 035023.}} \href{http://arxiv.org/abs/1203.3446}{{[\texttt arXiv:1203.3446].}}
\item G. Bélanger, C. B\oe{}hm, M. Cirelli, J. Da Silva and A. Pukhov. \textit{PAMELA and FERMI-LAT limits on the neutralino-chargino mass degeneracy}. \href{http://iopscience.iop.org/1475-7516/2012/11/028}{{\texttt JCAP \textbf{1211} (2012) 028.}} \href{http://arxiv.org/abs/1208.5009}{{[\texttt arXiv:1208.5009].}}
\item C. B\oe{}hm, J. Da Silva, A. Mazumdar and E. Pukartas. \textit{Probing the Supersymmetric Inflaton and Dark Matter link via the CMB, LHC and XENON1T experiments}. \href{http://prd.aps.org/abstract/PRD/v87/i2/e023529}{{\texttt Phys. Rev. \textbf{D87} (2013) 023529.}} \href{http://arxiv.org/abs/1205.2815}{{[\texttt arXiv:1205.2815].}}
\end{itemize}

\subsubsection*{Workshop contribution}

\begin{itemize}
\item G. Brooijmans, B. Gripaios, F. Moortgat, J. Santiago, P. Skands, \textit{et al. Les Houches 2011:
Physics at TeV Colliders New Physics Working Group Report}. \href{http://arxiv.org/abs/1203.1488}{{[\texttt arXiv:1203.1488].}}
\end{itemize}

\adjustmtc

\newpage\null\newpage

\mainmatter
\setcounter{page}{1}
\pagestyle{fancy}
\fancyhf{}
\renewcommand{\chaptermark}[1]{\markboth{\thechapter{} - #1}{}}
\renewcommand{\sectionmark}[1]{\markright{\thesection{} - #1}}
\lhead[\thepage]{\textsl{\rightmark}}
\rhead[\textsl{\leftmark}]{\thepage}
\headsep = 25pt

\chapter*{Introduction}
\addcontentsline{toc}{chapter}{Introduction}
\markboth{Introduction}{}

\section*{English version}

With the discovery of a particle whose characteristics match well those of the Standard Model Higgs boson and with the precise results of cosmological experiments, the standard models of particle physics and cosmology are now clearly successful. Nevertheless some observations are not completely understood in these theoretical fameworks. One of the most well-known is the Dark Matter problem. The fact that the particle content of the Standard Model of particle physics only explain a fraction of the matter content of the standard cosmological model is an important issue to solve. There are also some theoretical issues in the foundations of these models. The fine-tuning problem in the Higgs sector of the Standard Model of particle physics that can be related to the fact that no symmetry protects the mass of the Higgs boson in this model gives also motivations to look for New Physics. A lot of new models were built to respond to the issues of these two standard models and one of the most well-known which is considered in this thesis is Supersymmetry. Defining such a symmetry between two type of particles, the fermions and the bosons, gives a response to the fine-tuning problem and introduce new particles within which we can find a Dark Matter candidate. A nice feature of Supersymmetry is that several methods can be used to test its predictions : they span a large array of experiments and energies. Several of them will be considered to constrain the supersymmetric models analysed in this thesis.\\

This thesis has the following structure :\\

\begin{itemize}
\item In part \ref{part1} we will present a review of the content, the successes and the issues of the Standard Model of particle physics (chapter \ref{chap:revpart}) and the cosmological standard model (chapter \ref{chap:revcosmo}). We will then introduce Supersymmetry, its simplest version and the experimental constraints that will be considered in this thesis (chapter \ref{chapter:SUSY}).
\item In part \ref{part2} we will present studies of supersymmetric models considering the lightest neutralino as a Dark Matter candidate. In chapter \ref{chapter:NUHM2} we will analyse the Minimal Supersymmetric Standard Model within which cosmic inflation candidates can be found. In chapter \ref{chapter:ID} constraints from Indirect Detection of Dark Matter will be used on this model. To close this part we will analyse in chapter \ref{chapter:NMSSM} collider constraints on coloured supersymmetric particles within a singlet extension of the Minimal Supersymmetric Standard Model.
\item In part \ref{part3} the study of $U(1)$ extensions of the Minimal Supersymmetric Standard Model will be developed. In chapter \ref{chapter:UMSSM} we will introduce such extensions and their characteristics. Chapter \ref{chapter:RHsneu} will be devoted to the study of a scalar Dark Matter candidate in this model, the Right-Handed sneutrino. Finally more constraints will be considered especially coming from low energy observables and we will take into account recent results on the Higgs boson study (chapter \ref{chapter:B_Higgs_UMSSM}).
\end{itemize}

\section*{Version fran\c{c}aise}

Avec la d\'ecouverte d'une particule dont les caract\'eristiques correspondent de mieux en mieux \`a celles du boson de Higgs du Mod\`ele Standard et avec les r\'esultats d'une pr\'ecision in\'egal\'ee des exp\'eriences en cosmologie, les mod\`eles standards de la physique des particules et de la cosmologie ont d\'esormais clairement un grand succ\`es. Cependant certaines observations ne sont pas clairement expliqu\'ees dans ces cadres th\'eoriques. L'un des plus c\'el\`ebres est le probl\`eme de la Mati\`ere Noire. Le fait que les constituants du Mod\`ele Standard de la physique des particules expliquent seulement une fraction du contenu en mati\`ere de mod\`ele standard cosmologique est un point important \`a r\'esoudre. Il y a en outre des soucis th\'eoriques dans la construction de ces mod\`eles. Le probl\`eme de r\'eglage fin dans le secteur de Higgs du Mod\`ele Standard de la physique des particules qui peut \^{e}tre reli\'e au fait qu'aucune sym\'etrie ne prot\`ege la masse du boson de Higgs dans ce mod\`ele donne aussi des motivations dans la recherche de Nouvelles Physiques. Un grand nombre de mod\`eles ont \'et\'e d\'evelopp\'es afin de r\'epondre aux probl\`emes de ces deux mod\`eles standards et l'un des plus connus qui est \'etudi\'e tout au long de cette th\`ese est la Supersym\'etrie. D\'efinir une telle sym\'etrie entre deux sortes de particules, les fermions et les bosons, donne une solution au probl\`eme de r\'eglage fin et introduit de nouvelles particules au sein desquelles nous pouvons trouver un candidat \`a la Mati\`ere Noire. Une caract\'eristique int\'eressante de la Supersym\'etrie est que de nombreuses m\'ethodes peuvent \^{e}tre employ\'ees afin de tester ses pr\'edictions : ces outils couvrent une large gamme d'exp\'eriences et d'\'energies. Plusieurs d'entre eux seront utilis\'es pour contraindre les mod\`eles supersym\'etriques \'etudi\'es dans cette th\`ese.\\

Cette th\`ese poss\`ede la structure suivante :\\

\begin{itemize}
\item Dans la partie \ref{part1} nous pr\'esenterons le contenu, les succ\`es et les probl\`emes du Mod\`ele Standard de la physique des particules (chapitre \ref{chap:revpart}) et du mod\`ele standard cosmologique (chapitre \ref{chap:revcosmo}). Nous introduirons alors la Supersym\'etrie, sa version la plus simple et les contraintes exp\'erimentales qui seront consid\'er\'ees dans cette th\`ese (chapitre \ref{chapter:SUSY}).
\item Dans la partie \ref{part2} nous pr\'esenterons diff\'erentes \'etudes de mod\`eles supersym\'etriques en consid\'erant le neutralino le plus l\'eger comme candidat \`a la Mati\`ere Noire. Dans le chapitre \ref{chapter:NUHM2} nous analyserons le Mod\`ele Standard Supersym\'etrique Minimal au sein duquel des candidats \`a l'inflation cosmique peuvent \^{e}tre \'etudi\'es. Dans le chapitre \ref{chapter:ID} des contraintes provenant de la D\'etection Indirecte de Mati\`ere Noire seront utilis\'ees dans ce mod\`ele. Pour clore cette partie nous analyserons dans le chapitre \ref{chapter:NMSSM} des contraintes en collisionneur sur des particules supersym\'etriques color\'ees dans le cadre d'une extension singulet du Mod\`ele Standard Supersym\'etrique Minimal.
\item Dans la partie \ref{part3} l'\'etude d'extensions $U(1)$ du Mod\`ele Standard Supersym\'etrique Minimal sera d\'evelopp\'ee. Nous introduirons dans le chapitre \ref{chapter:UMSSM} ce genre d'extensions et leurs caract\'eristiques. Le chapitre \ref{chapter:RHsneu} sera consacr\'e \`a l'\'etude d'un candidat scalaire \`a la Mati\`ere Noire dans ce mod\`ele, le sneutrino droit. Pour finir davantage de contraintes seront consid\'er\'ees, tout particuli\`erement ceux provenant d'observables de basses \'energies et nous prendrons en compte les r\'ecents r\'esultats dans l'\'etude du boson de Higgs (chapitre \ref{chapter:B_Higgs_UMSSM}).
\end{itemize}

\adjustmtc

\part{Status of particle physics and cosmology ... and beyond}
\label{part1}

\chapter{From the infinitely small : the Standard Model of particle physics ...}
\label{chap:revpart}

\minitoc\vspace{1cm}

\newpage

\section{Building of the model : gauge sector}
\label{sec:1.gauge}

From the discovery of the nuclear structure of the atoms by Rutherford in the early 20th century \cite{Rutherford:1911zz} to the final hunting of the Higgs boson in 2012 at the Large Hadron Collider (LHC) with the ATLAS and the CMS collaborations \cite{Chatrchyan:2012ufa,Aad:2012tfa}, progress in particle physics were so strong that just a few theoretical models are still relevant to explain the behaviour of elementary particles. In this context the Standard Model of particle physics (SM) is the current most reliable description of the main fundamental interactions between the elementary particles\footnote{Gravitation can be safely neglected at this scale and even at the largest energies obtained both at colliders and in cosmic rays although there is no successful description of quantum gravity.} : the electromagnetic, the weak and the strong nuclear interactions. Symmetry considerations were essential to build this theory. Starting from the interactions between the known matter particles, namely quarks and leptons, the gauge theory developed by Yang and Mills \cite{Yang:1954ek} tells us that these particles interact through the exchange of spin 1 particles, the vector gauge bosons. The full Lagrangian of the theory, which contains the masses and interactions between the particles, is invariant under a non-abelian gauge symmetry that reads $SU(3)_c  \otimes  SU(2)_L  \otimes  U(1)_Y$; it introduces 12 gauge bosons. The strong interaction is described by the \textit{Quantum ChromoDynamics} (QCD) theory whose gauge symmetry is the exact $SU(3)_c$ where the subscript \textit{c} denote the colour charge. This interaction is mediated by eight massless vector gauge bosons, the gluons, whose first strong evidence of their existence was obtained by the observation of the \textit{three-jets} events at the PETRA accelerator \cite{Brandelik:1979bd}. The remaining part $SU(2)_L  \otimes  U(1)_Y$ corresponds to the unification of the \textit{ElectroMagnetic} (EM) interaction, based on the \textit{Quantum ElectroDynamics} (QED) theory, and the weak interaction into the same framework introduced by Glashow, the \textit{Electro-Weak} (EW) model \cite{Glashow:1961tr}. This gauge symmetry predicts, in addition to the massless EM propagator called photon, new massive vector gauge bosons, the $W^\pm$ and $Z$ bosons discovered by the UA1 collaboration at the CERN SPS collider \cite{Arnison:1983rp,Arnison:1983mk}. 

The self-interaction of gauge bosons and the kinetic terms are encoded in the Yang-Mills expression :
\beq \mathscr{L}_{YM} = -\frac{1}{4} \sum_{i=1}^3 F^i_{\mu\nu}F^{i \mu\nu}, \eeq
where
\beq \begin{split}
F^i_{\mu\nu} & = \p_\mu F^i_\nu - \p_\nu F^i_\mu + g_i c_{ijk} F^j_\mu F^k_\nu.
\end{split} \eeq
$\mu$ and $\nu$ are Lorentz indices, the metric being that of \textit{Special Relativity} (SR), the Minkowski one $\eta_{\mu\nu} = \mathrm{diag}(1,-1,-1,-1)$. The general covariant derivative for the $SU(3)_c  \otimes  SU(2)_L  \otimes  U(1)_Y$ symmetry reads
\beq \label{eq:1.cov} D_\mu = \p_\mu +i \sum_{i=1}^3 k_i g_i F^i_\mu T_i. \eeq
We use the following definitions for each symmetry :
\begin{itemize}
\item $i$ = 1 stands for the $U(1)_Y$ symmetry whose coupling constant is $g_1 = g_Y$. There is one gauge field associated $F^1_\mu = B_\mu$ and one generator $T_1 = Y/2$ with the hypercharge $Y$. The $U(1)_Y$ structure constant $c_{ijk}$ is equal to zero. Thus there is no self-interaction of the field $B_\mu$ : this corresponds to an abelian gauge group. $k_1 = 1$ which means that all matter particles of the SM couple through $U(1)_Y$.
\item $i$ = 2 stands for the $SU(2)_L$ symmetry whose coupling constant is $g_2$. There are three gauge fields $F^2_\mu = \{ W^a_\mu, a =1,2,3 \}$ and three associated generators, $T_2~\in~\{I_a = \s^a/2, a =1,2,3\}$ with $\s^a$ the Pauli matrices and $I_a$ the isospin. The $SU(2)_L$ structure constants $c_{ijk} = \e_{ijk}$, with $i,j,k \in \{1,2,3\}$, are the antisymmetric Levy-Civita symbols in three dimensions. Finally $k_2 = 0$ and $1$ respectively for the covariant derivative acting on a singlet and a doublet of $SU(2)_L$.
\item $i$ = 3 stands for the $SU(3)_c$ symmetry whose coupling constant is $g_3 = g_s$. There are eight gauge fields $F^3_\mu = \{ G^a_\mu, a =1,...,8 \}$ and eight associated generators $T_3 \in \{\l^a/2, a =1,...,8 \}$. The $SU(3)_c$ structure constants $c_{ijk} = f_{ijk}$, with $i,j,k \in~\{1,...,8\}$, are antisymmetric. To finish we have $k_3 = 0, 1$ and $-1$ respectively for a singlet, a triplet and an antitriplet of $SU(3)_c$. 
\end{itemize}
In the framework of EW unification, the electrical charge $\mathcal{Q}$ of a particle, characterized by its hypercharge $Y$ and $I_3$ which is the third component of the $SU(2)_L$ isospin, is given by the Gell-Mann Nishijima formula
\beq \label{eq:Gm-N} \mathcal{Q} = I_3 + \frac{Y}{2}. \eeq

 The main issue of this model is to explain how some of these gauge bosons become massive and how to avoid the explicit breaking of the EW symmetry. One simple way to give mass to the particles of the SM is to break the $SU(2)_L  \otimes  U(1)_Y$ symmetry introducing a doublet of scalar fields that has non-zero vacua as we will show in section~\ref{sec:1.higgs}. When we decide to choose one vaccum the symmetry is spontaneously broken to the QED symmetry $U(1)_{em}$ and the particles of the SM get a mass in an elegant mathematical way. This mechanism, developed in the 60's \cite{Englert:1964et,Higgs:1964ia,Higgs:1964pj,Guralnik:1964eu,Higgs:1966ev,Kibble:1967sv}, is the \textit{Brout-Englert-Higgs} mechanism and it gives a physical state, the massive scalar \textit{Brout-Englert-Higgs}\footnote{In the rest of this thesis we will simplify this naming by using the much more used designations \textit{Higgs mechanism} and \textit{Higgs boson}.} boson which seems to correspond to the boson discovered by the LHC experiments in 2012. The Higgs mechanism was successfully incorporated into the EW theory by Weinberg and Salam \cite{Weinberg:1967tq,Salam:1968rm} and the predictiveness of the theory was finally established by ’t Hooft who demonstrated that the SM theory of the EW interaction is renormalizable \cite{'tHooft:1971rn}. For a more complete overview of the SM theory see \cite{Peskin:1995ev,Halzen-Martin,Weinberg:1996kr}. In the following section we look at the matter sector of the SM.

\section{Matter sector}
\label{sec:1.matter}

The matter sector of the SM consists of 24 spin 1/2 particles called fermions and their corresponding antiparticles\footnote{Same mass and spin but the other quantum numbers are opposite such as the electric and colour charges.}. This sector is divided into two parts, depending on the sensitivity of this particles to the strong interaction : the leptons and the quarks. These fermions are described by the fundamental representation of the group of space-time transformations, the Lorentz group, and the object we use to define these fields are the four complex components Dirac spinor. This framework allows to put in one mathematical object the fermion, the corresponding antifermion and their two helicities\footnote{Projection of the spin onto the direction of momentum of the particle considered.}. One useful decomposition of the fermions is linked to the chiral symmetry. A Dirac spinor $\psi$ can be rewritten as a combination of its left and right chiralities $\psi_L$ and $\psi_R$ :
\beq \psi = P_L \psi_L + P_R \psi_R, \eeq
where the projectors are defined using the Dirac matrix $\g_5$
\beq P_L = \frac{1-\g_5}{2}, \ P_R = \frac{1+\g_5}{2}.\eeq
Using this decomposition we can define the fermion masses analysing the mixing of chiralities as we will see in section~\ref{sec:1.higgs}. The weak interaction only acts on the left part $\psi_L$ of the Dirac spinor and on the right one of the conjugate spinor : it breaks parity symmetry\footnote{Transformation between left and right chirality. The breaking of P is the reason why the symmetry group of the weak interaction is called $SU(2)_L$.} P and also both charge conjugation\footnote{Transformation between a particle and its antiparticle.} C and P symmetry.
When we classify the fermions with respect to the weak interaction, we then define doublets of chirality left and singlet of chirality right as detailed in sections \ref{subsec:1.lepton} and \ref{subsec:1.quark}.

\subsection{Leptons}
\label{subsec:1.lepton}

 Leptons are singlets of $SU(3)_c$; they do not interact through strong interaction. There are six leptons divided in three families or flavours in the SM. In this sector a family corresponds to a doublet of a charged (with an opposite electric charge with respect to that of the proton) and a neutral lepton with respect to the EM interaction. The most known charged lepton is the electron $e^-$ which is the lightest (its mass\footnote{We use the useful convention $\hbar = c = 1$ which allows us to write most of our high energy physics observables in units of eV$^n$ with n $\in \mathbb{Z}$. 1 eV is the energy gained (or lost) by an electron which is moving on a lengh of 1 meter across an electric potential difference of 1 Volt; a typical ballpark is the mass of a nucleon $\sim$ 1 GeV = 10$^{9}$ eV.} $m_e \sim$ 511 keV) and is stable, namely its lifetime is bigger than the age of the Universe. The others charged leptons are the muon $\mu^-$ dicovered in 1936 by Anderson and Neddermeyer \cite{Anderson:1936zz} and the tau $\tau^-$ detected in 1975 by Perl and collaborators from the Stanford Linear Accelerator Center (SLAC) and the Lawrence Berkeley Laboratory (LBL) \cite{Perl:1975bf}. The neutral states called neutrinos were found interacting only by weak interaction with the other particles \cite{Cowan:1992xc,Danby:1962nd,Kodama:2000mp}. It results that the three neutrino flavours $\nu_e$, $\nu_\mu$ and $\nu_\tau$ are mostly Left-Handed (LH) and almost massless. We insist on the \textit{almost} statement since neutrino oscillations were observed in many different ways (flux of solar neutrino, atmospheric neutrino, neutrinos produced in particle accelerators or nuclear reactors). This shows that these particles have non-zero masses and gives hints on the existence of a Right-Handed (RH) component (see part \ref{part3} for examples of such extensions in the neutrino sector). Neutrino oscillation describes the possibility to detect a neutrino with a flavour $i \in\{e,\mu,\tau\}$ knowing that it was created with a flavour $j \neq i$ using the unitary transformation relating the flavor $\nu_i$ and mass $\nu_\a$ eigenstates
\beq \nu_i = \sum_{\a=1,2,3} U_{i \a} \nu_\a, \ i \in\{e,\mu,\tau\} \eeq
through the Pontecorvo-Maki-Nakagawa-Sakata (PMNS) unitary matrix \cite{Pontecorvo:1957qd,Pontecorvo:1967fh,Maki:1962mu}
\beq U = 
      \begin{pmatrix}
			c_{12} c_{13} & s_{12} c_{13} & s_{13} e^{-i\d} \\
		  - s_{12} c_{23} - c_{12} s_{23} s_{13} e^{i \d} & c_{12} c_{23} - s_{12} s_{23} s_{13} e^{i \d} & s_{23} c_{13} \\
			s_{12} s_{23} - c_{12} c_{23} s_{13} e^{i \d} & - c_{12} s_{23} - s_{12} c_{23} s_{13} e^{i \d} & c_{23} c_{13}
	  \end{pmatrix}
	  \begin{pmatrix}
			1 & 0 & 0 \\
			0 & e^{i\a_{21} / 2} & 0 \\
			0 & 0 & e^{i\a_{31} / 2}
	  \end{pmatrix}, 
\eeq
where $c_{ij} = \cos \t_{ij}$, $s_{ij} = \sin \t_{ij}$ ($\t_{ij}$ being mixing angles), $\d$ is the Dirac CP violation phase and $\a_{21}$ and $\a_{31}$ are two Majorana CP violation phases. These two last parameters are only relevant if neutrinos are Majorana fermions\footnote{The fermion is identical to its antifermion.} which would be the case if neutrinoless double beta decay (two neutrons decaying into two protons and two electrons) is observed. $\d$ is currently unknown but the angles are now well determined \cite{Beringer:1900zz} :
\beq \begin{split}
\sin^2 \t_{12} & = 0.306^{\,+0.018}_{\,-0.015} \\
\sin^2 \t_{23} & = 0.42^{\,+0.08}_{\,-0.03} \\
\sin^2 2\t_{13} & = 0.096 \pm 0.013.
\end{split} \eeq
The recent observation of non-zero $\t_{13}$ angle by the Daya Bay reactor neutrino experiment \cite{An:2012eh} turned to be a powerful constraint on neutrino models. If the masses of the neutrinos are sufficiently large, one of the interesting way to determine them is through the study of the Cosmic Microwave Background (CMB) (see chapter \ref{chap:revcosmo}). Depending on the cosmological scenario considered and on the combination of observations used, the upper limit on the combined mass of the three neutrinos was determined by the Planck collaboration to be down to 0.23 eV at the $2\s$ level \cite{Ade:2013lta}.

The leptons and antileptons can now be represented in terms of doublets and singlets of LH and RH fields :
\beq L = \left\{ \binom{\nu_e}{e}_L , \binom{\nu_\mu}{\mu}_L , \binom{\nu_\tau}{\tau}_L \right\}, \
e = \left\{ e_R, \mu_R, \tau_R \right\}. \eeq

\subsection{Quarks}
\label{subsec:1.quark}

The quarks are triplets of $SU(3)_c$ and interact through the three interactions of the SM. As for leptons we have six quarks but each of them exists in three colours, the quantum number linked to the $SU(3)_c$ symmetry. The SM contains three families of quark doublets. In each doublet we have an up-type quark with a fractionary electric charge of 2/3 in unit of the proton charge and a down-type\footnote{These names stem from those of the first quark family.} quark with a charge -1/3. The quarks are always\footnote{With the exception of the top quark which is not able to hadronise because its lifetime is too short; however its decay product hadronises.} detected in bound states because of the confinement property of the strong interaction. The bound states they form are the hadrons. These composite particles are colourless and have an integer electric charge. They are classified into two categories : the baryons like the nucleons, composed of three quarks, and the mesons composed of a quark and an antiquark.
 
The original quark model composed of the three lightest ones has been proposed independently by Gell-Mann \cite{GellMann:1964nj} and Zweig \cite{Zweig:1981pd,Zweig:1964zz} in 1964. The existence of these quarks, the up, down and strange quarks ($u$, $d$ and $s$) was proved in 1968 when evidence of the proton substructures was obtained using deep inelastic scattering at SLAC \cite{Bloom:1969kc,Breidenbach:1969kd}. To evade the issue of the theoretically possible existence of Flavour-Changing Neutral Current (FCNC) processes, still unobserved, Glashow, Iliopoulos and Maiani introduced in 1970 a fourth quark, the charm quark \cite{Glashow:1970gm}. The mechanism they developed that allows a flavour mixing through charged current is called the Glashow-Iliopoulos-Maiani (GIM) mechanism. The first hadron containing charm quarks, the $J/\psi$ meson, was discovered in 1974 by a SLAC team \cite{Augustin:1974xw} and a team from Brookhaven National Laboratory (BNL) \cite{Aubert:1974js}. This model was parameterized by only one parameter, the Cabibbo angle $\t_c$ \cite{Cabibbo:1963yz}. The discovery of CP violation due to weak interaction in the quark sector \cite{Christenson:1964fg} led to the need to define a quark sector made of three families that mix through charged current since this framework naturally introduce a CP violation phase. This work was done by Kobayashi and Maskawa in 1973 \cite{Kobayashi:1973fv} and the footprint of the bottom quark, the heaviest down-type quark, was obtained four years later by the E288 experiment at Fermilab \cite{Herb:1977ek}. The top quark was much more difficult to detect because of its large mass and its disability to hadronise. It was finally discovered also at Fermilab by the CDF and D\O \ collaborations in 1995 \cite{Abe:1995hr,Abachi:1994td}.

The quark flavours can mix to form mass eigenstates (denoted by a prime ') through charged current. The flavour mixing in the quark sector is parameterized by three angles and one CP phase as in the Dirac neutrino sector; we use a 3 $\times$ 3 unitary matrix to represent this mixing, the Cabibbo-Kobayashi-Maskawa matrix (CKM) :
\beq  
	  \begin{pmatrix}
			d' \\
			s' \\
			b'
	  \end{pmatrix} = V_\mathrm{CKM}
	  \begin{pmatrix}
			d \\
			s \\
			b
	  \end{pmatrix}, \ V_\mathrm{CKM} = 
	  \begin{pmatrix}
			V_{ud} & V_{us} & V_{ub} \\
			V_{cd} & V_{cs} & V_{cb} \\
			V_{td} & V_{ts} & V_{tb} 
	  \end{pmatrix}. 
\eeq
Using global fits, the CKM elements are now precisely determined in the framework of the SM and their magnitudes are \cite{Beringer:1900zz} :
\beq  
	  V_\mathrm{CKM} = 
	  \begin{pmatrix}
			0.97427 \pm 0.00015 & 0.22534 \pm 0.00065 & 0.00351^{\,+0.00015}_{\,-0.00014} \\
			0.22520 \pm 0.00065 & 0.97344 \pm 0.00016 & 0.0412^{\,+0.0011}_{\,-0.0005} \\
			0.00867^{\,+0.00029}_{\,-0.00031} & 0.0404^{\,+0.0011}_{\,-0.0005} & 0.999146^{\,+0.000021}_{\,-0.000046}
	  \end{pmatrix}. 
\eeq
The experimental observations based on charged current and CP violation like for example in the $B$-mesons sector can be crucial to determine the viability of theories aiming to go Beyond the Standard Model (BSM) as we will observe in parts \ref{part2} and \ref{part3}. 
As for leptons, we classify the quarks and antiquarks in doublets and singlets of $SU(2)_L$ :
\beq Q = \left\{ \binom{u}{d}_L , \binom{c}{s}_L , \binom{t}{b}_L \right\}, \
u = \left\{ u_R, c_R, t_R \right\}, \ 
d = \left\{ d_R, s_R, b_R \right\}. \eeq

\section{The Higgs mechanism}
\label{sec:1.higgs}

\begin{figure}[!htb]
\begin{center}
\includegraphics[width=9cm,height=5cm]{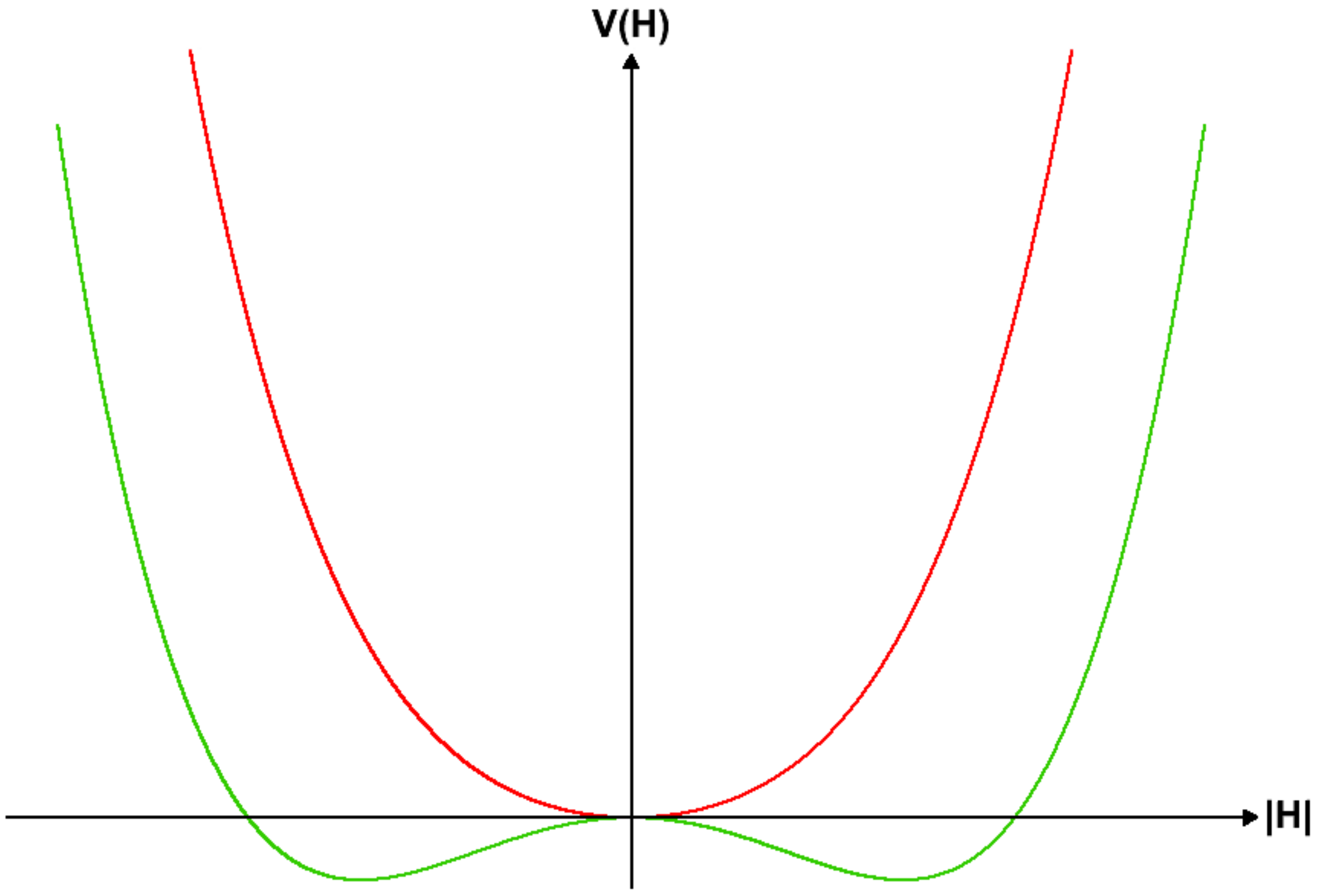} 
  \caption{Higgs potential for two cases with the same $|\mu^2|$ and $\l$ but with $\mu^2 > 0$ for the one represented by the red curve and $\mu^2 < 0$ for the other depicted by the green curve, \textit{mexican hat} shaped.}
\label{fig:higpot}
\end{center}
\end{figure}

Now let us look how to generate the mass of the particles we described in the SM. To do this an $SU(2)_L$ doublet of complex scalars with an hypercharge $Y$ = 1 is introduced~: 
\beq H = \binom{\phi^+}{\phi^0}. \eeq
The requirement of $SU(2)_L  \otimes  U(1)_Y$ invariance yields the following Lagrangian for the field H :
\beq \mathscr{L}_H = (D_\mu H)^\dag (D_\mu H) - V(H). \eeq 
H is an $SU(3)_c$ singlet; it implies that it has no tree level interaction with the gluon fields $G^a_\mu$ which in consequence are massless.
The gauge invariant potential chosen for the field H reads
\beq V(H) = \mu^2 H^\dag H + \l (H^\dag H)^2. \eeq
We then rewrite the Lagrangian $\mathscr{L}_H$ :
\beq \mathscr{L}_H = \left| \left\{ \p_\mu +\frac{i}{2} \left( g_Y Y B_\mu + g_2 \sum_{a=1}^3 \s^a W^a_\mu \right) \right\} H \right|^2 - \mu^2 H^\dag H - \l (H^\dag H)^2. \eeq 
To get a physical minimum for the potential implies that the parameter $\l$ has to be positive. Then the averaged expected value of the field H in the vacuum, also called Vacuum Expectation Value (VEV), is obtained by calculating the minimum of the potential $V(H)$ :
\beq \frac{\p V(H)}{\p |H|} = 0 = (2\mu^2 +4\l |H|^2)|H|, \eeq
which leads to the VEV
\beq \langle 0|H|0 \rangle = H_0 = \sqrt{\frac{-\mu^2}{2\l}} \equiv \pm \frac{v}{\sqrt{2}}. \eeq
We are thus left with two types of potential as depicted in figure~\ref{fig:higpot} :
\begin{itemize}
\item Either $\mu \geq 0$; we get a trivial minimum $\langle 0|H|0 \rangle$ = 0,
\item or $\mu < 0$; two minima are obtained : $\langle 0|H|0 \rangle = \pm \frac{v}{\sqrt{2}}$. 
\end{itemize}
The latter is the most interesting : by choosing one specific minimum $+ \frac{v}{\sqrt{2}}$, the $SU(2)_L  \otimes  U(1)_Y$ symmetry is spontaneously broken. Using the relation \ref{eq:Gm-N} we can check that this procedure does not break the EM symmetry : $SU(2)_L  \otimes  U(1)_Y$ breaks into the exact $U(1)_{em}$ symmetry :
\beq \mathcal{Q} H_0 = 0. \eeq

Expanding the field H around the chosen vacuum we have
\beq H_0 = \binom{\phi^+_0}{\phi^0_0} = \binom{G^+}{\frac{1}{\sqrt{2}} (v+h^0+iG^0) }. \eeq
A gauge transformation, the unitary gauge, allows us to get rid of the non-physical degrees of freedom $G^\pm$ and $G^0$, the Goldstone bosons. The new expression
\beq \label{eq:Hvev} H_0 = \frac{1}{\sqrt{2}}\binom{0}{v+h^0} \eeq
is now used to determine how the interaction between $H_0$ and the gauge fields gives their mass. 
First note that the only physical state remaining in the Higgs sector gets a mass :
\beq m^2_{h^0} = -2\mu^2 = 2 \l v^2.\eeq
 From the expression of $\mathscr{L}_H$ we analyse the term
\beq \label{SM:boson_m_terms} \begin{split} \left| \frac{i}{2} \left( g_Y Y B_\mu + g_2 \sum_{a=1}^3 \s^a W^a_\mu \right) H_0 \right|^2 = \ & 
 \frac{1}{8}  \left| \begin{pmatrix}
			g_2 W^3_\mu + g_Y Y B_\mu & g_2 ( W^1_\mu - i W^2_\mu) \\
			g_2 ( W^1_\mu + i W^2_\mu) & - g_2 W^3_\mu + g_Y Y B_\mu
	  \end{pmatrix}  \binom{0}{v+h^0} \right|^2 \\
= \ & \frac{1}{8} g_2^2 (v+h^0)^2 \left[ W^1_\mu W^{1 \mu} + W^2_\mu W^{2 \mu} \right] \\
&+\frac{1}{8} (v+h^0)^2 (g_Y B_\mu -g_2 W^3_\mu) (g_Y B^\mu -g_2 W^{3 \mu}).
\end{split} \eeq 
Let us look more precisely at the mass terms. Using 
\beq W^\pm_\mu \equiv \frac{1}{\sqrt{2}} (W^1_\mu \mp i W^2_\mu), \eeq
we find the mass term corresponding to the $W$ bosons :
\beq M^2_W W^+_\mu W^{- \mu} \quad \mathrm{with} \quad M_W = \frac{1}{2}g_2 v. \eeq
The remaining mass term reads
\beq \frac{1}{8} v^2  \begin{pmatrix} W^3_\mu & B_\mu \end{pmatrix} 
	  \begin{pmatrix}
			g_2^2 & -g_2 g_Y\\
			-g_2 g_Y & g_Y^2
	  \end{pmatrix} \binom{W^{3 \mu}}{B^\mu}. \eeq
The diagonalisation of this mass matrix gives two new physical states, the massless photon $A_\mu$ and the $Z$ boson described by the field $Z_\mu$. Their corresponding mass terms are 
\beq \label{SM:ZA_m_terms} \frac{1}{2} (M^2_Z Z_\mu^2 + M^2_A A_\mu^2) \eeq
with
\beq \label{SM:masseigen} \begin{split}
A_\mu & = \frac{g_Y W^3_\mu + g_2 B_\mu}{\sqrt{g_Y^2 + g_2^2}}, \quad M_A = 0,\\
Z_\mu & = \frac{g_2 W^3_\mu - g_Y B_\mu}{\sqrt{g_Y^2 + g_2^2}}, \quad M_Z = \frac{1}{2}\sqrt{g_Y^2 + g_2^2} v.
\end{split} \eeq
$g_Y$ and $g_2$ can be reexpressed in terms of the $U(1)_{em}$ coupling constant $e$ and a mixing angle called the Weinberg angle $\t_W$ to get a simpler relation between the physical states ($A_\mu$, $Z_\mu$) and the gauge eigenstates ($W^3_\mu$, $B_\mu$) :
\beq g_Y = \frac{e}{\cos \t_W}, \quad g_2 = \frac{e}{\sin \t_W} \ \RA \ \left\{\begin{array}{rl}
A_\mu = \sin \t_W W^3_\mu + \cos \t_W B_\mu \\
Z_\mu = \cos \t_W W^3_\mu - \sin \t_W B_\mu \end{array}\right. .
\eeq
The mixing between the $W^3_\mu$ and $B_\mu$ fields explains the different masses for the $Z$ and $W$ bosons :
\beq \frac{M_W}{M_Z} = \cos \t_W. \eeq
Interactions involving the exchange of $Z$ and $W$ bosons are respectively put into neutral and charged current interactions. The relative strength between these two types of interaction is encoded in the $\rho$ parameter defined as
\beq \label{eq:rhoSM} \rho = \frac{M^2_W}{M_Z^2 \cos^2 \t_W} = 1.\eeq
Note that this is the tree level definition of $\rho$; higher order corrections could lead to small deviation of $\rho$ from the unity which would be a way to probe New Physics (NP). More generally the experimental study of the EW sector imply strong constraints on BSM models especially those characterized by an extension of the gauge sector as we will see in part \ref{part3}. \\

In the matter sector the mass terms are generated through the mixing of the fermion chiralities represented by the Yukawa terms\footnote{Neutrinos are not considered here; the mechanism to generate their masses is not yet well established because we observe only one neutrino chirality.}. Recalling the singlets and doublets of $SU(2)_L$ defined in section~\ref{sec:1.matter}, the Yukawa Lagrangian reads :
\beq \label{eq:YukSM} \mathscr{L}_Y = \sum_{i,j=1}^3 y_u^i\bar{Q}_i \bar{H} V_{ij} u_j + y_d^i\bar{Q}_i H d_i + y_e^i\bar{L}_i H e_i + \textrm{h.c}\footnote{Hermitian coujugate of the expression. Note that most of the time the flavour and colour indices will be omitted.}, \quad \bar{H}=i\s^2 H^* \eeq 
where $i$ and $j$ stand for the fermion families, $V_{ij}$ are the CKM matrix elements and the $y$'s are the Yukawa couplings. The EW spontaneous breaking allows to derive the expression of the mass of the fermions. Using the eq.~\ref{eq:Hvev} and neglecting the effect of the CKM elements, the mass of a fermion $f$ of the i\textit{th} family reads
\beq m_{f_i} = \frac{y_f^i v}{\sqrt{2}}.\eeq

\section{Full standard picture}

The last part of the SM Lagrangian is the one that contains the interactions between the matter and gauge sectors :
\beq \mathscr{L}_{MG} = \bar{\Psi} \cancel{D} \Psi, \quad \cancel{D} = \g^\mu D_\mu, \quad \Psi \in \{Q,\bar{u},\bar{d},L,\bar{e}\}. \eeq
The full SM Lagrangian then reads
\beq \begin{split} 
\mathscr{L}_{SM} = & \ \mathscr{L}_{YM} + \mathscr{L}_{H} + \mathscr{L}_{Y} + \mathscr{L}_{MG} \\
= & -\frac{1}{4} \left( B_{\mu\nu}B^{\mu\nu} + \sum_{i=1}^3 W^i_{\mu\nu}W^{i \mu\nu} + \sum_{i=1}^8 G^i_{\mu\nu}G^{i \mu\nu} \right) \\
& + (D_\mu H)^\dag (D_\mu H) - \mu^2 H^\dag H - \l (H^\dag H)^2 \\
& + \sum_{i,j=1}^3 y^u_{i}\bar{Q}_i \bar{H} V_{ij} u_j + y^d_{i}\bar{Q}_i H d_i + y^e_{i}\bar{L}_i H e_i + \textrm{h.c} \\
& + \sum_{i=1}^3 \bar{Q}_i \cancel{D} Q_i + \bar{u}_i \cancel{D} u_i + \bar{d}_i \cancel{D} d_i + \bar{L}_i \cancel{D} L_i + \bar{e}_i \cancel{D} e_i.
\end{split} \eeq
It follows that the SM contains 19 parameters : 9 Yukawa couplings, 3 CKM angles and a CP phase, 3 coupling constants, the $\mu$ and the $\l$ parameters from the Higgs potential and the QCD vacuum angle $\t_{\mathrm{QCD}}$ linked to the strong CP problem that we will tackle in section~\ref{sec:1.SMpbs}. By considering massive neutrinos we must also add new terms : neutrino masses and PMNS elements.

\begin{figure}[!htb]
\begin{center}
\includegraphics[width=10.3cm,height=6.5cm]{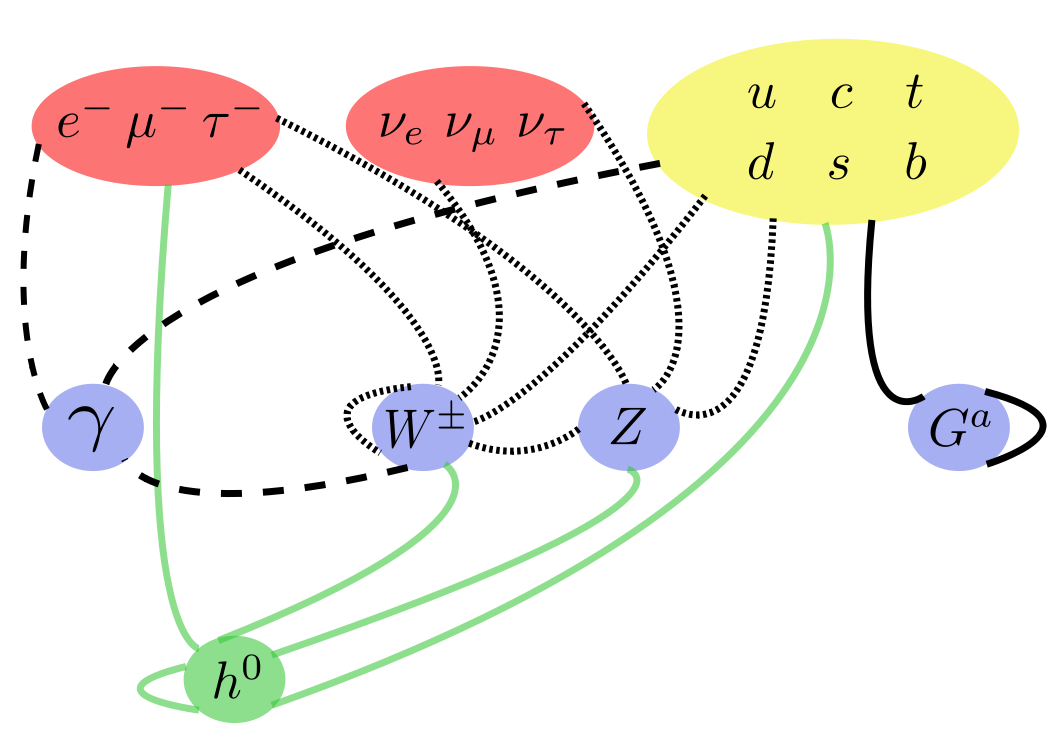} 
  \caption[Picture representing the tree-level interactions between the SM particles.]{Tree-level interactions between the SM particles. Lines represent interactions with gluons (plain black), $\g$ (dashed), $Z$ and $W$ (dotted) and Higgs boson (green).}
\label{fig:SMint}
\end{center}
\end{figure}

The SM interactions are depicted in figure~\ref{fig:SMint} and table \ref{tab:SM} summarizes the main characteristics of the SM particles. In this table the mass of the Higgs boson is the weighted mean of the mass measurement derived by the ATLAS collaboration with integrated luminosities of about $4.8$~fb$^{-1}$ at a center-of-mass energy of proton-proton collision at the LHC of $\sqrt{s} = 7$~TeV and $20.7$~fb$^{-1}$ at $\sqrt{s} = 8$~TeV ($m_{h^0} = 125.5 \pm 0.2^{\,+\,0.5}_{\,-\,0.6}$~GeV, \cite{ATLAS:2013mma}) and the measurement done by the CMS collaboration with integrated luminosities up to $5.1$~fb$^{-1}$ at $\sqrt{s} = 7$~TeV and up to $19.6$~fb$^{-1}$ at $\sqrt{s} = 8$~TeV ($m_{h^0} = 125.7 \pm 0.3 \pm 0.3$~GeV, \cite{CMS:yva}).

\renewcommand{\arraystretch}{1.2}
\begin{table}[!htb]
\begin{center}
\begin{tabular*}{1.\textwidth}{  c  c  c  c  c  }
       \hline \hline
      \multicolumn{2}{c}{\textbf{Name}} & \textbf{Mass} & $\mathbf{\mathcal{Q}}$ & $\mathbf{SU(3)_c, SU(2)_L, U(1)_Y}$ \\ \hline \hline
      \multicolumn{5}{c}{\textbf{Fermions}} \\ \hline \hline
          & $u$ & $2.3^{\,+0.7}_{\,-0.5}$ MeV & & \\ 
Up-type quarks & $c$ & $1.67 \pm 0.07$ GeV & 2/3 & ($\textbf{3}$, $\textbf{2}$, $\frac{1}{3}$)$_\mathrm{LH}$; ($\bar{\textbf{3}}$, $\textbf{1}$, -$\frac{4}{3}$)$_\mathrm{RH}$ \\ 
      	  & $t$ & $173.5 \pm 0.6 \pm 0.8$ GeV & & \\ \hline
          & $d$ & $4.8^{\,+\,0.7}_{\,-\,0.3}$ MeV & & \\ 
Down-type quarks & $s$ & $95 \pm 5$ GeV & -1/3 & ($\textbf{3}$, $\textbf{2}$, $\frac{1}{3}$)$_\mathrm{LH}$; ($\bar{\textbf{3}}$, $\textbf{1}$, $\frac{2}{3}$)$_\mathrm{RH}$\\ 
      	  & $b$ & $4.78 \pm 0.06$ GeV & & \\ \hline 
          & $e^-$ & 5.10998928 keV & & \\ 
Charged leptons & $\mu^-$ & 105.6583715 MeV & -1 & ($\textbf{1}$, $\textbf{2}$, -1)$_\mathrm{LH}$; ($\textbf{1}$, $\textbf{1}$, 2)$_\mathrm{RH}$ \\ 
      	  & $\tau^-$ & $1776.82 \pm 0.16$ MeV & & \\ \hline
          & $\nu_e$ & & & \\ 
Neutrinos & $\nu_\mu$ & $\sum m_{\nu} < 0.23$ eV \cite{Ade:2013lta} & 0 & ($\textbf{1}$, $\textbf{2}$, -1)$_\mathrm{LH}$ \\ 
      	  & $\nu_\tau$ & & & \\ \hline \hline
      \multicolumn{5}{c}{\textbf{Vector bosons}} \\ \hline \hline
Gluons & $G^a$ & - & 0 & ($\textbf{8}$, $\textbf{1}$, 0) \\ \hline
$W$ bosons & $W^\pm$ & 80.385 $\pm$ 0.015 GeV & $\pm$ 1 & $W^i \in (\textbf{1}, \textbf{3}, 0)$ \\
$Z$ boson & $Z$ & 91.1876 $\pm$ 0.0021 GeV & 0 & $B \in (\textbf{1}, \textbf{1}, 0)$ \\
Photon & $\gamma$ & $< 1 \times 10^{-18}$ eV & 0 & \\ \hline \hline
      \multicolumn{5}{c}{\textbf{Scalar boson}} \\ \hline \hline
Higgs boson & $h^0$ & $125.63^{\,+\,0.33}_{\,-\,0.35}$ GeV \cite{ATLAS:2013mma,CMS:yva} & 0 & ($\textbf{1}$, $\textbf{2}$, 1) \\ \hline \hline

\end{tabular*}
\caption[Summary of the characteristics of the SM particles.]{\label{tab:SM}Summary of the characteristics of the SM particles. If there is no citation, the mass comes from \cite{Beringer:1900zz}. The uncertainties are neglected if there are too small. Note that here the $c$ and $b$ masses are combinations of pole mass determinations and the mass of the top quark comes from an average of top mass measurements, all this done in \cite{Beringer:1900zz}.}
\end{center}
\end{table}

\section{Successes of the SM}
\label{sec:1.success}

Since the discovery of a boson whose properties (spin \cite{ATLAS:2013mla,CMS:yva}, parity \cite{CMS:yva}, couplings to other particles \cite{ATLAS:2013sla,CMS:yva}) match more or less those of the only scalar boson of the SM, the Higgs boson, all the particles predicted by the SM are now experimentally studied and the results reveal an excellent agreement between the predictions and the measurements. The observation of the massive gauge bosons and the success of the EW precision tests are especially an impressive achievement of the theory with $SU(2)_L  \otimes  U(1)_Y$ symmetry breaking \cite{LEP:2003aa}. 

Rare decays of mesons are an interesting way to test the viability of the SM since they are characterized by a suppression of the SM contributions. For instance the branching ratios $\bsmu$ and $\bsg$ are known as important observables to probe NP. However, their recent measurements show that they stand close to their SM expectation.

\section{SM issues}
\label{sec:1.SMpbs}

Despite the strong evidences of the viability of the SM, several problems remain unsolved.

\subsection{Theoretical problems}

One of the main argument that might indicate that the SM is not the fundamental theory of particle physics is that it does not include gravitation. Moreover there is no reliable theory of quantum gravity, namely the quantisation of the gravitational field is an open topic. Such a theory would have to be reliable when quantum gravity becomes non-negligible which is the case at the Planck scale defined by the Planck mass $M_{Pl}$
\beq M_{Pl} = \sqrt{\frac{\hbar c}{G_N}} = 1.220~93(7) \times 10^{19} \ \mathrm{GeV},\eeq
with the Newtonian gravitational constant $G_N = 6.673~8(8) \times 10^{-11} \ \mathrm{m}^2 \; \mathrm{kg}^{-1} \; \mathrm{s}^{-2}$.

Another issue concerning the unification of the fundamental interactions is linked to the unification at higher scale of the three SM interactions into one single interaction. These theories are called Grand Unified Theories (GUT). The running of the three SM coupling constants with the energy scale of the process studied has long been observed. This motivates the idea of unifying the coupling constants of the SM at a given energy scale. However, the theoretical computation of this running shows that these coupling constants do not converge to a common value at an energy scale $M_{\mathrm{GUT}} \sim 10^{16}$ GeV. Nevertheless, as we will see later, NP effects can restore the unification.

Another non-explained issue in the SM arises through the $\t_{\mathrm{QCD}}$ parameter. In the SM, there is no reason why the strong interaction could not also break the CP symmetry. It implies that the SM should include another CP violation term that looks like
\beq \mathscr{L}_{SU(3)_c}^{\cancel{\mathrm{CP}}} \propto \t_{\mathrm{QCD}} \sum_{a=1}^8 G^a_{\mu\nu} \e^{\mu\nu\rho\s} G^a_{\rho\s} \eeq
where $\e^{\mu\nu\rho\s}$ are the Levy-Civita symbols in four dimensions.
Experimental constraints on the value of $\t_{\mathrm{QCD}}$ coming from neutron electric dipole moment studies lead to an upper bound of $|\t_{\mathrm{QCD}}| < 10^{-11}$ \cite{Kim:2008hd}, thus raising a fine-tuning issue that BSM considerations could solve.

Another more severe fine-tuning problem appearing in the SM is related to the hierarchy problem. When we consider quantum corrections to the SM Higgs mass through a fermionic one-loop contribution, it yields a divergent integral. To solve this issue we introduce a cut-off $\Lambda$ corresponding to the limit of validity of the theory. We are then left with a correction
\beq \label{eq:1.higcor} \d m^2_{h^0} = \frac{y_f^2}{16 \pi^2} \left( -2\Lambda^2 +6m_f^2 \ln\frac{\Lambda}{m_f} + ...\right).\eeq
The problem here is the choice of the scale at which particle physics cannot be described only by the SM. The higher scales we know are the GUT scale and the Planck scale which implies that corrections to the Higgs boson mass are extremely large. It follows that the compensation between the bare mass and these corrections has to be really precise to get a Higgs boson mass around 125 GeV. This fine-tuning issue could be avoided in BSM models where for instance $\Lambda$ corresponds to a lower scale or new contributions cancel the SM corrections.

As we saw in section \ref{subsec:1.lepton} the neutrino sector is not well understood. The fact that these fermions are massive even if we found only one kind of neutrino chirality tells us either that the mass generation in the SM is not completely understood or that we must add at least one new field to try to solve this issue. Furthermore the Dirac or Majorana nature of the neutrinos is not established.

Finally the SM gives us no fundamental explanation for the origin of some other features like the number of families, the hierarchy between the fermion masses as can be seen in table \ref{tab:SM}, the separation between quarks and leptons or simply the number of parameters.

\subsection{Experimental discrepancies}

Some aspects of the Higgs sector could lead to discrepancies with the SM expectations. As an example the signal strength corresponding to the decay of the Higgs boson into two photons could be such an interesting observable. Table~\ref{tab:hgaga} shows the observed signal strength $\mathbf{\mu_{\g\g}}$ for different production modes and for fixed $m_{h^0}$ values.  They were obtained by the authors in \cite{Dumont:2013wma} from the Moriond 2013 results released by the ATLAS and CMS collaborations. As can be seen slight deviations from unity are currently observed (SM expectation : $\mu_{\g\g}^{\mathrm{SM}} \equiv 1$). If these discrepancies are confirmed it would allow to probe NP.
\renewcommand{\arraystretch}{1.2}
\begin{table}[!htb]
\begin{center}
\begin{tabular*}{1.\textwidth}{c|c|c|ccccc}
\hline \hline
\textbf{Channel} & \textbf{Signal strength} $\mathbf{\mu_{\g\g}}$ & $\mathbf{m_{h^0}}$ \textbf{(GeV)} & \multicolumn{5}{c}{\textbf{Production mode}} \\
& & & \textbf{ggF} & \textbf{VBF} & \textbf{WH} & \textbf{ZH} & \textbf{ttH} \\
\hline \hline
\multicolumn{8}{c}{ATLAS 4.8 fb$^{-1}$ at 7 TeV + 20.7 fb$^{-1}$ at 8 TeV~\cite{ATLAS:2013oma,ATLAS:2013mma}} \\
\hline
$\mu_{\g\g}^{\rm ggF+ttH}$ & $1.60 \pm 0.41$ & 125.5 & 100\% & -- & -- & -- & -- \\
$\mu_{\g\g}^{\rm VBF+VH}$ & $1.94 \pm 0.82$ & 125.5 & -- & 60\% & 26\% & 14\% & -- \\
\hline
\multicolumn{8}{c}{CMS 5.1 fb$^{-1}$ at 7 TeV + 19.6 fb$^{-1}$ at 8 TeV~\cite{CMS:ril}} \\
\hline
$\mu_{\g\g}^{\rm ggF+ttH}$ & $0.49 \pm 0.39$ & 125 & 100\% & -- & -- & -- & -- \\
$\mu_{\g\g}^{\rm VBF+VH}$ & $1.65 \pm 0.87$ & 125 & -- & 60\% & 26\% & 14\% & -- \\
\hline \hline
\end{tabular*}
\caption[Signal strength $\mu_{\g\g}$ at a given $m_{h^0}$ for different Higgs boson production modes at the LHC.]{\label{tab:hgaga}Signal strength $\mu_{\g\g}$ at a given $m_{h^0}$ for different Higgs boson production modes at the LHC : gluon-gluon fusion (ggF), vector boson fusion (VBF), associated production with an EW gauge boson $V=W,Z$ (\textit{Higgs Strahlung}, VH) and associated production with a $t\bar{t}$ pair (ttH). Adapted from \cite{Dumont:2013wma}.}
\end{center}
\end{table}

The motion of a charged lepton in an external EM field gives one of the most precisely measured experimentally and calculated theoretically quantities in particle physics, the lepton anomalous magnetic moment. In QED the interaction of a photon field $A_\nu(x)$, $x$ being the position in space-time, with a Dirac field, here the lepton $\psi_\ell(x)$ is given by the Lagrangian density of interaction
\beq
\mathscr{L}^{\mathrm{QED}}_{\mathrm{int}}(x)= e j^\nu_{em}(x)A_\nu(x),\quad
j^\nu_{em}(x)=- \bar{\psi}_\ell(x)\gamma^\nu \psi_\ell(x),
\eeq
where $j^\nu_\mathrm{em}(x)$ is the EM current and $\gamma^\nu$ are the Dirac matrices.

The intrinsic angular momentum or spin $\vec{S}$ of the lepton is responsible for the intrinsic magnetic dipole moment $\vec{\mu}$ which is, with the usual convention $\hbar = c = 1$ :
\beq
\vec{\mu}=g_\ell \frac{e}{2m_{\ell}} \vec{S}.
\eeq
The Dirac theory predicts a \textit{$g$--factor} for the leptons of $g_\ell=2$ \cite{Dirac:1928hu,Dirac:1928ej}. The \textit{anomalous} contribution to the magnetic moment of the lepton, defined as
\beq
a_\ell= \frac12 (g_\ell-2),
\eeq
is given by a form factor defined by the matrix element $\langle{\ell^-(p')}| j_\mathrm{em}^\nu(0) |{\ell^-(p)\rangle}$ where $|{\ell^-(p)\rangle}$ is a lepton state of momentum $p$. The leading QED contribution to $a_\ell$ was obtained at the end of the 1940's \cite{Schwinger:1948iu} and reads
\beq
a^{\mathrm{QED}(1 \, \mathrm{loop})}_\ell = \frac{\alpha_{_{em}}}{2\pi}.
\eeq
with $\alpha_{_{em}}$ the fine-structure constant. Higher order contributions and the experimental value are now determined with an impressive accuracy. Nevertheless the SM contributions to the muon anomalous magnetic moment do not fit well the measured value. The current discrepancy between the experimental measurement \cite{Roberts:2010cj} and the theoretical calculation of this observable shows a deviation of about 3$\s$ :
\beq
\amu = a^{\mathrm{exp}}_\mu - a^{\mathrm{SM}}_\mu = (249 \pm 86) \times 10^{-11}.
\eeq

Finally, the BaBar collaboration reported last year \cite{Lees:2012xj} a combined 3.4$\s$ deviation of the measured ratios involving $B$ and $D$ mesons to their SM prediction,
\beq
\mathscr{R}(D^{(*)}) = \frac{\mathscr{B}(\bar{B} \ra D^{(*)}\tau^- \bar{\nu}_\tau)}{\mathscr{B}(\bar{B} \ra D^{(*)}\ell^- \bar{\nu}_\ell)}. \quad l = \{e,\mu\}
\eeq
These decay modes may give important clues on BSM physics.

\subsection{Cosmological connexion}

Perhaps the most important open issue in the SM is that it does not provide a Dark Matter (DM) candidate.
This specific kind of matter is required for the viability of the cosmological standard model called Lambda Cold Dark Matter ($\mathbf{\Lambda}$CDM). Since it is thought that the Cold Dark Matter (CDM) is made of heavy, stable and neutral particles, the SM could naively include such a candidate with the neutrinos. Nevertheless, as we will analyse in chapter \ref{chap:revcosmo}, the calculated total density of neutrinos in the Universe does not match the DM density.

Tle $\mathbf{\Lambda}$CDM model tells us that a period of the early Universe was characterized by an important annihilation of baryons and antibaryons into photons. A slight baryon excess implied that the SM matter component of the Universe is mainly composed of baryons. The baryon asymmetry in the Universe is not explained in the SM : this observation cannot be described only by SM CP violation processes.

\chapter{... To the infinitely large : the Lambda Cold Dark Matter model}
\label{chap:revcosmo}

\minitoc\vspace{1cm}
\newpage

\section{Theoretical framework}
\label{sec:2.th}

As the study of the smallest-scale structures, the study of the largest-scale structures of the Universe has developed greatly in the 20th century. The construction of the standard cosmological picture was done in the context of the Einstein’s \textit{General Relativity} (GR), the current most reliable theory of gravitation \cite{Einstein:1916vd}. As in SR, the distance $ds$ between two points in the spacetime depends on the spacetime metric $g_{\mu\nu}$ and the spacetime coordinates defined as $x^\mu$ where $\mu$ is a Lorentz index :
\beq \label{eq:ds} ds^2 = g_{\mu\nu} dx^\mu dx^\nu.\eeq
In GR, $g_{\mu\nu}$ is defined as a deformation of the Minkowski metric $\eta_{\mu\nu}$ parameterized by $h_{\mu\nu}$ :
\beq g_{\mu\nu}(x^\rho) = \eta_{\mu\nu} + h_{\mu\nu}(x^\rho).\eeq

\subsection{Cosmological principle and its consequences}

 By considering the idea that on a sufficiently large scale, the properties of the Universe are the same for all \textit{observers}, standard cosmology yields the \textit{cosmological principle} : the Universe is homogeneous and isotropic. Using these symmetry considerations the spacetime distance of eq.~\ref{eq:ds} reads, using the spacetime signature $(+,-,-,-)$, as
\beq \label{eq:RW} ds^2 = dt^2 - a^2(t) \left[ \frac{dr^2}{1-kr^2} + r^2 (d\t^2 + \sin^2 \t \, d\phi^2) \right],\eeq
where ($r,\t,\phi$) are the spherical coordinates, the curvature parameter $k$ is defined for three different cases, namely for an open ($k = -1$), flat ($k = 0$) and closed ($k = +1$) spacetime and $a(t)$ is the scale factor of the Universe. This metric is called the Robertson-Walker (RW) metric. The function $a(t)$ depends only on the time coordinate $t$ and is then used to determine the evolution of the Universe. This function is calculated using the Einstein equation of GR which links the geometry of the Universe represented by the Einstein tensor $G_{\mu\nu}$ and the energy-momentum tensor $T_{\mu\nu}$ describing its matter and energy content :
\beq \label{eq:Ein_eq} G_{\mu\nu} = R_{\mu\nu} -\frac12 Rg_{\mu\nu} = 8\pi G_N T_{\mu\nu},\eeq
where $R_{\mu\nu}$ and $R$ are respectively the Ricci tensor and the Ricci scalar which depend on the metric $g_{\mu\nu}$.
Assuming that the matter-energy content of the Universe behaves as a perfect homogenous and isotropic fluid with a total energy density $\rho$ and a pressure $p$, the energy-momentum tensor is rewritten as
\beq \label{eq:perf} T_{\mu\nu} = (p+\rho)u_\mu u_\nu - pg_{\mu\nu},\eeq
with $u = (1,0,0,0)$ the velocity vector of the fluid. The eqs. \ref{eq:RW}, \ref{eq:Ein_eq} and \ref{eq:perf} gives the so-called Friedmann-Lema\^{i}tre (FL) equations :
\begin{align}
H^2 & = \left(\frac{\dot{a}}{a}\right)^2 = \frac{8\pi G_N}{3}\rho - \frac{k}{a^2}, \label{eq:FL1} \\
\frac{\ddot{a}}{a} & = -\frac{4\pi G_N}{3} (\rho + 3p),\label{eq:FL2} 
\end{align}
where the dots correspond to time derivatives and $H$ is called the Hubble parameter. The first equation gives the expansion rate of the Universe while the second one determines if the expansion is accelerated or decelerated, both at a given time $t$. Using eqs. \ref{eq:FL1} and \ref{eq:FL2} a third fundamental relation is obtained (it can also be found through energy-momentum conservation or simply through the first law of thermodynamics) :
\beq \label{eq:conserv} \dot{\rho} = - 3H(\rho + p).\eeq
Different types of fluid composed the Universe and they are characterized by their equation of state which connect $p$ and $\rho$ :
\beq \omega = \frac{p}{\rho}.\eeq
Using the eq.~\ref{eq:conserv}, the evolution of the energy density with respect to the scale factor can be approximated, for a constant value of $\omega$ with respect to time, as
\beq \label{eq:rho} \rho \propto a^{-3(1+\omega)}.\eeq
Two well-known cases are then obtained :
\begin{itemize}
\item For a Universe composed of matter\ie for non-relativistic particles, $p=p_m$ is negligible; $\omega = 0$. It follows that $\rho=\rho_m \propto a^{-3} \equiv V^{-1}$ where $V$ describes the total volume. In other words the matter dilutes linearly as the total volume of the Universe expands.
\item For relativistic particles with $p=p_r$ and $\rho=\rho_r$, which mainly corresponds to the radiation, $p_r=\rho_r c^2_s=\frac{1}{3}\rho_r$, with $c_s$ the speed of sound in this fluid. It implies that $\rho=\rho_r \propto a^{-4}$ : radiation dilutes quicker than matter.
\end{itemize}
By looking at eq.~\ref{eq:rho} one peculiar case is found for $w=-1$ : the energy density is constant with respect to the scale factor, it does not dilute. This case appears if a positive integration constant $\Lambda$, called the cosmological constant, is considered in the Einstein eq.~\ref{eq:Ein_eq} and is interpreted as a vacuum energy. Using eq.~\ref{eq:FL2} it follows that the scale factor is an exponential function of time; the expansion of a Universe dominated by vacuum energy is accelerating. More generally the cosmological constant is a special case of fluids implying an acceleration of the expansion of the Universe called \textit{Dark Energy} (DE). The parameter $\omega_{_\mathrm{DE}}$ of the equation of state of DE is not necessarily constant. 

\subsection{Cosmological parameters}

The eq.~\ref{eq:FL1} is used to determine the critical density $\rho_c$\ie the total density in a flat spacetime :
\beq \rho_c = \frac{3H^2}{8\pi G_N}.\eeq
Using this definition it is more common to define for each species the density parameters
\beq \Om_i = \frac{\rho_i}{\rho_c},\eeq
where $i$ stands for radiation ($r$), matter ($m$) and DE. Note that for the specific cases of the curvature and the cosmological constant the densities read
\beq \Om_k = -\frac{k}{a^2H^2}, \quad \Om_\Lambda = \frac{\Lambda}{3H^2}.\eeq
By rewriting eq.~\ref{eq:FL1}, it follows that
\beq \Om_r + \Om_m + \Om_{_\mathrm{DE}} = \Om_\mathrm{tot} = 1 - \Om_k.\eeq
The determination of the density parameters (see section \ref{sec:2.success}) allows to safely neglect the present values of the radiation and curvature densities. It follows from the eq.~\ref{eq:FL2} that a current acceleration of the expansion of the Universe needs
\beq \rho_{_{\mathrm{DE},0}} (1+3 \, \omega_{_{\mathrm{DE},0}}) + \rho_{m,0} < 0,\eeq
where the subscript \textit{0} refers to the present value of the quantities considered. 
Using the density parameters and the fact that $\Om_{m,0} + \Om_{_{\mathrm{DE},0}} \sim 1$ the condition for a current acceleration of the expansion of the Universe reads
\beq \omega_{_{\mathrm{DE},0}} \lesssim -\frac{1}{3 \, \Om_{_{\mathrm{DE},0}}},\eeq
which is satisfied for the cosmological constant.
Instead of using the time variable when describing the evolution of the Universe, it is more convenient to handle an other variable directly linked to the scale factor of the RW metric, the cosmological redshift $z$
\beq 1+z = \frac{a_0}{a}.\eeq
It is then possible to link the cosmological parameters at different times by modifying again eq.~\ref{eq:FL1}. The Hubble parameter $H_z$ at a redshift $z$ reads
\beq H_z^2 = H^2_0 \left[ \Om_{r,0}(1+z)^4 + \Om_{m,0}(1+z)^3 + \Om_{k,0}(1+z)^2 + \Om_{_{\mathrm{DE},0}} f(z)\right]\eeq
where
\beq f(z) = \exp\left(3\int_0^z \frac{1+\omega_{_{\mathrm{DE}}}(z')}{1+z'} dz'\right).\eeq
For more details on the construction of the physical cosmology we refer the reader for instance to the textbooks \cite{Weinberg1,Kolb-Turner,Weinberg2} and the reviews \cite{CervantesCota:2011pn,Trodden:2004st}.

\section{Cosmological observations}
\label{sec:2.success}

\subsection{Methods}

Different types of observations are used to probe the evolution of the Universe. Since the work of Hubble \cite{Hubble:1929ig} we know that the Universe is expanding. The remaining question about this expansion was whether it accelerates or not. The luminosity curve of type Ia supernov\ae \ (SNIa) is well known and calibrated with precision since their physics are quite similar. SNIa are then used as \textit{standard candles} to determine the characteristics of the expansion of the Universe. With this method the acceleration of the expansion of the Universe was determined by the Supernova Search Team \cite{Riess:1998cb} and the Supernova Cosmology Project \cite{Perlmutter:1998np} collaborations in 1998.

The study of Large Scale Structures (LSS) is also an important tool to probe cosmological features. The standard cosmological model predicts Baryon Acoustic Oscillations (BAO) in the matter power spectrum of LSS which thus could give informations about the early Universe. The first BAO signal was detected in 2005 by the Sloan Digital Sky Survey (SDSS) collaboration \cite{Eisenstein:2005su}.

The best way to understand with accuracy the evolution of the Universe is through the study of the CMB radiation. It also contains the most solid arguments in support of the standard cosmological model. The CMB is the relic of the time when the Universe had sufficiently cooled down (around 4~000 K) to allow the photons to decouple from the most standard form of matter observed, the baryonic matter, and to fill the Universe even nowadays. Predicted with a black-body spectrum by Gamow and collaborators in the 1940s \cite{Gamow:1946eb,Alpher:1948ve} and with a low temperature around 5 K due to the expansion of the Universe \cite{Alpher:1950zz}, this radiation was finally detected by Penzias and Wilson in 1965 with an almost perfect black-body spectrum as forecast \cite{Penzias:1965wn,Dicke:1965zz}. Several alternative cosmological models were thus ruled out. Its temperature is now well measured at 2.72548~$\pm$~0.00057~K~\cite{Fixsen:2009ug}.

\subsection{Success of the $\mathbf{\Lambda}$CDM model}

\begin{figure}[!htb]
\begin{center}
\subfloat[]{\includegraphics[width=8cm,height=5cm]{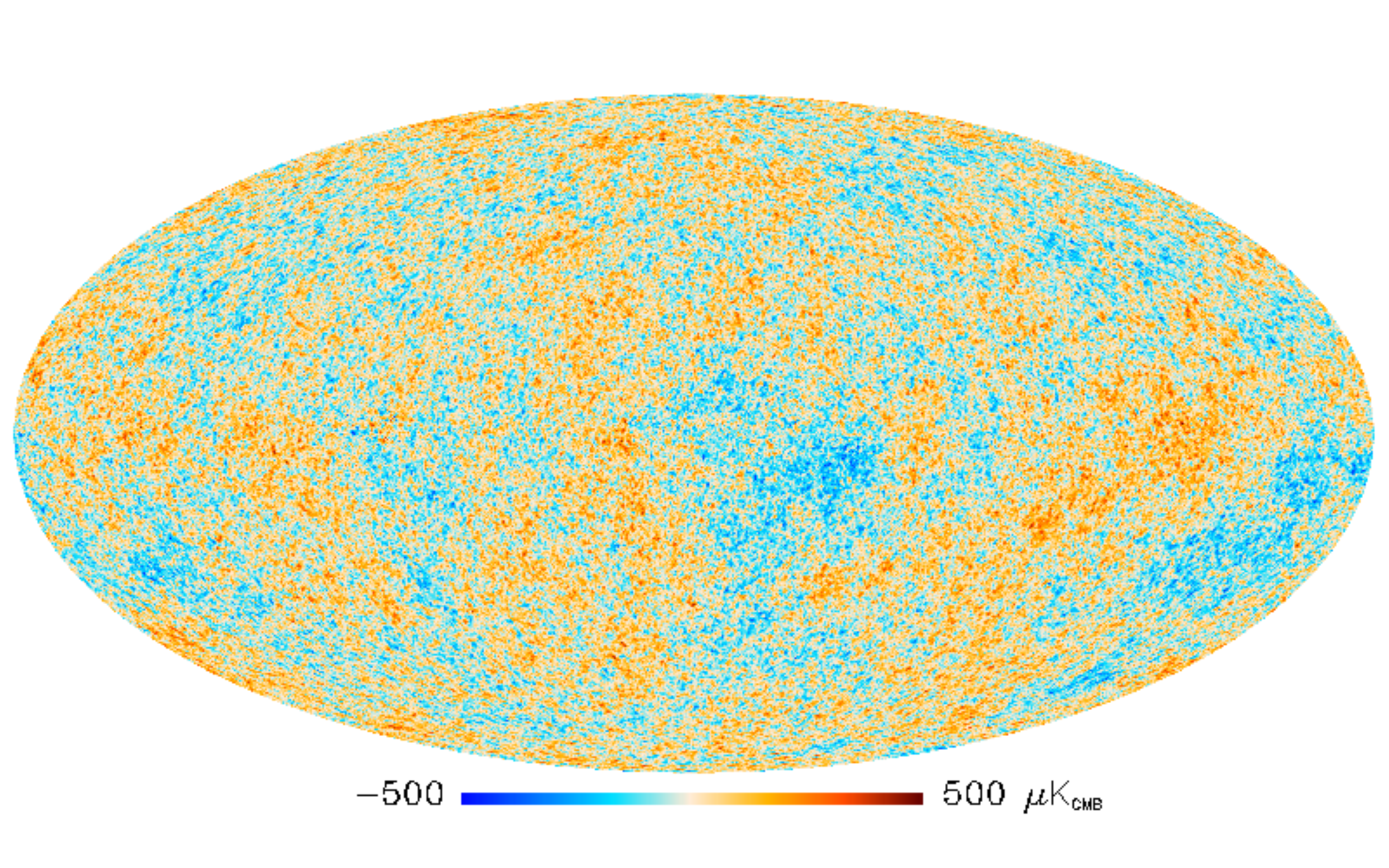}} 
\subfloat[]{\includegraphics[width=8cm,height=5cm]{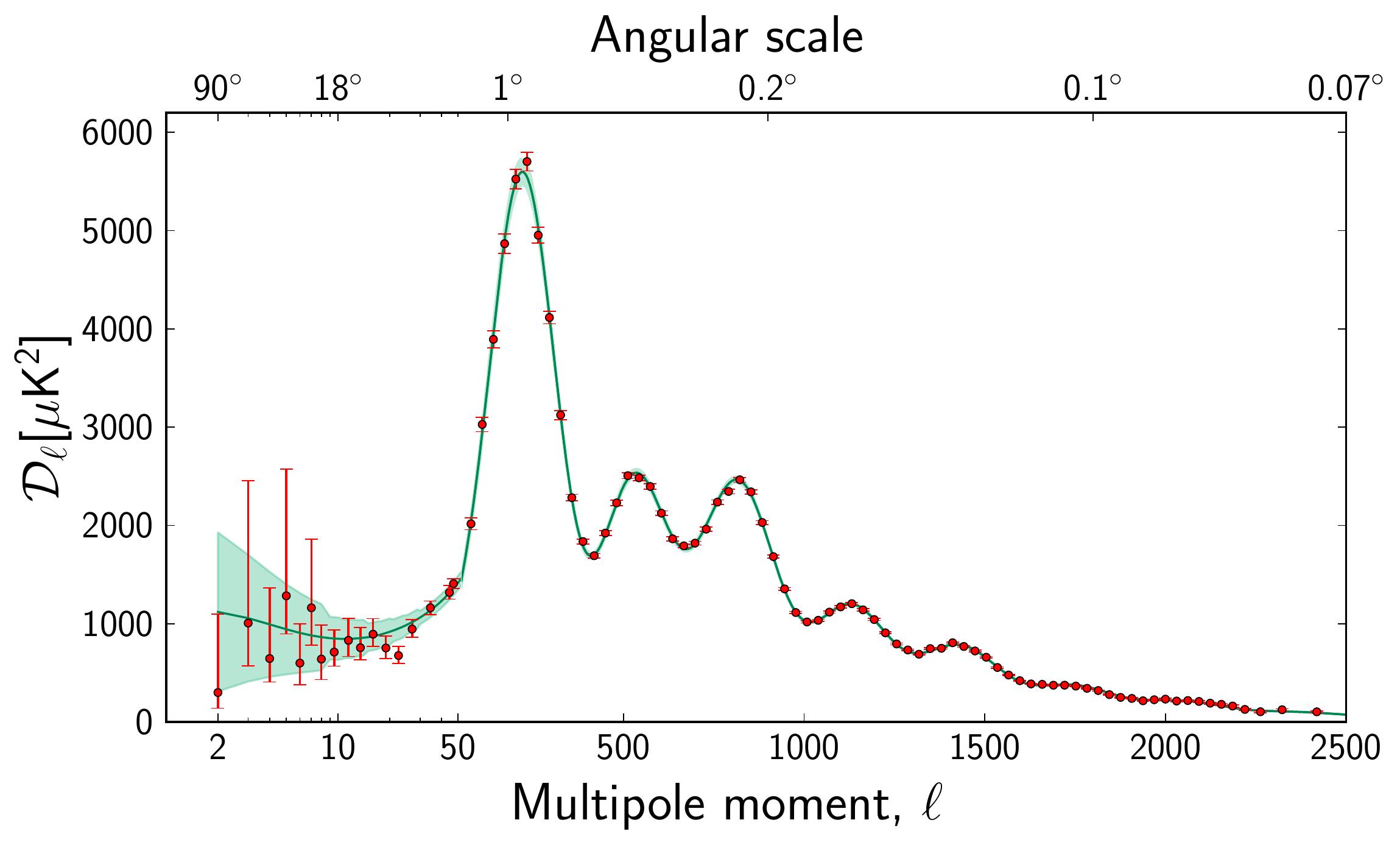}} 
  \caption[Temperature fluctuations of the CMB (a) and its temperature angular power spectrum compared to a simple $\mathbf{\Lambda}$CDM model (b) obtained by the Planck collaboration.]{Temperature fluctuations of the CMB (a) and its temperature angular power spectrum compared to a simple $\mathbf{\Lambda}$CDM model (b) obtained by the Planck collaboration. The vertical scale of (b) is defined as $\mathcal{D}_\ell[\mu K^2] = \ell(\ell+ 1)C_\ell/2\pi$. Figures taken from \cite{Ade:2013xsa}.}
\label{fig:CMBmappow}
\end{center}
\end{figure}

Cosmological observations are now reaching a high-level of precision in the determination of the cosmological parameters. It results that very small anisotropies are detectable in the CMB : it is not a perfect homogeneous and isotropic radiation. The temperature anisotropies are studied through an $Y_{lm}(\t,\phi)$ spherical harmonics expansion
\beq \frac{\d T}{T}(\t,\phi) = \sum_{\ell=2}^{\infty} \sum_{m=-\ell}^\ell a_{\ell m} Y_{\ell m}(\t,\phi).\eeq
The power spectrum of the CMB is then defined as the sum of the multipole moment $C_\ell$ defined as
\beq C_\ell = \frac{1}{2\ell +1} |a_{\ell m}|^2.\eeq
The best measurements of these anisotropies came with the 2013 data release of the Planck space observatory \cite{Ade:2013lta} with the CMB map and the CMB power spectrum shown in figure~\ref{fig:CMBmappow}. 

\begin{figure}[!htb]
\begin{center}
\includegraphics[width=5cm,height=5cm]{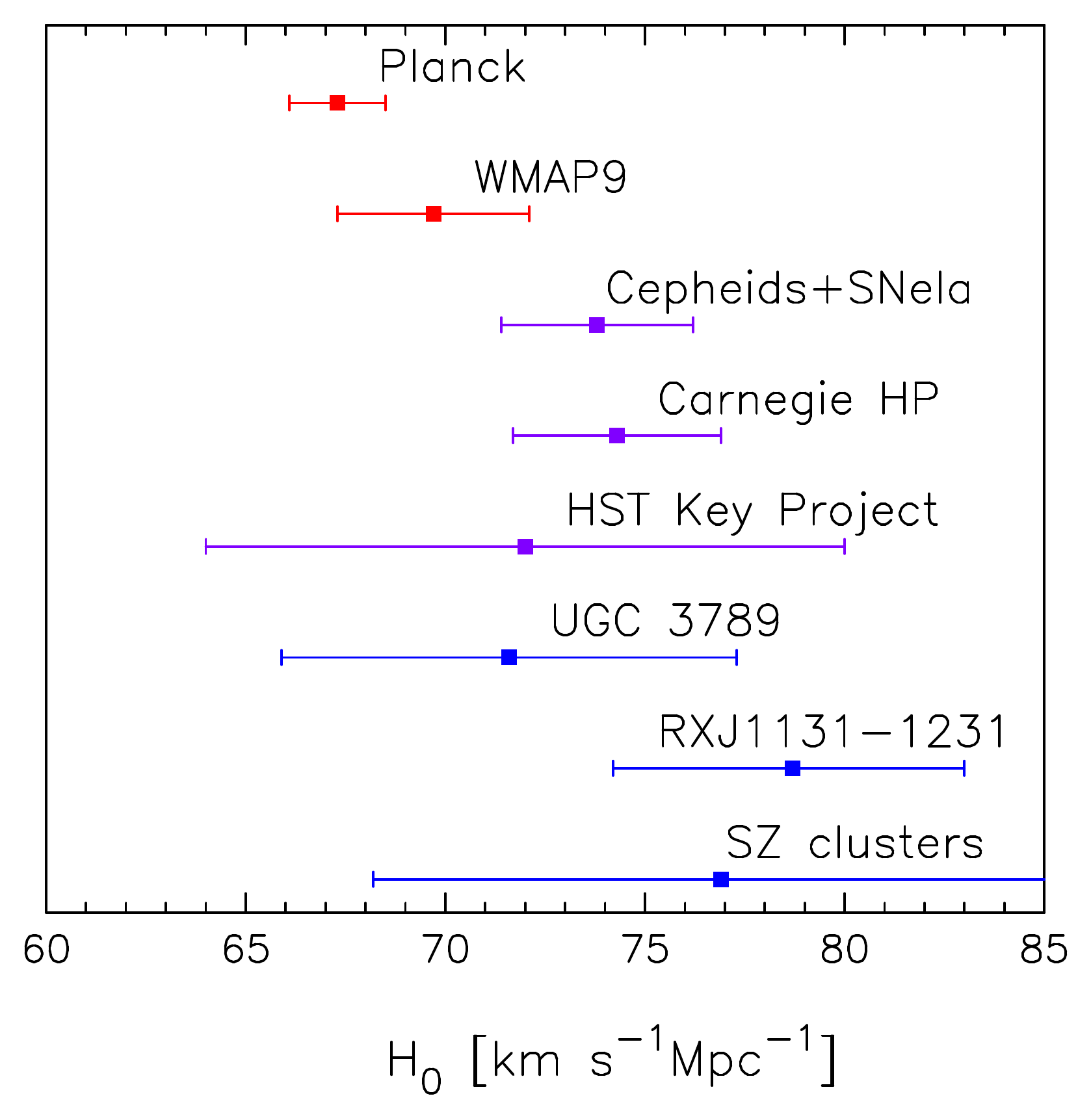} 
  \caption[Comparison of some $H_0$ measurements at 1$\s$ obtained by different collaborations.]{Comparison of some $H_0$ measurements at 1$\s$ obtained by different collaborations. Figure obtained from \cite{Ade:2013lta}.}
\label{fig:H0evol}
\end{center}
\end{figure}

To demonstrate how well the results were improved, figure~\ref{fig:H0evol} shows the evolution in time of the measurement of the Hubble constant $H_0$ using different techniques. 

Without going into the details of the CMB analysis figure~\ref{fig:CMBmappow}b shows that the most reliable form of cosmological model is the $\mathbf{\Lambda}$CDM model. The standard Big Bang scenario is fully described by only six parameters. We will talk about the three most relevant ones for the rest of this chapter, namely : 
\begin{itemize}
\item The scalar spectral index $n_s$ which enters in the definition of the primordial fluctuations (whose standard explanation is obtained in the context of cosmic inflation, see section~\ref{sec:2.inflation}) that give the CMB inhomogeneities;
\end{itemize}
and two parameters that characterized the matter content of the Universe :
\begin{itemize}
\item The physical baryon density $\Om_{_{b,0}} h^2$;
\item The physical Cold Dark Matter (CDM) density $\Om_{_{c,0}} h^2$;
\end{itemize}
where $h$ is the reduced Hubble constant defined using the relation $H_0 = 100 \, h \, \mathrm{km} \, \mathrm{s}^{-1} \, \mathrm{Mpc}^{-1}$ with 1 Mpc $\sim 3.1 \times 10^{22}$ m. 

Table \ref{tab:cosmo_param} summarizes the main results of the cosmological observations combining several data and will be discussed throughout this chapter. Note first that the radiation density is not included here since it is not significant today : $\Om_{_{r,0}} h^2 = 2.47 \times 10^{-5}$~\cite{Beringer:1900zz}. The curvature of the Universe is, to a good approximation, zero and the Universe is mainly filled by DE which drives the acceleration of its expansion. Moreover, the DE equation of state looks like that of a cosmological constant. The main interesting result for us here is the fact that the matter content of the Universe corresponds mainly to a non-standard, yet unknown form of matter, the \textit{Dark Matter} (DM) which interacts mainly gravitationally with the photons of the CMB. We will speak more about this component of the Universe in sections \ref{sec:2.DM} and \ref{sec:2.drawback}.

\renewcommand{\arraystretch}{1.2}
\begin{table}[!htb]
\begin{center}
\begin{tabular*}{0.706\textwidth}{ c c c }
       \hline \hline
      & \multicolumn{2}{ c }{\textbf{Planck+WP+highL+BAO}} \\ \cline{2-3}
     \textbf{Parameter} & \textbf{Best fit} & \textbf{1$\s$ uncertainty} \\ \hline \hline
     $\Om_{_{b,0}} h^2$ & 0.022161 & 0.02214 $\pm$ 0.00024 \\
     $\Om_{_{c,0}} h^2$ & 0.11889 & 0.1187 $\pm$ 0.0017 \\
     $n_s$ & 0.9611 & 0.9608 $\pm$ 0.0054 \\ 
     $\Om_{_{\mathrm{DE},0}}$ & 0.6914 & 0.692 $\pm$ 0.010 \\
     $H_0 \, (\mathrm{km} \, \mathrm{s}^{-1} \, \mathrm{Mpc}^{-1})$ & 67.77 & 67.80 $\pm$ 0.77 \\ 
     Age of the Universe (Gyr) & 13.7965 & 13.798 $\pm$ 0.037 \\  \cline{2-3}
      & \textbf{Best fit} & \textbf{2$\s$ uncertainty} \\ \hline
     $\Om_{_{k,0}}$ & 0.0009 & -0.0005$^{\,+\,0.0065}_{\,-\,0.0066}$ \\
     $\omega_{_{\mathrm{DE},0}}$ & -1.109 & -1.13$^{\,+\,0.23}_{\,-\,0.25}$ \\ \hline \hline
\end{tabular*}
\caption[Value for some cosmological parameters combining Planck data with WMAP polarization data (WP), high-$\ell$ CMB (highL) and BAO data.]{\label{tab:cosmo_param}Value for some cosmological parameters combining Planck data with WMAP polarization data (WP), high-$\ell$ CMB (highL) and BAO data. Adapted from \cite{Ade:2013lta}.}
\end{center}
\end{table}

\section{Dark Matter}
\label{sec:2.DM}

The DM postulate was initiated by Oort in 1932 to account for the motion of the stars in the Milky Way (MW). One year later Zwicky observed, through the study of the velocity distribution of galaxies in the Coma cluster, that luminous mass was not sufficient to explain the total mass responsible of the dynamic of the cluster \cite{Zwicky:1933gu}. He then also introduced DM, a specific type of matter that does not emit light, to explain this observation. By analysing the Virgo cluster, Smith reached the same conclusion in 1936 but the lack of astrophysical knowledge at this time prevented to emphasize these observations. 

\subsection{DM evidences}

\begin{figure}[!htb]
\begin{center}
\includegraphics[width=6cm,height=5cm]{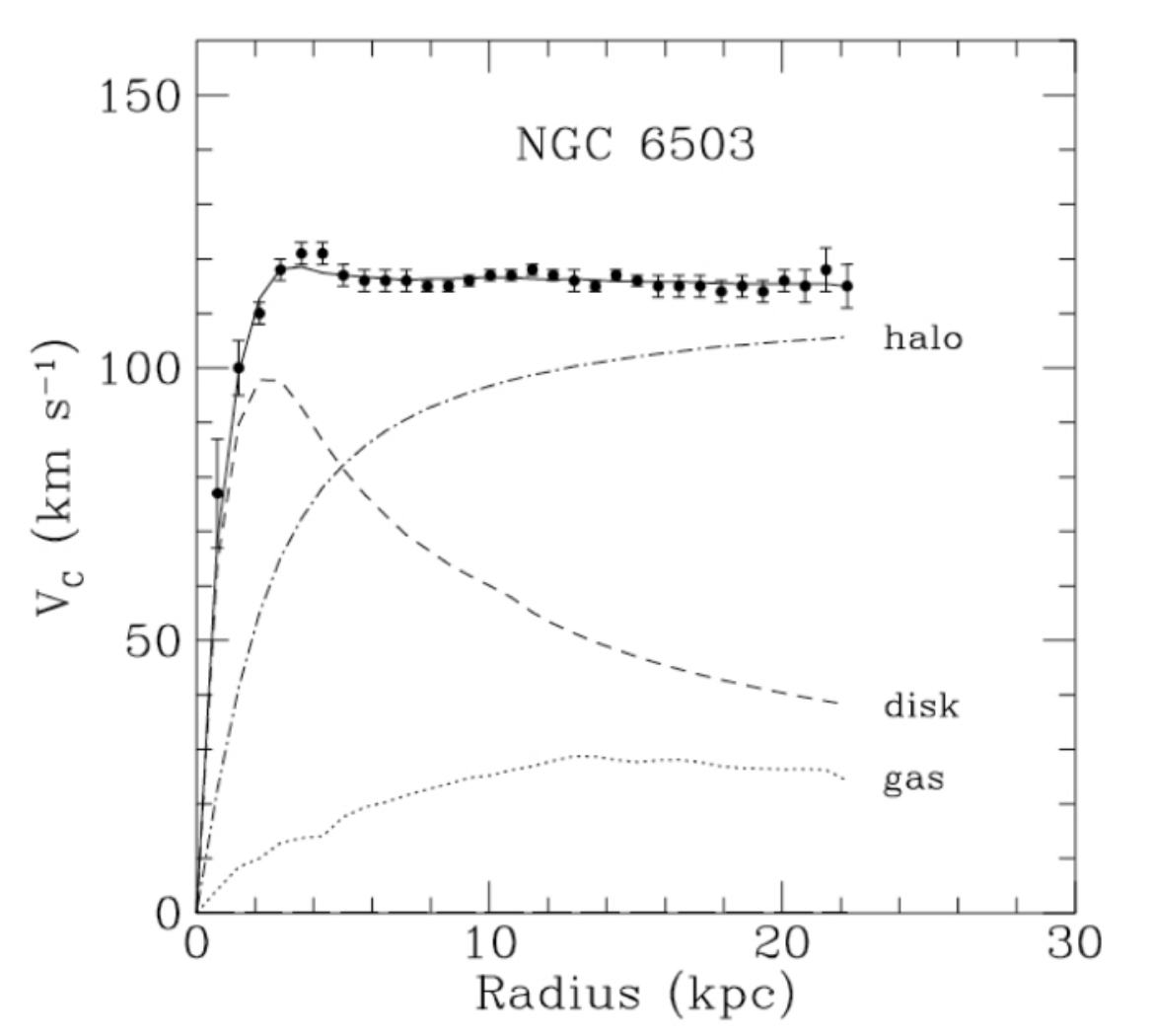} 
  \caption[Rotation curve of the galaxy NGC 6503, plotting its main contributions.]{Rotation curve of the galaxy NGC 6503, plotting its main contributions. Figure taken from \cite{Begeman:1991iy}.}
\label{fig:rotcurv}
\end{center}
\end{figure}

The first strong evidence came with the study of rotation curves of some spiral galaxies by Rubin and collaborators \cite{Rubin:1970zza,Rubin:1980zd,Rubin:1985ze}. These measurements made at the galactical scale consist in calculating the rotational velocity $v(r)$ at a distance $r$ of the center of a given galaxy using spherical symmetry assumption and Newtonian dynamics :
\beq v(r) = \sqrt{\frac{G_N M(r)}{r}},\eeq
where $M(r)$ is the mass inside the sphere of radius $r$.
Outside the luminous mass of the galaxy, $M(r)$ is clearly expected to be constant, which leads to a velocity $v(r) \propto 1/\sqrt{r}$. However this does not correspond to the observations : the velocity distribution is approximately flat far away from the center of the galaxy, as shown in figure~\ref{fig:rotcurv}. This is well explained if a DM halo is considered around the galaxy. 

\begin{figure}[!htb]
\begin{center}
\includegraphics[width=7.5cm,height=5cm]{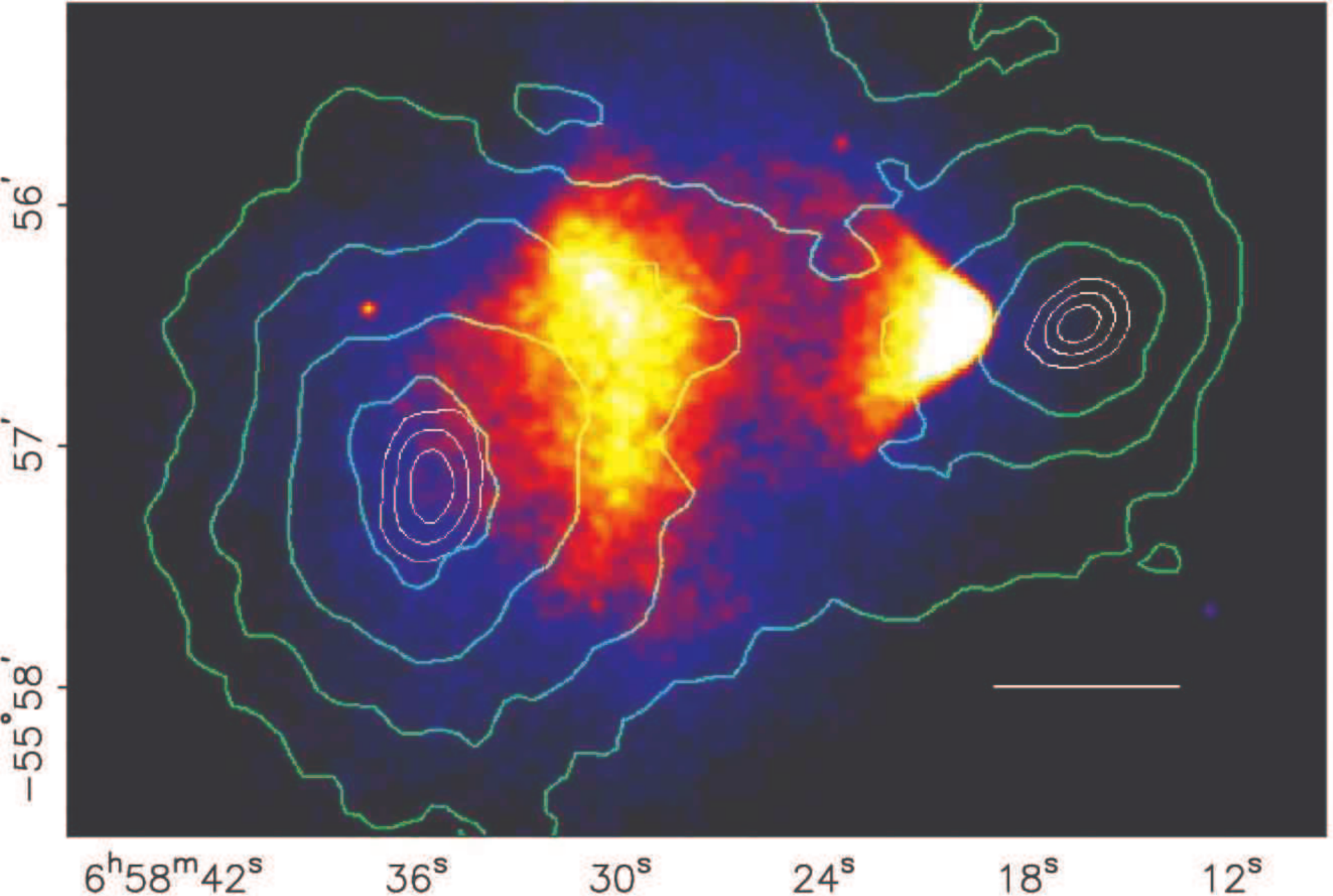} 
  \caption[Matter distribution in X-rays (colours) and using the gravitational lensing (contours) of the Bullet cluster.]{Matter distribution in X-rays (colours) and using the gravitational lensing (contours) of the Bullet cluster. Figure obtained from \cite{Clowe:2006eq}.}
\label{fig:bullet}
\end{center}
\end{figure}

Gravitational lensing gives the most powerful argument in favour of DM at the scale of galaxy clusters. Indeed their gravitational potential can \textit{bend} the light of an object behind them, which gives informations about the total matter distribution in the clusters that can be compared to their gas/dust distribution. The study of the Bullet cluster 1E 0657-56 \cite{Clowe:2006eq} shows the power of this method. This object is actually made of two smaller clusters that are colliding. Tracing the gas distribution and the distribution derived from lensing effects presents in figure~\ref{fig:bullet} a clear separation between them. It shows exactly what we expect from the DM scenario : since DM interacts mainly through gravitation it is less slowed down than luminous matter during the collision inside the cluster.

Finally evidences of DM are also obtained at cosmological scales. All these arguments show that DM interacts essentially through gravitation with its environment. N-body simulations of LSS formations, the direct study of the LSS and the CMB as developed in section~\ref{sec:2.success} tells us that the quantity of baryons in the Universe is not sufficiently high to explain our present Universe as presented in table~\ref{tab:cosmo_param} : the possibility of the presence of MAssive Compact Halo Objects (MACHOs) made of baryonic matter in galaxy halos is then not sufficient to realize DM. The question that follows is the problem of the nature of its constituents. If we look at the relativistic or non-relativistic nature of the DM it turns out that relativistic particles that are candidates to Hot Dark Matter (HDM) fail to match N-body simulation predictions and LSS observations. However some scenarios can fulfill DM constraints. The most reliable scenario that is considered in the standard cosmological model and that will be considered throughout this thesis is CDM. More precisely we will mainly focus on DM made of one specie of weakly interacting, massive, neutral and stable\footnote{At least on cosmological time scales, namely with a greater lifetime than the age of the Universe.} particles, the WIMPs. The exact nature of DM is a major problem of the $\mathbf{\Lambda}$CDM model and we will develop this point later in section~\ref{sec:2.drawback}. Before that to understand the behaviour and to compute the abundance of this type of matter we must look at the interactions between CDM and SM particles in the early and hot Universe.

\subsection{Equilibrium}

First of all, to be able to describe the physical processes in the hot Universe that corresponded to a thermal bath with a temperature $T$, we must determine the thermal distribution of each particle in this bath. The distribution function for a particle $\chi$ in thermal equilibrium at a temperature $T_\chi = T$, with an energy $E_\chi$ and a chemical potential $\mu_\chi$ reads
\beq f_\chi(\vec{p},\vec{x},t) = \frac{1}{\exp\left(\frac{E_\chi-\mu_\chi}{T_\chi}\right)\pm 1}.\eeq
It is defined for the Fermi-Dirac (+) and the Bose-Einstein (-) statistics. Since we start from the cosmological principle and the RW metric we have $f_\chi(\vec{p},\vec{x},t) \equiv f_\chi(|\vec{p}|,t)$. Knowing that each particle $\chi$ has $g_\chi$ internal degrees of freedom it is now possible to define the evolution of the number density $n_\chi$, the energy density $\rho_\chi$ and the pressure $p_\chi$ of particles $\chi$ :
\beq \begin{split}
n_\chi(t) & = \frac{g_\chi}{(2\pi)^3}\int f_\chi(|\vec{p}|,t) d\vec{p}, \\
\rho_\chi(t) & = \frac{g_\chi}{(2\pi)^3}\int E_\chi(|\vec{p}|)f_\chi(|\vec{p}|,t) d\vec{p}, \\
p_\chi(t) & = \frac{g_\chi}{(2\pi)^3}\int \frac{|\vec{p}|^2}{3E_\chi(|\vec{p}|)}f_\chi(|\vec{p}|,t) d\vec{p}.
\end{split} \eeq
The evolution of these particles in the thermal bath is determined using the Boltzmann equation 
\beq {\bf L}[f_\chi] = {\bf C}[f_\chi],\eeq
where ${\bf L}$ is the Liouville operator, that reflects the effects of the metric chosen, and ${\bf C}$ is the collision operator, describing the interactions of $\chi$ with the thermal bath.
In the case we are interested in\ie a massive particle, it is worthwhile to look at the evolution of the number density in the non-relativistic regime. Using the RW metric the Liouville operator then reads
\beq {\bf L}[f_\chi] = E_\chi \frac{\p f_\chi}{\p t} -H|\vec{p}|^2 \frac{\p f_\chi}{\p E_\chi}.\eeq
Integrating over momentum gives
\beq \frac{g_\chi}{(2\pi)^3} \int \frac{{\bf L}[f_\chi]}{E_\chi}d\vec{p} = \dot{n}_\chi + 3Hn_\chi = \frac{1}{a^3}\frac{d}{dt}(n_\chi a^3).\eeq
Without collision we then find an expected dilution with the expansion with $n_\chi \propto a^{-3}$.

More assumptions are needed to compute the collision operator. First, dealing with a simple $i+j \ra k+l$ process with $\chi=i$, we have
\beq \begin{split}
\frac{g_i}{(2\pi)^3} \int \frac{{\bf C}[f_i]}{E_i}d\vec{p} = - \int & d\pi_i d\pi_j d\pi_k d\pi_l (2\pi)^4 \d(\vec{p}_i+\vec{p}_j-\vec{p}_k-\vec{p}_l) \d(E_i+E_j-E_k-E_l) \\
& \sum_{spins} \left[\left|\mathcal{M}_{i+j \ra k+l}\right|^2 f_if_j(1\pm f_k)(1\pm f_l) - (i+j \leftrightarrow
 k+l)\right]
\end{split} \eeq
with $d\pi = \frac{d\vec{p}}{(2\pi)^3 2E}$, $\left|\mathcal{M}_{i+j \ra k+l}\right|$ the matrix element of the interaction with the sum describing the average over initial states spins and the addition of the final states spins and $(i+j \leftrightarrow k+l)$ corresponds to the addition of the same terms but with the exchange between ($i,j$) and ($k,l$).
Then the following assumptions are considered :
\begin{itemize}
\item Statistical factors are neglected since the scattering takes place at an energy $E \gg T$ : the Fermi-Dirac and the Bose-Einstein distributions are replaced by the Maxwell-Bolzmann distribution;
\item CP invariance is assumed, which means that  $\left|\mathcal{M}_{i+j \ra k+l}\right| = \left|\mathcal{M}_{k+l \ra i+j}\right| = \left|\mathcal{M}\right|$;
\item Assuming that $k$ and $l$ are in kinetic and chemical equilibrium leads to $f_kf_l = f_k^{eq}f_l^{eq}$ where \textit{eq} depicts the quantities at the equilibrium;
\item Assuming that $i$ and $j$ stay in kinetic equilibrium even after chemical equilibrium is lost;
\item In equilibrium we have $\frac{d}{dt}(n_i a^3) = 0$, which implies $f_i^{eq}f_j^{eq} = f_k^{eq}f_l^{eq}$.
\end{itemize}
Then, looking at the interaction of $\chi$ and its antiparticle $\bar{\chi}$ with particles form the thermal bath $X$ ($\bar{X}$), and assuming $n_\chi = n_{\bar{\chi}}$, the integrated Boltzmann equation is rewritten as 
\beq \dot{n}_\chi + 3Hn_\chi = \langle \s_{\chi \bar{\chi} \leftrightarrow X \bar{X}}v \rangle (n_\chi^{eq \, 2} - n_\chi^{2}),\eeq
where the thermally averaged cross section times velocity $\langle \s_{ij \leftrightarrow kl}v \rangle$ is given by
\beq \begin{split}
\langle \s_{ij \leftrightarrow kl}v \rangle = \frac{1}{n_i^{eq}n_j^{eq}} \int & d\pi_i d\pi_j d\pi_k d\pi_l (2\pi)^4 \d(\vec{p}_i+\vec{p}_j-\vec{p}_k-\vec{p}_l) \d(E_i+E_j-E_k-E_l)\\
& \left|\mathcal{M}\right|^2 e^{-\frac{E_i+E_j}{T}},
\end{split} \eeq
and the Maxwell-Boltzmann approximation shows that the number density of $\chi$ decreases with the temperature of the thermal bath : 
\beq n_\chi^{eq} = g_\chi \left(\frac{m_\chi T}{2\pi}\right)^{3/2} e^{-m_\chi/T}. \eeq
To get rid of the Hubble parameter a new variable $Y$, the comobile density, is used :
\beq Y = \frac{n_\chi}{s},\eeq
where $s$ is the entropy density $s=h_{\rm{eff}} \frac{2\pi^2 T^3}{45}$ and $h_{\rm{eff}}$ parameterizes the number of relativistic degrees of freedom in the Universe. Then entropy conservation in comoving volume gives $\dot{n}_\chi+3Hn_\chi=s\dot{Y}$ which implies
\beq \dot{Y}=-\langle \s_{\chi \bar{\chi} \leftrightarrow X \bar{X}}v \rangle s\left(Y^{eq \, 2} - Y^2\right). \eeq
Introducing the variable $x = \frac{m_\chi}{T}$ the equation is rewritten as
\beq \frac{d Y}{dx} = \frac{1}{3H}\frac{ds}{dx} \langle \s_{\chi \bar{\chi} \leftrightarrow X \bar{X}}v \rangle \left(Y^{eq \, 2}-Y^2\right).\eeq

\subsection{Freeze-out}
\label{subsec:2.freezeout}

When the temperature becomes negligible comparing to the mass the particle $\chi$, its comobile density decreases ($X$ and $\bar{X}$ cannot anymore annihilate into $\chi$ and $\bar{\chi}$ : $\langle \s_{\chi \bar{\chi} \leftrightarrow X \bar{X}}v \rangle = \langle \s_{ann}v \rangle$) but at a given temperature its interaction rate with the thermal bath drops below the expansion rate of the Universe : $\chi$ is decoupled from the thermal bath, it is the \textit{freeze-out} of $\chi$. Its comobile density becomes approximately constant from the decoupling to today an after some assumptions like a constant averaged cross section the current $\chi$ yield reads
\beq Y_0 \approx \sqrt{\frac{45 G_N}{\pi g_*}} \frac{x_F}{m_\chi} \frac{1}{\langle \s_{ann}v \rangle},\eeq
where $g_*$ is another parameterization of the number of degrees of freedom and the subscript \textit{F} denotes quantities at the time of the $\chi$ freeze-out. The present density of the relic $\chi$ is therefore
\beq \Om_{\chi,0} h^2 = \frac{\rho_{\chi,0}h^2}{\rho_c} = \frac{m_\chi n_{\chi,0}h^2}{\rho_c} = \frac{m_\chi s_0 Y_0 h^2}{\rho_c}.\eeq

For a WIMP mass at the weak scale, namely for hundreds of GeV, $\langle \s_{ann}v \rangle \approx 10^{-26} \mathrm{cm}^3 \, \mathrm{s}^{-1}$. Using an approximation of the relic density :
\beq \Om_{\chi,0} h^2 = \frac{3\times 10^{-27} \mathrm{cm}^3 \, \mathrm{s}^{-1}}{\langle \s_{ann}v \rangle},\eeq
it is clear that a WIMP could get the expected DM relic density shown in the table~\ref{tab:cosmo_param} : this is called the \textit{WIMP miracle}. 

\subsection{Precise calculation}

The calculation of $\langle \s_{ann}v \rangle$ could be approximated by an expansion in $v$, but this choice is not accurate in several cases : 
\begin{itemize}
\item Annihilation of $\chi$ through an s-channel resonance in which the propagator $\phi$ is characterized by $2m_\chi \approx m_\phi$;
\item Close to the threshold of the annihilation channel $\chi \bar{\chi} \rightarrow \phi \bar{\phi}$ with $m_\chi \approx m_\phi$;
\item Coannihilation with another particle $\chi'$ which could improve the annihilation rate through processes like $\chi \chi' \rightarrow X X'$ or $\chi' \bar{\chi'} \leftrightarrow X \bar{X}$.
\end{itemize}
The detailed calculation of the relic density and the discussion of these cases can be found in \cite{Gondolo:1990dk,Griest:1990kh}. 

In the case of coannihilation, an important point in some of the analyses that we will develop in part~\ref{part2}, let us consider $N$ particles $\chi_i$ ($i=1,\ldots,N$) with masses $m_{\chi_{_i}}$ ordered like $m_{\chi_{_1}} \leq m_{\chi_{_2}} \leq \cdots \leq m_{\chi_{_{N-1}}} \leq m_{\chi_{_N}}$, where the lightest particle is assumed stable, and with the internal degrees of freedom $g_i$. The thermal averaged cross section was defined by Edsj{\"o} and Gondolo \cite{Edsjo:1997bg} as
\beq \label{eq:sigmavefffin2}
  \langle \s_{ann}v \rangle = \frac{\int_0^\infty
  dp_{\rm{eff}} p_{\rm{eff}}^2 W_{\rm{eff}} K_1 \left( \frac{\sqrt{s}}{T} \right)}
  { m_{\chi_{_1}}^4 T \left[ \sum_i \frac{g_i}{g_1} \frac{m_{\chi_{_i}}^2}{m_{\chi_{_1}}^2} K_2 \left(\frac{m_{\chi_{_i}}}{T}\right) \right]^2}.
\eeq
where $K_{i}$ are the modified Bessel functions of the second kind and of order $i$, $p_{\rm{eff}} = p_{11}$ and $s = 4(p_{\rm{eff}}^2 + m_{\chi_{_1}})$. The quantity $W_{\rm{eff}} $ is defined as
\beq \label{eq:weff}
  W_{\rm{eff}} = \sum_{ij}\frac{p_{ij}}{p_{11}}
  \frac{g_ig_j}{g_1^2} W_{ij},
\eeq
where $ W_{ij} = 4 E_{i} E_{j} \sigma_{ij} v_{ij}$ and $p_{ij}$ is the momentum of the particle $\chi_i$ (or $\chi_j$) in the center-of-mass frame of the pair ($\chi_i,\chi_j$). Note that
\beq \sigma_{ij} = \sum_{X,\bar{X}} \sigma (\chi_i \chi_j \rightarrow X \bar{X}) \eeq
is the total annihilation rate for $\chi_i \chi_j$ annihilations into the particles $X$ ($\bar{X}$) and the relative particle velocity $v_{ij}$ reads 
\beq v_{ij} = \frac{\sqrt{(\vec{p_i} \cdot \vec{p_j})^2-m_{\chi_{_i}}^2 m_{\chi_{_j}}^2}}{E_i E_j}. \eeq

All these details and special cases are now taken into account in numerical codes like \DSUSY \cite{Gondolo:2004sc}, \SisoR \cite{Arbey:2009gu,Arbey:2011zz} and \micro \cite{Belanger:2001fz,Belanger:2004yn,Belanger:2006is,Belanger:2008sj,Belanger:2010gh,Belanger:2013oya}. The last one was used throughout this thesis and all results were based on it.

\section{Cosmic inflation}
\label{sec:2.inflation}

The $\mathbf{\Lambda}$CDM model faces several issues regarding the understanding of the physics of the very early Universe. Some of them can be solved introducing a phase of extremely rapid expansion of the Universe, the inflation area. 

\subsection{Cosmological puzzles}
\label{sec:2.inflation_issues}

The small inhomogeneities observed in the CMB grew with time to give the current structures of the Universe. One would expect that they were even smaller before the CMB formation. The problem that arises here is linked to the notion of causal connexion. Two photons separated by a distance greater than their comoving horizon $d_H(t)$, defined as
\beq d_H(t) = 2 a(t) \int_{t_i}^t \frac{dt'}{a(t')},\eeq
where $t_i$ corresponds to the time of the initial singularity in the Big Bang model, cannot exchange information. This quantity increases with time, which means that it was very small at the time of CMB formation. While all regions of the CMB seem to be causally linked, the size of the CMB horizon was determined to be only of the order of 1°. This causality problem is called the horizon problem.

The evolution of the curvature density in time reads
\beq \label{eq:flatness} |\Om_k(t)| = \left|\frac{k}{a(t)^2H(t)^2}\right| \propto \dot{a}(t)^{-2}.\eeq
Since radiation dominated and matter dominated Universe evolve respectively like $t^{1/2}$ and $t^{2/3}$, it follows that the flatness of the Universe reduce with time. However the present curvature density was determined to be very small, as shown in table~\ref{tab:cosmo_param}. Calculating $|\Om_k(t)|$ in the very early Universe and especially at the Planck time $t_{Pl} \approx 5.391 \times 10^{-44} $ s gives an upper bound on $|\Om_k(t_{Pl})|$ of the order of $10^{-60}$. This fine-tuning issue of the curvature density is called the flatness problem.

The last puzzle we present here is the monopole problem. One important issue of GUT is that the breaking of the associated symmetry into the SM gauge symmetry could predict the formation of topological defects in the early Universe, the monopoles. These objects have non-standard features : a mass of the order of the GUT scale and a magnetic charge. We still not have detected such objects in the Universe and it is therefore expected that their number density is extremely low. The problem is that if no monopole annihilation occurs, the calculations show that one monopole is expected per nucleon which is clearly in disagreement with the observations.

\subsection{Inflationary Universe}

The standard solution of the issues raised in section~\ref{sec:2.inflation_issues} is the introduction of a simple scalar field, the inflaton $\phi$ with its potential $V(\phi)$, whose energy density and pressure are defined as
\beq \rho_\phi = \frac12 \dot{\phi}^2 + V(\phi), \quad p_\phi = \frac12 \dot{\phi}^2 - V(\phi),\eeq
that fills the very early Universe. The FL equations yield the useful expressions for the Hubble parameter :
\beq \begin{split}
H^2 & = \frac{8\pi G_N}{3}\left(\frac{1}{2}\dot{\phi}^2 + V(\phi)\right), \\
\dot{H} & = -4\pi G_N \dot{\phi}^2.
\end{split} \eeq
Motivated by the condition of a strong acceleration of the expansion of the Universe, we must have
\beq |\dot{H}| \ll H^2,\eeq  
which leads to
\beq \dot{\phi}^2 \ll V(\phi).\eeq
The necessary requirement to have a viable inflationary model is thus that $V(\phi)$ has to be large but almost flat to allow the slow-roll motion of the inflaton field, which implies
\beq |\ddot{\phi}| \ll H |\dot{\phi}|.\eeq  
These expressions are called the \textit{slow roll} conditions. There are often rewritten in terms of the slow roll parameters
\beq \label{eq.2:inflaparams} \e_\phi = \frac{1}{16\pi G_N}\left(\frac{V'(\phi)}{V(\phi)}\right)^2, \quad \eta_\phi = \frac{1}{8\pi G_N}\frac{V''(\phi)}{V(\phi)},\eeq
where $V'(\phi) = \p V(\phi)/\p\phi$ and with the condition $\{\e_\phi,\eta_\phi\} \ll 1$ during inflation. At the end of the inflation area\ie when $\{\e_\phi,\eta_\phi\} \ra 1$, the inflaton field decays into relativistic particles and especially into photons : the temperature of the Universe increases until a maximum called the reheating temperature $T_R$. 

The flatness problem is then easily solved : a positive acceleration gives an increasing scale factor which leads, using eq.~\ref{eq:flatness}, to an increase of the flatness of the Universe.
As it was precised, the inflaton fills the entire Universe and then the exponential expansion takes place : the regions of the Universe that were causally related can therefore no longer be. Since the decay of the inflaton occurs identically in all regions of the Universe the horizon problem is solved.
Finally note that with this rapid expansion, topological defects are diluted which erases the monopole problem. Note that the energy scale of the inflaton is of the order of the GUT scale. 

\subsection{Cosmological perturbations and constraints}

\begin{figure}[!htb]
\begin{center}
\includegraphics[width=9cm,height=5cm]{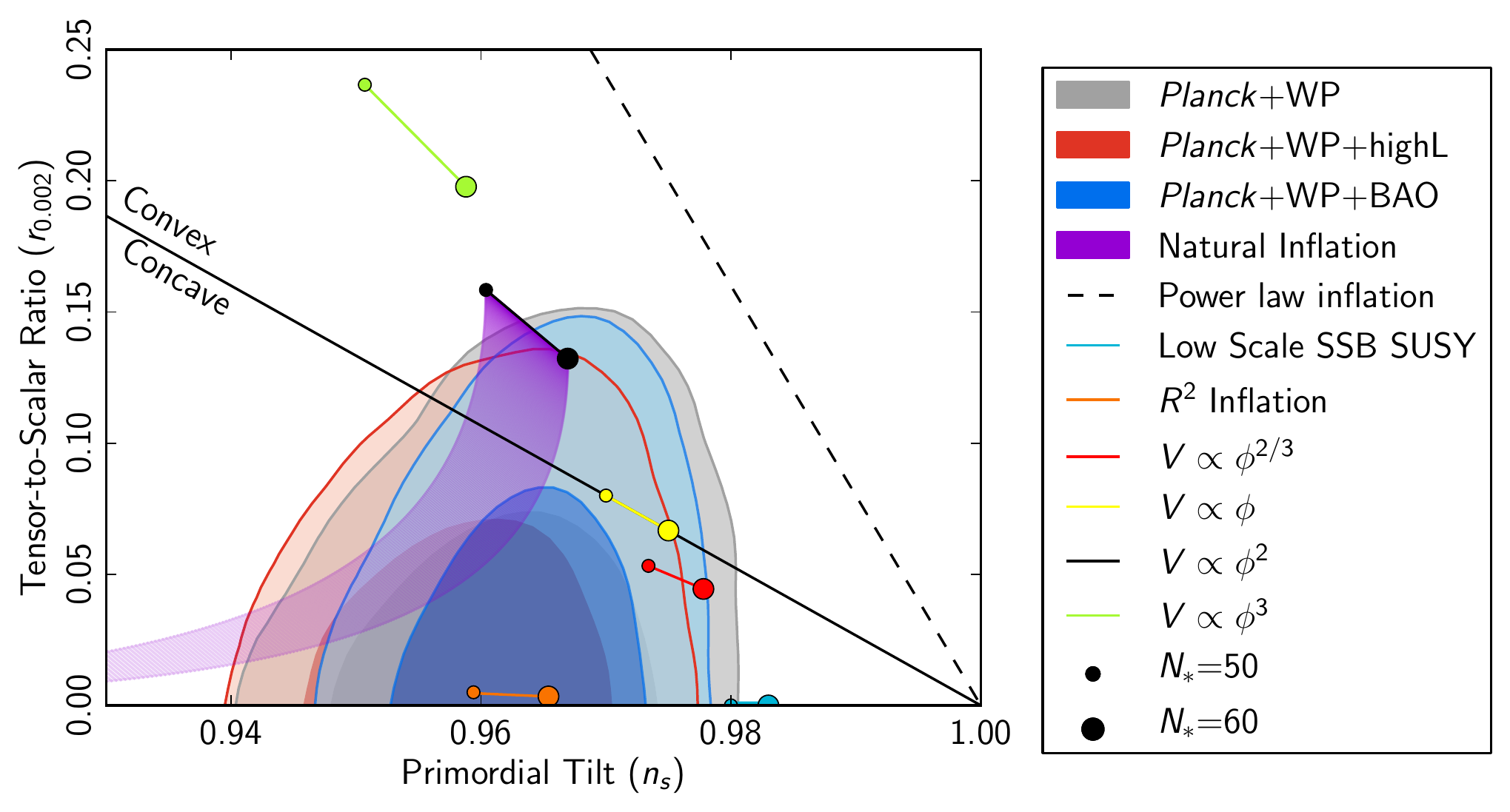} 
  \caption[Constraints on inflationary models in the plane (tensor-to-scalar ratio, scalar spectral index).]{Constraints on inflationary models in the plane (tensor-to-scalar ratio, scalar spectral index). Figure taken from \cite{Ade:2013uln}.}
\label{fig:inflation}
\end{center}
\end{figure}

We know that the CMB is not completely homogeneous and isotropic. The small perturbations measured give a lot of informations about the evolution of the Universe till today but it also brings us to the question of the origin of these small perturbations. Inflation models predict that the current structures of the Universe stem from initial quantum fluctuations created during the inflation area. By looking at simple scalar fluctuations\footnote{Inflation scenarios also predict tensor fluctuations in the RW metric. Their amplitude is parameterized by the tensor-to-scalar ratio whose Planck+WP+highL+BAO combination gives an upper bound of 0.111 at 2$\s$.}, the spectrum of the perturbations is fully determined by only two parameters : its amplitude $\d_H$ and the scalar spectral index $n_s$ which defines the scale-dependence of the power spectrum. Inflation models have the specificity to predict slight deviations from the scale invariance case $n_s=1$. The spectral index obtained by the Planck collaboration and shown in table~\ref{tab:cosmo_param} shows a more than 5$\s$ deviation from the scale invariance which is a strong support to inflationary models as shown in figure~\ref{fig:inflation}.   
The expansion factor of the Universe during the inflation is defined by the number of $e$-folds from the time when the observationally relevant perturbations are generated (whose scale factor is $a_i$) to the end of inflation $a_f$ and reads
\beq N_* = \ln\frac{a_f}{a_i} = \int_{t_i}^{t_f} H dt.\eeq
Depending on the model of inflation chosen we have $N_* \sim 40 - 60$.\\
For more informations on inflationary models see \cite{Baumann:2009ds,Martin:2013tda}.

\section{Thermal history of the Universe in the $\mathbf{\Lambda}$CDM model}

As explained in section~\ref{sec:1.SMpbs}, our description of the early Universe can at most go back to the physics of the Planck epoch, namely when the gravitational interaction becomes non-negligible comparing to the other interactions. Starting from this point the $\mathbf{\Lambda}$CDM model is characterized by the following steps :

\begin{itemize} 
\item When the temperature of the Universe was around $T \simeq 10^{16}$~GeV, an yet unknown symmetry associated to a GUT breaks into the SM gauge symmetry. This period also corresponds to the end of the accelerated expansion of the Universe with the decay of the inflaton field into SM particles. Depending on the inflationary model chosen, this happens before or after GUT breaking. 

\item At $T \sim 10^{2}$~GeV, the EW symmetry breaking takes place : the SM gauge symmetry breaks into $SU(3)_c \otimes U(1)_\mathcal{Q}$. The baryon asymmetry in the Universe is supposed to appear around this energy or even before. The mechanism that introduce this asymmetry, the baryogenesis, has to fulfill the following conditions given by Sakharov \cite{Sakharov:1967dj} : violation of the baryonic number, of C and of CP and the interactions must occur out of thermal equilibrium to ensure that the opposite reaction do not compensate those introducing the asymmetry.

\item At $T \sim 10^1 - 10^3$~GeV, the most interesting CDM candidate, the WIMP, is decoupled from the thermal bath filling the Universe, it is the DM freeze-out.

\item At $T \sim 0.3$~GeV, QCD phase transition, namely the confinement of quarks and gluons into hadrons, occurs : we move from a quark-gluon plasma to an hadron plasma. This phase is important in the DM relic density computation since it affects the calculation of the effective number of relativistic degrees of freedom at the freeze-out temperature as shown in \cite{Hindmarsh:2005ix}. However since the DM freeze-out temperature $T_F$ is of the order of $m_\chi/20$ these effects are mainly relevant for light DM with $m_\chi \lesssim $~7 GeV which is mainly ignored in this thesis.   
 
\item At $T \sim 1$~MeV, neutrinos decouple from the thermal bath : the neutrino freeze-out occurs. When $T$ falls below the mass of the electron, photons have not sufficient energy to annihilate into $e^+ e^-$ pairs. Baryogenesis provides enough excess of electrons to account for the current electron density observed.

\item At $T \sim 100$~keV, photons energy becomes too small to prevent formation of light atoms : it is the standard Big Bang Nucleosynthesis (BBN), which gives stringent constraints on the $\mathbf{\Lambda}$CDM model, as the determination of the Helium abundance in the Universe, which strengthens the CMB argument of a low current baryon density \cite{Iocco:2008va}.

\item At $T \sim 1$~eV, the Universe is now dominated by matter instead of radiation because of the faster radiation dilution mentioned in section~\ref{sec:2.th}. Large scale structures start to form.

\item At $T \sim 0.4$~eV, the photon decoupling from the matter, which gives the CMB, happens.

\item At $T \sim 10^{-3}$~eV, DE becomes the main component of the Universe.

\item Finally the current Universe is at $T \simeq 2.72548$~K $\sim 2 \times 10^{-4}$~eV with approximately 68.3\% of DE, 26.8\% of DM and 4.9\% of standard matter, mainly baryonic.
\end{itemize}

\section{$\mathbf{\Lambda}$CDM drawbacks}
\label{sec:2.drawback}

The main theoretical issue of the cosmological standard model is related to the cosmological constant. Cosmological observations implies an energy density related to the cosmological constant of the order of $10^{-29} \, \mathrm{g} \, \mathrm{cm}^{-3}$. Meanwhile, the total energy density coming from vacuum fluctuations is approximately 120 orders of magnitute higher when assuming a cut-off beyond which quantum field theory as to be modified, commonly the Planck scale \cite{Carroll:2000fy}. This fundamental theoretical problem is still open \cite{Martin:2012bt}.

As said in section~\ref{sec:1.SMpbs} CP violation processes in the SM are not sufficient to account for the baryogenesis which then needs NP. 

Some deeper motivations could be looked for in the case of the inflationary postulate, like the origin and the construction of the inflaton field which could be given in extension of the SM. See \cite{Martin:2013tda} for a review on inflationary scenarios. 

Neutrinos could be viable DM candidates since we saw in chapter~\ref{chap:revpart} that they interact weakly with the other SM particles. Nevertheless the neutrino relic density is given by 
\beq \Om_\nu h^2 = \sum_{i=e,\mu,\tau} \frac{m_{\nu_i}}{93 \, \mathrm{eV}},\eeq
which, using limits on neutrino masses like the ones obtained with the Planck satellite \cite{Ade:2013lta}, exclude neutrinos as the main component of DM. Moreover, since neutrinos are relativistic particles, they only could form HDM which is disfavoured as explained in section~\ref{sec:2.DM}. Thus no SM particles fit the DM hypothesis. 

One other issue facing the CDM idea is that it predicts more small scale structures, like dwarf galaxies, than observed \cite{Klypin:1999uc}. Alternatives could be considered like an intermediate situation between HDM and CDM : Warm Dark Matter (WDM) \cite{deVega:2011xh,deVega:2013ysa}. Other conflicts between some $\mathbf{\Lambda}$CDM predictions and astrophysical measurements are noted \cite{Famaey:2011kh}. 

$\mathbf{\Lambda}$CDM is not the only model able to give right predictions. For instance the MOdified Newtonian Dynamics (MOND) theory proposed by Milgrom in 1983 \cite{Milgrom:1983ca} is able to predict the right rotation curves of the galaxies, but it has several drawbacks. The most notable one is that, even within its relativistic extension called Tensor-Vector-Scalar gravity (TeVeS), this model is not able to reproduce the correct CMB power spectrum and in particular the third acoustic peak of the CMB \cite{Skordis:2009bf}.

\section{Some solutions to the $\mathbf{\Lambda}$CDM and SM issues}
\label{sec.3:sol}

Several proposals offer the possibility to solve drawbacks of the two standard models presented in chapters \ref{chap:revpart} and \ref{chap:revcosmo}. The main BSM theory studied in the literature as well as the main topic of this thesis, \textit{Supersymmetry} (SUSY), which is able to address these questions, will be developed in chapter~\ref{chapter:SUSY}. 

To solve the hierarchy problem of the Higgs boson mass, some models propose that the fundamental Planck scale could be much lower than expected, and especially at the TeV scale. This is done introducing extra spatial dimensions whose volume enters in the calculation of the Planck scale \cite{ArkaniHamed:1998rs,Randall:1999ee}. This type of model implies new particles, Kaluza-Klein (KK) excitations of SM particles, among which DM candidates can be found. As an example, in the Universal Extra Dimensions (UED) model \cite{Appelquist:2000nn} the introduction of a new parity, called KK parity, makes the lightest KK particle stable. Thus, if it weakly interacts with SM particles, it is able to fit the DM constraints \cite{Servant:2002aq,Belanger:2010yx}. Moreover these models give motivation for the number of families \cite{Dobrescu:2001ae}.

The strong CP problem linked to the $\t_{\mathrm{QCD}}$ fine-tuning can be solved introducing a global $U(1)$ symmetry, the Peccei-Quinn (PQ) symmetry \cite{Peccei:1977hh,Peccei:1977ur}. This symmetry preserves CP invariance raising the QCD parameter as a dynamic field. The $U(1)_{\mathrm{PQ}}$ symmetry breaking predicts the existence of a light pseudoscalar\footnote{A scalar that changes sign under P.} boson, the axion \cite{Weinberg:1977ma,Wilczek:1977pj}. Astrophysical constraints give an upper bound on its mass of approximately 16 meV \cite{Raffelt:2006cw,Beringer:1900zz}. It is supposed that axions interact very weakly with SM particles, which implies that they can be viable non-thermal CDM candidates even though they must be very light.

Several solutions involve new neutrino fields and especially in GUT as we will see in part \ref{part3}. Adding massive RH Majorana neutrinos gives a means to explain the LH neutrino masses through the \textit{seesaw mechanism} \cite{Minkowski:1977sc}. This could imply a leptonic asymmetry which possibly gives baryogenesis \cite{Fukugita:1986hr}. As another example, sterile neutrinos at the keV scale are also interesting WDM candidates \cite{Dodelson:1993je,deVega:2011xh,deVega:2013ysa}. A more complete list of DM candidates can be found in \cite{Bertone:2004pz}.

\chapter{Supersymmetry}
\label{chapter:SUSY}

\setcounter{minitocdepth}{3} %Numbered subsubsections in TOC
\minitoc\vspace{1cm}
\newpage

\section{SUSY responses to SM problems}

Before looking at its theoretical construction, we briefly summarize the three main points that motivate the phenomenological study of SUSY.

The main theoretical problem that SUSY is able to solve is the hierarchy problem. First of all it is important to note that every particle gets quantum corrections to its mass. For instance the self-energy of particles like the photon or the electron could give important corrections to their mass which are known to be very small, especially for the massless photon. The crucial point here is that these masses are protected by symmetries. QED symmetry is exact and if the amplitude of the QED process, here the photon self-energy, is defined as $\mathcal{M} = \e_\mu(k) \mathcal{M}^\mu(k)$ where $\e_\mu(k)$ is the polarization vector of the photon and $k$ the momentum then the $U(1)_{em}$ conservation gives the Ward identity
\beq k_\mu \mathcal{M}^\mu(k) = 0,\eeq
so that all QED corrections to the photon mass vanish. In the case of the electron it is the chiral symmetry that stabilizes the mass of this fermion. However since the electron is not massless chiral symmetry is not exact. Fortunately the correction remains negligible : it is logarithmic and proportional to the measured electron mass. In the case of the Higgs boson of the SM there is no symmetry that protects its mass as shown in chapter~\ref{chap:revpart}. SUSY, a symmetry that links fermions and bosons, can play such a role. New scalars added by SUSY give new loop corrections to the Higgs boson mass. If $\l_s$ parameterizes the coupling of a scalar with a mass $m_s$ to the Higgs boson the new contribution to the Higgs boson mass reads
\beq \d m^2_{h^0} = \frac{\l_s}{16 \pi^2} \left( \Lambda^2 -2m_s^2 \ln\frac{\Lambda}{m_s} + ...\right).\eeq
Bringing together this correction and the SM one in eq.~\ref{eq:1.higcor}, assuming that each SM fermion is associated with two supersymmetric complex scalars, and that $y_f^2 = \l_s$ it follows that the quadratic divergences cancel each other. The remaining correction is logarithmic and proportional to the mass difference between the SM fermions and the new scalars. Providing that SUSY is exact (equal mass of fermion and scalar partners) or at least that the mass degeneracy between these fermions and scalars particles is below the TeV scale the EW scale is stabilized.

Another argument in favour of SUSY is the unification of coupling constants at the GUT scale. The evolution of the coupling constants with the energy scale is given by Renormalization Group Equations (RGEs). At one-loop they are defined as
\beq \frac{d g_i}{dt} = \frac{1}{16 \pi^2} b_i g_i^3, \quad i \in \{1,2,3\},\eeq
where $t=\ln(Q/Q_0)$, $Q$ being the energy scale considered and $Q_0$ the GUT scale. $g_1 = \sqrt{5/3}g_Y$ and the other coupling constants are given in section~\ref{sec:1.gauge}. Depending on the content of the model,\eg number of generations, number of Higgs doublets or number of scalars, the $b_i$ coefficients can be drastically modified. As we will see in section~\ref{subsubsec:1.chiral} the Minimal Supersymmetric Standard Model (MSSM) studied in this thesis needs two Higgs doublets. Moreover, as mentioned above, SUSY contains more particles than the SM. As a result the coefficients $b_i$ differ :
\beq \begin{split}
\mathrm{SM : \ } (b_1,b_2,b_3) & = (41/10, -19/6, -7), \\
\mathrm{MSSM :\ } (b_1,b_2,b_3) & = (33/5, 1, -3). 
\end{split} \eeq
\begin{figure}[!htb]
\begin{center}
\includegraphics[width=6cm,height=5cm]{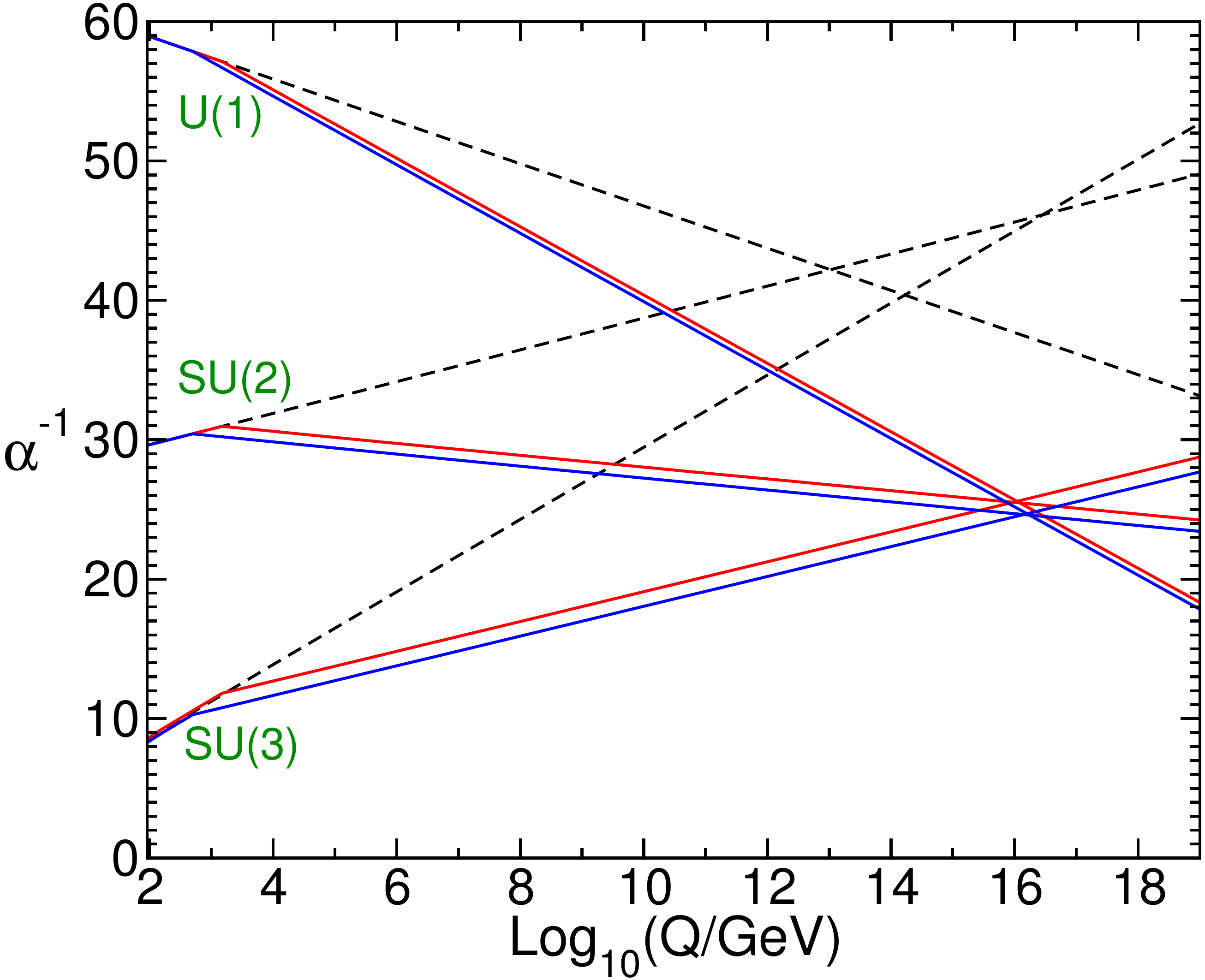} 
  \caption[Two-loop renormalization group evolution of $\a_i^{-1}$  in the SM (dashed lines) and the MSSM (solid lines).]{Two-loop renormalization group evolution of $\a_i^{-1}$  in the SM (dashed lines) and the MSSM (solid lines). Figure obtained from \cite{Martin:1997ns}.}
\label{fig:unification}
\end{center}
\end{figure}

Defining $\a_i = g_i^2/4\pi$, figure~\ref{fig:unification} compares the evolution of coupling constants in the SM and the MSSM and shows the possibility of unifying the three interactions in the context of SUSY.

SUSY is also able to link particle physics and cosmology through the possibility of getting supersymmetric DM candidates. We will note in section~\ref{subsec:3.MSSM} that an exact discrete $\mathbb{Z}_2$ symmetry of SUSY, called $R$-parity, gives a stable Lightest Supersymmetric Particle (LSP) which could, if it weakly interacts with the other particles, be a WIMP. For a review on supersymmetric DM candidates see \cite{Jungman:1995df}. Note that SUSY could also gives inflaton candidates, as we will show in chapter~\ref{chapter:NUHM2}. 

There are also \textit{\ae{}sthetic} motivations for SUSY as in the search of extensions of the Poincar\'e algebra which is presented in next section. If SUSY is formulated as a local symmetry a spin 2 particle corresponding to the graviton, the hypothetical particle that mediates gravity, is introduced. Then the supersymmetric models of gravity called supergravity have the elegant feature to link the SM fundamental interactions with gravity \cite{VanNieuwenhuizen:1981ae}.

\section{Elements on the theoretical construction of exact SUSY}

The Poincar\'e group contains all the symmetries of SR : in addition to space-time tranformations that characterized the Lorentz group\ie the rotations and the boosts encoded in the generator $M_{\mu \nu}$, it includes the translations represented by the generator $P_\mu$. The Poincar\'e algebra, which is a Lie algebra, is defined by the following commutation relations :
\beq \begin{split}
[P_\mu,P_\nu] & = 0, \\
[M_{\mu \nu},P_\rho] & = i (\eta_{\nu \rho} P_\mu - \eta_{\mu \rho} P_\nu), \\
[M_{\mu \nu},M_{\rho \s}] & = i (\eta_{\mu \s} M_{\nu \rho} - \eta_{\mu \rho} M_{\nu \s} + \eta_{\nu \rho} M_{\mu \s} - \eta_{\nu \s} M_{\mu \rho}).
\end{split} \eeq
In 1967, Coleman and Mandula \cite{Coleman:1967ad} demonstrated a \textit{no-go theorem} : there is no non-trivial extension of the Poincar\'e algebra by an usual Lie algebra, namely defined only by commutation relations. The search of possible extensions leads to the need to include a more general algebra that also contains anticommutation relations. A graded Lie algebra called super-Poincaré algebra was then shown by Haag, Lopuszanski and Sohnius to be the most general extension of the Poincar\'e algebra \cite{Haag:1974qh}.

\subsection{Super-Poincaré algebra}

This superalgebra is characterized by fermionic generators. An interesting implication is the definition of a supersymmetric tranformation : these operators transfrom a fermion into a boson and vice versa. Its most simple version that we consider in this thesis contains only $\mathcal{N} = 1$ pair of fermionic generators $\mathscr{Q}$
\beq \mathscr{Q} \equiv \binom{\Qa}{\Qda}, \quad \{\a,\dot{\a}\} \in \{1,2\},\eeq
where $\Qa$ and $\Qda$ correspond to two-component spinors, the Weyl spinors. This version is the most phenomenologically interesting one since it is the only case which allows to define properly chiral fermions. Actually the supermultiplets\footnote{Irreductible representations of the superalgebra.} of the supersymmetric versions with $\mathcal{N} > 1$ can contain both left and right chiralities with the same gauge tranformations, which is clearly not observed experimentally especially in the case of the weak interaction. In addition to the commutation relations of the Poincar\'e algebra, this superalgebra adds the following commutation and anticommutation relations :
\begin{align}
\{\Qa, \Qdb \} & = 2(\s^\mu)_{\a \dot{\b}} P_\mu, \label{eq:susycommut1} \\
\{\Qa, \Qb \} & = \{ \Qda, \Qdb \} = 0, \label{eq:susycommut2} \\
[ P_\mu, \Qa ] & = [ P_\mu, \Qda ] = 0, \label{eq:susycommut3} \\
[ M_{\mu \nu} , \Qa] & = i (\s_{\mu \nu})^\b_\a \Qb, \label{eq:susycommut4} \\
[ M_{\mu \nu} , \Qda] & = i (\bar{\s}_{\mu \nu})^{\dot{\a}}_{\dot{\b}} \Qdb, \label{eq:susycommut5} \\
[ \Qa, R ] & = \Qa, \label{eq:susycommut6} \\
[ \bar{Q}_{\dot{\a}}, R ] & = -\bar{Q}_{\dot{\a}}. \label{eq:susycommut7} 
\end{align}
The $\s$ matrices are obtained using the relations :
\beq \begin{split}
(\s^{\mu \nu})^\b_\a & = \frac14 \left[(\s^\mu)_{\a \dot{\g}} (\bar{\s}^\nu)^{\dot{\g} \b} - (\s^\nu)_{\a \dot{\g}} (\bar{\s}^\mu)^{\dot{\g} \b}\right], \\
(\bar{\s}^{\mu \nu})^{\dot{\a}}_{\dot{\b}} & = \frac14 \left[(\bar{\s}^\mu)^{\a \dot{\g}} (\s^\nu)_{\dot{\g} \b} - (\bar{\s}^\nu)^{\a \dot{\g}} (\s^\mu)_{\dot{\g} \b}\right], \\
(\s^\mu)_{\dot{\a} \a} & = (\mathbb{I}_2,\s^i)_{\dot{\a} \a}, \\
(\bar{\s}^\mu)^{\dot{\a} \a} & = \e^{\dot{\a}\dot{\b}} \e^{\a \b}(\s^\mu)_{\b \dot{\b}} = (\mathbb{I}_2,-\s^i)^{\dot{\a} \a},
\end{split} \eeq
where $\mathbb{I}_2$ is the $2 \times 2$ identity matrix, the $\s^i, i \in \{1,2,3\}$ are the Pauli matrices and the antisymmetric tensors $\e$ are
\beq \e^{\a \b} = \e^{\dot{\a}\dot{\b}} =  \begin{pmatrix} 0 & 1 \\ -1 & 0 \end{pmatrix}, \quad
     \e_{\a \b} = \e_{\dot{\a}\dot{\b}} =  \begin{pmatrix} 0 & -1 \\ 1 & 0 \end{pmatrix}. \eeq
Here $R$ is the generator of a global $U(1)$ symmetry of SUSY called $R$-symmetry. Without going into the details it could be considered as a kind of generalization of the chiral symmetry. The Nelson-Seiberg theorem \cite{Nelson:1993nf} shows that its presence is a necessary condition for SUSY breaking. This $R$-symmetry could also be an explanation of the $R$-parity. SUSY breaking and $R$-parity will be developed respectively in sections \ref{subsec:3.SSB} and \ref{subsec:3.MSSM}. 

If we look at eq.~\ref{eq:susycommut3} it results that fermionic generators commute with the mass operator $P^2 = P_\mu P^\mu$. Therefore in each supermultiplet we find particles with the same mass. As mentioned before they also have same gauge transformations which implies same electric and colour charges. However eqs. \ref{eq:susycommut4} and \ref{eq:susycommut5} imply that the supermultiplets are made of particles with different spin. Another interesting characteristic of these supermultiplets is that they are constituted by an equal number of fermionic and bosonic degrees of freedom. It is then tempting to put in a same multiplet bosons and fermions of the SM which have the same electric charge, like a set with the photon and a neutrino and another with the $W$ boson and the electron \cite{Fayet:1974pd}. Of course this needs a SUSY breaking mechanism to explain the mass difference between particles of a same multiplet. Nevertheless the problem of assigning coloured particles in supersymmetric multiplets remains. This led Fayet to define in each supermultiplet a SM particle characterized either by a bosonic or a fermionic statistic and its associated \textit{superpartner} with the other statistic \cite{Fayet:1976et,Fayet:1977yc}. $\mathcal{N} = 1$ SUSY is described by two types of supermultiplets : chiral and gauge supermultiplets. A nice way to define them is going to the \textit{superspace} and the \textit{superfield} formalisms. Here we decide not to present these concepts and we refer the reader to the textbooks \cite{Weinberg:2000cr,Baer:2006rs,Ryder:1985wq,Gates:1983nr} and the reviews \cite{Bilal:2001nv,Sohnius:1985qm,Quevedo:2010ui}. The supersymmetric Lagrangian will now be introduced. For details on the demonstration of the relations that will be presented, see \cite{Martin:1997ns}.

\subsection{Chiral supermultiplet}
\label{subsubsec:1.chiral}
    
The fermions and the scalar of SM are defined using the simplest supermultiplet of the $\mathcal{N} = 1$ SUSY called chiral supermultiplet. It consists of one LH Weyl spinor\footnote{For simplicity we now omit Weyl indices.} $\psi$ and one complex scalar field $\phi$. If the spinor refers to a SM fermion, $\phi$ is called a \textit{scalar fermion} or more succinctly a \textit{sfermion} whereas the supersymmetric partner of the Higgs boson is an \textit{higgsino}. The simplest model constructed with this supermultiplet containing massless and non-interacting particles is the Wess-Zumino model \cite{Wess:1974tw}. Its Lagrangian reads
\beq \mathscr{L}_{WZ} = -\p^\mu \phi^* \p_\mu \phi + i\bar{\psi} \bar{\s}^\mu \p_\mu \psi + F^* F, \eeq
where $F$ is a complex scalar \textit{auxiliary} field without kinetic term necessary for $\mathscr{L}_{WZ}$ to be invariant under SUSY even if the equation of motion $\bar{\s}^\mu \p_\mu \psi = 0$ is not satisfied, namely for off-shell particles. The corresponding SUSY transformations are parameterized by an infinitesimal and anticommuting Weyl fermion object $\e$. Here we confine ourselves to global SUSY; it results that $\e$ does not depend on spacetime coordinates\ie $\p_\mu \e = 0$. The transformations for each fields are given by the following relations :
\beq \begin{split}
\d \phi & = \e \psi, \qquad \qquad \qquad \, \, \, \d \phi^* = \bar{\e} \bar{\psi}, \\
\d \psi & = -i \s^\mu \bar{\e} \p_\mu \phi + \e F, \quad \d \bar{\psi} = i \e \s^\mu \p_\mu \phi^* + \bar{\e} F^*, \\
\d F & = -i \bar{\e} \bar{\s}^\mu \p_\mu \psi, \qquad \quad \, \d F = i \p_\mu \bar{\psi} \bar{\s}^\mu \e.
\end{split} \eeq
If several chiral supermultiplets defined by the set $\Psi_i = (\phi_i,\psi_i,F_i)$ are considered, interactions between them must be analysed. The most general Lagrangian terms for the interactions between these supermultiplets which is renormalisable and invariant under SUSY transformation has the following form:
\beq \mathscr{L}_{\mathrm{int}} = -\frac12 \mathcal{W}^{ij} \psi_i \psi_j  + \mathcal{W}^i F_i + \textrm{h.c.} \eeq
Here $\mathcal{W}_i$ and $\mathcal{W}_{ij}$ are derivatives with respect to the scalar fields $\phi_i$ or $\phi_j$ of an important function in SUSY, the \textit{superpotential} $\mathcal{W}$ :
\beq \begin{split}
\mathcal{W} & = L^i \phi_i + {1\over 2} M^{ij} \phi_i \phi_j + {1\over 6} y^{ijk} \phi_i \phi_j \phi_k , \\
\mathcal{W}^i & = \frac{\p \mathcal{W}}{\p \phi_i} , \quad \mathcal{W}^{ij} = \frac{\p^2 \mathcal{W}}{\p \phi_i \p \phi_j}.
\end{split} \eeq
$M^{ij}$ represents the symmetric mass matrix for the Weyl fermions while $y^{ijk}$ is the Yukawa coupling between a scalar $\phi_k$ and two Weyl fermions $\psi_i$ and $\psi_j$. $L^i$ is only allowed if there is a gauge singlet $\phi_i$ in the SUSY model considered, which is not the case in the MSSM. A crucial point to emphasize is that for the sake of $\mathscr{L}_{\mathrm{int}}$ invariance under SUSY transformations, $\mathcal{W}$ must be holomorphic and then does not depend on the complex conjugates $\phi^*_i$. It follows that the Yukawa term $y^u_{i}\bar{Q}_i \bar{H} V_{ij} u_j$ that appears in the SM Lagrangian in eq.~\ref{eq:YukSM} cannot exist in SUSY. To address this issue two Higgs doublets are introduced in the MSSM : one gives mass to up-type quarks and the other gives mass to down-type quarks and charged leptons. When adding $\mathscr{L}_{WZ}$ and $\mathscr{L}_{\mathrm{int}}$ the auxiliary fields are eliminated using their equations of motion that lead to
\beq F_i = -\mathcal{W}^*_i, \quad F^i = -\mathcal{W}^i.\eeq
The chiral Lagrangian can now be written as
\beq \mathscr{L}_{\mathrm{chiral}} = \sum_{\Psi_i} -\p^\mu \phi^{* i} \p_\mu \psi_i + i\bar{\psi}^i \bar{\s}^\mu \p_\mu \psi_i -\frac12 \left(\mathcal{W}^{ij} \psi_i \psi_j + \mathcal{W}^*_{ij} \bar{\psi}^i \bar{\psi}^j\right)  - \mathcal{W}^i \mathcal{W}^*_i. \eeq

\subsection{Gauge supermultiplet}

Gauge fields of the SM are defined in gauge supermultiplets. They contain a massless gauge boson field $F^a_\mu$, a Weyl spinor $\l^a$ called \textit{gaugino} and a real bosonic auxiliary field $D^a$ which is introduced for the same argument as the field $F$ in the chiral supermultiplet. The superscript $a$ depends on the gauge symmetry considered and we then refer to the definitions given in section \ref{sec:1.gauge} for the objects used here. The gauge and SUSY invariant Lagrangian for gauge supermultiplets reads
\beq \mathscr{L}_{\mathrm{gauge}} = \sum_{\mathcal{G}} -\frac14 F^a_{\mu\nu}F^{a \mu\nu} +i\bar{\l}^a \bar{\s}^\mu \nabla_\mu \l^a + \frac12 D^a D^a,\eeq
where we sum over the gauge symmetries\ie $\mathcal{G} \in \left\{SU(3)_c, SU(2)_L, U(1)_Y\right\}$ and the covariant derivative acting on gauginos has the following form :
\beq \nabla_\mu \l^a = \p_\mu \l^a + g c_{abc} F^b_\mu \l^c.\eeq
The final step in the construction of the supersymmetric Lagrangian is the introduction of couplings between the two types of supermultiplet considered. For the sake of renormalizability, only three interaction terms that look like $(\phi^* T^a \psi)\l^a$, $\bar{\l}^a(\bar{\psi} T^a\phi)$ and $(\phi^* T^a \phi)D^a$, where $T^a$ are the generators of the gauge symmetry considered, are possible. Finally the complete Lagrangian is written as
\beq \mathscr{L} = \mathscr{L}_{\mathrm{chiral}} + \mathscr{L}_{\mathrm{gauge}} - \sum_{\mathcal{G}, \Psi_i} \sqrt2 g(\phi^{*i} T^a \psi_i)\l^a +g\bar{\l}^a(\bar{\psi}^i T^a\phi_i) - g(\phi^{*i} T^a \phi_i)D^a. \eeq
Combining the expressions that include $D^a$ shows that this auxiliary field depends only on scalar fields :
\beq D^a = - g(\phi^{*i} T^a \phi_i).\eeq
Since $F_i$ is also a function of scalar fields, the scalar potential of SUSY is
\beq V(\phi,\phi^*) = \sum_{\mathcal{G}, \phi_i} F^{*i} F_i + \frac12 D^a D^a = \sum_{\mathcal{G}, \phi_i} \mathcal{W}^i \mathcal{W}^*_i + \frac12 g^2(\phi^{*i} T^a \phi_i)^2.\eeq

\section{SUSY breaking}
\label{subsec:3.SSB}

Exact SUSY predicts that SM particles have the same mass as their superpartners, which is experimentally excluded; for example no selectron is observed around 511 keV. Thus if SUSY is realized it must be broken to give heavier masses to the superpartners. Spontaneous SUSY breaking can occur if at least one of the auxiliary fields $F_i$ and $D^a$ acquires a VEV. There are two mechanisms : the Fayet-Iliopoulos mechanism which deals with the $D$ term \cite{Fayet:1974jb} and the O'Raifeartaigh mechanism which concerns the $F$ term \cite{O'Raifeartaigh:1975pr}. However these mechanisms do not give a sufficiently high supersymmetric mass spectrum. Therefore the main scheme studied to explain these mass differences is as follows : SUSY may be broken in a \textit{hidden sector} which couples very weakly to the \textit{visible sector}\ie usually the MSSM. Three scenarios of mediation of the breaking from the hidden to the visible sector are mainly studied in the literature :
\begin{itemize}
\item The \textit{gravity-mediated} scenario, where the mediation is depicted by gravitational interactions between the two sectors \cite{Chamseddine:1982jx,Barbieri:1982eh,Hall:1983iz,Nilles:1983ge}. The VEV $\langle F \rangle$ of the SUSY breaking in the hidden sector is connected to the superpartners mass terms by
\beq m \sim \langle F \rangle / M_{Pl}.\eeq
Therefore to get mass terms in the GeV-TeV range SUSY breaking must happen at $\sqrt{\langle F \rangle} \sim 10^{10} - 10^{11}$ GeV. We will analyse in chapter~\ref{chapter:NUHM2} a version of gravity-mediated model, the Non-Universal Higgs Mass model (NUHM);
\item The \textit{Gauge-Mediated Supersymmetry Breaking} (GMSB) scenario, where the mediation occurs through gauge interactions. An intermediate scale that looks like a messenger scale interacts with the hidden and visible sectors and gives the mass terms of the superpartners through loop diagrams involving messenger particles \cite{Dine:1981gu,Nappi:1982hm,AlvarezGaume:1981wy,Dine:1993yw,Dine:1994vc,Dine:1995ag}. To obtain experimentally relevant mass terms and assuming the messenger scale is of the order of the SUSY breaking scale then $\sqrt{\langle F \rangle} \sim 10^4$ GeV. A major problem of this scenario in the context of this thesis is that its minimal version does not lead to a possible WIMP DM;
\item The \textit{Anomaly-Mediated Supersymmetry Breaking} (AMSB) scenario \cite{Randall:1998uk,Giudice:1998xp} also based on gravity mediation has in its minimal version a non-negligible drawback : sleptons get negative squared masses. An example of possible solution is a combination of AMSB with GMSB \cite{Kaplan:2000jz,Sundrum:2004un}.
\end{itemize}
These scenarios refer to a \textit{top-down approach} : we look at high-scale motivations like GUT unification and through RGEs we determine the low energies supersymmetric parameters. Universality assumptions at GUT scale allow to shrink the supersymmetric parameter space down to a few free parameters, typically less than ten. The other approach is called \textit{bottom-up approach} : starting from the SM and the low energies supersymmetric parameters we derive theoretical implications at GUT scale. This approach is characterized by many more free parameters since we use explicit SUSY breaking terms at low energy. These terms will be detailed in the next section. Several public codes are available to compute the supersymmetric spectra using RGEs : \Spect \cite{Djouadi:2002ze}, \ISA \cite{Paige:2003mg}, \SPno \cite{Porod:2003um} and \SoSUSY \cite{Allanach:2001kg}. The latter is used in chapters \ref{chapter:NUHM2} and \ref{chapter:ID}. 

An interesting consequence of these scenarios is that the Electro-Weak Symmetry Breaking (EWSB) is realized in a more natural way than in the SM. We saw in chapter~\ref{chap:revpart} that this symmetry breaking is realized in the SM introducing a negative mass squared term, which has no real explanation. In SUSY the Higgs masses squared for the two Higgs doublets are positive at the GUT scale but their evolution to low scales especially with (s)top contribution to the RGEs lead one of them to be negative which then implies EWSB : it is the \textit{radiative EWSB}. Nevertheless the issue of symmetry breaking is postponed to the higher scale where an unknown SUSY breaking mechanism takes place.

\section{The Minimal Supersymmetric Standard Model}
\label{subsec:3.MSSM}

The MSSM is the supersymmetric extension that contains the minimum number of fields. As mentioned above the chiral supermultiplet contains a LH Weyl spinor; it implies that sfermions are associated either to LH or RH fermions. We then denote sfermions by the subscripts \textit{L} and \textit{R} even though they do not carry chirality. As we saw two Higgs doublets are present in this minimal version : $H_u$ which gives mass to up-type quarks and $H_d$ which concerns down-type quarks and charged leptons. Using the notations given in chapter~\ref{chap:revpart} the content of the MSSM is represented in table~\ref{tab:MSSM}. For an overview of this model see~\cite{Drees:2004jm}.
\renewcommand{\arraystretch}{1.4}
\begin{table}[!htb]
\begin{center}
\begin{tabular*}{0.87\textwidth}{  c  c  c  c  c  }
       \hline \hline
      \multicolumn{5}{c}{\textbf{Chiral supermultiplets}} \\ \hline \hline
      \multicolumn{2}{c}{\textbf{Name}} & \textbf{spin 0} & \textbf{spin 1/2} & $\mathbf{SU(3)_c, SU(2)_L, U(1)_Y}$ \\ \hline 
      squarks, quarks & $\widetilde{Q}, Q$ & ($\tilde{u}_L$ $\tilde{d}_L$) & ($u_L$ $d_L$) & ($\textbf{3}$, $\textbf{2}$, $\frac{1}{3}$) \\ 
      	 (3 families) & $\bar{u}$ & $\tilde{u}^*_R$ & $\bar{u}_R$ & ($\bar{\textbf{3}}$, $\textbf{1}$, $-\frac{4}{3}$) \\ 
      		      & $\bar{d}$ & $\tilde{d}^*_R$ & $\bar{d}_R$ & ($\bar{\textbf{3}}$, $\textbf{1}$, $\frac{2}{3}$) \\ \hline 
    sleptons, leptons & $\widetilde{L}, L$ & ($\tilde{\nu}_L$ $\tilde{e}_L$) & ($\nu_L$ $e_L$) & ($\textbf{1}$, $\textbf{2}$, $-1$) \\ 
      	 (3 families) & $\bar{e}$ & $\tilde{e}^*_R$ & $\bar{e}_R$ & ($\bar{\textbf{1}}$, $\textbf{1}$, $2$) \\ \hline 
    Higgs, higgsinos  & $H_u$ & ($H^+_u$ $H^0_u$) & ($\widetilde{H}^+_u$ $\widetilde{H}^0_u$) & ($\textbf{1}$, $\textbf{2}$, $1$) \\ 
      		      & $H_d$ & ($H^0_d$ $H^-_d$) & ($\widetilde{H}^0_d$ $\widetilde{H}^-_d$) & ($\textbf{1}$, $\textbf{2}$, $-1$) \\ \hline \hline
      \multicolumn{5}{c}{\textbf{Gauge supermultiplets}} \\ \hline \hline
      \multicolumn{2}{c}{\textbf{Name}} & \textbf{spin 1/2} & \textbf{spin 1} & $\mathbf{SU(3)_c, SU(2)_L, U(1)_Y}$ \\ \hline 
      \multicolumn{2}{c}{gluinos, gluons} & $\widetilde{G}^a$ & $G^a$ & ($\textbf{8}$, $\textbf{1}$, 0) \\ \hline
      \multicolumn{2}{c}{winos, $W$'s} & $\widetilde{W}^\pm$ $\widetilde{W}^3$ & $W^\pm$ $W^3$ & ($\textbf{1}$, $\textbf{3}$, 0) \\ \hline
      \multicolumn{2}{c}{bino, $B$} & $\widetilde{B}$ & $B$ & ($\textbf{1}$, $\textbf{1}$, 0) \\ \hline \hline
\end{tabular*}
\caption{\label{tab:MSSM}MSSM supermultiplets and their gauge properties.}
\end{center}
\end{table}

\subsection{Lagrangian at low energy}

The parameterization of SUSY breaking at low energies is written in terms of explicit breaking terms called \textit{soft terms}. This name stems from some important requirements : 
\begin{itemize}
\item No quadratic divergences must be reintroduced to remain the EW scale stable;
\item Gauge invariance has to be preserved;
\item The theory must stay renormalisable\ie the new parameters in the Lagrangian must have positive mass dimension;
\item Baryon number $\mathcal{B}$ and lepton number $\mathcal{L}$ must be conserved.
\end{itemize}
It results that the Soft SUSY Breaking (SSB) Lagrangian reads
\beq \begin{split} \label{eq:Lsoft}
\mathscr{L}_{\mathrm{MSSM}}^{\mathrm{soft}} = & -\frac{1}{2} \left(\sum_{a=1}^{8} M_3\widetilde{G}^a\widetilde{G}_a + \sum_{i=1}^{3} M_2\widetilde{W}^i\widetilde{W}_i + M_1\widetilde{B}\widetilde{B} + \textrm{h.c.}\right) \\
			    & -(\tilde{u}^*_R \mathbf{a_u} \widetilde{Q}H_u - \tilde{d}^*_R \mathbf{a_d} \widetilde{Q}H_d - \tilde{e}^*_R \mathbf{a_e} \widetilde{L}H_d + \textrm{h.c.}) \\
			    & -\widetilde{Q}^\dag \mathbf{m^2_{\widetilde{Q}}}\widetilde{Q} -\widetilde{L}^\dag \mathbf{m^2_{\widetilde{L}}}\widetilde{L} -\tilde{u}^*_R \mathbf{m^2_{\tilde{u}_R}} \tilde{u}_R -\tilde{d}^*_R \mathbf{m^2_{\tilde{d}_R}}\tilde{d}_R -\tilde{e}^*_R \mathbf{m^2_{\tilde{e}_R}}\tilde{e}_R \\
			    & -m^2_{H_u} H^\dag_u H_u -m^2_{H_d} H^\dag_d H_d - (b H_u H_d + \textrm{h.c.}),
\end{split} \eeq
where $M_1, M_2$ and $M_3$ are respectively bino, wino and gluino mass terms. The sfermion fields appearing in $\mathscr{L}_{\mathrm{MSSM}}^{\mathrm{soft}}$ and defined in table~\ref{tab:MSSM} are vectors in the family space. The bold terms in the second and the third line of eq.~\ref{eq:Lsoft} are respectively trilinear coupling and soft sfermion mass term matrices. This introduce a lot of new parameters. However several phenomenological constraints allow to reduce this free parameter space. Physics of the $K$ mesons as well as experimental limits on the measurements of electric dipole moments force us to assume universality for the first and second generation of squark soft terms and Minimal Flavour Violation (MFV). Generalizing these assumptions to all sfermions and to the trilinear couplings we finally consider the previously defined matrices as diagonal in the family space. This phenomenological MSSM (pMSSM) will be studied in chapter~\ref{chapter:ID}. Last line of eq.~\ref{eq:Lsoft} contains new terms that will enter in the Higgs potential.

The MSSM superpotential is given by
\beq \label{eq.3:W_MSSM} \mathcal{W}_{\mathrm{MSSM}} = \tilde{u}^*_R\mathbf{y_u}\widetilde{Q}H_u - \tilde{d}^*_R\mathbf{y_d}\widetilde{Q}H_d - \tilde{e}^*_R\mathbf{y_e}\widetilde{L}H_d + \mu H_uH_d,\eeq
where $\mu$ is a supersymmetric mass term in the Higgs sector and the Yukawa matrices $\mathbf{y_f}$, which are in one-to-one correspondence with the trilinear couplings, are diagonal. As postulated before the conservation of baryon and lepton number is maintained. Actually there are terms which could be added in the MSSM superpotential that fulfill all the conditions for SUSY except $\mathcal{B}$ and $\mathcal{L}$ conservations. Nevertheless they lead to features non-observed experimentally such as a fast proton decay, for instance with the decay in positron and neutral pion $p \ra e^+ \pi^0$. It is here that $R$-parity is introduced \cite{Farrar:1978xj} :
\beq P_R = (-1)^{3(\mathcal{B}-\mathcal{L}) + 2S},\eeq
where $S$ is the spin of the particle considered. It results that $P_R = +1$ for particles from the SM and $P_R = -1$ for their superpartners. The $R$-parity conservation prevents baryon and lepton number violation but it has other important consequences :
\begin{itemize}
  \item The LSP is stable. Moreover if it has no electric and colour charge, in other words if it interacts weakly with the other particles it could be a viable WIMP DM;
  \item The LSP can only be produced in pairs at colliders;
  \item Each superpartners, except the LSP, can only decay into an odd number of other supersymmetric particles. Thus we usually expect only one LSP at the end of the decay chain.
\end{itemize}
As we have seen in the gauge sector of the SM presented in chapter~\ref{chap:revpart}, mixing effects can differentiate gauge and mass eigenstates. It is also the case in SUSY.

\subsection{Higgs sector}
\label{subsec:3.higgs}

Since we have two doublets of complex fields, the Higgs sector contains eight scalar degrees of freedom. The scalar potential of the two Higgs doublets $H_2 \equiv H_u = (H_u^+, H_u^0)$ and $H_1 \equiv H_d = (H_d^0, H_d^-)$ is a sum of $F$, $D$ and SSB terms and reads
\beq \begin{split} \label{eq:3.potential}
V_{\mathrm{MSSM}} = & \, V_{\mathrm{MSSM}}^F + V_{\mathrm{MSSM}}^D + V_{\mathrm{MSSM}}^{\mathrm{soft}}\\
= & \, (|\mu|^2 + m^2_{H_u}) (|H_u^0|^2 + |H_u^+|^2)
+ (|\mu|^2 + m^2_{H_d}) (|H_d^0|^2 + |H_d^-|^2) \\ 
& + [b(H_u^+ H_d^- - H_u^0 H_d^0) + \textrm{h.c.}] \\
& + {1\over 8} (g_2^2 + g_Y^2) (|H_u^0|^2 + |H_u^+|^2 - |H_d^0|^2 - |H_d^-|^2 )^2
+ \frac12 g_2^2 |H_u^+ H_d^{0*} + H_u^0 H_d^{-*}|^2,
\end{split} \eeq
where the terms proportional to the coupling constants come from $D$ terms, those proportional to $|\mu|^2$ stem from $F$ terms and the others from SSB terms.
The EWSB leads to one VEV for each neutral component of the Higgs doublets
\beq \langle H_u^0 \rangle = \frac{v \sin \b }{\sqrt{2}},\qquad \langle H_d^0 \rangle = \frac{v \cos \b }{\sqrt{2}}, \eeq
with $v$ the SM VEV and
\beq \label{eq.3:simpl} v \sin \b \equiv v_u \equiv v_2,\qquad v \cos \b \equiv v_d \equiv v_1.\eeq
The ratio of the VEVs then reads
\beq \tan \b = \frac{v_u}{v_d}.\eeq
With the EWSB, three of the degrees of freedom ($G^0, G^\pm$) become the longitudinal modes of the $W$ and $Z$ bosons. We are then left with five physical Higgs scalars : two of them are neutral and CP-even ($h^0$ and $H^0$), one is neutral and CP-odd ($A^0$) and there are two charged Higgs bosons $H^\pm$. The relations between the gauge and mass eigenstates are parameterized by two mixing angles $\a$ and $\b$ and read
\begin{align}
\sqrt{2}\binom{H_u^0}{H_d^0} & = \binom{v_u}{v_d} +  \begin{pmatrix} \cos \a & \sin \a \\ -\sin \a & \cos \a \end{pmatrix} \binom{h^0}{H^0} + i\begin{pmatrix} \sin \b & \cos \b \\ -\cos \b & \sin \b \end{pmatrix} \binom{G^0}{A^0}, \\
\binom{H_u^+}{H_d^{-*}} & = \begin{pmatrix} \sin \b & \cos \b \\ -\cos \b & \sin \b \end{pmatrix} \binom{G^+}{H^+}.
\end{align}
The minimization condition of the potential derived from $\frac{\p V_{\mathrm{MSSM}}}{\p H^0_u} = \frac{\p V_{\mathrm{MSSM}}}{\p H^0_d} = 0$ gives the tree-level masses in the Higgs sector :
\beq \begin{split}
m^2_{A^0} & = \frac{2b}{\sin 2\b} = 2|\mu|^2 + m^2_{H_u} + m^2_{H_d},\\
m^2_{h^0, H^0} & = \frac{1}{2} \left(m^2_{A^0} + M^2_{Z} \mp \sqrt{(m^2_{A^0} -  M^2_{Z})^2 + 4 M^2_{Z} m^2_{A^0} \sin^2 (2\b) }\right),\\
m^2_{H^\pm}	& = m^2_{A^0} + M^2_W,
\end{split} \eeq
where by convention $h^0$ is lighter than $H^0$. Note that the tree-level value of the mixing angle $\a$ is determined by :
\beq \frac{\sin 2\a}{\sin 2\b} = -\frac{m^2_{H^0}+m^2_{h^0}}{m^2_{H^0}-m^2_{h^0}}, \quad \frac{\tan 2\a}{\tan 2\b} = \frac{m^2_{A^0}+M^2_{Z}}{m^2_{A^0}-M^2_{Z}}.\eeq
Since these relations give a stringent upper bound on the lightest Higgs mass which is widely excluded by experimental searches ($m_{h^0} <  M_{Z} |\cos 2\b| $), radiative corrections are crucial for the validity of the model. These corrections, mainly coming from top (s)quark loops, are taken into account in codes like \SoSUSY that we will use in part~\ref{part2}.

\subsection{Sfermion sector}

Mixing also occurs between the supersymmetric partners of the LH and RH components of a fermion $f$. The mass matrix for a sfermion $Y$ of the i\textit{th} family whose soft mass terms are $m^2_{X_i}$ for the LH part and $m^2_{x_i}$ for the RH one is usually defined as\footnote{Note that from now on we will sometimes use the simplications $\cos \b = \cb, \ \sin \b = \sb, \ \tan \b = \tb, \ \cos \t_W = \cw, \ \sin \t_W = \sw, \ \tan \t_W = t_W$ and the notations given in eq.~\ref{eq.3:simpl}.}
\beq \label{eq:3.sfer} \mathbf{M^2_{Y_i}} =\begin{pmatrix}
    m^2_{X_i} + m^2_f + M^2_Z c_{2\b} \left(I^3_f - \mathcal{Q}_f s^2_W\right) & m_f\left(A_f - \mu (\tb)^{-2 I^3_f}\right)\\
    m_f\left(A_f - \mu (\tb)^{-2 I^3_f}\right) & m^2_{x_i} + m^2_f + M^2_Z c_{2\b} \left(I^3_f - \mathcal{Q}_f s^2_W\right)
  \end{pmatrix} \eeq
where $I^3_f$ is the third component of the $SU(2)_L$ isospin of $f$, $\mathcal{Q}_f$ its electric charge and $m_f$ its mass. This mass matrix is then only relevant when we consider superpartners of heavy SM fermions. It results that the soft mass terms and the trilinear couplings $A_f$ are here determined only for $i=3$ with $Y_3 \in \{\tilde{t}_{L,R}, \tilde{b}_{L,R}, \tilde{\tau}_{L,R}\}$, $X_3 \in \{\widetilde{Q}_3, \widetilde{L}_3\}$ and $x_3 \in \{\tilde{u}_3, \tilde{d}_3, \tilde{e}_3\}$. After diagonalisation, the physical states $\tilde{t}_1, \tilde{t}_2, \tilde{b}_1, \tilde{b}_2, \tilde{\tau}_1,  \tilde{\tau}_2$ are ordered in mass,\ie $m_{\tilde{f}_1} < m_{\tilde{f}_2}$ with $\tilde{f} \in \{\tilde{t}, \tilde{b}, \tilde{\tau}\}$. For the other sfermions their mass simply reads
\beq m_{Y_i}^2 = m^2_{X_i} + M^2_Z c_{2\b} (I^3_f - \mathcal{Q}_f s^2_W), \eeq
with $i \in \{1,2\}$, $Y_1 \in \{\tilde{u}_L, \tilde{u}_R, \tilde{d}_L, \tilde{d}_R, \tilde{e}_L, \tilde{e}_R\}$, $Y_2 \in \{\tilde{c}_L, \tilde{c}_R, \tilde{s}_L, \tilde{s}_R, \tilde{\mu}_L, \tilde{\mu}_R\}$ and $X_2 = X_1 \in \{\widetilde{Q}_1, \widetilde{L}_1, \tilde{u}_1, \tilde{d}_1, \tilde{e}_1\}$. For the sneutrinos $i \in \{1,2,3\}$ : $Y = (\tilde{\nu}_{e L}, \tilde{\nu}_{\mu L}, \tilde{\nu}_{\tau L})$ and $X = (\widetilde{L}_1, \widetilde{L}_2, \widetilde{L}_3)$.

\subsection{Gaugino and higgsino sector}

The sector of the supersymmetric partners of gauge bosons and Higgs fields is divided into three parts : first the supersymmetric partners of the gluons, second the charged gauginos and higgsinos and finally the most interesting SUSY particles in the context of this thesis, the neutral gauginos and higgsinos that form the neutralino sector.

\subsubsection{Gluinos and charginos}

The eight supersymmetric partners of the gluons, the \textit{gluinos} $\tilde{g}$, are Majorana fermions with a mass term simply equal to $|M_3|$.

The two charged winos ($\tilde{W}^\pm$) as well as the two charged higgsinos ($\tilde{H}^+_u$ et $\tilde{H}^-_d$) mix to form two charged Dirac fermions called \textit{charginos} $\chi^\pm_i$ ($i \in \{1,2\}$), where $m_{\chi^\pm_1} < m_{\chi^\pm_2}$. The mass matrix in the gauge-eigenstate basis $(\widetilde W^+, \widetilde H_u^+, \widetilde W^-, \widetilde H_d^-)$ is
\beq \label{eq.3:chimass} \mathbf{M}_{\chi^\pm} = \begin{pmatrix} 0 &  \mathbf{X}^T \\  \mathbf{X} & 0 \end{pmatrix}, \eeq
with
\beq \mathbf{X} = \begin{pmatrix} M_2 & \sqrt{2} M_W \sin \b \\ \sqrt{2} M_W \cos \b & \mu \end{pmatrix}.\eeq
The relation between mass and gauge eigenstates is given by two unitary 2$\times$2 matrices $\mathbf{Z_v}$ and $\mathbf{Z_u}$ with
\beq \binom{\chi^+_1}{\chi^+_2} = {\mathbf{Z_v}}\binom{\widetilde W^+}{\widetilde H_u^+}, \quad
     \binom{\chi^-_1}{\chi^-_2} = {\mathbf{Z_u}}\binom{\widetilde W^-}{\widetilde H_d^-}.
\eeq

\subsubsection{Neutralinos}
\label{subsubsec.3:neut}

Finally the mixing of the four neutral gauge eigenstates $(\widetilde{B}, \widetilde{W}^3, \widetilde{H}_u^0, \widetilde{H}_d^0)$ gives rise to four Majorana mass eigenstates called \textit{neutralinos} $\chi^0_i$ ($i \in \{1,2,3,4\}$) with $m_{\chi^0_1} < m_{\chi^0_2} < m_{\chi^0_3} < m_{\chi^0_4}$. It is a crucial sector of this thesis since these particles are neutral, massive and they weakly interact with the rest of the MSSM sector. The lightest one is the most commonly WIMP DM candidate studied in the literature and this candidate will be considered in the work presented in part~\ref{part2}. In the basis $\psi^0 = (\widetilde{B}, \widetilde{W}^3, \widetilde{H}^0_d, \widetilde{H}^0_u)$ the neutralino Lagrangian involving mass terms has the following form :
\beq \mathscr{L}_{\chi^0} = -\frac{1}{2} (\psi^0)^T \mathbf{M_{\chi^0}} \psi^0 + \textrm{h.c.},\eeq 
with the neutralino mass matrix given by
\beq
  \mathbf{M_{\chi^0}}=\begin{pmatrix}
    M_1 & 0 & -M_Z \cb \sw & M_Z \sb \sw \\
    0 & M_2 & M_Z \cb \cw & -M_Z \sb \cw \\
    -M_Z \cb \sw & M_Z \cb \cw & 0 & -\mu\\
    M_Z \sb \sw & -M_Z \sb \cw & -\mu & 0
  \end{pmatrix}.
\eeq
This matrix is diagonalised by a 4$\times$4 unitary matrix $\mathbf{Z_n}$ which links mass and gauge eigenstates :
\beq \chi^0_i = Z_{n ij} \psi^0_j\textrm{,} \qquad i\textrm{,}j \in \{1,2,3,4\}.\eeq

Thence several scenarios are considered for the LSP neutralino when we want to match its computed relic density using the \micro code with the measured DM relic density. For large values of the $\mu$ term and assuming universality of the gaugino mass terms at the GUT scale the lightest neutralino $\chi^0_1$ is mostly bino,\ie the bino fraction $Z^2_{n 11} \sim 1$. This state interacts too weakly with the SM particles; its total annihilation cross section calculated in the early Universe is too small and this candidate mostly overclose the Universe. For some scenarios this problem is avoided. For instance when $m_{\chi_1^0} \approx m_{A^0}/2$ the DM annihilation into $b\bar{b}$ or $\tau^+\tau^-$ is enhanced through an $s$-channel pseudo-scalar Higgs boson exchange : this corresponds to the \textit{funnel regions}. 

For small $\mu$ comparing to the gaugino masses, the LSP is mostly higgsino, \ie $Z^2_{n 13} + Z^2_{n 14} \sim 1$. Non-negligible mass degeneracy with  $\chi_2^0$ and the lightest chargino leads to an enhancement of the DM annihilation into gauge bosons through $t$-channel $\chi_2^0/\chi_1^\pm$ exchange. This could imply the opposite problem in comparison to bino LSP : the predicted $\chi^0_1$ relic density is too low to satisfy the cosmological constraints. Taking larger $\mu/m_{\chi_1^0}$ values solve this issue as we will show in chapter~\ref{chapter:NUHM2}. 

Loosing the universality assumption as in chapter~\ref{chapter:ID} gives the opportunity to study wino LSP ($Z^2_{n 12} \sim 1$). This candidate can be strongly constrained by DM observables since it is charged under $SU(2)_L$ : its annihilation rate into SM particles both in the early Universe and presently can be too large.

\section{Constraints on SUSY}

One of the nice features of SUSY is that it offers possible signatures in different types of experiments. Thus the model can be explored and constrained by very different types of observations. These observational data are separated into two main categories : astroparticle and cosmological constraints on supersymmetric DM candidates and constraints on SUSY coming from collider experiments.

\subsection{Cosmological and astroparticle constraints}

The current amount of DM in the Universe gives the most stringent constraint on supersymmetric WIMP DM candidates. Assuming a thermal production and the freeze-out mechanism, section~\ref{sec:2.DM} gives the method to compute the supersymmetric WIMP DM relic density which is compared to the precise measurement of the DM relic density. The recent result, given in table~\ref{tab:cosmo_param}, obtained combining Planck data with WMAP polarization data, high-$\ell$ CMB and BAO data reads
\beq \Om_{\mathrm{WIMP}} h^2 = 0.1187 \pm 0.0017.\eeq
In the case of SUSY this calculation becomes tricky and can involve a large number of processes when coannihilations give the dominant contribution to the WIMP relic density. Numerical computations appear to be essential hence the use of codes like the \micro code. This point will be particularly elaborated in chapters \ref{chapter:NUHM2} and \ref{chapter:ID} where coannihilation processes involving a slightly heavier supersymmetric particle, the chargino, have consequences on DM observables. Note that the particle which is the second lightest superpartner is called the Next-to-LSP (NLSP). In this thesis DM relic density lower bound is sometimes relaxed\footnote{Which is not the case for the upper bound since the DM candidate chosen should not overclose the Universe.}. Either we assume that the WIMP considered does not account for the total DM density, in this case it will also impact the calculation of DM detection observables developed below. Another possibility for models with low DM relic density is to regenerate the correct DM density through more sophisticated mechanisms, for example the freeze-in mechanism \cite{Hall:2009bx,Chu:2011be,Williams:2012pz}. 

To probe the DM candidate the most powerful method is simply to detect it. Two main strategies are then used. 

\subsubsection{DM Direct Detection}
\label{subsubsec:3.DD}

It is usually assumed that the WIMP halo inside which the MW stands has a local density\footnote{A recent estimate assuming either an Einasto \cite{Graham:2005xx} or a Navarro-Frenk-White \cite{Navarro:1996gj} DM density profile leads to a local density of $0.39 \pm 0.03$~GeV cm$^{-3}$ \cite{Catena:2009mf,Beringer:1900zz}.}\ca 0.3-0.4 GeV cm$^{-3}$ and local circular and galactic escape velocities of a few hundreds km s$^{-1}$. This implies that many of these WIMPs should pass through the Earth and then weakly interact with our planet. The idea that follows is simply to detect WIMP scattering on nuclei through the measurement of the recoil energy of this normal matter. This DM Direct Detection (DD) proposal was initiated in 1984 by Goodman and Witten \cite{Goodman:1984dc} and from there a lot of underground\footnote{Which is mandatory because of background issues and the low number of events expected.} detectors were built to find this DM footprint. 

The simple elastic scattering on a nucleus leads to recoil energies of the order of dozens of keV, which is detectable by current experiments. The cross section of the process is expressed in terms of Spin-Dependent (SD) and Spin-Independent (SI) scattering. The SD interaction comes from the coupling of the WIMP to the spin content of a nucleon and the total SD cross section on a nucleus depends only on its unpaired nucleons. Meanwhile the SI scattering results from a scalar coupling between the WIMP and a nucleon; the SI cross section on a nucleus increases with the total number of nucleons $A$ inside this nucleus. As a result SD scattering is competitive with SI scattering only for light nuclei. The best constraints are then expected in the case of SI WIMP-nucleon scattering. Moreover SD scattering is clearly irrelevant for scalar WIMPs which will be considered in chapter~\ref{chapter:RHsneu}. 

\begin{figure}[!htb]
\begin{center}
\includegraphics[width=10cm,height=7cm]{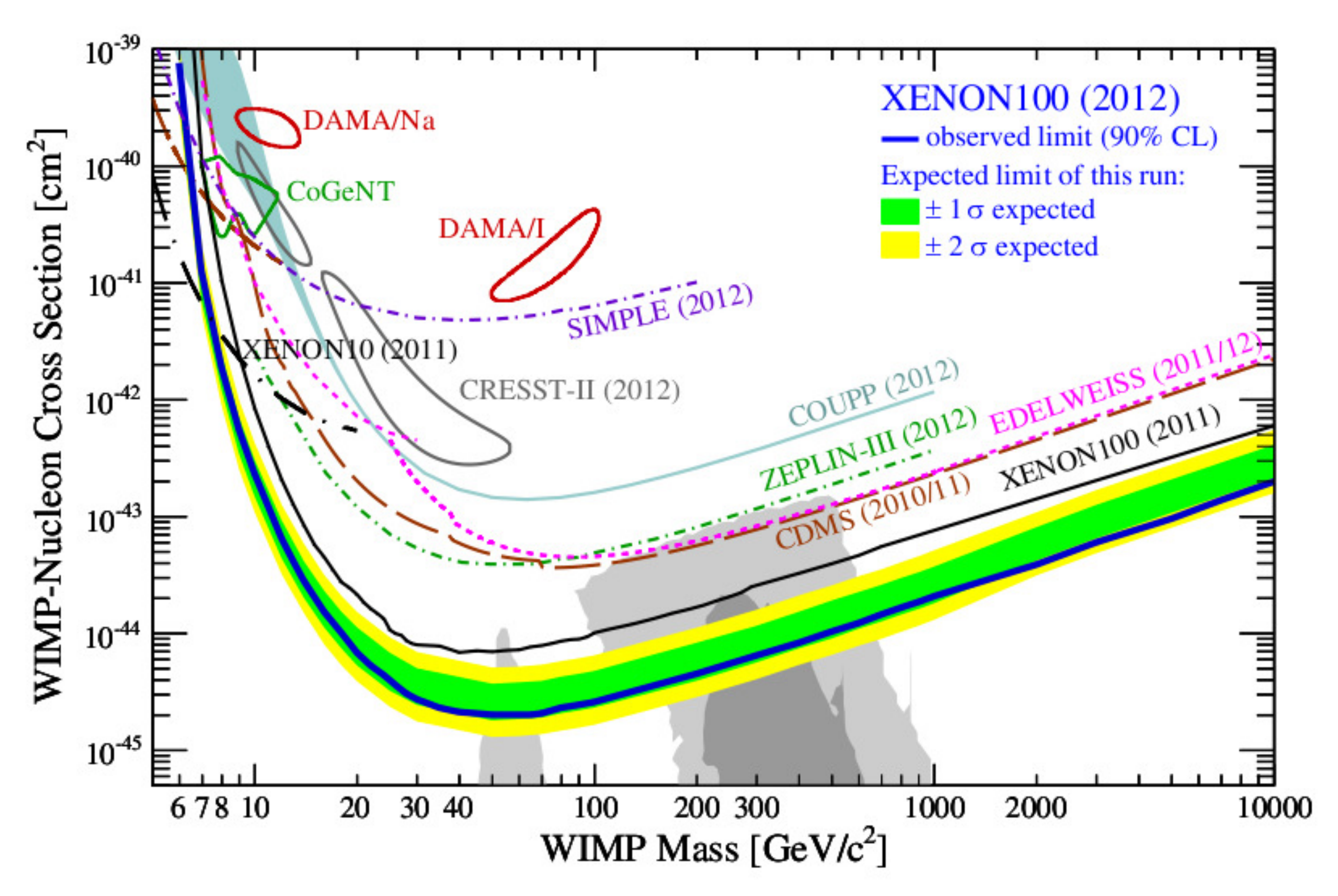} 
  \caption[Results on the SI WIMP-nucleon cross section versus the WIMP mass from various experiments compared to some supersymmetric predictions (grey regions).]{Results on the SI WIMP-nucleon cross section versus the WIMP mass from various experiments compared to some supersymmetric predictions (grey regions). Figure taken from~\cite{Aprile:2012nq}.}
\label{fig:XENON100}
\end{center}
\end{figure}

Figure~\ref{fig:XENON100} shows the current best exclusion limits for SI scattering that come from the XENON100 experiment~\cite{Aprile:2012nq}. For instance it is able to probe cross sections down to $2 \times 10^{-45}$ cm$^2$ $= 2 \times 10^{-9}$ pb for a WIMP mass of 55 GeV and thus constrains quite significantly the predictions from SUSY. This figure also shows that some collaborations like DAMA \cite{Bernabei:2008yi}, CoGeNT \cite{Aalseth:2010vx}, CRESST \cite{Angloher:2011uu} or recently CDMS \cite{Agnese:2013rvf} have detected anomalous features that could be interpreted as signals of DM. However these results are in tension with the stringent bounds put by the XENON100 and the XENON10 \cite{Angle:2011th} collaborations. In the different studies considered in this thesis we have decided to only take into account the negative search from XENON100.

These limits on the WIMP-nucleon cross section are obtained assuming that the interactions WIMP-proton and WIMP-neutron are the same. Since it is not necessary the case, the normalized cross-section on a point-like nucleus $N$ with a mass $m_N$ with $Z$ protons and $A-Z$ neutrons is considered. It reads
\beq \sigma^{\rm SI}_{\chi N}= \frac{4 \mu_{\chi N}^2}{\pi}\frac{\left( Z f_p+ (A-Z)f_n\right)^2}{A^2 } \eeq 
with $\mu_{\chi N} = \frac{m_{\chi} m_N}{m_{\chi} + m_N}$. To determine the nucleon amplitudes $f_p$ and $f_n$, respectively for the protons and the neutrons, the WIMP-parton\footnote{Quarks and gluons.} scattering must be computed. The calculation of the DD cross section is done using the \micro code. For a scalar interaction resulting for instance from Higgs boson exchange the WIMP-nucleon process depends on the quark content of the nucleon. This is determined by ratios of light quark masses $m_u/m_d$ and $m_s/m_d$ and by two parameters extracted from lattice QCD calculations : the light-quark sigma term $\s_{\pi N} =~(m_u+m_d) \langle N|\bar{u}u + \bar{d}d|N \rangle$ and the strange quark content of the nucleon $\s_s =~m_s \langle N|\bar{s}s|N \rangle$. These quantities have a non-negligible uncertainty which then impacts the calculation of the cross section. The main values used in this thesis are summarized in table~\ref{contentnucl}. Note that a compilation of recent lattice results leads to different mean values for $\s_{\pi N}$ and $\s_s$ and especially to lower uncertainties, namely $\s_{\pi N} = 34 \pm 2$~MeV and $\s_s = 42 \pm 5$~MeV \cite{Belanger:2013oya}.
\begin{table}[!htb]
\begin{center}
\begin{tabular}{c c}\hline \hline
\textbf{Parameter} & \textbf{Value}\\ \hline \hline
$m_u/m_d$ & 0.553 $\pm$ 0.043 \cite{Leutwyler:1996qg}\\ 
$m_s/m_d$ & 18.9 $\pm$ 0.8 \cite{Leutwyler:1996qg}\\ 
$\s_{\pi N}$ & 44 $\pm$ 5 MeV \cite{Thomas:2012tg}\\ 
$\s_s$ & 21 $\pm$ 7 MeV \cite{Thomas:2012tg}\\ \hline \hline
\end{tabular}
\caption{\label{contentnucl} Parameters determining the quark content of the nucleon in the calculation of WIMP-nucleon processes.}
\end{center}
\end{table}

\subsubsection{DM Indirect Detection}
\label{subsubsec:3.ID}

As explained in section~\ref{subsec:2.freezeout}, WIMP annihilation into SM particles is expected to be a rare event today. However in some regions of the space the rate of DM annihilation could be sufficiently large to give anomalous features in cosmic rays searches such as gamma-rays, neutrinos, positrons and anti-protons. This is the principle of the Indirect Detection (ID) of DM that was proposed a few decades ago \cite{Gunn:1978gr,Stecker:1978du,Zeldovich:1980st,Ellis:1988qp,Silk:1984zy,Stecker:1985jc,Rudaz:1987ry,Stecker:1988fx,Turner:1989kg}. Furthermore N-body simulations show that the DM distribution in the Universe seems to be inhomogeneous, thus leading to enhancement of DM annihilation. 

This interesting method to probe DM has however a significant drawback : it requires a detailed knowledge of the astrophysical sources. To claim the discovery of an anomalous feature the choice has to be made between two methods : remove the known or modelled background in order to show a signal that could come from DM or see a clear feature difficult to mimic by considering only astrophysical sources. Possible anomalies discovered that could give hints on DM include the positron excess observed by the \PAMELA \cite{Adriani:2008zr,Adriani:2010ib}, \Fermi-LAT \cite{FermiLAT:2011ab} and recently AMS \cite{Aguilar:2013qda} collaborations or the two possible $\g$-ray lines at 111 and 129 GeV seen in \Fermi-LAT data \cite{Bringmann:2012vr,Weniger:2012tx,Tempel:2012ey,Rajaraman:2012db,Su:2012ft,Hektor:2012kc,Su:2012zg}. Nevertheless the origin of these anomalies is not yet clearly established. There are also a lot of astrophysical data that fit well with the modelling of astrophysical background sources in the usual WIMP mass range, which is able to constrain DM models. Chapter~\ref{chapter:ID} will be devoted to $\bar{p}$ and $\g$-ray constraints on a neutralino DM which mainly annihilates into $W$ bosons.

Note that another detection method to look for DM, which is currently in development, could be promising. Summarizing quickly this method, called directional detection, it is based on the motion of the Solar System in the MW which imply that the motion of the Earth relatively to the galactic halo could gives hints on DM \cite{Spergel:1987kx}. 

\subsection{Collider constraints}

High energy colliders as the LHC could also give hints on supersymmetric DM, although it is not in their scope to give informations on whether this particle is stable at a cosmological scale or not. Nevertheless important constraints can be applied to SUSY since no superpartners have been discovered yet.

\subsubsection{Bounds on supersymmetric particles}

\begin{figure}[!htb]
\begin{center}
\includegraphics[width=6.2cm,height=4.2cm]{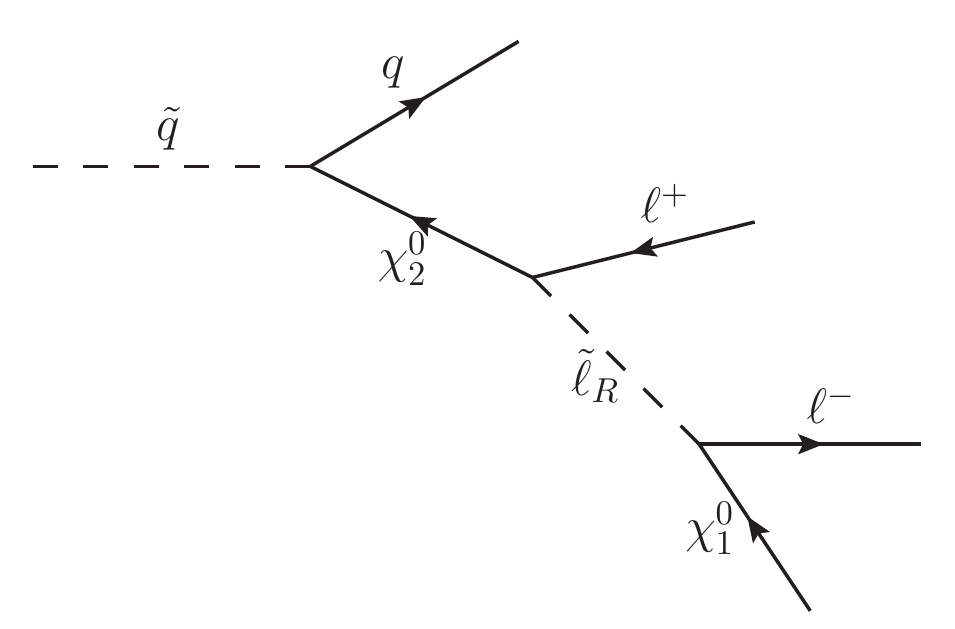} 
  \caption[Example of a cascade decay as a result of the $R$-parity conservation in the MSSM.]{Example of a cascade decay as a result of the $R$-parity conservation in the MSSM : a squark $\tilde{q}$ decays into the second lightest neutralino $\chi^0_2$ and a quark which will decay or hadronise. Then $\chi^0_2$ decays to a charged lepton plus a RH slepton which finally decays into another lepton and the LSP, here the lightest neutralino $\chi^0_1$.}
\label{fig:cascade}
\end{center}
\end{figure}

The LEP collider puts stringent lower bounds on the mass of several charged sparticles. For example it constrains the $\chi^\pm_1$ mass to be above $\sim 103$~GeV, except for small mass degeneracies with the LSP. Direct accelerator lower bounds on sparticles from the LEP collider for the chargino, the sneutrino and the RH charged slepton masses are implemented in the \micro code. Since the start of the LHC the limits on coloured sparticles are much improved and some searches are able to exclude supersymmetric particles with a mass beyond 1 TeV. Nevertheless these bounds have to be taken with care since there are derived under several assumptions. The first one, justified within the context of this thesis is the conservation of $R$-parity. As we saw in section~\ref{subsec:3.MSSM} this implies that a cascade decay of a SUSY particle leading to SM particles and the LSP, which appears as missing energy in the detectors\footnote{As with neutrinos.}, is one of the main method used to probe SUSY (see figure~\ref{fig:cascade}). 

\begin{figure}[!htb]
\begin{center}
\centering \noindent
\includegraphics{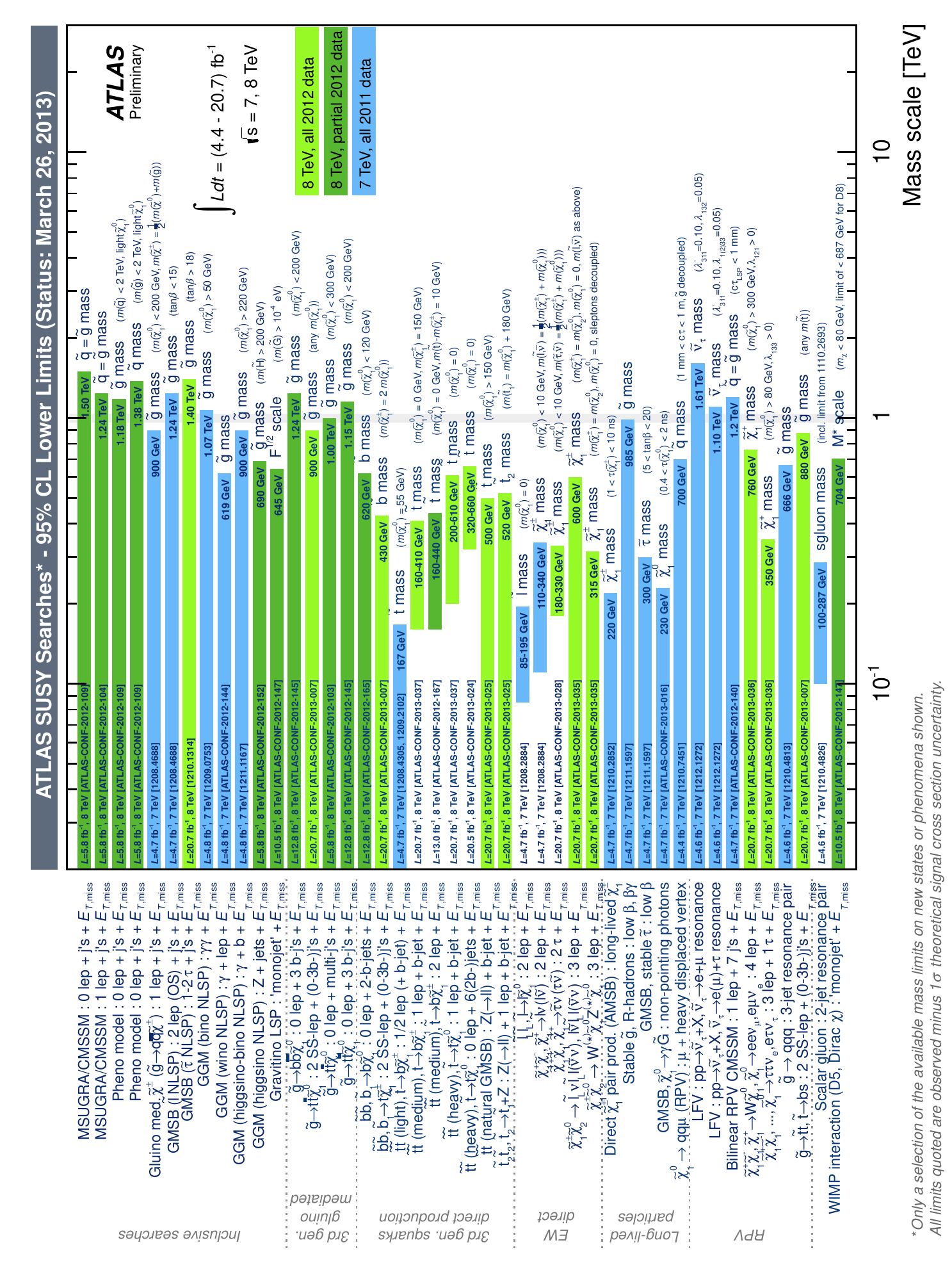} 
  \caption[Summary of ATLAS searches for SUSY at Moriond 2013.]{Summary of ATLAS searches for SUSY at Moriond 2013. Figure taken from \cite{atlassusy}.}
\label{fig:atlassusy}
\end{center}
\end{figure}

Actually experimentalists have obtained bounds on SUSY using many more assumptions. This implies that some bounds are not so stringent for the analysis developed in this thesis, especially when scenarios going beyond the MSSM are studied. Many limits were derived in specific MSSM scenarios GUT motivated, in most cases within the simplest gravity-mediated SUSY breaking scenario, the minimal SUperGRAvity/Constrained MSSM (mSUGRA/CMSSM). As an example figure~\ref{fig:atlassusy} summarizes ATLAS bounds on supersymmetric particle masses shown during the \textit{Rencontres de Moriond} conference in 2013 where, depending on the analysis done, stronger assumptions are considered like an equal mass for gluinos and first and second generation of squarks. Other analyses have been done within \textit{simplified} models with specific decay chains studied \cite{Chatrchyan:2013sza}. For most of the studies presented in this thesis these limits are avoided assuming that squarks and sleptons are at the TeV scale, except for third generation squarks which are more weakly constrained. Nonetheless it is possible to check the validity of these experimental limits on non-minimal supersymmetric models. Chapter~\ref{chapter:NMSSM} will be devoted to this type of study within a typical non-minimal model, the Next-to-MSSM (NMSSM). 

Note that the knowledge of SM particle masses have also consequences on the studies. For example the top quark mass influences the SUSY spectra in GUT scale models as well as provides large corrections to the SM-like Higgs boson mass.

\subsubsection{Low energy observables}

SUSY can also be detected in the measurement of rare processes or in the precise determination of some EW quantities. $B$-physics observables are used throughout this thesis. We will be mainly interested in the observables $\btau$, $\bsmu$, $\Delta M_s$, $\Delta M_d$ and $\bsg$. All these branching ratios and mass differences get supersymmetric contributions that can be large. As an example, the $\bsmu$ in the MSSM which for large $\tan \b$ values grows like $\tan^6 \b$. This allows to exclude some regions of the supersymmetric parameter space since no major discrepancies are obtained between the experimental measurements and the SM expected values. Other low energy observables such as the anomalous magnetic moment of the muon $\amu$ were considered. In part~\ref{part2} we use the \micro functions that calculate these observables. Part~\ref{part3} is dedicated to a supersymmetric model with an extended gauge symmetry, the UMSSM. Since $B$-physics observables and $\amu$ were adapted to this model for this thesis they will be detailed there.

\part{Neutralino Dark Matter in the (N)MSSM}
\label{part2}

\chapter{Unification with non-universal Higgs masses and the supersymmetric inflaton}
\label{chapter:NUHM2}

\setcounter{minitocdepth}{3} %Numbered subsubsections in TOC
\minitoc\vspace{1cm}
\newpage

This chapter is mainly based on the article \cite{Boehm:2012rh} with some figures added/updated and the scanning method carefully described.

\section{Introduction}

The primordial inflation dilutes all matter except the quantum fluctuations which we see in the CMB radiation. The last phases of inflation are commonly embedded within a BSM sector where the inflaton can directly excite quarks and leptons. This is why we consider in this chapter a supersymmetric model, the NUHM2 : within its scalar potential containing $F$ and $D$ terms two $D$-flat directions, $\widetilde{u}\widetilde{d}\widetilde{d}$ and $\widetilde{L}\widetilde{L}\widetilde{e}$, are viable inflaton candidates. By looking at constraints on the neutralino DM candidate, collider constaints such as $B$-physics observables as well as cosmological constraints on the inflaton candidates, indications on the inflaton mass can be obtained.

\section{Gravity-mediation of SUSY breaking}

As explained in section~\ref{subsec:3.SSB}, one of the method to mediate SUSY breaking is through gravity. The most popular gravity-mediated SUSY breaking model is the mSUGRA/CMSSM. The \textit{minimal} qualifier in the mSUGRA name stems from the fact that universal scalar masses at GUT scale is assumed. Only five free parameters are needed to study its phenomenological consequences at low scale :
\begin{itemize}
\item $m_{0}$, the common scalar mass at GUT scale;
\item $m_{1/2}$, the common gaugino mass at GUT scale;
\item $A_{0}$, the common trilinear coupling at GUT scale;
\item $\tan \beta$, the ratio of the two Higgs doublets VEV;
\item $sign(\mu)$, the sign of the superpotential Higgs mass term.
\end{itemize}\smallskip
However this universality in the scalar sector has some disadvantages when considering moderate values for the common parameters\footnote{High values for common parameters can respect physical constraints as shown in \cite{Akula:2011aa,Feng:2012jfa}, although these scenarios may not be accessible at the LHC scale.}. The LSP neutralino is for instance mainly bino in this model. This implies that finding regions of the parameter space where the neutralino relic density matches or is below the measured DM density is difficult. For instance DM constraints are usually fulfilled in narrow regions, the funnel regions. In addition it has been shown that these universality assumptions make it difficult, especially for low $\tan \beta$ values, to get the SM Higgs boson mass around 125 GeV \cite{Baer:2011ab,Arbey:2011ab,Ellis:2012aa}.

These universality conditions can be relaxed by removing the degeneracy between Higgs boson and sfermion masses at GUT scale, thus solving the previous issues. This type of models, the NUHM models \cite{Ellis:2002iu,Baer:2005bu}, are divided into two categories : NUHM1 and NUHM2. In the NUHM1, inspired by GUT models where the two Higgs doublets appear in the same multiplet, degeneracy is maintained in the Higgs sector : $m^2_{0} \neq m^2_{H_u} = m^2_{H_d}$. Thereby a new free parameter, defined at GUT scale, is added : $sign(m^2_{H_{u,d}}) \sqrt{|m^2_{H_{u,d}}|}$.

\subsection{The NUHM2 model}

In the NUHM2, the model we will study here, Higgs scalar masses at GUT scale are not equal; it is assumed that the Higgs doublets appear in different multiplets at GUT scale. Hence two free parameters are added to those of mSUGRA/CMSSM, $m^2_{H_u}$ and $m^2_{H_d}$. Following the EWSB conditions the following relations are obtained (see \cite{Ellis:2002iu}) :
\begin{align}
m_{H_d}^2(1+ \tan^2 \beta) = & \: m^2_{A^0} \tan^2 \beta - \mu^2 (\tan^2 \beta + 1 -\Delta_\mu^{(H_u)} ) - (c_{H_d} + c_{H_u} + 2 c_\mu)\tan^2 \beta \nonumber \\
& \: - \Delta_A \tan^2 \beta - \frac{1}{2} M_Z^2 (1 - \tan^2 \beta) - \Delta_\mu^{(H_d)} \quad \textrm{and} \\
m_{H_u}^2(1+ \tan^2 \beta) = & \: m^2_{A^0} - \mu^2 (\tan^2 \beta + 1 + \Delta_\mu^{({H_u})} ) - (c_{H_d} + c_{H_u} + 2 c_\mu) \nonumber \\
& \: - \Delta_A + \frac{1}{2} M_Z^2 (1 - \tan^2 \beta) + \Delta_\mu^{(H_d)},
\end{align}
where the $\Delta$ and $c$ terms are respectively loop and radiative corrections. These relations link the Higgs GUT scale parameters to the $\mu$ term and the pseudoscalar mass $m_{A^0}$ which are taken as new free parameters. Therefore the set of free parameters of NUHM2 is : 
\begin{center}
$m_{0}$, $m_{1/2}$, $A_{0}$, $\tan \beta$, $\mu$ and $m_{A^0}$.
\end{center}
To find the regions of NUHM2 which are compatible with colliders and DM constraints two methods, both relying on the \micro code and the \SoSUSY spectrum calculator, are used. In a first step benchmark points which satisfy all these requirements will be identified. Then, the model will be studied by performing a Markov Chain Monte Carlo (MCMC) analysis of the NUHM2 parameter space.

\subsection{Benchmark points with neutralino DM in the NUHM2}
\label{subsec:4.benchmark}

For the method consisting on identifying benchmark points the following requirements are imposed :

\begin{itemize}
\item The LSP must be the neutralino;
\item The relic density of the neutralino must be compatible with the measured DM abundance from the WMAP 7-year $+$ BAO $+ \, H_0$ mean value \cite{Komatsu:2010fb} :\beq 0.1088<\Om_{DM} h^2<0.1158;\eeq
\item The LEP2 bound on the mass of the lightest chargino must be satisfied\ie $m_{\chi^\pm_1}>103.5$ GeV \cite{Nakamura:2010zzi};
\item The mass of the lightest Higgs boson must be around $m_{h^0}=125$~GeV. 
\end{itemize}
Scans are done in the ranges $0 \leq m_0 \leq 3000$ GeV and $0 \leq m_{1/2} \leq 5000$~GeV and choosing specific values of $\mu$, $\tan\beta$ and $m_{A^0}$. The mass of the top quark is set at $m_t=173.2$ GeV \cite{Lancaster:2011wr} and the limits on the following branching ratios $\bsmu<4.5\times10^{-9}$ \cite{Aaij:2012ac} and $\bsg=(3.55\pm0.26)\times10^{-4}$ \cite{Asner:2010qj} are used. 

None of the scenarios found can explain the measured value of the anomalous magnetic moment of the muon $\amu$; the additional contributions in this model are indeed too small \cite{Baer:2011ab}. We assume in the rest of the study that as long as the contribution of a given scenario is not greater than the measured value, the configuration is valid.

The same observation and assumption are made when we consider the branching ratio $\btau$, using the average $\btau=(1.67\pm0.39)\times10^{-4}$ \cite{Asner:2010qj}. 
\begin{figure}[!htb]
\begin{center}
\centering
\subfloat[]{\includegraphics[width=8cm,height=9cm]{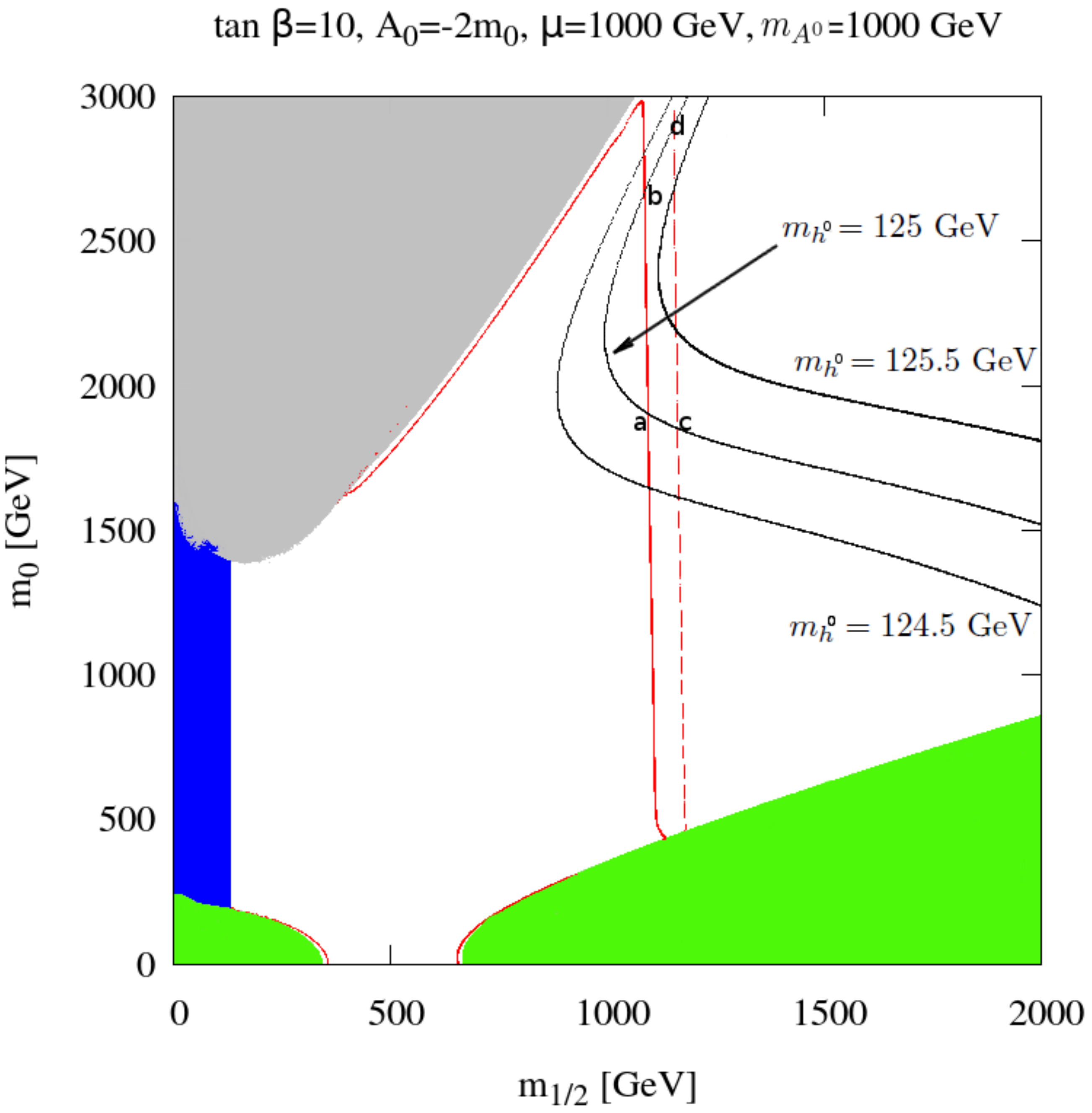}}
\subfloat[]{\includegraphics[width=8cm,height=9cm]{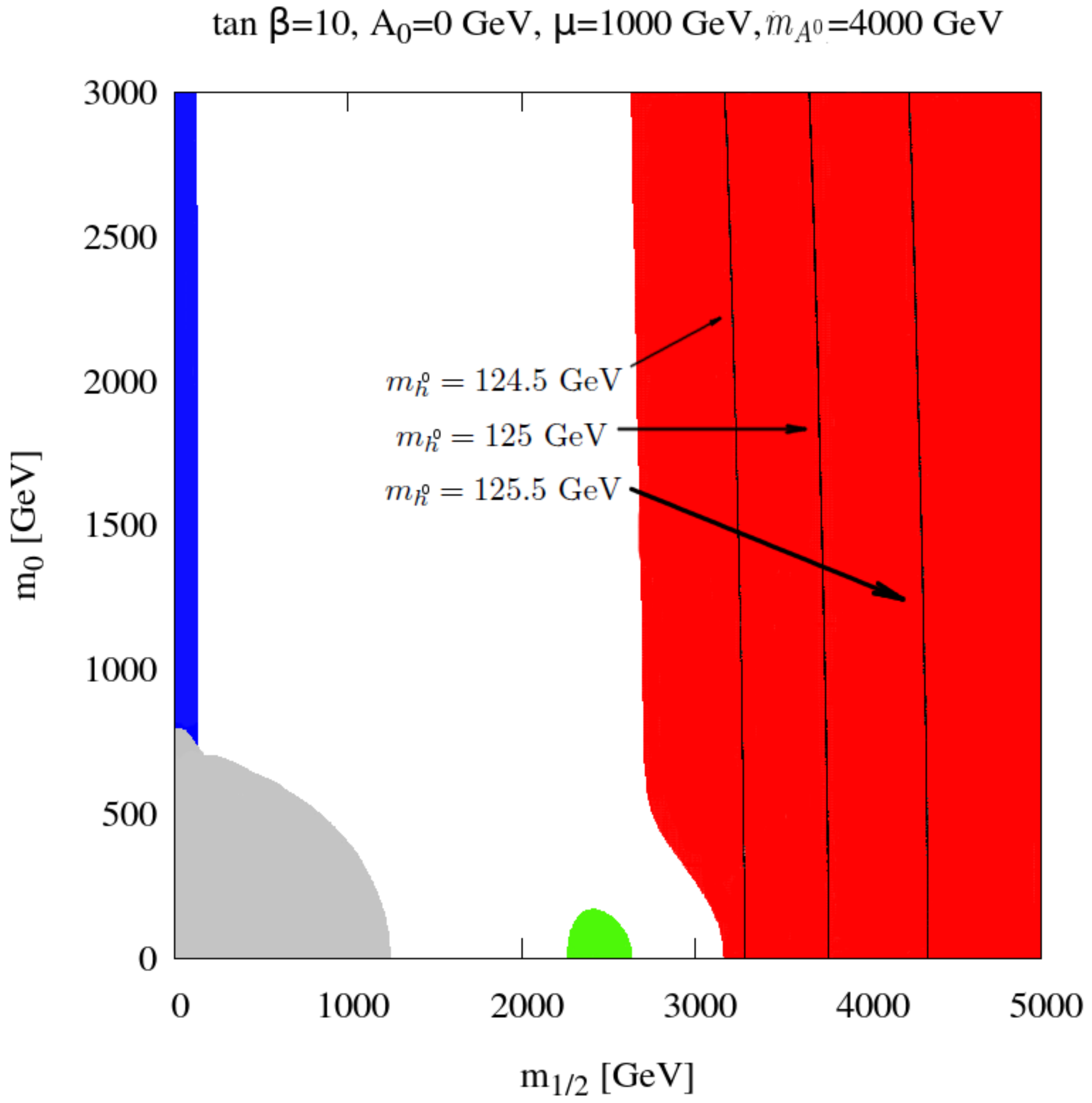}}
\caption[$(m_0,m_{1/2})$ plane for the NUHM2 model : Panel (a) is for $A_0=-2m_0$, $m_{A^0}=1$ TeV. Panel (b) is for $A_0=0$ TeV and $m_{A^0}=4$ TeV. Both panels share $\tan\beta=10$ and $\mu=1$ TeV.]{$(m_0,m_{1/2})$ plane for the NUHM2 model : Panel (a) is for $A_0=-2m_0$, $m_{A^0}=1$ TeV. Panel (b) is for $A_0=0$ TeV and $m_{A^0}=4$ TeV. Both panels share $\tan\beta=10$ and $\mu=1$ TeV. WMAP allowed regions are drawn in red while black lines show the interesting Higgs boson mass regions. Green regions are characterized by stau LSP. LEP2 bounds on the chargino mass is not satisfied in blue regions and non physical configurations are found in grey regions.}
\label{fig:4.SUSY_points}
\end{center}
\end{figure}

We show two examples of scan in figure~\ref{fig:4.SUSY_points}. Regions where the LSP is not a neutralino but a stau are coloured in green and regions excluded by LEP2 limits on the chargino mass are represented in blue. Grey regions correspond to non physical configurations (in particular the stop is tachyonic in most of these regions). In panel (a) of this figure the regions of the parameter space where the neutralino relic density is in agreement with the WMAP observations is represented by a red strip while panel (b) present a wider WMAP allowed region. 

Since we are looking for points which satisfy both the Higgs boson mass and DM constraints, benchmark scenarios are defined as 
the points which lie at the intersection between the red area representing the interesting DM relic density and the black line corresponding to a Higgs boson mass of $m_{h^0}=125$~GeV. 

In panel (a) of figure~\ref{fig:4.SUSY_points} configurations intersecting WMAP allowed and $m_{h^0}\approx 125$ GeV lines are found for $m_0 > 1.5$ TeV and $m_{1/2} > 1$ TeV. In this part of the plot allowed by relic density constraint, the LSP is mostly bino since $\mu$ is fixed at 1~TeV. To explain the observed abundance, given that binos interact too weakly with SM particles and then usually overclose the Universe, the neutralino mass must be such that annihilation via a CP-odd Higgs s-channel exchange is enhanced. This leads to the relation $m_{\chi_1^0}\approx m_{A^0}/2$ and thus implies that the neutralino mass is about $m_{\chi_1^0}\approx 500$ GeV for $m_{A^0}=1$ TeV. The allowed region is actually referred to as the funnel region. The mass of a bino LSP is roughly equal to $M_1$, which is related at the EW scale to $m_{1/2}$ via $M_1\approx 0.42 m_{1/2}$. Thus $m_{\chi_1^0}\approx M_1\approx 500$~GeV corresponds to $m_{1/2} \approx 1.1$~TeV. Between the two red strips, the relic density falls below the observed DM abundance because the annihilation process becomes resonant and reduces too much the relic density. Four benchmark points are identified for $m_{h^0}=125$ GeV. They are given by $m_0 = 1897,~ 2668,~ 1847,~ 2897$~GeV with $m_{1/2}\approx 1.1$~TeV and correspond respectively to the benchmark points `a',~`b',~`c' and `d'. These points match well the $B$-physics constraints; furthermore they are close to the SM expectations : $\bsmu_{a,b,c,d} \approx 3.1\times10^{-9}$ and $\bsg_{a,b,c,d} \approx 3.25\times10^{-4}$. Note that the choice of the Higgs boson mass for defining our benchmark points is crucial. Indeed this plot shows that increasing $m_{h^0}$ by just 500 MeV takes away half the resonance region to the interesting configurations. It gives only two benchmark points with $m_0 > 2$~TeV and $m_{1/2}\approx 1.1$~TeV.

In panel (b) of figure~\ref{fig:4.SUSY_points} the situation is completely different. Since $m_{A^0}=4$ TeV, no funnel region is expected until large $m_{1/2}$ values, typically around 4~TeV. Moreover since $\mu = 1$~TeV it is expected in this scenario to have a mainly higgsino-like lightest neutralino of a mass slightly higher than 1 TeV for sufficiently large $m_{1/2}$. It implies that there is a large degeneracy between the two lightest neutralinos and the lightest chargino. Thus if the neutralino is the LSP the t-channel chargino/neutralino exchange process and the neutralino-neutralino and neutralino-chargino coannihilation mechanisms contribute significantly to the LSP relic density. The consequence is that, within the explored range of $m_0$, any scalar masses at the GUT scale is compatible with $m_h=125.0\pm0.5$ GeV and the observed DM abundance, which leads to many more benchmark points than in the previous scenario. As a result it is clear that a broader scan of the parameter space is expected to give mainly higgsino LSP for the interesting configurations.

\subsection{A broader scan of the parameter space}

In the previous subsection we have identified a set of parameters for which the Higgs mass coincided with $m_{h^0}\approx$ 125 GeV, and simultaneously lead to a DM relic density compatible with WMAP observations. We now want to check whether predictions associated with these benchmark points are generic or not.

\subsubsection{A \textit{Markov Chain Monte Carlo} inspired algorithm}

\begin{figure}[!htb]
\begin{center}
\centering \noindent
\includegraphics{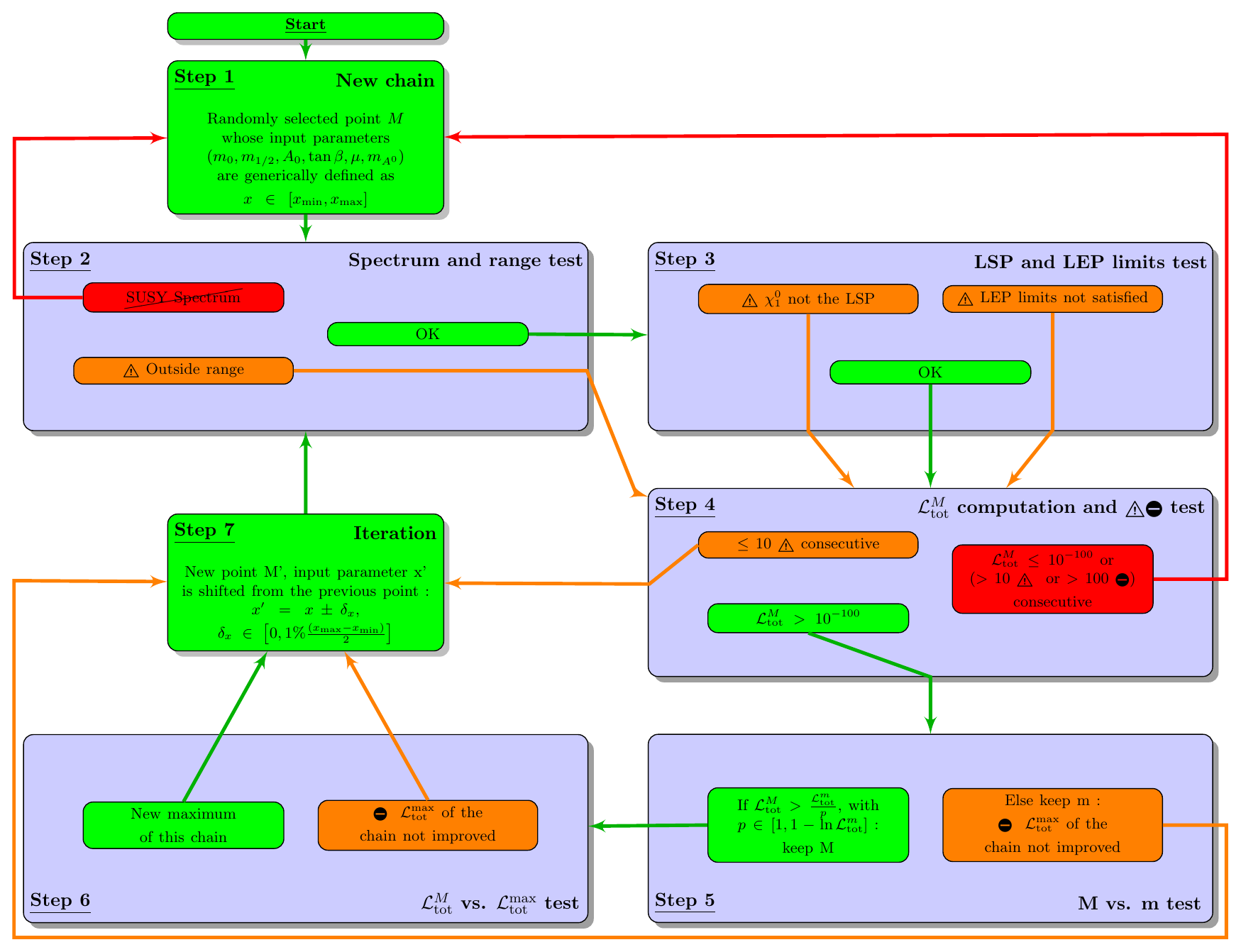}
\caption[Representation of the overall MCMC procedure used in this work.]{Representation of the overall MCMC procedure used in this work. Green paths represent the ideal paths while orange paths represent warning or \textit{stuck} signal paths. Red paths represent the rejection of a chain.}
\label{fig:4.MCMC}
\end{center}
\end{figure}

The multidimensional parameter space of the model $\{m_{0}, m_{1/2}, A_{0}, \tan \beta, \mu, m_{A^0}\}$ is explored by performing a random walk corresponding to a Markov Chain Monte Carlo (MCMC) analysis. This exploration based on the \micro code follows the procedure of a Metropolis-Hastings algorithm \cite{Metropolis:1953am} and is described by the following steps :
\begin{enumerate}
\item \label{MCMC:1} Let us start with a point $M(m_{0}, m_{1/2}, A_{0}, \tan \beta, \mu, m_{A^0})$ in a given location of a chain whose free parameters are selected inside a defined range $[x_\mathrm{min}, x_\mathrm{max}]$ (randomly if it is the first point of the chain, otherwise see next steps) where $x_\mathrm{min}$ ($x_\mathrm{max}$) is the lower (upper) bound of the range for a given free parameter $x$. We assume flat priors for all free parameters considered;
\item \label{MCMC:2} The SUSY spectrum is then computed using {\tt SOFTSUSY}. If it corresponds to a non physical configuration (tachyonic sfermion, Landau pole below the GUT scale,...), the point (and the corresponding chain) is rejected and we go back to step \ref{MCMC:1}. If this point is not the first one of the chain considered it was slightly shifted from the previous point. Then it might go beyond one of the intervals for the free parameters defined at step \ref{MCMC:1}. If it is the case we assign a warning signal to this point and we move to step \ref{MCMC:4};
\item \label{MCMC:3} After this we check if the point $M$ is characterized by a neutralino LSP and is allowed by LEP limits on sparticle masses. If it does not fulfill one of these conditions we assign a warning signal to this point;
\item \label{MCMC:4} Before discussing about the likelihood calculations and if a warning signal has been imposed on the point either in step~\ref{MCMC:2} or step \ref{MCMC:3}, we check that at most ten consecutive warning signals have been obtained in this chain. If it is the case we move directly to step \ref{MCMC:7} with the assignment $N=M$ where $N$ will be the point considered. Another signal that will be introduced at the step~\ref{MCMC:6}, the \textit{stuck} signal, is defined for the points that do not correspond to the highest total likelihood function of the chain. This signal allows to not stay closed in a region of the parameter space which was sufficiently analysed. We then check whether no more than one hundred consecutive \textit{stuck} signals were obtained in this chain. If it is the case we continue this step. Then we compute the total likelihood function $\mathcal{L}_\mathrm{tot}^M$ which is equal to the product of the likelihood functions associated with each observable that we will define in section~\ref{subsubsec:4.scancarac}. We continue to step \ref{MCMC:5} if it satisfies the quite conservative criterion $\mathcal{L}_\mathrm{tot}^M > 10^{-100}$. Otherwise if the point fails to fulfill one of the three tests presented above we reject the chain and we go back to step \ref{MCMC:1}.
\item \label{MCMC:5} We now look at the main part of the algorithm. If $M$ is the first point of the chain we automatically keep the point that we note $N=M$ and we move to step~\ref{MCMC:7}. Otherwise we compare $M$ to the previous point kept in this chain that we will call $m$. To do this we define a new parameter $p$ which is taken randomly in the range $[1, 1-\ln \mathcal{L}_{\mathrm{tot}}^m]$ and we compare $\mathcal{L}_\mathrm{tot}^M$ to $\mathcal{L}_{\mathrm{tot}}^m/p$. If $\mathcal{L}_{\mathrm{tot}}^M > \mathcal{L}_{\mathrm{tot}}^m/p$, M and all its characteristics like the SUSY spectrum or its associated values for each observable considered are kept and $N=M$. If not we keep the previous point, which means that $N=m$, and we move to step \ref{MCMC:7};
\item \label{MCMC:6} The last important step for the accepted point $M$ is to verify if it gives the highest total likelihood function of the chain. If it is not the case we assign a \textit{stuck} signal to this point (which is automatically assigned at the point $m$ when $\mathcal{L}_{\mathrm{tot}}^M \leq \mathcal{L}_{\mathrm{tot}}^m/p$); 
\item \label{MCMC:7} We now generate, starting from $N$, a new point $M'$. Its corresponding free parameters $x'$ are shifted from those of $N$ as $x' = x + \delta_x$ where $\delta_x$ is randomly selected inside the range $\left[-1\% \frac{(x_\mathrm{max}-x_\mathrm{min})}{2}, 1\% \frac{(x_\mathrm{max}-x_\mathrm{min})}{2}\right]$. Finally we close the loop procedure returning to step \ref{MCMC:2} and doing the same tests for the point $M'$.
\end{enumerate}
This algorithm is also described in figure~\ref{fig:4.MCMC}. We could think to use a more stringent criterion than $\mathcal{L}_\mathrm{tot}^M > 10^{-100}$. Nevertheless we observed that the code converge quite quickly to interesting configurations,\ie $\mathcal{L}_\mathrm{tot}^M > 10^{-1}$, thanks to the range chosen for the parameter $p$, $[1, 1-\ln \mathcal{L}_{\mathrm{tot}}^m]$. Actually we also tested $p \in [1, 10]$ but the result was less interesting : regions which are fine-tuned were left unexplored. The reason why the first method is more efficient in the exploration of fine-tuned regions is that for high $\mathcal{L}_\mathrm{tot}^m$ it is almost mandatory to have $\mathcal{L}_{\mathrm{tot}}^M \gtrsim \mathcal{L}_{\mathrm{tot}}^m$. Thus it is possible to fully explore fine-tuned regions, avoiding the previous problem. 

Note that this algorithm does not completely exclude regions where the LSP is not the neutralino and where some sparticle masses are below LEP bounds using the warning signals. The reason is that we do not want to get rid of the possibility that the corresponding point is near a region where LSP-NLSP coannihilation mechanism contribute to the relic density of DM or that the chargino mass is for instance close to LEP limits. 
  
Since we were not interested in characterising how statistically relevant the points that we found are but wanted instead to determine the full range of configurations that are possible,  we do not consider the number of occurrence of a given scenario. The drawback of such a method is that we cannot determine how likely a region of the parameter space is with respect to other regions. The advantage as explained above is that very small (fine-tuned) configurations are kept in the analysis. We imposed between $10^6$ and $10^7$ loops which led to more than $10^5$ interesting configurations for each scan.

\subsubsection{Characteristics of the scan}
\label{subsubsec:4.scancarac}

Now let us look at the physics part of this algorithm, the computation of the likelihood functions associated with each observable. We want to identify regions of the parameter space which lead to a SM Higgs boson mass within the range that was not excluded by the LHC experiments in early 2012,\ie $[115.5, 127]$ GeV~\cite{ATLAS:2012ae,Chatrchyan:2012tx}, and a neutralino relic density within the WMAP measurements, namely $\Omega_{\chi^0_1} h^2 \in [0.1088,0.1158]$. Note that we also considered a scan where the neutralino is only one of the components of DM, asking $\Omega_{\chi^0_1} h^2 \in [0.01123, 0.1123]$. We will make a quick comment on this point at the end of section~\ref{subsubsec:4.results}. Following the lines described in \cite{Vasquez:2010ru}, we use three kind of likelihood functions :
\begin{itemize}
\item For the Higgs boson mass and the DM relic density, we define the corresponding likelihood as a function $\mathcal{L}_1$ which decays exponentially at the edges of the $[x_{min}, x_{max}]$ range, according to 
\beq \begin{split} \label{eq:4.L1}
\mathcal{L}_1(x, x_{min}, x_{max}, \sigma)  = & \: e^{-\frac{\left( x - x_{min}\right)^2}{2\sigma ^2}} \: \textrm{if} \: x < x_{min},\\
= & \: e^{-\frac{\left( x - x_{max}\right)^2}{2\sigma ^2}} \: \textrm{if} \: x > x_{max},\\
= & \: 1 \: \textrm{for} \: x \in [x_{min}, x_{max}],
\end{split} \eeq
with a tolerance $\s$ and $x$ is the observable which corresponds in that case to either the Higgs boson mass or the LSP relic density;   
\item For all the other observables, we will use two types of likelihood :
\begin{itemize}
\item For an observable with a preferred value $\mu$ and an error $\s$, we use a Gaussian distribution $\mathcal{L}_2$ :
\beq \label{eq:4.L2} \mathcal{L}_2(x, \mu, \sigma) = e^{-\frac{\left(x-\mu\right)^2}{2\sigma ^2}}.\eeq
Note that the $\s$ of the $\bsg$ observable is calculated adding quadratically theoretical and experimental errors;
\item For an observable with a lower or upper bound (set experimentally), we will take the function $\mathcal{L}_3$ with a positive or negative variance $\s$ :
\beq \label{eq:4.L3} \mathcal{L}_3(x, \mu, \sigma) = \frac{1}{1+e^{-\frac{x-\mu}{\sigma}}}.\eeq
\end{itemize}
\end{itemize}
Here we do not implement limits on sparticle masses from the LHC since squark masses that we consider are mostly above present limits. The known constraints that we impose are summarized in table~\ref{tab:4.constraints} and the range that we consider for the different free parameters of the NUHM2 model is given in table~\ref{tab:4.range}. Note that we also took into account some EW observables although they were found to not constrain scenarios obtained since they mainly apply for light LSP (below\ca 50 GeV). They are the invisible decay width of the $Z$ boson and the cross section $\sigma_{e ^+ e ^- \rightarrow \chi^0_1 \chi^0_{2,3}} \times \mathscr{B}(\chi^0_{2,3} \rightarrow Z \chi^0_1)$. Note also that the $\Delta \rho$ parameter, which determines the deviation of the $\rho$ parameter from unity (recall eq.~\ref{eq:rhoSM}), did not constrain the parameter space considered.

\begin{table*}[!htb]
\begin{tabular*}{1.\textwidth}{  c  c  c  c  }
\hline \hline \textbf{Constraint} & \textbf{Value/Range} & \textbf{Tolerance $\s$} & \textbf{Likelihood} \\ 
\hline \hline
       $m_{h^0}$ (GeV) \cite{ATLAS:2012ae,Chatrchyan:2012tx} & [115.5, 127] & 1 & $\mathcal{L}_1$ \\ \hline 
       $\Om_{\chi^0_1} h^2$ \cite{Komatsu:2010fb} & [0.1088, 0.1158] & 0.0035 & $\mathcal{L}_1$ \\ 
       Relaxing constraint on $\Om_{\chi^0_1} h^2$ & [0.01123, 0.1123] & 0.0035 & $\mathcal{L}_1$ \\\hline
       $\bsg$ $\times$ $10^{4}$ & 3.55 & exp : 0.24, 0.09 & $\mathcal{L}_2$ \\  \cite{Asner:2010qj,Misiak:2006zs} && th : 0.23 & \\ \hline
       $\amu$ $\times$ $10^{10}$ \cite{Davier:2010nc} & 28.7 & -8 & $\mathcal{L}_3$ \\ \hline 
       $\bsmu$ $\times$ $10^{9}$ \cite{Aaij:2012ac} & 4.5 & -0.045 & $\mathcal{L}_3$ \\ \hline        
       $\Delta \rho$ & 0.002 & -0.0001 & $\mathcal{L}_3$ \\ \hline 
       $R_{B^\pm \to \tau^\pm \nu_\tau} (\frac{\mathrm{NUHM2}}{\mathrm{SM}})$ \cite{Charles:2011va} & 2.219 & -0.5 & $\mathcal{L}_3$ \\ \hline 
       $Z \ra \chi^0_1 \chi^0_1$ (MeV) & 1.7 & -0.3 & $\mathcal{L}_3$ \\ \hline 
       $\sigma_{e ^+ e ^- \rightarrow \chi^0_1 \chi^0_{2,3}}$ & 1 & -0.01 & $\mathcal{L}_3$ \\ 
$\times \mathscr{B}(\chi^0_{2,3} \rightarrow Z \chi^0_1)$ (pb) \cite{Abbiendi:2003sc} &&& \\ 
\hline \hline
\end{tabular*}
\caption[Constraints imposed in the MCMC for the NUHM2 model.]{\label{tab:4.constraints}Constraints imposed in the MCMC, from \cite{Nakamura:2010zzi} unless noted otherwise.}
\end{table*}

\begin{table}[!htb]
\begin{center}
\begin{tabular}{cc}\hline \hline
\textbf{Parameter} & \textbf{Range} \\ \hline \hline
$m_0$ & [0, 4] TeV\\ 
$m_{1/2}$ & [0, 4] TeV\\ 
$A_0$ & [-6, 6] TeV\\ 
$\tan \beta$ & [2, 60]\\
$\mu$ & [0, 3] TeV\\ 
$m_{A^0}$ & [0, 4] TeV\\ \hline \hline
\end{tabular}
\caption{Range chosen for the free parameters in the NUHM2 model. \label{tab:4.range}}
\end{center}
\end{table}

\subsubsection{Results}
\label{subsubsec:4.results}

\begin{figure}[!htb]
\begin{center}
\subfloat[]{\includegraphics[width=8cm,height=5cm]{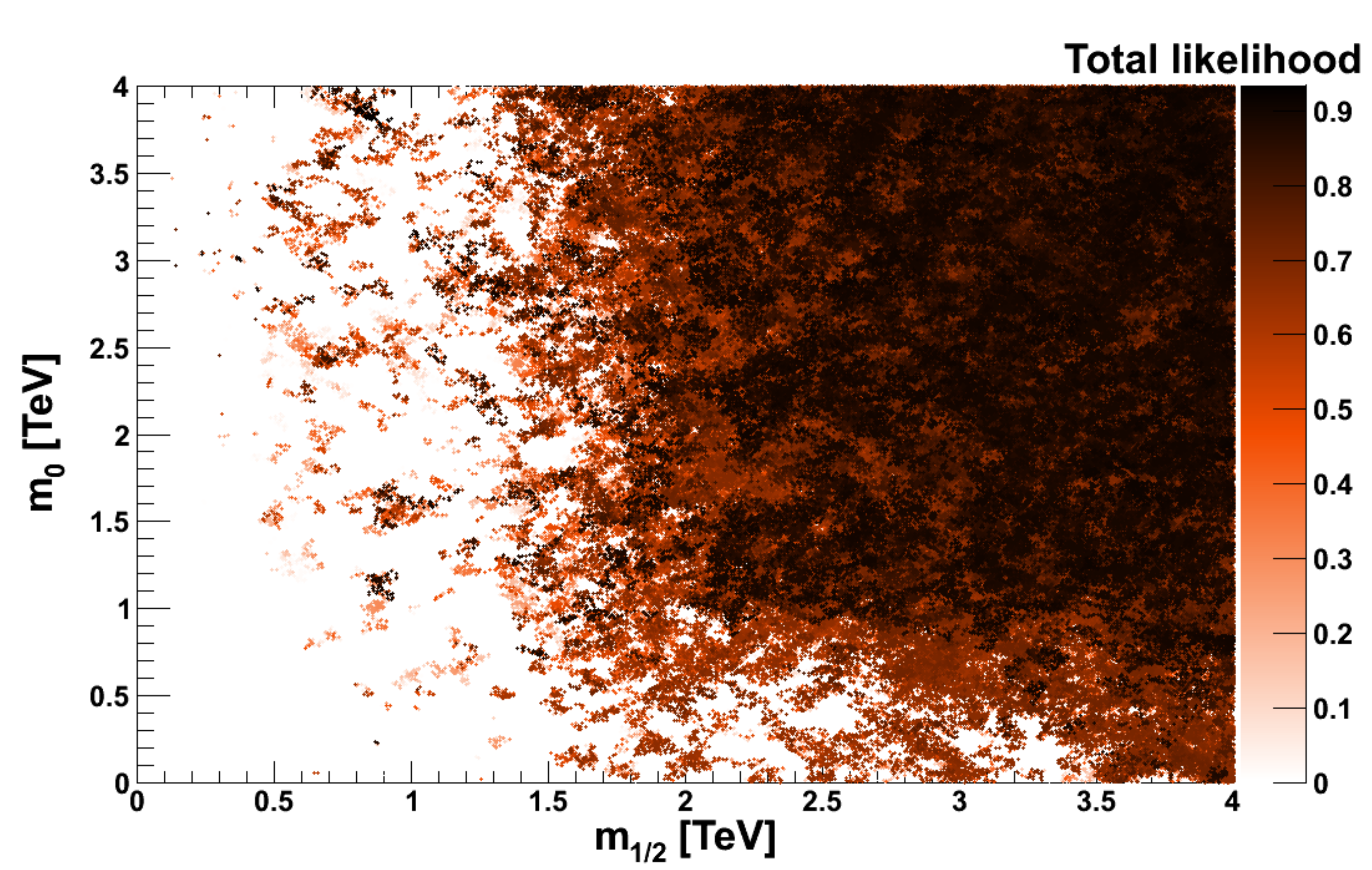}}
\subfloat[]{\includegraphics[width=8cm,height=5cm]{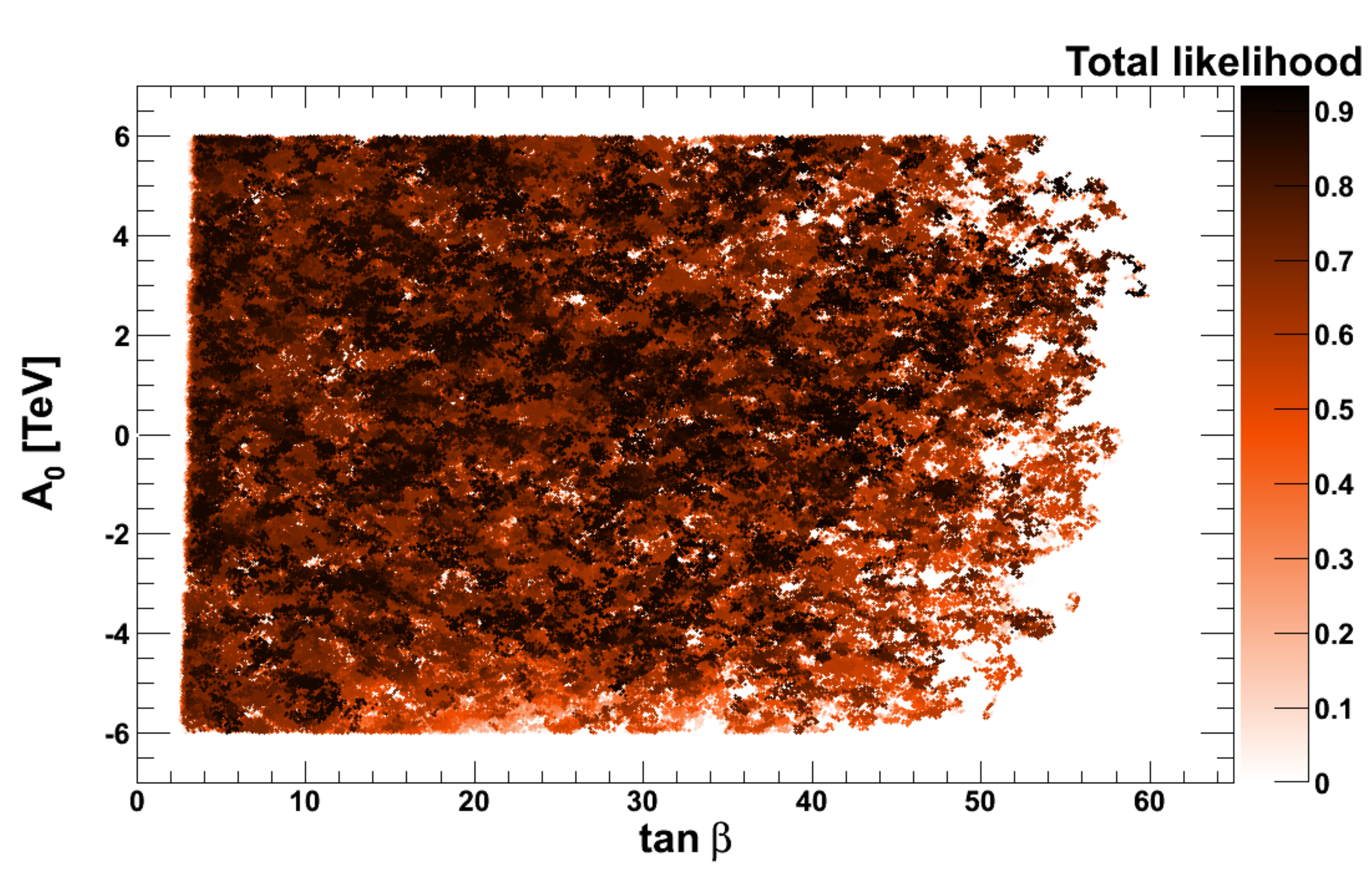}}\\
\subfloat[]{\includegraphics[width=8cm,height=5cm]{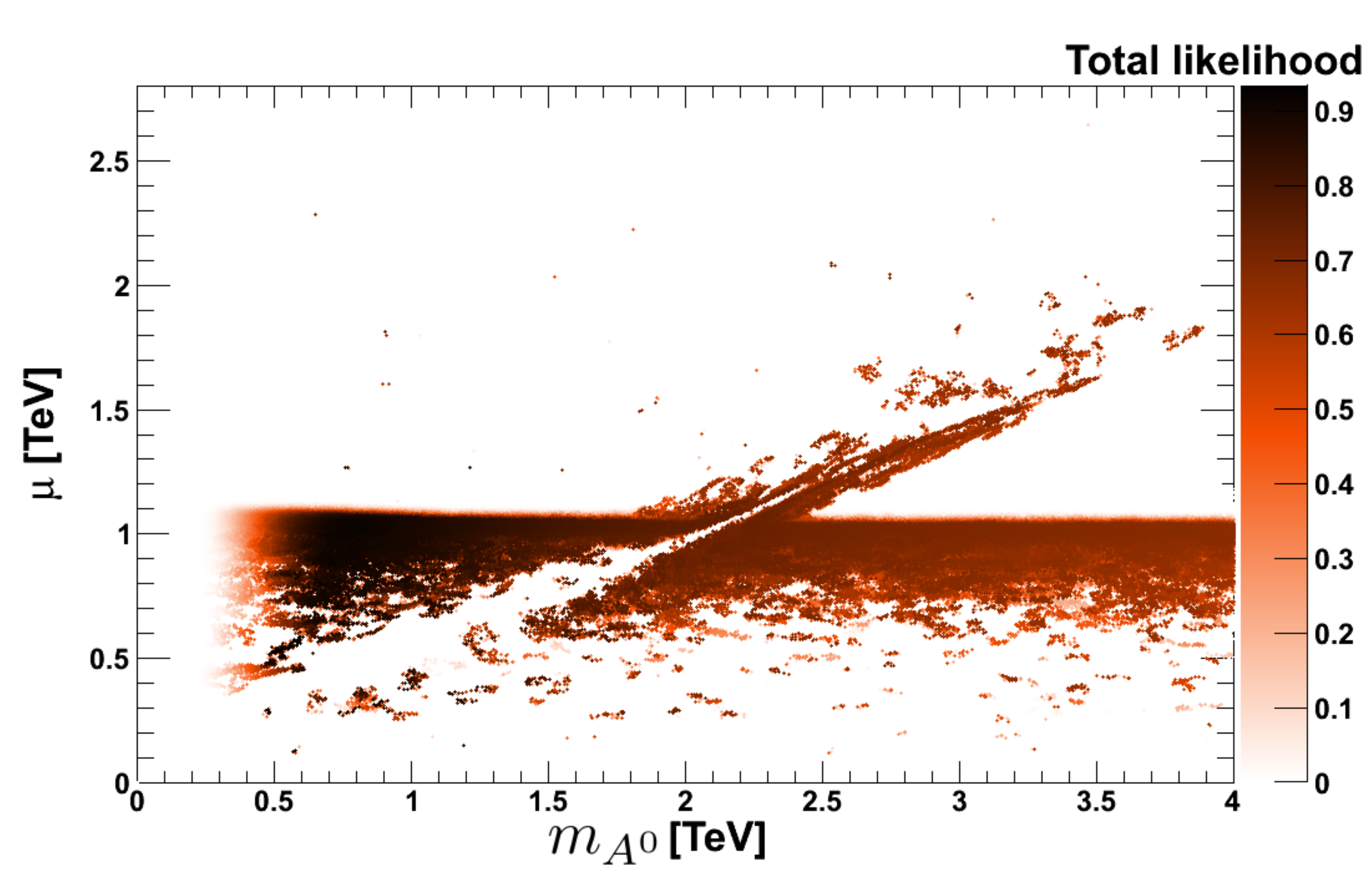}}
\caption[Plot of the allowed parameter space in the $(m_0,m_{1/2})$ (a), $(A_0,\tan \beta)$ (b) and $(\mu,m_{A^0})$ (c) planes.]{Plot of the allowed parameter space in the $(m_0,m_{1/2})$ (a), $(A_0,\tan \beta)$ (b) and $(\mu,m_{A^0})$ (c) planes. We use the total likelihood of the points as colour code. The darkest points have the highest likelihood. However they may not be statistically significant.}
\label{fig:4.freeparams}
\end{center}
\end{figure}

Panel (a) of figure~\ref{fig:4.freeparams} shows that most of the scenarios found by the MCMC involve TeV scale values of $m_0$ and $m_{1/2}$, but no real feature emerges from the plot. Panel (b) illustrates one of the main advantages of the NUHM2 model over the mSUGRA/CMSSM model : low values of $\tan \beta$ down to\ca 2 can give configurations with $m_{h^0} \gtrsim 115$~GeV. Note however that $m_{h^0} \simeq 125$~GeV cannot be reached for such low $\tan \beta$ values. Finally panel (c) shows that the $\mu$ term must be around the TeV scale to have viable scenarios. We will see later the main reason of this observation.

\begin{figure}[h]
\centering
\includegraphics[width=8cm,height=5cm]{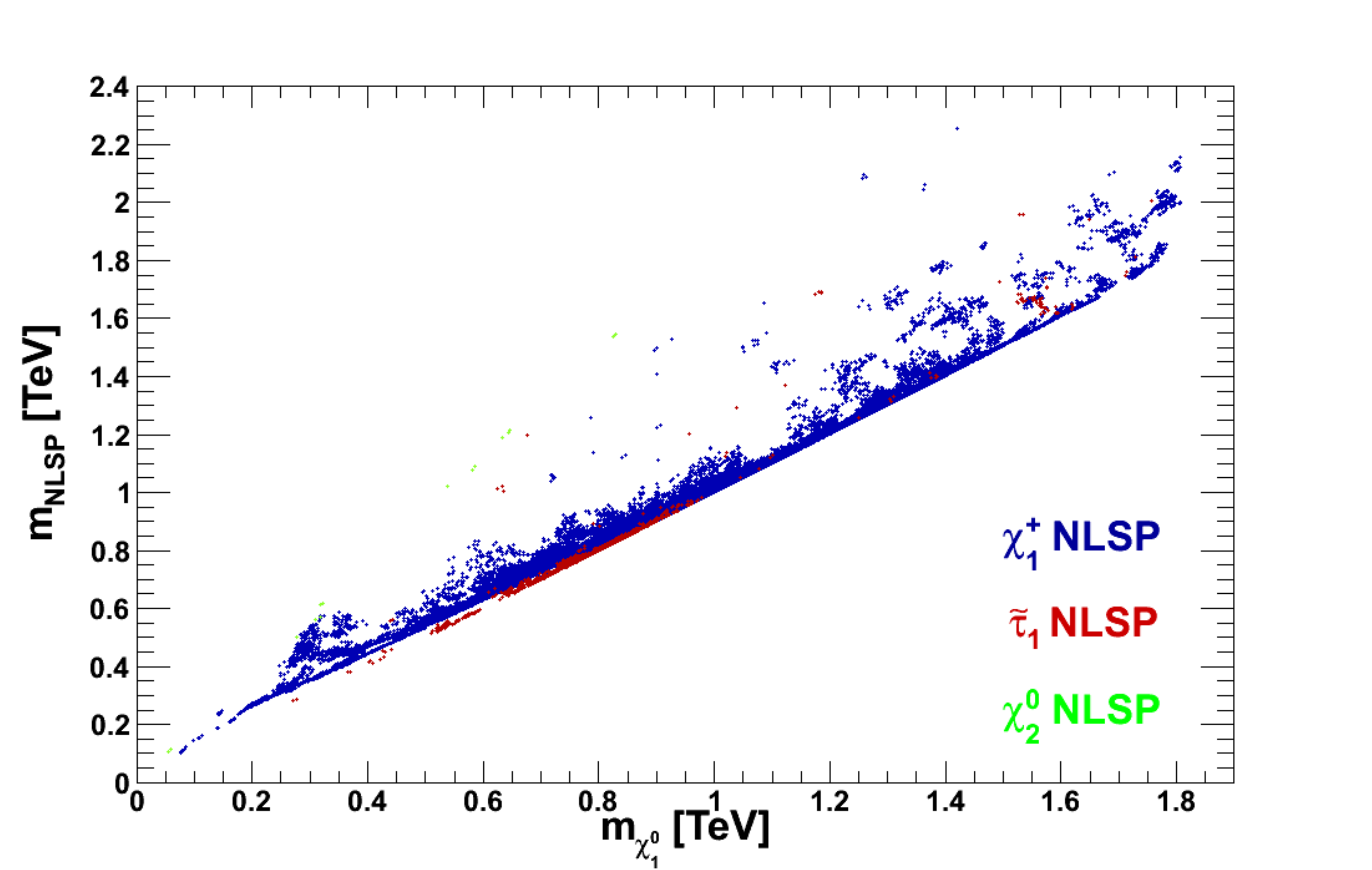}
\caption[Mass of the LSP\vs the mass of the NLSP, depending on the nature of the NLSP.]{Mass of the LSP\vs the mass of the NLSP, depending on the nature of the NLSP. Only points with large total likelihood were considered.}
\label{4.mLSP_mNLSP_mcmc}
\end{figure}

As illustrated in figure~\ref{4.mLSP_mNLSP_mcmc}, there is a very strong correlation between the mass of the LSP and that of the NLSP, suggesting that the neutralino relic density either relies on the coannihilation mechanism or a $t$-channel exchange of the NLSP (or both). The NLSP is found to be mostly a chargino and sometimes a neutralino or a stau. The $A^0$-pole resonance corresponding to the benchmark points `a',`b',`c',`d' requires however a certain amount of fine-tuning (precisely because it requires $m_{\chi^0_1} \simeq m_{A^0}/2$) and is not the most represented configuration found by the MCMC. 

\begin{figure}[h]
\centering
\includegraphics[width=8cm,height=5cm]{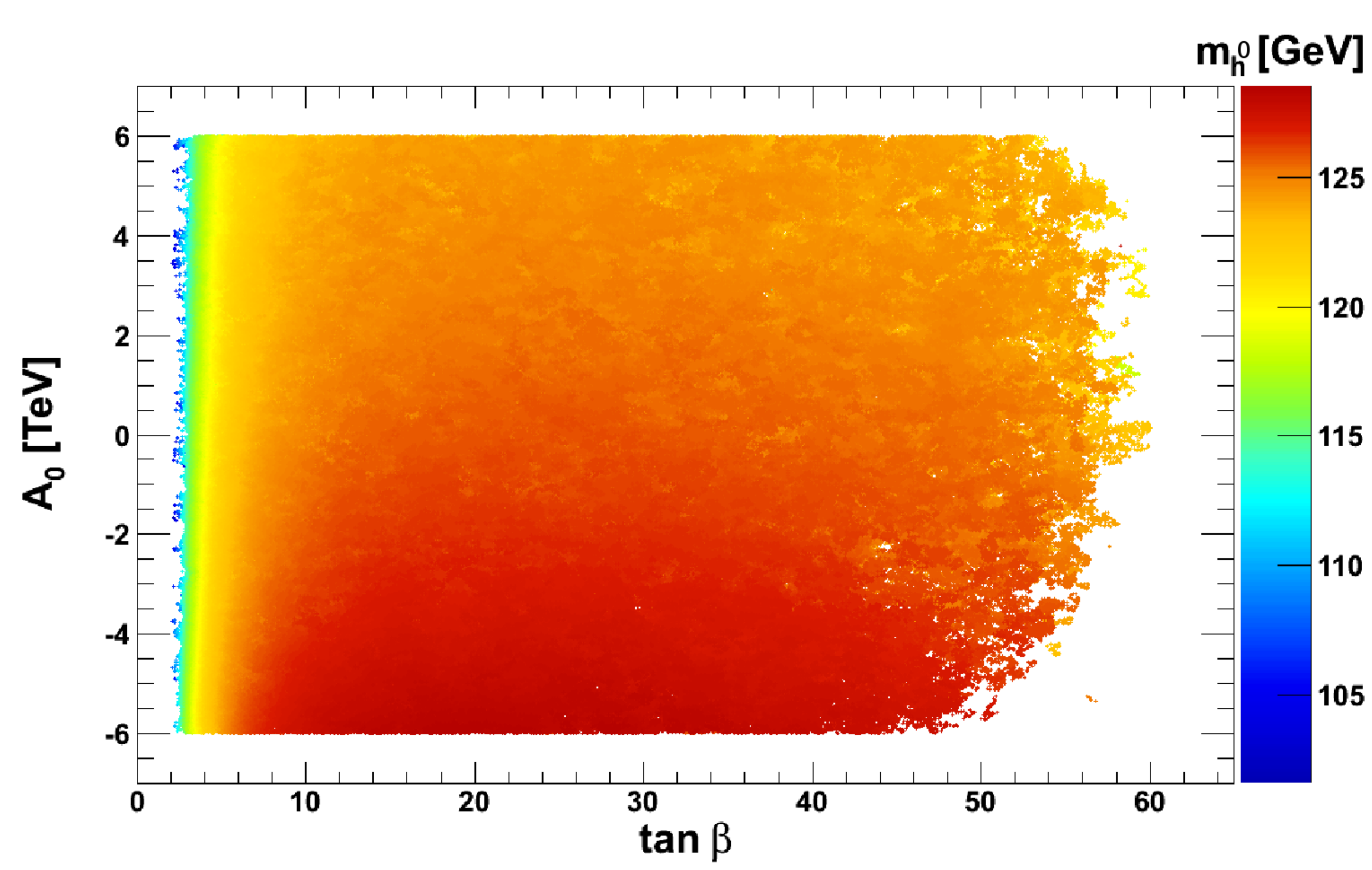}
\caption[SM Higgs boson mass in the $ (A_0,\tan \beta)$ plane.]{SM Higgs boson mass in the $ (A_0,\tan \beta)$ plane. Light Higgs boson can be found whatever the value of the trilinear coupling $A_0$, provided that $\tan \beta$ is small.} 
\label{4.a0_vs_tb_vs_mh}
\end{figure}

The predominance of scenarios in which charginos are mass degenerated with neutralinos\footnote{Note that we did not compute the amount of EW fine-tuning in our NUHM2 scenarios. It was shown, for instance in \cite{Ellis:2007by} that some NUHM2 benchmark points wherein $A_0$ = 0 TeV give non-negligible EW fine-tuning.} can be understood by inspecting figure~\ref{4.a0_vs_tb_vs_mh}. For the configurations with $A_0 = -2 m_0$, the Higgs boson mass $m_{h^0}$ tends to exceed the upper experimental bound unless one decreases the value of $\tan \beta$.

\begin{figure}[!htb]
\begin{center}
\subfloat[]{\includegraphics[width=8cm,height=5cm]{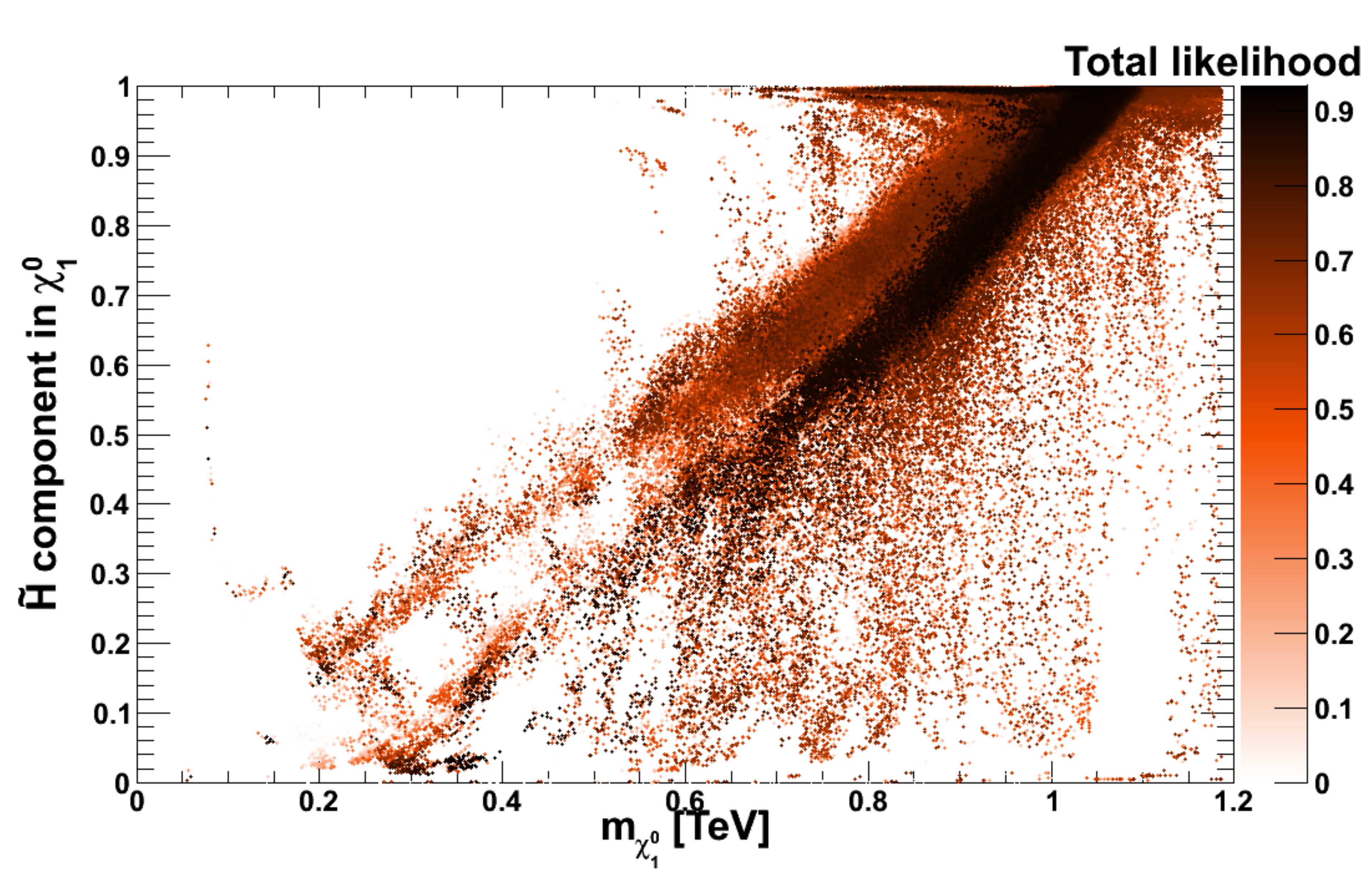}}
\subfloat[]{\includegraphics[width=8cm,height=5cm]{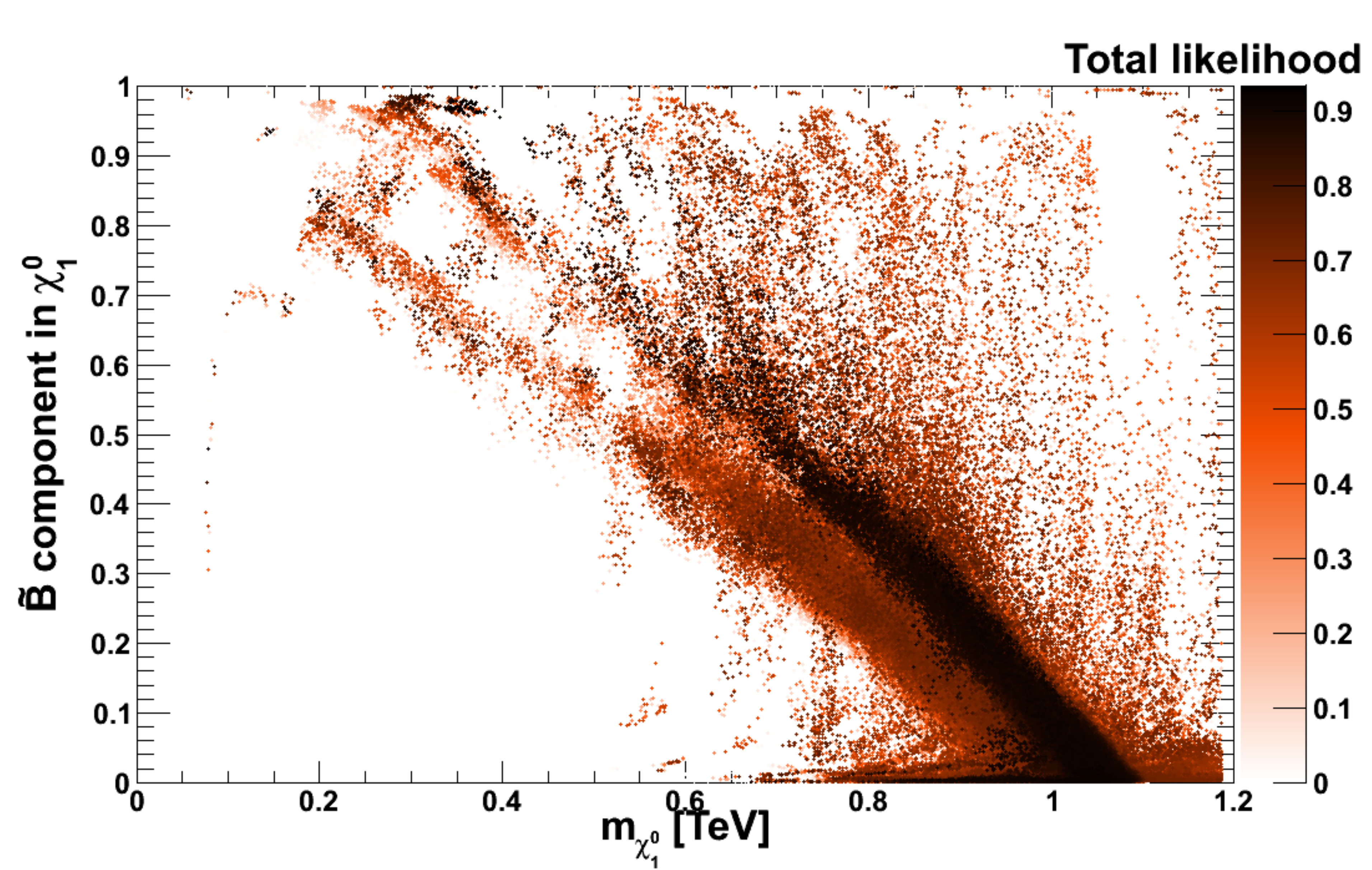}}
\caption[Neutralino composition. Panel (a) shows the higgsino content\vs the neutralino mass while panel (b) shows the bino fraction.]{Neutralino composition. Panel (a) shows the higgsino content\vs the neutralino mass while panel (b) shows the bino fraction. The colour coding corresponds to the likelihood of these points.} 
\label{4.neutralino_composition}
\end{center}
\end{figure}

For such configurations, sfermion masses are generally too large for the sfermion-neutralino coannihilation channels to reduce the relic density significantly and both the neutralino and chargino have a significant higgsino fraction as represented in panel (a) of figure~\ref{4.neutralino_composition}. As a result, the possible channels to reduce the neutralino relic density either involve CP-odd Higgs portal annihilations or neutralino-chargino mass degeneracies. 

The exchange of a pseudoscalar Higgs is actually significant when $m_{\chi^0_1} \sim m_{A^0}/2$ (as found for the benchmark points `a',`b',`c',`d') but neutralino-chargino coannihilation or chargino t-channel exchange are dominant when the higgsino fraction is very large. In fact, among the configurations with a non-negligible higgsino fraction, 
the larger the bino fraction, the more favoured the $A^0$-pole since small neutralino couplings to the Higgs can be compensated by having $m_{\chi^0_1}$ closer to $m_{A^0}/2$. The distribution of points depending on their bino fraction is represented in the panel (b) of figure~\ref{4.neutralino_composition}. Clearly scenarios with bino-like neutralinos are under represented, illustrating how fine-tuned they are. This was expected after looking at the benchmark scenarios in section~\ref{subsec:4.benchmark}. 

Note that the panel (a) of figure~\ref{4.neutralino_composition} illustrates also the point that heavy neutralinos with a mass $m_{\chi^0_1} \geq 0.6$ TeV tend to have a large higgsino fraction, thus suggesting even more dominant coannihilations with charginos (or annihilations through chargino exchange) when the neutralino becomes fairly heavy. Interestingly though, for most of these scenarios, the value of the $\mu$ parameter varies between $500$ GeV and $1.5$ TeV but the  values which correspond to the highest likelihood are about $m_{\chi^0_1} \simeq \mu \simeq 1$~TeV, which is indeed consistent with a dominant higgsino fraction.   

\begin{figure}[h]
\centering
\includegraphics[width=8cm,height=5cm]{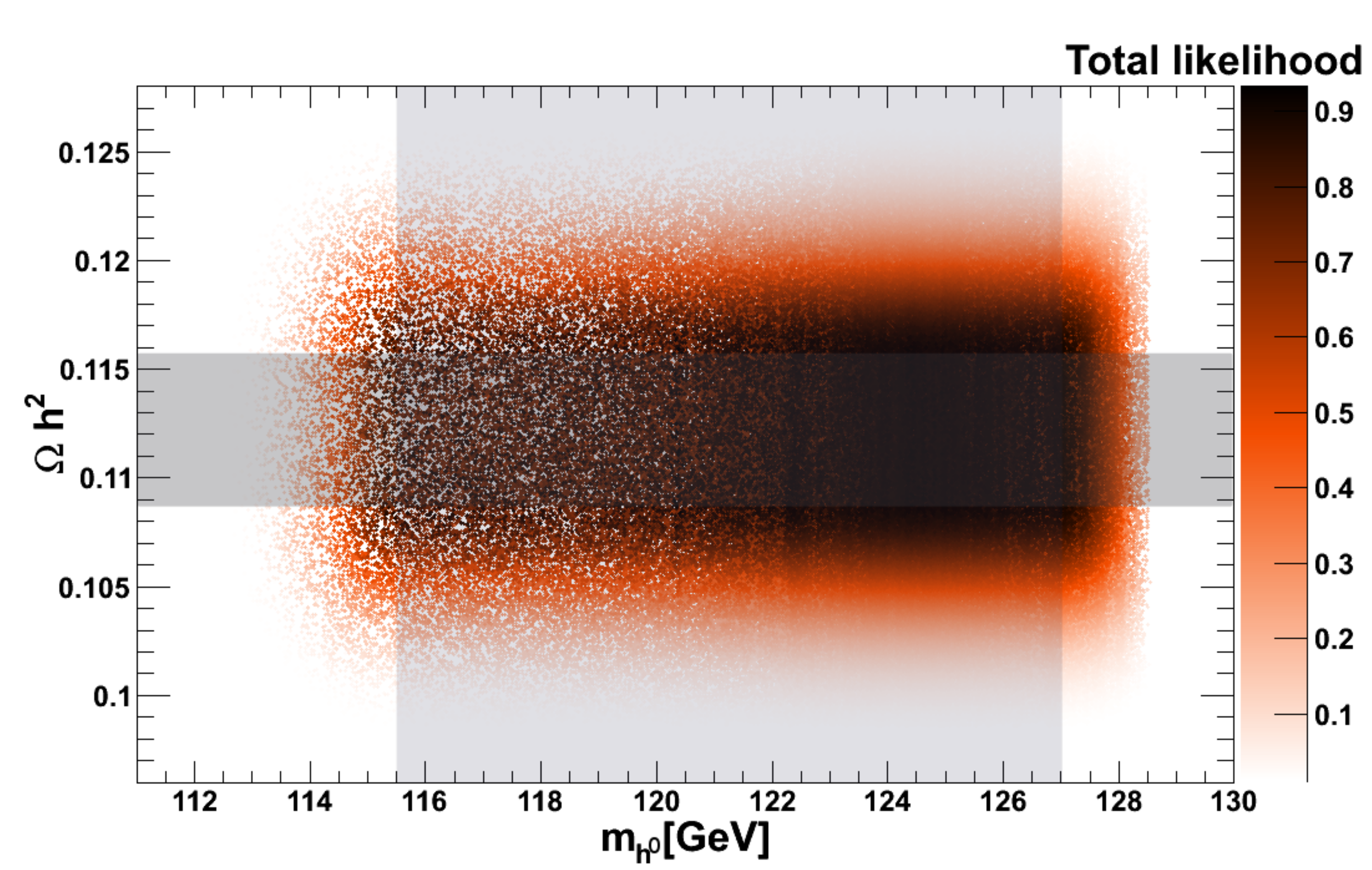}
\caption[Neutralino relic density\vs the mass of the lightest Higgs boson.]{Neutralino relic density\vs the mass of the lightest Higgs boson. Same colour code as in figure~\ref{4.neutralino_composition}.}
\label{4.RDHiggs}
\end{figure}

We can now investigate the distribution of points which satisfy constraints on the Higgs boson mass and the DM relic density, figure~\ref{4.RDHiggs}. Points with high likelihood are \textit{smoothly} distributed within the observed relic density and Higgs boson mass range. However there is a higher concentration of points corresponding to Higgs boson masses above 122 GeV, indicating that a Higgs boson mass of the NUHM2 model which is around the expected SM Higgs boson mass is in principle a better match to the particle physics constraints in such scenarios.

\begin{figure}[!htb]
\begin{center}
\subfloat[]{\includegraphics[width=8cm,height=5cm]{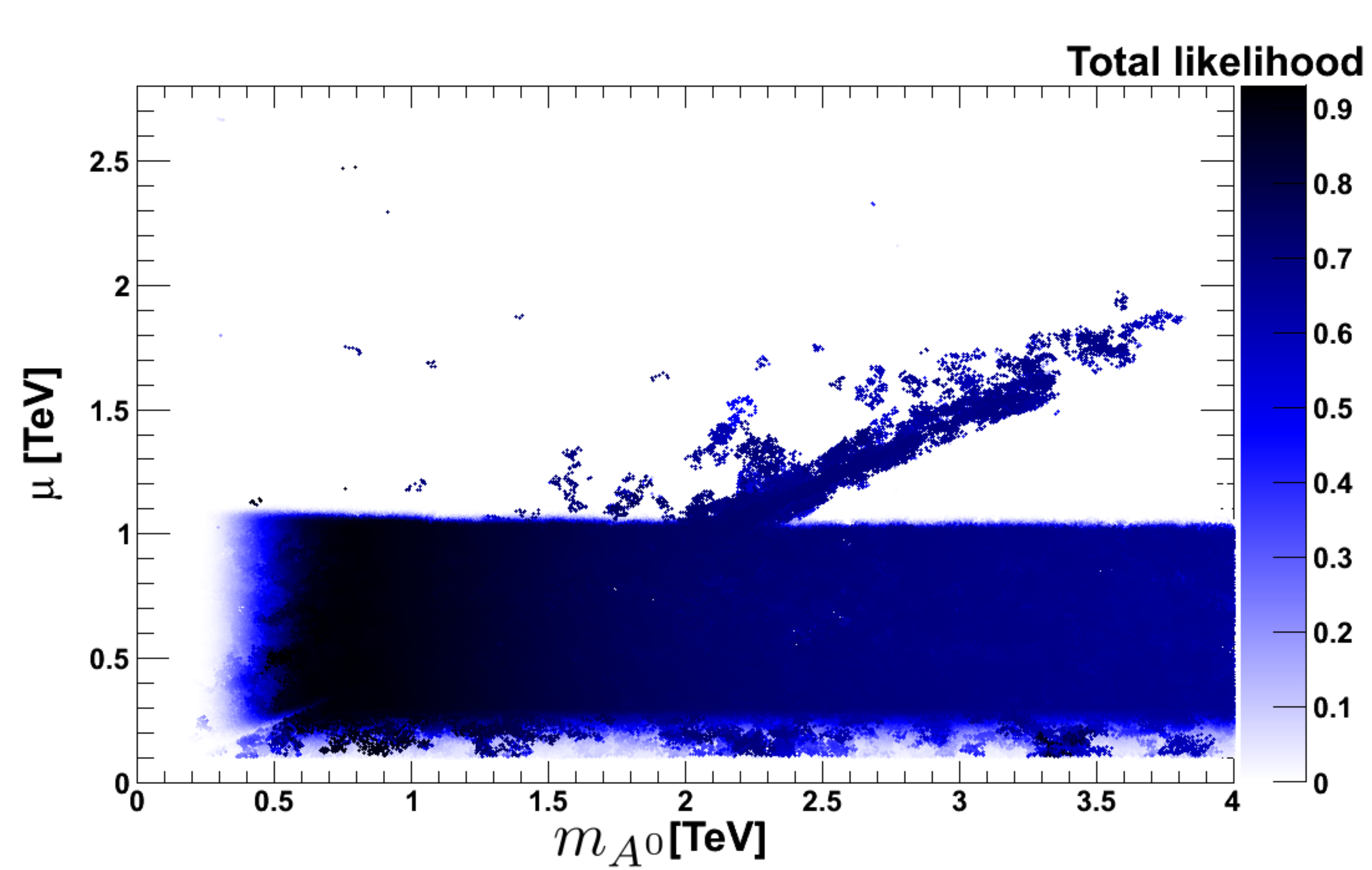}}
\subfloat[]{\includegraphics[width=8cm,height=5cm]{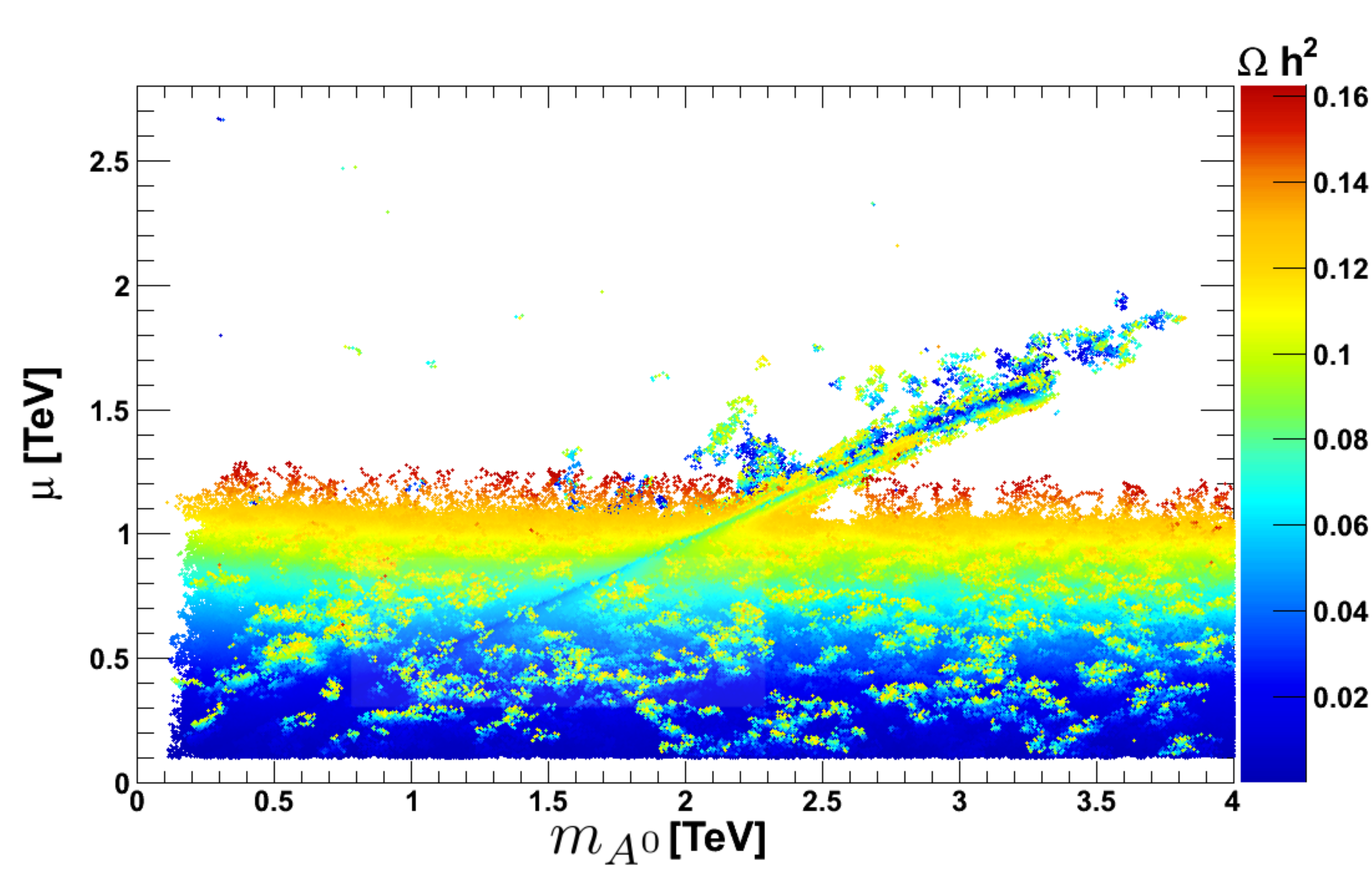}}\\
\subfloat[]{\includegraphics[width=8cm,height=5cm]{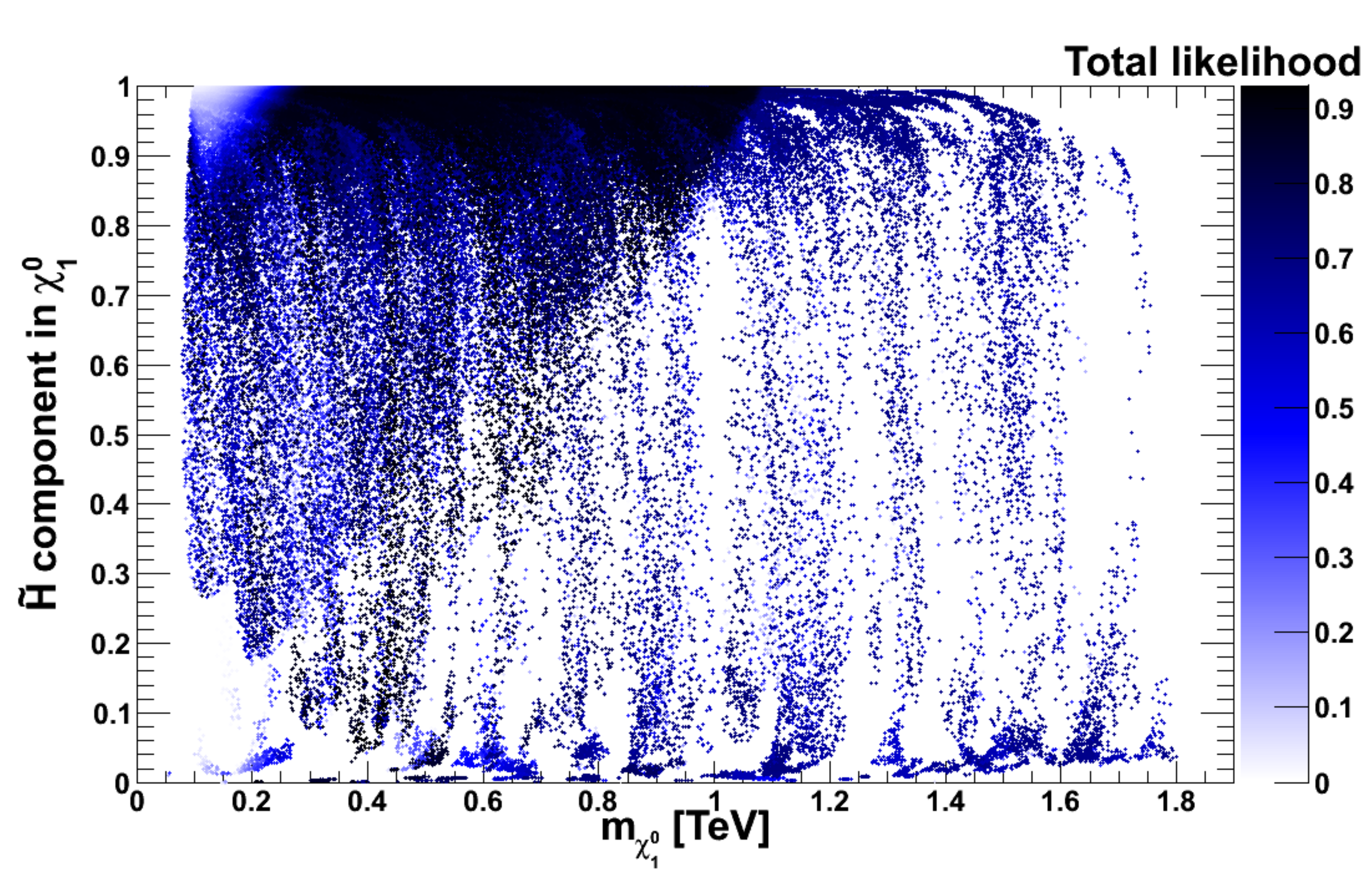}}
\subfloat[]{\includegraphics[width=8cm,height=5cm]{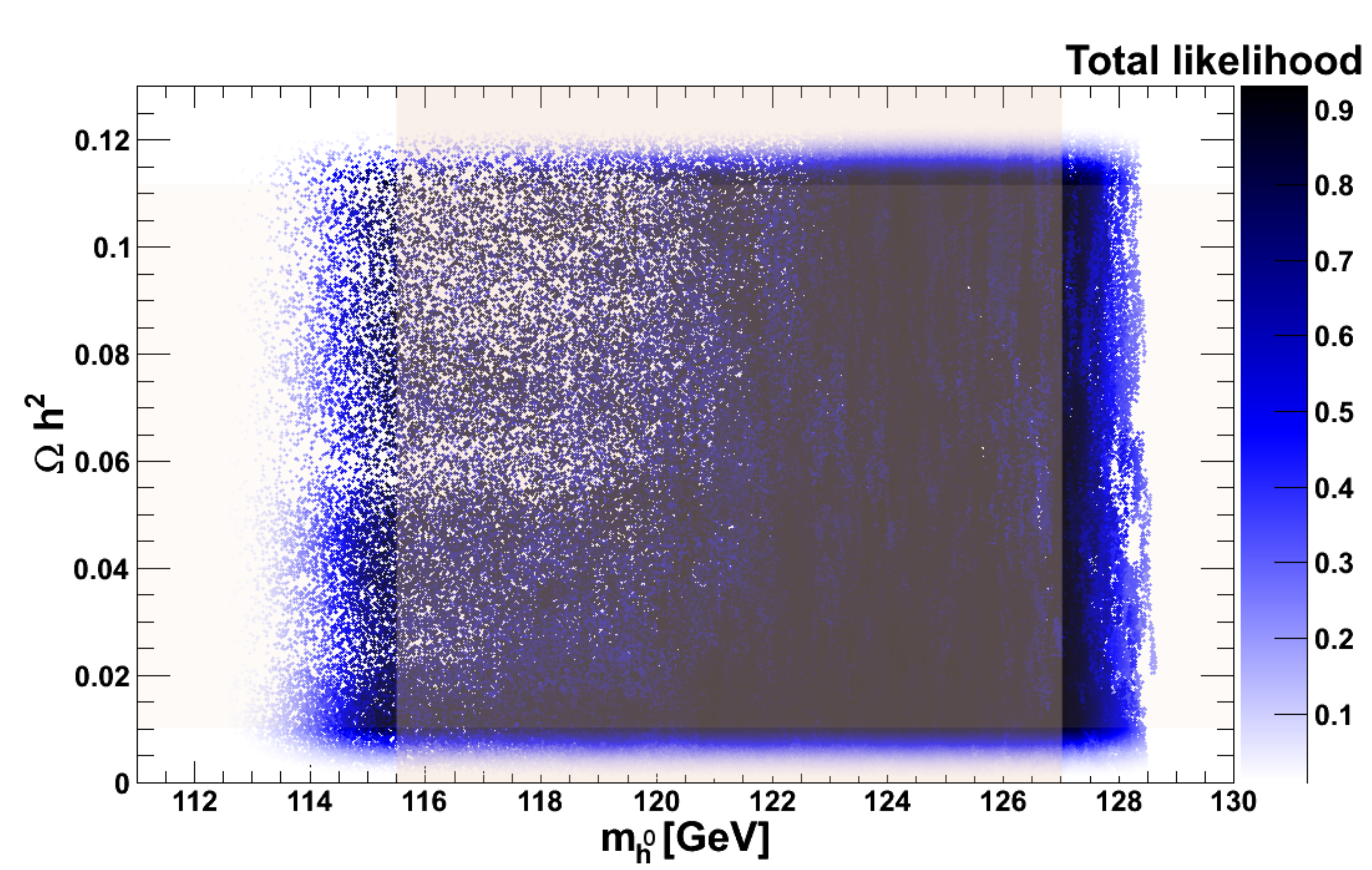}}
\caption[Characteristics of the scan with low relic densities allowed.]{Characteristics of the scan with low relic densities allowed. Panels (a, b) show the $(\mu,m_{A^0})$ plane with either the total likelihood of the points or the corresponding LSP relic density as colour code. Panel (c) shows the higgsino fraction and panel (d) present the neutralino relic density\vs the mass of the lightest Higgs boson where the colour coding corresponds to the likelihood of the points.} 
\label{4.lowRD}
\end{center}
\end{figure}

Relaxing the constraint on the DM relic density and allowing neutralinos to constitute only a fraction of the total DM density does not change the above features. The main effect in fact is to allow lower values of $\mu$, since a lighter higgsino LSP is characterized by a higher annihilation cross section and then by a lower relic density. As expected the degeneracy between the LSP and the NLSP, mainly chargino, is still present and larger values of the Higgs boson mass still give a higher likelihood. Figure~\ref{4.lowRD} summarizes the characteristics of this scan with relaxing constraint on $\Om_{\chi^0_1} h^2$.

We now look at the study of inflaton candidates in the framework of this supersymmetric model.

\section{Supersymmetric inflaton}

The primordial inflation must explain the seed perturbations for the CMB radiation, and after the end of inflation the coherent oscillations of the inflaton must excite SM quarks and leptons at temperatures sufficiently high to realize baryons and DM in the current Universe~\cite{Mazumdar:2010sa,Mazumdar:2011zd}. In this respect, it is vital that the last phase of primordial inflation must end in a vacuum of BSM physics which can solely excite the relevant degrees of freedom required for the success of BBN, see~\cite{Pospelov:2010hj} for a review. 

Inflation needs a potential which remains sufficiently flat along which the slow-roll inflation can take place in order to generate the observed temperature anisotropy in the CMB. Low scale SUSY not only provides a DM candidate and a testable framework for BSM physics, but also guarantees the flatness of such flat directions at a perturbative and a non-perturbative level (for a review see~\cite{Enqvist:2003gh}).

The flat directions of the MSSM provide nearly 300 gauge-invariant $F$-and $D$-flat directions 
~\cite{Gherghetta:1995dv,Dine:1995kz}, which are all charged under the SM gauge group. Out of these flat directions, there are particularly 2 $D$-flat directions, $\widetilde{u}\widetilde{d}\widetilde{d}$ and $\widetilde{L}\widetilde{L}\widetilde{e}$, which carry SM charges and can be ideal inflaton candidates~\cite{Allahverdi:2006iq,Allahverdi:2006we,Allahverdi:2006cx}. Here $\widetilde u$ and $\widetilde d$ correspond to the right handed squarks, $\widetilde L$ corresponds to the left handed sleptons, and $\widetilde e$ corresponds to the right handed charged sleptons. Both the inflaton candidates provide an \textit{inflection point} in their respective potentials where inflation can be driven for sufficiently large e-foldings of inflation to explain the current Universe and explain the seed perturbations for the temperature anisotropy in the CMB. 

The inflaton in this case only decays into the MSSM degrees of freedom which thermalize the Universe with a temperature $T_R\sim 10^{8}$~GeV~\cite{Allahverdi:2011aj}. This temperature is sufficient to excite the degrees of freedom which are needed for the LSP to get a relic density that matches observations. It is then natural to ask whether there exists any parameter space, where both successful inflation and thermal DM abundance can be explained simultaneously~\cite{Allahverdi:2007vy,Allahverdi:2010zp}.

Since in our case the inflaton candidates are gauge invariant, by using the RGEs at one loop level one can evaluate the mass of the inflaton $m_\phi$ from the scale of inflation to the scale of LHC. This eventually will enable us to relate the inflaton mass with the DM parameter space and the lightest CP-even Higgs boson mass.

\subsection{Inflaton candidates : flat directions of squarks and sleptons}
\label{MSSMI}

\begin{figure}[t]
\centering
\includegraphics[width=8cm,height=5cm]{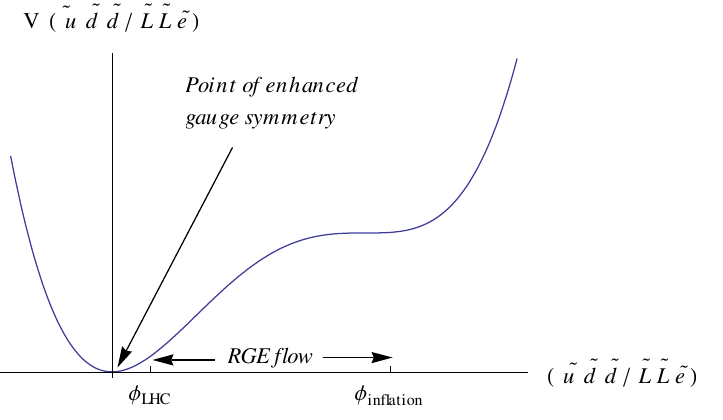}
\caption[A schematic drawing of inflationary potential for either $\widetilde u\widetilde d\widetilde d$ or $\widetilde L\widetilde L\widetilde e$ as shown in eq.~\ref{scpot}.]{A schematic drawing of inflationary potential for either $\widetilde u\widetilde d\widetilde d$ or $\widetilde L\widetilde L\widetilde e$ as shown in eq.~\ref{scpot}. Inflation happens near the \textit{inflection point} point as shown by $\phi_{\rm inflation}=\phi_0$. Inflation ends at the point of {\it enhanced gauge symmetry}, where the entire (MS)SM gauge symmetry is recovered. The physical mass and couplings at high scale $\phi_{0}$ and $\phi_{\mathrm{LHC}}$ are related via RGEs described by eqs.~\ref{rgudd} and \ref{rglle}.
}\label{potential}
\end{figure}

In references~\cite{Allahverdi:2006iq,Allahverdi:2006we,Allahverdi:2007vy,Allahverdi:2010zp} it was shown that the two $D$-flat directions $\widetilde{u}\widetilde{d}\widetilde{d}$ and $\widetilde{L}\widetilde{L}\widetilde{e}$ can be the ideal inflaton candidates, because both flat directions are lifted (they are approximately flat) by higher order superpotential terms of the following form\footnote{Note that to simplify a number of following equations we use the reduced Planck mass $m_{Pl} = M_{Pl}/\sqrt{8\pi}$.} which would provide non-vanishing $A$-term in the potential even at large VEVs~\footnote{Note that the $R$-parity is still conserved, both the superpotential terms ${\bf udd}$ and ${\bf LLe}$ do not appear at the renormalizable level, they are instead lifted by non-renormalizable operators. Further note that both the operators vanish in the vacuum which is shown at $\phi=0$ in figure~\ref{potential}.}: 
\beq \label{supot} W \supset {\lambda \over 6}{\Phi^6 \over m^3_{Pl}},\eeq
 where $\lambda \sim {\cal O}(1)$~\footnote{~The exact value of $\lambda$ is irrelevant for the CMB analysis, as it does not modify the CMB predictions. However it is possible to extract its value by integrating out the heavy degrees of freedom. If the origin of these operators arise from either $SU(5)$ or $SO(10)$, then the typical value is of order $\lambda \sim {\cal O}(10^{-2})$ for $SO(10)$ and $\lambda \sim {\cal O}(1)$ for $SU(5)$, as shown in \cite{Allahverdi:2007vy}.}. The scalar component of the $\Phi$ superfield, denoted by $\phi$, is given by\footnote{The representations for the flat directions are given by: $\widetilde u^{\alpha}_i=\frac1{\sqrt{3}}\phi\,,~\widetilde d^{\beta}_j=\frac1{\sqrt{3}}\phi\,,~\widetilde d^{\gamma}_k=\frac{1}{\sqrt{3}}\phi .$ Here $1 \leq \alpha,\beta,\gamma \leq 3$ are colour indices, and $1\leq i,j,k \leq 3$ denote the quark families. The flatness constraints require that $\alpha \neq \beta \neq \gamma$ and $j \neq k$.
 
In the case of sleptons we have $\widetilde L_i=\frac1{\sqrt{3}}\binom{0}{\phi},~\widetilde L_j=\frac1{\sqrt{3}}\binom{\phi}{0},~\widetilde e_k=\frac{1}{\sqrt{3}}\phi\,,$ where $1 \leq i,j,k \leq 3$ denote the lepton families. The flatness constraints require that $i \neq j \neq k$. Note that the cosmological perturbations do not care which combination arises, as gravity couples universally.}

\beq \phi = {\widetilde{u} + \widetilde{d} + \widetilde{d} \over \sqrt{3}}, \quad \phi = {\widetilde{L} + \widetilde{L} + 
\widetilde{e} \over \sqrt{3}},\eeq
for the $\widetilde{u}\widetilde{d}\widetilde{d}$ and $\widetilde{L}\widetilde{L}\widetilde{e}$ flat directions respectively.
After minimizing the potential along the angular direction $\theta$ ($\Phi$ = $\phi e^{i \theta}$), we can determine the real part of $\phi$ by rotating it to the corresponding angles $\theta_{\rm min}$. The scalar potential is then found to be~\cite{Allahverdi:2006iq,Allahverdi:2006we}
\beq \label{scpot}
V(\phi) = {1\over2} m^2_\phi\, \phi^2 - A {\lambda\phi^6 \over 6\,m^{3}_{Pl}} + \lambda^2 {{\phi}^{10} \over m^{6}_{Pl}}\,,
\eeq
where $m_\phi$ and $A$ are the soft breaking mass and the $A$-term respectively ($A$ is a positive quantity since its phase is absorbed by a redefinition of $\theta$ during the process). The masses for $\widetilde{L}\widetilde{L}\widetilde{e}$ and $\widetilde{u}\widetilde{d}\widetilde{d}$ are given by:
\beq \begin{split}\label{masses}
m^2_{\phi_{\widetilde{L}\widetilde{L}\widetilde{e}}}& =\frac{m^2_{\widetilde L}+m^2_{\widetilde L}+m^2_{\widetilde e}}{3},\\
m^2_{\phi_{\widetilde{u}\widetilde{d}\widetilde{d}}}& =\frac{m^2_{\widetilde u}+m^2_{\widetilde d}+m^2_{\widetilde d}}{3}.
\end{split} \eeq
These masses are now VEV dependent,\ie $m^2(\phi)$. The inflationary perturbations will be able to constrain 
the inflaton mass only at the scale of inflation,\ie $\phi_0$, while LHC will be able to constrain the masses at the LHC scale. However both the physical quantities are related to each other via RGEs as we will discuss below.
For
\beq {A^2 \over 40 m^2_{\phi}} \equiv 1 - 4 \a^2,\eeq
where $\a^2 \ll 1$~\footnote{The value of $\a$ during inflation could be small,\ie $\a \sim 10^{-10}$, but it runs dynamically from the GUT scale where $A^2=40 m_{\phi}^2$ to the required value at scale of inflation via RGEs. For a detailed discussion see \cite{Allahverdi:2010zp}.}, there exists a point of inflection ($\phi_0$) in $V(\phi)$, where
\beq \begin{split}
\phi_0^4 & = {m_\phi m^{3}_{Pl}\over \lambda \sqrt{10}} + {\cal O}(\alpha^2),\\
V^{\prime \prime}(\phi_0) & = 0,\\
V(\phi_0) & = \frac{4}{15}m_{\phi}^2\phi_0^2 + {\cal O}(\alpha^2),\\
V'(\phi_0) & = 4 \alpha^2 m^2_{\phi} \phi_0 \, + {\cal O}(\alpha^4),\\
V^{\prime \prime \prime}(\phi_0) & = 32\frac{m_{\phi}^2}{\phi_0} + {\cal O}(\alpha^2).
\end{split} \eeq

From now on we only keep the leading order terms in all expressions. Note that inflation occurs within an interval\footnote{For a low scale inflation, setting the initial condition is always challenging. However in the case of a MSSM or string theory landscape where there are many false vacua at high and high scales, then it is conceivable that earlier phases of inflation could have occurred in those false vacua. This large vacuum energy could lift the flat direction condensate either via quantum fluctuations~\cite{Allahverdi:2007wh}, however see also the challenges posed by the quantum fluctuations~\cite{Enqvist:2011pt}, or via classical initial condition which happens at the level of background without any problem, see~\cite{Allahverdi:2008bt}.}
\beq \label{plateau}
\vert \phi - \phi_0 \vert \sim {\phi^3_0 \over 60 m^2_{Pl}} ,
\eeq
in the vicinity of the point of inflection, within which the slow roll parameters $\e_\phi$ and $\eta_\phi$ given in eq.~\ref{eq.2:inflaparams} are smaller than 1. The Hubble expansion rate during inflation is given by
\beq \label{hubble} H_{\mathrm{inf}} \simeq \frac{2}{\sqrt{45}}\frac{m_{\phi}\phi_0}{m_{Pl}}. \eeq
In order to obtain the flat potential, it is crucial that the $A(\phi_0)$-term ought to be close to $m_{\phi}(\phi_0)$ in the above potential eq.~\ref{scpot}. This can be obtained within two particular scenarios : \textit{Gravity Mediation}~\cite{Nilles:1983ge} and \textit{Split SUSY}, where the scale of SUSY is high and sfermions are very heavy~\cite{Babu:2005ui}.

Keeping low scale and high scale SUSY breaking scenarios in mind here we will consider a large range of $(m_{\phi},~\phi_0)$ to match the cosmological observations.

\subsection{Gaussian fluctuations and tensor to scalar ratio}

\begin{figure}[!htb]
\begin{center}
\subfloat[]{\includegraphics[width=8cm,height=6cm]{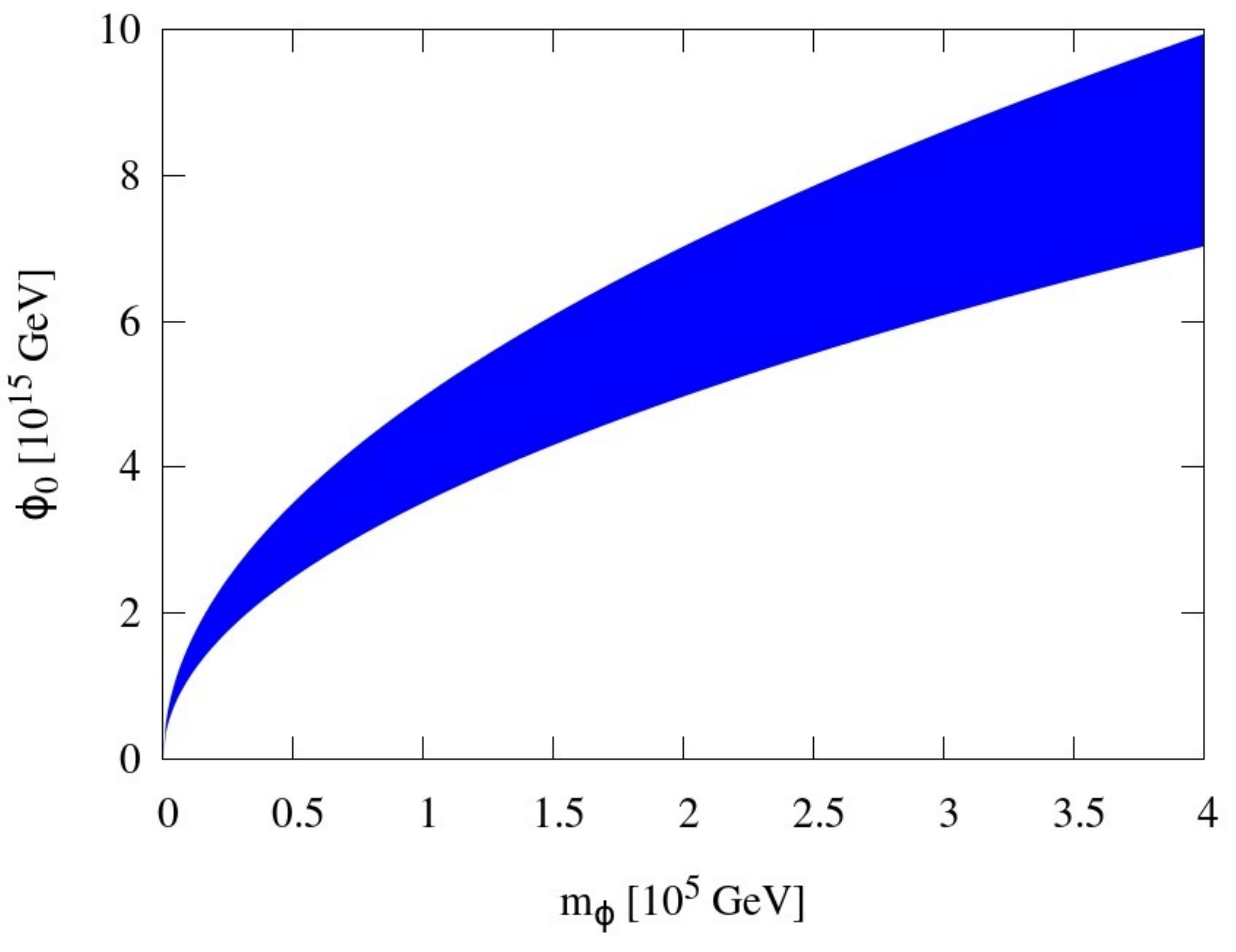}}
\subfloat[]{\includegraphics[width=8cm,height=6cm]{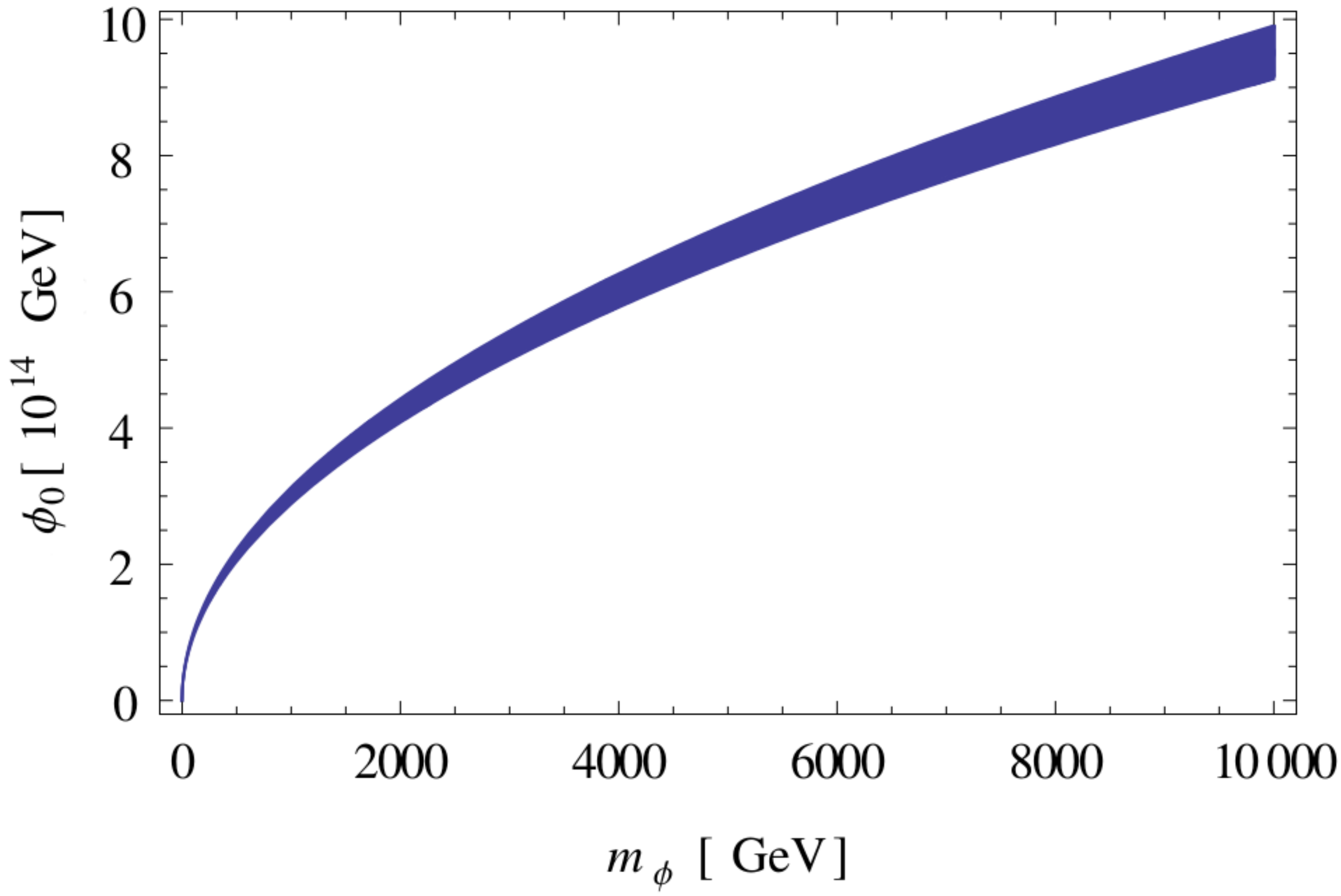}}
\caption[$(\phi_0,m_{\phi})$ plane in which inflation is in agreement with the cosmological observations of the temperature anisotropy of the CMB fluctuations.]{$(\phi_0,m_{\phi})$ plane in which inflation is in agreement with the cosmological observations of the temperature anisotropy of the CMB fluctuations. The blue regions show the inflaton energy scale and inflaton mass which are compatible with the central value of the amplitude of the seed perturbations and the allowed range of spectral tilt. For panel (a) we have $\delta_H=1.91\times 10^{-5}$ and $0.934 \leq n_s\leq 0.988$ at 2$\s$~\cite{Komatsu:2010fb}. The panel (b), taken from \cite{Wang:2013hva}, use post-Planck results : $\delta_H=1.87\times 10^{-5}$ and $n_s = 0.9603 \pm 0.0073$ at 1$\s$. Note that we restricted ourselves to inflaton VEVs $\phi_0$ below the GUT scale.}\label{Fig-0}
\end{center}
\end{figure}

The above potential eq.~\ref{scpot} has been studied extensively in \cite{Allahverdi:2006we,Bueno-Sanchez:2006xk,Enqvist:2010vd}. The amplitude of density perturbations $\delta_H$ and the scalar spectral index $n_s$ are given respectively by :
\beq \label{ampl}
\delta_H = {8 \over \sqrt{5} \pi} {m_{\phi} m_{Pl} \over \phi^2_0}{1 \over \Delta^2}
\sin^2 \left[{\cal N}_{\rm COBE}\sqrt{\Delta^2}\right], \eeq
and
\beq \label{tilt}
n_s = 1 - 4 \sqrt{\Delta^2} \cot \left[{\cal N}_{\rm COBE}\sqrt{\Delta^2}\right], \eeq
where
\beq \label{Delta}
\Delta^2 \equiv 900 \alpha^2 {\cal
N}^{-2}_{\rm COBE} \left({m_{Pl} \over \phi_0}\right)^4. \eeq
In the above, ${\cal N}_{\rm COBE}$ is the number of e-foldings between the time when the observationally relevant perturbations are generated till the end of inflation and follows : 
${\cal N}_{\rm COBE} \simeq 66.9 + (1/4) {\rm ln}({V(\phi_0)/ m^4_{Pl}}) \sim 50$. 
Since the perturbations are due to a single field, one does not expect large non-gaussianity from this model ($f_{NL}\leq 1$, see~\cite{Maldacena:2002vr}).

Panel (a) of figure~\ref{Fig-0} shows the exploration of a wide range of the inflaton mass, $m_{\phi}$, where inflation can explain the observed temperature anisotropy in the CMB with the right amplitude, $\delta_H=1.91\times 10^{-5}$, and the tilt in the power spectrum, $0.934 \leq n_s\leq 0.988$~\cite{Komatsu:2010fb}. This figure represents the inflation energy scale versus the mass of the inflaton. The configurations which fit the observed values of $\delta_H$ and $n_s$ are shown in blue. The main effect of last release of the results from the Planck collaboration \cite{Ade:2013uln} is a narrowing of this blue region because of the much more precise measurement of $n_s$ as can be shown in panel (b) of figure~\ref{Fig-0}, taken from \cite{Wang:2013hva}\footnote{Note that the scales on this plot are not the same than in the previous one. Nevertheless this representation looks always the same at any scale if we use fixed cosmological constraints.}. Although we have restricted ourselves to VEV values below the GUT scale, the model does provide negligible running in the tilt which is well within the observed limit.
 
Here we have allowed for a wide range of $m_{\phi}$ and $\phi_0$ values because ultimately we want to show that inflation can happen within low-scale SUSY scenarios from high-scale SUSY breaking soft masses as in the split-SUSY scenario~\cite{Babu:2005ui}. 

Here we mostly consider scenarios where the scale of inflation is low enough that one would not expect any observed tensor perturbations in any future CMB experiments. To obtain large observable tensor to scalar ratio $r$ one would have to embed these inflaton candidates within $\mathcal{N}=1$ supergravity. This would modify the potential with a large vacuum energy density besides providing supergravity corrections to mass and A-term~\cite{Mazumdar:2011ih,Hotchkiss:2011gz}. One could then obtain $r\sim 0.05$ for both inflaton flat directions: $\widetilde u\widetilde d\widetilde d$ and $\widetilde L\widetilde L\widetilde e$ as shown in \cite{Hotchkiss:2011gz} which then could be probed by the Planck experiment \cite{Ade:2013uln}.

\subsection{Renormalization Group Equations}

Since the inflaton carries SM charges and they are fully embedded within the MSSM, it is possible to probe various regions of the parameter space for inflation. The CMB fluctuations probe the inflaton potential at the inflationary scale. At low energies the inflaton properties can be probed by the LHC from the masses of the squarks and sleptons.

The inflaton mass and the non-renormalizable $A$ term in the inflationary potential are both scale dependent quantities, and they can be tracked down to lower energies by using the RGEs. In \cite{Allahverdi:2006we,Allahverdi:2007vy,Allahverdi:2010zp}, it was shown that the RGEs at one loop level for the relevant flat direction are, for $\widetilde u\widetilde d\widetilde d$ :
\beq \begin{split}
\label{rgudd}
\hat{\mu} \frac{dm^2_\phi}{d\hat{\mu}} & =-\frac{1}{6\pi^2}(4M_3^2g_3^2+\frac{2}{5}M_1^2g_1^2),\\
\hat{\mu} \frac{dA}{d\hat{\mu}} & =-\frac{1}{4\pi^2}(\frac{16}{3}M_3g_3^2+\frac{8}{5}M_1g_1^2).
\end{split} \eeq
where  $\hat{\mu} = \hat\mu_0=\phi_0$ is the VEV at which inflation occurs. We have for $\widetilde{L}\widetilde{L}\widetilde{e}$ :
\begin{equation}
\begin{aligned}
\label{rglle}
&\hat\mu\frac{dm^2_\phi}{d\hat\mu}=-\frac{1}{6\pi^2}(\frac{3}{2}M_2^2g_2^2+\frac{9}{10}M_1^2g_1^2),
\\&\hat\mu\frac{dA}{d\hat\mu}=-\frac{1}{4\pi^2}(\frac{3}{2}M_2g_2^2+\frac{9}{5}M_1g_1^2).
\end{aligned}
\end{equation}
$M_1$, $M_2$ and $M_3$ are equal to $m_{1/2}$ at the unification scale. To solve these equations, one needs to take into account the running of the gaugino masses and coupling constants as given in \cite{Nilles:1983ge}. So every point in the $(m_0,m_{1/2})$ plane can now be mapped onto the $(\phi_0,m_\phi)$ plane.

\subsection{Indirect detection of the inflaton at LHC}

\begin{figure}[!htb]
\begin{center}
\centering
\subfloat[]{\includegraphics[width=8cm,height=9cm]{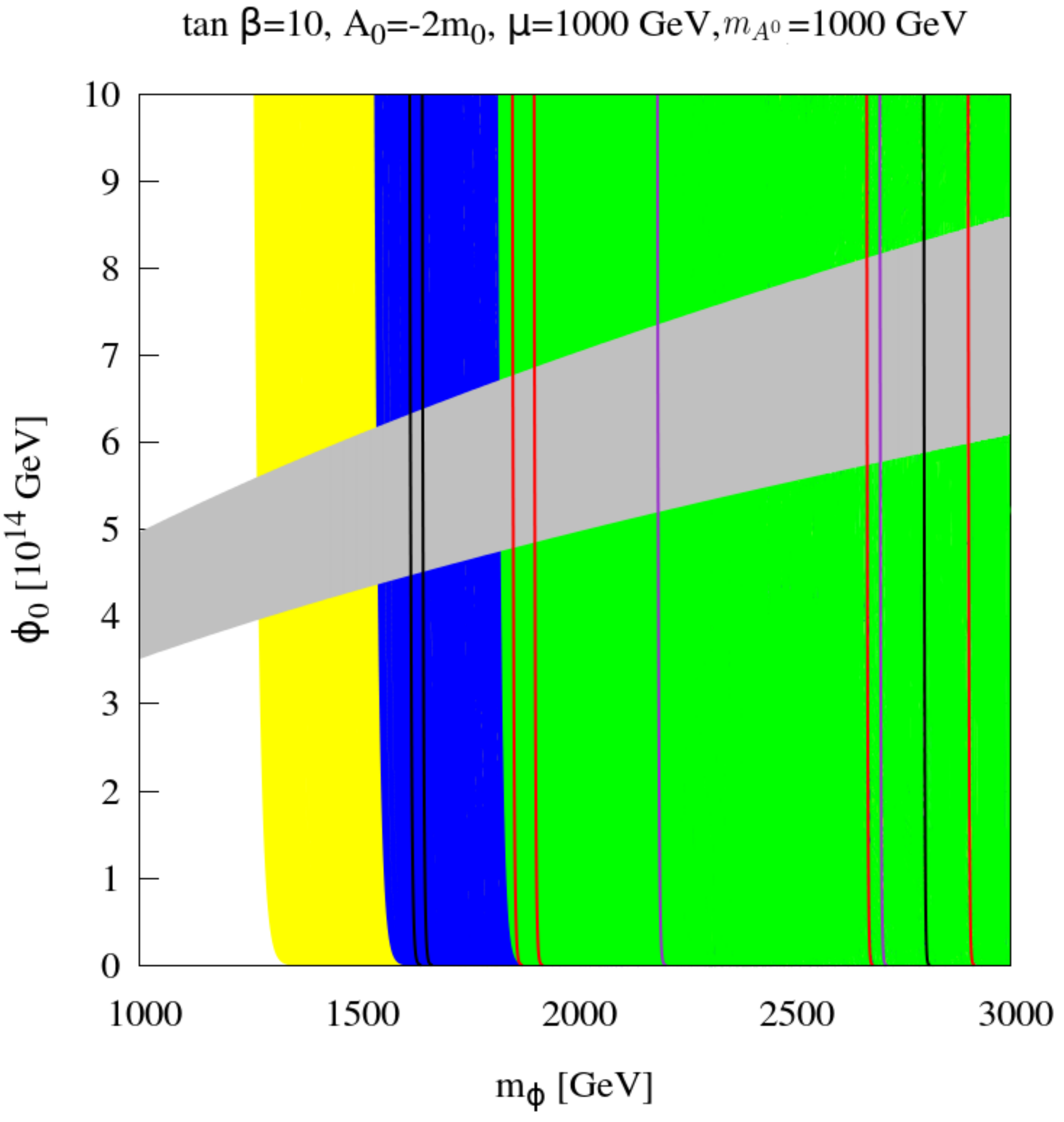}}  
\subfloat[]{\includegraphics[width=8cm,height=9cm]{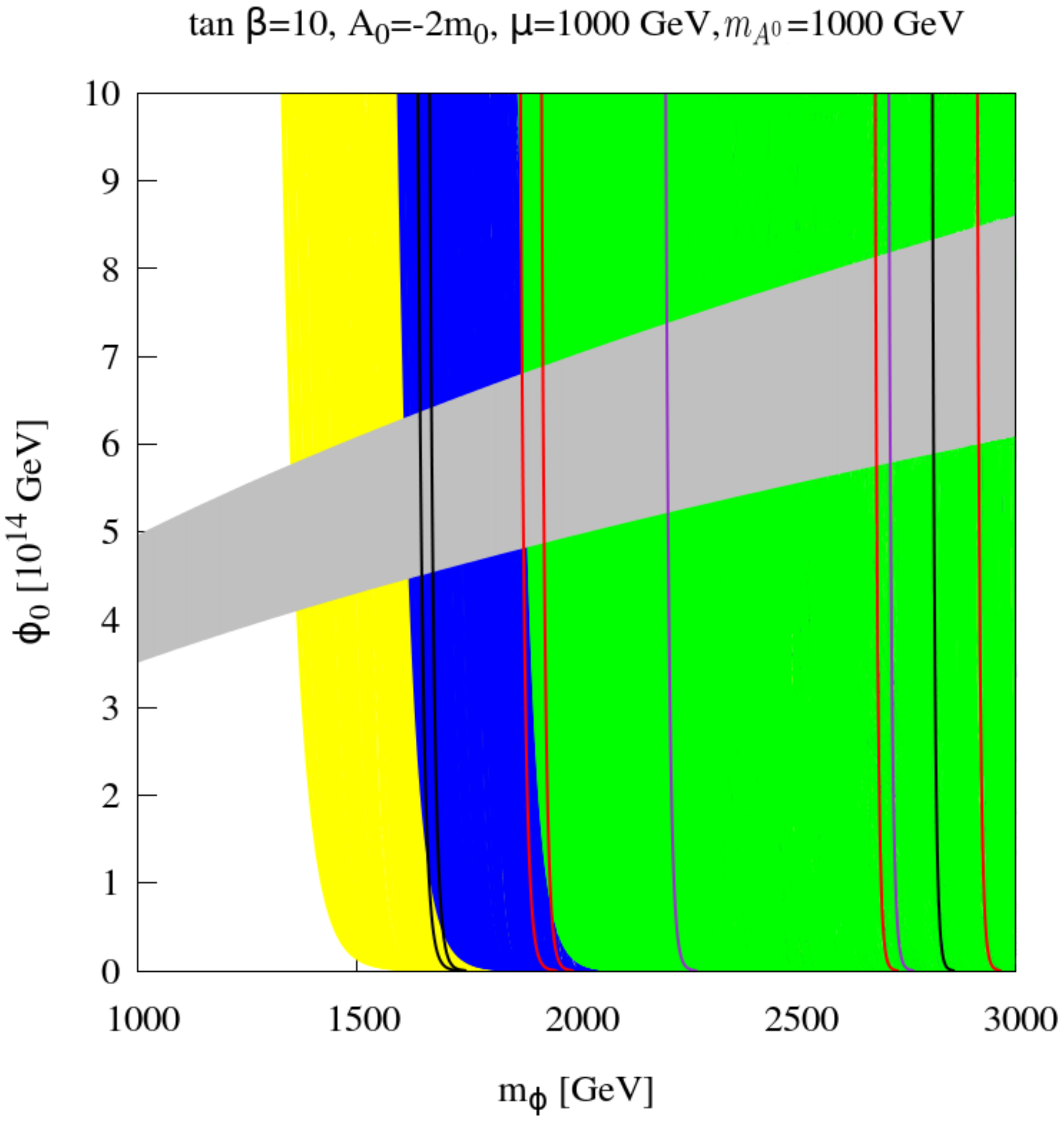}}
\caption[($\phi_0$,$m_\phi$) plane for $\widetilde{L}\widetilde{L}\widetilde{e}$ (a) and $\widetilde{u}\widetilde{d}\widetilde{d}$ (b) flat direction inflatons respectively, where $\tan\beta=10$, $A_0=-2m_0$ and 
$\mu=m_{A^0}=1$ TeV.]{($\phi_0$,$m_\phi$) plane for $\widetilde{L}\widetilde{L}\widetilde{e}$ (a) and $\widetilde{u}\widetilde{d}\widetilde{d}$ (b) flat direction inflatons respectively, where $\tan\beta=10$, $A_0=-2m_0$ and 
$\mu=m_{A^0}=1$ TeV. The green region is the region where $m_{h^0}$ can go up to 125.5 GeV while the blue region corresponds to $m_{h^0} \leq 125$~GeV. In the same way the yellow region is characterized by $m_{h^0} \leq 124.5$~GeV. Black, red and violet lines correspond respectively to $m_{h^0} = 124.5,125,125.5$~GeV and simultaneously a DM abundance in the range $0.1088<\Omega_{\chi^0_1} h^2<0.1158$. }
\label{fig3}
\end{center}
\end{figure}

In the previous section we have verified the validity of our benchmark points and could measure how fine-tuned they are with respect to other configurations. In particular, we have seen that scenarios with large scalar masses $m_0$ require small values of $\tan \beta$ in order not to exceed the upper limit on the Higgs boson mass and lead to scenarios in which the neutralino has generally a non-negligible higgsino fraction. 

We can now determine the inflation energy scale and mass of the inflaton for these benchmark points. We will follow a similar approach as in \cite{Allahverdi:2010zp} in order to estimate the inflaton mass which is compatible with the temperature anisotropy of the CMB data.

\subsubsection{Inflaton mass for benchmark points}

In figure~\ref{fig3} the regions of figure~\ref{fig:4.SUSY_points}a are mapped onto the $(\phi_0,m_\phi)$ plane. Panel (a) is for the $\widetilde{L}\widetilde{L}\widetilde{e}$ case and panel (b) is for the $\widetilde{u}\widetilde{d}\widetilde{d}$ case. The red lines show regions where both the right DM abundance and a Higgs boson mass of 125 GeV are obtained, which corresponds to the four benchmark points `a',~`b',~`c' and `d'. The grey shaded region shows where the NUHM2 inflation can explain the CMB observations from WMAP. From these figures we see that in the case of a $m_{h^0}=125$ GeV Higgs boson mass, inflation should happen around $\phi_0\approx(4.8-6.8)\times10^{14}$~GeV for the `a' and `c' benchmark points, yielding $m_{\phi_{\widetilde{L}\widetilde{L}\widetilde{e}}} \approx 1.9$~TeV and $m_{\phi_{\widetilde{u}\widetilde{d}\widetilde{d}}} \approx 2.2$~TeV. Another two possibilities correspond to the benchmark points `b' and `d'. For `b' point we have inflation happening in a range of $\phi_0\approx(5.7-8)\times10^{14}$ GeV with  $m_{\phi_{\widetilde{L}\widetilde{L}\widetilde{e}}} \approx 2.7$~TeV and $m_{\phi_{\widetilde{u}\widetilde{d}\widetilde{d}}} \approx 2.9$~TeV and similarly for `d' we have range of $\phi_0\approx(6-8.1)\times10^{14}$ GeV with $m_{\phi_{\widetilde{L}\widetilde{L}\widetilde{e}}} \approx 2.9$~TeV and $m_{\phi_{\widetilde{u}\widetilde{d}\widetilde{d}}} \approx 3.1$~TeV. From the cosmological point of view, the heavier Higgs boson is, the more of the parameter space for inflation which become compatible with the CMB observations we have. In general for the $\widetilde{u}\widetilde{d}\widetilde{d}$ inflaton, we have a larger running than in the $\widetilde{L}\widetilde{L}\widetilde{e}$ case, essentially because of the running of $g_3$. However, it is hard to appreciate this running visibly by comparing panel (a) and (b) of figure~\ref{fig3}, because of the large range of $m_\phi$ we have plotted.

\subsubsection{LHC predictions and Inflaton mass}

\begin{figure}[h]
\centering
\subfloat[]{\includegraphics[width=8cm,height=5cm]{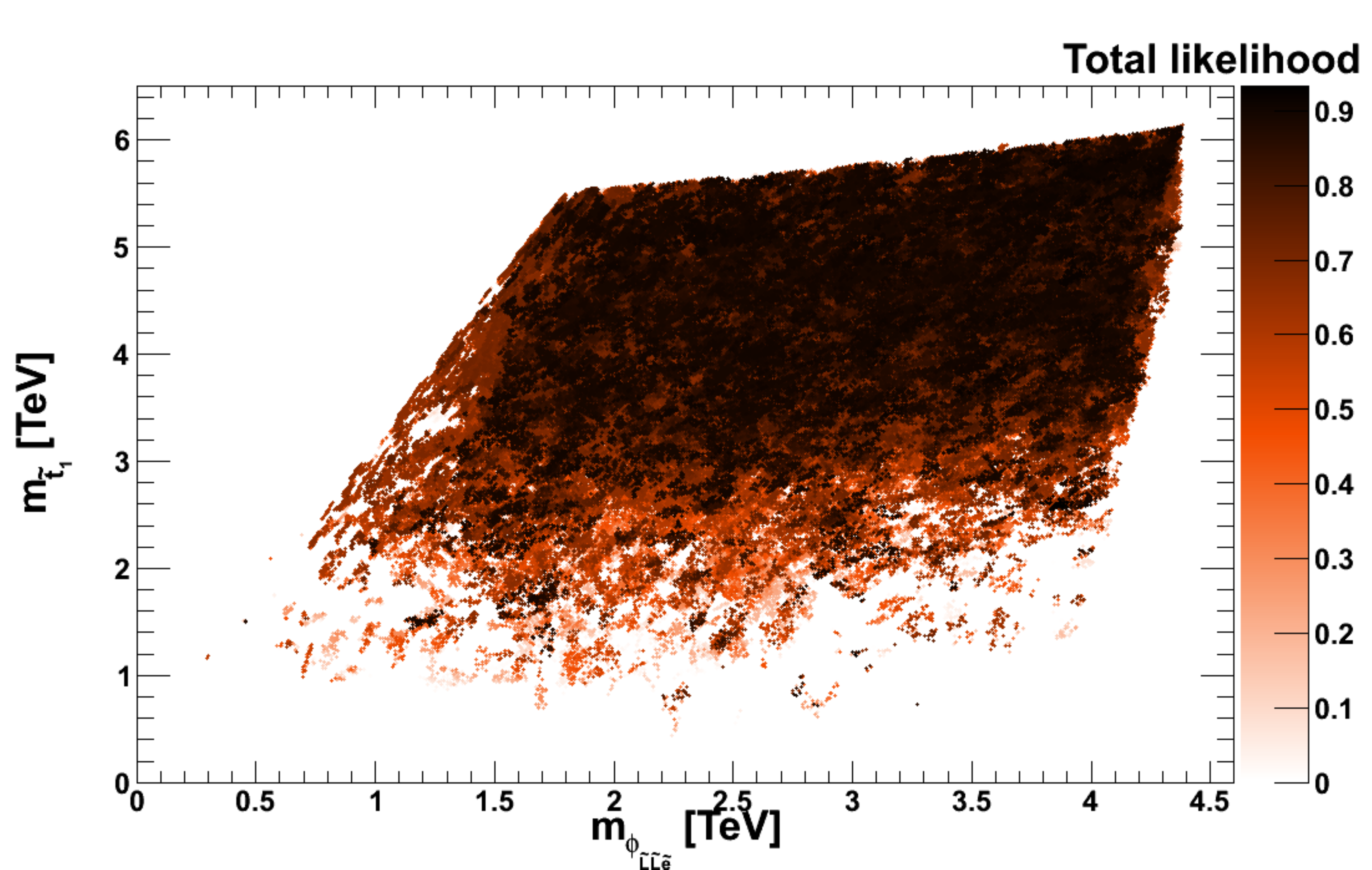}}
\subfloat[]{\includegraphics[width=8cm,height=5cm]{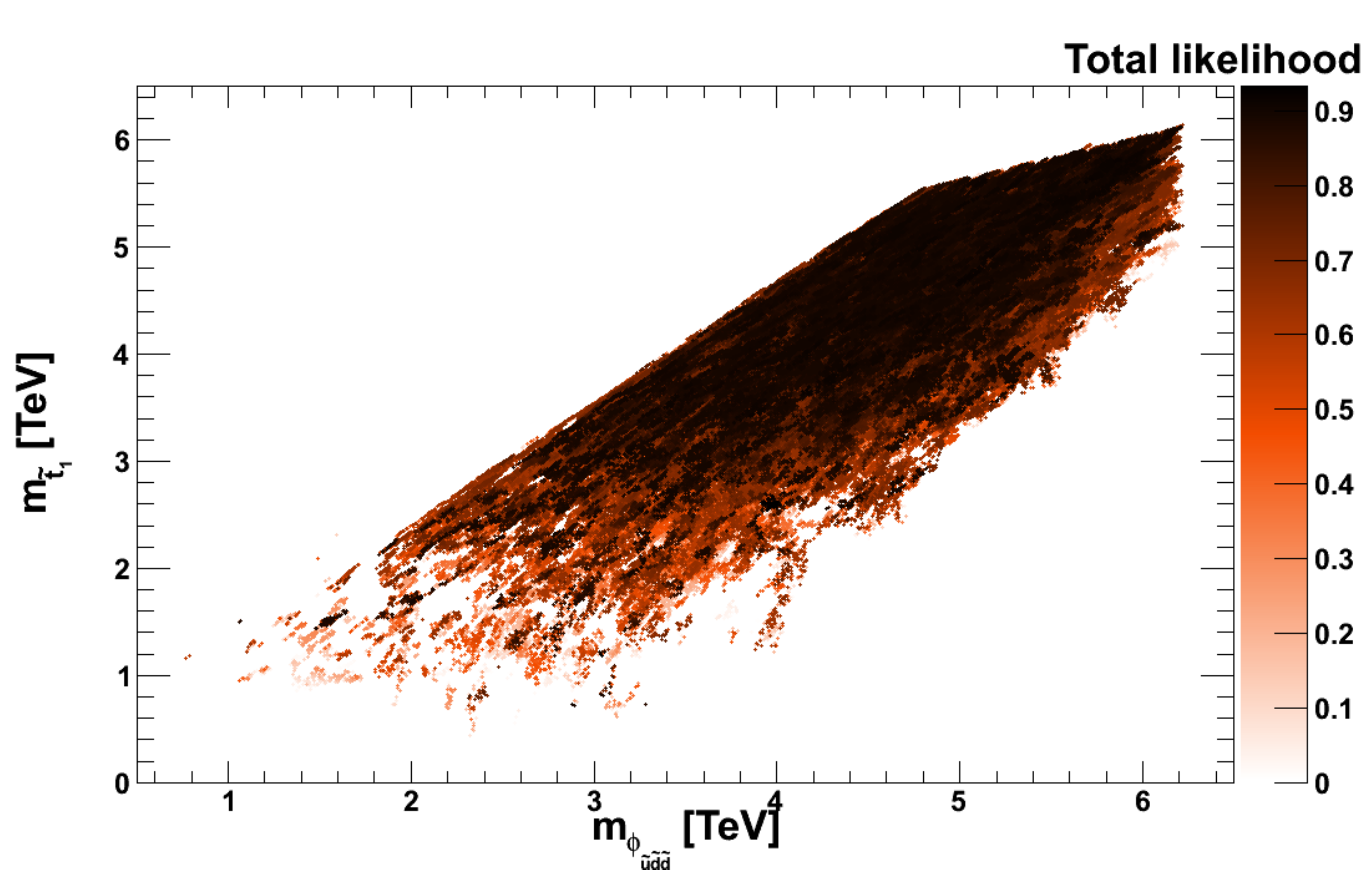}}
\caption[The lightest stop mass $m_{\tilde{t}_1}$ versus the inflaton masses for $\widetilde L\widetilde L\widetilde e$ (a) and $\widetilde u\widetilde d\widetilde d$ (b), see eq.~\ref{masses}.]{The lightest stop mass $m_{\tilde{t}_1}$ versus the inflaton masses for $\widetilde L\widetilde L\widetilde e$ (a) and $\widetilde u\widetilde d\widetilde d$ (b), see eq.~\ref{masses}. Same colour code as in figure~\ref{4.neutralino_composition}.}
\label{mstopmphi}
\end{figure}

Our previous scans of the NUHM2 parameter space have selected neutralinos with a high higgsino fraction when the neutralino mass falls within the 0.6 and 1.2 TeV range. It is now interesting to check the prediction for the stop mass depending on the inflaton mass at TeV scale, see figure~\ref{mstopmphi}. 
We find that in both inflation scenarios, the inflaton mass is above $500$ GeV and is associated with a very massive stop. For the $\widetilde{u}\widetilde{d}\widetilde{d}$ combination, the lightest stop mass is constrained to be within $m_{\phi} > m_{\tilde{t}_1} > m_{\phi}/3$. Scenarios with the lightest stops (namely $m_{\tilde{t}_1} \lesssim 2$ TeV) may offer a chance to probe the NUHM2 parameter space and thus a mean to determine the inflaton mass.

\begin{figure}[h]
\centering
\subfloat[]{\includegraphics[width=8cm,height=5cm]{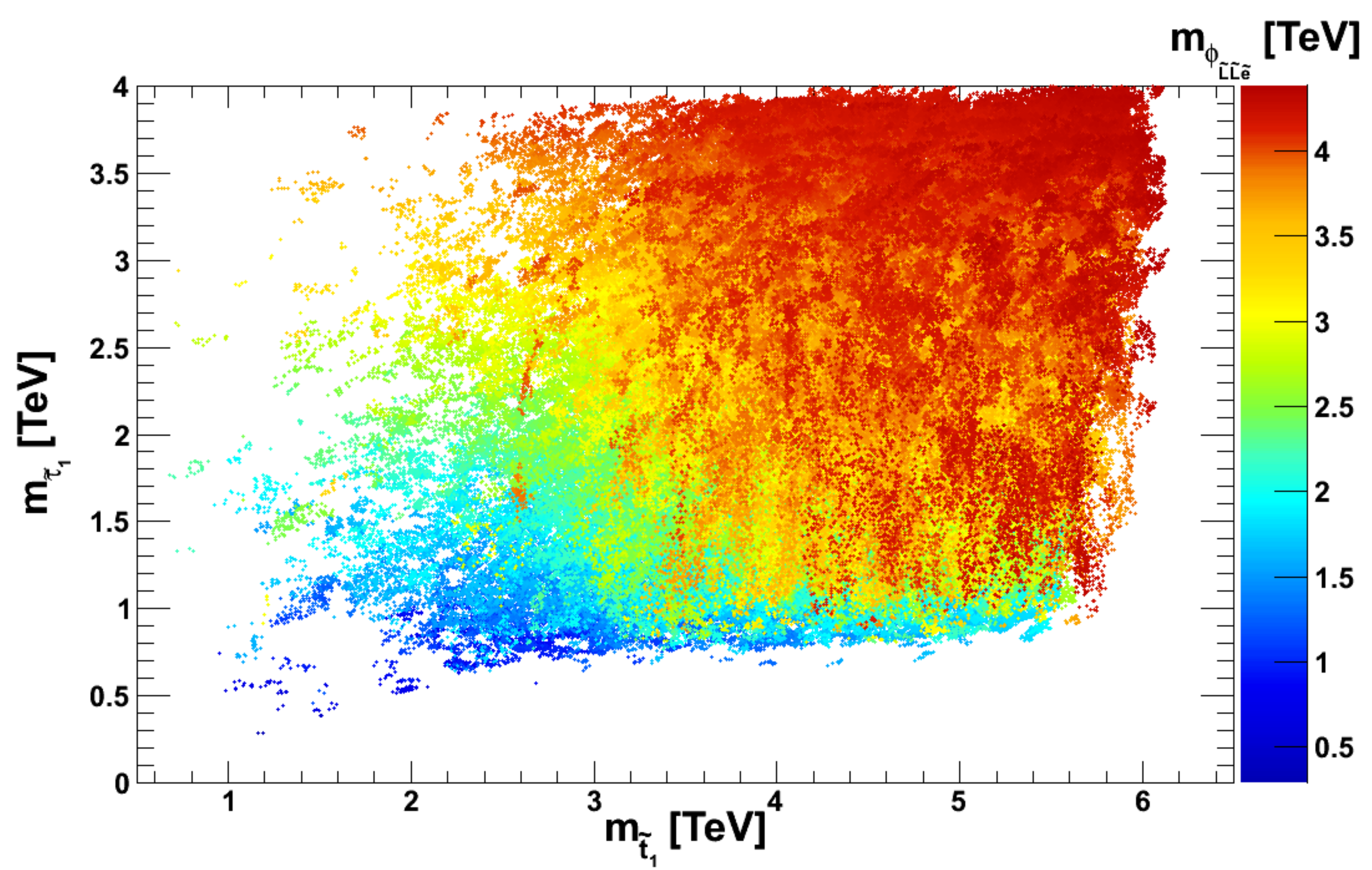}}
\subfloat[]{\includegraphics[width=8cm,height=5cm]{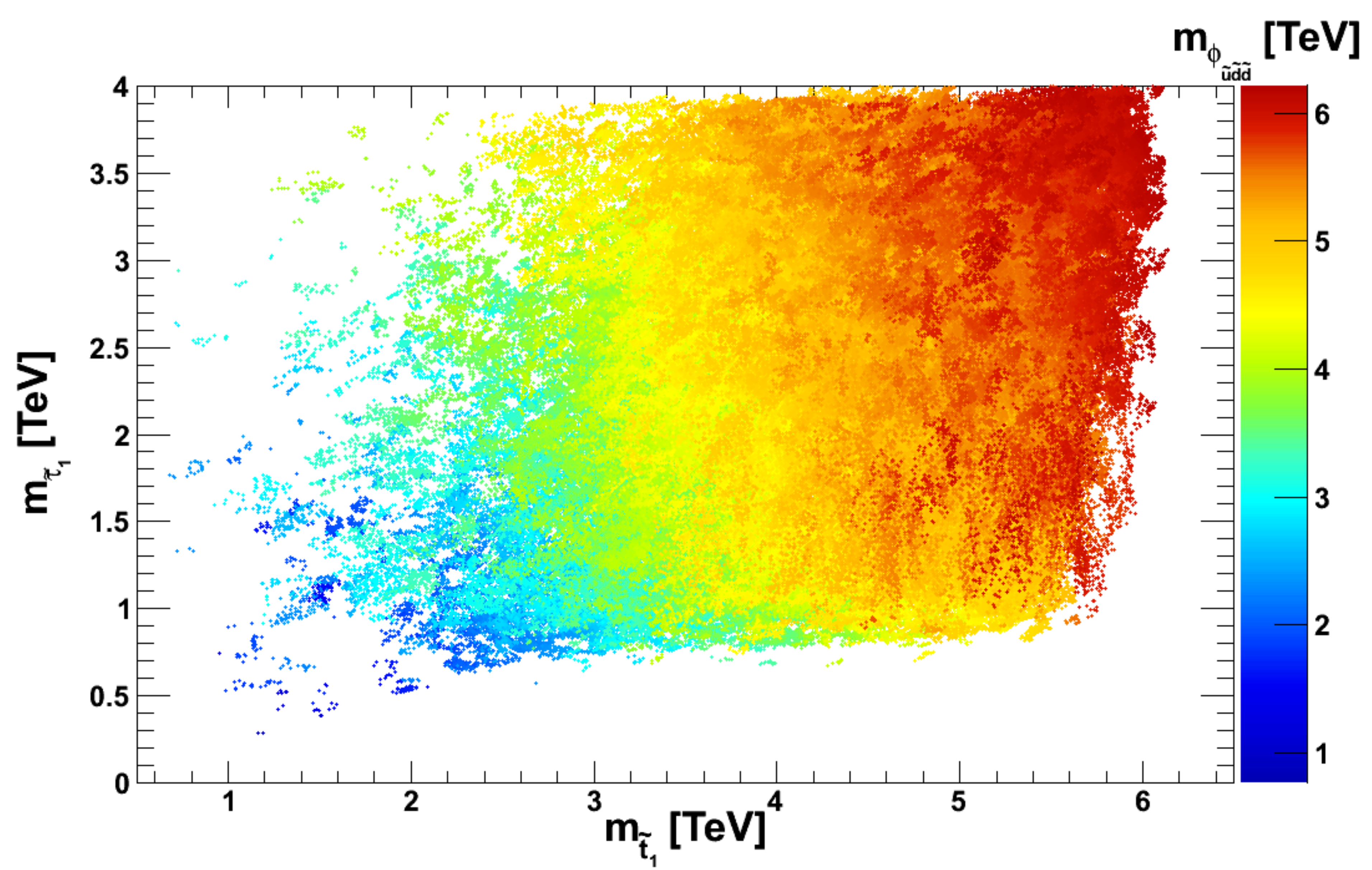}}  
\caption{The correlation between stau mass $m_{\tilde\tau_{1}}$ and stop mass $m_{\tilde t_1}$. The colour coding corresponds to the inflaton masses for $\widetilde L\widetilde L\widetilde e$ (a) and $\widetilde u\widetilde d\widetilde d$ (b).}
\label{staumasses}
\end{figure}

Such predictions can also depend on other parameters, such as the stau mass, see figure~\ref{staumasses}. The prediction differs depending on whether the inflaton correspond to the $\widetilde{u}\widetilde{d}\widetilde{d}$ or $\widetilde{L}\widetilde{L}\widetilde{e}$ inflation mechanism. For the $\widetilde{L}\widetilde{L}\widetilde{e}$ case shown in the panel (a) of figure~\ref{staumasses}, one finds that scenarios with \textit{light} inflaton (\ie with a mass lower than 2 TeV) correspond to staus lighter than 2 TeV and stops lighter than 2-3 TeV. More generally there is a relation between the inflaton and the stau masses, whatever the value of the stop mass. This correlation between the stau and the $\widetilde{L}\widetilde{L}\widetilde{e}$ inflaton mass can be understood because the inflaton is of leptonic origin.  Similarly, for the $\widetilde{u}\widetilde{d}\widetilde{d}$ case plotted in the panel (b) of figure~\ref{staumasses}, the inflaton mass is related to the stop mass but there is no constraint on the stau. Although such a feature can be easily understood given the nature of the inflaton, using LHC observables and searches for sparticles could provide a way to distinguish between the $\widetilde{u}\widetilde{d}\widetilde{d}$ and $\widetilde{L}\widetilde{L}\widetilde{e}$ scenarios. In addition, we find that staus in both scenarios can be lighter than $1$ TeV, thus offering another possible window for probing this model at LHC. Discovering a relatively light stau at LHC together with a specific stop mass would constrain the parameters of the model and thus provide a determination of the inflaton mass.

\begin{figure}[h]
\centering
\includegraphics[width=8cm,height=5cm]{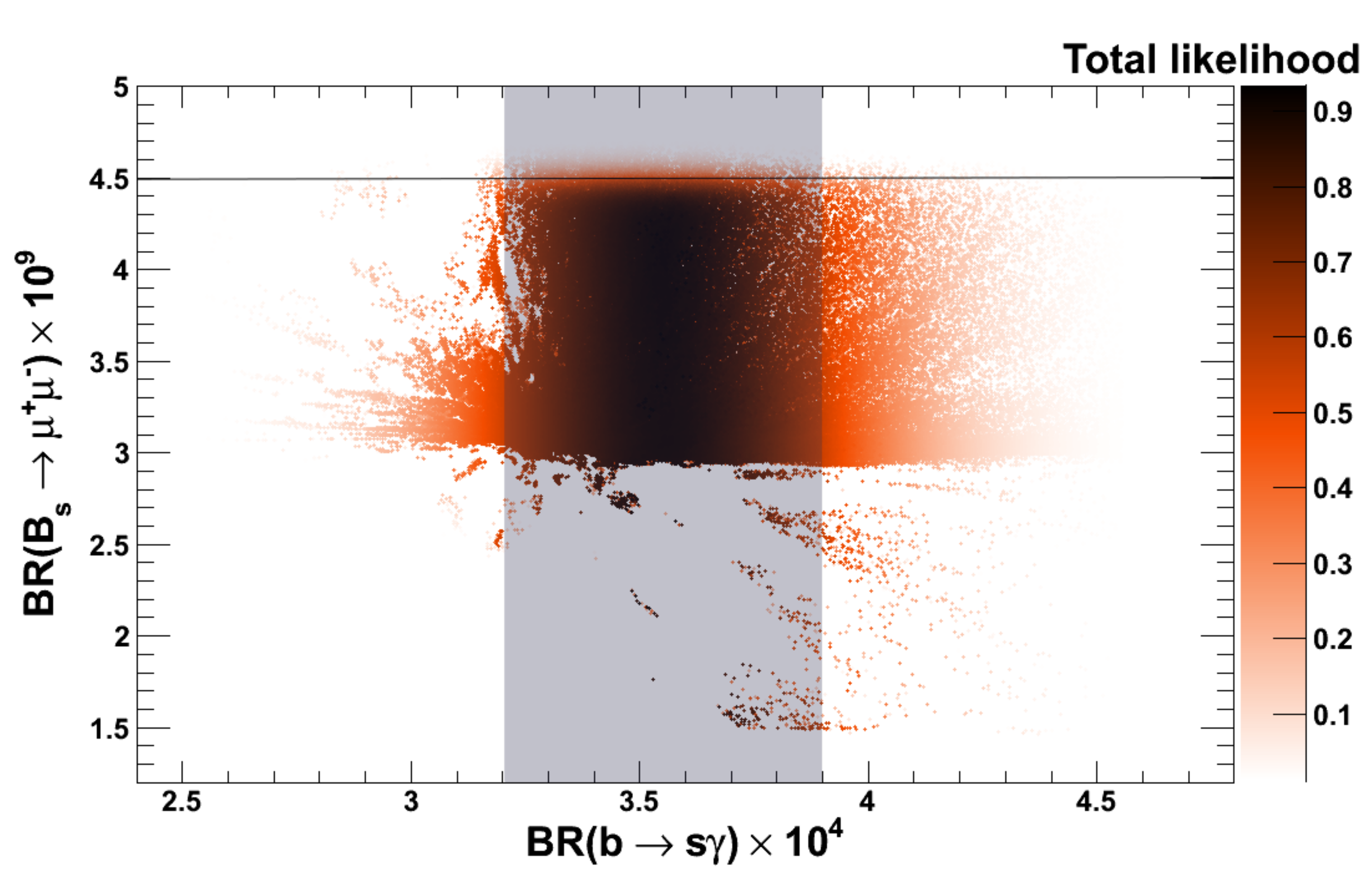}
\caption{The branching ratios of $B_s \rightarrow \mu^+ \mu^-$  and $b\rightarrow s \gamma$ are shown with the colour coding corresponding to the likelihood. The shaded region shows points within $b\rightarrow s \gamma$ experimental and theoretical error bars. }
\label{bsmumu}
\end{figure}

Specific observables such as $\bsmu$ and $\bsg$ are also interesting to consider. In particular, in figure~\ref{bsmumu}, one can see that most of the scenarios which fall within the observed range of the $b \ra s \gamma$ decay rate lead to a relatively large $B_s \ra \mu^+ \mu^-$ branching ratio, basically within $3 \times 10^{-9}$ and $4.5 \times 10^{-9}$. Some scenarios are nevertheless excluded (\ie with a contribution larger than $4.5 \times 10^{-9}$) and the latest LHCb results \cite{:2012ct} exclude now the predictions $\bsmu < 2 \times 10^{-9}$. This provides additional scope for detecting such scenarios at LHC since most scenarios are within the sensitivity of the LHCb experiment.

Finally, for completeness, we display the expected SI elastic scattering cross section associated with these scenarios in a Xenon-based experiment. We superimpose on this plot the limits obtained by the XENON100 experiment in 2011~\cite{Aprile:2011hi} as well as the improvements obtained in 2012 \cite{Aprile:2012nq} which are extremely robust regarding the relative scintillation efficiency $L_{eff}$ at this mass scale\footnote{Even though it may be affected by astrophysical uncertainties, see~\cite{McCabe:2011sr,Frandsen:2011gi} and uncertainties on quark coefficients of the nucleon that were taken as the default values in {\tt micrOMEGAs}.}~\cite{Davis:2012vy}. We also represent the predicted limit for the XENON1T experiment.  

As one can see, most of the scenarios presented in this chapter regarding NUHM2 are well below the limit set by the 2011 data of the XENON100 experiment. Nevertheless the 2012 data already probe the allowed parameter space quite significantly, especially for LSP masses below 800 GeV. The projected sensitivity for XENON1T indicates that it may be possible to probe the whole NUHM2 parameters in the future if not already ruled out by the LHC.

\begin{figure}[h]
\centering
\includegraphics[width=8cm,height=5cm]{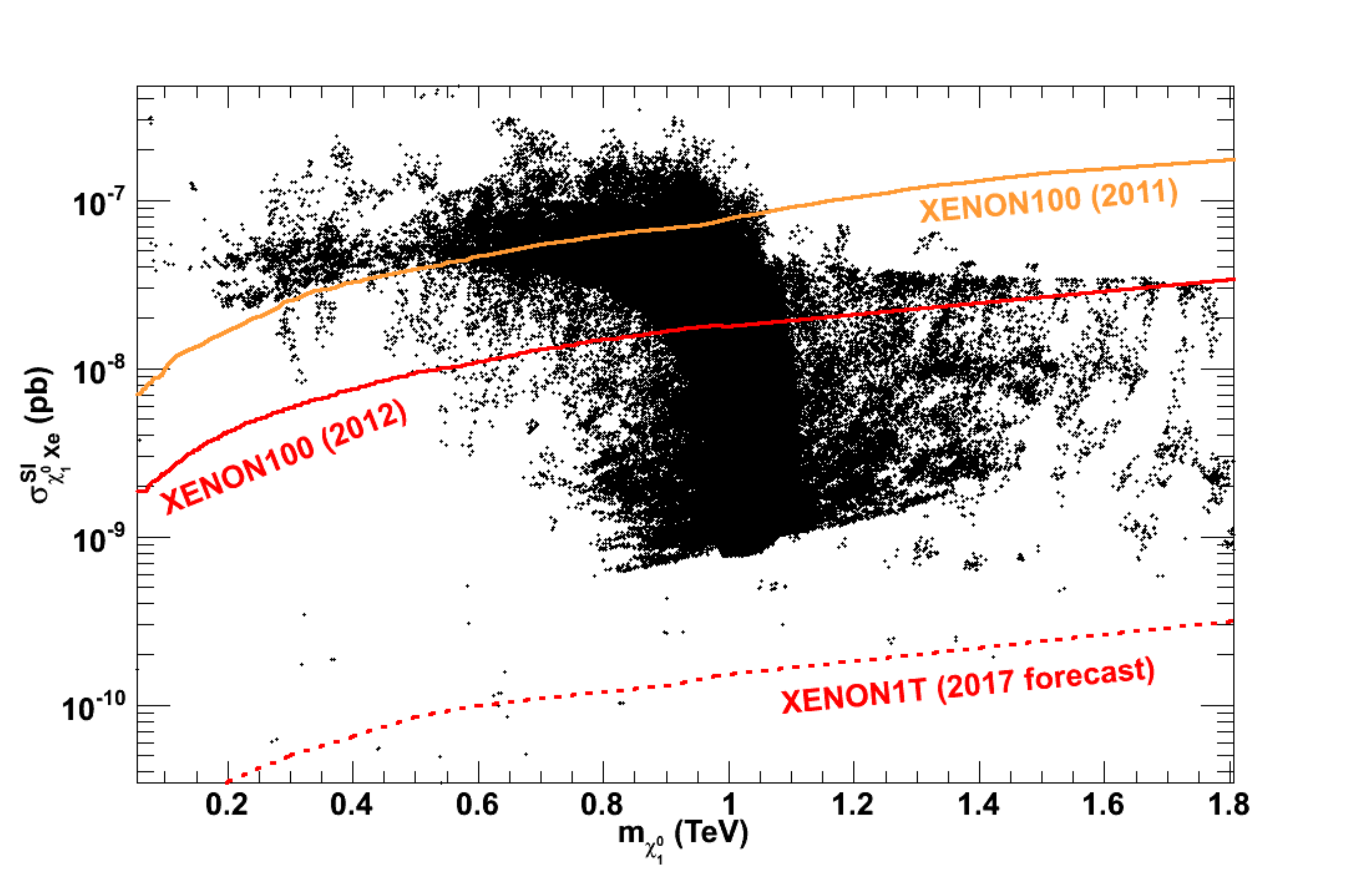}
\caption{The expected and observed limits from Xenon-based experiments and the allowed NUHM2 parameter space in the plan neutralino SI cross section\vs neutralino mass.}
\end{figure}

\section{Conclusions}

In this study we have identified the regions of the NUHM2 parameter space which are compatible with the observed DM abundance (assuming that the neutralino is the DM candidate), the Higgs boson mass constraints from LHC, and the constraints set on the inflationary potential to match the CMB constraints.

We have considered two inflaton candidates ($\widetilde{u}\widetilde{d}\widetilde{d}$ and $\widetilde{L}\widetilde{L}\widetilde{e}$) for which the \textit{high} scale of inflation $\phi_0$ is intimately tied up to the low scale physics at the LHC scale via the RGEs, and which are compatible with the amplitude of the perturbations, $\delta_H=1.91\times 10^{-5}$ and the $2\sigma$ tilt in the power spectrum $0.934 \leq n_s\leq 0.988$. 

We used two methods. One consists in finding benchmark points and the other one in performing a more complete scan of the parameter space by using a MCMC code. Our main conclusion is that for most configurations the neutralino DM is mainly higgsino-like and nearly degenerate with an NLSP chargino. The $\widetilde{u}\widetilde{d}\widetilde{d}$ inflaton appears to be \textit{fairly light} but still heavier than 1 TeV while the $\widetilde{L}\widetilde{L}\widetilde{e}$ inflaton can be as light as 500 GeV. In both cases however it is possible to find configurations in which both the staus and the stops are potentially within the reach of the LHC, thus indicating that sparticle searches at LHC could actually provide a mean to constrain the inflaton mass for some subset of the NUHM2 parameter space. These NUHM2 scenarios will have to be confronted to new measurements of the branching ratios $\bsmu$ and $\bsg$. Finally LHC constraints or potential hints could be enhanced by the results of the forthcoming DM Direct Detection experiments such as the XENON1T experiment.

As can be seen from figure~\ref{fig3}, hints of a TeV scale inflaton together with the precise measurement of the Higgs boson mass would actually narrow down the scale of inflation. Combined with the Planck satellite precise measurements, one should actually be able to pinpoint both the scale of inflation $\phi_0$ and the corresponding mass $m_{\phi}$ at the scale of inflation, thus providing a window on extremely high energy physics which also complements the current observations from the CMB radiation.

\chapter{The phenomenological MSSM confronting Indirect Detection of Dark Matter}
\label{chapter:ID}

\minitoc\vspace{1cm}
\newpage

This chapter is mostly based on the article \cite{Belanger:2012ta} with slight modifications on the MCMC study and additional details.
\section{Introduction}

As described in section~\ref{subsubsec:3.ID}, despite some drawbacks the ID of DM can be used as a powerful tool to probe DM scenarios. This is what we will analyse in this chapter. For instance, the \Fermi-LAT\ collaboration has been able to set stringent limits on the DM pair annihilation cross section into SM particles using the diffuse $\gamma$-ray emission in dwarf Spheroidal (dSph) galaxies~\cite{Ackermann:2011wa} and also in the Milky Way~\cite{Ackermann:2012rg,Ackermann:2012qk}. It has ruled out DM candidates with a total annihilation cross section of $\langle \sigma v \rangle = 3 \times  10^{-26} \ \rm{cm^3/s}$ if $m_{\rm DM} \lesssim 30$ GeV. 
This constituted a remarkable milestone as such a value corresponds to the one suggested by the thermal freeze-out scenario, which is generally considered as a strong argument in favour of WIMPs. 

These limits nevertheless weaken at higher DM masses, therefore allowing for heavier DM candidates with a larger pair annihilation cross section. For example, for $m_{\rm{DM}} =$ 100 GeV the limit relaxes to $\langle \sigma v \rangle \lesssim 10^{-25} \ \rm{cm^3/s}$ while for $m_{\rm DM} =$ 500 GeV, it reads $\langle \sigma v \rangle \lesssim 3 \times 10^{-25} \ \rm{cm^3/s}$, which is one order of magnitude higher than the \textit{thermal} cross section.

Discovering such a configuration with large values of the pair annihilation cross section would invalidate the thermal freeze-out model and either point towards the existence of non-thermal processes in the early Universe or potentially call for mechanisms such as freeze-in and regeneration. Explaining the observed DM relic density may remain nevertheless  challenging. For example, in \cite{Williams:2012pz}, it was shown that candidates with a total annihilation cross section exceeding $\langle \sigma v \rangle =  10^{-24} \ \rm{cm^3/s}$ (corresponding to a thermal relic density smaller than $3 \%$) would be ruled out by the \Fermi-LAT experiment if they were regenerated at 100$\%$. 

The measurement of the galactic $\bar{p}$ flux presented by the \PAMELA\ collaboration~\cite{Adriani:2008zq,Adriani:2010rc} is an interesting tool to probe DM. While extensive work was done to explain the electron/positron excesses in terms of DM annihilations (or decays), the implications of the absence of anomalies in the $\bar{p}$ spectrum has remained relatively unexploited. Indeed only a relatively small number of works~\cite{Cirelli:2008pk,Donato:2008jk,Boehm:2009vn,Evoli:2011id,Garny:2011ii,
Asano:2011ik,Garny:2012eb} have dealt with it and shown that large DM annihilation cross sections can be constrained by the  \PAMELA\ data.

In this chapter we will see a more systematic analysis of these general anti-proton constraints on the DM annihilation cross section, including paying attention to the uncertainties associated with DM and astrophysical predictions, which will demonstrate that these measurements can actually constrain the properties of specific DM scenarios. To illustrate this, we will work within a \textit{simplified} version of the pMSSM~\cite{Djouadi:1998di} in which all sfermion masses are set to 2 TeV, except for the stop and sbottom masses. The soft masses for the stop are allowed to be much lighter to obtain a Higgs boson mass around 125 GeV. In this scenario the only particles with masses below the TeV threshold are therefore the neutralino, chargino, the supersymmetric Higgs bosons and the lightest stop and sbottom. Such a configuration with a large mass splitting between gauginos and sfermions may actually seem unusual from a supersymmetric point of view (albeit close to split SUSY~\cite{ArkaniHamed:2004fb}) but it is supported by the unfruitful searches for squarks and gluinos at LHC. 

We will investigate scenarios where the neutralino pair annihilation cross section into $W^+ \, W^-$  gauge bosons is enhanced (due in particular to the chargino exchange diagram). Such a large annihilation cross section gives both a significant anti-proton and diffuse gamma ray flux, together with a gamma ray line, and is therefore potentially constrained by the \PAMELA\ and \Fermi-LAT data. In the pMSSM, where non-universality for the gaugino masses can be assumed, such an enhancement is realized when the LSP neutralino is mass degenerated with the chargino,\ie when the neutralino has a significant wino component. Assuming total regeneration of the neutralino DM, the combination of both \Fermi-LAT and \PAMELA\ data is therefore expected to constrain the wino fraction of the LSP. Note that constraints on the neutralino composition are also expected to be obtained in presence of a lower sfermion mass spectrum. However the effect of the chargino-neutralino mass degeneracy on $\gamma-$ray and $\bar{p}$ production would be much harder to characterize. Hence our choice in favour of a heavy sfermion mass spectrum.

\section{Anti-proton and $\g$-ray bounds on  $\sigma_{{\rm DM}  \ {\rm DM} \  \ra \ W^+ W^-}$}
\label{constraints}

In this section we discuss how anti-proton and gamma ray data impose generic constraints on the DM pair annihilation cross section into $W^+ W^-$ as a function of the DM mass.

\subsection{Generic bounds on $\sigma_{{\rm DM} \ {\rm DM} \ \ra \ W^+ W^-}$ from $\bar{p}$}
\label{pbarconstraints}

$W^\pm$ production in space leads to abundant anti-proton production as the $W^\pm$'s decay products hadronise. The flux of anti-protons thus produced by DM annihilations into a pair of $W^{\pm}$ gauge bosons in the MW and collected at Earth is therefore determined by the DM pair annihilation cross section into $W^+ W^-$, the DM mass and the DM halo profile. It also depends on the anti-proton propagation parameters which are being considered. Here it is assumed that the DM halo profile is described by an Einasto profile (other choices make a small difference) and the standard three sets of propagation parameters (`MIN', `MED', `MAX'), summarized in table~\ref{tab:proparam}, are considered. The anti-protons fluxes used are given in~\cite{Cirelli:2010xx}. 

\begin{table}[t]
\center
\begin{tabular}{c|cccc} \hline \hline
 &  \multicolumn{4}{c}{\textbf{Antiproton parameters}}  \\
Model  & $\delta$ & $\mathcal{K}_0$ [kpc$^2$/Myr] & $V_{\rm conv}$ [km/s] & $L$ [kpc]  \\
\hline \hline 
MIN  &  0.85 &  0.0016 & 13.5 & 1 \\
MED &  0.70 &  0.0112 & 12 & 4  \\
MAX  &  0.46 &  0.0765 & 5 & 15 \\ \hline \hline 
\end{tabular}
\caption[Propagation parameters for anti-protons in the galactic halo.]{Propagation parameters for anti-protons in the galactic halo (from~\cite{Delahaye:2007fr,Donato:2003xg}). $\delta$ and $\mathcal{K}_0$ are the index and the normalization of the diffusion coefficient, $V_{\rm conv}$ is the velocity of the convective wind and $L$ is the thickness of the diffusive cylinder. 
\label{tab:proparam}}
\end{table}

In order to constrain the annihilation cross section, the \PAMELA\ data \cite{Adriani:2010rc} above an anti-proton energy of 10 GeV (to avoid the uncertainty related to solar modulation) will be considered. Both the predicted energy spectrum and the flux depend on the DM mass that is being assumed. For each value $m_{\rm{DM}}$, the sum of the astrophysical background flux and predicted anti-protons flux originating from DM will be compared with the \PAMELA\ data. Given the uncertainties on the astrophysical background, we will see two different procedures to derive meaningful limits :
\begin{itemize}
\item One can be regarded as aggressive since it assumes a fixed background. For obtaining the corresponding limits (referred to as \textit{fixed background} in the following). The standard flux of astrophysical (secondary) anti-protons from~\cite{Bringmann:2006im} is considered and the DM anti-protons flux is added. The result is then compared with the \PAMELA\ data and a 95\% C.L. limit is derived by imposing that the global $\chi^2$ of the background $+$ DM flux does not exceed by more than 4 units the $\chi^2$ of the background only hypothesis.

\item The other one is more conservative given that it considers an adjusted background within the uncertainties. For obtaining conservative limits (hereafter referred to as \textit{marginalized background}), the standard form of the background spectrum predicted is again considered except that now the normalisation of the background spectrum $A$ and the spectral index $p$ are allowed to vary within 40$\%$ and $\pm 0.1$ respectively (for each value of the DM mass and pair annihilation cross section into $W^+ \, W^-$). The standard description of the background spectrum is multiplied by a factor $A \, (T/T_0)^p$, where $T$ is the anti-proton kinetic energy, $T_0 = 30$ GeV is a pivot energy with $0.6 < A < 1.4$ and $-0.1 < p < +0.1$. This point allow to include the uncertainty predicted in~\cite{Bringmann:2006im}. Then the DM contribution is added and the pair of parameters $A$ and $p$ which minimizes the global $\chi^2$ with the \PAMELA\ data for a point in the plane $(\langle \s_{ann}v \rangle, m_{\rm DM})$ is determined. The 95\% C.L. limit is then imposed by requiring that the marginalized global $\chi^2$ does not exceed 4 units with respect to the null hypothesis (which has been marginalized consistently). 
\end{itemize}

The latter procedure allows to increase the gap between the expected $\bar{p}$ background and the actual \PAMELA\ data which in consequence leaves more space for a possible DM injection of anti-protons and leads to weaker limits on the DM pair annihilation cross section.
\begin{figure}[!htb]
\begin{center}
\centering
\subfloat[]{\includegraphics[width=5cm,height=6cm]{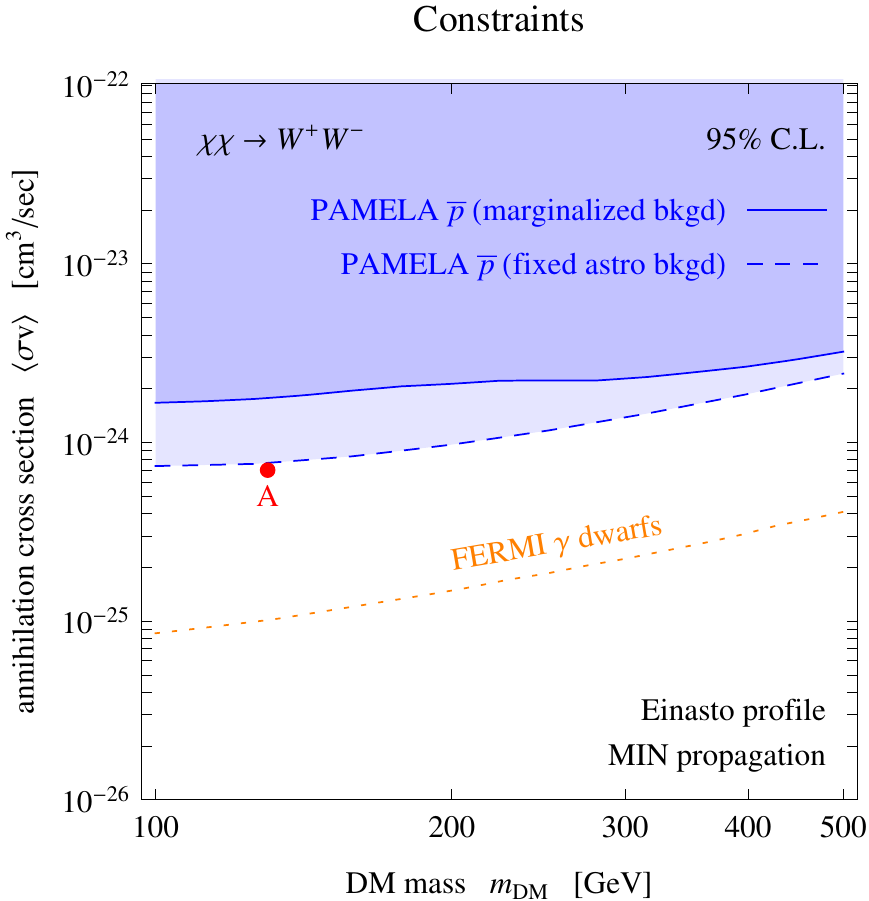}}  
\subfloat[]{\includegraphics[width=5cm,height=6cm]{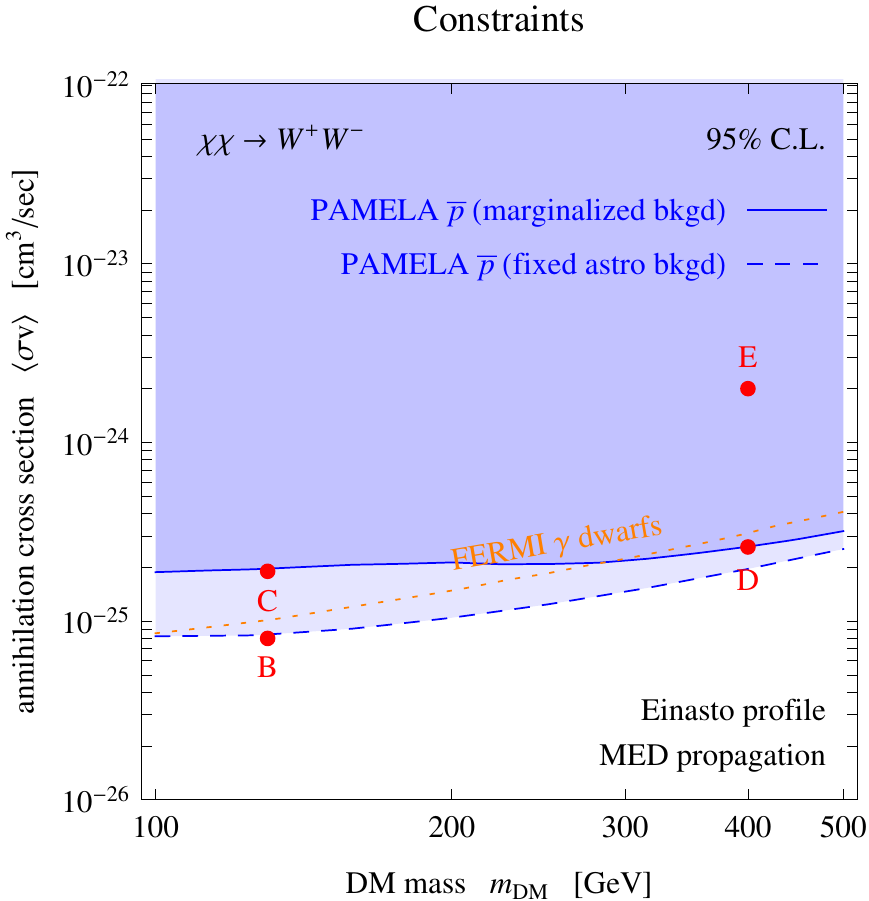}}
\subfloat[]{\includegraphics[width=5cm,height=6cm]{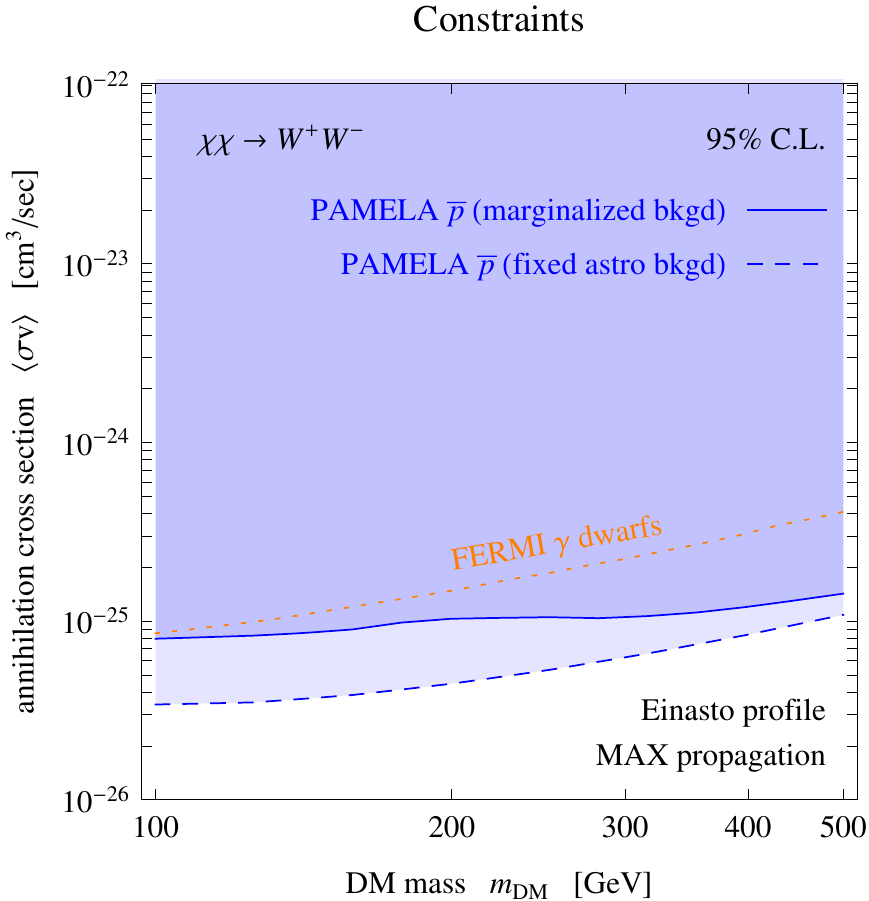}}
\caption[Anti-proton constraints on DM annihilation into $W^+W^-$, respectively for the `MIN' (a), `MED' (b) and `MAX' (c) propagation parameters.]{\label{antip_crosssection} Anti-proton constraints on DM annihilation into $W^+W^-$, respectively for the `MIN' (a), `MED' (b) and `MAX' (c) propagation parameters. The constraints obtained by the \Fermi-LAT collaboration from satellite dwarf galaxies are superimposed. Five benchmark points are also displayed.}
\end{center}
\end{figure}

The derived constraints are displayed in figure~\ref{antip_crosssection} for the `MIN', `MED' and `MAX' set of parameters. As expected, the \textit{conservative} limits are slightly less constraining than the \textit{aggressive} ones. The choice of propagation parameters has also a big impact on the type of constraints that can be set : in terms of cross sections, the difference between the `MIN' and `MAX' limits exceeds a factor 10.

\begin{figure}[!htb]
\begin{center}
\centering
\subfloat[]{\includegraphics[width=5cm,height=5.5cm]{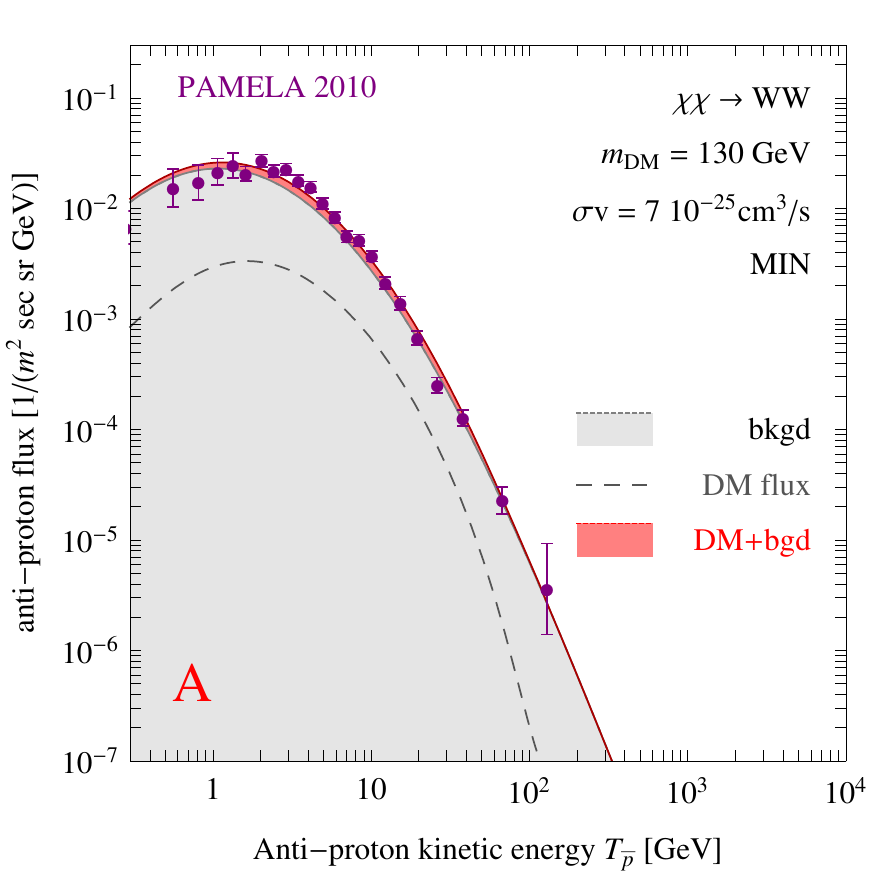}}  
\subfloat[]{\includegraphics[width=5cm,height=5.5cm]{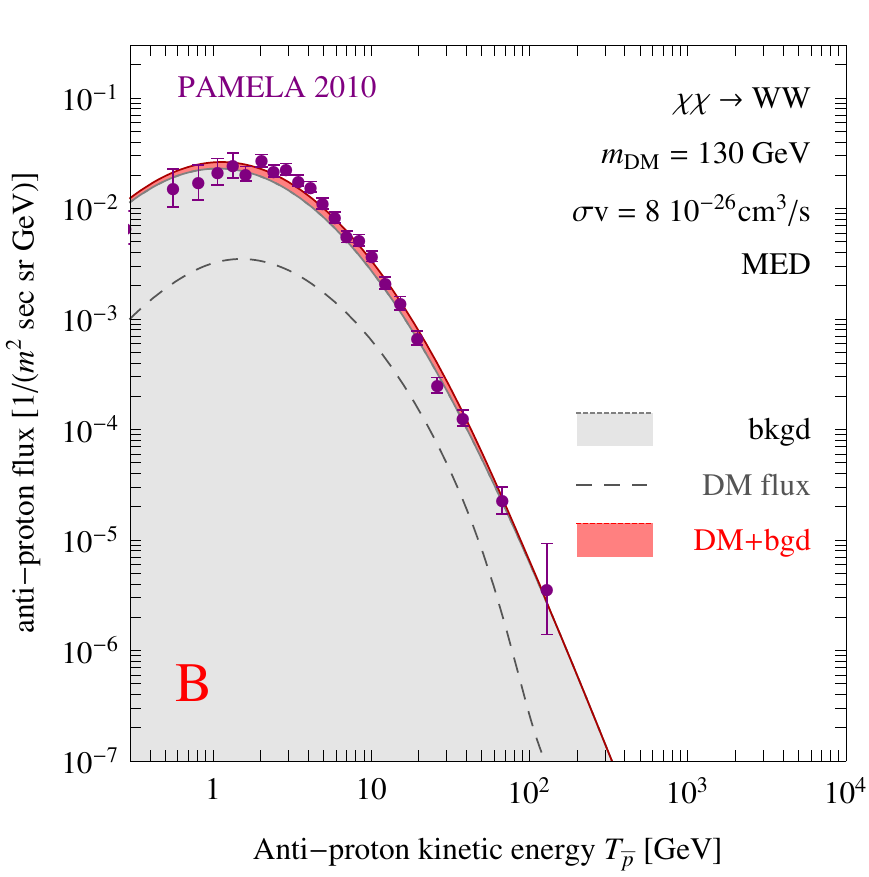}}
\subfloat[]{\includegraphics[width=5cm,height=5.5cm]{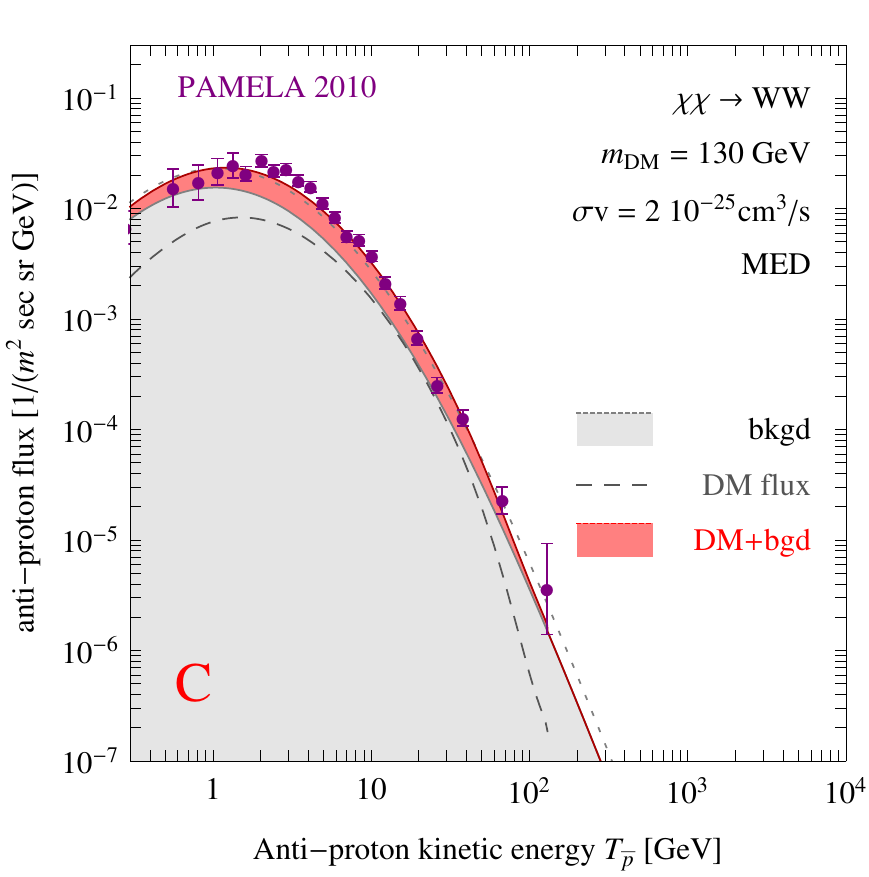}}\\
\subfloat[]{\includegraphics[width=5cm,height=5.5cm]{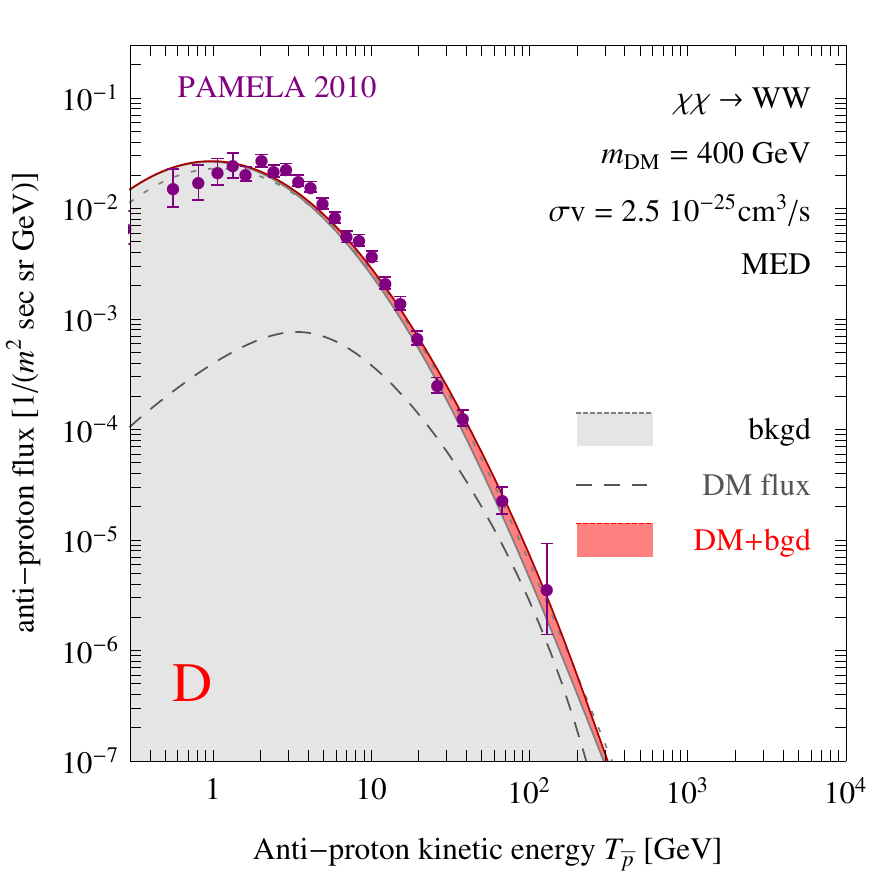}}
\subfloat[]{\includegraphics[width=5cm,height=5.5cm]{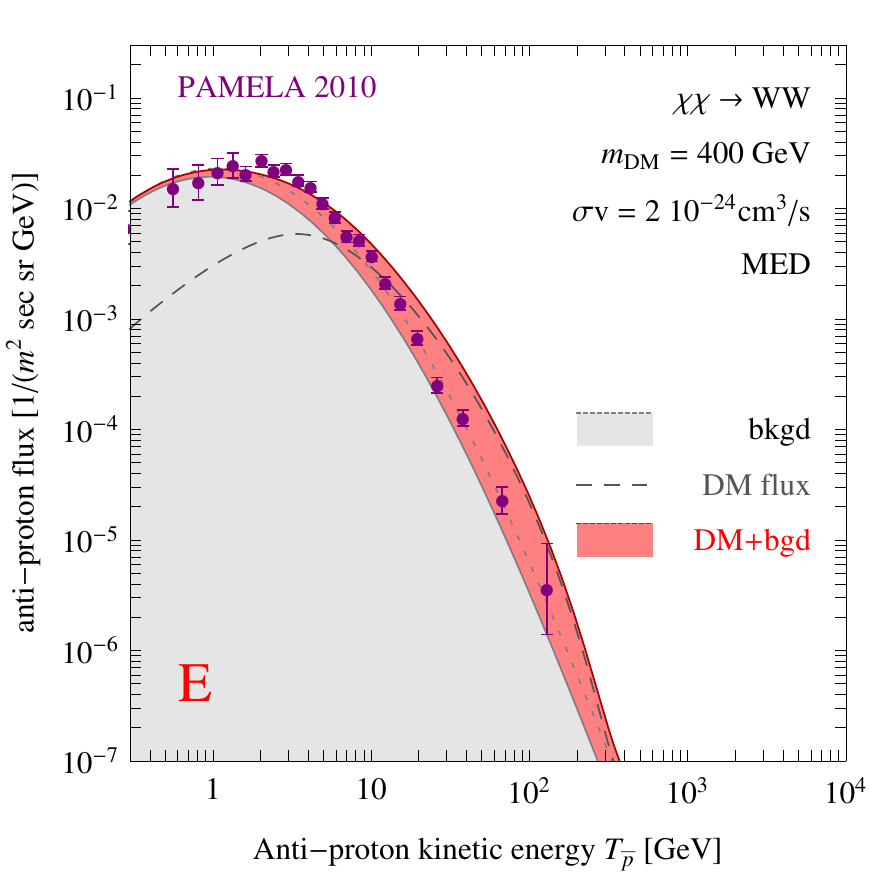}}
\caption[Examples of the fluxes of anti-protons (astrophysical background and DM-produced) compared with the data from the \PAMELA\, experiment for the sample points `A' to `E' as defined in figure~\ref{antip_crosssection}.]{\label{examples} Examples of the fluxes of anti-protons (astrophysical background and DM-produced) compared with the data from the \PAMELA\, experiment for the sample points `A' to `E' as defined in figure~\ref{antip_crosssection}. In each panel the assumed parameters (DM mass, annihilation cross section and propagation scheme) are reported.}
\end{center}
\end{figure}

To understand more precisely how these constraints work, 5 benchmark points are defined (hereafter referred to as `A',`B',`C',`D',`E'), corresponding to different DM masses, cross sections, propagation parameters and constraint procedures. The corresponding fluxes are plotted in figure~\ref{examples}. As can be seen, benchmark points `A' to `D' correspond to \textit{borderline} scenarios where the total $\bar{p}$ flux (i.e. the sum of the expected flux from DM and astrophysical background) is not significantly exceeding the data. Point `E', on the other hand, displays how \textit {badly} the data is violated inside the excluded region.

The first apparent feature from figure~\ref{examples} is that one can actually exclude a small excess in anti-protons produced by relatively light DM particles because the \PAMELA\ data set have very small error bars at energies below 100 GeV, hence the strength of the constraints. It is then instructive to compare case `A' and `B' : these two scenarios refer to the same DM mass and constraint procedure; they also predict a very similar flux, as can be seen in panel (a) and (b) of figure~\ref{examples}, but have a different annihilation cross section. The latter is  much larger for `A' than for `B'. This is because the propagation scheme was assumed to be `MIN' for the former and `MED' for the latter. With the `MIN' propagation set, the yield of anti-protons is about one order of magnitude smaller than with `MED' (since the galactic diffusion zone is much smaller in the former case) and therefore the constraint on the annihilation cross section is about one order of magnitude looser than for the `MED' case. On the other hand the constraint obtained for `MAX' (which is not shown here) is stronger than for `MED'. 

The comparison between points `B' and `C' shows the impact of the constraint procedures. Although both `B' and `C' have the same DM mass and propagation scheme, we find that the value of the annihilation cross section that is allowed for `C' is larger than for `B'. The reason is that `C' corresponds to the scenario in which the limit is obtained by using the \textit{marginalized background} procedure so there is more room for DM while `B' corresponds to a \textit{fixed background} scenario so the associated constraints are stronger.  

Finally, the comparison between `C' and `D' enables one to understand why the \textit{marginalized background} constraints are rather independent of the DM mass, despite the fact that the error bars in the \PAMELA\ data become larger at larger energies. For a large DM mass (case `D') the $\bar{p}$ flux is shifted towards larger energies and rather negligible at $\sim$ 10 GeV with respect to the astrophysical background; there is thus little room to reduce the background (which alone has to fit the data at low energy) and consequently there is little room left for a DM contribution at large energies. As a result, the bound remains stringent. 

We will then use in the study of the pMSSM the derived limit from the \textit{marginalized background} procedure obtained using the `MED' set of anti-protons propagation parameters.

\subsection{Generic bounds on $\sigma_{{\rm DM} \ {\rm DM} \ \ra \ W^+ W^-}$ from gamma-rays}
\label{gammaconstraints}

In DM scenarios, the $W^\pm$ production is associated with gamma-ray emission through (i) the decay and hadronisation of the decay products of the $W^{\pm}$ bosons, (ii) the radiation of a photon from the internal and/or final states associated with ${\rm DM} \ {\rm DM}  \ra W^+ W^-$  (iii) DM annihilations into $\gamma\gamma$ and $\gamma Z$ (which can be seen as a higher order process based on ${\rm DM} \ {\rm DM} \ra W^+ W^-$). The first case leads to a \textit{continuum} spectrum of $\gamma$-rays; the second leads to \textit{sharp features} in the $\gamma$-ray continuum spectrum and the third to \textit{$\gamma$-ray lines}. The resulting fluxes from these processes have to be compared with the gamma-ray flux measurements from the MW or from other nearby galaxies. Therefore we now look at the current $\gamma$-ray constraints derived in the literature (mainly from \Fermi-LAT analyses), paying particular attention to that derived from the $W^+ W^-$ channel since this is the main focus of our analysis.

\medskip 

\subsubsection{Continuum}

The \Fermi-LAT collaboration has recently published two different analyses of the continuum diffuse gamma-ray emission from the Milky Way halo~\cite{Ackermann:2012rg,Ackermann:2012qk}. Since no clear DM signal has been found, these have been used to set upper limits on the DM pair annihilation cross-section into various channels as $b\bar{b},gg,W^+W^-,e^+e^-,\mu^+\mu^-$ and $\tau^+\tau^-$. However the most stringent limits on the DM annihilation cross section have actually been obtained from another \Fermi-LAT analysis based on the diffuse $\gamma$-ray emission from dSph galaxies. These DM dominated objects indeed represent a good target for DM searches. In the present analysis we will use the results from~\cite{Ackermann:2011wa}. Figure~\ref{antip_crosssection} shows the comparison between the dSph galaxies limits and the \PAMELA\ anti-proton bounds that were derived in section~\ref{pbarconstraints}. Depending on the propagation scheme that has been chosen for the anti-protons, the dSph galaxies $\gamma$-ray bounds is somewhat more stringent or looser than the constraints from the anti-proton data. For example, for the `MED' case and \textit{marginalized background}, $\bar{p}$ limits become more constraining than $\gamma$-ray bounds when $m_{\rm DM} \gtrsim 290$ GeV. However they are stronger than the $\gamma$-ray limits whatever the value of $m_{\rm DM}$ (assuming $m_{\rm DM}> 100$ GeV) for the \textit{fixed background} procedure. Since nevertheless the $\bar{p}$ and $\gamma$-ray limits are basically of the same order of magnitude, we will include both constraints in our study.

\subsubsection{Internal bremsstrahlung and final state radiation}  

Gamma rays produced directly as radiation from an internal line or a final state are in general suppressed by the fine structure constant $\alpha_{_{em}}$. However, for a $t$-channel diagram, the associated cross section can be enhanced when the intermediate particle is almost mass degenerated with the DM. Typically the enhancement factor is about $m_{\rm DM}^2/(M_I^2-m_{\rm DM}^2)$ where $M_I$ is the mass of the intermediate particle (\ie a chargino for neutralino pair annihilation into a $W^{\pm}$ pair). These processes are model dependent and cannot be constrained generically but they will be included in our $\gamma$-ray estimates when we investigate the neutralino pair annihilations into $W^+ W^-$ in the pMSSM. Section~\ref{subsec:fsr} will be dedicated to the study of their impact on the pMSSM scenarios considered.

\subsubsection{Line(s)}  

Annihilations directly into $\gamma\gamma$ or $\gamma Z$ occur at one-loop level (since DM particles do not couple directly to photons) and are therefore generically suppressed. However they lead to a distinctive signature, namely a mono-energetic gamma-ray line at an energy $E=m_{\rm DM}$ or $E = m_{\rm DM} \, (1- M_Z^2/(4 m_{\rm DM}^2))$ which can be looked for. Since the purpose of this study is to set constraints on the DM properties we will disregard the possible evidences for two gamma-ray lines at 129 and 111 GeV. We only consider the constraints which were reported by the \Fermi-LAT collaboration on line searches in the MW~\cite{Ackermann:2012qk}. Since the status of these searches is not definite, we made the choice to not include these constraints to perform the scans over the pMSSM parameter space. However we do check that the scenarios which survive the $\bar{p}$ and $\gamma$-ray constraints are not killed by these line searches.

\section{Chargino-neutralino mass degeneracy}

Now that we have obtained the maximal value of the DM pair annihilation cross section into $W^+ W^-$ that is observationally allowed as a function of the DM mass, we can focus on the pMSSM and investigate the impact of this generic limit on the neutralino DM parameter space.

\subsection{Neutralino pair annihilations into $W^+ W^-$}

\begin{figure}[h]
\begin{center}
\includegraphics[scale=0.8]{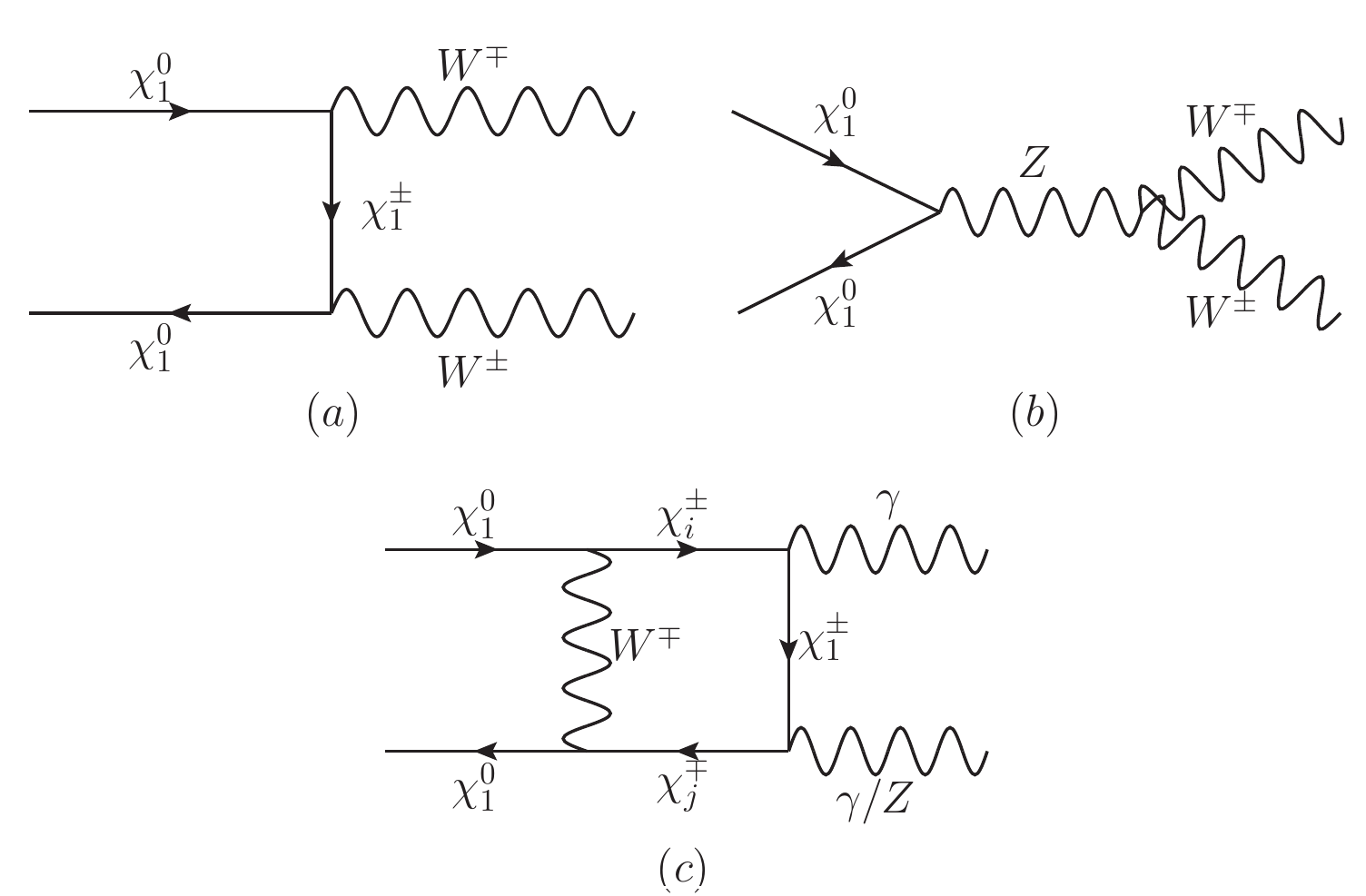}
\end{center}
\caption[Dominant neutralino pair annihilation diagrams into $W^+ \, W^-$, $\gamma \, \gamma$ and $\gamma \, Z$ for this analysis.]{Dominant neutralino pair annihilation diagrams into $W^+ \, W^-$, $\gamma \, \gamma$ and $\gamma \, Z$ for this analysis, with $i,j \in \{1,2\}$. \label{pMSSM_diagrams}}
\end{figure}

In a scenario where all the sfermions are very heavy, the dominant neutralino annihilation channels are expected to be mostly into gauge bosons, more specifically into $W^+W^-$ pairs. All loop-induced $W^{\pm}$ production diagrams which involve sfermions are expected to be suppressed. Hence the processes which are expected to lead to a significant $W^+ W^-$ production in the pMSSM only involve charginos and Z boson. The corresponding diagrams are displayed in figure~\ref{pMSSM_diagrams}. Since they correspond to $s-$ and $t-$channel diagrams, we typically expect resonant or enhanced annihilations when $m_{\chi^0_1} \sim  M_{Z}/2$ or $m_{\chi^0_1} \sim m_{\chi^\pm_1}$ (\ie when the neutralino and chargino are mass degenerated). These ultimately enhance the neutralino pair annihilations into $\gamma \gamma$ \cite{Bergstrom:1997fh,Boudjema:2005hb} and $\gamma Z$ \cite{Boudjema:2005hb,Ullio:1997ke} through in particular the box diagram displayed in figure~\ref{pMSSM_diagrams}c. 

The questions that we want to address in the next subsections are : i) which part of the SUSY parameter space is excluded by the $\bar{p}$ limits and do these limits exclude more allowed configurations than the $\gamma$-rays bounds ? ii) which values of the neutralino-chargino mass degeneracy are actually constrained by astrophysical data ?

\subsection{Exploring the supersymmetric parameter space}

To answer this, we explore the pMSSM parameter space using the same approach based on MCMC as in chapter \ref{chapter:NUHM2}. 

\begin{table}
\begin{center}
\begin{tabular}{cc|cc}\hline \hline
\textbf{Free parameter} & \textbf{Range} & \textbf{Nuisance parameter} & \textbf{Range} \\ \hline \hline
$M_1$ & [10, 500] GeV & $m_u/m_d$ & [0.424,0.682]\\ 
$M_2$ & [100, 1000] GeV & $m_s/m_d$ & [16.5,21.3]\\ 
$\mu$ & [-2000, 2000] GeV & $\s_{\pi N}$ & [29,59] MeV\\ 
$\tan \beta$ & [2, 75] & $\s_s$ & [0,42] MeV\\ 
$m_{\tilde{Q}_3}, m_{\tilde{u}_3}$ & [100, 3000] GeV & $m_t$ & [170.5,175.9] GeV\\
$A_t$ & [-8000, 8000] GeV\\ \hline \hline
\end{tabular}
\caption{Range chosen for the pMSSM free parameters and nuisance parameters. \label{tab:5.range}}
\end{center}
\end{table}

Our free parameters and their corresponding range are summarized in table~\ref{tab:5.range}. These include the soft mass terms associated with the squarks of the third generation (\ie here $m_{\tilde{Q}_3}$ and $m_{\tilde{u}_3}$) and the trilinear coupling $A_t$. To obtain sfermion masses at the TeV scale, we set all the soft masses to 2 TeV. In addition, we set the other trilinear couplings to 0 TeV and the CP-odd Higgs boson mass to 2 TeV. In this framework, the bino mass $M_1$ does not exceed 500 GeV; our choice for the other parameters indeed ensures that the neutralinos and charginos are light and the mass splitting between the neutralinos and charginos remains relatively small. 

On top of these free parameters, we had to include some nuisance parameters over which we will marginalize~\cite{Dumont:2012ee}. These are related in particular to the quark content of the nucleons (since they have a non-negligible impact on the computation of the DM-nucleon scattering cross section as explained in section \ref{subsubsec:3.DD} with the corresponding values in table~\ref{contentnucl}) and the top quark mass \cite{Lancaster:2011wr} (since it has an impact on the Higgs sector). All of them are allowed to vary in the range [$N_\mathrm{exp}$ -3$\s_\mathrm{err}$, $N_\mathrm{exp}$ +3$\s_\mathrm{err}$], with $N_\mathrm{exp}$ ($\s_\mathrm{err}$) the corresponding value (error) as shown in table~\ref{tab:5.range}. The same scanning method is used for both nuisance and free parameters; the multidimensional parameter space of the model is then 
\begin{center}
$\{M_1, M_2, \mu, \tan \beta, m_{\tilde{Q}_3}, m_{\tilde{u}_3}, A_t, m_u/m_d, m_s/m_d, \s_{\pi N}, \s_s, m_t\}.$
\end{center}
A slight difference with respect to the free parameters is that we will define a likelihood function for the nuisance parameters since we want to vary their values strictly around their bounds at 1$\s$. 

We also require that the lightest Higgs boson mass only varies within the range allowed by the ATLAS and CMS experiments in july 2012 \cite{Aad:2012tfa,Chatrchyan:2012ufa} plus a small theoretical error of 1 GeV, namely $m_{h^0} = 125.9 \pm 2.0$~GeV. Nevertheless note that the theoretical uncertainty on the calculation of the lighter scalar Higgs boson in the MSSM can reach 5 GeV as shown in \cite{Allanach:2004rh}. By precaution, we checked that the scenarios which seemed allowed were compatible with the \HB code~\cite{Bechtle:2008jh,Bechtle:2011sb} (even though the most recent LHC results on the Higgs boson~\cite{Aad:2012tfa,Chatrchyan:2012ufa} are not included in this version). Note that we did not add any requirement about the Higgs boson signal strength to perform the scans.

The neutralino relic density $\Omega_{\chi^{0}_{1}} h^2$ is allowed to vary in $[1\%\textrm{ WMAP7}, \textrm{ WMAP7}]$ with $\Omega_{\rm WMAP7} h^2 = 0.1123 \pm 0.0035$,  using WMAP 7-year $+$ BAO $+$ $H_0$ and the \texttt{RECFAST 1.4.2} code~\cite{Komatsu:2010fb}. We do not consider smaller values of the relic density as these correspond to DM scenarios with very large values of the annihilation cross section and ultimately overproduce gamma-rays in the galaxy (\ie are excluded) if their relic density is entirely regenerated, see \cite{Williams:2012pz}. 

We did not implement LHC limits on sfermion masses because our requirement of a sfermion spectrum at the TeV scale should ensure that they are satisfied. However updates on direct searches for relatively light stop and sbottom would be useful to implement to further constrain the parameter space.

For each scenario (corresponding to a point in the pMSSM parameter space), we then calculate the total likelihood function as in chapter \ref{chapter:NUHM2}. The likelihood functions associated with each observable and nuisance parameter are defined as follows :

\begin{itemize}
\item To $m_{h^0}$, $\Omega_{\chi^{0}_{1}} h^2$ and all nuisance parameters, we associate the likelihood function  $\mathcal{L}_1$ defined in eq.~\ref{eq:4.L1}. The tolerance $\s$ concerning the nuisance parameters is equal to $\s_\mathrm{err}/10$ to prevent too large deviation especially for the strange quark content of the nucleon $\s_s$ which has a large uncertainty.
\item As in chapter \ref{chapter:NUHM2}, we use the Gaussian likelihood function $\mathcal{L}_2$ defined in eq.~\ref{eq:4.L2} for the $\bsg$ observable. This observable is important as it receives a potentially large contribution from chargino/stop loops when either one of these particles is light. This contribution can be compensated by the charged Higgs/top diagram but the latter is however suppressed when the charged Higgs mass is at the TeV scale.
\item We also include the likelihood function $\mathcal{L}_3$ defined in eq.~\ref{eq:4.L3} for the 2012 XENON100 limits \cite{Aprile:2012nq} to ensure that the scans do not select too large values of the DM-nucleon scattering cross section.  
In fact we also associate $\mathcal{L}_3$ to regions of the parameter space where ${\sigma v}_{{\chi^{0}_{1}\chi^{0}_{1}\ra W^{+}W^{-}}} $ is greater than $10^{-27}$~${\rm cm}^3/{\rm s}$ to render the scan more efficient. As in chapter \ref{chapter:NUHM2} we note that some experimental measurements are very discrepant with the SM expectations as the example of $\amu$. These observables receive additional contributions from particles in the pMSSM but they are too small to explain the observations. Therefore we associate $\mathcal{L}_3$ to them so that the likelihood is equal to unity if the predictions are much below the measured value. 
\end{itemize}
The set of constraints that we use is summarized in table~\ref{tab:5.constraints}. 

\begin{table*}[!htb]
\begin{tabular*}{1.\textwidth}{  c  c  c  c  }
\hline \hline \textbf{Constraint} & \textbf{Value/Range} & \textbf{Tolerance} & \textbf{Likelihood} \\ 
\hline \hline
       $m_{h^0}$ (GeV) \cite{Aad:2012tfa,Chatrchyan:2012ufa} & [123.9, 127.9] & 0.1 & $\mathcal{L}_1$ \\ \hline 
       $\Om_{\chi^0_1} h^2$ \cite{Komatsu:2010fb} & [0.001123, 0.1123] & 0.0035 & $\mathcal{L}_1$ \\ \hline
       $\bsg$ $\times$ $10^{4}$ & 3.55 & exp : 0.24, 0.09 & $\mathcal{L}_2$ \\ 
        \cite{Asner:2010qj,Misiak:2006zs} & & th : 0.23 & \\ \hline
       $\s^{SI}_{\chi^0_1 {\rm Xe}}$ (pb) & ($\s_N$, $m_{\rm DM}$) plane & -$\s_N(m_{\rm DM})$/100 & $\mathcal{L}_3$ \\
       &  from \cite{Aprile:2012nq} & & \\ \hline
       ${\sigma v}_{^{\chi^{0}_{1}\chi^{0}_{1}\ra W^{+}W^{-}}}$ & 1 & 0.01 & $\mathcal{L}_3$ \\
       ($10^{-27}$~${\rm cm}^3/{\rm s}$) & & & \\ \hline
       $\amu$ $\times$ $10^{10}$ \cite{Davier:2010nc} & 28.70 & -0.287 & $\mathcal{L}_3$ \\ \hline 
       $\bsmu$ $\times$ $10^{9}$~\cite{Aaij:2012ac} & 4.5 & -0.045 & $\mathcal{L}_3$ \\ \hline        
       $\Delta \rho$ & 0.002 & -0.0001 & $\mathcal{L}_3$ \\ \hline 
       $R_{B^\pm \to \tau^\pm \nu_\tau} (\frac{\rm pMSSM}{\rm SM})$ \cite{Charles:2011va} & 2.219 & -2.219$\times10^{-2}$ & $\mathcal{L}_3$ \\ \hline 
       $Z \ra \chi^0_1 \chi^0_1$ (MeV) & 1.7 & -0.3 & $\mathcal{L}_3$ \\ \hline 
       $\sigma_{e ^+ e ^- \ra \chi^0_1 \chi^0_{2,3}}$ & 1 & -0.01 & $\mathcal{L}_3$ \\ 
       $\times \mathscr{B}(\chi^0_{2,3} \ra Z \chi^0_1)$ (pb) \cite{Abbiendi:2003sc} &&& \\ \hline \hline 
\end{tabular*}
\caption[Constraints imposed in the MCMC for the pMSSM.]{\label{tab:5.constraints} Constraints imposed in the MCMC, from \cite{Nakamura:2010zzi} unless noted otherwise.}
\end{table*}

\section{Results}
\label{results}

We now look at the results of the scans regarding the ID constraints.

\subsection{Bounds on the NLSP-LSP mass splitting}

\begin{figure}[!htb]
\begin{center}
\centering
\includegraphics[width=9cm,height=6cm]{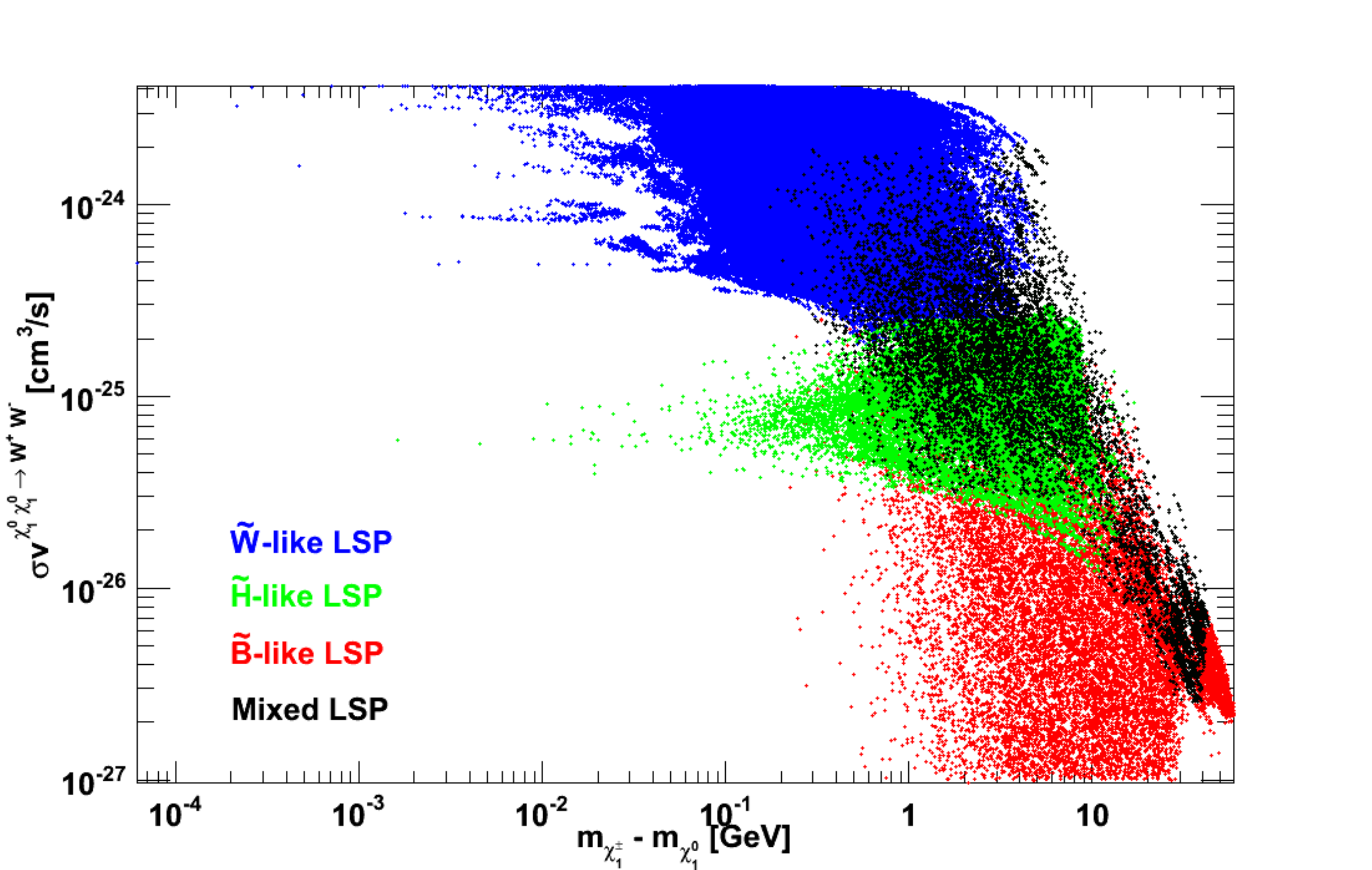}
\caption[Plot of the neutralino pair annihilation cross section into $W^+ W^-$ as a function of the chargino-neutralino mass splitting. The colour coding indicates the LSP composition.]{\label{diff-vcsWW-compoLSP}Plot of the neutralino pair annihilation cross section into $W^+ W^-$ as a function of the chargino-neutralino mass splitting. The colour coding indicates the LSP composition : in blue it is a wino, in green an higgsino, in red a bino and in black we have a mixed LSP.}
\end{center}
\end{figure}

The first result of the scan is shown in figure~\ref{diff-vcsWW-compoLSP}. The neutralino pair annihilation cross section into $W^+ W^-$ as a function of the mass degeneracy between the neutralino LSP and the chargino NLSP and in terms of the neutralino composition is displayed. 

This figure indicates the neutralino composition which maximises the $W^{\pm}$ production. As can be seen scenarios where $\sigma v_{\chi^0_1 \chi^0_1 \ra W^+ W^-}$ is the largest and the neutralino-chargino mass splitting is the smallest correspond to neutralinos with a very large wino fraction. Large values of both $\sigma v_{\chi^0_1 \chi^0_1 \ra W^+ W^-}$ and the $\chi_1^0-\chi_1^\pm$ mass splitting correspond to wino-dominated neutralinos but with a non negligible higgsino component. For these two types of wino-dominated configurations the neutralino and chargino mass degeneracy is small enough to make the $t-$channel (chargino) exchange diagram very large. As the wino fraction decreases, the mass splitting becomes larger and the $t-$channel chargino exchange diagram contribution decreases. However it remains large till the higgsino fraction which ensures large values of the $\chi_1^0 - \chi_1^\pm - W^\mp$ coupling remain significant (\ie dominate over the bino fraction).

\begin{figure}[!htb]
\begin{center}
\centering
\subfloat[]{\includegraphics[width=8cm,height=6cm]{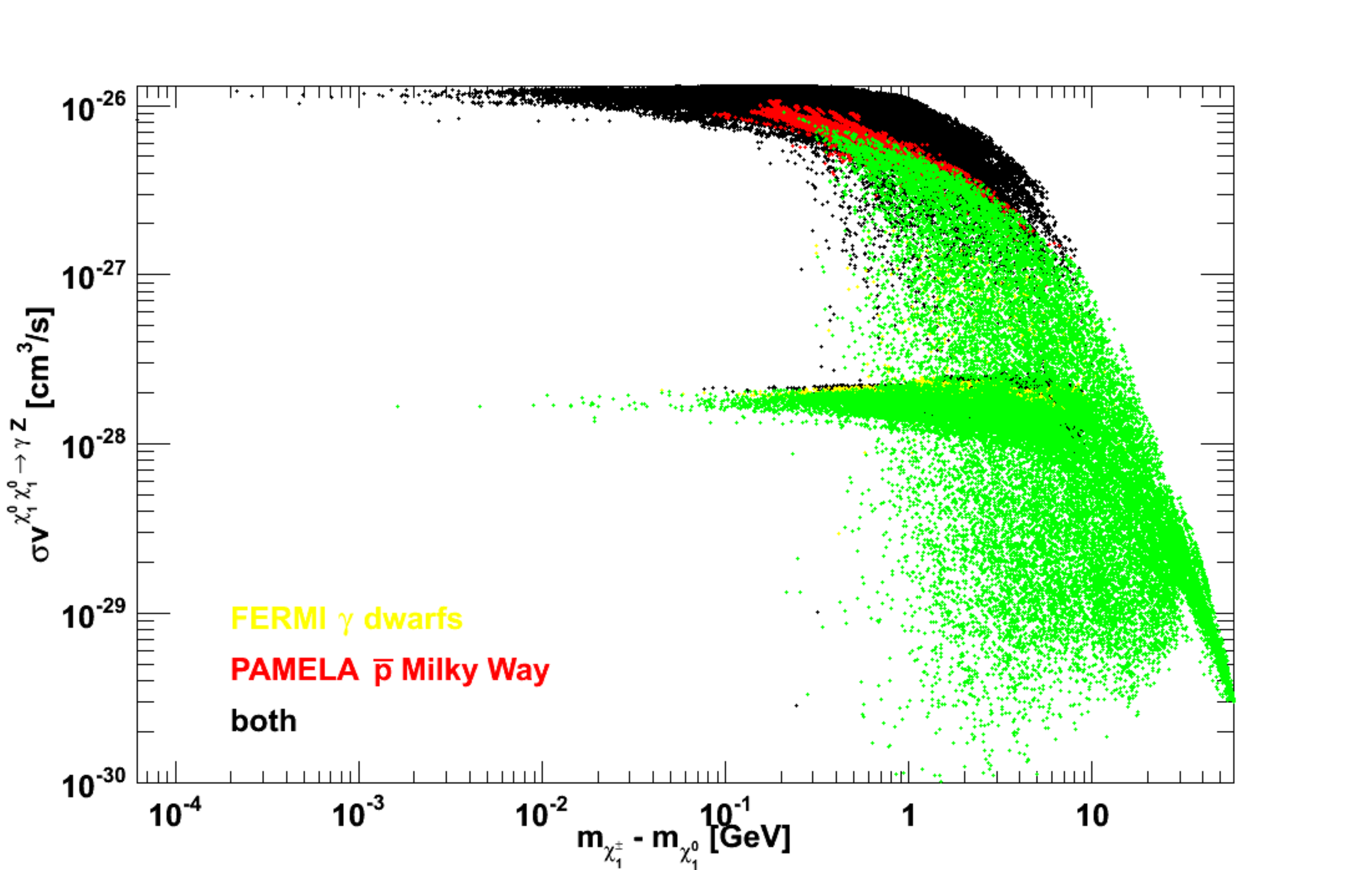}}
\subfloat[]{\includegraphics[width=8cm,height=6cm]{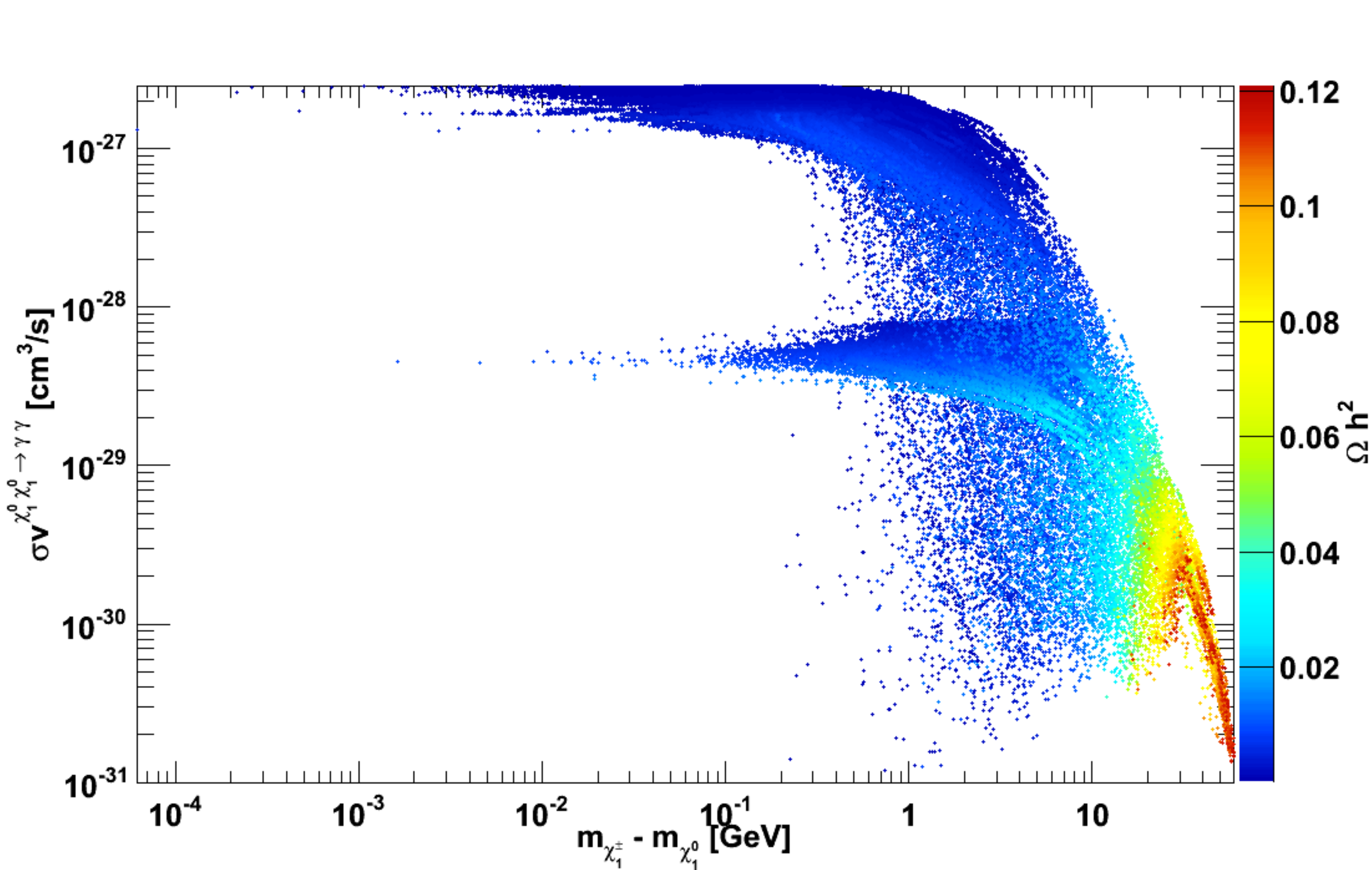}}
\caption[Plots of the neutralino pair annihilation cross section into $\gamma \gamma$ (panel (a)) and into $\gamma Z$ (panel (b)) as a function of the chargino-neutralino mass splitting.]{\label{diff-vcsg}Plots of the neutralino pair annihilation cross section into $\gamma \gamma$ (panel (a)) and into $\gamma Z$ (panel (b)) as a function of the chargino-neutralino mass splitting. In panel (a) the colour coding shows the point in the pMSSM parameter space which are not excluded by ID constraints in green and those which are excluded by \Fermi-LAT (yellow), \PAMELA\ (red) or both (black). The freeze-out relic density is displayed in panel (b).}
\end{center}
\end{figure}

In figure~\ref{diff-vcsg} we show the pair annihilation cross section into $\gamma Z$ (panel (a)) and $\gamma \gamma$ (panel (b)) as a function of the neutralino-chargino mass degeneracy $\Delta m = m_{\chi_1^\pm} - m_{\chi_1^0}$. Panel (a) shows which values of the neutralino pair annihilation cross section into $\gamma Z$ are excluded by astrophysical data as a function of the neutralino-chargino mass splitting. A similar plot is shown for $\gamma \gamma$ but the colour code now illustrates the relation between the different values of this cross section and the neutralino \textit{thermal} relic density. As one can see the shape of the scenario distribution for $\gamma \gamma$ and $\gamma Z$ is essentially the same in the $(\s v, \Delta m)$ plane. However the $\gamma Z$ cross section is approximately 10 times larger than that for $\gamma \gamma$ for every scenario. Hence combining these two figures actually gives an information about the relic density of the scenarios which are excluded by astrophysical data.

In the $\gamma Z$ plot the points excluded by the \Fermi-LAT dSph continuum $\gamma$-ray data are displayed in yellow. Those correspond, by construction, to scenarios where there is a very large $W^{\pm}$ production (and thus a large contribution to the continuum $\gamma$-ray spectrum). The regions which are excluded by the \PAMELA\ data are shown in red. The black points correspond to scenarios excluded by both the \PAMELA\ and the \Fermi-LAT data while those in green represent the points allowed by these two types of constraints. 

As one can see from the distribution of black points the largest values of the annihilation cross sections into $\gamma Z$ (and therefore $W^+ W^-$) are excluded by both measurements. Since these scenarios correspond to a small (or relatively small) chargino-neutralino mass splitting and thus large values of the $t-$channel chargino exchange diagram, we can conclude that both \PAMELA\ and \Fermi-LAT data are relevant to constrain wino-dominated neutralinos. A small number of these configurations is however constrained by only one of the \PAMELA\ or \Fermi-LAT dataset but this does not affect the maximal value of the $\chi_1^0-\chi_1^\pm$ mass splitting that can be excluded by using astrophysical considerations.

By inspecting where the neutralino pair annihilations into $\gamma Z$, $\gamma \gamma$ and $W^+ W^-$ are significant in these plots, one also finds that higgsino-dominated scenarios are constrained by both \PAMELA\ and \Fermi-LAT data because diagram~\ref{pMSSM_diagrams}a still generates a large $W^{\pm}$ production. In fact, for such an LSP, the annihilation cross section into $ZZ$ also becomes non-negligible compared to that into $W^+W^-$. Since the expected $\gamma$-ray and $\bar{p}$ spectra from $W^{\pm}$ and $Z$ production are very similar, we accounted for them both when we made the comparison with the \PAMELA\ and \Fermi-LAT data. Finally the green points which pass all the constraints have a non-negligible bino component. This reduces the chargino exchange diagram contribution and thus enables to decrease the $W^{\pm}$ (and therefore anti-proton and $\gamma$-ray) production. For these bino-like configurations one expects the stop and sbottom exchange to be relevant, leading to quarks in the final state and possibly (in particular for $b \bar{b}$) an overproduction of gamma-rays. Note however that such process would become non-negligible especially near pseudo-scalar Higgs boson resonances for heavy neutralinos. Since we put $m_{A^0}$ at 2 TeV these scenarios are not expected in this scan.

\begin{figure}[!htb]
\begin{center}
\centering
\subfloat[]{\includegraphics[width=8.3cm,height=6cm]{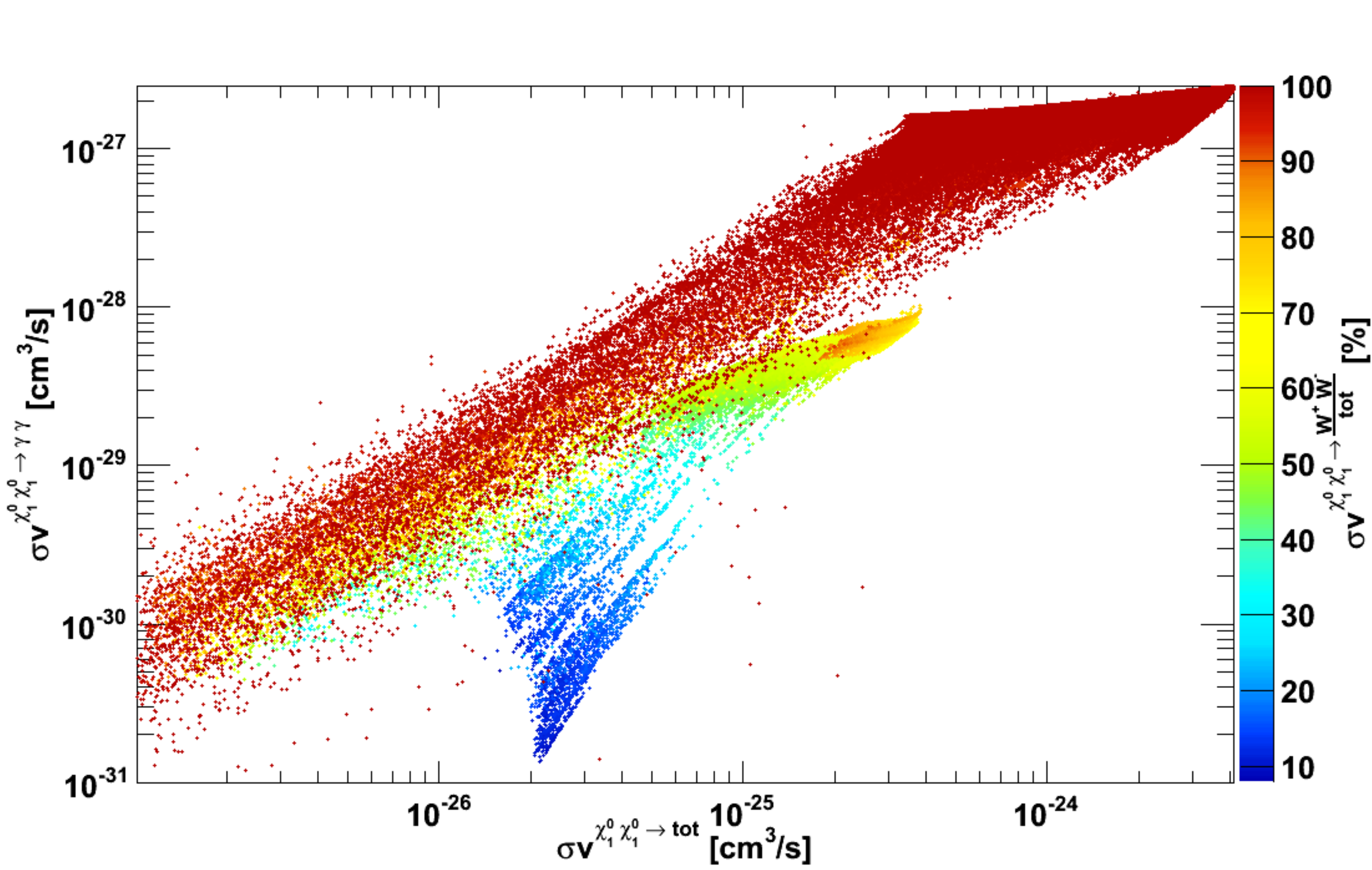}}
\subfloat[]{\includegraphics[width=8.3cm,height=6cm]{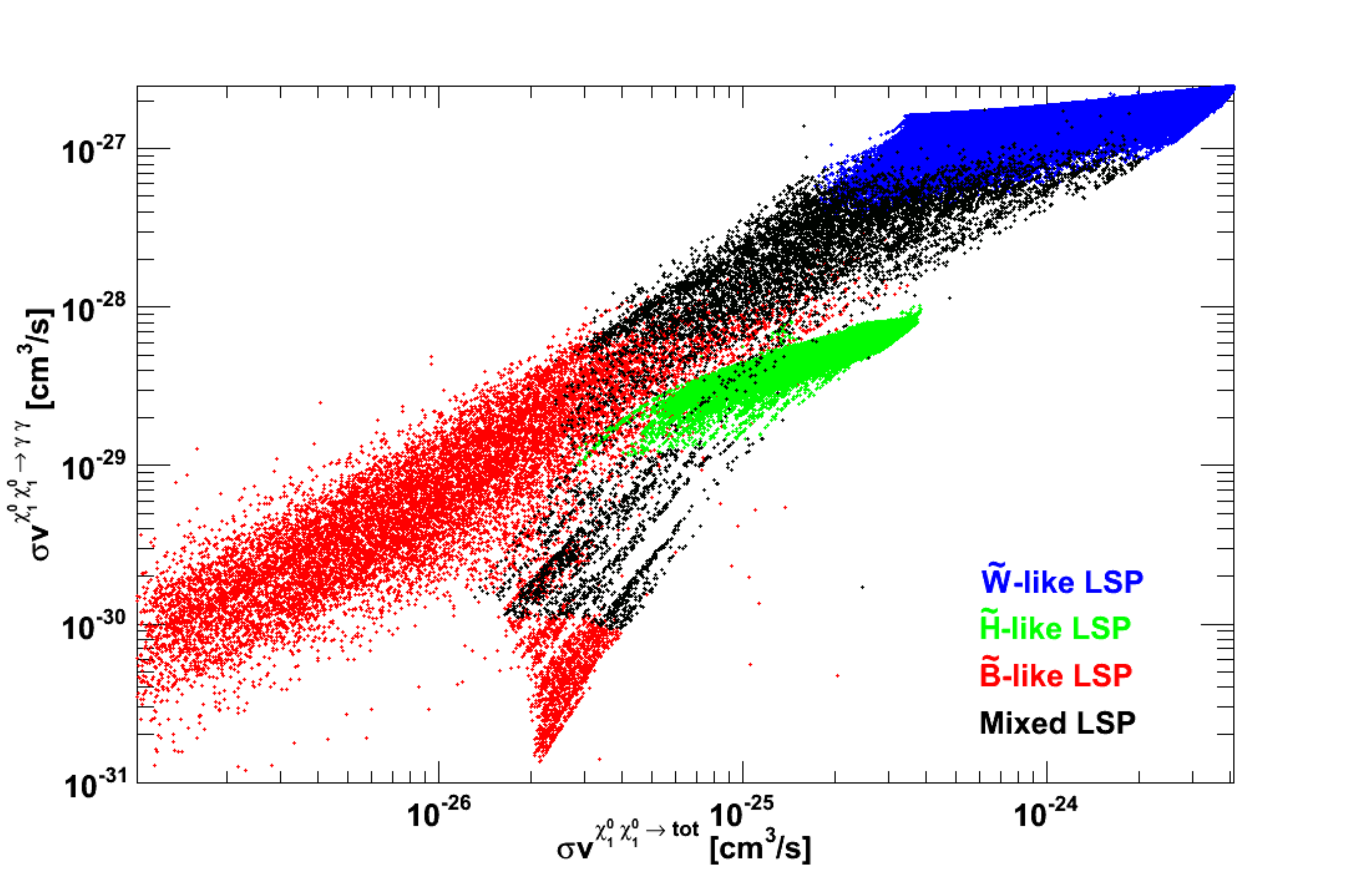}}\\
\subfloat[]{\includegraphics[width=8.3cm,height=6cm]{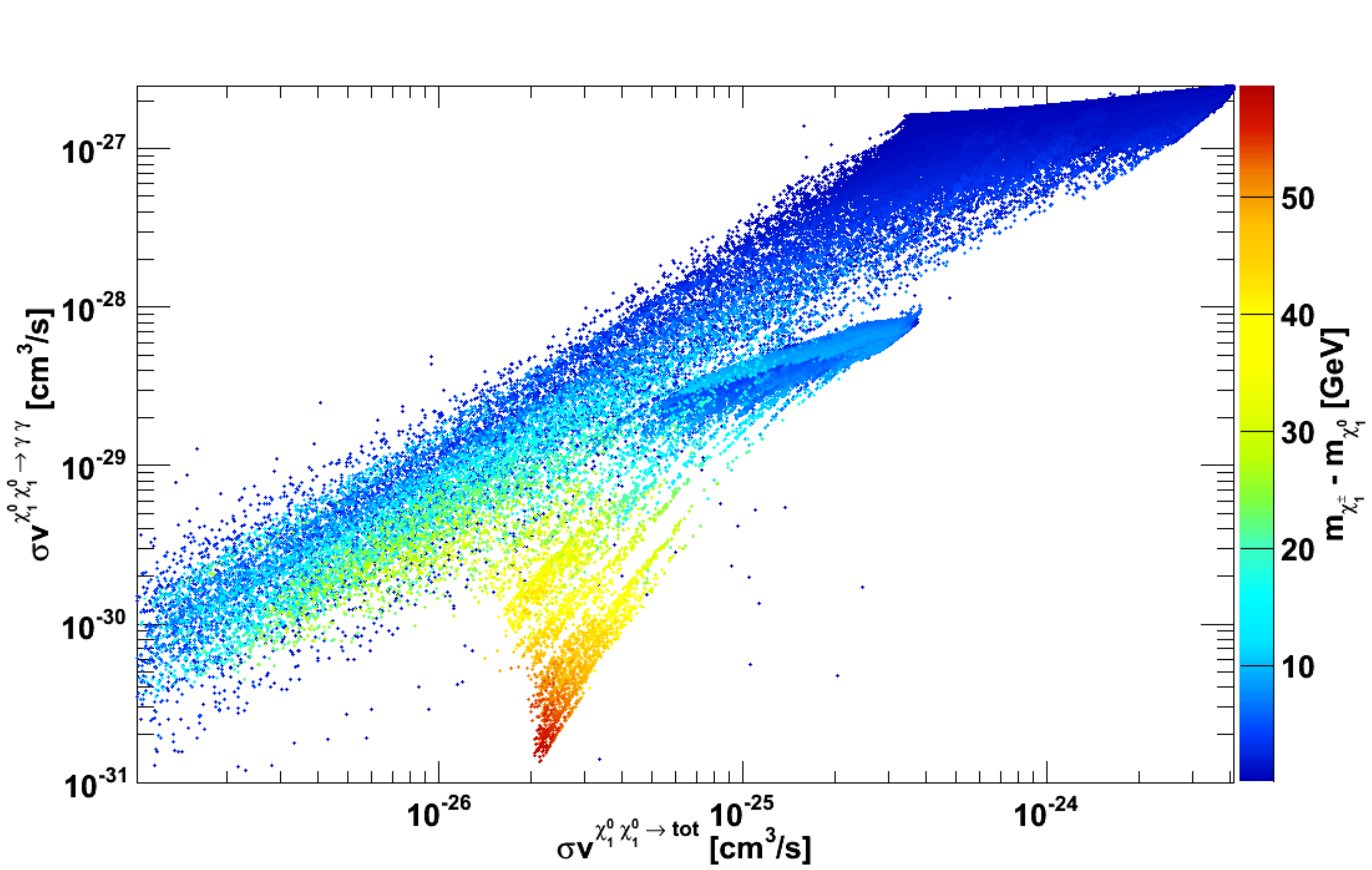}}
\caption[The neutralino pair annihilation cross section into $\gamma \gamma$ as a function of the total annihilation cross section, considering three types of colour code.]{\label{vcstot-vcsgg}The neutralino pair annihilation cross section into $\gamma \gamma$ as a function of the total annihilation cross section. The colour code of panel (a) present the fraction of annihilation cross section into $W^+ W^-$ whereas panel (b) has the same colour code as in figure~\ref{diff-vcsWW-compoLSP}. Panel (c) shows the mass splitting $\Delta m$ as colour code.}
\end{center}
\end{figure}

We then see that it is not easy in the pMSSM to always get a large branching ratio of the neutralino annihilation cross section into $W^{\pm}$. This characteristic is depicted in figure~\ref{vcstot-vcsgg} where we show the neutralino pair annihilation cross section into $\gamma \gamma$ as a function of the total annihilation cross section. For panel (a) the colour code corresponds to the fraction of annihilation cross section into $W^+ W^-$ and panel (b) represents the neutralino composition. As explained above higgsino LSP gives a non-negligible LSP annihilation into $Z$ bosons which then decreases the $W^+W^-$ branching ratio. Note that some scenarios are characterized by really small annihilation cross section into $W^{\pm}$ down to 10\%. These cases correspond to an LSP which is mostly bino and with a non-negligible higgsino component. Panel (c) shows that the lower is the fraction of annihilation cross section into $W^+ W^-$, the higher is the mass splitting $\Delta m$. Actually there are characterized by an increasing fraction of annihilation cross section into $t \bar{t}$ which becomes dominant for the largest $\Delta m$. However this region in the parameter space corresponds to low cross sections.

\begin{figure}[!htb]
\begin{center}
\centering
\includegraphics[width=9cm,height=6cm]{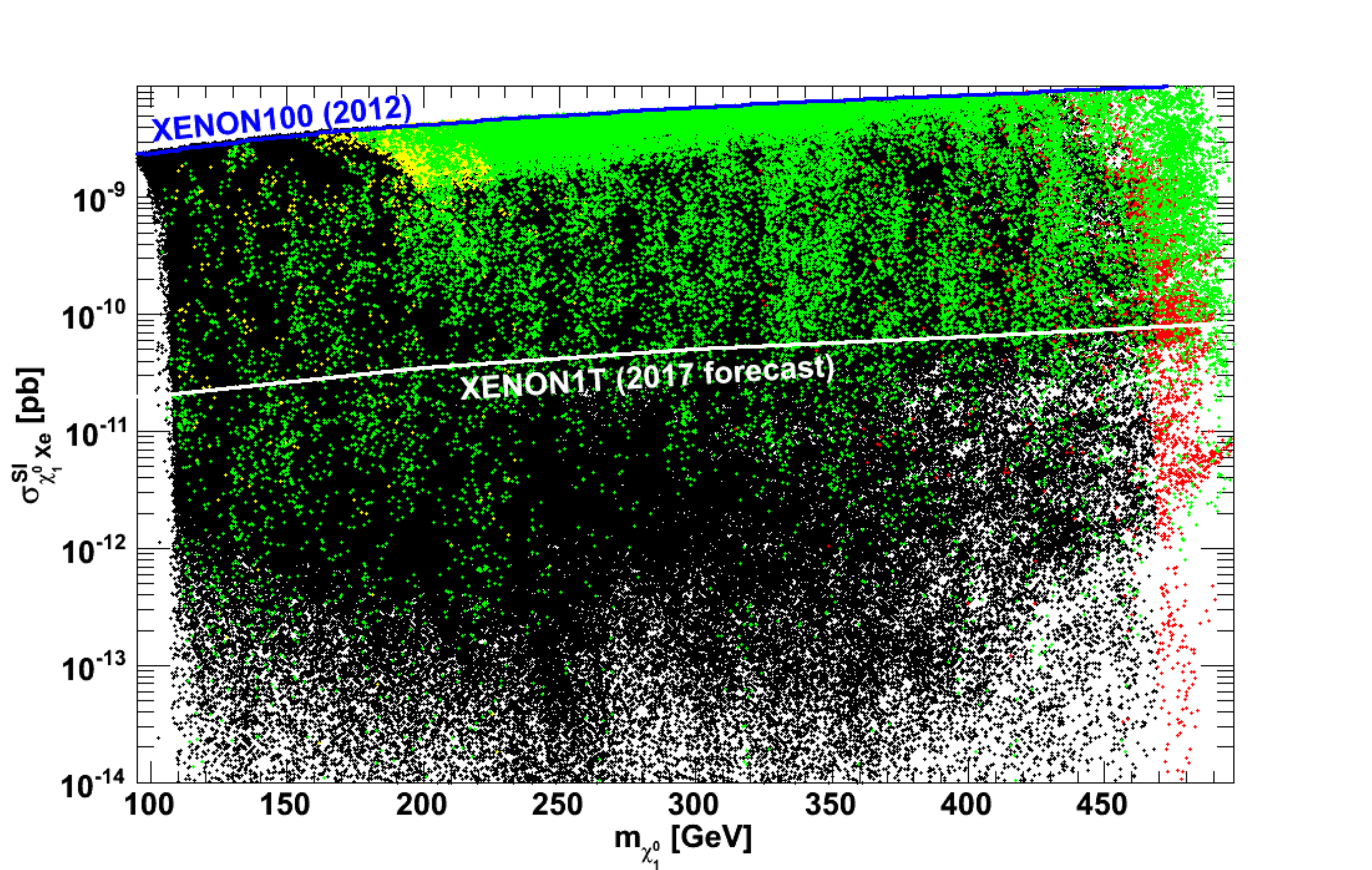}
\caption[SI DM-nucleon cross section as a function of the DM mass together with the XENON 2012 limit.]{\label{csSI}SI DM-nucleon cross section as a function of the DM mass together with the XENON 2012 limit. Same colour code as in figure~\ref{diff-vcsg}a.}
\end{center}
\end{figure}

We give in figure~\ref{csSI} the forecasted SI elastic scattering cross section as a function of the neutralino mass for a Xenon-based experiment. Again in green are the points which are astrophysically allowed, in black the points which are excluded by both \PAMELA\ and \Fermi-LAT data and in red or yellow the points which are either excluded by the \PAMELA\ or \Fermi-LAT experiments respectively. Clearly one can see that the combination of both the \PAMELA\ and \Fermi-LAT astrophysical constraints surpass the latest exclusion limit set by the XENON100 experiment. In fact in general the astrophysical constraints discussed in this chapter even have a stronger exclusion power than the forecasted XENON1T limit, illustrating how important adding astrophysical knowledge is in this specific scenario. 

Even though many configurations are excluded by the \PAMELA\ and \Fermi-LAT data, we do find scenarios which are neither excluded by the XENON100 2012 limit nor by the astrophysical constraints discussed in this chapter. Hence the XENON100 experiment could still discover evidence for relatively light pMSSM neutralinos ($m_{\chi^0_1} < 500$ GeV) if these particles indeed exist. We note nevertheless that in \cite{Davis:2012hn}, a constraint as strong as the XENON100 2012 limit was obtained by using the XENON100 2011 data and a Bayesian analysis where the full information available in the $(S_1,S_2)$ scintillation plane was exploited. It is therefore likely that the XENON100 experiment can improve its present exclusion limit with the 2012 data and rule out some of the configurations shown here in green. Note that the scan gives mainly LSP mass above 100 GeV because of the requirement imposed ${\sigma v}_{{\chi^{0}_{1}\chi^{0}_{1}\ra W^{+}W^{-}}} > 10^{-27}$~${\rm cm}^3/{\rm s}$. Moreover the three-body processes involving an off-shell $W$ boson in the DM annihilation are not taken into account in the \micro version used in this work\footnote{Note that the latest \micro version \cite{Belanger:2013oya} allows to include three- and four-body final states involving one or two off-shell $W$ or $Z$ gauge bosons in the DM observable calculations.}. Then the diagram~\ref{pMSSM_diagrams}b does not contribute here to the neutralino pair annihilation. 

\begin{figure}[t]
\includegraphics[width=\textwidth]{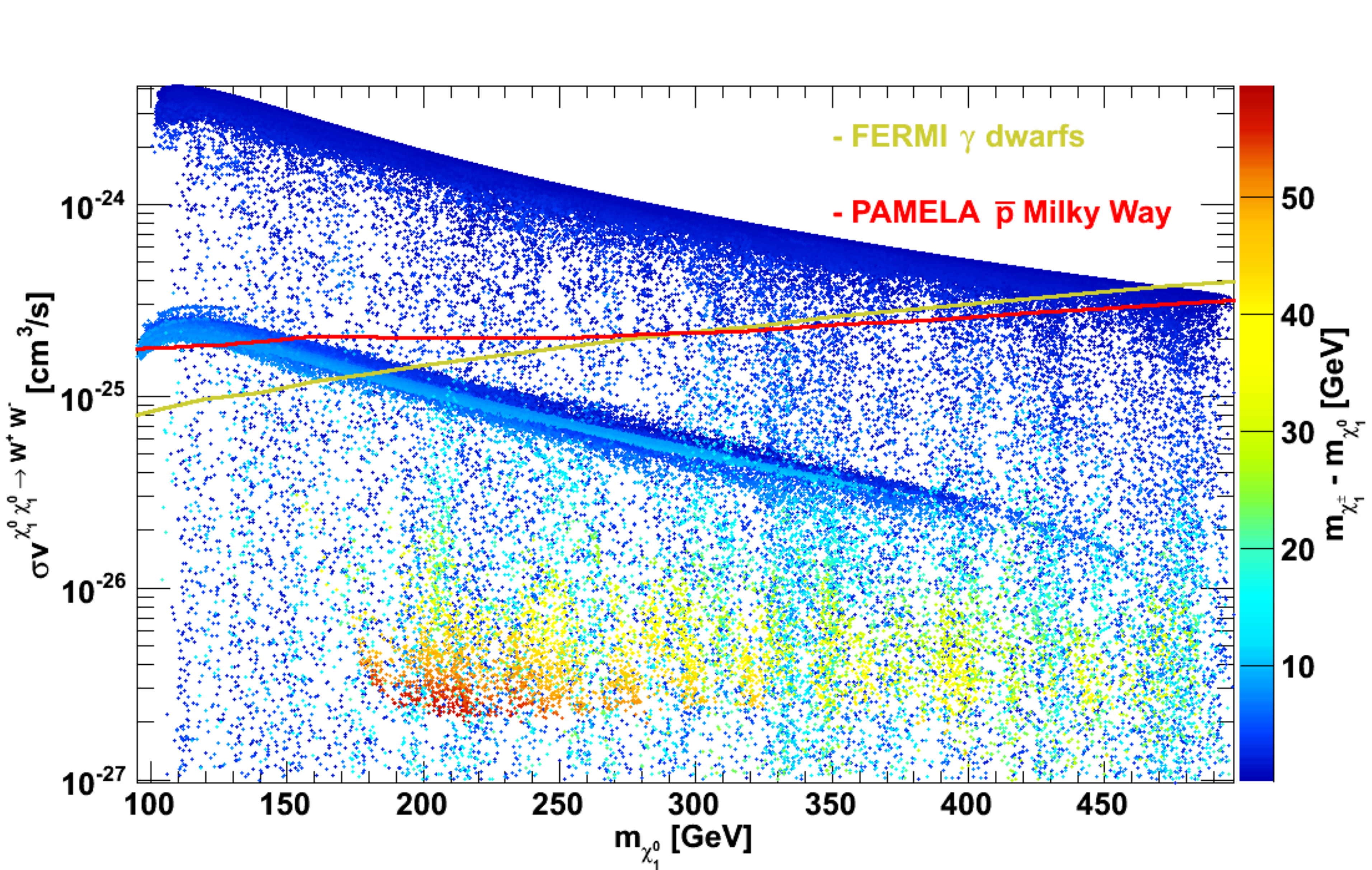}
\caption{\label{WW}Annihilation cross section into $W^+ W^-$ as a function of the neutralino DM mass. The chargino NLSP-neutralino LSP mass splitting is shown as colour code.}
\end{figure}

Finally, in figure~\ref{WW}, we show the annihilation cross section into $W^+ W^-$ as a function of the neutralino mass and superimpose the \PAMELA\ (for the `MED' set of propagation parameters and \textit{marginalized background},\ie the conservative limits) and \Fermi-LAT limits (red and yellow lines respectively). The colour code indicates the different values of the neutralino-chargino mass splitting. As can be seen from this plot, the \PAMELA\ and \Fermi-LAT constraints are actually complementary. The \Fermi-LAT limit excludes more configurations below 300 GeV than the \PAMELA\ bound but it assumes that the observations are independent of the DM energy distribution in dSph galaxies, which can be debated~\cite{Hooper:2003sh,Charbonnier:2011ft}. In contrast, the anti-proton limit excludes a bit more configurations than the gamma-rays above 300 GeV. This is reassuring since it is set by observations \textit{within} the galaxy but the drawback is that it relies on a specific choice of propagation parameters and knowledge of astrophysical sources. In any case, the fact that both limits exclude similar configurations enables us to validate the exclusion region that we found. 

Hence the main information that one can read from this plot, combined with that displayed in figures~\ref{diff-vcsWW-compoLSP} and \ref{diff-vcsg}, is that :
\begin{itemize} 
\item one can rule out neutralino-chargino mass splitting up to $\sim$ 13 GeV if $m_{\chi_1^0} \lesssim 150$ GeV and the neutralino is at 80\% a mixture of wino and higgsino;
\item one can exclude all scenarios in which the 80\% wino LSP-chargino mass difference is smaller than 0.25 GeV for $m_{\chi^0_1} < 500$ GeV, thanks to both \PAMELA\ and \Fermi-LAT data. 
\end{itemize}

\subsection{Final state radiation in the pMSSM} 
\label{subsec:fsr}

\begin{figure}[!htb]
\begin{center}
\centering
\subfloat[]{\includegraphics[width=7.5cm,height=7cm]{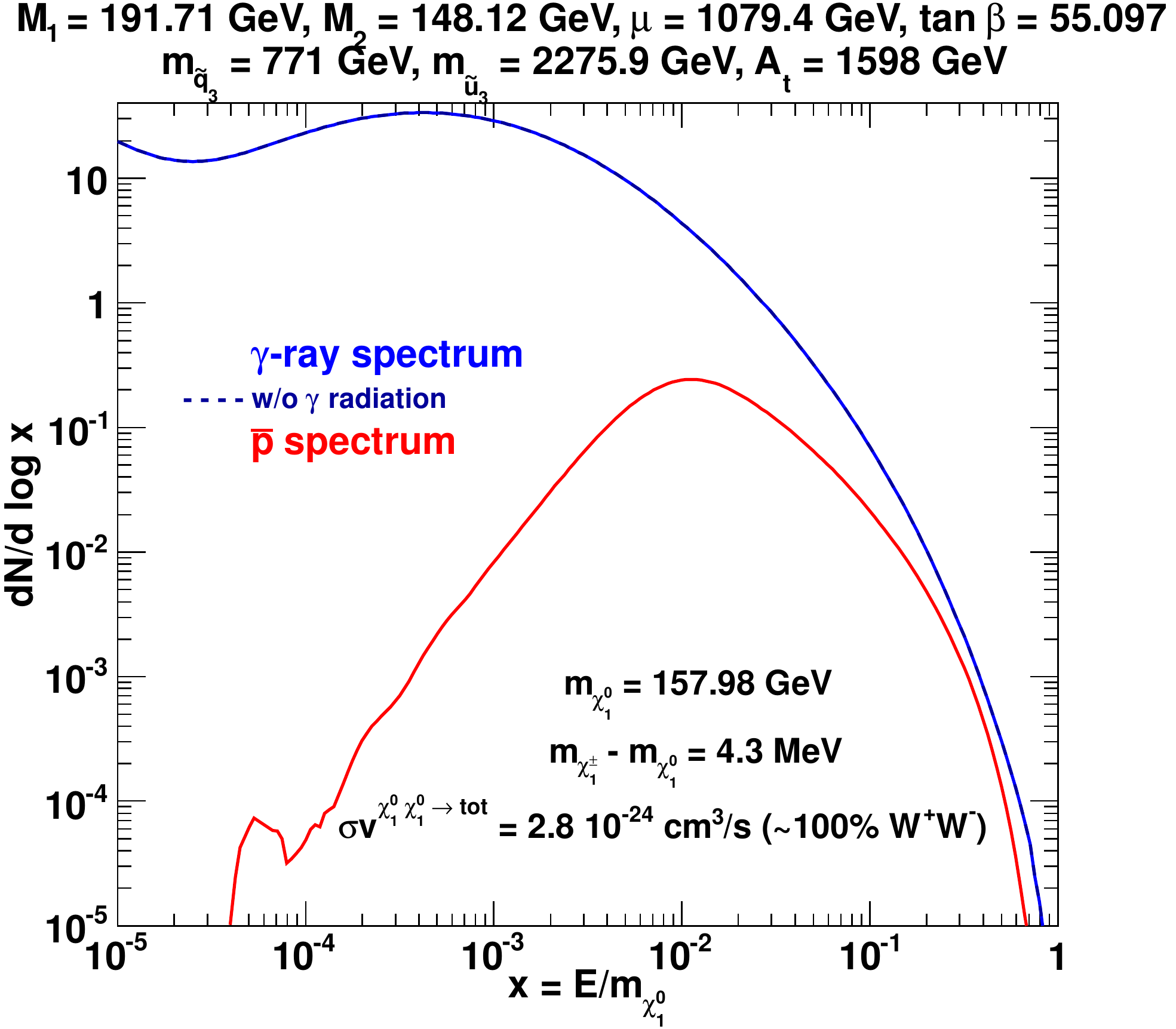}} \quad \,
\subfloat[]{\includegraphics[width=7.5cm,height=7cm]{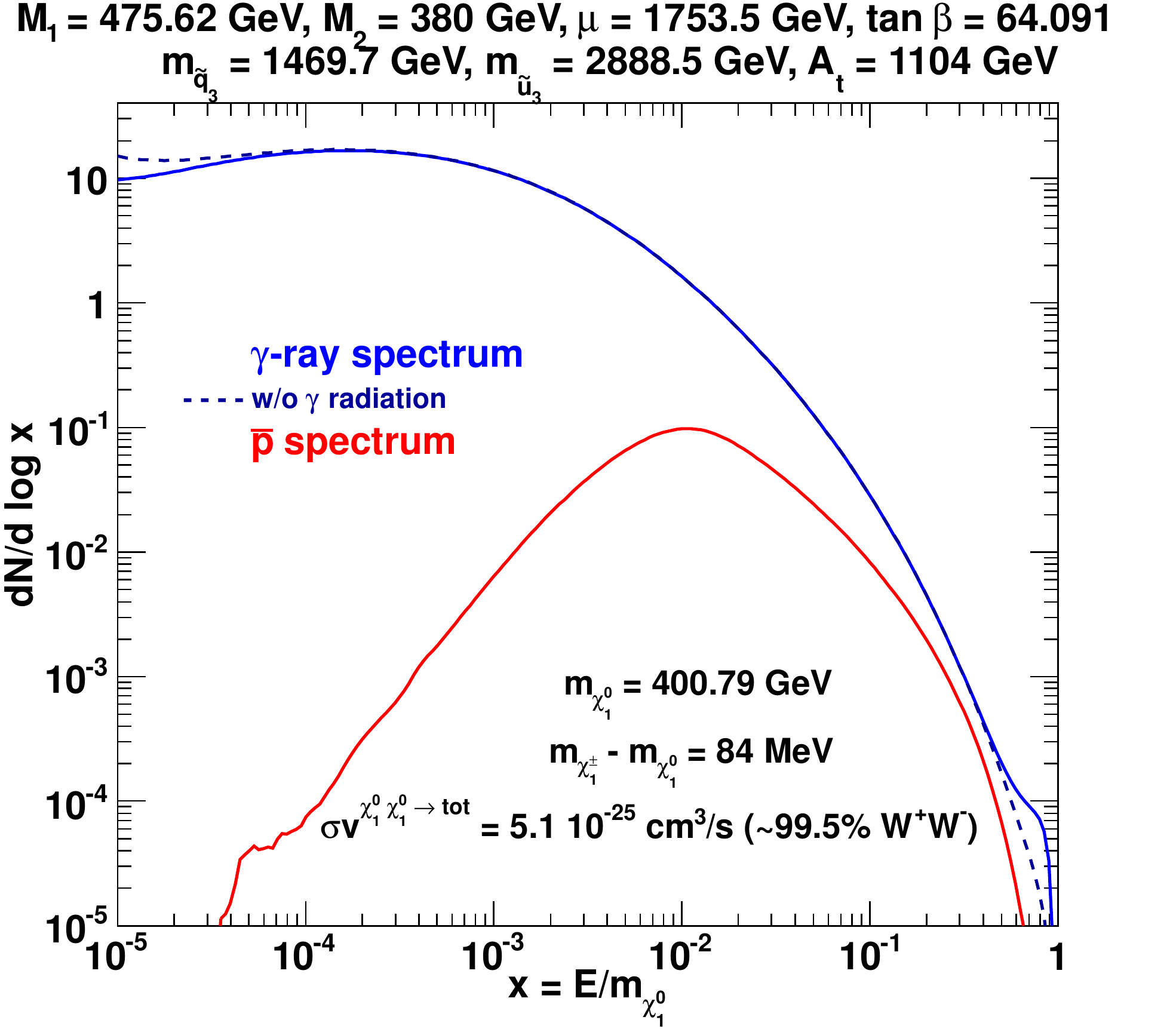}}
\caption[Photon and anti-proton distributions as a function of the energy fraction $x=E/m_{\chi^0_1}$ for two pMSSM cases.]{\label{fig:5.spec}Photon and anti-proton distributions as a function of the energy fraction $x=E/m_{\chi^0_1}$. The blue plain (dark blue dashed) lines show the $\gamma$-ray spectrum with (without) photon radiation while the red lines show the $\bar{p}$ spectrum. Panel (a) presents a pMSSM case with a light DM candidate ($m_{\chi^0_1} \approx 158$~GeV) which is excluded by ID : ${\sigma v}_{^{\chi^{0}_{1}\chi^{0}_{1}\ra W^{+}W^{-}}} \gg {\sigma v}_\mathrm{\Fermi}$ for this $m_{\chi^0_1}$. Panel (b) shows the spectrum obtained with a large DM mass ($m_{\chi^0_1} \approx 400$~GeV) which is just above the ID limits considered here.}
\end{center}
\end{figure} 

We also analysed the impact of final state radiations on $\gamma$-ray constraints. Indeed, in principle, the regime of quasi-degeneracy is the one where it can be most important~\cite{Bringmann:2007nk} and, since we decided to only use the constraints from \cite{Ackermann:2011wa}, it is important to check that it is a good choice. We explicitly computed using the \micro code the $\gamma$-ray spectrum including photon radiation for a number of cases from the MCMC scans\footnote{Note obviously that we always considered photon radiation for the DM annihilation cross section calculations in the scans.} and we show two examples in figure~\ref{fig:5.spec}. The blue plain (dark blue dashed) lines show the $\gamma$-ray spectrum with (without) photon radiation while as an example we plot the $\bar{p}$ spectrum in red. For small DM masses as shown in panel (a) the modification to the $\gamma$-ray spectrum is generally too small to have an effect : the two blue curves are merged. On the other hand, for large DM masses as shown in panel (b), the inclusion of final state radiation starts being noticeable. However, it creates a peak at an energy fraction\ca 1 which falls beyond the range of energies on which \Fermi\ dwarf bounds are based.

\subsection{130 GeV line} 

\begin{figure}[!htb]
\begin{center}
\centering
\includegraphics[width=9cm,height=6cm]{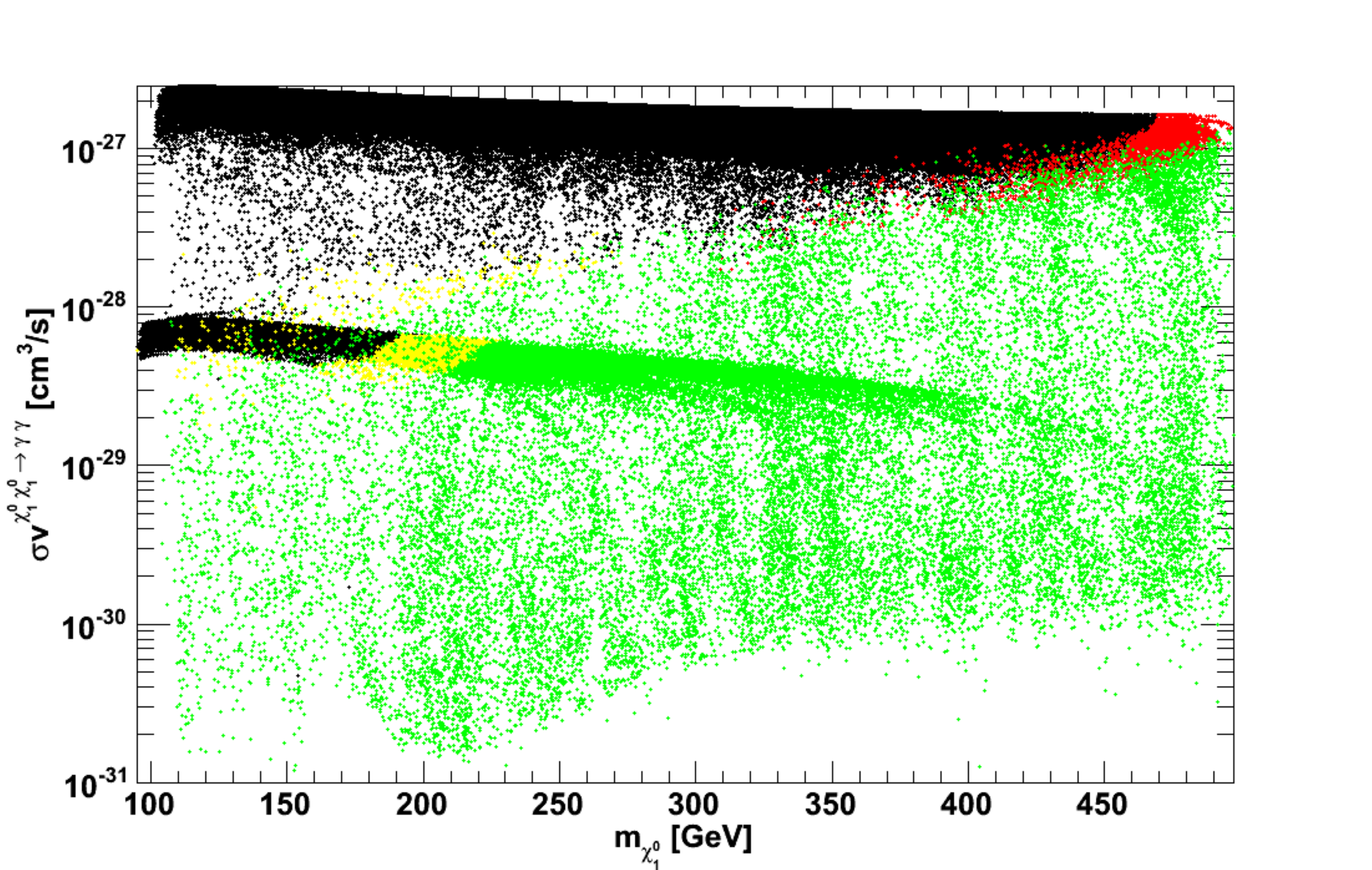}
\caption[The neutralino pair annihilation cross section into $\gamma \gamma$ as a function of the LSP mass.]{\label{130GeV}The neutralino pair annihilation cross section into $\gamma \gamma$ as a function of the LSP mass. The colour coding shows the points in the pMSSM parameter space which are not excluded by ID constraints in green and those which are excluded by \Fermi-LAT (yellow), \PAMELA\ (red) or both (black).}
\end{center}
\end{figure}

As a side comment regarding the so-called \textit{130 GeV line} : we do find scenarios where $\sigma v_{\chi^0_1 \chi^0_1 \ra \gamma \gamma} \simeq  10^{-27} \ \rm{cm^3/s}$, which is the value of the cross section that is required to explain the feature in the spectrum. These configurations predict a neutralino-chargino mass splitting greater than $\sim 0.2$ GeV. However none of the points corresponding to neutralinos with a mass of about 130 GeV are allowed by the \PAMELA\ and the \Fermi-LAT data as displayed in figure~\ref{130GeV}. Hence, our results suggest that one cannot explain the \textit{130 GeV line} in our simplified version of the pMSSM, which is in agreement with~\cite{Cohen:2012me,Buchmuller:2012rc}. Indeed, due to the anti-proton limit, scenarios with $\sigma v_{\chi^0_1 \chi^0_1  \ra \gamma \gamma} \simeq  10^{-27} \ \rm{cm^3/s}$ rather correspond to neutralinos with a mass of about 450 GeV. In fact, for the same reason, all allowed points with $\sigma v_{\chi^0_1 \chi^0_1  \ra \gamma \gamma}>2 \times 10^{-28} \ \rm{cm^3/s}$ correspond to configurations where $m_{\chi^0_1} >$ 250 GeV. Finally note that in the pMSSM the existence of 130 GeV neutralinos should give rise to a second $\gamma$-ray line at $\sim$ 111  GeV (on top of that at 130 GeV),  corresponding to the neutralino pair annihilation into $\gamma Z$. Given our prediction for $\gamma Z$ and $\gamma \gamma$, the flux associated with this 111 GeV line should be about ten times larger than that corresponding to the 130 GeV line, which is in conflict with the observations.

\subsection{The case of no DM regeneration}

\begin{figure}[!htb]
\begin{center}
\centering
\includegraphics[width=9cm,height=6cm]{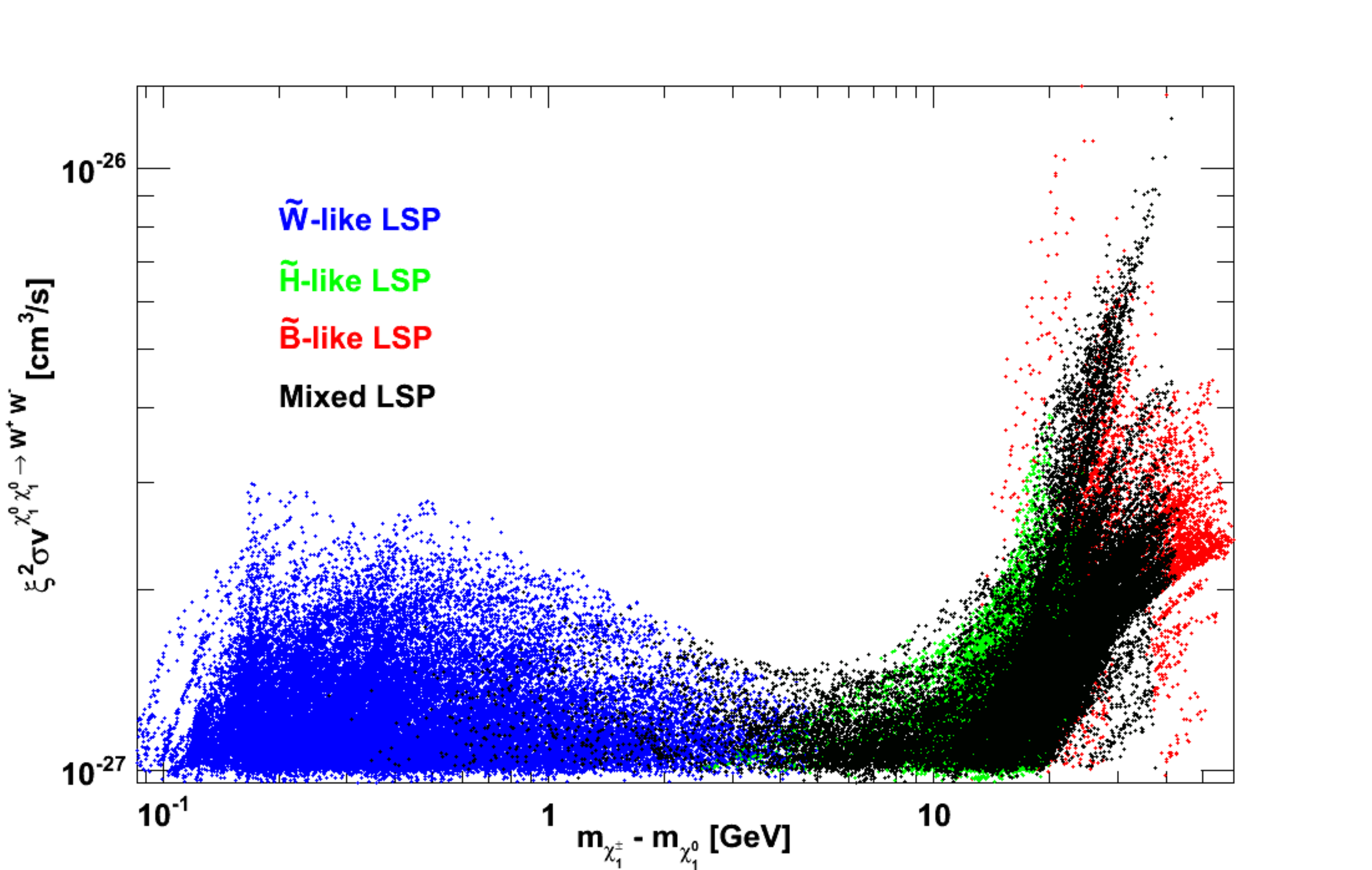}
\caption{\label{resc}Plot of the rescaled neutralino pair annihilation cross section into $W^+ W^-$ as a function of the chargino-neutralino mass splitting in the scan where no DM regeneration is assumed. Same colour code as in figure~\ref{diff-vcsWW-compoLSP}.}
\end{center}
\end{figure}

In this study we have assumed that the relic density was regenerated at 100 $\%$ for candidates with a total annihilation cross section much larger than the \textit{thermal} one (\ie with a suppressed freeze-out relic density). This way we could ensure a fair comparison between theoretical expectations and the limits set by the \Fermi-LAT and XENON100 experiments. Looking at figure~\ref{diff-vcsg}b one sees that invoking regeneration is needed to obtain a DM relic density in agreement with the WMAP values for all scenarios with a chargino-neutralino mass splitting smaller than\footnote{For larger values of the mass splitting, no regeneration assumption is required but the annihilation cross sections into $\gamma \gamma$ and $\gamma Z$ are strongly suppressed. In particular $\s v_{\chi^0_1 \chi^0_1 \ra \gamma \gamma}$ is much below $10^{-29} \ \rm{cm^3/s}$.} $\sim$ 20 GeV. Assuming that all these candidates have the correct relic density, we could indeed exclude scenarios with a neutralino-chargino mass splitting up to $\sim$ 10 GeV and values of $\s v_{\chi^0_1 \chi^0_1 \ra \gamma Z}$ down to $10^{-28} \ \rm{cm^3/s}$ (see figure~\ref{diff-vcsg}a), corresponding to $\s v_{\chi^0_1 \chi^0_1 \ra W^+ W^-} > 10^{-25} \ \rm{cm^3/s}$ and $\Om h^2 \ll 0.06$. However, relaxing the regeneration assumption would completely relax the exclusion regions and therefore the bound on the mass splitting. It is what is presented in figure~\ref{resc}. This plot comes from a scan where the neutralino LSP is allowed to constitute only a fraction of the total DM density. The rescaling parameter is defined as $\xi = \Om_{\chi^0_1}/\Om_{\mathrm{WMAP},1\s}$, where we take the lower 1$\s$ bound from WMAP7, namely $\Om_{\mathrm{WMAP},1\s} h^2 = 0.1088$. As shown the rescaled neutralino pair annihilation cross section into $W^+ W^-$ is far below the \Fermi-LAT and \PAMELA\ limits, between $10^{-27} \ \rm{cm^3/s}$ and $10^{-26} \ \rm{cm^3/s}$ : all points are safe from current ID constraints. However in this scenario the DM problem is not solved.

\section{Conclusions}

In this chapter we examined the constraints on the ${\rm DM} {\rm DM} \ra W^+ W^-$ annihilation cross section by using the \PAMELA\ anti-proton data and we analysed the choice of propagation parameters and uncertainties on the astrophysical background. These results are independent of the so-called \PAMELA\ positron excess and are obtained for two different (fixed vs marginalized) choices of the background spectrum.

We then compared these bounds with the most stringent gamma-ray limits which have been derived using the \Fermi-LAT\ measurements of the gamma-ray continuum spectrum from dSph galaxies, for the same DM annihilation channel and DM mass range. We found that the anti-proton constraints appear to be very competitive with the gamma-ray bounds. More precisely, choosing the `MED' propagation scheme, the $\bar p$ constraints are slightly weaker than the $\gamma$-ray ones when $m_{\rm DM} \lesssim  300$ GeV and slightly stronger for $m_{\rm DM} \gtrsim 300$ GeV.  On the other hand, the anti-proton constraints are stronger if we assume the 'MAX' set of propagation parameters and less powerful if we assume the `MIN' set. We also recall that the gamma ray limits themselves may be subject to some uncertainties related to the modelling of the DM profile in dSph  galaxies. 

Finally we applied as fiducial limits the \Fermi\ dwarf bounds and the $\bar p$ constraints relative to `MED' and the marginalized astrophysical background to the neutralino LSP in a simplified version of the pMSSM, where we set all the sfermion masses (apart from that of the third generation) to the TeV scale. We found that the fiducial \PAMELA\ anti-proton and \Fermi-LAT gamma-ray limits rule out small but non negligible neutralino-chargino mass splittings. In particular for $m_{\chi_1^0} \lesssim 150$ GeV, one can rule out mass splittings up to 13 GeV. Our results also suggest that pure wino or wino-like neutralinos are excluded if they are lighter than 450 GeV. Overall, this limit surpasses the bounds that can be set by using the XENON100 data and even in fact the projected XENON1T limit. 

Note that the main difference of this MCMC study comparing to that of the paper \cite{Belanger:2012ta} is that no significant $\chi^0_1 \chi^0_1 \ra b \bar{b}$ annihilation cross section is found. The reason is that we decided here to put the CP-odd Higgs boson mass at 2 TeV which then exclude the possibility to have an enhancement of the $b \bar{b}$ final state due to the $s$-channel pseudo-scalar Higgs boson exchange. This has also the consequence to avoid some scenarios with a bino-wino LSP and a mass splitting around 20 GeV which could be probe by ID, hence the lower bound of 13 GeV found here.
 
Hence from this work, we conclude that present ID data already enable one to exclude regions of the parameter space where the neutralino-chargino mass splitting is small but non-negligible. Since these regions are difficult to probe directly at the LHC, these findings show that \Fermi-LAT and \PAMELA\ data constitute modern tools to explore the supersymmetric parameter space and even beat LHC (and also in fact DD) searches on their own territory even though, on the negative side, they assume a regeneration of the relic density for neutralinos with a very large annihilation cross section.

\chapter{Direct SUSY searches at LHC and a singlet extension of the MSSM}
\label{chapter:NMSSM}

\minitoc\vspace{1cm}
\newpage

\section{Going beyond the minimal supersymmetric scenario}

As shown in the previous chapters the MSSM can solve several problems encountered in particle physics and cosmology. Nevertheless this minimal version of SUSY has some drawbacks.

\subsection{The \textit{$\mu$-problem}}

As explained in chapter~\ref{chapter:SUSY}, two Higgs $SU(2)_L$-doublets have to be defined in a supersymmetric extension of the SM. As we saw in eq.~\ref{eq.3:W_MSSM} the superpotential of the MSSM contains a mass term for $H_u$ and $H_d$, the $\mu$ term. Since this parameter is SUSY preserving, \textit{natural} values are $\mu=0$ or very large,\eg $\mu \sim M_{Pl}$. With the EWSB, the minimization condition of the MSSM potential allows to determine $M_Z^2$ and $\tan \beta$ from the MSSM Lagrangian parameters $|\mu|^2, b, m^2_{H_u}$ and $m^2_{H_d}$ with the relations :
\beq \begin{split}
\sin 2\beta & = \frac{2b}{m^2_{H_u} + m^2_{H_d} + 2|\mu|^2},\\
M_Z^2 & = \frac{|m^2_{H_d} - m^2_{H_u}|}{\sqrt{1-\sin^2 2\beta}} - m^2_{H_u} - m^2_{H_d} -  2|\mu|^2.
\end{split} \eeq
It follows that if we do not want to consider a fine-tuned cancellation to obtain the expected $Z$ boson mass, the supersymmetric mass term squared ought to be at most within (10 - 100)$\times M_Z^2$. This is the case for the SUSY breaking parameters $b, m^2_{H_u}$ and $m^2_{H_d}$, but as said above it was not expected for the $\mu$ term. Moreover $\mu=0$ is not compatible with the limits on the chargino mass and the mass matrix in eq.~\ref{eq.3:chimass} which imply that $\mu \gtrsim 100$~GeV. The fact that the MSSM does not account for a $\mu$ term around the EW scale is called the \textit{$\mu$-problem} \cite{Kim:1983dt}.

\subsection{MSSM limitations}

Besides this theoretical problem, some experimental results can be difficult to explain within the MSSM.

As we saw at the end of section~\ref{subsec:3.higgs} and in chapter~\ref{chapter:NUHM2}, the simplest scenarios of the MSSM have difficulties to generate the expected mass for the SM-like Higgs boson because its tree-level value is too low. However it is still possible in the MSSM to get $m_{h^0} \sim 125$~GeV, but it requires appreciable fine-tuning \cite{Hall:2011aa}. 

Non-standard signal strength in some Higgs decay channels are difficult to explain in the MSSM. As an example, the ATLAS colaboration reported in early 2012 an excess at 2.8$\s$ in the $h^0 \ra \g\g$ decay channel, consistent with a Higgs boson mass $m_{h^0}=126$~GeV \cite{ATLAS:2012ad}. This signal was larger than expected from a SM Higgs boson (with a signal strength\footnote{However as noted in table~\ref{tab:hgaga} this excess is now less pronounced.} $\s/\s_{\mathrm{SM}}=2\pm 0.8$). An excess in the $\g\g$ channel was also reported by the CMS experiment \cite{Chatrchyan:2012twa} but at a slightly lower mass $m_{h^0}=124$~GeV. Assuming that the Higgs boson is produced by gluon fusion, we can define the signal strength by the ratio 
\beq \label{eq:6.R} R_{ggXX} =  \frac{\sigma(gg\rightarrow h^0)_{\mathrm{BSM}} \mathscr{B}(h^0\rightarrow X X)_{\mathrm{BSM}}}{\sigma(gg\rightarrow h^0)_{\mathrm{SM}} \mathscr{B}(h^0\rightarrow X X)_{\mathrm{SM}}},\eeq
where here we are interested on $X=\g$. It is generally expected that, after applying other constraints on the MSSM, $R_{gg \g\g} $ is at most as large as unity~\cite{Arbey:2011aa,Brooijmans:2012yi,AlbornozVasquez:2011aa} except in a corner of the parameter space where heavily mixed staus can lead to an increase in the $h^0 \ra \g\g$ partial decay width~\cite{Carena:2011aa}.

As shown in chapters \ref{chapter:NUHM2} and \ref{chapter:ID} the lightest neutralino can be a viable DM candidate in the MSSM. However if the signals in DD experiments presented in section~\ref{subsubsec:3.DD}, indicative of a light DM\ie below\ca 15 GeV, are confirmed it will be difficult to explain them within the MSSM and we would need to go beyond the minimal model. Indeed such a neutralino is strongly constrained because it leads to too much DM.

\subsection{The Next-to-MSSM}

To solve the $\mu$-problem, a simple method consists in generating an effective $\mu$ term in a similar way that the generation of quark and lepton masses. A Yukawa coupling $\l$ of $H_u$ and $H_d$ to a new scalar field $S$ is added, and the VEV of $S$, 
\beq \langle S \rangle = \frac{v_s}{\sqrt{2}},\eeq
which is induced by new soft SUSY breaking terms, is of the order of the EW scale. Given that the $\mu$ term of the MSSM carries no SM gauge symmetry quantum numbers, the new scalar field has to be a singlet of $SU(3)_c  \otimes  SU(2)_L  \otimes  U(1)_Y$. Since we are in SUSY, we define this new field in a new chiral supermultiplet $\mathbf{S}$; then the complex scalar field $S$ is complemented by a fermionic partner, the singlino $\widetilde{S}$. The MSSM superpotential is thus modified as
\beq \mathcal{W}_{\mathrm{MSSM}} \ra \mathcal{W}_{\mathrm{MSSM}}|_{\mu = 0} + \lambda S H_uH_d,\eeq
and the effective $\mu$ term reads
\beq \mu_\mathrm{eff} = \l \frac{v_s}{\sqrt{2}}.\eeq

However the introduction of the singlet supermultiplet has the drawback to add an additional $U(1)$ (PQ) global symmetry which gives, as we saw in section~\ref{sec.3:sol}, an axion. It results that experimental limits on axion searches allow rather small values of $\l$, namely $|\l|<10^{-7}$. In order to get an effective $\mu$ term around the EW scale we then need to fine-tune $\langle S \rangle$. To solve this we introduce an additional parameter $\kappa$ which breaks the PQ symmetry through a cubic self-coupling term $\frac{1}{3}\kappa S^3$ : it is in this framework that the NMSSM is defined (see\eg \cite{Ellwanger:2009dp,Maniatis:2009re} for reviews). In the literature, the NMSSM model which is mostly considered is the one which has the minimal number of new parameters to solve the $\mu$-problem. Its superpotential is defined as
\beq \label{W_NMSSM} \mathcal{W}_{\mathrm{NMSSM}} = \mathcal{W}_{\mathrm{MSSM}}|_{\mu = 0} + \lambda S H_uH_d + \frac{1}{3}\kappa S^3.\eeq
The SSB Lagrangian of the NMSSM is
\beq \mathscr{L}_{\mathrm{NMSSM}}^{\mathrm{soft}} = \mathscr{L}_{\mathrm{MSSM}}^{\mathrm{soft}}|_{b = 0} - m^2_S |S|^2 - (\l A_\l S H_uH_d + \frac{1}{3}\kappa A_\kappa S^3 + \textrm{h.c.}),\eeq
with two trilinear couplings added, $A_\l$ and $A_\kappa$, and the singlet mass term $m_S$.

The addition of a complex scalar field leads to a different Higgs sector with respect to the MSSM. The NMSSM contains seven physical Higgs scalars : two charged Higgs bosons $H^\pm$ as in the MSSM but three CP-even Higgs bosons $h_i, i \in \{1,2,3\}$ and two CP-odd Higgs bosons $a_i, i \in \{1,2\}$. Moreover the SM Higgs boson mass gets a pure NMSSM positive contribution which then improves the possibility to get a SM-like Higgs boson around 125 GeV. If $h_1$ is this one its tree-level mass squared upper bound reads
\beq m^2_{h_1} \lesssim M_Z^2 \cos^2 2\b + \frac{\l^2}{2}v^2 \sin^2 2\b.\eeq
The new contribution is largest for large $\l$ and small $\tan \b$. Note also that important doublet-singlet mixing can occurs in the Higgs sector. This leads to specific NMSSM features as we will see in sections \ref{sec.6:Rgggaga1} and \ref{sec.6:Rgggaga2}.

Another point to highlight in the NMSSM is that a new field enters in the neutralino sector, the singlino $\widetilde{S}$. Therefore the neutralino mass matrix defined in section~\ref{subsubsec.3:neut} is modified here. In the basis $\psi^0 = (\widetilde{B}, \widetilde{W}^3, \widetilde{H}^0_d, \widetilde{H}^0_u, \widetilde{S})$ the neutralino mass matrix is given by
\beq
  \mathbf{M_{\chi^0}}=\begin{pmatrix}
    M_1 & 0 & -M_Z  \cb \sw & M_Z \sb \sw & 0\\
    0 & M_2 & M_Z \cb \cw & -M_Z \sb \cw & 0\\
    -M_Z \cb \sw & M_Z \cb \cw & 0 & -\mu_\mathrm{eff} & -\l \frac{v_u}{\sqrt{2}}\\
    M_Z \sb \sw & -M_Z \sb \cw & -\mu_\mathrm{eff} & 0 & -\l \frac{v_d}{\sqrt{2}}\\
    0 & 0 & -\l \frac{v_u}{\sqrt{2}} & -\l \frac{v_d}{\sqrt{2}} & \frac{2\kappa \mu_\mathrm{eff}}{\l}
  \end{pmatrix}.
\eeq
Its diagonalisation by a 5$\times$5 unitary matrix $\mathbf{Z_n}$ gives the neutralino mass eigenstates :
\beq \chi^0_i = Z_{n ij} \psi^0_j\textrm{,} \qquad i\textrm{,}j \in \{1,2,3,4,5\}.\eeq
Because of its singlet nature the singlino can be very light, hence providing a potential light DM candidate. Nevertheless, scenarios with such a DM candidate are strongly constrained by Indirect Detection and Direct Detection of DM \cite{Vasquez:2010ru,AlbornozVasquez:2011js,AlbornozVasquez:2012px}. Keeping the scenarios that fulfill DM constraints, we could check whether pure NMSSM signatures with such a DM candidate can be probed at the LHC. This is what we will see in section~\ref{sec.6:squglu}.

The work that will be presented in the next sections constitutes a part of the article \cite{Vasquez:2012hn}. Section \ref{sec.6:squglu} contains additional material. 

\section{Previous scans on the NMSSM parameter space}

In order to make sure that the scenarios that we consider are all relevant, the analysis that we will show is based on \cite{Vasquez:2010ru,AlbornozVasquez:2011js} in which the NMSSM parameter space was explored in light of particle physics and astroparticle physics constraints. In these studies, based on MCMC analyses, the LSP neutralino was required to be relatively light (\ie with a mass below 15 GeV) motivated by hints of a signal in DD experiments~\cite{Aalseth:2010vx,Angloher:2011uu,Bernabei:2010mq}. It was also required that the LSP relic density does not exceed the WMAP5 observed value $\Om_{\rm WMAP5} h^2=0.1131\pm 0.0034$~\cite{Komatsu:2008hk}, but it can be much lower than this, hence calling for another type of particles to solve the DM problem. Limits from $B$-physics, $\amu$, as well as LEP and Tevatron limits on the Higgs boson and SUSY particles were also taken into account. LHC limits on the Higgs sector computed with NMSSMTools~\cite{Ellwanger:2005dv} were also included. Additional constraints such as DD limits from XENON100~\cite{Aprile:2011hi}, $\g$-rays from dSph galaxies probed by Fermi-LAT~\cite{Strigari:2006rd} and the radio emission in the MW and in galaxy clusters \cite{Boehm:2002yz,Boehm:2010kg} were superimposed on the parameter space selected by the MCMC.

\begin{figure}[!htb]
\begin{center}
\centering
\subfloat[]{\includegraphics[width=8cm,height=5cm]{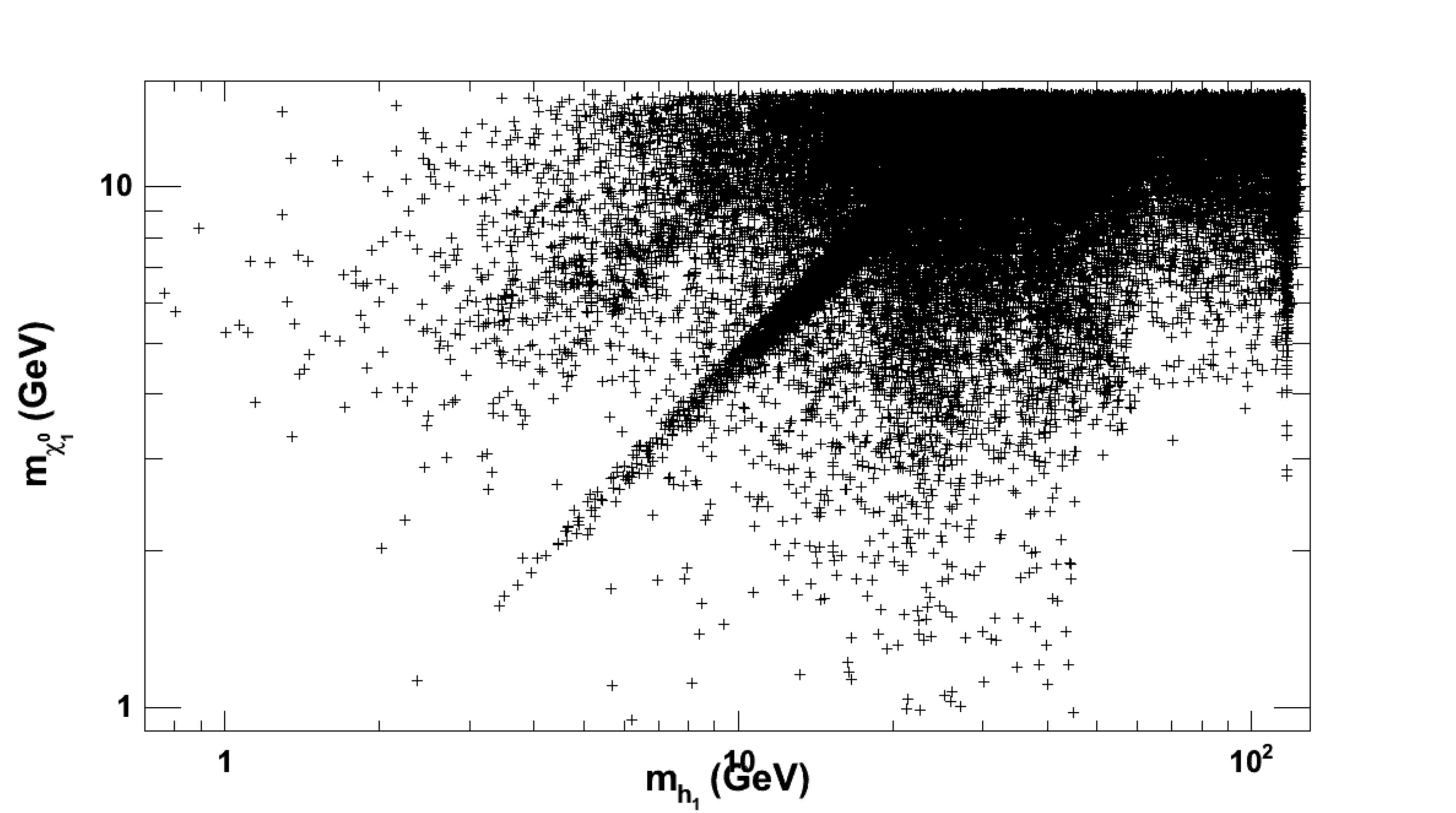}}
\subfloat[]{\includegraphics[width=8cm,height=5cm]{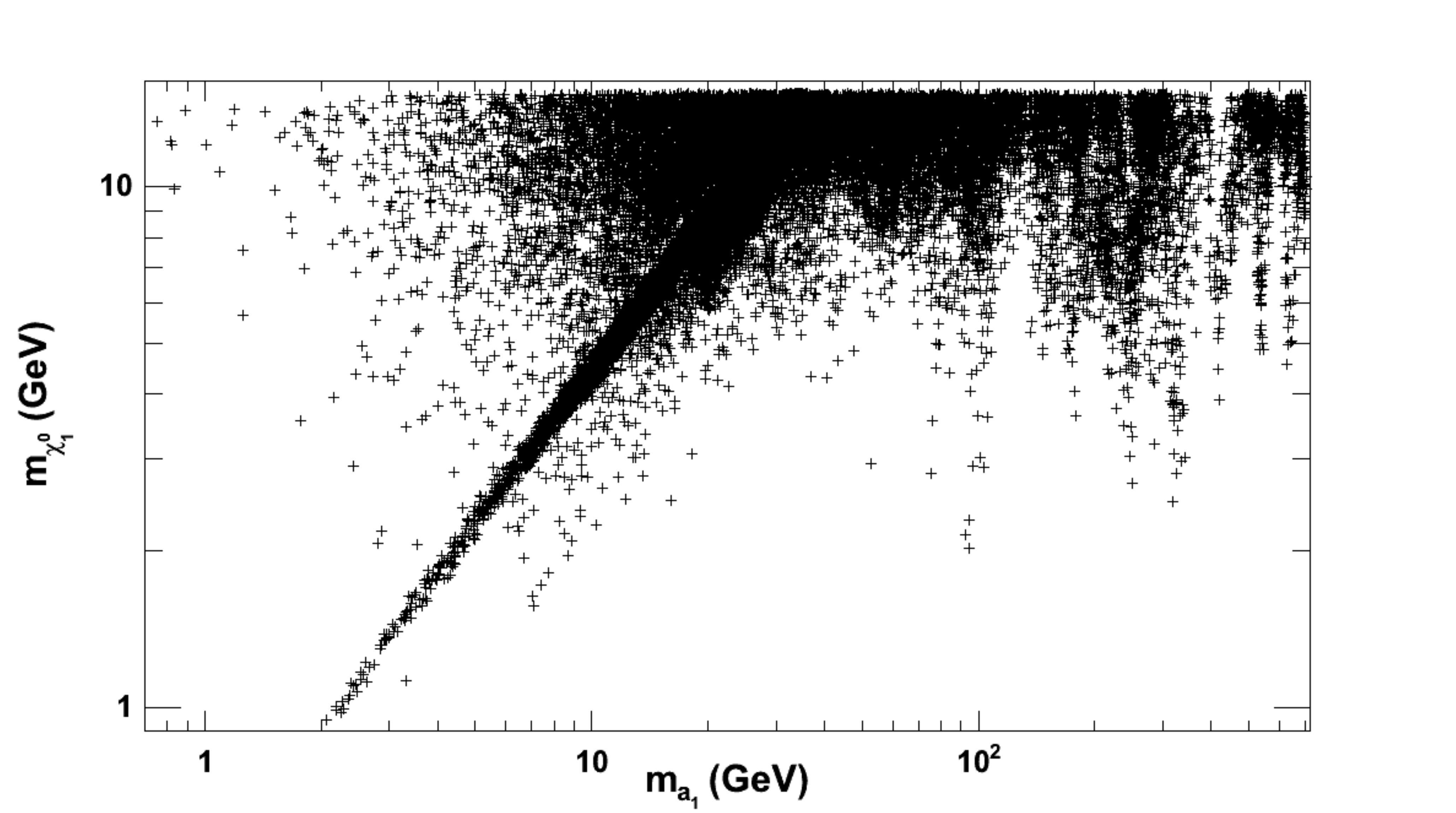}}
\caption{\label{neutmh1ma1}LSP mass\vs the mass of the lightest CP-even (CP-odd) Higgs boson, in panel (a) (panel (b)) for the light LSP scenario.}
\end{center}
\end{figure}

Since neutralinos lighter than 15 GeV can either be binos or singlinos, the easiest way to ensure significant annihilations  is through resonant exchange of a light scalar or pseudoscalar Higgs boson. Hence scenarios with light singlet-like $a_1$ or $h_1$, namely $m_{h_1,a_1} < 30$ GeV, are preferred as shown in figure~\ref{neutmh1ma1}. However, when the neutralino mass is large enough, annihilation mechanisms through $Z$-exchange or light sleptons can also be efficient and a very light Higgs boson singlet is no longer important.

\section{Squarks and gluinos searches at the LHC}
\label{sec.6:squglu}

To make sure that SUSY searches at the LHC does not exclude these interesting configurations, we decided to look at the LHC limits on sparticles. We took into account the exclusion limit coming from the ATLAS $1.04 \ \rm{fb}^{-1}$ search for squarks and gluinos via jets and missing transverse momentum \cite{Aad:2011ib}. For each relevant SUSY point, namely in the range which is excluded by this search in simplified SUSY models with first and second generation squarks lighter than $\sim 0.6$ TeV when the gluinos are much more heavy and gluinos lighter than $\sim 0.5$~TeV for heavy squarks, signal events were generated\footnote{Note that the matrix elements used by all event generators for the hard production of two SUSY particles are accurate only to leading order in perturbative QCD. It is therefore desirable to supplement the resulting signal cross-section with an NLO K-factor for the production process, obtained in a separate calculation. In the MSSM, Prospino~\cite{Beenakker:1996ed} is commonly used; unfortunately for the NMSSM there is no automated calculation of NLO cross-sections publicly available. Exclusion calculated without the K-factor ($\mathcal{O}$(1-3) in the MSSM) is therefore slightly conservative.} using the Monte Carlo event generator \textsf{Herwig++~2.5.1}~\cite{Bahr:2008pv,Gieseke:2011na}. Experimental cuts of each search channel were then applied using \textsf{RIVET 1.5.2}~\cite{Buckley:2010ar}. Then, the ATLAS jets and missing $E_T$ searches at  $1.04 \ \rm{fb}^{-1}$ which are included~\cite{Grellscheid:2011ij} in the \textsf{RIVET} package are used to set constraints on the NMSSM scenarios considered. Of course current limits are much more stronger with coloured sparticles possibly excluded up to $1.5$~TeV \cite{ATLAS:2012ona} and could probe a larger part of our interesting points.

\subsection{Relevant NMSSM region in light of ATLAS jets $+$ $\cancel{E_T}$ searches}

Analysing the mass of the squarks and gluinos of the interesting NMSSM configurations, we found one region of the parameter space where the ATLAS limits on coloured sparticles could be relevant : it is characterized by squarks in the range $[0.3-1.1]$~TeV and gluino masses around $5.3-5.5$~TeV.

Using the tools described above we can extract for each NMSSM point the final production cross section in the different signal regions of the jets + $E_T$ searches in \cite{Aad:2011ib}, namely the \textit{$\geq$ two-jets}, \textit{$\geq$ three-jets}, \textit{$\geq$ four-jets} with $m_\mathrm{eff}\footnote{As defined in \cite{Aad:2011ib}, $m_\mathrm{eff}$ is the sum of the missing transverse energy and the magnitudes of the transverse momenta of the two, three or four highest transverse momentum jets used to define the signal region.} >$ 0.5 TeV, \textit{$\geq$ four-jets} with $m_\mathrm{eff} >$ 1 TeV and the \textit{high mass} (stringent requirements on the transverse momentum and $m_\mathrm{eff}$) channels. We then compare the results to the excluded values at 2$\s$ in each signal region given in \cite{Aad:2011ib}.

\subsection{Light squark masses}

\begin{figure}[!htb]
\begin{center}
\centering
\subfloat[]{\includegraphics[width=8cm,height=5cm]{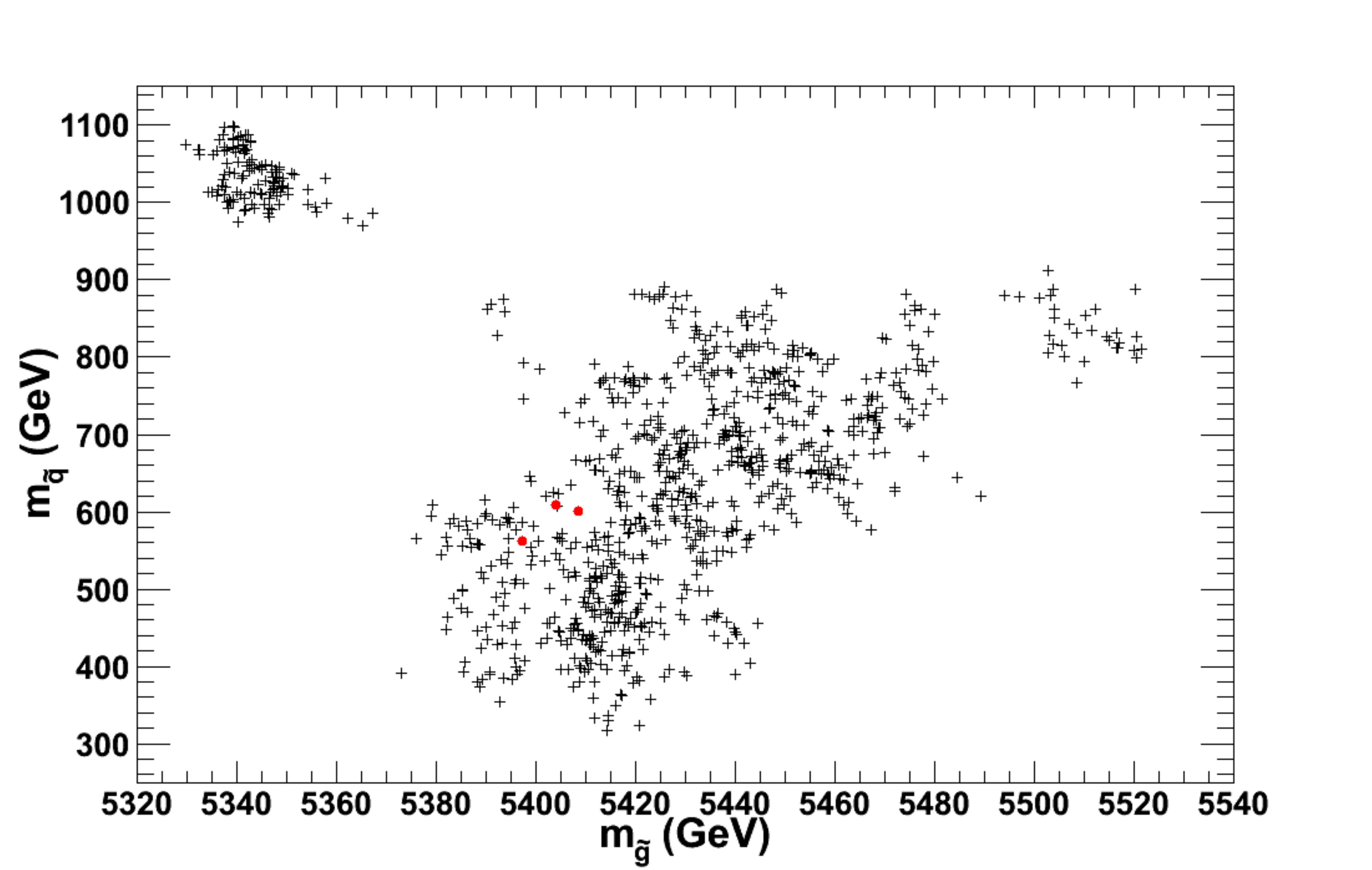}}
\subfloat[]{\includegraphics[width=8cm,height=5cm]{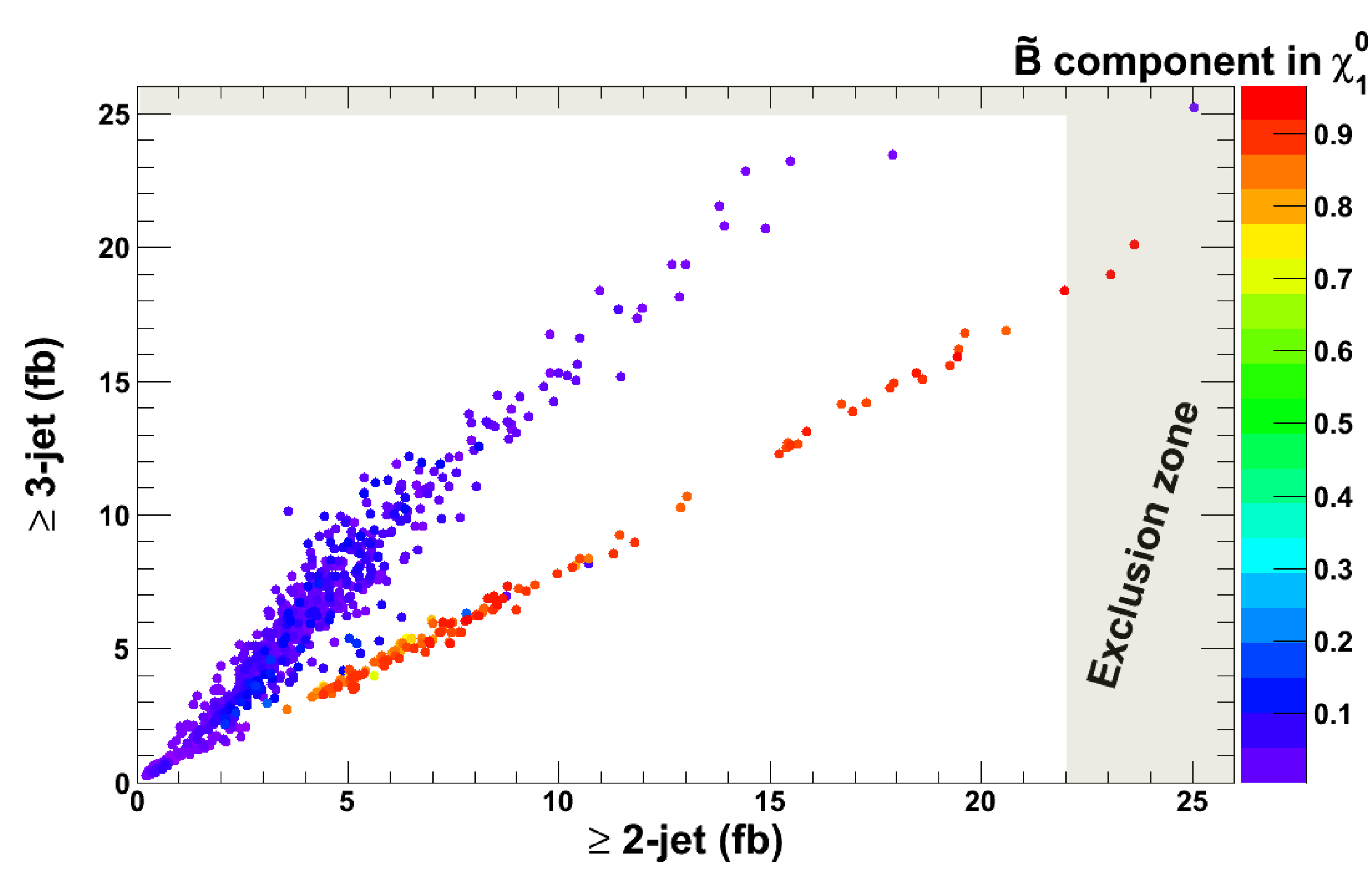}}    
\caption[Representation of the points in the plane ($m_\mathrm{squark}, m_\mathrm{gluino}$) (panel (a)) and in the plane (\textit{$\geq$ three-jets}, \, \textit{$\geq$ two-jets}) (panel (b)). In panel (a) the three excluded points are plotted in red. The colour coding in panel (b) indicates the LSP composition in each scenarios.]{\label{binosinglino} Representation of the points in the plane ($m_\mathrm{squark}, m_\mathrm{gluino}$) (panel (a)) and in the plane (\textit{$\geq$ three-jets}, \, \textit{$\geq$ two-jets}) (panel (b)). In panel (a) the three excluded points are plotted in red. The colour coding in panel (b) corresponds to the bino component of the LSP neutralino in each scenarios : red points have a bino-like LSP whereas blue points have a singlino-like LSP. The exclusion zone ($\sigma_{2j}>22~\text{fb},\:\sigma_{3j}>25~\text{fb}$) is shaded.}
  \end{center}
\end{figure}

This region with light squarks is displayed in the panel (a) of figure~\ref{binosinglino}. Panel (b) of figure~\ref{binosinglino} illustrates the most sensitive signal regions. The exclusion limits at 2$\s$ are respectively $22$~fb for the cross section in the \textit{$\geq$ two-jets} channel and $25$~fb for the cross section in the \textit{three-jets} channel. We are able to exclude three points in this region. What is striking when we look at figure~\ref{binosinglino}a is that these excluded points, plotted in red, are not those corresponding to the lightest squarks. To understand the reason of this unusual feature we must look at the characteristics of the LSP in these scenarios. We recall that the limit of $m_{\tilde{q}} \sim 0.6$~TeV is obtained in MSSM scenarios with large branching ratios of gluinos and squarks into jets and the neutralino LSP, mostly bino. 

Two of the excluded points can be easily understood. They correspond to the points only excluded by the \textit{$\geq$ two-jets} channel limit as shown in figure~\ref{binosinglino}b. Actually, their main characteristic is that the LSP is bino-like. Thus the usual $\tilde{q}\rightarrow q \chi_1^0$ decay takes place and the familiar jets + missing $E_T$ exclusion is observed for $m_{\tilde{q}} \sim 0.6$ TeV.

\begin{figure}[!htb]
\begin{center}
\centering
\includegraphics[width=9cm]{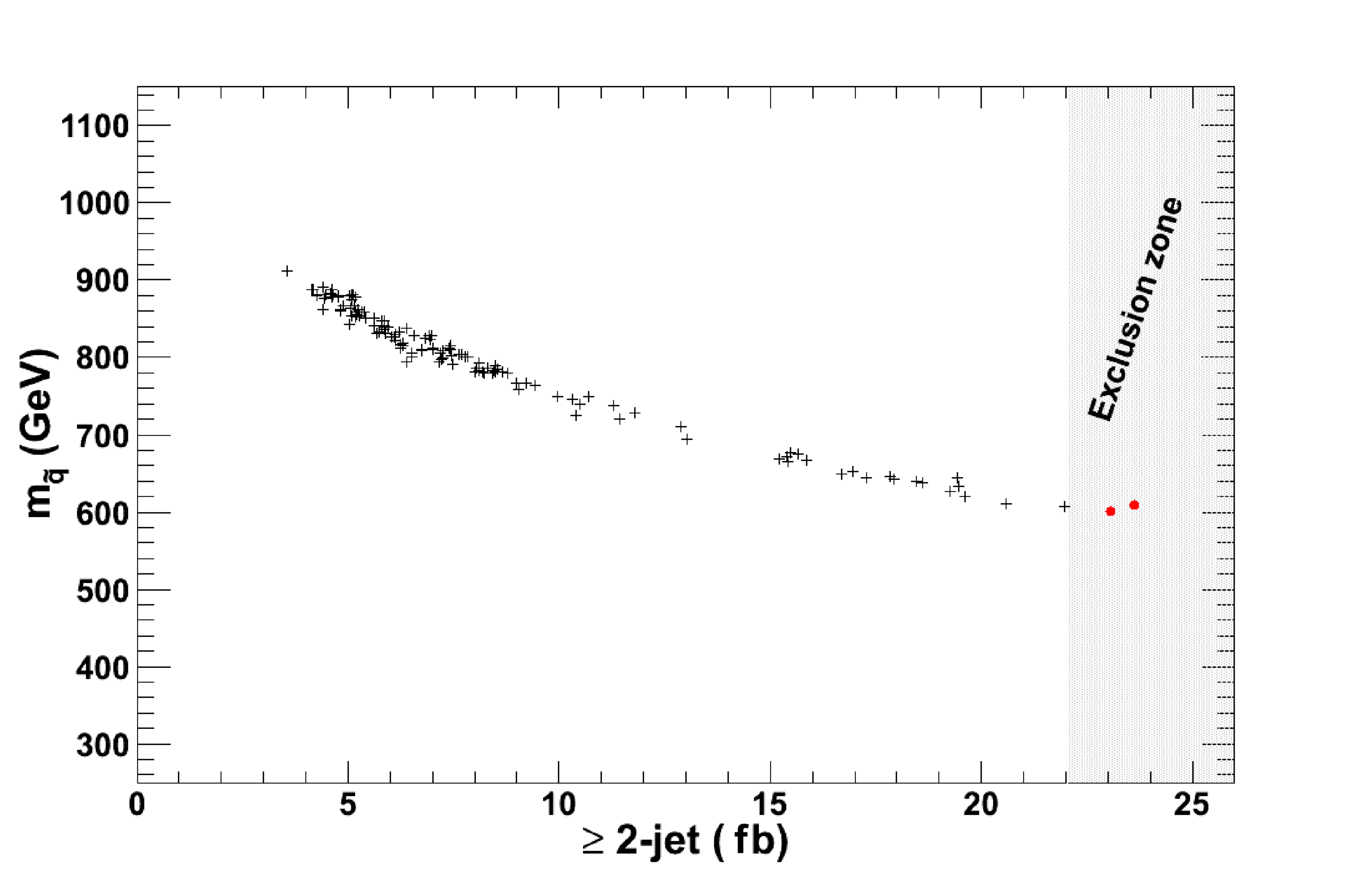}
\caption{\label{binochannel}The squark mass\vs the most sensitive signal region for scenarios with bino-like LSP. The grey zone corresponds to the excluded region and the two excluded points are plotted in red.}
\end{center}
\end{figure}

This is confirmed by figure~\ref{binochannel} which only plots the bino-like cases of figure~\ref{binosinglino}b : these two points are characterized by the lowest squark masses, then with the highest acceptance in the \textit{$\geq$ two-jets} channel. 

\begin{figure}[!htb]
\begin{center}
\centering
\includegraphics[width=9cm]{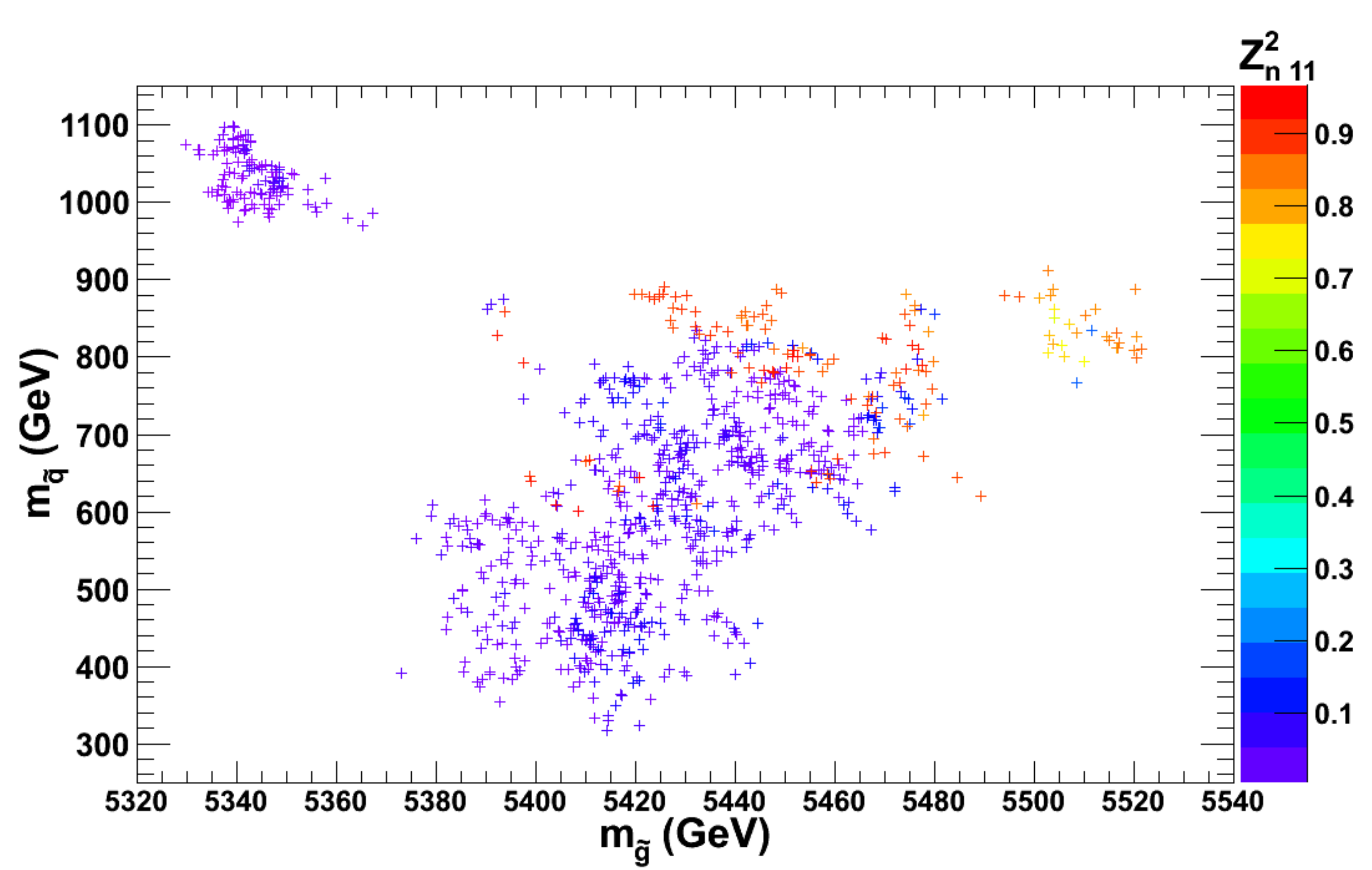}
\caption{\label{binosinglino2} Points in the plane ($m_\mathrm{squark}, m_\mathrm{gluino}$) where the colour coding is the same as in figure~\ref{binosinglino}b.}
\end{center}
\end{figure}

\begin{figure}[!htb]
\begin{center}
\centering
\subfloat[]{\includegraphics[width=8cm,height=5cm]{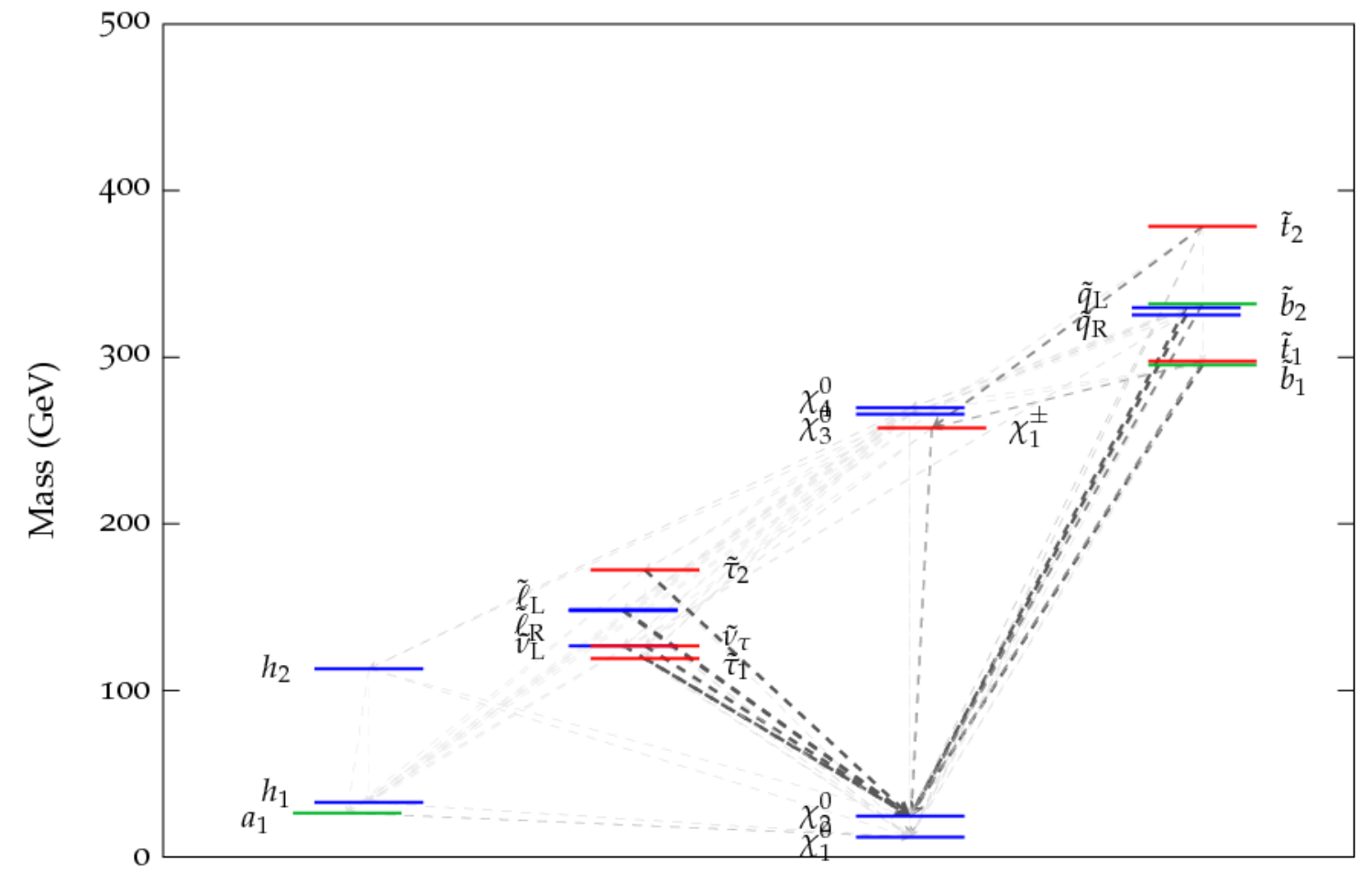}}
\subfloat[]{\includegraphics[width=8.5cm,height=5.5cm]{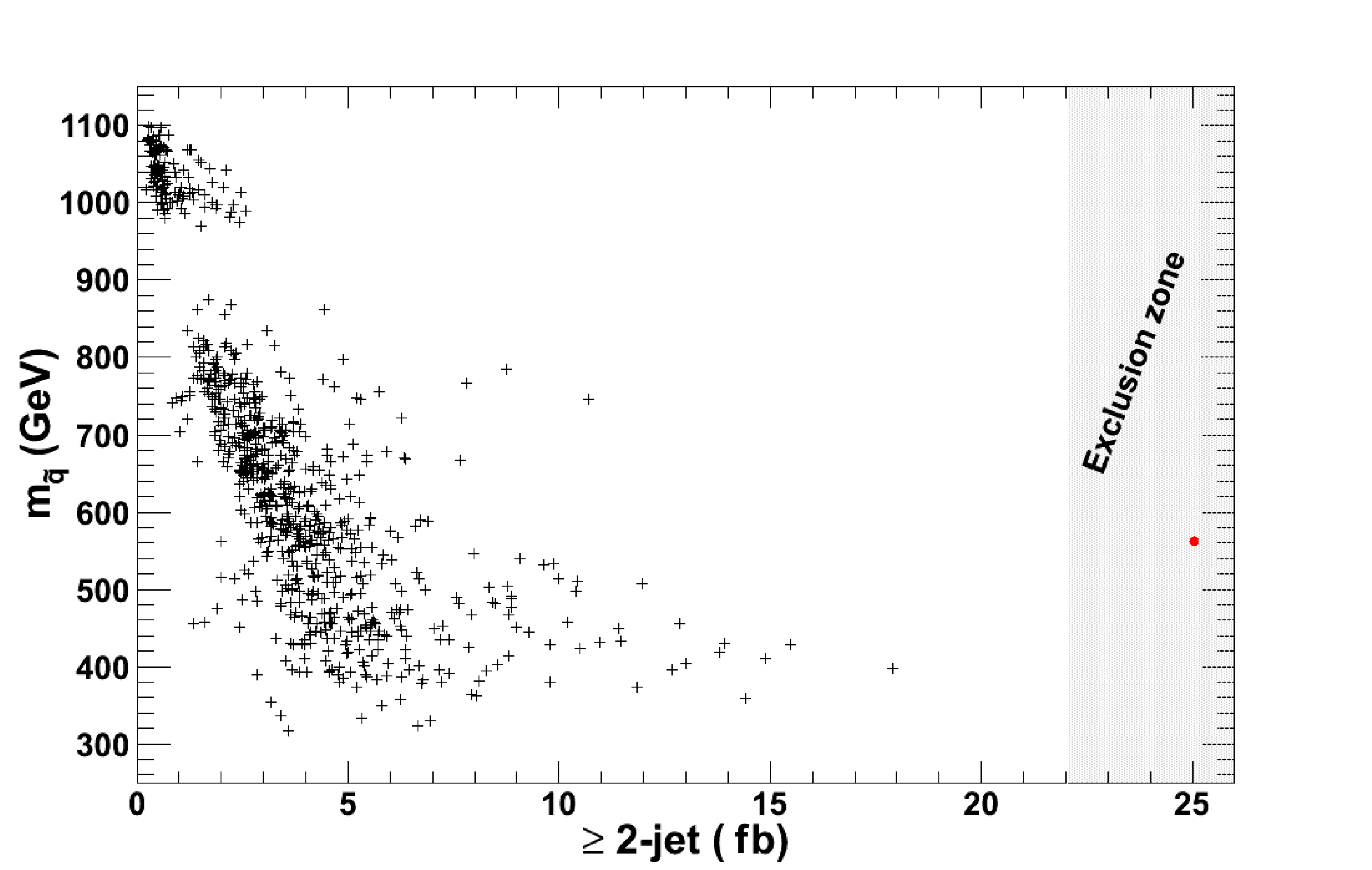}}    
\caption[The case of singlino-like LSP : panel (a) shows the branching ratios and masses of supersymmetric particles and Higgs bosons in the case of a non-excluded point. Panel (b) represents the squark mass\vs the best sensitive signal region for these scenarios with singlino-like LSP.]{\label{singlinoexple}The case of singlino-like LSP : panel (a) shows the branching ratios and masses of supersymmetric particles and Higgs bosons in the case of a non-excluded point with singlino-like LSP, where we restrict ourselves to sparticles and Higgs bosons below $0.5$~TeV. The darker the dotted arrow the higher the corresponding branching ratio. Panel (b) represents the squark mass\vs the best sensitive signal region for scenarios with singlino-like LSP. The grey zone correspond to the excluded region and one excluded point is plotted in red.}
\end{center}
\end{figure}

We may ask why points with squarks down to $0.3$~TeV are allowed in figure~\ref{binosinglino}a. The colour code in figure~\ref{binosinglino2} gives the key to answer to that. In red we have points with a bino-like LSP and in blue a singlino-like LSP. We clearly see that all points with squarks below $0.6$~TeV contain singlino-like LSP. When the LSP is purely singlino, the squarks and gluinos cannot decay to this LSP directly but must do via an intermediate particle, frequently the second-lightest neutralino as illustrated in figure~\ref{singlinoexple}a. As noted in \cite{Das:2012rr} this reduces the acceptance into jets + missing $E_T$ search channels, as the extra step reduces the missing $E_T$ and may result in leptons\footnote{SUSY searches with leptons would have in fact greater sensitivity but they do not compensate for the loss of sensitivity in the 0-lepton search \cite{Das:2012rr}.}. Then these limits are not so efficient for singlino-like LSP as shown in figure~\ref{singlinoexple}b. 

It is also interesting to note that the relative acceptance into different search channels can help to distinguish a bino-like LSP from a singlino-like LSP : the latter produces a higher average number of jets in the cascade as illustrated in figure~\ref{binosinglino}b. The graph also shows the low level of exclusion of singlino-like LSP points for the reasons discussed above. The bulk of bino-like LSP points is not excluded simply because squarks are too heavy and out of reach.

\begin{figure}[!htb]
  \begin{center}
    \includegraphics[width=9cm]{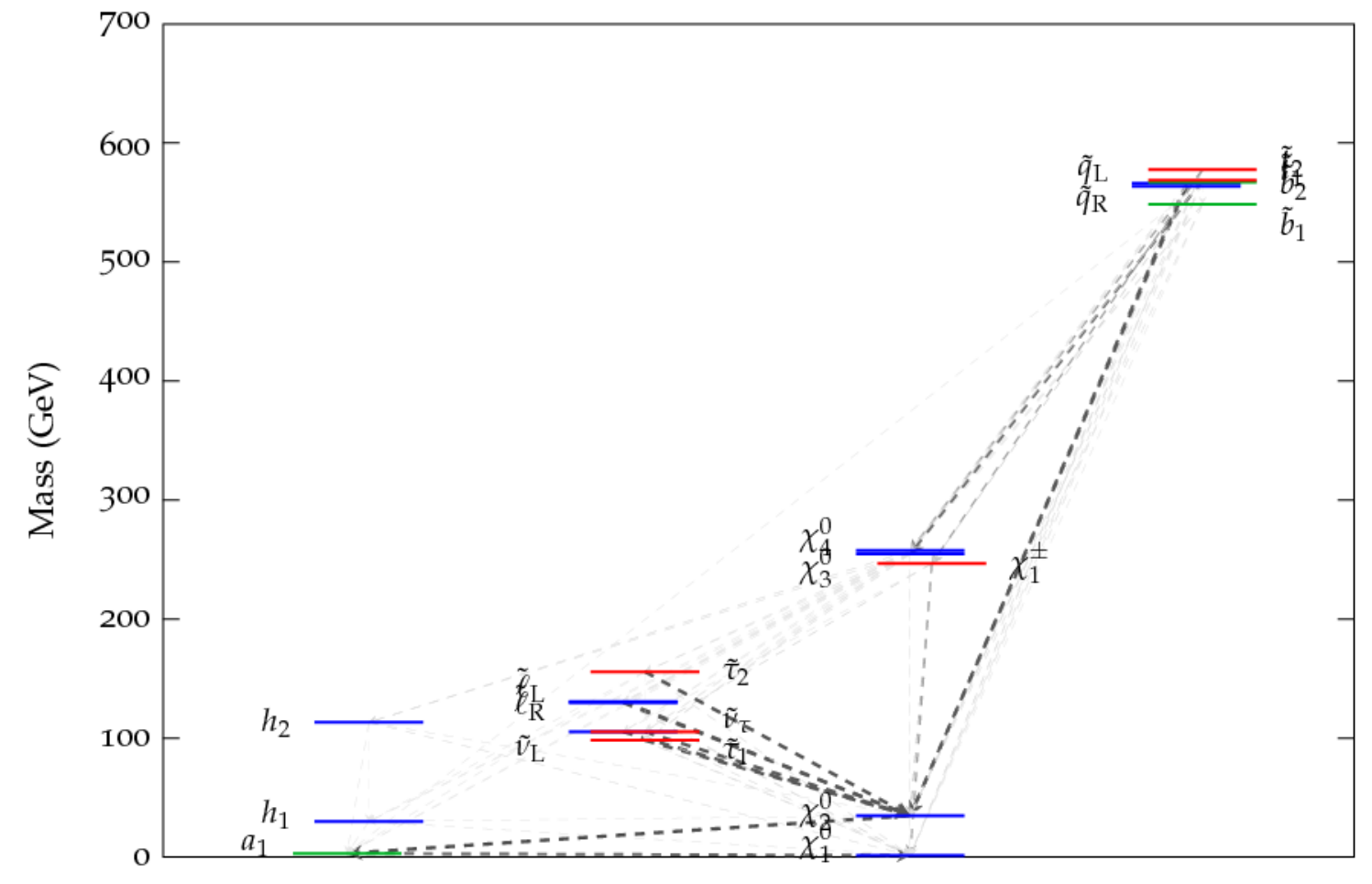}
    \caption[Branching ratios and masses of supersymmetric particles and Higgs bosons in the case of the excluded point with singlino-like LSP.]{\label{singlinoexcl}Branching ratios and masses of supersymmetric particles and Higgs bosons in the case of the excluded point with singlino-like LSP. Here we restrict ourselves to sparticles and Higgs bosons below $0.7$~TeV.}
  \end{center}
\end{figure}

However figure~\ref{singlinoexple}b presents also an interesting exception since one point with singlino-like LSP and squark masses above $0.5$~TeV is excluded by this ATLAS limits. This point is described in figure~\ref{singlinoexcl}. In this case we see that the lightest pseudoscalar Higgs boson is really light : $m_{a_1} \simeq 3$~GeV. Thus the bino-like $\chi^0_2$ decays into the singlino-like LSP plus $a_1$. This explain why this point is excluded : since this pseudoscalar Higgs boson has a mass below the $b\bar{b}$ and $\tau^+\tau^-$ pair thresholds it decays fully invisibly, then giving large missing $E_T$.

\section{Higgs boson signal strength with light LSP}
\label{sec.6:Rgggaga1}

\begin{figure}[!htb]
\begin{center}
\centering
\subfloat[]{\includegraphics[width=8cm]{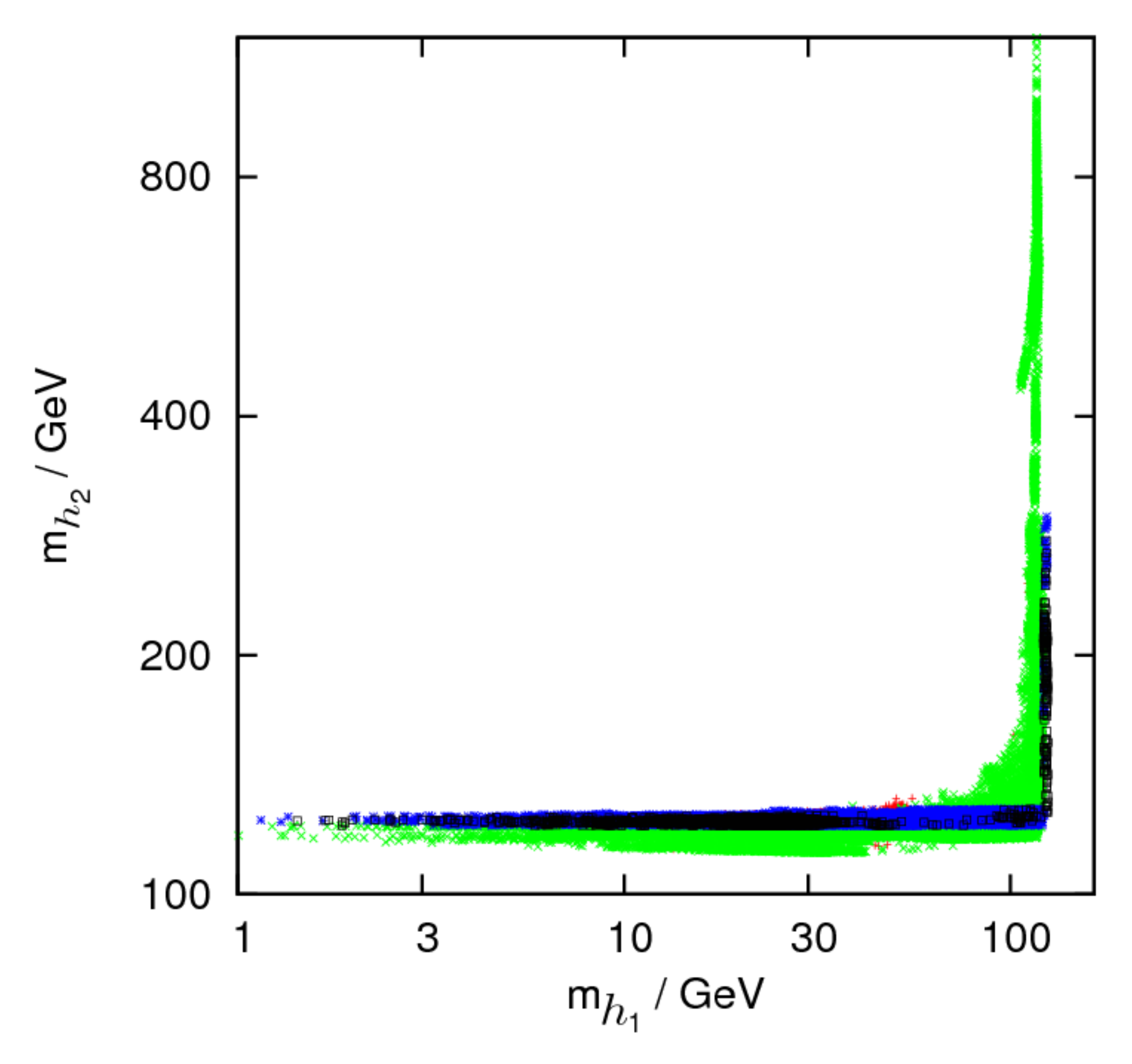}}
\subfloat[]{\includegraphics[width=8cm]{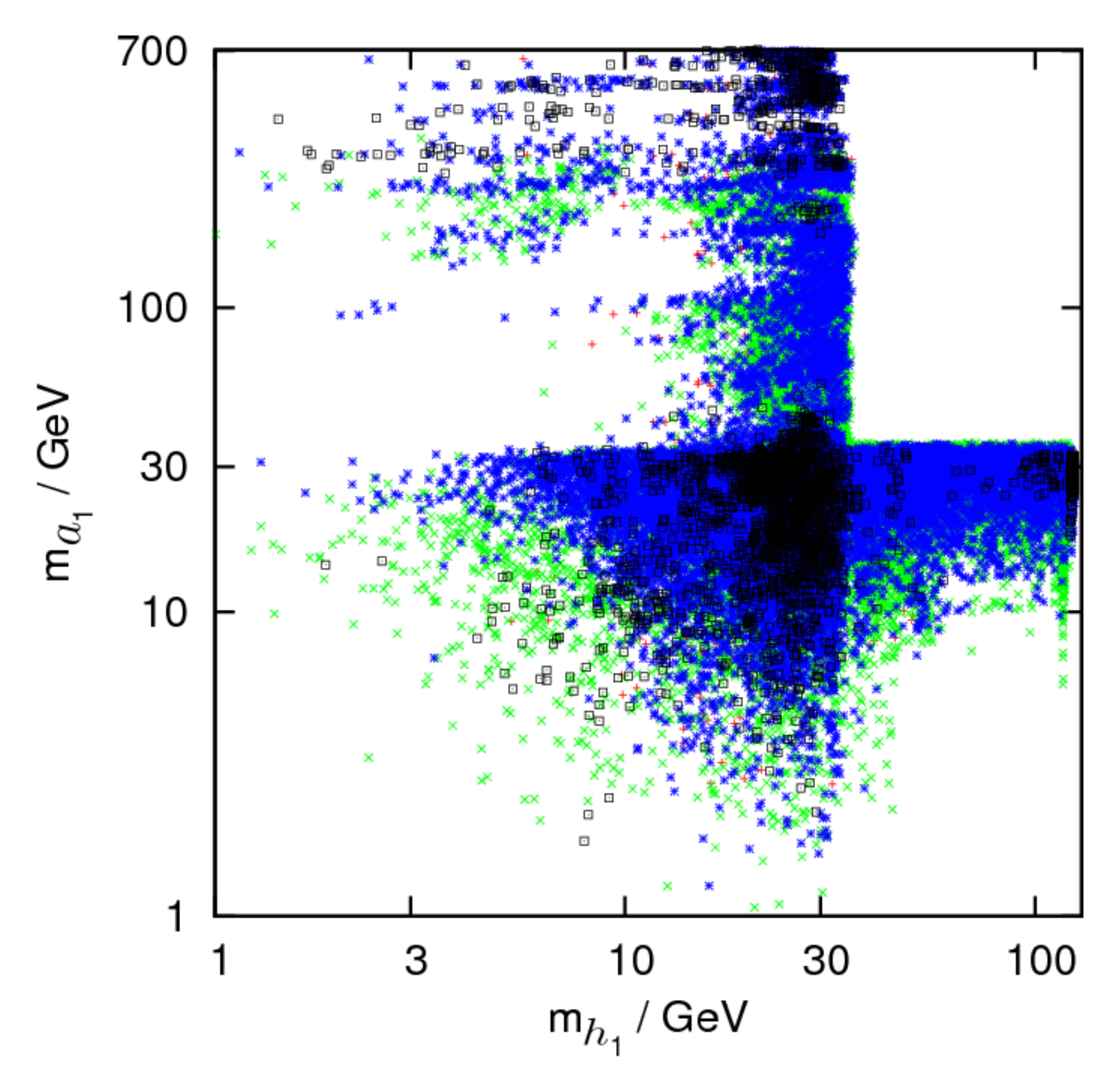}}    
\caption[Plots in the plane ($m_{h_2},m_{h_1}$) (panel (a)) and ($m_{a_1},m_{h_1}$) (panel (b)) where the colour coding indicates scenarios excluded or not by SUSY or Higgs boson search and gives informations on the Higgs boson signal strength.]{\label{fig:mh12}Plots in the plane ($m_{h_2},m_{h_1}$) (panel (a)) and ($m_{a_1},m_{h_1}$) (panel (b)). Red points are ruled out either by {\tt HiggsBounds-3.6.1beta\;} constraints or the ATLAS $1\text{fb}^{-1}$ jets and missing $E_T$ SUSY search. Green points have no Higgs boson with a mass in $122-128$~GeV, blue points have a Higgs boson ($h_1$ and/or $h_2$) within this mass range, and black points have such a Higgs boson with ${R_{gg\g\g}}>0.4$.}
\end{center}
\end{figure}

For a model to be compatible with the data, it is required that the SM-like Higgs boson be in the observed mass range (say  [122-128] GeV) and the signal strength be consistent with the data. As a criteria for $R_{gg\gamma\gamma}$, we choose a 2$\s$ error bar around the central value determined by ATLAS~\cite{ATLAS:2012ad}, thus $0.4<R_{gg\g\g}<3.6$. This is also compatible with the CMS results and the SM expectations (where $R_{gg\g\g}^{\mathrm{SM}} \equiv 1$). 

As illustrated in figure~\ref{neutmh1ma1} and highlighted by figure~\ref{fig:mh12}a, $h_1$ is typically much below the EW scale, thus with a large singlet component, when $h_2$ is SM-like. When $h_2$ is much heavier than 125 GeV, $h_1$ is SM-like (reaching at most up to $\sim$~122 GeV) and the pseudoscalar $a_1$ is basically a light singlet, see figure~\ref{fig:mh12}b. It is nevertheless possible for both scalars to be heavily mixed and have a mass around 100-130~GeV (in this case $a_1$ has to be light). In figure~\ref{fig:mh12} blue and black points show the scenarios with at least one of the scalars within the range preferred by ATLAS and CMS. As can be seen, this is generally $h_2$ since $m_{h_1}$ barely exceeds 122~GeV. Note that this do not take into account theoretical uncertainties that can reach a few GeV's.

\begin{figure}[!htb]
\begin{center}
\centering
\subfloat[]{\includegraphics[width=8cm]{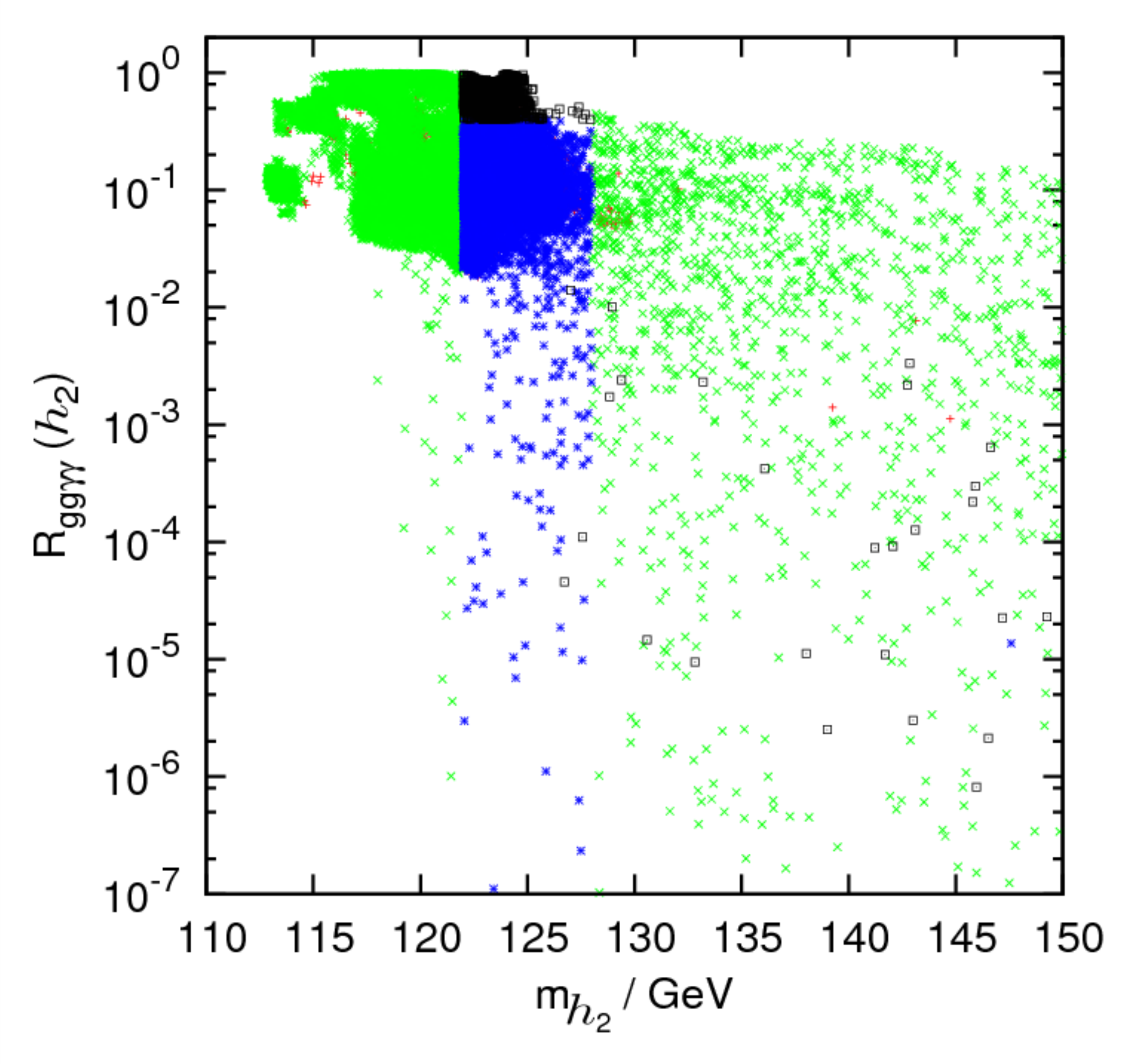}}
\subfloat[]{\includegraphics[width=8cm]{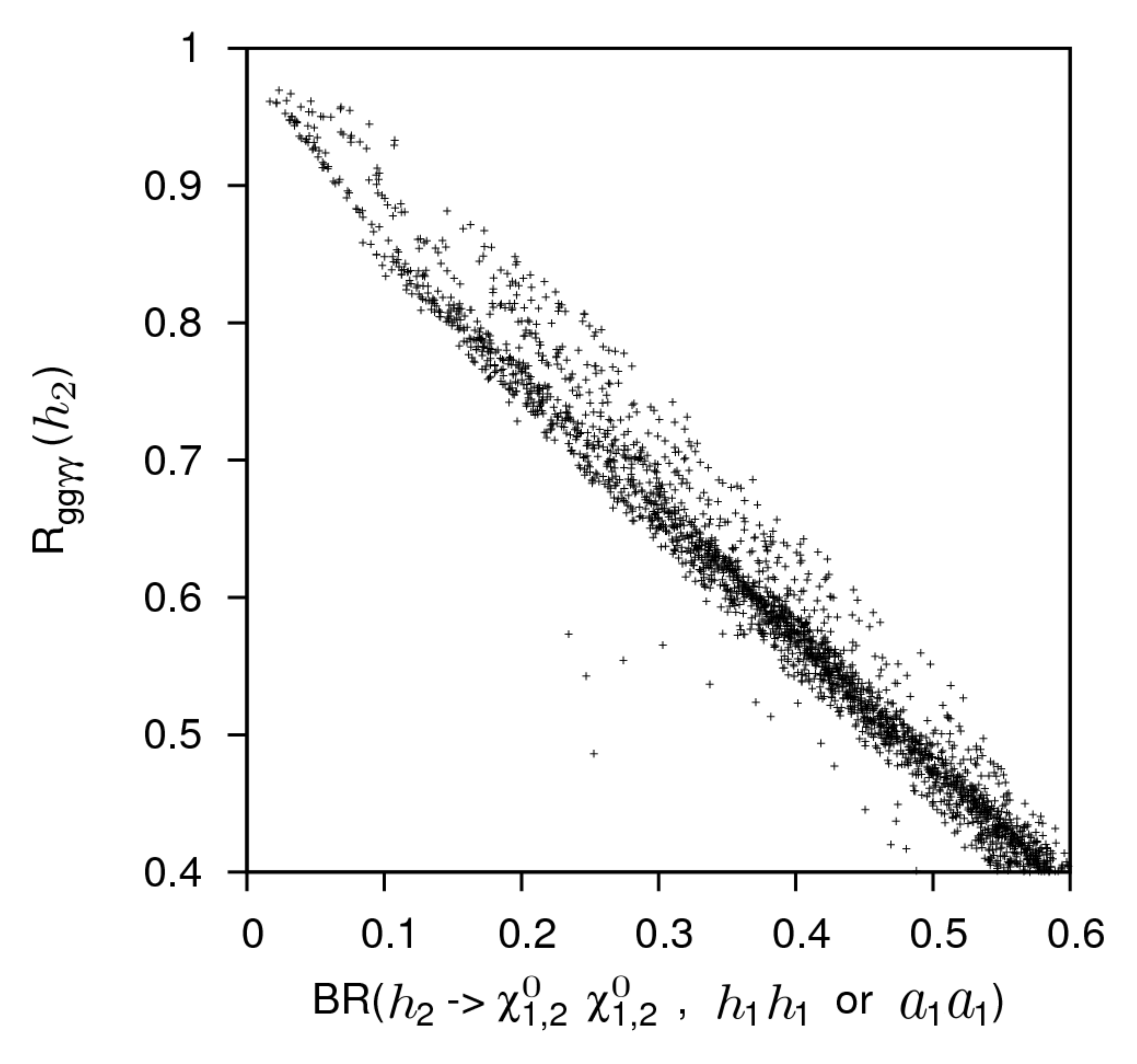}}
\caption{\label{fig:gggg_nmssm}$R_{gg\g\g}$ as a function of the mass of $h_2$ (panel (a)) and as a function of new BSM decays. Same colour code as in figure~\ref{fig:mh12}.}
\end{center}
\end{figure}

The predictions for $R_{gg\g\g}$ as a function of the $h_2$ mass are displayed in figure~\ref{fig:gggg_nmssm}a. We only display the region where this channel is relevant, that is when $m_{h_2} < 150$ GeV. We clearly note that all the configurations that were selected by the MCMC have $R_{gg\g\g} < 1 $. An explanation is that, although $h_2$ couplings are usually SM-like, large suppressions in $R_{gg\g\g}$ are possible because the width of the Higgs is enhanced by many new non-standard decay channels~\cite{Cao:2011re}. In particular, $h_2$ can decay into two neutralinos, two light scalar Higgs bosons ($h_1$) or two light pseudoscalar Higgs bosons ($a_1$) which reduces significantly the branching ratio into two photons.

In figure~\ref{fig:gggg_nmssm}a, all the points which do not satisfy either the {\tt HiggsBounds-3.6.1beta\;} limits or the SUSY searches in jets plus missing $E_T$ are coloured in red. As mentioned above very few of these points are excluded by SUSY searches. The points which fall within the Higgs boson observed mass range are highlighted in blue. Scenarios where the strength of the signal in $\g\g$ is also compatible with the 2$\s$ range reported by ATLAS ($R_{gg\g\g}>0.4$) are represented by black squares.

The effect of these non-standard decays on the signal strength is shown in figure~\ref{fig:gggg_nmssm}b.
Clearly, too large branching ratios into non-standard modes such as $\chi_i^0 \chi_j^0 \ (i,j=1,2), h_1 h_1$ and $a_1 a_1$ would render the two-photon mode invisible or suppressed with respect to the SM prediction~\cite{Cao:2011re}. In fact, in order for the signal strength  $R_{gg\g\g} $ to be compatible with $R_{gg\g\g} > 0.4$, the branching ratio $\mathscr{B}(h_2 \ra \rm{invisible})$ must be lower than $\sim 60\%$.

The existence of decay modes such as $h_2\ra h_1 h_1$ or $ a_1 a_1$, with the singlet Higgs boson further decaying into SM particles remains nevertheless interesting because such modes give a distinctive signature which could be searched for at LHC and would constitute evidence for NP if they are found. Extraction of this signal from background via jet substructure techniques has been studied for $h \ra 2a \ra 4\tau$ in \cite{Englert:2011iz}, $h \ra 2a \ra \tau^+ \tau^- \mu^+ \mu^-$ in \cite{Lisanti:2009uy}, and $h \ra 2a \ra 4g$ (less relevant for SUSY due to $\tan\beta$ suppression) in \cite{Falkowski:2010hi,Chen:2010wk}. These decays with an intermediate scalar ($h_1$) instead of an intermediate pseudoscalar ($a_1$) give the same signal.

\section{The case of heavy LSP}
\label{sec.6:Rgggaga2}

\begin{figure}[!htb]
\begin{center}
\centering
\subfloat[]{\includegraphics[width=8cm]{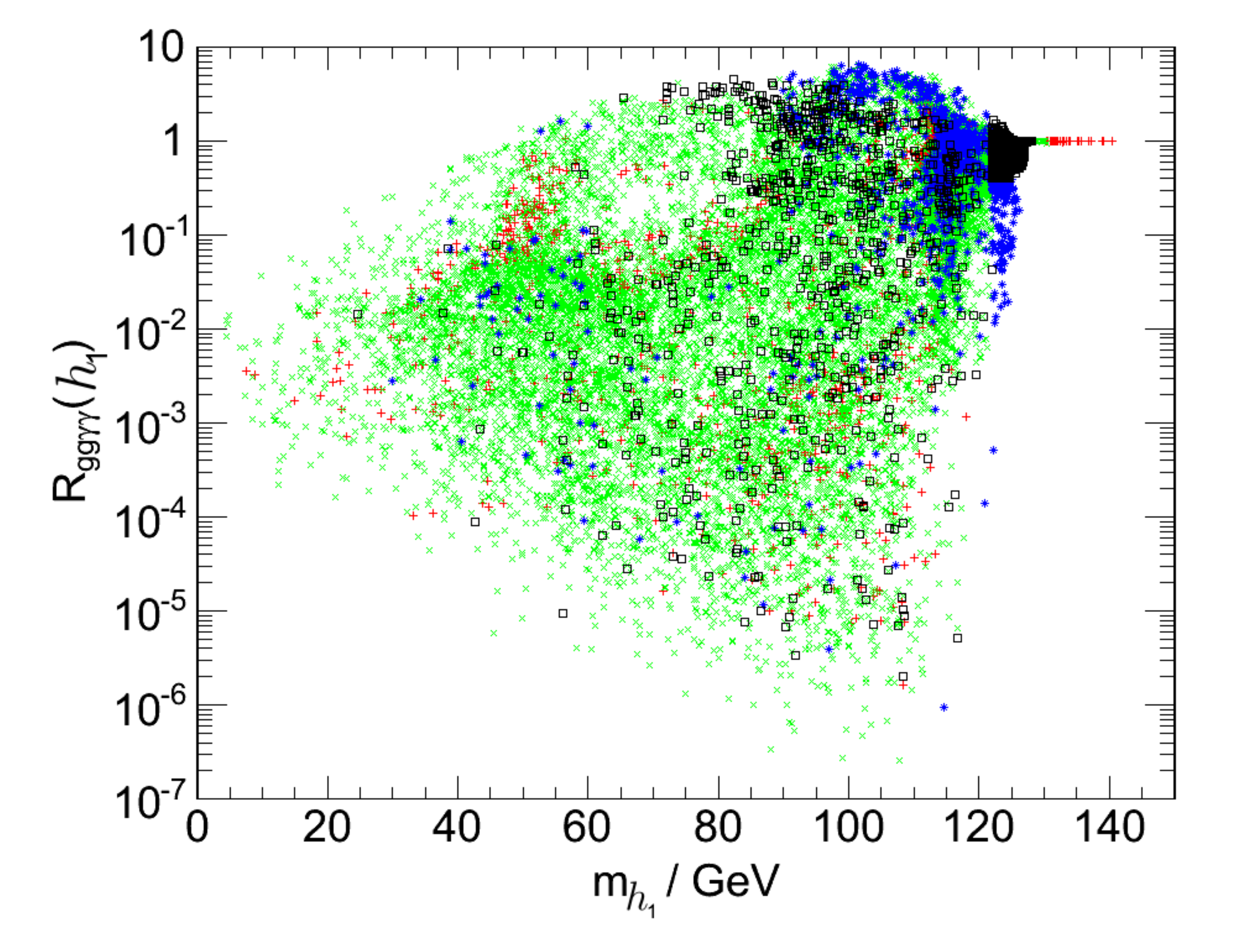}}
\subfloat[]{\includegraphics[width=8cm]{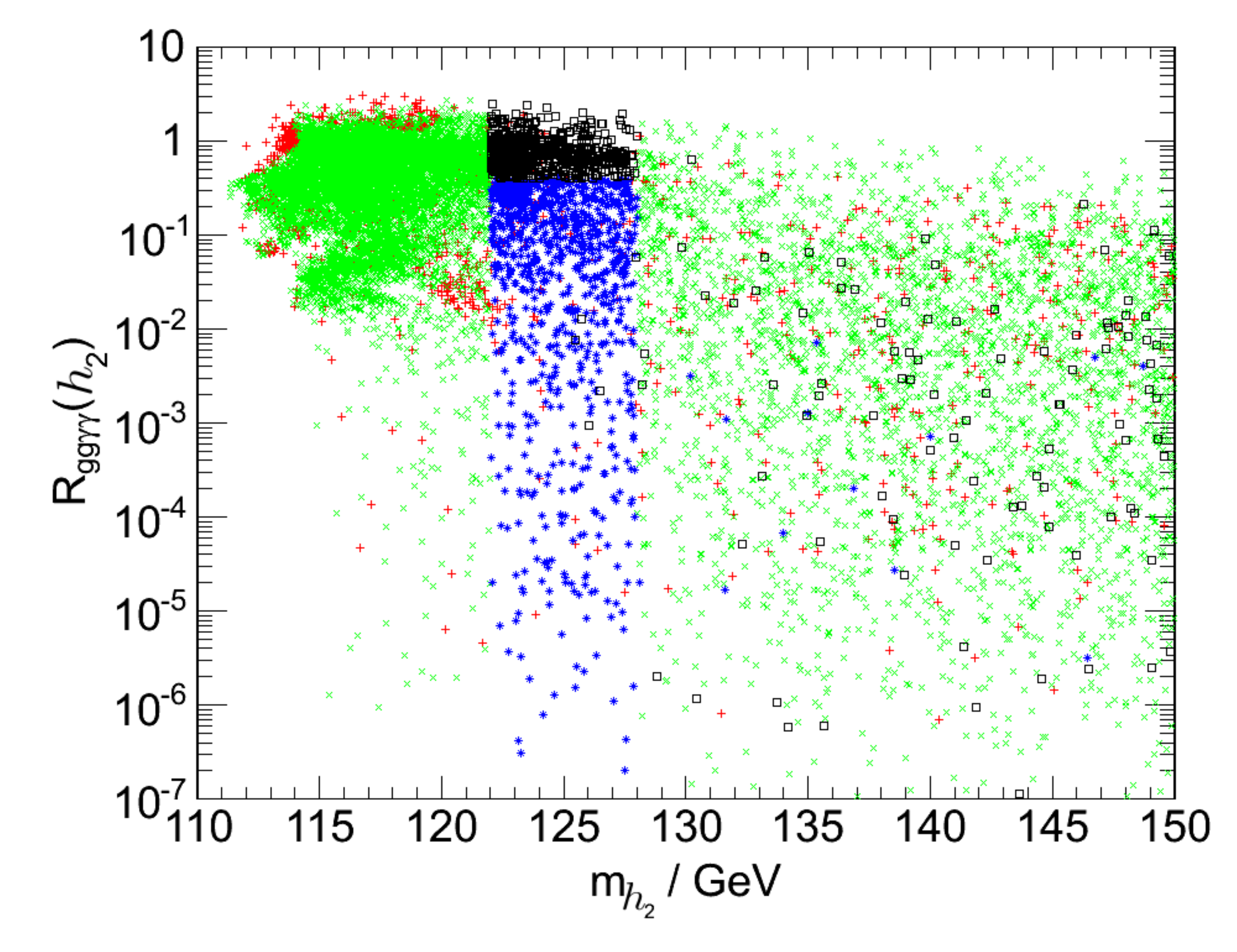}}    
\caption[$R_{gg\g\g}$ as a function of the mass of $h_1$ (panel (a)) and of $h_2$ (panel (b)) in the arbitrary neutralino LSP model.]{\label{fig:gggg_all}$R_{gg\g\g}$ as a function of the mass of $h_1$ (panel (a)) and of $h_2$ (panel (b)) in the arbitrary neutralino LSP model. Same colour code as in figure~\ref{fig:mh12} except red points are denoting only exclusion by {\tt HiggsBounds-3.6.1beta\;}.}
\end{center}
\end{figure}

As a side comment note that the analyses were repeated for another MCMC scan in which there is no $m_{\chi_1^0} < 15$ GeV requirement. Event though this study encompass all LSP masses, a separate analysis with a prior on the neutralino mass
was needed to ensure a complete coverage of the parameter space of the $m_{\chi_1^0} < 15$ GeV scenarios. Then as opposed to the previous case with a light neutralino the Higgs boson in the mass region preferred by the LHC is a SM-like $h_1$. Indeed, without very light neutralinos, a very light singlet sector is not needed for resonant annihilations. Thus the associated values for $R_{gg\g\g}$ are naturally of order unity (see figure~\ref{fig:gggg_all}a).  Nevertheless cases where $R_{gg\g\g}<0.4$ are possible, when invisible decay modes (such as $h_1 \ra \chi_1^0 \chi_1^0$ or $h_1 \ra a_1 a_1$) are kinematically accessible.  

$h_2$ masses extend over a wide range (all the way to several TeV's) and include some points in the mass region preferred by the LHC. The values of $R_{gg\g\g}$ for $h_2$ are displayed in figure~\ref{fig:gggg_all}b for the range of masses where  the two-photon search mode is relevant. In this region the signal strength reaches values as high as $R_{gg\g\g}=2$.
This enhancement with respect to the SM expectations is found when $h_2$ has some singlet component and a suppressed partial width to $b\bar{b}$ ($h_1$, conversely, has an enhanced $b\bar{b}$ partial width and a reduced signal strength $R_{gg\g\g}$). 

Note however that for most of these points with an enhanced $R_{gg\g\g}$ the neutralino would form only a fraction of the observed DM. If a lower limit on the relic density is used these points characterized by a non-negligible higgsino component are then excluded. Note however that $R_{gg\g\g} > 1$ and a strict lower bound on the DM relic density is still possible by looking thoroughly at regions with $\tan \beta \sim 2$ and $\l \sim 1$ \cite{Ellwanger:2011aa}.

\section{Conclusions}

In this chapter we have investigated the Higgs signal strength and direct SUSY search in the context of the NMSSM. Many scenarios where found with either $h_1$ or $h_2$ had a mass in the range [122-128] GeV and a signal strength compatible with the SM. We also found scenarios where the signal strength in the two-photon mode was as large as the excess reported by ATLAS and CMS.  

When insisting on a light neutralino, we found that the most promising configurations favour a SM-like $h_2$ rather than a $h_1$ SM-like Higgs boson and therefore predict the existence of a light Higgs boson dominantly singlet. The possibility of observing a second light Higgs boson provides a distinct signature of the NMSSM Higgs sector.

Furthermore we note that the traditional jets + missing $E_T$ signature of squarks and gluinos search in the MSSM can be modified in the NMSSM when the LSP is a singlino. These sparticles decay into quarks and the second-lightest neutralino $\chi_2^0$ bino-like which then decays into the singlino-like $\chi_2^0$ and moslty leptons. As a result the missing $E_T$ is reduced and the usual limits on squark and gluino masses are weakened for these scenarios.

 Note nevertheless that the limits have greatly improved since the study we made. For instance assuming equal squark and gluino masses the ATLAS $5.8 \ \rm{fb}^{-1}$ search for squarks and gluinos via jets and missing transverse momentum at $\sqrt{s}=8$~TeV center-of-mass energy of proton-proton collision \cite{ATLAS:2012ona} can exclude such coloured sparticles up to $1.5$~TeV (see figure~\ref{fig:atlassusy}) in the MSSM. As a prospect it could then be interesting to check if our favourite NMSSM points with singlino-like LSP still survive LHC constraints on squark and gluino masses which could give a possibility to derive, starting from mSUGRA/CMSSM limits, indirect bounds on these NMSSM scenarios.

\part{$\mathbf{U(1)}$ extensions of the MSSM}
\label{part3}

\chapter{The UMSSM}
\label{chapter:UMSSM}

\minitoc\vspace{1cm}
\newpage

\section{Another solution to the \textit{$\mu$-problem}}

In chapter~\ref{chapter:NMSSM} we analysed some phenomenological consequences of an extension of the MSSM, the NMSSM. This model has in particular the nice feature to solve the $\mu$-problem of the MSSM. However this approach has drawbacks that can be debated. Actually the superpotential of the NMSSM defined in eq.~\ref{W_NMSSM} is invariant under a discrete $\mathbb{Z}_3$ symmetry. As shown in \cite{Vilenkin:1984ib} the spontaneous breaking of a discrete symmetry at a phase transition of the early Universe can generate topological defects, more precisely two-dimensional objects called \textit{domain walls}. Then as denoted in \cite{Abel:1995wk} the $\mathbb{Z}_3$ symmetry breaking at the EW phase transition leads to cosmological domain walls. These objects can dominate the energy density of the early Universe : they therefore can be viewed as a source of DE \cite{Friedland:2002qs,Bucher:1998mh}. Nevertheless the equation of state of domain walls is predicted to be $\omega = -2/3$ which is excluded by CMB observations, especially recent Planck results with $\omega_{_{\mathrm{DE}}} = -1.13^{\,+\,0.23}_{\,-\,0.25}$ at the 2$\s$ level, see table~\ref{tab:cosmo_param}. Several solutions have been proposed for the NMSSM~\cite{Abel:1996cr,Panagiotakopoulos:1998yw}.

Another possibility to solve the $\mu$-problem avoiding the domain walls problem is, when adding the chiral supermultiplet $\mathbf{S}$, to promote the $U(1)_{\mathrm{PQ}}$ global symmetry to a new $U(1)$ Abelian gauge symmetry. Then the extra $S^3$ term of eq.~\ref{W_NMSSM} and the usual $\mu$ term of the MSSM superpotential in eq.~\ref{eq.3:W_MSSM} are forbidden by this new gauge symmetry. It follows that the VEV of the scalar singlet field $S$ will break the new Abelian symmetry and its massless pseudoscalar component becomes the longitudinal mode of a new gauge boson which then gets a mass, the $Z'$ vector boson \cite{Cvetic:1997ky}. It is this type of extensions of the MSSM that will be considered throughout the last part of this thesis.

\section{An $E_6$ inspired model}

Models with extended gauge symmetries are well motivated within the context of BSM models. They occur in GUT scale models \cite{Langacker:1980js,London:1986dk}, extra-dimension motivations \cite{Masip:1999mk}, superstring models \cite{Cvetic:1995rj,Cvetic:1996mf,Cleaver:1997jb,Cleaver:1998gc,Cleaver:1998sm}, strong dynamics models \cite{Hill:2002ap}, little Higgs models \cite{ArkaniHamed:2001nc,ArkaniHamed:2002qy,Han:2003wu} or through the Stueckelberg \cite{Stueckelberg:1900zz} mechanism \cite{Kors:2004ri,Kors:2005uz,Feldman:2006wb}.

Looking at supersymmetric models, several Abelian extensions are considered in the literature. An example is the $U(1)_{\mathcal{B}-\mathcal{L}}$ scenario \cite{Malinsky:2005bi,DeRomeri:2011ie}, and specific supersymmetric $U(1)_{\mathcal{B}-\mathcal{L}}$ inflationary scenarios were studied in \cite{Allahverdi:2007wt}. Note also that in supersymmetric models with extended $U(1)$ symmetry a $\l S H_u H_d$ interaction allows for the strong first order phase transition that is needed for EW baryogenesis~\cite{Ahriche:2010ny}. One of the most analysed $U(1)$ extension originates from a string-inspired $E_6$ grand unified gauge group \cite{Hewett:1988xc,Langacker:1998tc,King:2005jy,Howl:2007zi,Hall:2011yb}. Since the rank of the $E_6$ group is 6 the breaking of the $E_6$ symmetry gives models based on rank-5 or rank-6 gauge groups. As a consequence these models in general lead to a low-energy gauge symmetry with one or two additional $U(1)$ symmetries plus the SM gauge symmetry. $E_6$ contains $SO(10) \otimes U(1)_{\psi}$ while $SO(10)$ can be decomposed into $SU(5) \otimes U(1)_{\chi}$. Using the Hosotani mechanism \cite{Hosotani:1983vn} and the fact that $SU(5)$ contains the SM gauge symmetry $E_6$ can be broken directly into $SU(3)_c  \otimes  SU(2)_L  \otimes  U(1)_Y \otimes U(1)_{\psi} \otimes U(1)_{\chi}$ which is of rank 6. We assume that at low energy the gauge symmetry is\footnote{For more details see \cite{King:2005jy}.} \begin{center}$SU(3)_c  \otimes  SU(2)_L  \otimes  U(1)_Y \otimes U(1)'$,\end{center} where $U(1)'$ is a linear combination of $U(1)_{\psi}$ and $U(1)_{\chi}$. This combination is parameterized by an angle $\te6$ with $\te6 \in [-\pi/2, \pi/2]$ and we have
\beq U(1)' = \cos \te6 U(1)_{\chi} + \sin \te6 U(1)_{\psi}.\eeq
The $U(1)'$ charge for each field of the model is then defined as 
\beq \mathcal{Q}' = \cos \te6 \mathcal{Q}'_{\chi} + \sin \te6 \mathcal{Q}'_{\psi},\eeq
where the charges $\mathcal{Q}'_{\chi}$ and $\mathcal{Q}'_{\psi}$ are given in \cite{Langacker:1998tc,Barger:2007nv} for all fields of the $E_6$ model.

The matter sector of the $E_6$ model contains, in addition to the supermultiplets containing the SM fermions, three families of RH neutrinos, three families of Higgs doublets ($H_u, H_d$), three singlets, three families of extra colour $SU(3)_c$ (anti)triplets and finally an additional Higgs-like $SU(2)_L$ doublet and anti-doublet. The complete matter sector is needed for anomaly cancellations but for simplicity we will assume that these exotic fields, with the exception of the RH neutrinos (one of the supersymmetric partners will be considered as a possible DM candidate in chapter~\ref{chapter:RHsneu}), one family of Higgs doublets and one Higgs singlet, are above a few TeV's. We will assume that they play no role in the phenomonology of the rest of the matter sector and neglect them in the following (noted as $\cal{O}(\mathrm{TeVs})$ in the forthcoming equations). It is what we will call the UMSSM model. 

\renewcommand{\arraystretch}{1.4}
\begin{table}[!htb]
\begin{center}
\begin{tabular*}{0.95\textwidth}{  c  c  c  c  c  }
       \hline \hline
      \multicolumn{5}{c}{\textbf{Chiral supermultiplets}} \\ \hline \hline
      \multicolumn{2}{c}{\textbf{Name}} & \textbf{spin 0} & \textbf{spin 1/2} & $\mathbf{SU(3)_c, SU(2)_L, U(1)_Y, U(1)'}$ \\ \hline 
      squarks, quarks & $\widetilde{Q}, Q$ & ($\tilde{u}_L$ $\tilde{d}_L$) & ($u_L$ $d_L$) & ($\textbf{3}$, $\textbf{2}$, $\frac{1}{3}$, $\mathcal{Q}'_Q$) \\ 
      	 (3 families) & $\bar{u}$ & $\tilde{u}^*_R$ & $\bar{u}_R$ & ($\bar{\textbf{3}}$, $\textbf{1}$, $-\frac{4}{3}$, $\mathcal{Q}'_u$) \\ 
      		      & $\bar{d}$ & $\tilde{d}^*_R$ & $\bar{d}_R$ & ($\bar{\textbf{3}}$, $\textbf{1}$, $\frac{2}{3}$, $\mathcal{Q}'_d$) \\ \hline 
    sleptons, leptons & $\widetilde{L}, L$ & ($\tilde{\nu}_L$ $\tilde{e}_L$) & ($\nu_L$ $e_L$) & ($\textbf{1}$, $\textbf{2}$, $-1$, $\mathcal{Q}'_L$) \\ 
      	 (3 families) & $\bar{\nu}$ & $\tilde{\nu}^*_R$ & $\bar{\nu}_R$ & ($\bar{\textbf{1}}$, $\textbf{1}$, $0$, $\mathcal{Q}'_\nu$) \\ 
      	  & $\bar{e}$ & $\tilde{e}^*_R$ & $\bar{e}_R$ & ($\bar{\textbf{1}}$, $\textbf{1}$, $2$, $\mathcal{Q}'_e$) \\ \hline 
    Higgs, higgsinos  & $H_u$ & ($H^+_u$ $H^0_u$) & ($\widetilde{H}^+_u$ $\widetilde{H}^0_u$) & ($\textbf{1}$, $\textbf{2}$, $1$, $\mathcal{Q}'_{H_u}$) \\ 
      		      & $H_d$ & ($H^0_d$ $H^-_d$) & ($\widetilde{H}^0_d$ $\widetilde{H}^-_d$) & ($\textbf{1}$, $\textbf{2}$, $-1$, $\mathcal{Q}'_{H_d}$) \\
      		      & $\mathbf{S}$ & $S$ & $\widetilde{S}$ & ($\textbf{1}$, $\textbf{1}$, $0$, $\mathcal{Q}'_S$) \\ \hline \hline
      \multicolumn{5}{c}{\textbf{Gauge supermultiplets}} \\ \hline \hline
      \multicolumn{2}{c}{\textbf{Name}} & \textbf{spin 1/2} & \textbf{spin 1} & $\mathbf{SU(3)_c, SU(2)_L, U(1)_Y, U(1)'}$ \\ \hline 
      \multicolumn{2}{c}{gluinos, gluons} & $\widetilde{G}^a$ & $G^a$ & ($\textbf{8}$, $\textbf{1}$, 0, 0) \\ \hline
      \multicolumn{2}{c}{winos, $W$'s} & $\widetilde{W}^\pm$ $\widetilde{W}^3$ & $W^\pm$ $W^3$ & ($\textbf{1}$, $\textbf{3}$, 0, 0) \\ \hline
      \multicolumn{2}{c}{bino, $B$} & $\widetilde{B}$ & $B$ & ($\textbf{1}$, $\textbf{1}$, 0, 0) \\ \hline
      \multicolumn{2}{c}{bino', $B'$} & $\widetilde{B'}$ & $B'$ & ($\textbf{1}$, $\textbf{1}$, 0, 0) \\ \hline \hline
\end{tabular*}
\caption{\label{tab:UMSSM}UMSSM supermultiplets and their gauge properties. The $U(1)'$ charges are given in table~\ref{tab:Ucharge}.}
\end{center}
\end{table}

\begin{table}[!htb]
\begin{center}
\begin{tabular}{cccccccccc}
       \hline \hline
       & $\mathcal{Q}'_Q$ & $\mathcal{Q}'_u$ & $\mathcal{Q}'_d$ & $\mathcal{Q}'_L$ & $\mathcal{Q}'_\nu$ & $\mathcal{Q}'_e$ & $\mathcal{Q}'_{H_u}$ & $\mathcal{Q}'_{H_d}$ & $\mathcal{Q}'_S$ \\ \hline \hline
      $\sqrt{40}\mathcal{Q}'_{\chi}$ & $-1$ & $-1$ & $3$ & $3$ & $-5$ & $-1$ & $2$ & $-2$ & $0$ \\ 
      $\sqrt{24}\mathcal{Q}'_{\psi}$ & $1$ & $1$ & $1$ & $1$ & $1$ & $1$ & $-2$ & $-2$ & $4$ \\
      $2\sqrt{15}\mathcal{Q}'_{\eta}$ & $-2$ & $-2$ & $1$ & $1$ & $-5$ & $-2$ & $4$ & $1$ & $-5$ \\ \hline  \hline
\end{tabular}
\caption{\label{tab:Ucharge} $U(1)'$ charges for the matter content considered in the UMSSM.}
\end{center}
\end{table}

\section{Description of the UMSSM}
\label{sec:7.descUMSSM}

\begin{figure}[!htb]
\begin{center}
\centering
\includegraphics[width=9cm]{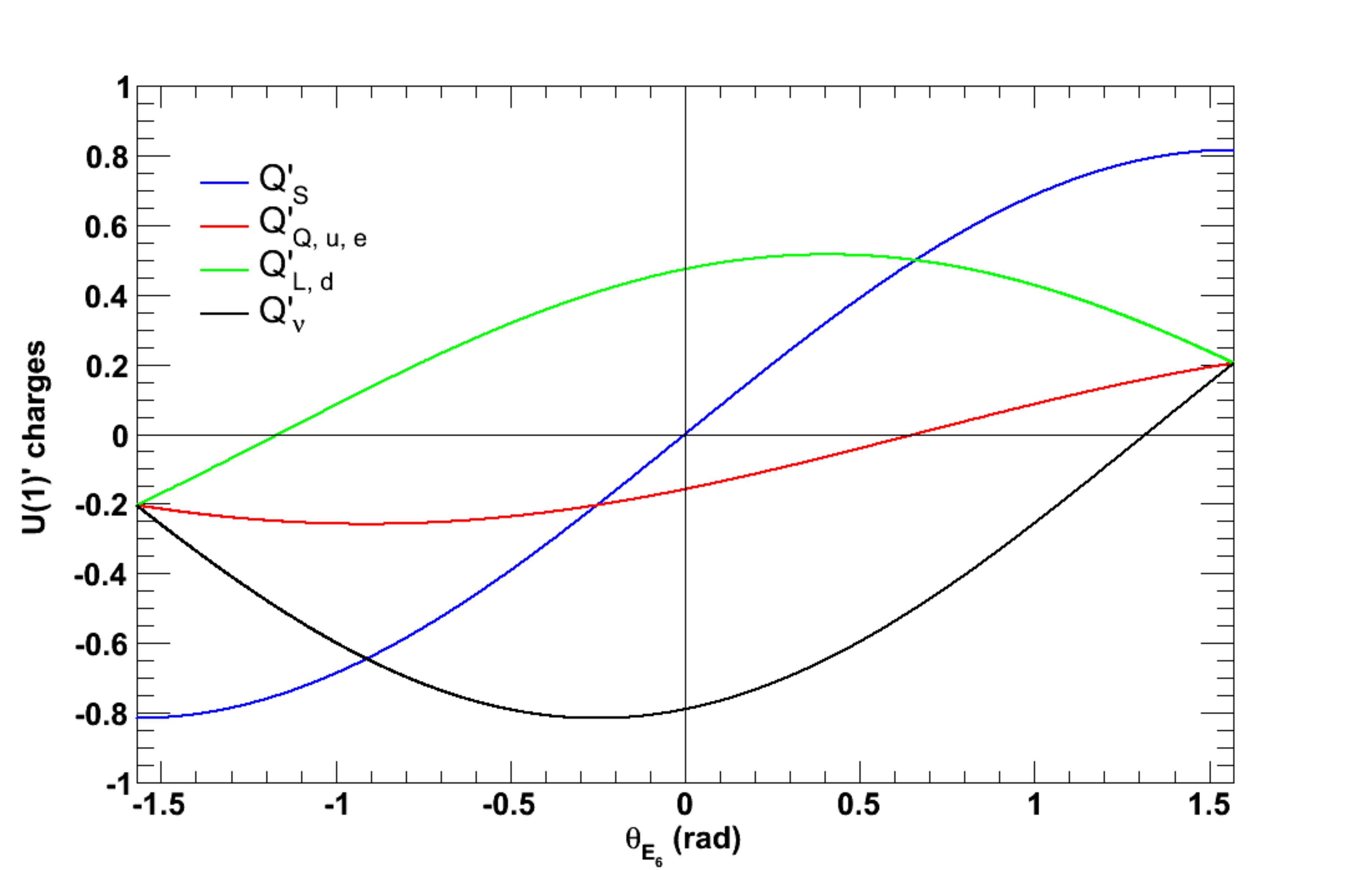}
\caption{\label{fig:7.zprime_evol}$U(1)'$ charges of some fields of the UMSSM as a function of $\te6$.}
\end{center}
\end{figure}

Table~\ref{tab:UMSSM} summarizes the supermultiplet and particle content of the UMSSM whereas table~\ref{tab:Ucharge} gives the $\mathcal{Q}'_{\chi}$ and the $\mathcal{Q}'_{\psi}$ charges for each matter field in the model as well as a combination often used in the literature corresponding to $U(1)_\eta$ with $\te6 = -\arctan \sqrt{5/3}$. In addition to the MSSM supermultiplets, the UMSSM model has a new vector supermutiplet, containing a new boson $B'$ and the corresponding Majorana gaugino $\widetilde{B'}$, and two types of chiral supermultiplets : the one with the singlet $S$ and the singlino $\widetilde{S}$ as defined in the NMSSM and another which gives three RH neutrino $\sum_{i=e,\mu,\tau} \nu_{i R}$ and their supersymmetric partners the RH sneutrinos $\sum_{i=e,\mu,\tau} \tilde{\nu}_{i R}$. Note that, as depicted with the black curve in figure~\ref{fig:7.zprime_evol}, the RH (s)neutrinos decouple completely of the rest of the UMSSM sector,\ie $\mathcal{Q}'_\nu = 0$ when $\te6=\arctan \sqrt{15}$. This choice of $U(1)'$ charge is referred in the literature to as the $U(1)_N$ choice and is used in several phenomenological studies of $E_6$ inspired models \cite{King:2005jy,Kalinowski:2008iq,Hall:2012mx,Athron:2012sq,Athron:2013ipa}. This case will not be considered in next chapter which considers the RH sneutrino as thermal DM candidate. The superpotential is the same as in the MSSM with $\mu=0$ but has an additional term involving the singlet :
\beq \label{W_UMSSM} \mathcal{W}_{\mathrm{UMSSM}} = \mathcal{W}_{\mathrm{MSSM}}|_{\mu = 0} + \l S H_u H_d + \tilde{\nu}^*_R \mathbf{y_{\nu}} \widetilde{L}H_u + \cal{O}(\mathrm{TeVs}),\eeq
where $\mathbf{y_{\nu}}$ is the Yukawa matrix associated to $\bar{\nu}$ which is diagonal in the family space and the VEV of $S$, $\langle S \rangle = \frac{v_s}{\sqrt{2}}$ breaks the $U(1)'$ symmetry and induces a $\mu$ term
\beq \mu= \l \frac{v_s}{\sqrt{2}}.\eeq
For $\te6=0$, where $\mathcal{Q}'_S=0$, this symmetry cannot be broken by the singlet field. Note also that the invariance of the superpotential under $U(1)'$ imposes a condition on the Higgs sector, namely $\mathcal{Q}'_{H_u} + \mathcal{Q}'_{H_d} + \mathcal{Q}'_S = 0$. The SSB Lagrangian of the UMSSM is
\beq \begin{split}
\mathscr{L}_{\mathrm{UMSSM}}^{\mathrm{soft}} = \, & \mathscr{L}_{\mathrm{MSSM}}^{\mathrm{soft}}|_{b = 0} - \left( \frac{1}{2} M'_1\widetilde{B'}\widetilde{B'} + \tilde{\nu}^*_R \mathbf{a_{\nu}}\widetilde{L}H_u + \textrm{h.c.} \right) - \tilde{\nu}^*_R \mathbf{m^2_{\tilde{\nu}_R}}\tilde{\nu}_R \\
& - m^2_S |S|^2 - (\l A_\l S H_uH_d + \textrm{h.c.}) + \cal{O}(\mathrm{TeVs}),
\end{split} \eeq
with the trilinear coupling $A_\l$, the $\widetilde{B'}$ mass term $M'_1$, the singlet mass term $m_S$ and the trilinear coupling and soft sneutrino mass term matrices $\mathbf{a_{\nu}}$ and $\mathbf{m^2_{\tilde{\nu}_R}}$  which are also diagonal in the family space. We now describe how each sector of the model is modified with respect to the MSSM.

\subsection{Gauge bosons}
\label{subsec:7.boson}

As we saw in table~\ref{tab:UMSSM} the UMSSM contains a new Abelian gauge boson $B'$. Since the two Higgs doublets and the Higgs singlet are charged under $U(1)'$ this new boson gets a mass through the breaking of the $SU(2)_L  \otimes  U(1)_Y$ and $U(1)'$ symmetries. We start from the interaction between the Higgs sector and the gauge sector :
\beq \mathscr{L}_{\mathrm{UMSSM}}^{\mathrm{Higgs}} = (D_\mu H_u)^\dag (D_\mu H_u) + (D_\mu H_d)^\dag (D_\mu H_d) + (D_\mu S)^* (D_\mu S) - V_{\mathrm{UMSSM}},\eeq
where $V_{\mathrm{UMSSM}}$ will be detailed in section~\ref{subsec:7.higgs} and the general covariant derivative for the SM gauge symmetry is defined in eq.~\ref{eq:1.cov}. For the $U(1)'$ symmetry we have
\beq D_\mu = \p_\mu +i g'_1 \mathcal{Q}'_X B'_\mu,\eeq
when this covariant derivative acts on a matter field $X$ and we choose the normalization of the coupling constants such that $g'_1=\sqrt{5/3} g_Y$ where $g_Y$ and $g'_1$ are the coupling constants of $U(1)_Y$ and $U(1)'$ respectively. Now if we look at the gauge boson mass terms after EWSB the corresponding expression in the context of the SM that is included in eq.~\ref{SM:boson_m_terms} is rewritten in the context of the UMSSM for each VEV as
\begin{center} \label{UMSSM:hud_m_terms} 
\beq \sum_{i=u,d} \left| \frac{i}{2} \left(g_2 \sum_{a=1}^3 \s^a W^a_\mu + \kappa_i g_Y B_\mu + 2 g'_1 \mathcal{Q}'_{H_i} B'_\mu\right) \frac{1}{\sqrt{2}}\binom{v_i \d^i_d}{v_i \d^i_u} \right|^2 + \left| i g'_1 \mathcal{Q}'_S B'_\mu \frac{v_s}{\sqrt{2}} \right|^2,\eeq 
\end{center}
where $\kappa_u=1$ and $\kappa_d=-1$. As in the SM we recover the $W$ boson mass term $M_W = \frac{1}{2}g_2 v$ with $v^2 = v_u^2 + v_d^2$. Nevertheless instead of a 2$\times$2 matrix with leads to the definition of the photon and the $Z$ boson as in the case of the SM and the (N)MSSM the remaining terms read
\begin{center} \label{UMSSM:m_terms}
\beq \sum_{i=u,d}\frac{1}{8} v_i^2  \begin{pmatrix} W^3_\mu & B_\mu & B'_\mu \end{pmatrix} 
	  \begin{pmatrix}
			g_2^2 & -g_2 g_Y & -2 \kappa_i g_2 g'_1 \mathcal{Q}'_{H_i}\\
			-g_2 g_Y & g_Y^2 & 2 \kappa_i g_Y g'_1 \mathcal{Q}'_{H_i}\\
			-2 \kappa_i g_2 g'_1 \mathcal{Q}'_{H_i} & 2 \kappa_i g_Y g'_1 \mathcal{Q}'_{H_i} & 4 \gp2 \mathcal{Q}'^2_{H_i}
	  \end{pmatrix}  \begin{pmatrix} W^{3 \mu} \\ B^\mu \\ B'^\mu \end{pmatrix} \nonumber \eeq
\beq + \frac{1}{2} \gp2 \mathcal{Q}'^2_{S} v_s^2 B'_\mu B'^\mu.\eeq
\end{center}	  
In the upper left part of the 3$\times$3 matrix we recognize the usual SM part which leads to the definition of the photon and the $Z$ boson given in eqs.~\ref{SM:ZA_m_terms} and \ref{SM:masseigen}. The remaining terms are new :
\beq \frac{1}{2} g'_1 \sqrt{g_2^2 + g_Y^2} \left(\mathcal{Q}'_{H_d} v_d^2 - \mathcal{Q}'_{H_u} v_u^2\right)Z_\mu B'^{\mu} + \frac{1}{2} \gp2\left(\mathcal{Q}'^2_{H_u} v_u^2 + \mathcal{Q}'^2_{H_d} v_d^2 + \mathcal{Q}'^2_{S} v_s^2\right)B'_\mu B'^{\mu}.\eeq
The second term gives the mass term to the new Abelian gauge boson $B'=Z'$ :
\beq \label{eq:7.MZp} M_{Z'}^2  = \gp2\left(\mathcal{Q}'^2_{H_u} v_u^2 + \mathcal{Q}'^2_{H_d} v_d^2 + \mathcal{Q}'^2_{S} v_s^2\right),\eeq
and the mixing between the two massive Abelian gauge bosons is parameterized by the term $\Delta_Z^2$ :
\beq \label{eq:7.delta} \Delta_Z^2 = \frac{1}{2} g'_1 \sqrt{g_2^2 + g_Y^2} \left(\mathcal{Q}'_{H_d} v_d^2 - \mathcal{Q}'_{H_u} v_u^2\right).\eeq
Note that the two neutral massive gauge bosons $Z$ and $Z'$ can also mix both through kinetic mixing~\cite{Cvetic:1997wu,Langacker:2008yv,Kalinowski:2008iq}. Here we simplify the situation by neglecting the kinetic mixing. The mass matrix of the $Z$ and $Z'$ boson reads
\beq \label{eq:7.matZZp} M^2_{ZZ'} =
  \left( \begin{array}{cc}
   M^2_Z & \Delta_Z^2\\
   \Delta_Z^2 & M_{Z'}^2
  \end{array}\right),\eeq
with $M_Z$ given in eq.~\ref{SM:masseigen}. The diagonalisation of the mass matrix leads to two mass eigenstates $Z_1$ and $Z_2$ :
\beq \begin{split}
Z_1 & = \cos\azz Z + \sin\azz Z',\\
Z_2& = - \sin\azz Z + \cos\azz Z',
\end{split} \eeq
where the mixing angle $\azz$ is defined as
\beq \label{eq:7.sinazz} \sin 2 \azz =\frac{2 \Delta_Z^2}{M^2_{Z_1} - M^2_{Z_2}}, \eeq
and the masses of the physical fields are
\beq M^2_{Z_1,Z_2}=\frac{1}{2} \left( M^2_Z + M_{Z'}^2 \mp \sqrt{\left(M^2_{Z}+M_{Z'}^2\right)^2+4\Delta_Z^4}  \right).\eeq
The mixing angle is constrained from precise measurements of $Z$ properties to be of the order or smaller than $10^{-3}$ radians~\cite{Leike:1991if,Erler:2009jh}, the new gauge boson $Z_2$ will therefore have approximately the same properties as the $Z'$. The input parameters are the physical masses, $M_{Z_1}=91.1876$~GeV (see table~\ref{tab:SM}), $M_{Z_2}$ and the mixing angle $\azz$.
Then the mass of the $Z$ and $Z'$ states are computed using
\beq \label{eq:7.MZZp}\begin{split}
  M^2_{Z} & = M^2_{Z_1}\cos^2 \azz + M^2_{Z_2}\sin^2 \azz, \\
  M^2_{Z'} & = M^2_{Z_1}\sin^2 \azz + M^2_{Z_2}\cos^2 \azz.
\end{split} \eeq
From these together with the coupling constants, we extract both the value of $\tan\beta=v_u/v_d$ and the 
 value of $v_s$. For the latter we use eqs~\ref{eq:7.MZp} and \ref{eq:7.MZZp}. Using eqs.~\ref{eq:7.delta} and \ref{eq:7.sinazz} the ratio of the VEVs of the two Higgs doublets is obtained with
\beq \label{eq:7.c2b} \cos^2\beta = \frac{1}{\mathcal{Q}'_{H_d} + \mathcal{Q}'_{H_u}}\left( \frac{\sin 2 \azz(M^2_{Z_1} - M^2_{Z_2})}{ v^2 g'_1 \sqrt{g_Y^2 + g_2^2}} + \mathcal{Q}'_{H_u}\right).\eeq
For each $U(1)'$ model, the value of $\tan\beta$ can be strongly constrained as a consequence of the requirement $\cos^2\beta>0$. For example for the $U(1)_\psi$ case with $\sin \azz>0$ and $M_{Z_2} > M_{Z_1}$, the value of $\tan\beta$ has to be below 1. This is because for this $\te6$ choice we have
\beq \Delta_Z^2  = g'_1 \sqrt{\frac{g_Y^2 + g_2^2}{24}} (\tan^2\beta-1) v^2_d < 0.\eeq
One might think that small values of $\tan\beta$ are problematic for the Higgs boson mass, however as we will see below additional corrections to the light Higgs boson mass can bring it at an experimentally preferred value.

\subsection{Higgs sector}
\label{subsec:7.higgs}

The Higgs sector of the UMSSM consists of three CP-even Higgs bosons $h_i, i \in \{1,2,3\}$, two charged Higgs bosons $H^\pm$ and one CP-odd Higgs boson $A^0$. As explained in the beginning of this chapter the reason for having only one CP-odd field is that the pseudoscalar part of $S$ is absorbed by the new gauge boson $Z'$. The tree-level mass terms are obtained after minimization of the potential $V_{\mathrm{UMSSM}}$ that reads
\beq V_{\mathrm{UMSSM}} = V_{\mathrm{MSSM}}|_{\mu,b = 0} + V_{\mathrm{UMSSM}}^F + V_{\mathrm{UMSSM}}^D + V_{\mathrm{UMSSM}}^{\mathrm{soft}} \eeq
where
\beq \begin{split}
V_{\mathrm{UMSSM}}^F & = |\l H_u\cdot H_d|^2 + |\l S|^2 \left(|H_d|^2+|H_u|^2 \right), \\
V_{\mathrm{UMSSM}}^D & = \frac{\gp2}{2}\left(\mathcal{Q}'_{H_d} |H_d|^2 + \mathcal{Q}'_{H_u} |H_u|^2 + \mathcal{Q}'_S |S|^2\right)^2,\\
V_{\mathrm{UMSSM}}^{\mathrm{soft}} & = m_s^{2}|S|^2 + \left( \l A_\l S H_u\cdot H_d + h.c. \right).
\label{eq:7.potential}
\end{split} \eeq
where $H_u \cdot H_d$ is defined as $H_u \cdot H_d = \e_{ij} H_u^i H_d^j$, with the sum over $i,j \in \{1,2\}$ and $ V_{\mathrm{MSSM}}$ is defined in eq.~\ref{eq:3.potential}. We expand the Higgs fields as
\begin{align}
H_d^0 & = \frac{1}{\sqrt{2}} \left( v_d + \phi_d + i \varphi_d \right),\\
H_u^0 & = \frac{1}{\sqrt{2}} \left( v_u + \phi_u + i \varphi_u \right),\\
S     & = \frac{1}{\sqrt{2}} \left( v_s + \s + i \xi \right),
\end{align} 
while the charged Higgs sector is the same as that of the MSSM :
\beq
H_d^- = -\cos\beta G_{W^-} + \sin\beta H^-, \quad
H_u^+ = \sin\beta G_{W^+} + \cos\beta H^+,
\eeq
with $G_{X}$ the Goldstone boson associated to $X$.\\

Following \cite{Barger:2006dh}, the minimization conditions of $V_{\mathrm{UMSSM}}$ are
\beq \begin{split} \label{eq:minLO}
\left(m_{H_d}\right)^2 = & - \frac{1}{2}\left[\frac{g_Y^2 + g^2_2}{4} + \mathcal{Q}'^2_{H_d} \gp2\right]v_d^2 + \frac{1}{2}\left[\frac{g_Y^2 + g^2_2}{4} - \l^2- \mathcal{Q}'_{H_d} \mathcal{Q}'_{H_u} \gp2\right]v_u^2 \\
& - \frac{1}{2}\left[\l^2 + \mathcal{Q}'_{H_d} \mathcal{Q}'_S \gp2\right]v^2_s + \frac{\l A_\l v_s v_u}{v_d \sqrt{2}},\\
\left(m_{H_u}\right)^2 = & \quad \, \frac{1}{2}\left[\frac{g_Y^2 + g^2_2}{4} - \l^2- \mathcal{Q}'_{H_d} \mathcal{Q}'_{H_u} \gp2\right]v_d^2 - \frac{1}{2}\left[\frac{g_Y^2 + g^2_2}{4} + \mathcal{Q}'^2_{H_u} \gp2\right]v_u^2 \\
& - \frac{1}{2}\left[\l^2 + \mathcal{Q}'_{H_u} \mathcal{Q}'_S \gp2\right]v^2_s + \frac{\l A_\l v_s v_d}{v_u \sqrt{2}},\\
\left(m_S\right)^2 = & - \frac{1}{2}\left[\l^2 + \mathcal{Q}'_{H_d} \mathcal{Q}'_S \gp2\right]v_d^2  - \frac{1}{2}\left[\l^2 + \mathcal{Q}'_{H_u} \mathcal{Q}'_S \gp2\right]v_u^2  - \frac{1}{2} \mathcal{Q}'^2_S \gp2 v^2_s + \frac{\l A_\l v_u v_d}{v_s \sqrt{2}},
\end{split} \eeq 
and we can write the tree-level mass-squared matrices for the CP-even $(\mathcal{M}_+^0)$ and the CP-odd $(\mathcal{M}_-^0)$ Higgs bosons in the basis $\{H^0_d, H^0_u, S\}$ using the relations
\beq
\left( {\mathcal{M}}_{+}^{0}\right)_{ij}= \left. \frac{\partial ^{2}V_{\mathrm{UMSSM}}}{\partial \phi _{i}\partial \phi_{j}} \right|_0 ,\qquad \left( {\mathcal{M}}_{-}^{0}\right) _{ij}= \left. \frac{\partial ^{2}V_{\mathrm{UMSSM}}}{\partial \varphi_{i}\partial  \varphi _{j}} \right|_0,
\eeq
where $(\phi_1,\phi_2,\phi_3) \equiv (\phi_d,\phi_u,\s)$ and $(\varphi_1,\varphi_2,\varphi_3) \equiv (\varphi_d,\varphi_u,\xi)$.
For the neutral CP-even Higgs bosons the relations are
\beq \begin{split} \label{treeUMSSM_even}
\left({\mathcal{M}_{+}^0}\right)_{11} & =  \left[\frac{({g_Y}^2 + g^2_2)^2}{4} +  \mathcal{Q}'^2_{H_d} \gp2\right] v_d^2 + \frac{\l A_\l v_s v_u}{\sqrt{2} v_d},\\
\left({\mathcal{M}_{+}^0}\right)_{12} & = -\left[\frac{({g_Y}^2 + g^2_2)^2}{4} - \l^{2} - \mathcal{Q}'_{H_d} \mathcal{Q}'_{H_u} \gp2\right] v_u v_d - \frac{\l A_\l v_s}{\sqrt{2}},\\
\left({\mathcal{M}_{+}^0}\right)_{13} & =  \left[\l^{2} + \mathcal{Q}'_{H_d} \mathcal{Q}'_{S} \gp2\right] v_s v_d - \frac{\l A_\l v_u}{\sqrt{2}},\\
\left({\mathcal{M}_{+}^0}\right)_{22} & =  \left[\frac{({g_Y}^2 + g^2_2)^2}{4} + \mathcal{Q}'^2_{H_u} \gp2\right] v_u^2 + \frac{\l A_\l v_s v_d}{\sqrt{2} v_u},\\
\left({\mathcal{M}_{+}^0}\right)_{23} & =  \left[\l^{2} + \mathcal{Q}'_{H_u} \mathcal{Q}'_S \gp2\right] v_s v_u - \frac{\l A_\l v_d}{\sqrt{2}},\\
\left({\mathcal{M}_{+}^0}\right)_{33} & = \mathcal{Q}'^2_S \gp2 v^2_s + \frac{\l A_\l v_u v_d}{v_s \sqrt{2}}.\\
\end{split} \eeq
For the CP-odd sector we have
\beq \begin{split}\label{treeUMSSM_odd}
\left({\mathcal{M}_{-}^0}\right) & = \frac{\l A_\l}{\sqrt{2}} 
  \begin{pmatrix}
    \frac{v_s v_u}{v_d} & v_s & v_u\\
    v_s & \frac{v_s v_d}{v_u} & v_d\\
    v_u & v_d & \frac{v_u v_d}{v_s}
  \end{pmatrix}, 
\end{split} \eeq
which leads to
\beq\label{treeUMSSM_odd2} 
\left(m_{A^0}\right)^2 = \mu A_\l \left( \frac{v_d}{v_u} + \frac{v_u}{v_d} + \frac{v_u v_d}{v^2_s}\right).
\eeq
The charged Higgs mass at tree-level reads
\beq\label{treeUMSSM_charged}
\left(m_{H^\pm}\right)^2 = M^2_W + \frac{\l A_\l \sqrt{2}}{\sin 2 \beta} v_s - \frac{\l ^2}{2} v^2.  
\eeq
Then it is important to consider radiative corrections in the Higgs sector. To do that two method will be used. In chapter~\ref{chapter:RHsneu} we will just add radiative corrections in the unitary gauge through a Coleman-Weinberg potential \cite{Coleman:1973jx}. In chapter~\ref{chapter:B_Higgs_UMSSM} an effective Lagrangian approach following the method used in the NMSSM \cite{Belanger:2005kh} will be considered. 
The CP-even squared mass matrix is diagonalised by a 3$\times$3 unitary matrix $\mathbf{Z_h}$ which gives the mass eigenstates (ordered in mass) by $(h_1,h_2,h_3)^T=\mathbf{Z_h}^T (\phi_d,\phi_u,\s)^T$. As we saw there is only one pseudoscalar Higgs boson. The relation between the Goldstone of the $Z$ and $Z'$ bosons, $A^0$ and the basis of the Higgs fields is given by a 3$\times$3 unitary matrix $\mathbf{Z_A}$ : $(G_Z,G_{Z'},A^0)^T=\mathbf{Z_A}^T (\varphi_d,\varphi_u,\xi)^T$.

Because of the constraints on the new boson that we will presented in section~\ref{sec:7.new}, we typically find a Higgs spectrum which consists on a SM-like CP-even Higgs boson, a heavy mostly doublet scalar which is almost degenerate with the pseudoscalar and the charged Higgs, and finally a predominantly singlet scalar. Note that the mass of the pure singlet $m_S \approx g'_1  \mathcal{Q}'^2_{S} v_s$ and is therefore close to that of $M_{Z_2}$ when $v_s \gg v_u,v_d$. The hierarchy in the mass of the heavy doublet and the singlet depends on the parameters of the model. For large values of $A_\l$ and $\l$ (and therefore $\mu$) the mass of the heavy Higgs doublet increases and can exceed that of the singlet. Furthermore the Higgs mixing increases, however the singlet component of the light state is usually not large. Although the lightest scalar Higgs boson is usually SM-like, it can be significantly heavier than in the MSSM. Indeed the upper bound on the Higgs boson mass receives two types of additional contributions as compared to the MSSM, one proportional to $A_\l$ that is also found in the NMSSM and which enters in the SSB part of the potential in eq~\ref{eq:7.potential}, and the other from the introdution of the new symmetry which implies new $D$-terms. The upper bound on the lightest Higgs boson mass is thus raised to ${\cal O}(170)$~GeV~\cite{Barger:2006dh}.

\subsection{Sfermions}

The important new feature in the sfermion sector is that the $U(1)'$ symmetry induces some new $D$-term contributions to the sfermion masses which are added in the diagonal part of the sfermion matrix defined in eq.~\ref{eq:3.sfer}. These contributions read
\beq \label{eq:7.sferU} \Delta_f= \frac{1}{2} \gp2 \mathcal{Q}'_f \left( \mathcal{Q}'_{H_d} v_d^2 + \mathcal{Q}'_{H_u} v_u^2 + \mathcal{Q}'_S v_s^2 \right),\eeq 
where $f \in \{Q, u, d, L, e, \nu\}$.

The new $D$-term contribution can completely dominate the sfermion mass, especially for large values of $v_s$. Those are found in particular when $\te6 \approx 0$. The $D$-term contribution can induce negative corrections to the mass, so that light sfermions can be found even when the soft masses are set to 2 TeV. For $\te6>0$, $\Delta_\nu < \Delta_{Q,u,e}<\Delta_{L,d} $ so that a universal soft mass term for the sfermions at the weak scale naturally leads to a
RH sneutrino as the lightest sfermion, which is useful if we want to consider it as the LSP. Furthermore the NLSP will be the RH slepton or the stop if a large mixing decreases the mass of the lightest $\tilde{t}$. On the other hand for $\te6<0$, $\Delta_\nu >\Delta_{Q,u,e} > \Delta_{L,d}$ so the sneutrino cannot be the LSP with universal soft sfermion masses at the weak scale. However in general one expects non-universality in sfermion masses at the weak scale even if universality is imposed when the model is embedded in a GUT scale model. In particular the RH sleptons, whose RGEs are driven only by $U(1)$ couplings have the smallest soft terms at the weak scale. Therefore as long as $\Delta_\nu$ is not much larger than for other sfermions, it is still natural that the RH sneutrino be the lightest sfermion. Then in chapter~\ref{chapter:RHsneu} we analyse this sort of LSP as a thermal DM candidate.

\subsection{Neutralinos}

In the UMSSM the neutralino mass matrix in the basis $(\widetilde{B},\widetilde{W}^3,\widetilde{H}_d,\widetilde{H_u},\widetilde{S},\widetilde{B'})$ reads
\beq
\mathbf{M_{\chi^0}}=
  \left( \begin{array}{cccccc}
   M_1 	&   0    & -M_Z \cb \sw  &    M_Z \sb \sw  &   0   &   0\\  
   0 	&   M_2    & M_Z \cb \cw  & -M_Z \sb \cw  &    0   &   0\\  
   -M_Z \cb \sw &  M_Z \cb \cw  & 0 &  -\mu  & -\l \frac{v_u}{\sqrt{2}} & \mathcal{Q}'_{H_d} g'_1 v_d   \\  
    M_Z \sb \sw & -M_Z \sb \cw  & -\mu &   0 & -\l \frac{v_d}{\sqrt{2}} & \mathcal{Q}'_{H_u} g'_1 v_u \\  
   0 &  0 & -\l \frac{v_u}{\sqrt{2}} & -\l \frac{v_d}{\sqrt{2}} & 0 & \mathcal{Q}'_S g'_1 v_s\\  
   0 &  0 & \mathcal{Q}'_{H_d} g'_1 v_d  & \mathcal{Q}'_{H_u} g'_1 v_u  & \mathcal{Q}'_S g'_1 v_s  &  M'_1\\  
  \end{array}\right) \,.
\label{eq:sfermion}
\eeq
Diagonalisation by a 6$\times$6 unitary matrix $\mathbf{Z_n}$ leads to the neutralino mass eigenstates :
\beq \chi^0_i = Z_{n ij} \psi^0_j\textrm{,} \qquad i\textrm{,}j \in \{1,2,3,4,5,6\}.\eeq
Several studies were done to analyse the neutralino sector in the UMSSM \cite{Hesselbach:2001ri,Choi:2006fz}. As in the NMSSM $\widetilde{S}$ LSP was studied \cite{Franke:2001nx,Suematsu:2005bc,Nakamura:2006ht}, and more generally the neutralino LSP as a viable DM candidated in the UMSSM was obtained \cite{deCarlos:1997yv,Barger:2004bz}. The chargino sector is identical to that of the (N)MSSM.

\section{Constraints on the UMSSM}
\label{sec:7.new}

As in the case of the other supersymmetric models, the UMSSM is constrained using several observables : DM observables, Higgs boson mass bounds, constraints on sparticle masses as well as low energy observables. The first ones, focused here to the relic density and the Direct Detection of DM, will be used in the next two chapters while sparticle constraints will be divided into two categories : LEP constraints on sparticles will be always taken into account, whereas most LHC constraints will be avoided assuming that coloured sparticle masses are above current LHC limits on simplified models. For low energy observables, chapter~\ref{chapter:RHsneu} will only consider the $\Delta M_s$ and $\Delta M_d$ observables. In chapter~\ref{chapter:B_Higgs_UMSSM}, the effects of a more larger set of low energy observables in the context of the UMSSM, namely several $B$-physics observable, the anomalous magnetic moment of the muon and $\Delta \rho$ will be discussed. However the UMSSM scenario has also a specific feature that leads to important constraints. This feature, a new gauge symmetry, is reflected in a new gauge boson whose experimental and observational implications are now dicussed.

\subsection{Collider constraints on the $Z'$}
\label{sec.7:constraint}

The main constraint that will be used in the next two chapters comes from the direct collider searches for a $Z'$ resonance. The new Abelian gauge boson can be produced directly in $pp$ and $p\bar{p}$ collisions and is searched primarily using leptonic decay modes since they provide very clean signals without much background. The results from the LHC experiments are now pushing the mass of this particle well above the TeVs. 
\begin{figure}[!htb]
\begin{center}
\centering
\subfloat[]{\includegraphics[width=8cm,height=5cm]{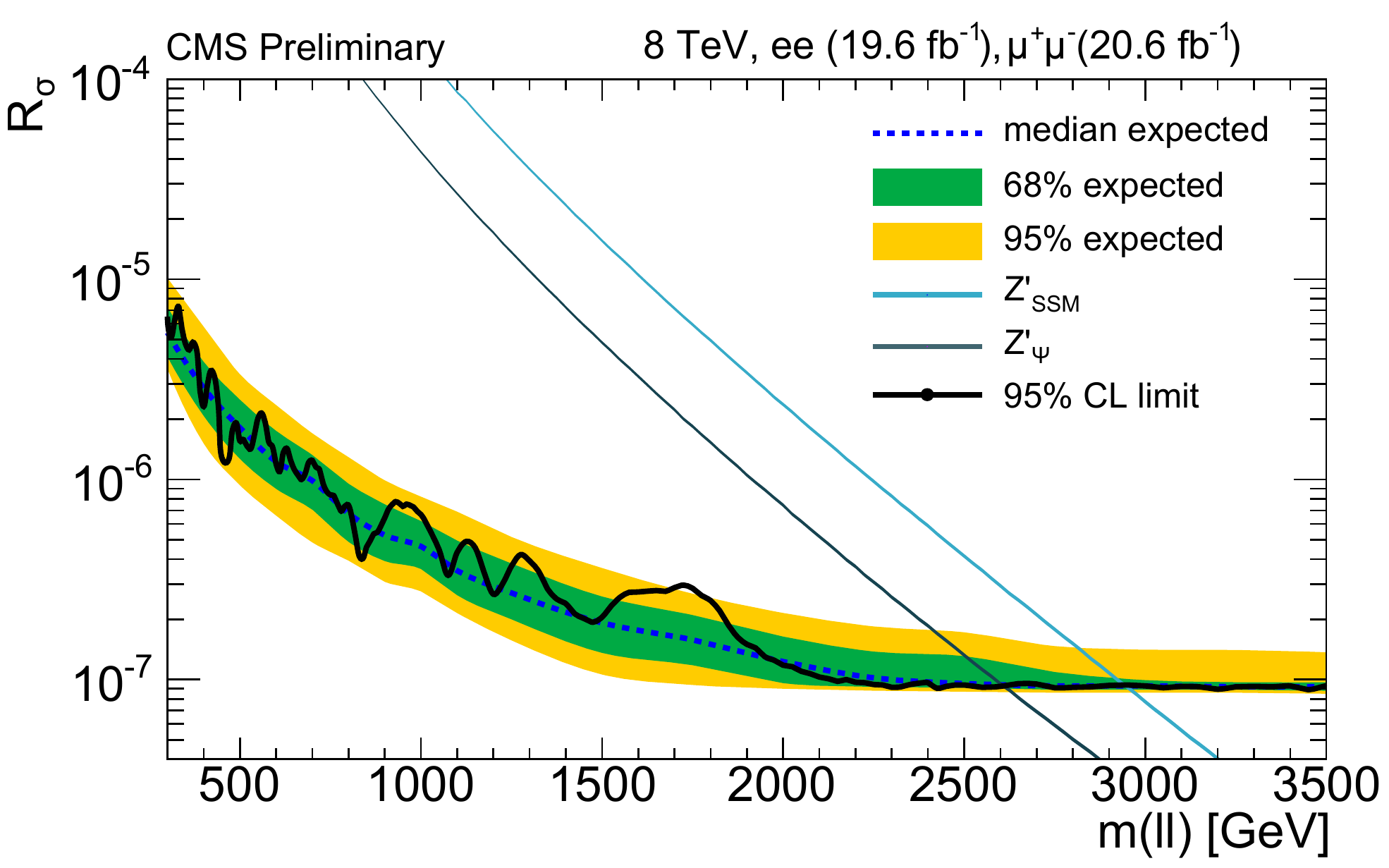}}
\subfloat[]{\includegraphics[width=8cm,height=5cm]{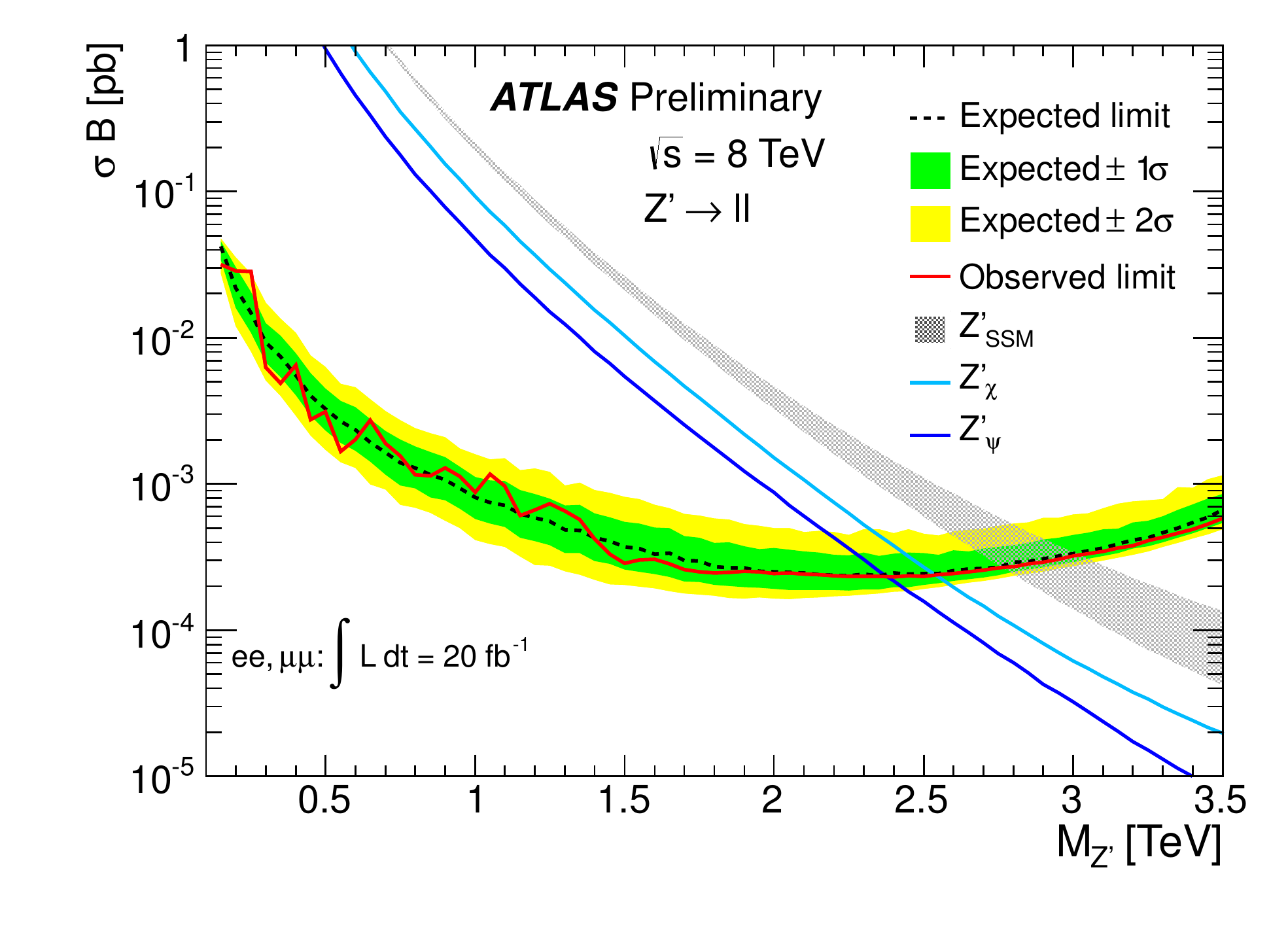}}
\caption[Limits on the mass of the $Z'$ boson obtained by the CMS and ATLAS collaborations in 2013.]{\label{zpLHC}Limits on the $Z'$ boson. Panel (a) presents the upper limit obtained by CMS on the ratio $R_\s$ of cross section times branching ratio into lepton pairs for a new gauge boson as compared to the same quantity for the $Z$ boson. Two cases are considered for the $Z'$ : the sequential SM and the $U(1)_{\psi}$ model. Panel (b) shows the upper limit obtained by ATLAS on the cross section times branching ratio into lepton pairs for a new gauge boson : the sequential SM as well as the $U(1)_{\psi}$ and the $U(1)_{\chi}$ models. Plots obtained respectively from \cite{CMS:Zprime} and \cite{ATLAS:2013jma}.}
\end{center}
\end{figure}

Figure~\ref{zpLHC} shows the current best limits obtained by the CMS and the ATLAS collaborations on specific $U(1)'$ models. CMS limits were derived using $20.6$~fb$^{-1}$ ($19.6$~fb$^{-1}$) of data in the $\mu^+\mu^-(e^+e^-)$ channels and ATLAS bounds were obtained using $20$~fb$^{-1}$ of data in the $\ell^+\ell^- = \mu^+\mu^-, \, e^+e^-$ channels, both at $\sqrt{s} = 8$~TeV. If we focus on the $U(1)_{\psi}$ model, CMS is able to exclude\footnote{Here we denote $Z'$ as the experimentally sought gauge boson; in our UMSSM model it corresponds to the $Z_2$ mass eigenstate.} a $Z'$ below $2.6$~TeV at 2$\s$ and ATLAS gives a 2$\s$ lower bound of $2.38$~TeV. This puts really stringent constraints on our model. Nevertheless these bounds were derived assuming the $Z'$ decays only into SM particles. In our case, the $Z_2$ can also decay into supersymmetric particles, into RH neutrinos and into Higgs bosons, thus reducing the branching ratio into leptons. The limits on the $Z_2$ mass are therefore weakened~\cite{Gunion:1986ky,Gherghetta:1996yr,Chang:2011be}. To take this effect into account we determined the modified leptonic branching ratio for each point in our studies and have re-derived the corresponding limits. It is also possible to significantly weaken these limits by considering through kinetic mixing a leptophobic $Z'$ \cite{Babu:1996vt,An:2012va,Cheung:2012pq}.

\subsection{Other constraints on $Z'$ physics}
\label{sec.7:constraint2}

There are other methods to constrain the $Z'$ mass and the mixing with the $Z$ boson. For instance from the modification of low energy neutral currents, experiments examining atomic parity violation due to $Z'$ interaction between the nucleus and the electrons could be competitive with collider limits for some $U(1)'$ models \cite{Rosner:2001ck,Diener:2011jt}.

EW precision measurements provide extremely relevant constraints in the context of the UMSSM. We saw in section~\ref{subsec:7.boson} that the $\azz$ mixing angle between $Z$ and $Z'$ has to be very small. The mixing between these two bosons gives also contributions to the $\Delta \rho$ parameter which can be non-negligible \cite{Babu:1996vt}.

BBN can provide the most stringent limits on the $Z'$ mass when the RH neutrinos are charged under $U(1)'$, as considered here. Actually such RH Dirac neutrinos can be produced by $Z'$ interactions prior to BBN leading to a faster expansion rate of the Universe and to too much $^4\mathrm{He}$. For example in \cite{Barger:2003zh} the lower bound on the $Z'$ mass, in some scenarios and assuming a given effective number of neutrinos, can go up to 4 TeV. With the latest results of the Planck experiment these limits could be updated.

For more informations on $Z'$ physics see \cite{ZpHuntersGuide}.

\chapter{The Right-Handed sneutrino as thermal Dark Matter in the UMSSM}
\label{chapter:RHsneu}

\minitoc\vspace{1cm}
\newpage

This chapter is largely based on the article \cite{Belanger:2011rs} with a few added details. Since this work was done two years ago a lot of new experimental data become available (Higgs boson discovery, $Z'$ searches, DM observables). A brief discussion on the impact of the new results will be given after the presentation of the original study. This would actually need a new study, including more constraints.

\section{Introduction}

The MSSM contains two neutral weakly interacting particles that could be DM candidates, the neutralino and the LH sneutrino. While the neutralino has been extensively studied and remains one of the favourite DM candidates~\cite{Goldberg:1983nd,Ellis:1983ew,Roszkowski:2004jc,Baer:2009bu,Ellis:2010kf}, the LH sneutrino faces severe problems. The sneutrino coupling to the $Z$ boson induces a cross section for elastic scattering off nuclei that can exceed the experimental limit by several orders of magnitude~\cite{Falk:1994es}. In particular the bounds from CDMS~\cite{Ahmed:2009zw} or XENON100~\cite{Aprile:2010um} cannot be satisfied even for a LH sneutrino mass above 1 TeV. Furthermore the sneutrino annihilation rate is usually too rapid to provide enough DM~\cite{Falk:1994es}.

The observation of neutrino oscillations indicative of massive neutrinos gives a natural motivation for adding a RH neutrino to the SM fields as we saw in section~\ref{subsec:1.lepton}. Extending the MSSM with RH neutrinos and their supersymmetric partners provides then an alternate DM candidate, the RH sneutrino (RHSN). The smallness of the neutrino masses is usually explained by introducing Majorana mass terms and making use of the see-saw mechanism. The natural scale for the RH neutrinos is generally around $10^{12}$~GeV so that RH neutrinos are too heavy to play a direct role in physics below the TeV scale and so are their supersymmetric partners. Note however that the inverse see-saw mechanism proposes scenarios with RH Majorana neutrinos at the TeV scale~\cite{Mohapatra:1986aw,Mohapatra:1986bd}. We will not consider these scenarios. 
It is also possible to generate neutrino masses through Dirac mass terms at the expense of introducing some large hierarchy among the fermions. In this case the supersymmetric partners of the neutrinos are expected to be, as for other
sfermions, at the SUSY breaking scale,\ie around or below 1 TeV. In this framework the RH sneutrino can be the LSP. This is the scenario we will consider here. 

To make a RHSN LSP a viable DM candidate requires special conditions. The RH sneutrino being sterile under SM gauge interactions cannot be brought into thermal equilibrium\footnote{Note however that non thermal mechanisms can make a mostly sterile sneutrino a good DM candidate~\cite{Asaka:2005cn,Asaka:2006fs,Gopalakrishna:2006kr,Yaguna:2008mi}.}. Nevertheless several proposals for sneutrino DM have emerged including mixed sneutrinos~\cite{ArkaniHamed:2000bq,Borzumati:2000mc,Hooper:2004dc,MarchRussell:2009aq,Kumar:2009sf,Belanger:2010cd,Dumont:2012ee},
RHSN in models with Dirac mass terms that result from the decay of thermal equilibrium MSSM particles~\cite{Asaka:2005cn,Asaka:2006fs}, DM from a RH sneutrino condensate~\cite{McDonald:2007mr}, sneutrinos in inverse see-saw models~\cite{Hall:1997ah,Arina:2008bb,An:2011uq,Bandyopadhyay:2011qm} or RHSN in extensions of the MSSM~\cite{Arina:2007tm,Long:2007ev,Arina:2008yh,Cerdeno:2008ep,Cerdeno:2009dv,Demir:2009kc,Cerdeno:2011qv,Kang:2011wb}. The RHSN Dark Matter was also considered in hybrid inflationnary models~\cite{Deppisch:2008bp}. Extending the gauge group provides another alternative as the RHSN can reach thermal equilibrium because it couples to new vector and/or scalar fields~\cite{Lee:2007mt,Langacker:2008yv,Bandyopadhyay:2011qm}. For example an additional $U(1)'$ gauge symmetry provides new couplings of the sneutrino with the $Z'$ as well as to new scalar fields. In this framework annihilation of pairs of RH sneutrino  
can be efficient enough to obtain $\Omega h^2\approx 0.1$. The annihilation is specially enhanced when the particle exchanged in the $s$-channel is near resonance. Furthermore the elastic scattering cross section of the RHSN is naturally suppressed by several orders of magnitude as compared to the MSSM sneutrino as on the one hand the couplings to the EW scale particles (the $Z$ and the light Higgs boson) are strongly suppressed and on the other hand the $Z'$ exchange is suppressed because its mass is above the TeV scale\footnote{Note that their signatures in cosmic rays~\cite{Allahverdi:2009ae,Demir:2009kc} or neutrino telescopes~\cite{Allahverdi:2009se} were also explored.}. This is the framework that we will study here, within the UMSSM model described in chapter~\ref{chapter:UMSSM}.

In the UMSSM, the DM candidate can either be the neutralino or the sneutrino. The neutralino was investigated in~\cite{Kalinowski:2008iq,deCarlos:1997yv,Barger:2007nv}. We rather concentrate on the RHSN LSP which is a possible thermal DM candidate when it is charged under $U(1)'$; in this case the RH neutrinos are Dirac neutrinos \cite{King:2005jy}.

In~\cite{Lee:2007mt} sneutrinos annihilation through $Z'$ exchange and $\widetilde{B'}$ $t$-channnel exchange into neutrinos were considered while the interactions through the Higgs sector were included in~\cite{Allahverdi:2009ae,Demir:2009kc}. We extend those analyses by including all possible annihilation and coannihilation channels and by performing a complete exploration of the parameter space allowing for different choices of the $U'(1)$ symmetry within the context of an $E_6$  inspired model. Furthermore we take into account early 2011 LHC results on the $Z'$, on supersymmetric particles and on the Higgs sector. Insisting on having a sneutrino LSP, to complement previous studies that had concentrated on the neutralino case, will lead to some constraint on the parameter space. In particular the parameters $\mu$ and $M_1,M_2$ have to be larger than the sneutrino mass in order to avoid an higgsino or a gaugino LSP. We will generically consider only cases with coloured sparticles above the TeV scale to easily avoid LHC limits on squarks. 

Within the UMSSM with a RHSN Dark Matter candidate, we found a variety of annihilation channels for the sneutrino, with a predominance of annihilations near a resonance. The main annihilation processes can be classified as
\begin{itemize}
\item{} Annihilation near a light Higgs boson resonance
\item{} Annihilation near a heavy Higgs boson resonance
\item{} Annihilation near a $Z_2$ resonance 
\item{} Annihilation into $W/Z$ pairs through Higgs boson exchange
\item{} Coannihilation with the NLSP, here the NLSP can be neutralino, chargino, charged slepton or any other sfermion.
\end{itemize}

Note that annihilation into neutrinos can occur either through $Z_2$ exchange or through $\widetilde{B'},\widetilde{S}$ exchange when the latter are near the mass of the LSP. The neutrino channels are however never the dominant ones. Finally
annihilation of sneutrinos into pairs of new gauge bosons could also contribute, we will not consider this channel here as
it is relevant only for sneutrino heavier than the $Z'$, that is above the TeV scale. For such a heavy sneutrino to be the LSP means that all soft parameters are also above the TeV scale. These configurations for supersymmetric particles have a restricted parameter space. For example, very large values of $\mu$ often lead to a light Higgs boson mass below 114~GeV. Furthermore when the sneutrino is heavier than the $Z'$, the neutralino can become the LSP. We therefore choose rather to concentrate on the subTeV scale sneutrinos. For each scenario we also include an analysis of the DD rate as the SI cross section of sneutrinos on nucleons turns out to pose severe constraints on a whole class of $U(1)'$ models. 

We will first describe the main particle physics constraints used in this work. Then the contributions to the relic abundance and the DD rate will be presented. Finally we will see the results for two sample $U(1)'$ models which have significant differences between their sneutrino DM candidate, the $U(1)_\psi$ and $U(1)_\eta$ models whose $U(1)'$ charges are given in table~\ref{tab:Ucharge}, as well as results from a global analysis of the parameter space, and we will conclude.

\section{Constraints imposed}
\label{sec:8.constraint}

The main constraints on the model arise from the gauge boson and Higgs sector. Since coloured supersymmetric particles do not play a direct role in sneutrino annihilation, we will generally assume that they are above the TeV scale thus evading the LHC constraints~\cite{Chatrchyan:2011zy,Aad:2011ib}.

We used the ATLAS exclusion limits published in~\cite{Collaboration:2011dca} obtained with an integrated luminosity of ${\cal L}=1.01(1.21) \, \mathrm{fb}^{-1}$ in the $e^+e^-(\mu^+\mu^-)$ channels. These limits are extracted using different $U(1)'$ models. In the two models we will consider for the phenomenological analysis the bounds\footnote{Note the impressive improved limits now obtained at the LHC as shown in figure~\ref{zpLHC}.} are similar, with $\mzp >1.49$~TeV for $U(1)_\psi$ and $\mzp >1.54$~TeV for $U(1)_\eta$. As explained in section~\ref{sec.7:constraint} these bounds were derived assuming the $Z_2$ decays only into SM particles and we then re-derived the limits since we have new decays of the $Z_2$ in the UMSSM. 

Because of EW precision measurements we will typically set for the study of some benchmark points the mixing angle $\azz$ to $|\azz|\leq0.001$.

The Higgs sector was constrained in the early 2011 as follows. The SM-like Higgs boson was constrained by LEP to be above 114 GeV~\cite{Barate:2003sz} and by LHC searches to be below 144 GeV~\cite{CMS:2011mpa}. These limits can be relaxed when the Higgs couplings to gauge bosons are reduced due to the mixing with the singlet. In the parameter space explored, the singlet component is small and the limit is not modified significantly. The upper limit can also be relaxed if the Higgs boson has a large branching fraction into invisible particles. The relic density constraint imposes that light sneutrinos have a mass very near $m_{h_1}/2$. Despite a phase space suppression factor, the contribution of the invisible mode to the light Higgs boson decay can in some cases reach 90\% for light sneutrinos, thus relaxing the constraint on the lightest Higgs boson. When imposing a limit on the light Higgs boson mass, we have folded in the effect of the invisible decay modes therefore allowing in a few cases a light Higgs boson heavier than 144 GeV. Note that we introduced radiative corrections to the Higgs boson masses to get relevant SM-like Higgs boson masses. 

Observables in the $B$ sector provide powerful constraints on the supersymmetric parameter space assuming MFV. The measured values in 2011 of the mass differences of $B$ mesons, $\Delta M_s$ and $\Delta M_d$, are respectively $17.63\pm 0.11 \, \mathrm{ps}^{-1}$~\cite{Aaij:2011qx} and $0.507\pm 0.004 \, \mathrm{ps}^{-1}$~\cite{Barberio:2006bi}, somewhat below SM predictions given in \cite{Domingo:2007dx} :
\begin{eqnarray}
\Delta M_s^\mathrm{SM}&=&20.5\pm 3.1 \, \mathrm{ps}^{-1},\nonumber\\
\Delta M_d^\mathrm{SM}&=&0.59\pm 0.19 \, \mathrm{ps}^{-1}.
\end{eqnarray}
Additional supersymmetric contributions of the same sign as the SM ones are therefore strongly constrained, even though there are large uncertainties in the SM prediction mainly due to the CKM matrix elements and hadronic parameters. Supersymmetric contributions include box diagrams with for instance charged Higgs/quark or squark/chargino loops~\cite{Bertolini:1990if} as well as double penguin diagrams with a neutral Higgs boson exchange. The latter give a significant contribution for large values of $\tan\beta$ and were not included in our analysis. The former are important at small values of $\tan\beta$ which are often found in the UMSSM. In particular the charged Higgs/quark box diagram contribution adds to the SM contribution inducing too large values for $\Delta M_s$. One of the dominant contributions, for small values of $\tan\beta$, is proportional to $ x \ln x (\cot^4\beta)$, where $x=m^2_t/m^2_{H^\pm}$. This observable thus constrains severely scenarios with $\tan\beta<1$. The computation of the mass difference is adapted from the routine provided in \NTools\cite{Domingo:2007dx}. We have also used the same estimate for the theoretical uncertainties. Note that we will more carefully look at this mass differences calculation in chapter~\ref{chapter:B_Higgs_UMSSM}.

Other observables such as $\bsg$, $\bsmu$ or $\btau$ are known to receive large supersymmetric contributions when $\tan\beta$ is large, the heavy Higgs doublet is light and/or the squarks are light. The scenario we will study have heavy squarks,  Higgs doublets above several hundred GeV's and features small to moderate values of $\tan\beta$, we therefore have not included these constraints. However it could be interesting to check if this choice is safe. As for $\Delta M_s$ and $\Delta M_d$, we will consider these observables in chapter~\ref{chapter:B_Higgs_UMSSM}.

\section{Relic abundance of sneutrinos}

\begin{figure}[h]
\begin{center}
\includegraphics[scale=0.8]{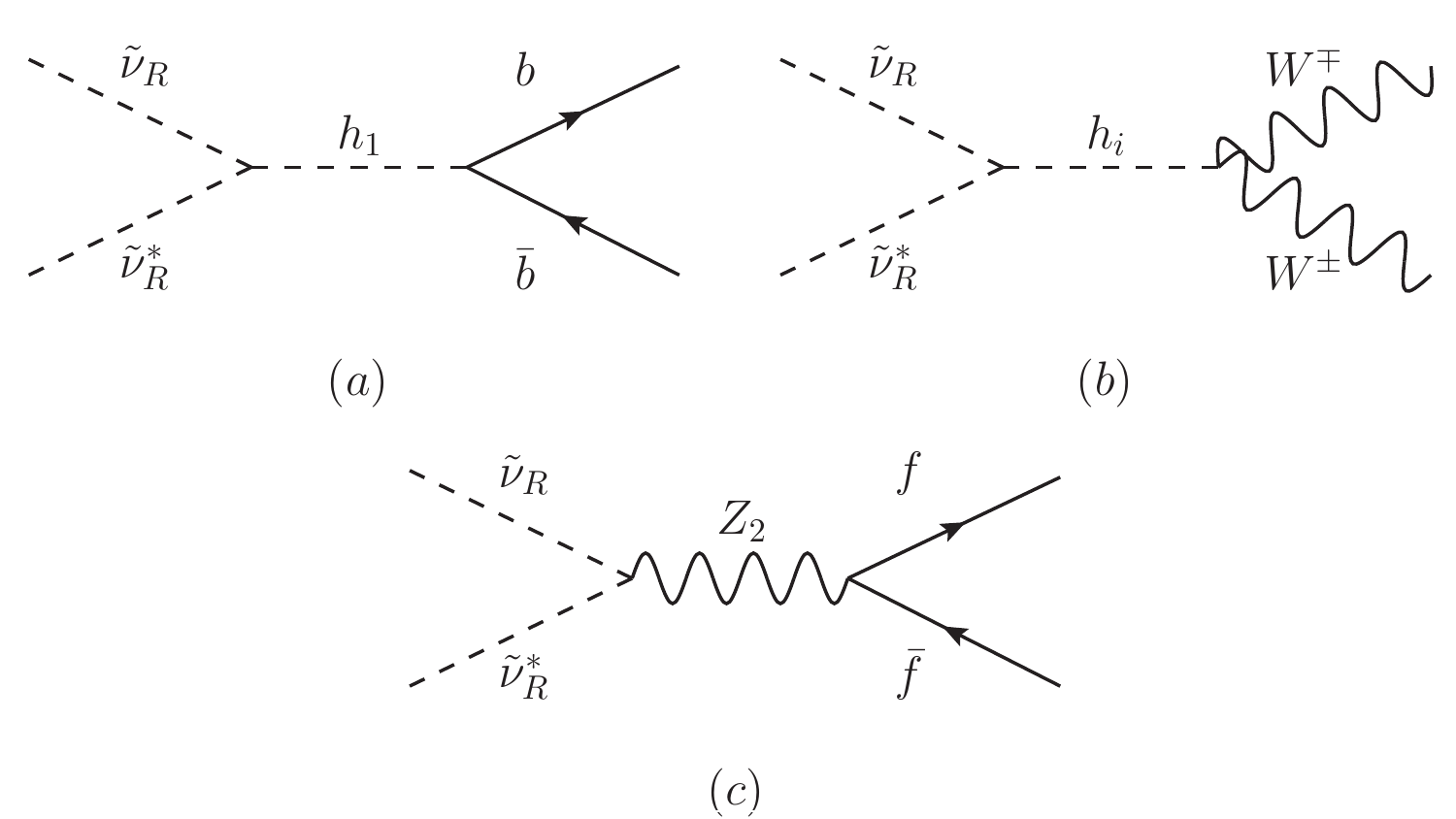}
\caption[Main annihilation processes for RH sneutrinos.]{\label{RD_UMSSM}Main annihilation processes for RH sneutrinos, with $i \in \{1,2\}$ for panel (b).}
\end{center}
\end{figure}

The RH sneutrino couples very weakly to the MSSM particles, its annihilation is therefore primarily through the particles of the extended sector, the new vector boson $Z_2$ (figure~\ref{RD_UMSSM}c) and new scalars (panel (a) and (b) of figure~\ref{RD_UMSSM}). The coupling of the RHSN to $Z_2$ is directly proportional to its $U(1)'$ charge, $\mathcal{Q}'_\nu$ :
\beq g_{Z_2\lsp\lsp^*}= g_1' \mathcal{Q}'_\nu\cos\azz,\eeq
with $\cos\azz\approx 1$. Annihilation of sneutrinos becomes efficient when $\mlsp\approx M_{Z_2}/2$ since the $s$-channel processes can benefit from resonance enhancement.

Sneutrino annihilation also become efficient when $m_{\tilde\nu_R}\approx  m_{h_i}/2$ where $h_i$ can be any neutral CP even Higgs field. The coupling of the sneutrino to neutral scalars reads
\beq
g_{h_i\lsp\lsp^*} = -\gp2 \mathcal{Q}'_\nu\left[ v_d \mathcal{Q}'_{H_d} Z_{h i1}  + v_u \mathcal{Q}'_{H_u} Z_{h i2} + v_s \mathcal{Q}'_S Z_{h i3} \right].\label{eq:hnunu} \eeq
Since $v_s\gg v$, the largest coupling will be to the predominantly singlet Higgs, for which $Z_{hi3}\approx 1$. Typically the singlet Higgs boson is the one that has a mass close to $Z_2$, resonant Higgs annihilation will therefore occur for roughly the same sneutrino mass as the resonant annihilation through $Z_2$. The light Higgs boson is dominantly doublet, nevertheless its coupling to sneutrinos receives contributions from all three terms in eq.~\ref{eq:hnunu} since $v_s\gg v$.
This coupling is generally sufficient to have a large cross section enhancement near $\mlsp\approx m_{h_1}/2$. In some cases efficient annihilation can occur away from the resonance, for example annihilation into $W/Z$ pairs through light Higgs boson exchange or, for heavier RHSN's, into light Higgs boson pairs or $t\bar{t}$ pairs through singlet exchange. Note that the sneutrino coupling to the lightest Higgs boson depends on $\l$ ($\mu$), an important parameter to determine the mixing of Higgs bosons. For some choice of parameters, the couplings of the mostly doublet $h_1$ to the RHSN can be strongly suppressed, not allowing a large enough enhancement on the annihilation cross section.  Specific examples will be presented in section~\ref{sec:8.results}. 

Sneutrinos can also annihilate in neutrino pairs through t-channel exchange of $\widetilde{B'}$ in addition to the usual $Z_2$  contributions. This process contributes mostly for light $\widetilde{B'}$ and is never the dominant channel. Finally it is also possible to reduce the abundance of sneutrinos through coannihilation processes, this occur when the masses of the NLSP and the LSP are within a few GeV's. Coannihilation can occur with neutralinos, charginos or other sfermions. Typically, because there is only weak couplings of the RH sneutrino to the rest of the MSSM particles, coannihilation processes involve self-annihilation of the NLSP's and/or NLSP/Next-to-NLSP annihilation. The neutralino NLSP will decay via $\chi^0_1 \ra \lsp \nu_R$ with a typical lifetime of $10^{-19} - 10^{-17}$s except when it is almost degenerate with the sneutrino LSP which can lead to an increase of lifetime. The NLSP decay will always occur much before BBN and will not spoil its predictions.

\section{Direct Detection}
\label{sec:8.DD}

\begin{figure}[h]
\begin{center}
\includegraphics[scale=0.8]{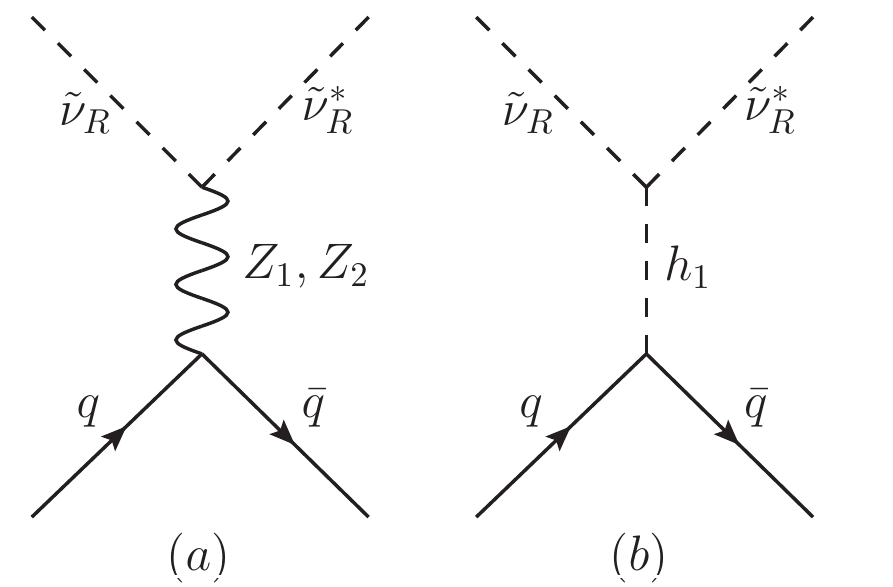}
\caption{\label{DD_UMSSM}Main scattering processes of the RH sneutrino.}
\end{center}
\end{figure}

The cross section for scattering of sneutrinos on nucleons will be purely Spin Independent as the sneutrino is a scalar particle. This process receives contributions from only two types of diagrams : exchange of gauge or Higgs bosons. The gauge boson contribution (figure~\ref{DD_UMSSM}a) depends on the vectorial coupling of the fermions $f$ to $Z_{1,2}$, with $Q_V^f= I_3^f -2\mathcal{Q}^f \sin^2\theta_W$ for $Z$ and $Q_V'^f=\mathcal{Q}'_f-\mathcal{Q}'_{\bar{f}}$ for $Z'$ where $f \in \{Q,L\}$ and $\bar{f} \in \{u,d,\nu,e\}$.
 The total gauge boson contribution to the SI cross section on point-like nuclei reads
\begin{eqnarray}
\sigma^{\rm SI,\,Z_{1,2}}_{\lsp N} = \frac{\mu_{\lsp N}^2}{\pi} (g'_1 \mathcal{Q}'_\nu)^2  
\left[\left(y(1-4\sin^2\theta_W)+ y'\right) Z + (-y+ 2y')(A-Z)\right]^2,
\label{eq:ddz}
\end{eqnarray}
where here $A$ and $Z$ are respectively the number of nucleons and protons inside the nucleus and
\beq \begin{split}
y & =  \frac{g_Y \sin \alpha_{Z} \cos \alpha_{Z}}{4 \sin\theta_W}\left(\frac{1}{M_{Z_2}^2}- \frac{1}{M_{Z_1}^2}\right),\nonumber\\
y'& = -\frac{g'_1}{2} Q_V'^d \left(\frac{\sin^2\alpha_Z}{M_{Z_1}^2} + \frac{\cos^2\alpha_Z}{M_{Z_2}^2}\right),
\end{split} \eeq
with $Q_V'^d=-4/\sqrt{40} \cos\te6$. The calculation is given in appendix~\ref{appen:sigmaZ1Z2}. The contribution in $1/M^2_{Z_2}$ in $y$ and the one proportional to $\sin^2\azz$ in $y'$ are always suppressed. From table~\ref{tab:Ucharge} one sees that all fermions have purely axial-vector couplings to $Z_\psi$ and that u-quarks also have purely axial vector couplings to $Z_\chi$. Therefore the $Z'$ contribution is solely dependent on its coupling to d-quarks, hence the term in $Q_V'^d$ in $y'$. This contribution is expected to be twice as large for neutrons than for protons, see eq.~\ref{eq:ddz}. 
In the model $U(1)_\psi$, $y'=0$ and the cross section will depend on the $Z_1$ exchange contribution. In this case the amplitude for protons is suppressed by a factor $1 -4\sin^2\theta_W$ as compared to that for neutrons. Furthermore the $Z_1$ contribution proceeds through the $Z'$ component so that the cross section is proportional to $\sin^2\azz$. When $\cos\te6 \neq 0$ the term in $y'$ usually dominates by about one order of magnitude for a TeV scale $Z_2$ and the mixing angle
$\azz=10^{-3}$. In these scenarios only a  weak dependence on $\alpha_Z$ is expected, the largest cross sections are expected for $\te6\simeq 0$ corresponding to the maximal value of $Q_V'^d$ as can be seen in figure~\ref{fig:7.zprime_evol} with $\mathcal{Q}'_Q$ in red and $\mathcal{Q}'_d$ in green.

The Higgs contribution leads to a cross section 
\begin{equation}
\sigma^{{\rm SI,}\,h_i}_{\lsp N} =\frac{\mu_{\tilde\nu_R N}^2 }{4\pi}  \sum_i \frac{g_{{h_i\lsp\lsp^*}}^2 } {m_{h_i}^4 m_{\lsp}^2} \left( (A-Z) \sum_q g_{{h_i}q\bar{q}} f_q^n m_n + Z \sum_q g_{{h_i}q\bar{q}} f_q^p m_p \right)^2,
\label{eq:ddh}
\end{equation}
where $g_{{h_i}q\bar{q}}=-eZ_{hi1}/(2 M_W s_W c_\beta)$ is the Higgs coupling to quarks after the quark mass has been factored out and $g_{h_i\lsp\lsp^*}$ is defined in eq.~\ref{eq:hnunu}. 

Because of the dependence on the Higgs boson mass and the fact that the Higgs coupling to quarks goes only through the doublet component, the lightest Higgs doublet generally gives the dominant contribution (figure~\ref{DD_UMSSM}b). Note that the Higgs contribution is inversely proportional to the square of the sneutrino mass and is therefore expected to dominate at low masses since the $Z_{1,2}$ contribution depends only weakly on the sneutrino mass through the effective mass, $\mu_{\tilde\nu_R N}$. Furthermore the Higgs contribution is roughly the same for neutrons and protons. The quark coefficients of the nucleon were taken to be the default values in \micro2.4 (with $f^p_{u,d,s}= 0.033,0.023,0.26 \; {\rm and} \; f^n_{u,d,s}=0.042,0018,0.26$)~\cite{Belanger:2008sj}. There can be large uncertainties in these coefficients and recent lattice results lead to lower values for the $f$'s, in particular with the parameters given in section~\ref{subsubsec:3.DD}. However the quark coefficients will have a significant impact only when the Higgs contribution is dominant, that is for sneutrinos below $\approx 100$~GeV.

The total SI cross section on point-like nucleus is obtained after averaging over the sneutrino and anti-sneutrino. Note that the interference between the $Z_{1,2}$ and Higgs boson exchange diagrams have opposite signs for $\tilde{\nu}_R N$ and $\tilde{\nu}_R^*N$. Here we assume equal numbers of sneutrino and anti-sneutrinos so that the total cross section is the sum of the $Z_{1,2}$ and Higgs boson contributions. To take into account the fact that the proton and neutron contributions are not necessarily equal, we compute the normalized cross-section on a point-like nucleus,  
\begin{equation}
\sigma^{\rm SI}_{\lsp N}= \frac{4 \mu_{\tilde\nu_R N}^2}{\pi}\frac{\left( Z f_p+ (A-Z)f_n\right)^2}{A^2 },
\end{equation}
where the average over $\lsp$ and $\lsp^*$ is assumed implicitly. This cross-section can be directly compared with the limits on $\sigma^{\rm SI}_{\chi \, p}$ that is extracted from each experiment.

We first implemented the model in \lhep\cite{Semenov:2008jy,Semenov:2010qt} in both unitary and Feynman gauges and checked gauge invariance for a large number of processes at tree-level. We then introduced radiative corrections to the
Higgs boson masses in the unitary gauge. The dominant radiative corrections from top quarks and stops are listed in appendix~\ref{appen:radcor}. This code then produces the model file for \chep\cite{Pukhov:2004ca,Belyaev:2012qa} and \micro which thus allows to study the DM properties in the UMSSM, namely the DD cross section as well as the relic abundance of RHSN. To check the gauge invariance involved two other sectors of the UMSSM : the gauge fixing sector and the ghost sector. They are detailed in appendix~\ref{appen:gold_and_ghost}.

\section{Results}
\label{sec:8.results}
The free parameters of the model are
$\mlsp, \mu, M_1,M_2,M'_1,A_\l, M_{Z_2}, \te6, \azz$, as well as all masses and trilinear couplings of sleptons and squarks. 
To reduce the number of free parameters we fix those that do not belong to the sneutrino or neutralino/Higgs sectors. We thus fix the soft masses of sleptons and squarks to 2 TeV and we take $A_f=0$ and $A_t=1$~TeV. We furthermore assume $M_1=M_2/2=M_3/6$ as dictated by universality at the GUT scale. Therefore the free parameter space considered is 
\begin{center}
$\{\mlsp, \mu, M_1, M_1', A_\l, M_{Z_2}, \azz, \te6\}.$
\end{center}
We first consider specific choices of $\te6$ before letting it be a free parameter. 

\subsection{The case of the $U(1)_\psi$ model}
\label{sec:psi}

As we have discussed above, we expect the relic abundance of the sneutrino DM to be generally too high. The only processes to bring the abundance within the range preferred by the WMAP experiment,\ie $\Om h^2 = 0.1123 \pm 0.0035$ \cite{Komatsu:2010fb}, will be to enhance the annihilation through a resonance effect or make use of coannihilation. We will first describe the typical behaviour of $\Omega h^2$ as a function of the $\lsp$ mass for fixed values of the free parameters. Once $\te6$ is fixed the parameters that have a strong influence on DM are $M_{Z_2}$ the mass of the new gauge boson, $A_\l$ that influence the Higgs spectrum, as well as $\mu$ and $M_1$ that determine the region where the sneutrino is LSP through their influence on the neutralino mass. The parameter $\mu$ also influences the Higgs mixing matrix and therefore the coupling of the Higgs bosons to sneutrinos. To limit the number of free parameters we will first fix $M_1=M_1'=1$~TeV and rather modify the property of the neutralino NLSP by varying $\mu$. Furthermore allowing to have   $\mu<M_1$ will cover the case of the higgsino NLSP which has much higher annihilation cross section than the bino and is  therefore more likely to play an important role in coannihilation channels.

\subsubsection{A case study  with $M_{Z_2}=1.6$~TeV}
\label{sec:psi:mz}

We first consider the case where the new gauge boson is just above the LHC exclusion limit, that is $M_{Z_2}=1.6$~TeV and we fix $\mu=1$~TeV, $\azz=-0.001$ and $A_\l=1.5$~TeV. For this choice of parameters, the mass spectrum is such that $h_2$ is dominantly singlet and has roughly the same mass as the $Z_2$ while the other heavy Higgs bosons ($h_3,A^0,H^\pm$) are nearly degenerate with a mass around 1.97~TeV. The light Higgs boson has a mass of 136.6~GeV and is safely above the LEP limit despite the small value of $\tan\beta=2.05$. The lightest neutralino has a mass of 957~GeV and is a mixture of bino and higgsino. The sneutrino is therefore the DM candidate when its mass ranges from a few GeV's up to the mass of the lightest neutralino. The value of $\Om h^2$ lies below the WMAP upper bound in two regions, the first when $60 \, {\rm GeV}\le m_{\lsp} \le 69 \, {\rm GeV}$ the second when $734 \, {\rm GeV}\le m_{\lsp} \le 805 \, {\rm GeV}$. In both regions the annihilation cross section is enhanced significantly by a resonance effect. In the first region the $h_1$ exchange is enhanced and the preferred annihilation channel is into $b{\bar b}$ pairs while in the second region the $h_2$ and $Z_2$ exchanges are enhanced with a dominant contribution from $h_2$. The preferred annihilation channels are into $W^+W^-,t\bar{t},h_1h_1,Z_1Z_1$ pairs, the dominant decay modes of $h_2$. Note that the decays into gauge bosons proceed through the small doublet component of $h_2$. For larger masses of the sneutrino, coannihilation with the lightest neutralino can take place. Coannihilation processes involving pairs of neutralinos annihilating through the heavy pseudoscalar Higgs boson can decrease $\Om h^2$ such that the WMAP upper bound is satisfied. This occurs when the NLSP-LSP mass difference drops below 60 GeV.

\begin{figure}[!htb]
\begin{center}
\centering
\subfloat[]{\includegraphics[width=8.25cm,height=6.5cm]{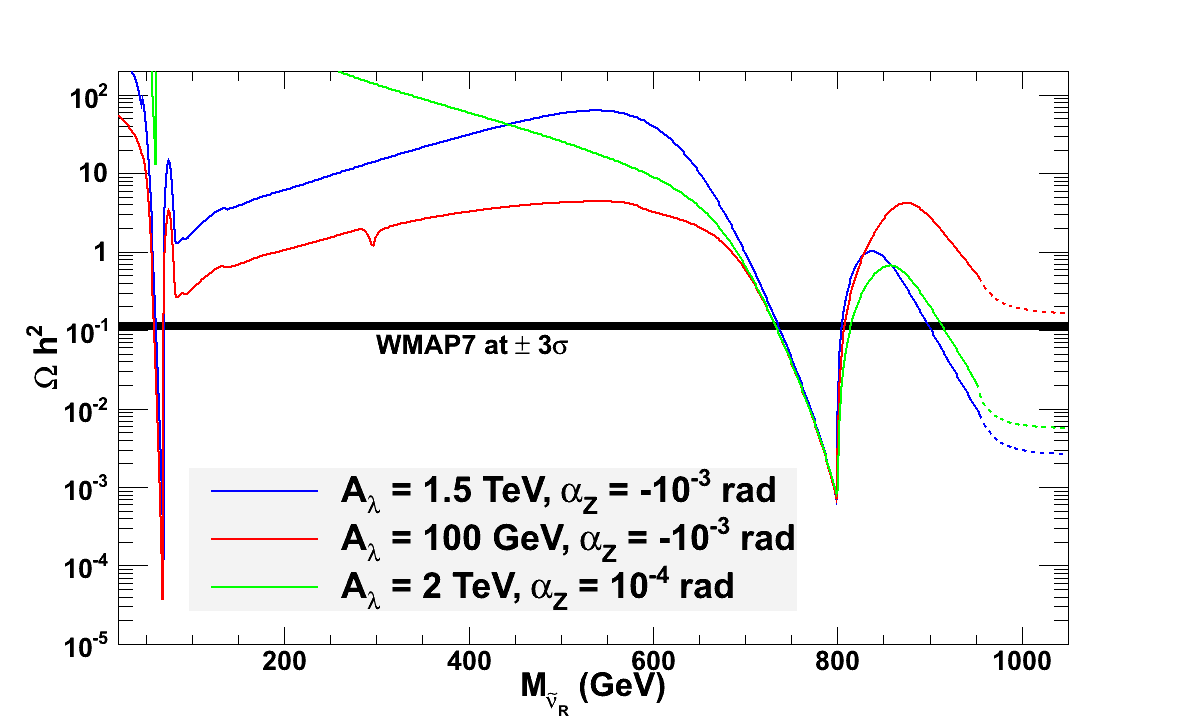}} 
\subfloat[]{\includegraphics[width=8.25cm,height=6.5cm]{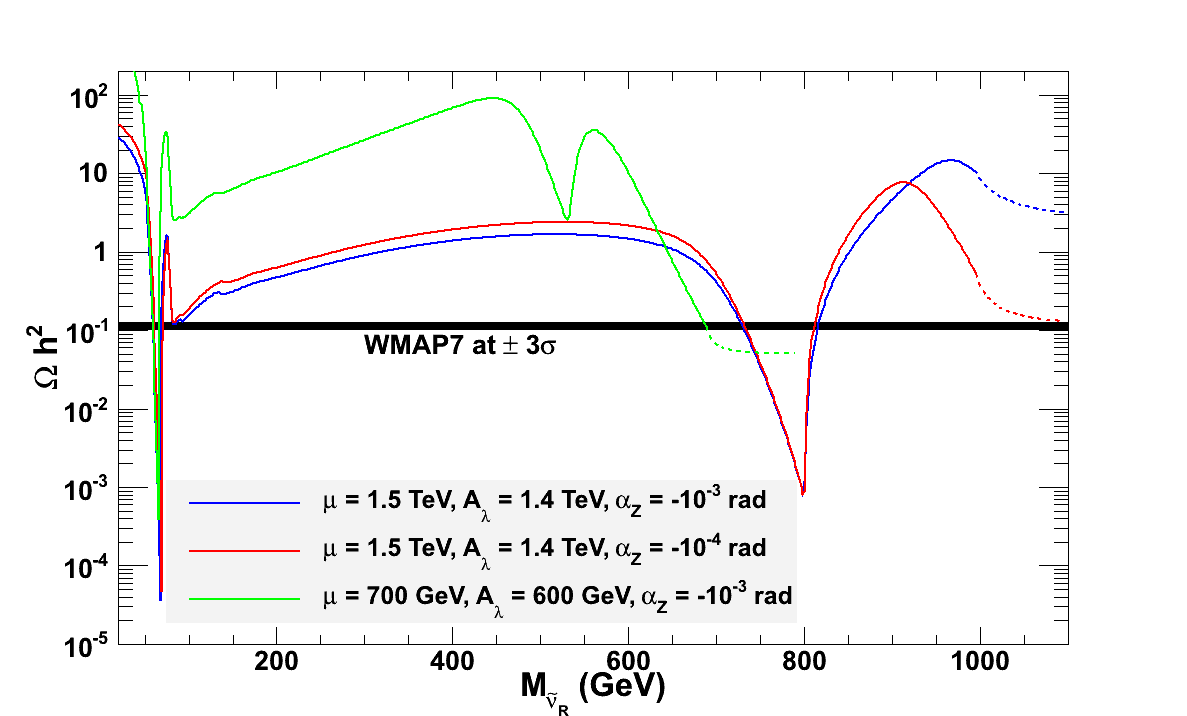}} 
\caption[$\Omega h^2$ as a function of the LSP mass for $\mzp=1.6$~TeV. Panel (a) has $\mu=1$~TeV, $(A_\lambda \,\mathrm{(TeV)}, \azz)=(1.5,-10^{-3})$, $(0.1,-10^{-3})$, $(2.,10^{-4})$ whereas panel (b) corresponds to $\mu=1.5$~TeV, $(A_\lambda \,\mathrm{(TeV)}, \azz)=(1.4,-10^{-3})$, $(1.4,-10^{-4})$ and $\mu=0.7$~TeV, $(A_\lambda \,\mathrm{(TeV)}, \azz)=(0.6,-10^{-3})$.]{$\Omega h^2$ as a function of the LSP mass for $\mzp=1.6$~TeV. Panel (a) has $\mu=1$~TeV, $(A_\lambda \,\mathrm{(TeV)}, \azz)=(1.5,-10^{-3})$, $(0.1,-10^{-3})$, $(2.,10^{-4})$ whereas panel (b) corresponds to $\mu=1.5$~TeV, $(A_\lambda \,\mathrm{(TeV)}, \azz)=(1.4,-10^{-3})$, $(1.4,-10^{-4})$ and $\mu=0.7$~TeV, $(A_\lambda \,\mathrm{(TeV)}, \azz)=(0.6,-10^{-3})$. Full (dash) lines correspond to the region where the LSP is the RHSN (neutralino). }
\label{fig:omega}
\end{center}
\end{figure}

This general behavior is somewhat influenced by our choice of $A_\l$, $\mu$ and $\azz$ since these parameters influence the masses and mixings of the Higgs doublets and the couplings of sneutrinos to Higgs bosons. For example for $A_\l=100$~GeV, the mass of the heavy doublet (which is now $h_2$) decreases to 592 GeV. The $h_2/A^{0}$ exchange contributes to sneutrino annihilation into top pairs thus leading to a decrease of the value of $\Om h^2$ as one approaches the Higgs resonance,
the drop is not significant enough to bring the relic density below the WMAP upper bound, see figure~\ref{fig:omega}a. As another example consider the case $A_\l=2$~TeV and $\azz=10^{-4}$. Here the $h_1\lsp\lsp^*$ coupling is suppressed,
$\Om h^2$ becomes very large and despite a resonance effect when $m_{\lsp} \approx m_{h_1}/2$ the WMAP upper bound can never be satisfied for a sneutrino lighter than 100 GeV, see the green line in figure~\ref{fig:omega}a. 

The parameter $\mu$ induces corrections to the light Higgs mass as well as shifts in the Higgs mixing matrix. In particular increasing $\mu$ (and therefore $\lambda$) to 1.5~TeV  increases the singlet mixing in the light Higgs boson and thus the $h_1\lsp\lsp^*$ coupling. This makes annihilation processes through Higgs exchange more efficient, and gives rise to 
a new region where $\Om h^2$ is below the WMAP upper bound when the sneutrino mass is just above the $W$ pair threshold, see figure~\ref{fig:omega}b with $A_\l=1.4$~TeV. Note that in this case the lightest neutralino has a large bino component and coannihilation is not very efficient. Lowering $\mu$ to 700 GeV changes the nature of the lightest neutralino which becomes dominantly higgsino with its mass determined by $\mu$. Thus the range of masses where the sneutrino is the LSP becomes narrower. In fact the singlet Higgs boson/$Z_2$ pole annihilation region cannot be reached when the sneutrino is the LSP, the region compatible with WMAP is rather one where higgsino coannihilation dominates. For the green line in figure~\ref{fig:omega}b, one can see a significant drop in $\Om h^2$ near the $h_2$ resonance, this is however not sufficient to have efficient annihilation of the RHSN.
 
\begin{figure}[!htb]
\begin{center}
\centering
\subfloat[]{\includegraphics[width=8.25cm,height=6.5cm]{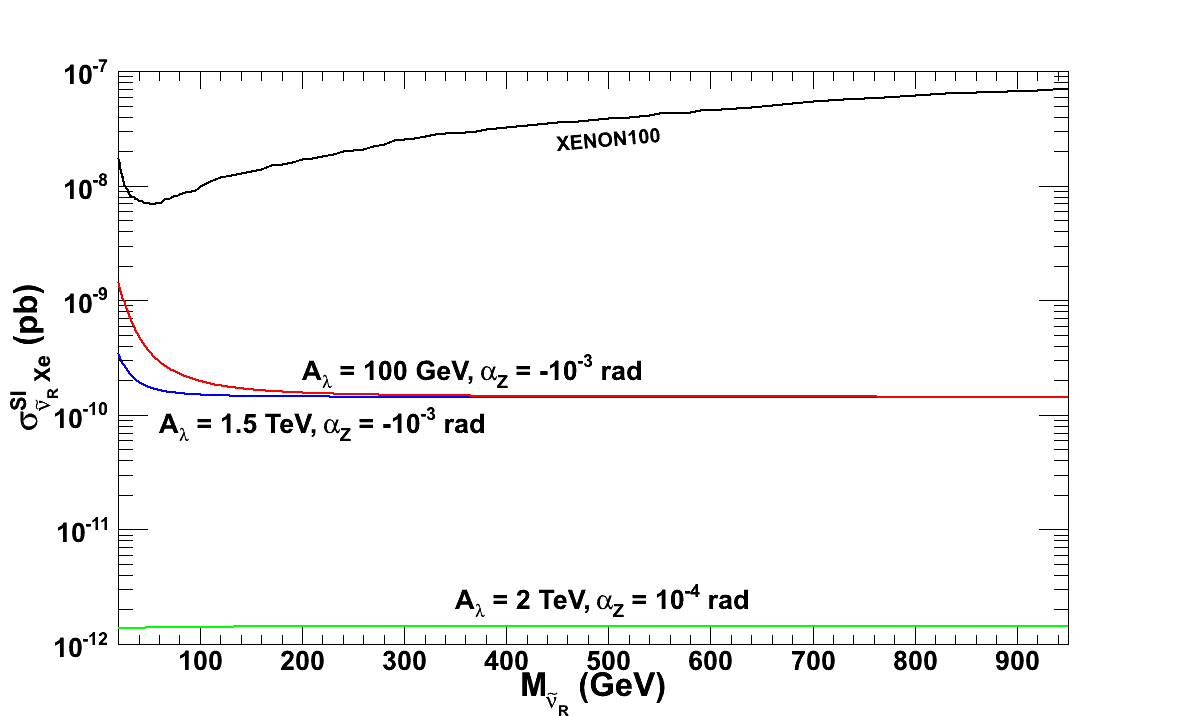}} 
\subfloat[]{\includegraphics[width=8.25cm,height=6.5cm]{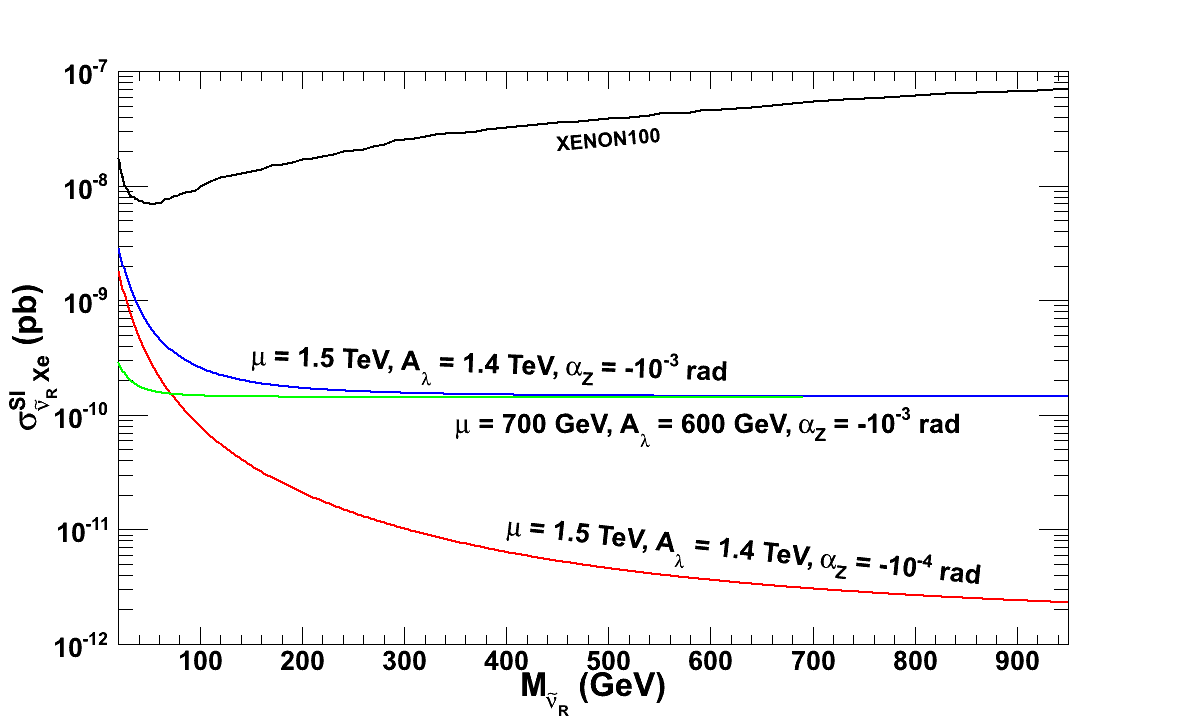}}  
\caption{$\sigma^{SI}_{\lsp Xe}$ as a function of the LSP mass for the same parameters as figure~\ref{fig:omega}. The XENON100 exclusion curve is also displayed.}
\label{fig:dd}
\end{center}
\end{figure}
 
The Direct Detection Spin Independent cross section receives contributions from both the light Higgs boson and $Z_1$ exchange.
In general for the $U(1)_\psi$ scenario, the cross-section is below the limits from XENON100~\cite{Aprile:2010um}. As we have argued above, this is because the $Z_1$ contribution is directly proportional to $\sin^2\azz$. This contribution nevertheless dominates for a mixing angle $|\azz|=10^{-3}$ except for small masses. It gives $\sigma^{SI}_{\lsp Xe}=1.4\times 10^{-10}$~pb for $\mlsp \geq 200$~GeV and is mostly independent of other input parameters, see figure~\ref{fig:dd}a,b. Furthermore the $Z_1$ contribution is much larger for neutrons than for protons. We take that into account by computing the cross section for scattering on point-like nucleons (here we use Xenon) normalized to one nucleon. For masses in the range $50-100$~GeV, the SI cross section is enhanced due to the light Higgs exchange contribution which increases at low sneutrino masses, although for this scenario the values are always below the experimental limit. The predictions are in the range $\sigma^{SI}_{\lsp Xe}=2\times 10^{-10}- 3\times 10^{-10}$~pb for $\azz=-10^{-3}$. The SI cross section drops rapidly with the mixing angle, for example for $\azz=|10^{-4}|$, $\sigma^{SI}_{\lsp Xe} \simeq 10^{-12}$~pb for $\mlsp \geq 200$~GeV, see figure~\ref{fig:dd}a. For small values of the mixing angle, the $Z_1$ contribution is suppressed and the Higgs boson contribution can become dominant even for masses of a few hundred GeV's, this however corresponds to a low overall cross section, see for example figure~\ref{fig:dd}b. The SI cross section can be further suppressed in the limit $\azz=0$ and for parameters for which the $h_1$ coupling to sneutrino is suppressed.

\subsubsection{Exploration of $U(1)_\psi$ parameter space}

After having described the general behavior of $\Omega h^2$ and of the DD rates for some choice of parameters, we now search for the region in the parameter space that are compatible with $0.1018 < \Omega h^2 <0.1228$ corresponding to the WMAP  $3\sigma$ range~\cite{Komatsu:2010fb} as well as with the DD limit of XENON100~\cite{Aprile:2010um}. We have further imposed the limits on the Higgs sector from LEP and the LHC and have taken into account the impact of the invisible decay mode of the Higgs boson. We have also used limits on sparticles from LEP as implemented in {\tt micrOMEGAs.\;}Note that since we are only considering the case of heavy sfermions, these limits as well as the LHC limits on sparticle masses do not come into play \cite{Aad:2011ib,Chatrchyan:2011zy}. Finally we have also imposed the constraints from $\Delta M_{d,s}$. We have performed random scans with $5\times 10^6$ points. The ranges for the parameters are listed in table~\ref{tab:e6}. As before we here fix all sfermion masses to 2 TeV and neglect all trilinear couplings with the exception of $A_t=1$~TeV. Note that we impose $\mu>0$ and therefore consider only positive values for $A_\l$, see eq.~\ref{treeUMSSM_odd2}.

\begin{table}[!htb]
\begin{center}
\begin{tabular}{cc}\hline \hline
\textbf{Parameter} & \textbf{Range} \\ \hline \hline
$\mlsp$ & [0, 1.5] TeV\\ 
$M_{Z_2}$ & [1.3, 3] TeV\\ 
$\mu$ & [0.1, 2] TeV\\ 
$A_\lambda$ & [0, 2] TeV\\
$\azz$ & [-0.003, 0.003] rad \\ 
$M_1$, $M'_1$ & [0.1, 2] TeV\\ \hline \hline
\end{tabular}
\caption{Range of the free parameters in the $U(1)'$ models \label{tab:e6}.}
\label{tab:range}
\end{center}
\end{table}

\begin{figure}[!htb]
\begin{center}
\centering
\subfloat[]{\includegraphics[width=8.25cm,height=6.5cm]{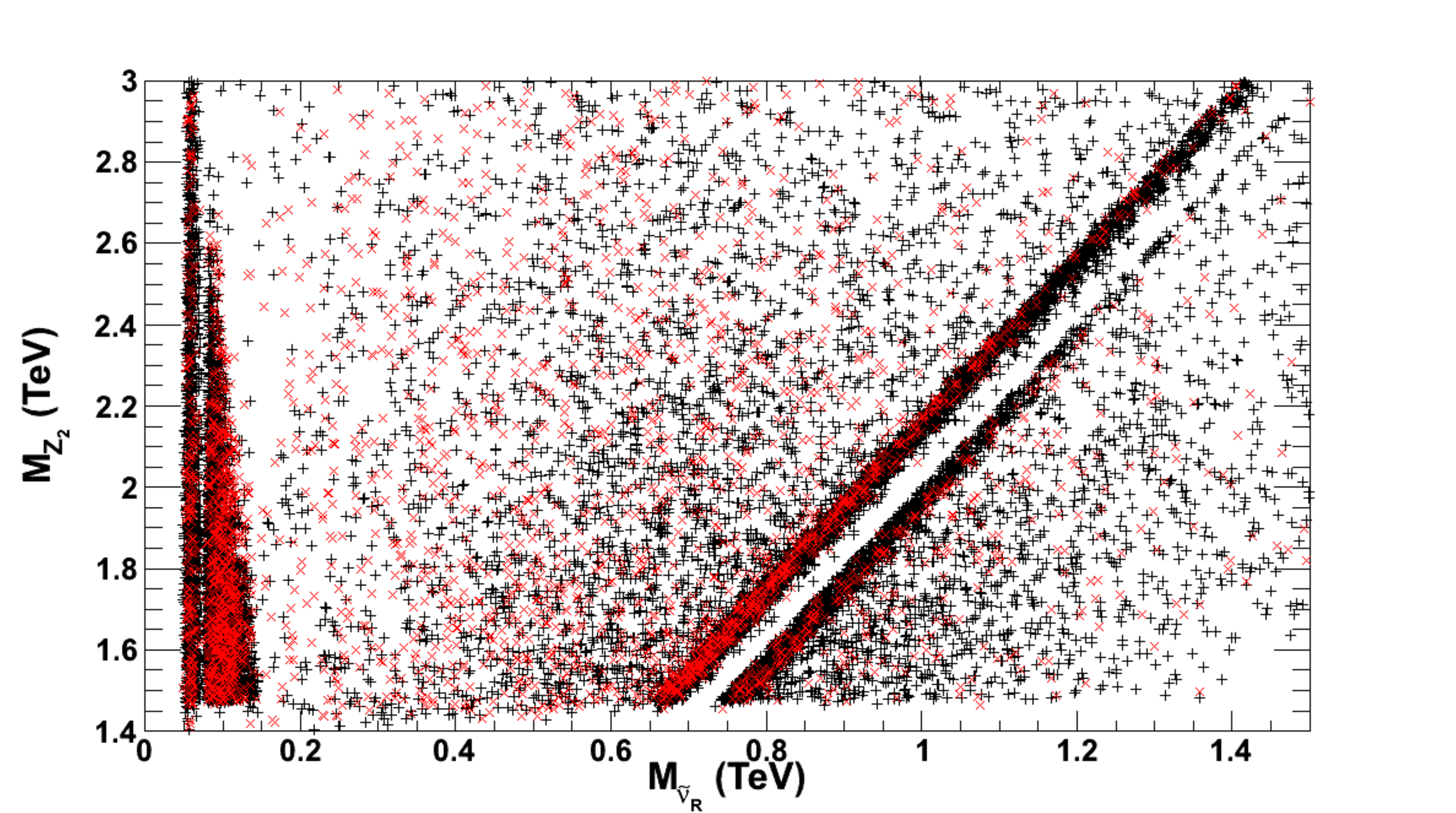}} 
\subfloat[]{\includegraphics[width=8.25cm,height=6.5cm]{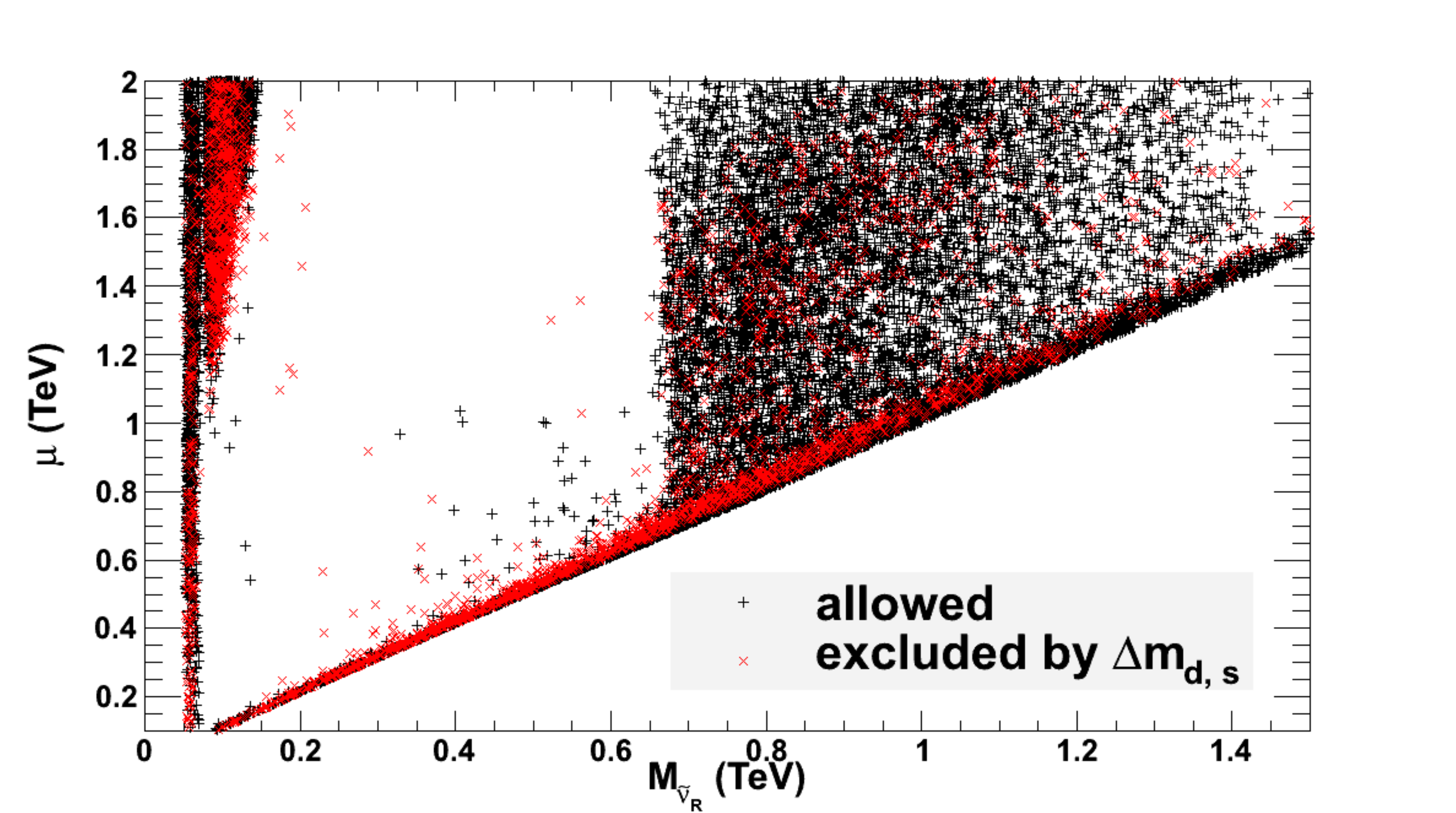}}
\caption{The allowed scenarios in the $M_{Z_2}$\vs $m_{\lsp}$ plane (panel (a)) and in the $\mu$\vs $m_{\lsp}$ plane (panel (b)). Points excluded by the $\Delta M_{d,s}$ constraint are displayed in red.}
\label{fig:random_psi}
\end{center}
\end{figure}

The results of the scan as displayed in figure~\ref{fig:random_psi} show that the allowed values for $m_{\lsp}$ cover a wide range from 55~GeV to the largest value probed. The allowed points in the plane $m_{\lsp}- M_{Z_2}$ are clustered in three regions around $m_{\lsp}\approx 60$~GeV, around $m_{\lsp} \approx M_{Z_2} \, -\delta \, (+\delta')$ with $\delta \approx 70$~GeV $(\delta' \approx 10$~GeV, see figure~\ref{fig:random_psi}a), and $m_{\lsp}\approx 90-100$~GeV. The first two are characterized by the main annihilation mechanism near a resonance. The latter corresponds to annihilation through a Higgs boson exchange. As discussed above, this requires a large value for $\mu$, see figure~\ref{fig:random_psi}b. The remaining allowed scenarios correspond to $m_{\lsp} \approx m_{\chi^0_1}$. In most cases the NLSP is an higgsino and $m_{\lsp}\approx \mu$, see figure~\ref{fig:random_psi}b, then important contributions from coannihilation processes involving neutral and charged higgsinos annihilating into fermion pairs are to be added to the sneutrino annihilation processes dominated by the channels into $W^+W^-,ZZ$ through Higgs boson exchange. Because $\mu$ is constrained by the LEP limit on charginos, the RHSN mass is in this case heavier than $\sim$ 90~GeV. A few cases where the NLSP has an important bino component are also found, those correspond to the points above the line $\mu \approx \mlsp$ in figure~\ref{fig:random_psi}b. Note that figure~\ref{fig:random_psi}a illustrates the possibility to relax the bounds on $M_{Z_2}$ in the UMSSM. Indeed several scenarios in the range $[1.40, 1.49]$~TeV are allowed when we re-derived the limits. The reason is that for the $U(1)_\psi$ model the $Z_2$ decays into supersymmetric particles can reach up to 20\% especially in cases where the sneutrino is light (hence neutralinos can be light as well) while decays into Higgs and gauge bosons are typically below 10\%.

In this model, the cross section for LSP scattering on nuclei is dominated by either Higgs or $Z_1$ exchange and is mostly independent of $M_{Z_2}$. The Direct Detection cross section can reach at most $\sigma^{SI}_{\lsp Xe} = 2 \times 10^{-9}$~pb when $m_{\lsp}\approx 50-60$~GeV and the light Higgs exchange dominates while the maximal value is only $\sigma^{SI}_{\lsp Xe} \approx 5\times 10^{-10}$~pb for sneutrinos above 200~GeV when $Z_1$ exchange is dominant. These predictions are much below the exclusion limits of XENON100~\cite{Aprile:2010um}. Values below $\sigma^{SI}_{\lsp Xe} = 10^{-13}$~pb can also be obtained, these occur when the $Z_1$ contribution is suppressed by the small mixing angle and the light Higgs coupling to the LSP is suppressed as well. Cases where coannihilation processes dominate can also lead to small cross sections.

\begin{figure}[!htb]
\begin{center}
\centering
\subfloat[]{\includegraphics[width=8.25cm,height=6.5cm]{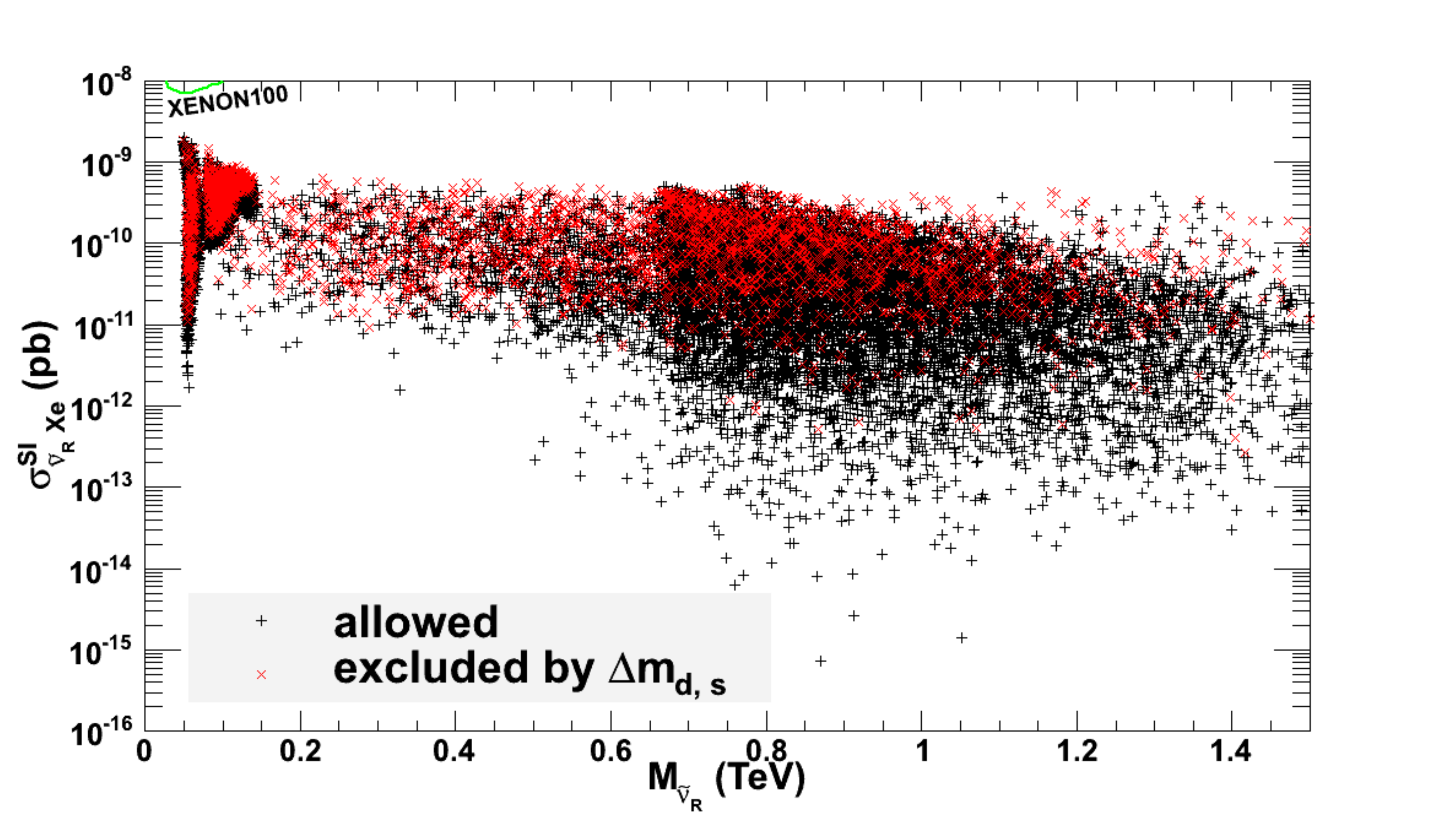}}
\subfloat[]{\includegraphics[width=8.25cm,height=6.5cm]{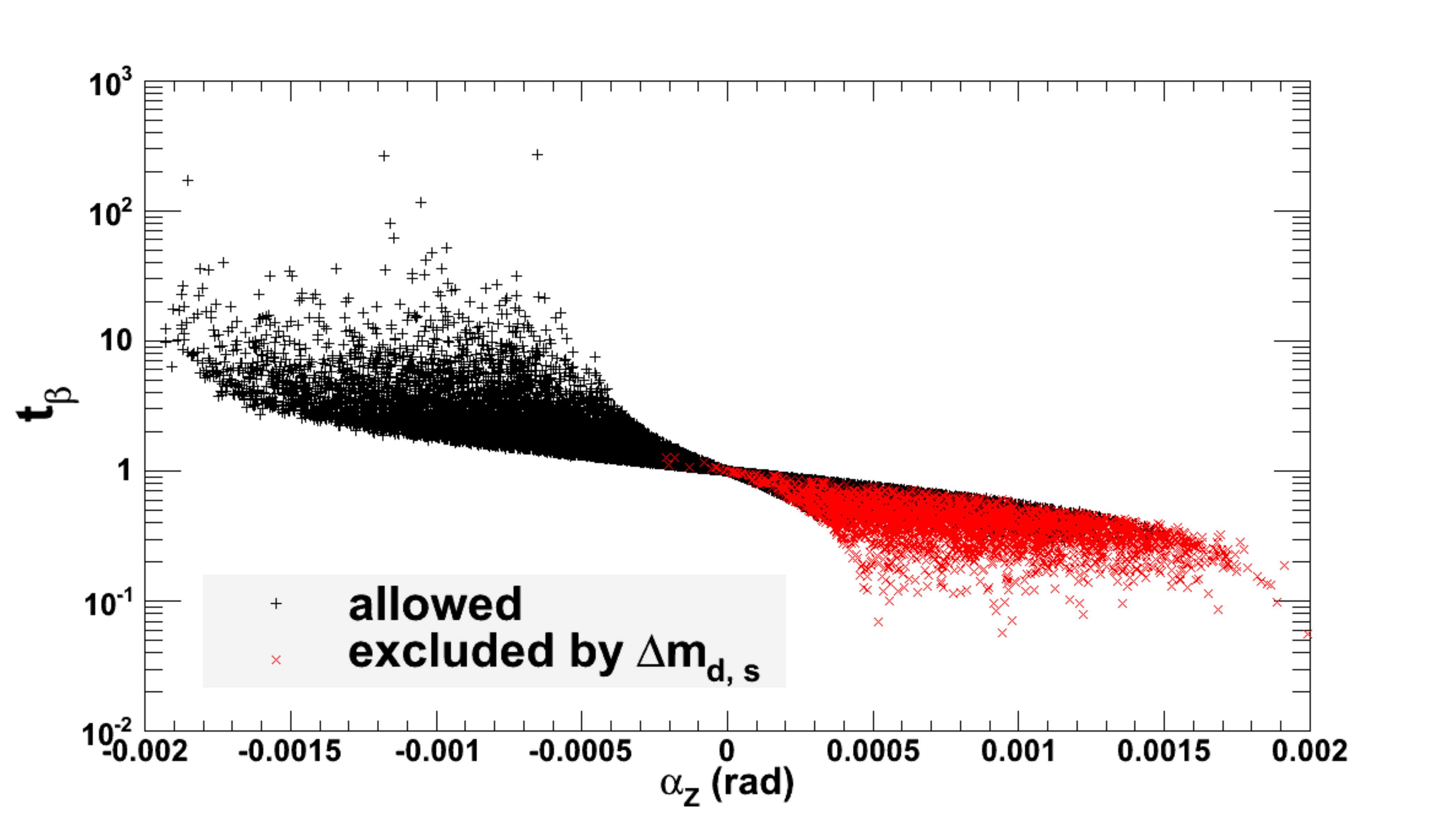}} 
\caption{(a) $\sigma^{SI}_{\lsp Xe}$ as a function of the LSP mass for the allowed scenarios in the $U(1)_\psi$ model.
(b) Allowed scenarios in the $\tan\beta -\azz$ plane. Points excluded by $\Delta M_s$ are displayed in red.  }
\label{fig:random_psi_direct}
\end{center}
\end{figure}

Constraints from $\Delta M_{d,s}$ rule out some parts of the parameter space, in particular the case where $\azz>0$ since it leads to values of $\tan\beta<1$. The allowed points in the $\tan\beta -\azz$ plane are displayed in figure~\ref{fig:random_psi_direct}b. As mentioned in section~\ref{sec:8.constraint}, when $\tan\beta$ is small the charged Higgs box diagram adds to the SM contribution and leads to value for $\Delta M_s$ that is too large. This constraint does not influence directly the range of predictions for the DD cross section presented in figure~\ref{fig:random_psi_direct}a, nor does it affect the range of allowed masses for the RHSN. 

\subsection{The case of the $U(1)_\eta$ model}

The properties of the RHSN Dark Matter are dependent on the choice of the $U(1)'$ charges. To illustrate some of the differences we present results for $\te6=-\arctan \sqrt{5/3}$ ($U(1)_\eta$) before discussing arbitrary values in the next section. We chose this model for its phenomenological properties. For this value of $\te6$, the coupling of the sneutrino to $Z_2$ is enhanced as compared to the previous example with $\mathcal{Q}'_\nu=-\sqrt{5}/2\sqrt{3}$. Furthermore the coupling of the RHSN to $h_1$ is typically enhanced as compared to the model $U(1)_\psi$.  Finally the vectorial couplings of $Z'$ to neutrinos and d-quarks are non-zero. Therefore both terms in eq.~\ref{eq:ddz} can contribute to the SI cross section, this implies a DD cross section that is much larger than found previously. Nevertheless the vectorial couplings of $Z_1$ are far enough from their maximal value that it is possible to find scenarios that satisfy Direct Detection limits as will be demonstrated below.

\subsubsection{A case study with $M_{Z_2}=1.6$~TeV}

\begin{figure}[!htb]
\begin{center}
\centering
\subfloat[]{\includegraphics[width=8.25cm,height=6.5cm]{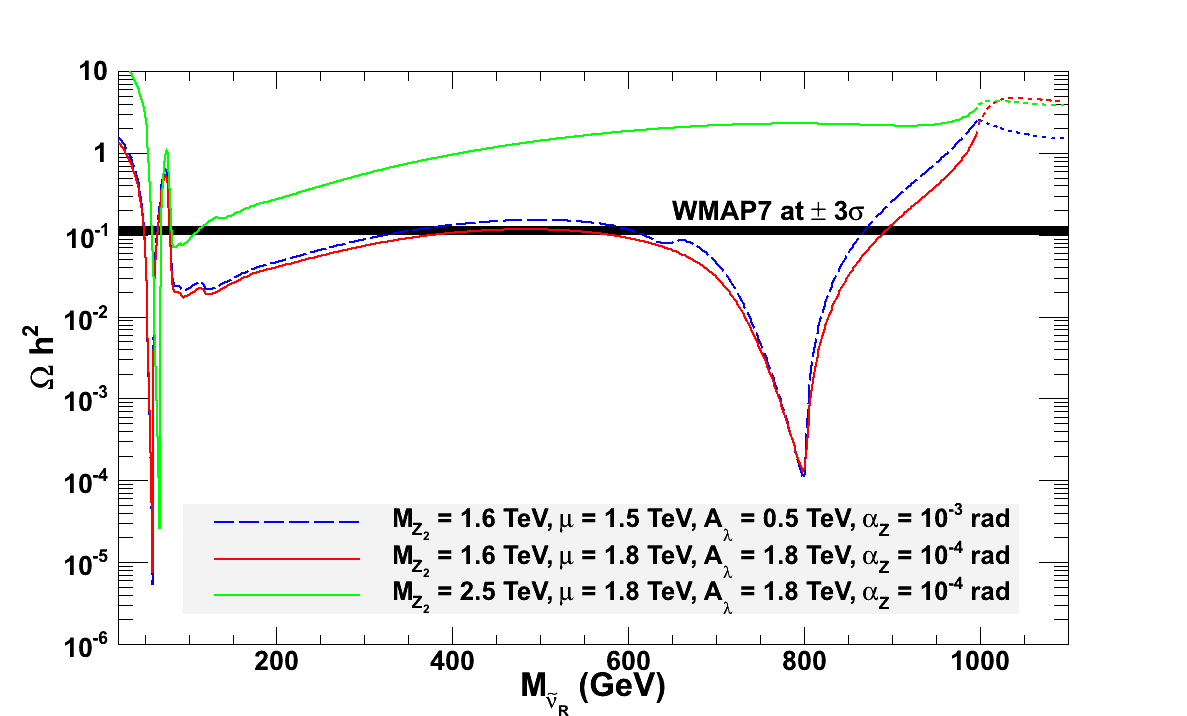}}
\subfloat[]{\includegraphics[width=8.25cm,height=6.5cm]{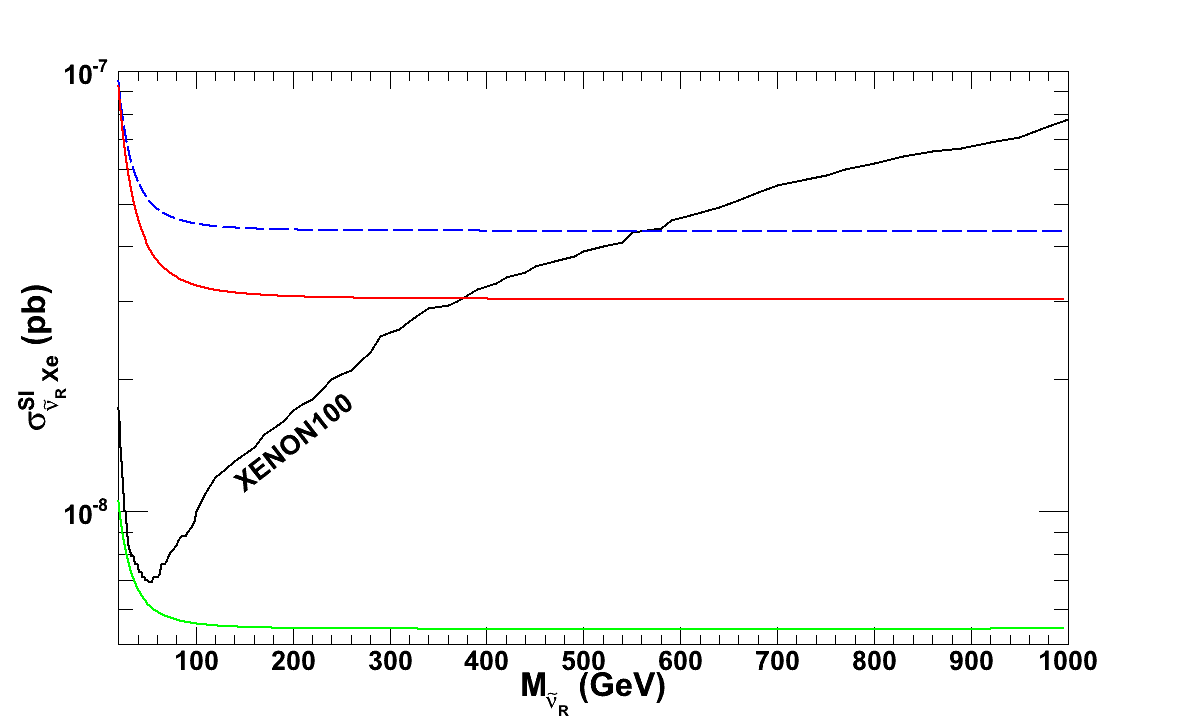}}
\caption[$\Omega h^2$ as a function of the LSP mass for (a) $\mzp=1.6$~TeV, $\mu=1.5$~TeV, $A_\lambda=0.5$~TeV, $\azz=10^{-3}$ and $\mzp=1.6,2.5$~TeV, $\mu=1.8$~TeV, $A_\lambda=1.8$~TeV, $\azz=10^{-4}$. (b) $\sigma^{SI}_{\lsp Xe}$ as a function of the LSP mass for the same choice of parameters.]{$\Omega h^2$ as a function of the LSP mass for (a) $\mzp=1.6$~TeV, $\mu=1.5$~TeV, $A_\lambda=0.5$~TeV, $\azz=10^{-3}$ and $\mzp=1.6,2.5$~TeV, $\mu=1.8$~TeV, $A_\lambda=1.8$~TeV, $\azz=10^{-4}$. The dotted lines correspond to the neutralino LSP. (b) $\sigma^{SI}_{\lsp Xe}$ as a function of the LSP mass for the same choice of parameters.}
\label{fig:omega_eta}
\end{center}
\end{figure}

To illustrate the behaviour of the relic density and the DD rate we choose $M_{Z_2}=1.6$~TeV, $\alpha_Z=0.001$, $\mu=1.5$~TeV and $A_\l=0.5$~TeV. For this choice of parameters $m_{h_1}=116.9$~GeV despite a small value of $\tan\beta=1.2$. The other doublet-like Higgs boson is around $1.3$~TeV while $h_3$ is nearly degenerate with $Z_2$. As for the model $U(1)_{\psi}$, the relic density satisfies the WMAP upper bound in the regions where annihilation into light Higgs boson
or singlet Higgs/$Z_2$ is enhanced by a resonance effect as well as in a region where annihilation via $h_1$ exchange is efficient without the benefit of a resonance enhancement. The latter region extends over a wide range of values for the LSP mass because of the large couplings of the sneutrinos to $h_1$. The preferred annihilation channels are typically into $W^+W^-$, $ZZ$ $t\bar{t}$ or $h_1h_1$ pairs as well as into $b\bar{b}$ for light sneutrinos. Note also that exchange of $h_2$ can contribute significantly to sneutrino annihilation, see the small dip at $\mlsp=650$~GeV in figure~\ref{fig:omega_eta}a. 
In this scenario the Direct Detection cross section is large ($\sigma^{SI}_{\lsp Xe} \approx 4.5\times 10^{-8}$~pb for $\mlsp \geq 100 {\rm GeV}$) and exceeds the XENON100 bound for a sneutrino LSP below 550 GeV, see figure~\ref{fig:omega_eta}b. This is because both the $Z_1$ and $Z_2$ contribute significantly to the cross section. Extending the region of validity of sneutrino DM over a larger mass range after considering DD limits thus requires decreasing the mixing angle $\azz$ and/or  increasing the mass of the $Z_2$. For example taking $\azz=10^{-4}$, $\mu=A_\l=1.8$~TeV has the effect of decreasing the relic density - so that it is below the WMAP upper bound for sneutrino masses in the range 90-900~GeV - while also decreasing the SI cross section. Yet the LSP is still constrained to be $\mlsp \geq 370$~GeV. In fact the Direct Detection rate decreases by less than a factor of two, this is because the contribution of the $Z_2$ exchange, the term proportional to $Q'^d_V$ in eq.~\ref{eq:ddz}, is not suppressed by $\sin\alpha_Z$. This means that to have sneutrino DM with a mass of a few hundred GeV's also requires increasing the mass of the $Z_2$. For example taking $M_{Z_2}=2.5$~TeV decreases the Direct Detection rate below the XENON100 exclusion for all masses of the LSP. However for this choice of parameters, only light sneutrinos also satisfy the relic density constraint since the heavy sneutrino LSP cannot annihilate through the singlet/$Z_2$ resonance.

Finally we comment on the coannihilation region. We have fixed $M_1=1$~TeV, so that the lightest neutralino has a mass  $\approx 1$~TeV and is dominantly bino. For values of $\mu \leq 1$~TeV, the lightest neutralino has a large higgsino component and coannihilation is efficient. For the benchmarks with $\mu=1.5$~TeV the coannihilation processes involving binos are not efficient enough to reduce $\Om h^2$ to a value compatible with WMAP.

\subsubsection{Exploration of $U(1)_\eta$ parameter space}

\begin{figure}[!htb]
\begin{center}
\centering
\subfloat[]{\includegraphics[width=8.25cm,height=6.5cm]{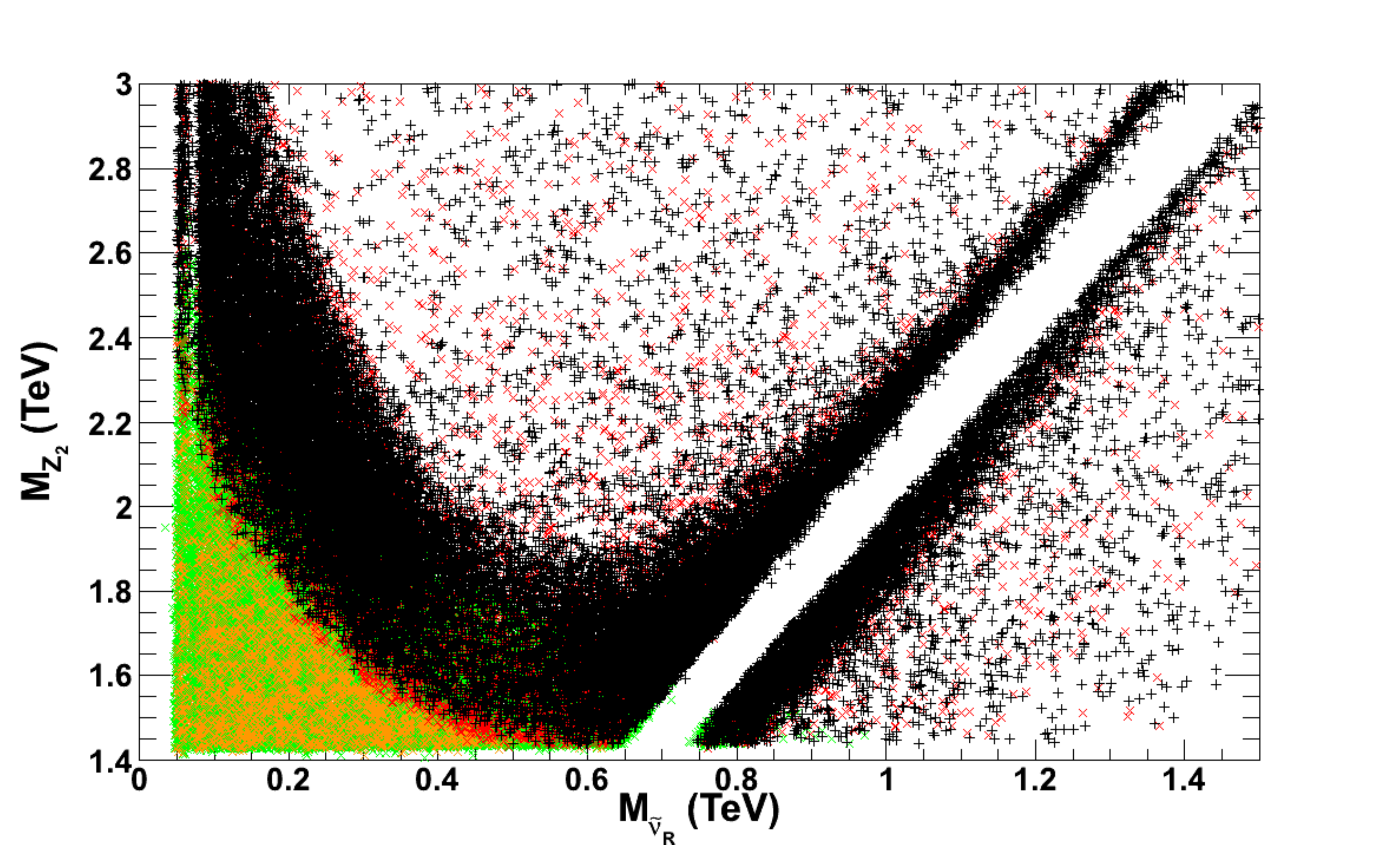}} 
\subfloat[]{\includegraphics[width=8.25cm,height=6.5cm]{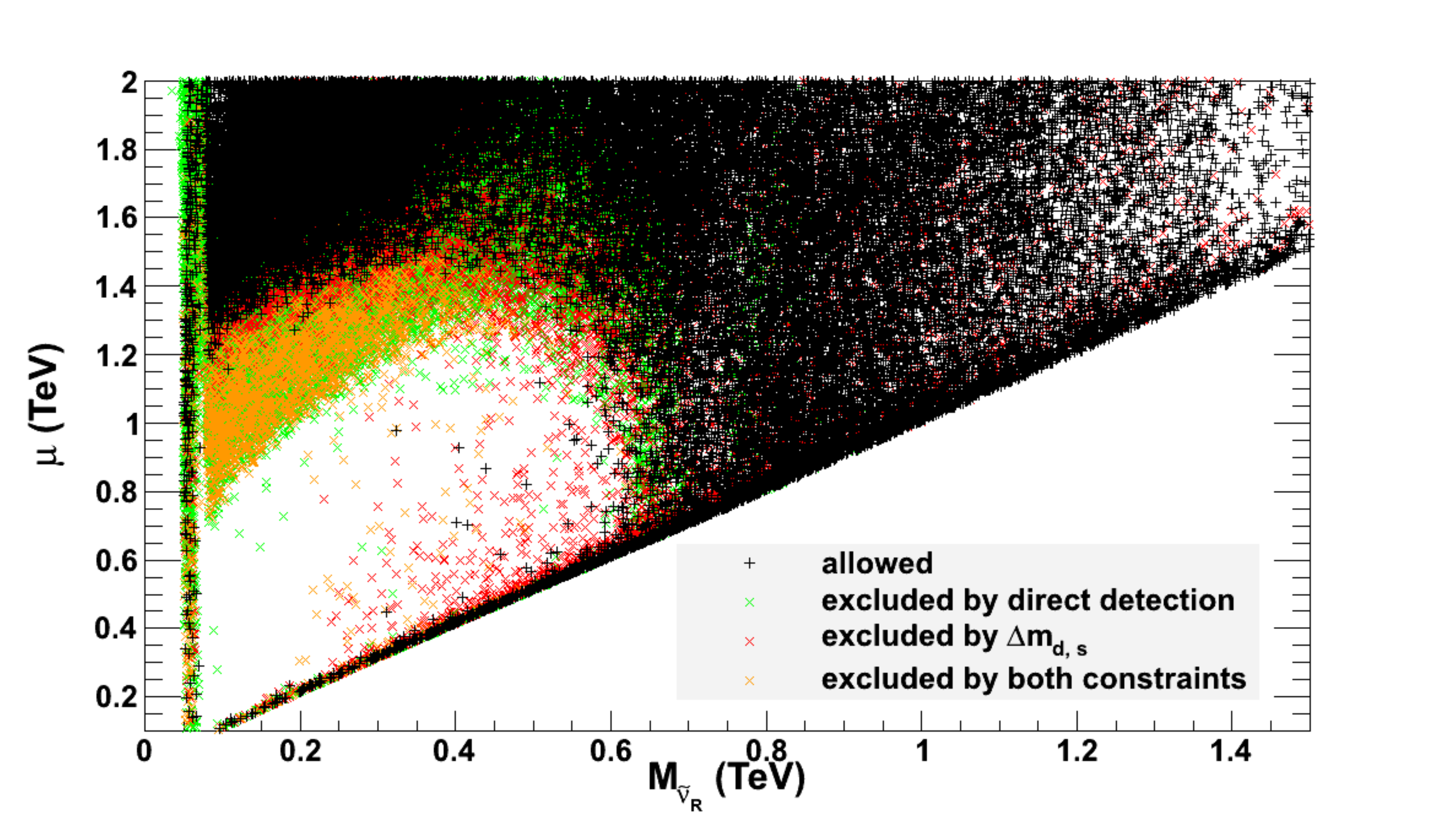}}
\caption{The allowed $U(1)_\eta$ scenarios in the (a) $M_{Z_2}$\vs $m_{\lsp}$ plane and (b) $\mu$\vs $m_{\lsp}$ plane. Points excluded by the $\Delta M_{d,s}$ constraints are plotted in red, those excluded by XENON100 are represented in green and those excluded by both constraints are shown in orange.}\label{fig:random_eta}
\end{center}
\end{figure}

\begin{figure}[!htb]
\begin{center}
\centering
\subfloat[]{\includegraphics[width=8.25cm,height=6.5cm]{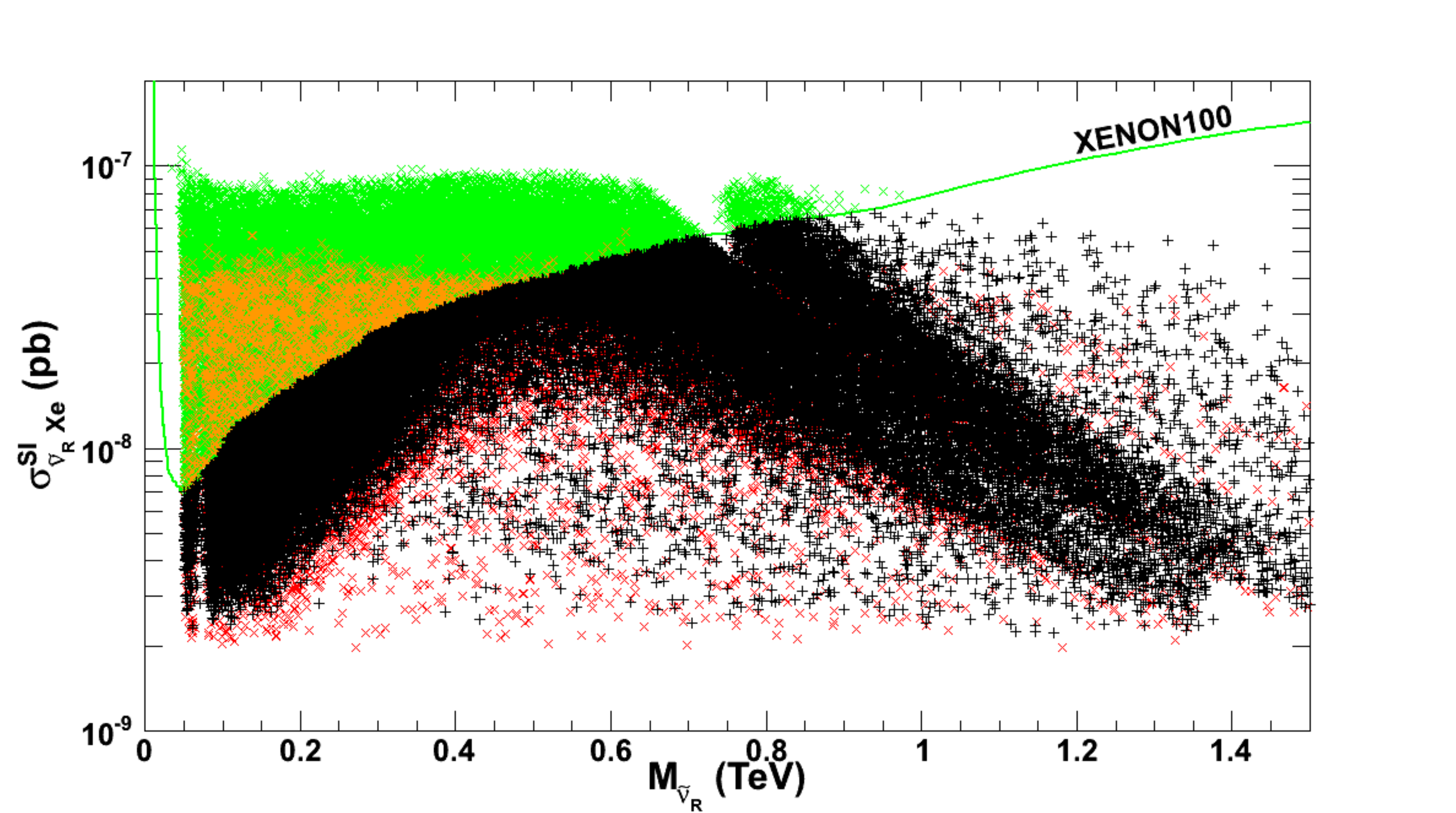}}
\subfloat[]{\includegraphics[width=8.25cm,height=6.5cm]{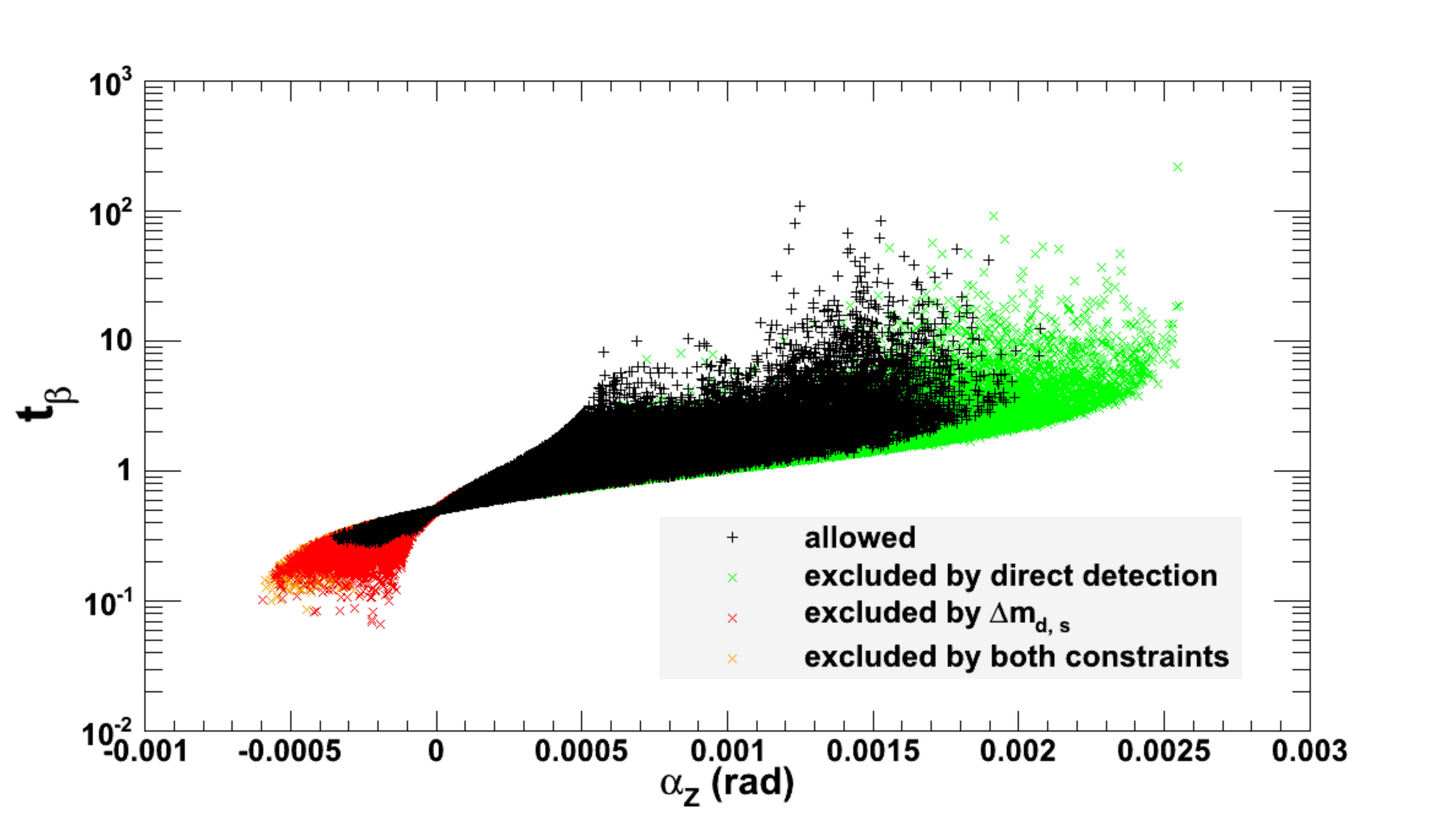}}
\caption{(a) $\sigma^{SI}_{\lsp Xe}$ as a function of the LSP mass for the allowed scenarios in the $U(1)_\eta$ model.
(b) Allowed scenarios in the $\tan\beta$ - $\alpha_Z$ plane, same colour code as figure~\ref{fig:random_eta}.  }
\label{fig:random_eta_direct}
\end{center}
\end{figure}

We also explore the parameter space of the $U(1)_\eta$ model, varying the parameters in the range shown in table~\ref{tab:range}.
As before, we impose in addition to limits on $\Omega h^2$ and on the SI cross section, the lower limit on the Higgs boson and $Z_2$ mass, and the limit on $\Delta M_{d,s}$. The results are displayed in figure~\ref{fig:random_eta} in the $M_{Z_2}-\mlsp$ plane, as well as in the $\mu-\mlsp$ plane. As we have discussed above, because Higgs boson annihilation is efficient enough, the sneutrinos are not confined to the $h_i,Z_2$ resonance region. Sneutrinos from 100 GeV to $M_{Z_2}/2$ can satisfy the relic density constraint, either through pair annihilation or through coannihilation (the region where $\mlsp\approx \mu$). In the former case large values of $\mu$ are preferred to have large enough couplings of the sneutrino LSP to $h_1$. The DD cross section is large with a maximum value near $\sigma^{SI}_{\lsp Xe} \approx 10^{-7}$~pb for the whole range of LSP masses (see figure~\ref{fig:random_eta_direct}). Because the experimental limits on the DD rate are more stringent for masses around 80 GeV, light sneutrinos are severely constrained especially when the $Z_2$ mass is lighter than 2 TeV. The lower bound on the cross section for $M_{Z_2}\leq 3$~TeV is $\sigma^{SI}_{\lsp Xe}=2\times 10^{-9}$~pb. The typically large Direct Detection rate is the main signature of this scenario. Several points are also constrained by $\Delta M_s$, in particular when $\azz<0$ which implies $\tan\beta<1$. This does not affect the range of values of sneutrino masses compatible with all constraints. As in the $U(1)_\psi$ case the limits on the $Z_2$ mass ($\mzp >1.54$~TeV) are weakened. In the $U(1)_\eta$ model, the decays into SM particles is even more suppressed, it never exceeds 65\%. This is mainly due to a large branching fraction into RH neutrinos (around 30\%) as well as into sparticles.

\subsection{A global scan of the parameter space}

Having illustrated the properties of the sneutrino DM in specific models we will next explore the complete parameter space of the model by keeping all parameters for the neutralino and gauge boson sector free while imposing the constraints from WMAP, Direct Detection as well as on Higgs boson masses. 

\begin{figure}[!ht]
\centering
\includegraphics[width=10cm]{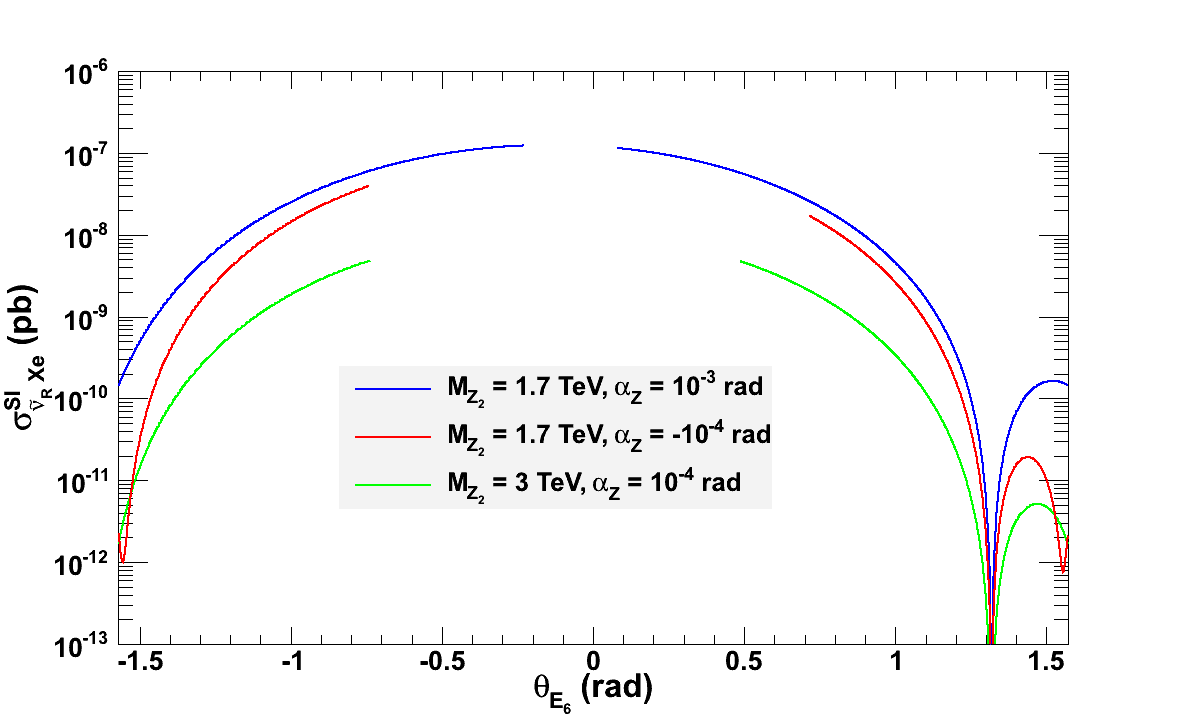}
\caption{$\sigma^{SI}_{\lsp Xe}$ as a function of $\te6$ for $(\mzp \, \mathrm{(TeV)},\azz)= (1.7,10^{-3})$, $(1.7,-10^{-4})$, $(3,10^{-4})$, $\mu=M_{Z_2}/2+0.5$~TeV, $A_\lambda=1$~TeV and $\mlsp=M_{Z_2}/2$.  }
\label{fig:sigmae6}
\end{figure}

The values of $\theta_{E_6}$ where the sneutrino is a good DM candidate are restricted. First for $\te6 \approx 0$ the value of $v_s$ becomes very large especially when $M_{Z_2}$ is large. This induces large negative corrections to sfermion masses and lead in particular to a charged LSP. For example, for soft terms at 2 TeV the values $-0.2<\theta_{E_6}<0.05$ are excluded when $M_{Z_2}=1$~TeV. Second the DD cross section is often too large when $|\te6|<0.5$. This is due mainly to the contribution of the $Z_2$ exchange to the Direct Detection cross section that is proportional to $\cos\te6$. To illustrate this we display the variation of the DD cross section as a function of $\te6$ for different values of $M_{Z_2}$ in figure~\ref{fig:sigmae6}. We choose $\mlsp=M_{Z_2}/2$ so as to guarantee that the WMAP upper bound is satisfied and fix $\mu=M_{Z_2}/2+0.5$~TeV in each case to ensure that the sneutrino is the LSP. We have also fixed for the other sfermions $m_{\tilde f}=2$~TeV. The DD bounds are easily satisfied near $\te6=\pm \pi/2$ because of the suppressed contribution of the $Z_{1,2}$ vectorial coupling, see figure~\ref{fig:sigmae6}. The mixing angle $\azz$ has only a moderate impact on the Direct Detection rate, while increasing the mass of the $Z_2$ reduces the SI cross section, except when $\te6=\theta_\psi$ as we have seen in section~\ref{sec:psi}. Note that for $\azz<0$ there is a dip in the cross section near $\te6=\pm \pi/2$, this is because of a cancellation between the $y$ and $y'$ contributions in eq.~\ref{eq:ddz} in the cross section on neutrons. As a side comment note that for $\te6 \simeq 0.42\pi$ the RH sneutrino is decoupled, which corresponds to the $U(1)_N$ model that was mentioned in section~\ref{sec:7.descUMSSM}.

\begin{figure}[!htb]
\begin{center}
\centering
\subfloat[]{\includegraphics[width=8.25cm,height=6.5cm]{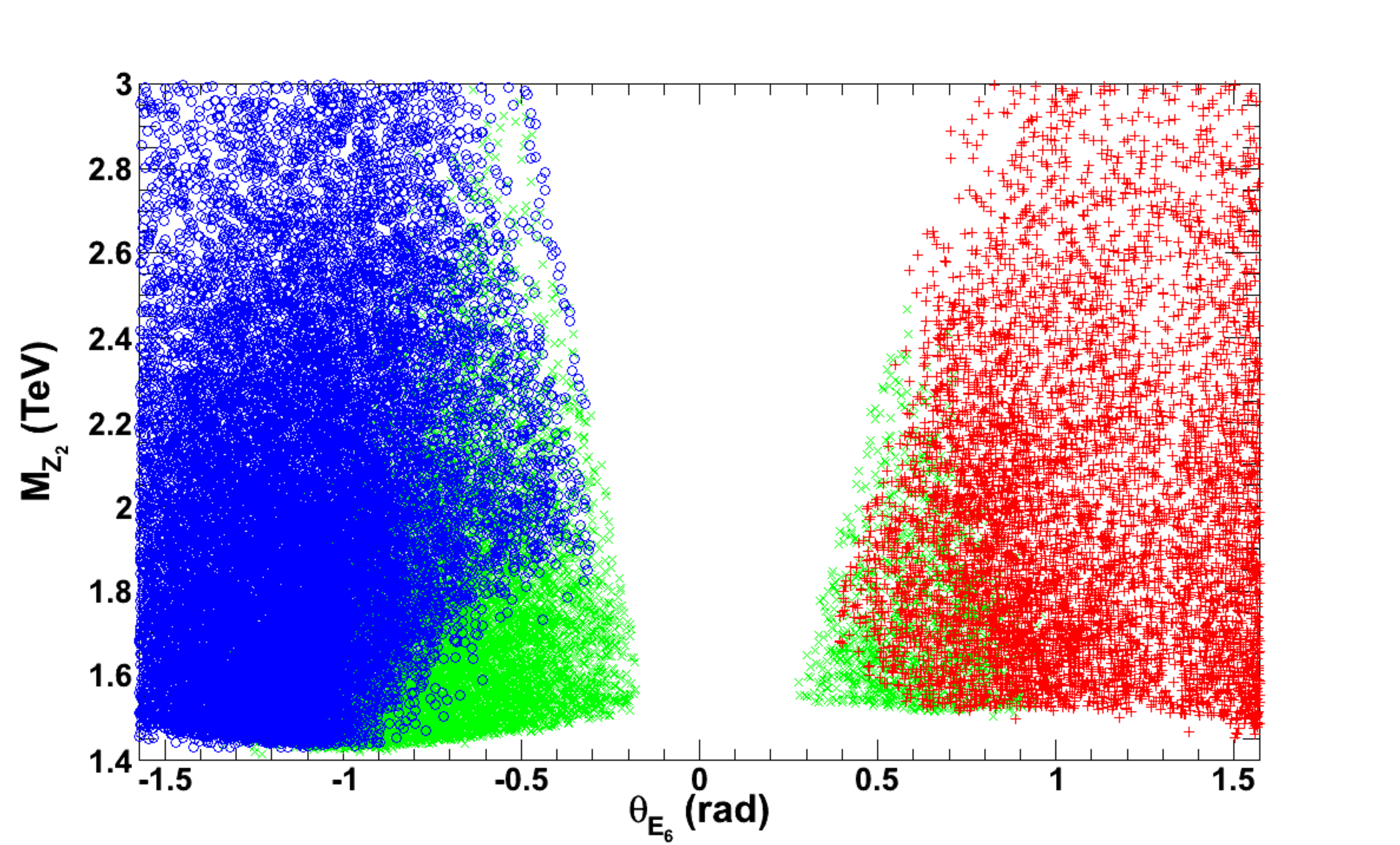}}  
\subfloat[]{\includegraphics[width=8.25cm,height=6.5cm]{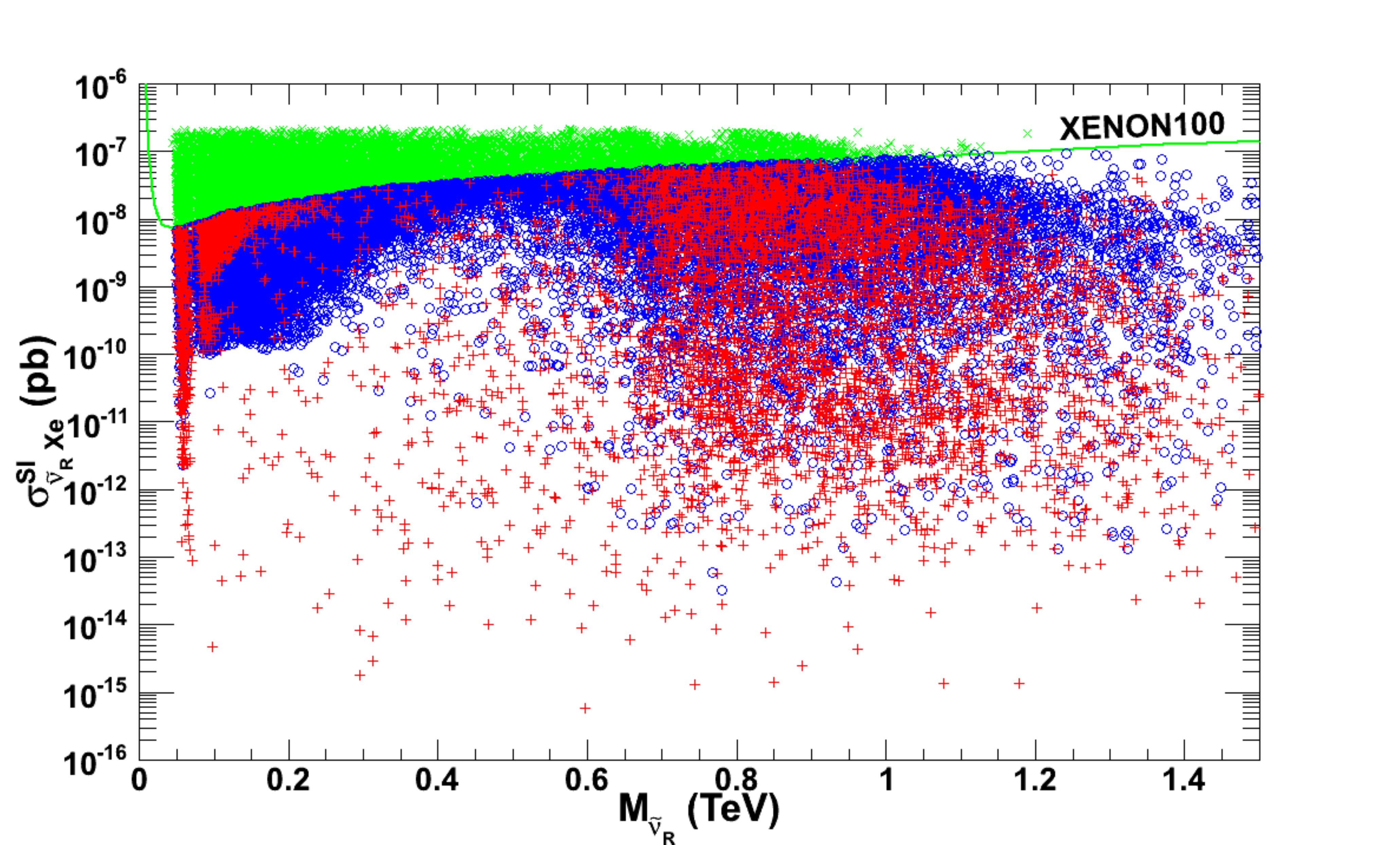}} 
  \caption{(a) Allowed scenarios in the $M_{Z_2}$\vs $\te6$ plane, the models above the XENON100 bound are in green. (b) $\sigma^{SI}_{\lsp Xe}$ as a function of $\mlsp$. In blue $\te6<0$ and in red $\te6>0$.}
\label{fig:te6:sigma}
\end{center}
\end{figure}

For the random scans we choose the same range as in table~\ref{tab:e6} with in addition $-\pi/2 \leq \te6 \leq \pi/2$. As before, we have applied the constraints on $M_{Z_2}$ from the LHC and on $\alpha_Z$. We have imposed the constraint from XENON100~\cite{Aprile:2010um} and from $\Delta M_s$ a posteriori to better illustrate the impact of these constraints.
The successful scenarios are displayed in figure~\ref{fig:te6:sigma}a in the $M_{Z_2}-\te6$ plane. The white region at small values of $\te6$ have a charged fermion LSP. Many scenarios with $-\pi/4<\te6<0 $ have too large a Direct Detection rate as was the case for the model $U(1)_\eta$. The value of the DD cross section span several orders of magnitude from less than $10^{-13}$~pb to $2\times 10^{-7}$~pb (see figure~\ref{fig:te6:sigma}b). In particular for sneutrino masses around 100 GeV there are many models which exceed the Direct Detection limit. These are scenarios with $\te6<0$ where the sneutrino coupling to light Higgs boson allows efficient annihilation even away from resonance as discussed in the case of $U(1)_\eta$. The enhanced couplings to the Higgs boson as well as the coupling to the $Z_{1,2}$ lead to a large DD rate. Models with $\te6>0$ have only light sneutrinos in that mass range when coannihilation plays a role and therefore tends to have much lower Direct Detection rate. In fact even for heavier masses models with $\te6>0$ predict smaller Direct Detection rates as discussed above.

\begin{figure}[!htb]
\begin{center}
\centering
\subfloat[]{\includegraphics[width=8.25cm,height=6.5cm]{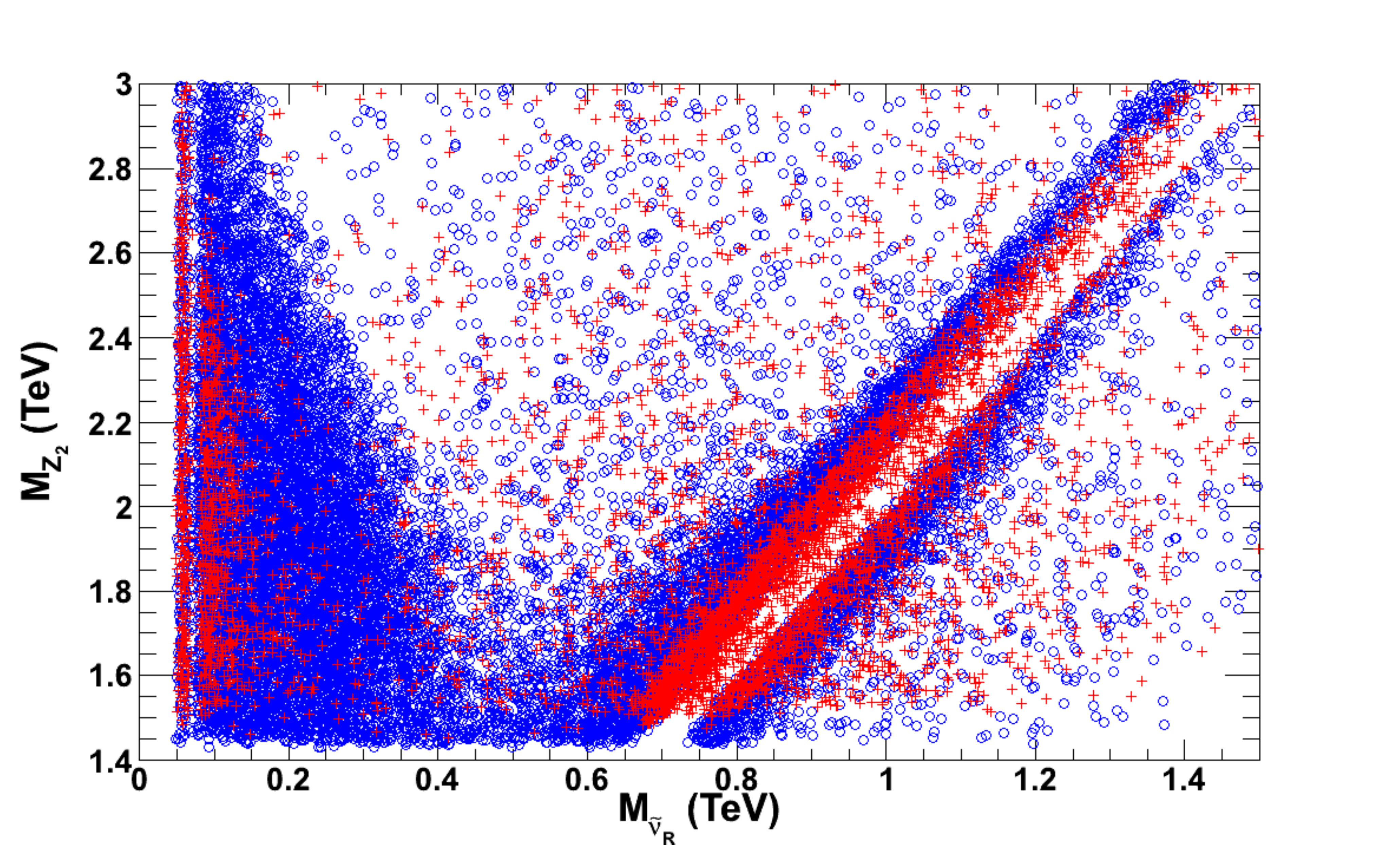}}
\subfloat[]{\includegraphics[width=8.25cm,height=6.5cm]{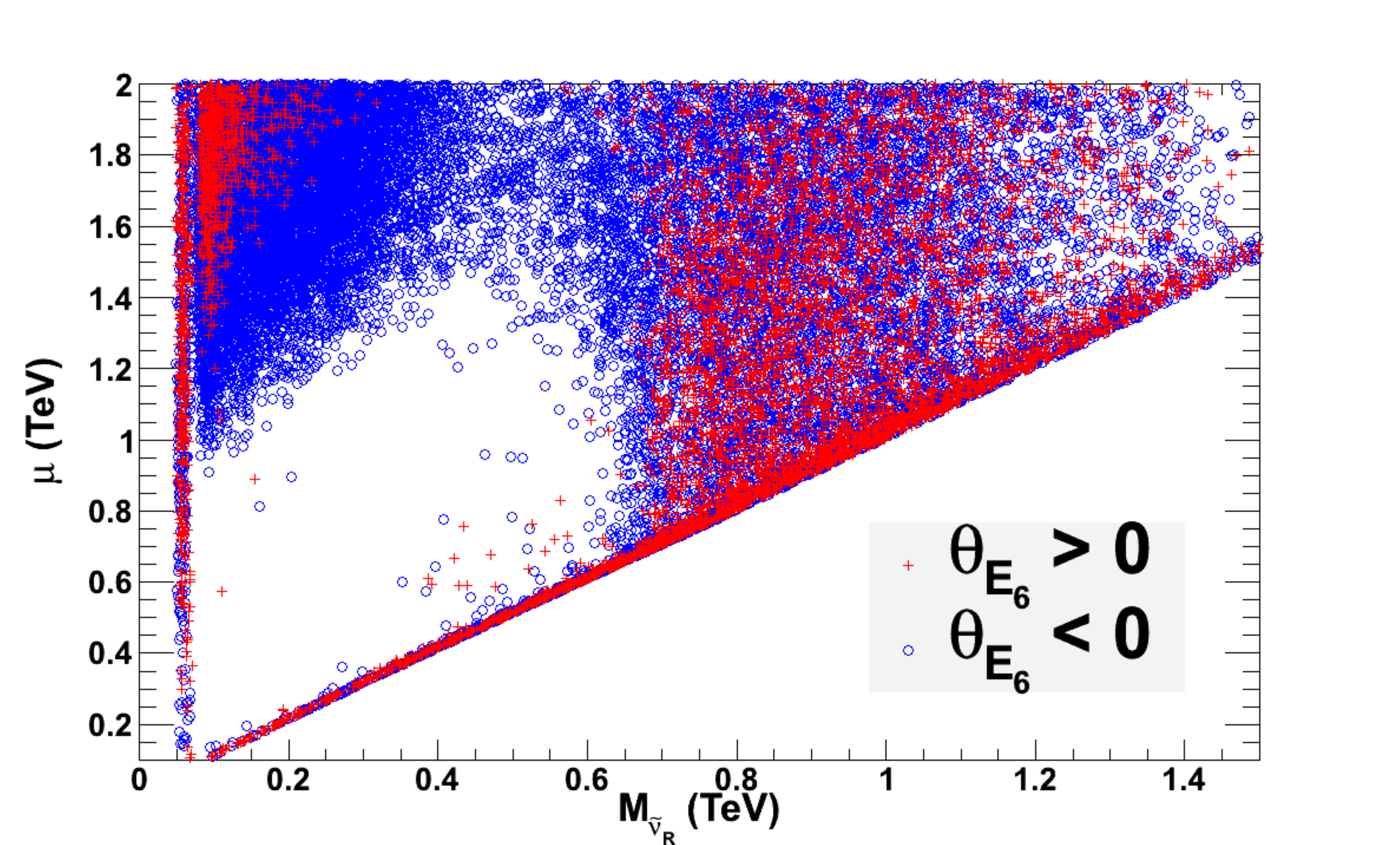}}
  \caption{The allowed scenarios in the (a) $M_{Z_2}$\vs $m_{\lsp}$ plane and (b) $\mu$\vs $m_{\lsp}$ plane after applying all constraints. Same colour code as in figure~\ref{fig:te6:sigma}.}
\label{fig:e6}
\end{center}
\end{figure}

\begin{figure}[!htb]
\begin{center}
\centering
\subfloat[]{\includegraphics[width=8.25cm,height=6.5cm]{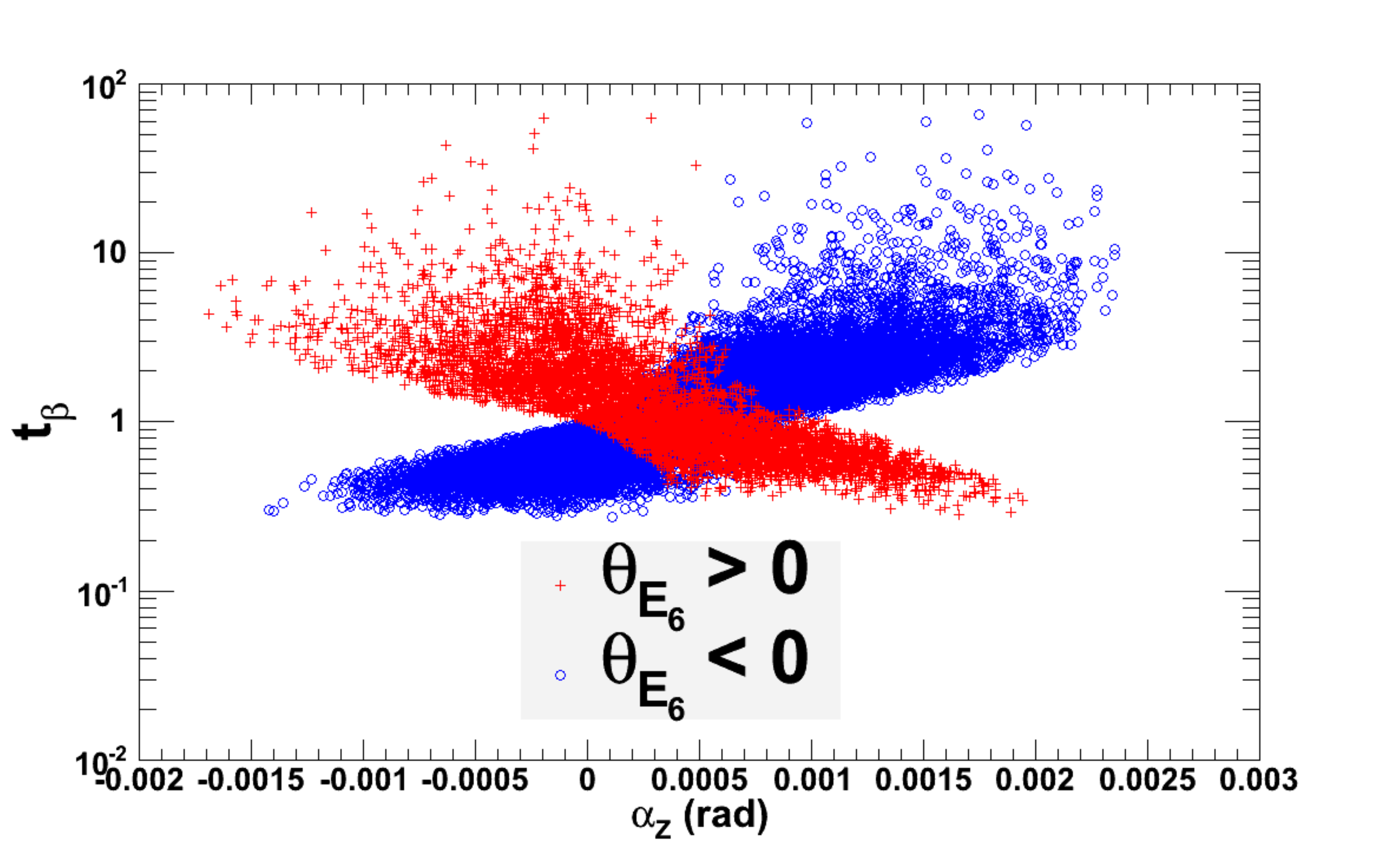}}\\
\subfloat[]{\includegraphics[width=8.25cm,height=6.5cm]{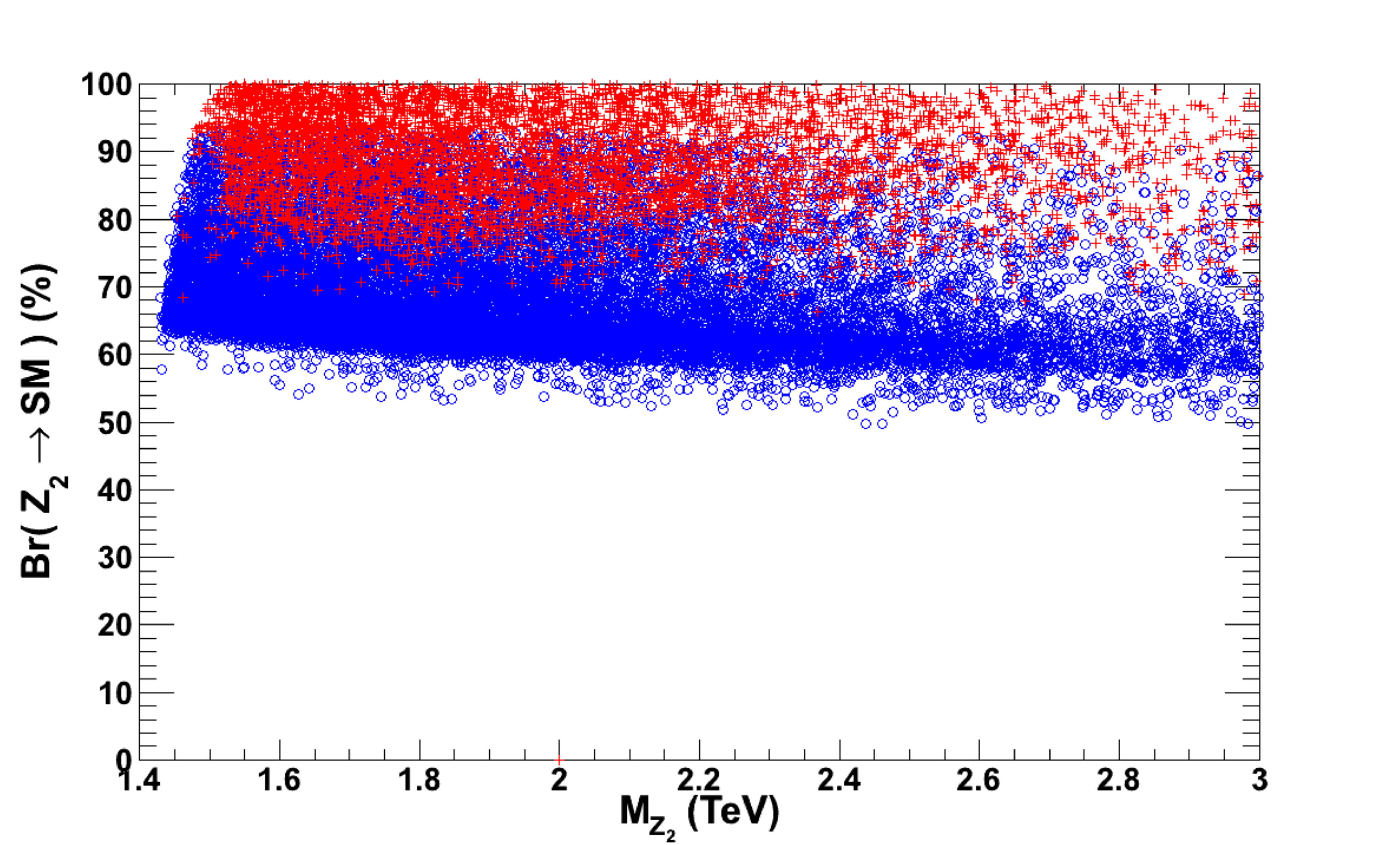}}
\subfloat[]{\includegraphics[width=8.25cm,height=6.5cm]{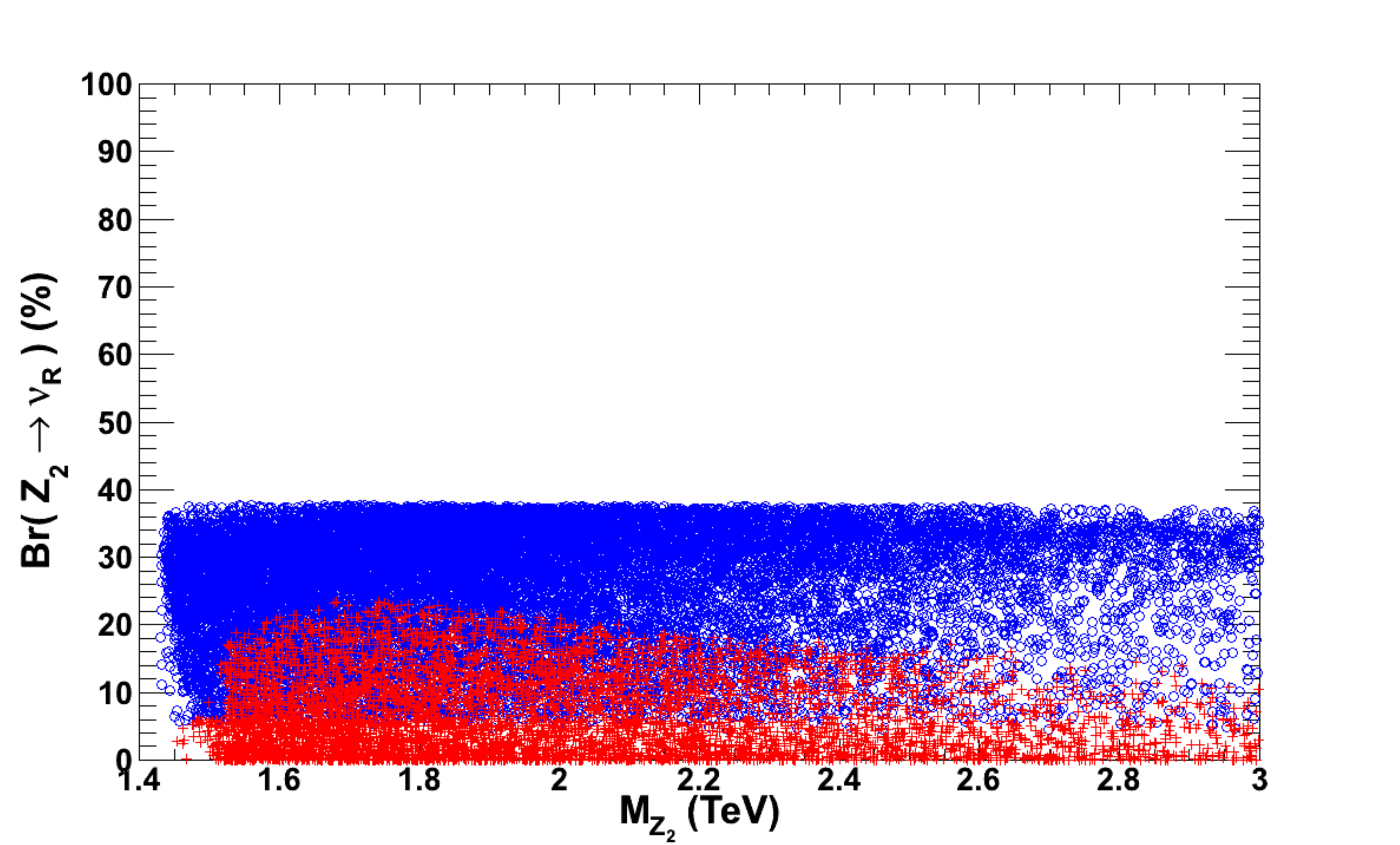}}
\caption{\label{fig:8.plus}(a) Allowed scenarios in the $\tan\beta$ - $\alpha_Z$ plane, (b) branching ratio of the $Z_2$ into SM particles and (c) branching ratio of the $Z_2$ into RH neutrinos as a function of $M_{Z_2}$. Same colour code as in figure~\ref{fig:te6:sigma}.}
\end{center}
\end{figure}

The processes that can contribute to $\Omega h^2$ were discussed both in the context of $U(1)_\psi$ and $U(1)_\eta$ models.
In the general case we find similar results. We find a predominance of annihilation near a $h_1$ or singlet Higgs/$Z_2$ resonance as well as annihilation into gauge boson pairs through $h_1$ exchange. The latter being confined to sneutrino masses just above the $W$ pair threshold when $\te6>0$. These regions have a high density of points in the plane $M_{Z_2}-\mlsp$ in figure~\ref{fig:e6}a. For $\te6>0$ the only other allowed scenarios have $\mlsp \approx
 \mu$ as displayed in figure~\ref{fig:e6}b. These are dominated by higgsino coannihilation. For $\te6<0$, sneutrinos of masses above 100 GeV can also annihilate efficiently through $h_1$ exchange provided their coupling to $h_1$ is large enough - this requires large values of $\mu$. In both figures we have imposed the $\Delta M_{d,s}$ constraint although we do not display explicitly its impact. As discussed in the previous section this constraint plays a role for small values of $\tan\beta$ for any values of the LSP mass. Figure~\ref{fig:8.plus}a displays the allowed scenarios in the $\tan\beta$ - $\alpha_Z$ plane. The dependence on the $\te6$ angle can be understood by looking at figure~\ref{fig:7.zprime_evol} and eq.~\ref{eq:7.c2b}. Actually we recall that the invariance of the UMSSM superpotential under $U(1)'$ leads to $\mathcal{Q}'_{H_d} + \mathcal{Q}'_{H_u} = -\mathcal{Q}'_S$. Then since figure~\ref{fig:7.zprime_evol} shows that $\mathcal{Q}'_S (-\te6) = -\mathcal{Q}'_S(\te6)$ it results that for the same set of parameters a $\te6$ of the opposite sigh needs to change the sign of $\azz$ too. Note finally that as seen in the cases of $U(1)_\psi$ and $U(1)_\eta$ the branching ratio of $Z_2$ into SM particles tends to be decreased for negative values of $\te6$ mainly due to the enhancement of $Z_2 \ra \nu_R \nu_R^*$ processes (see figure~\ref{fig:8.plus}b,c).

\section{Conclusions}

The RH sneutrino is a viable thermal DM candidates in $U(1)$ extensions of the MSSM. The allowed parameter space depends strongly on the value of the new $Z_2$ vector boson mass. Sneutrino annihilation is typically dominated by resonance annihilation, with in particular the dominant contribution from the Higgs sector rather than from the new gauge boson $Z_2$.
 For the light Higgs boson this requires fine tuning of the masses while for the heavy singlet Higgs boson the mass difference has to be within 15\% (of $m_{h_i}/2$). As in other supersymmetric models, coannihilation processes can also be important. For simplicity we have only discussed the case of one sneutrino Dark Matter candidate, in the case of complete degeneracy of three sneutrinos the relic density increases by a factor 3. Indeed the annihilation cross section is the same for all sneutrino flavours since there is never a significant contribution from annihilation into neutrino pairs, thus the increase in the number of channels is compensated by the increase in the degrees of freedom leading to a smaller effective annihilation cross section. This will imply a narrower range of masses for the LSP near $m_{h_i}/2$. 

The Direct Detection limit is very stringent for a whole class of models unless the mass of the $Z_2$ is above 2 TeV.
The scenarios with $\te6<0$ will be best tested with SI detectors with improved sensitivities. In particular with a factor of 2 better sensitivity the whole region with intermediate masses for the sneutrinos - when annihilation into $W$ pairs is dominant - could be probed. The scenarios with $\te6>0$ are more challenging to probe via DD. Another signature of the RHSN model is that of a new gauge boson at the LHC. In fact the results from searches for $Z_2$ gauge bosons at the LHC have had a significant impact on the parameter space of the model and pushed one of the favoured region for the sneutrino LSP to be near 800 GeV. A $Z_2$ up to several TeV's will be probed when the energy of the LHC is increased to 14 TeV. A negative result from such a search will imply a much reduced rate for Direct Detection searches as well. Indirect searches could also provide a good probe of the RHSN model, these will be investigated separately. 

New data on $Z'$ searches and DM observables put even more stringent constraints on this study. In particular the 2012 limits obtained by the XENON100 collaboration probe almost all the parameter space explored in the scan of the $U(1)_\eta$ parameter space. Nevertheless as said above with the increased lower bound on the $Z_2$ boson the DD cross-section would be reduced in this region of the parameter space. Now with the Higgs boson discovery more constraints can be used on the Higgs sector of the UMSSM. This will be analysed in next chapter.

In this analysis we have assumed that Dirac neutrinos were very light (sub-eV range). As said in section~\ref{sec.7:constraint2} such RH Dirac neutrinos could lead to a faster expansion rate of the Universe and predict too much $^4\mathrm{He}$. The resulting constraints on the mass of the new gauge bosons in the $E_6$ model were analysed in~\cite{Barger:2003zh} and compatibility with BBN resulted into lower limits on the $Z'$ mass in the multi-TeV range assuming the effective number of neutrinos was increased by 0.3. Nevertheless recent LHC bounds remain competitive when we consider a number of additional neutrinos $\Delta N_\nu \geq 0.5$ which is within the 2$\s$ error bars of the last combination of Planck data with WMAP polarization data, high-$\ell$ CMB and BAO data which gives an effective number of neutrino-like relativistic degrees of freedom of $N_\mathrm{eff} = 3.30^{\,+\,0.54}_{\,-\,0.51}$~\cite{Ade:2013lta}.

\chapter{The Higgs sector and low energy observables in the UMSSM}
\label{chapter:B_Higgs_UMSSM}

\minitoc\vspace{1cm}
\newpage

Motivated by the Higgs boson discovery which gives stringent constraints on BSM physics, we reconsider the analysis of Dark Matter properties in the UMSSM taking into account Higgs boson observables as well as new low energy observables and Direct Detection of Dark Matter. To this end we improve the computation of the Higgs boson mass and develop tools to compute low energy observables. We consider a much larger set of observables than in the previous chapter including those that are  known to be relevant in SUSY. The computation of the theoretical predictions for $B$-physics observables in this model is based on several routines contained in the \NTools code \cite{Domingo:2007dx}. We have modified and adapted this code to compute observables within the UMSSM.

\section{The Higgs sector in the UMSSM}

We first describe how the calculation of the corrected Higgs boson masses are modified as compared to chapter~\ref{chapter:RHsneu} and we will see how the Higgs bosons signal strengths are computed in this model. Recall that as mentioned in chapter~\ref{chapter:UMSSM} because of new contributions from $U(1)$ $D$-terms, in addition to those from the superpotential as in the NMSSM, a Higgs boson mass of 125 GeV is easily obtained in this model. 

\subsection{Radiative corrections through an effective potential}

For the study described in chapter~\ref{chapter:RHsneu} we implemented the UMSSM model in \lhep at tree-level in unitary and Feynman gauges and checked gauge invariance. Radiative corrections to the Higgs boson masses were just added in the unitary gauge through a Coleman-Weinberg potential \cite{Coleman:1973jx} and only the dominant (due to their large Yukawa couplings) one-loop contributions from the top quark and the stops loops were considered following \cite{Barger:2006dh}. At the time there was no need for a greater precision. However, after the discovery of a new boson and a precise measurement of its mass, it is important to make a precise theoretical prediction of at least the SM-like Higgs boson. We thus introduced the radiative corrections through the effective Lagrangian approach following the method used in the MSSM \cite{Boudjema:2001ii} and in the NMSSM \cite{Belanger:2005kh}. This ensures a gauge invariant treatment\footnote{Here a restricted number of operators was considered. See \cite{Chalons:2012qe} for an analysis of a general effective scalar potential involving two doublets and a singlet Higgs field.}.

The following effective potential is used :
\beq \begin{split}
V^\mathrm{eff}_\mathrm{UMSSM} = & \; \l_1 |H_d|^4/2 + \l_2 |H_u|^4/2
+ \l_3 |H_d|^2 |H_u|^2 \\
& + \l_4 |H_u \cdot H_d|^2
+ \l_5 ((H_u \cdot H_d)^2+(H_u \cdot H_d)^{\dag 2})/2 \\
& + (\l_6 |H_d|^2 + \l_7 |H_u|^2) ((H_u \cdot H_d) + (H_u \cdot H_d)^\dag) \\
& + \l_s (S H_u \cdot H_d + S^* (H_u \cdot H_d)^\dag).
\end{split} \eeq
Among the $\l$'s, $\l_s$ is the only dimensionful parameter. The $\l$'s contain some of the corrections that are considered in a recent update of the NMSSM model files in the \micro- \NTools interface \cite{Belanger:2013oya}, namely 1- and 2-loop (s)top and (s)bottom corrections, removing all the pure NMSSM terms not present in the UMSSM. In the present work we decide, as in the work done in \cite{Barger:2006dh}, to not consider pure UMSSM corrections as one-loop gauge contributions since the $U(1)'$ gauge coupling is small compared to the Yukawa coupling of the top quark.

In this framework, we have to modify the minimization conditions given in eq.~\ref{eq:minLO} in order to minimize the corrected Higgs potential :
\beq \begin{split}
\left(m^c_{H_d}\right)^2 = & \left(m_{H_d}\right)^2 + \l_s \frac{v_s v_u}{\sqrt{2} v_d} -\l_1 \frac{v_d^2}{2} -(\l_3+\l_4+\l_5) \frac{v_u^2}{2} + 3 \l_6 \frac{v_u v_d}{2} + \l_7 \frac{v_u^3}{2 v_d},\\
\left(m^c_{H_u}\right)^2 = & \left(m_{H_u}\right)^2 + \l_s \frac{v_s v_d}{\sqrt{2} v_u} -\l_2 \frac{v_u^2}{2} -(\l_3+\l_4+\l_5) \frac{v_d^2}{2} + \l_6 \frac{v_d^3}{2 v_u} + 3 \l_7 \frac{v_u v_d}{2},\\
\left(m^c_S\right)^2 = & \left(m_S\right)^2 + \l_s \frac{v_u v_d}{\sqrt{2} v_s}.
\end{split} \eeq \\
We are now able to rewrite the Higgs mass matrices. The corrected CP-even mass-squared matrix elements $(\mathcal{M}_+^c)_{ij}$ are
\beq \begin{split}
\left({\mathcal{M}_{+}^c}\right)_{11} & = \left({\mathcal{M}_{+}^0}\right)_{11} + \l_1 v_d^2 + \left(\l_s \frac{v_s}{\sqrt{2}} -3 \l_6 \frac{v_d^2}{2} + \l_7 \frac{v_u^2}{2}\right)\frac{v_u}{v_d},\\
\left({\mathcal{M}_{+}^c}\right)_{12} & = \left({\mathcal{M}_{+}^0}\right)_{12} +  (\l_3+\l_4+\l_5) v_u v_d -\frac{3}{2}(\l_6 v_d^2 + \l_7 v_u^2) -\l_s \frac{v_s}{\sqrt{2}},\\
\left({\mathcal{M}_{+}^c}\right)_{13} & = \left({\mathcal{M}_{+}^0}\right)_{13} -\l_s \frac{v_u}{\sqrt{2}},\\
\left({\mathcal{M}_{+}^c}\right)_{22} & = \left({\mathcal{M}_{+}^0}\right)_{22} + \l_2 v_u^2 + \left(\l_s \frac{v_s}{\sqrt{2}} + \l_6 \frac{v_d^2}{2} -3 \l_7 \frac{v_u^2}{2}\right)\frac{v_d}{v_u},\\
\left({\mathcal{M}_{+}^c}\right)_{23} & = \left({\mathcal{M}_{+}^0}\right)_{23} -\l_s \frac{v_d}{\sqrt{2}},\\
\left({\mathcal{M}_{+}^c}\right)_{33} & = \left({\mathcal{M}_{+}^0}\right)_{33} + \l_s \frac{v_u v_d}{\sqrt{2} v_s}.
\end{split} \eeq

The CP-odd elements read
\beq \begin{split}\label{eq:oddcorr}
\left({\mathcal{M}_{-}^c}\right)_{11} & = \left({\mathcal{M}_{-}^0}\right)_{11} + \left(\l_6 \frac{v_d^2}{2} + \l_7 \frac{v_u^2}{2} - \l_5 v_u v_d + \l_s \frac{v_s}{\sqrt{2}}\right)\frac{v_u}{v_d},\\
\left({\mathcal{M}_{-}^c}\right)_{12} & = \left({\mathcal{M}_{-}^0}\right)_{12} + \l_6 \frac{v_d^2}{2} + \l_7 \frac{v_u^2}{2} - \l_5 v_u v_d + \l_s \frac{v_s}{\sqrt{2}},\\
\left({\mathcal{M}_{-}^c}\right)_{13} & = \left({\mathcal{M}_{-}^0}\right)_{13} + \l_s \frac{v_u}{\sqrt{2}},\\
\left({\mathcal{M}_{-}^c}\right)_{22} & = \left({\mathcal{M}_{-}^0}\right)_{22} + \left(\l_6 \frac{v_d^2}{2} + \l_7 \frac{v_u^2}{2} - \l_5 v_u v_d + \l_s \frac{v_s}{\sqrt{2}}\right)\frac{v_d}{v_u},\\
\left({\mathcal{M}_{-}^c}\right)_{23} & = \left({\mathcal{M}_{-}^0}\right)_{23} + \l_s \frac{v_d}{\sqrt{2}},\\
\left({\mathcal{M}_{-}^c}\right)_{33} & = \left({\mathcal{M}_{-}^0}\right)_{33} + \l_s \frac{v_u v_d}{\sqrt{2} v_s}.
\end{split} \eeq
which allow us to write the corrected pseudoscalar mass-squared
\beq \begin{split}
\left(m^c_{A^0}\right)^2 = & \left(m_{A^0}\right)^2 + \frac{v^2}{v_s v_u v_d} \left\{ \sqrt{2} \l_s \left(v_s^2 + \left(\frac{v_u v_d}{v}\right)^2\right) - 2 \l_5 v_s v_u v_d + \l_6 v_s v_d^2 + \l_7 v_s v_u^2 \right.\\
& \left. + \sqrt{ \begin{aligned}
& \left[\sqrt{2} \l_s \left(v_s^2 + \left(\frac{v_u v_d}{v}\right)^2\right) - 2 \l_5 v_s v_u v_d + \l_6 v_s v_d^2 + \l_7 v_s v_u^2\right]^2\\
& - 4\sqrt{2} \l_s \left(\frac{v_u v_d}{v}\right)^2 (-2 \l_5 v_s v_u v_d + \l_6 v_s v_d^2 + \l_7 v_s v_u^2)\\
\end{aligned}
} \ \right\} .
\end{split} \eeq
Finally, the charged Higgs mass is corrected as
\beq
\left(m^c_{H^\pm}\right)^2 = \left(m_{H^\pm}\right)^2 + \left(\l_s \frac{v_s}{\sqrt{2}} + \l_6 \frac{v_d^2}{2} + \l_7 \frac{v_u^2}{2}\right) \left(\frac{v_u}{v_d} + \frac{v_d}{v_u}\right) -(\l_4+\l_5) \frac{v^2}{2}.  
\eeq

\subsection{Higgs bosons signal strengths in the UMSSM}
\label{subsec:8.H_coup_UMSSM}

\begin{figure}[!htb]
\begin{center}
\includegraphics[scale=0.74]{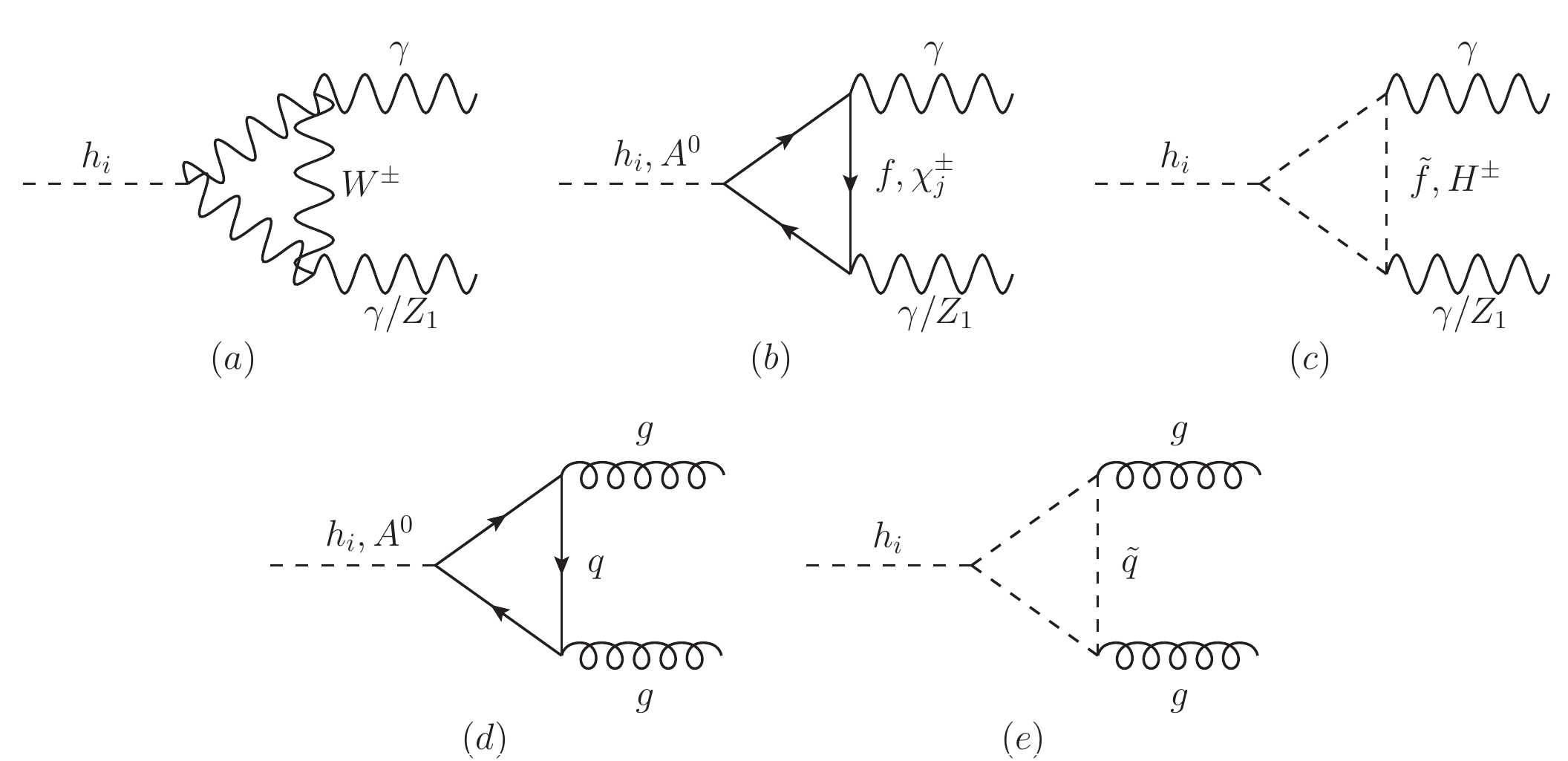} 
  \caption[Triangle one-loop contributions to the $\g\g$ and $\g Z_1$ reduced couplings (a, b, c) and to the $gg$ reduced couplings (d, e).]{Triangle one-loop contributions to the $\g\g$ and $\g Z_1$ reduced couplings (a, b, c) and to the $gg$ reduced couplings (d, e), with $i \in \{1,2,3\}$ and $j \in \{1,2\}$. $f$($\tilde{f}$) stands for (s)fermions and $q$($\tilde{q}$) stands for (s)quarks.}
\label{fig:hloop}
\end{center}
\end{figure}

To compute the signal strength as defined in eq.~\ref{eq:6.R} for the case of Higgs bosons decaying into photons when they are produced by gluon-gluon fusion, we adapted to the UMSSM the {\tt NMHDECAY\;}code contained in \NTools \cite{Ellwanger:2004xm,Ellwanger:2005dv}. Let us look at the determination of the Higgs bosons reduced couplings, relative to a SM Higgs boson with the same mass, which are obtained with tree-level interactions. The reduced Higgs couplings $C^{h_i, A^0}_f$ to a fermion $f$ relative to the SM Higgs coupling read :
\beq \begin{split} \label{eq:9.redcoupfer}
C^{h_i}_{q_d, \ell^\pm} & = \frac{Z_{h i1}}{\cos\beta}, \qquad C^{h_i}_{q_u} = \frac{Z_{h i2}}{\sin\beta},\\
C^{A^0}_{q_d, \ell^\pm} & = \frac{Z_{A 31}}{\cos\beta}, \qquad C^{A^0}_{q_u} = \frac{Z_{A 32}}{\sin\beta},
\end{split} \eeq
where $q_d \, (q_u)$ are down- (up-) type quarks and $\ell^\pm$ corresponds to charged leptons. The reduced couplings of CP-even Higgs bosons to $W$ bosons\footnote{Since the scalar-massive vector coupling is CP-even, there is no such tree-level coupling with a CP-odd Higgs boson.} read as in the NMSSM
\beq C^{h_i}_{W^\pm} = \cos\b Z_{h i1} + \sin\b Z_{h i2}. \eeq
However, unlike the (N)MSSM, the reduced couplings of CP-even Higgs bosons to $Z_1$ bosons are not exactly the same than for $W$'s here since there is a $Z'$ component in $Z_1$. The expression is
\beq \begin{split} C^{h_i}_{Z_1} = & \, \cos^2 \azz (\cos\b Z_{h i1} + \sin\b Z_{h i2})\\
& + 4 \cos \azz \sin \azz \sin \t_W \frac{g'_1}{g_Y} (\cos\b Z_{h i1} - \sin\b Z_{h i2})\\
& + 4 \sin^2 \azz \frac{\sin^2 \t_W}{v} \frac{\gp2}{g_Y^2} \left(\mathcal{Q}'^2_{H_d} v_d Z_{h i1} + \mathcal{Q}'^2_{H_u} v_u Z_{h i2} + \mathcal{Q}'^2_S v_s Z_{h i3} \right).
\end{split} \eeq
Since gluons and photons are massless particles, they do not couple directly to the different Higgs bosons. To get the reduced couplings of Higgs bosons to $gg$, $\g\g$ and $\g Z_1$, respectively $C^{h_i, A^0}_{g}, C^{h_i, A^0}_{\g}$ and $C^{h_i, A^0}_{\g Z_1}$, we must consider charged particle loops\footnote{For a review on the loop contributions to the Higgs boson decays into photons and gluons see \cite{Djouadi:2005gi} for the SM and \cite{Djouadi:2005gj} for the MSSM.}. To do that we again adapted the {\tt NMHDECAY\;}code to the UMSSM. We then consider the loop contributions listed in figure~\ref{fig:hloop}. We also automatically consider QCD radiative corrections for the quark contributions that are included in the {\tt NMHDECAY\;}code.

As can be seen there are contributions which imply the coupling of Higgs bosons to sfermions. Considering up-type squarks, we have the following couplings :
\begin{align}
\lambda_{\tilde{u}_L}^{h_i} = & -\frac{3g_2^2-g_Y^2}{12}\left(v_d Z_{h i1} - v_u Z_{h i2}\right)\nonumber\\
& -\gp2 \mathcal{Q}'_Q (\mathcal{Q}'_{H_d} v_d Z_{h i1} + \mathcal{Q}'_{H_u} v_u Z_{h i2} + \mathcal{Q}'_S v_s Z_{h i3}),\\
\lambda_{\tilde{u}_R}^{h_i} = & -\frac{g_Y^2}{3}\left(v_d Z_{h i1} - v_u Z_{h i2}\right)\nonumber\\
& -\gp2 \mathcal{Q}'_u (\mathcal{Q}'_{H_d} v_d Z_{h i1} + \mathcal{Q}'_{H_u} v_u Z_{h i2} + \mathcal{Q}'_S v_s Z_{h i3}).
\end{align}
where we recall that the $\mathcal{Q}$'s are given in table \ref{tab:Ucharge}. In the same way we have for down-type squarks :
\begin{align}
\lambda_{\tilde{d}_L}^{h_i} = & \, \frac{3g_2^2+g_Y^2}{12}\left(v_d Z_{h i1} - v_u Z_{h i2}\right)\nonumber\\
& -\gp2 \mathcal{Q}'_Q (\mathcal{Q}'_{H_d} v_d Z_{h i1} + \mathcal{Q}'_{H_u} v_u Z_{h i2} + \mathcal{Q}'_S v_s Z_{h i3}),\\
\lambda_{\tilde{d}_R}^{h_i} = & \, \frac{g_Y^2}{6}\left(v_d Z_{h i1} - v_u Z_{h i2}\right)\nonumber\\
& -\gp2 \mathcal{Q}'_d (\mathcal{Q}'_{H_d} v_d Z_{h i1} + \mathcal{Q}'_{H_u} v_u Z_{h i2} + \mathcal{Q}'_S v_s Z_{h i3}).
\end{align}
For the charged sleptons, the expressions read :
\begin{align}
\lambda_{\tilde{e}_L}^{h_i} = & \, \frac{g_2^2-g_Y^2}{4}\left(v_d Z_{h i1} - v_u Z_{h i2}\right)\nonumber\\
& -\gp2 \mathcal{Q}'_Q (\mathcal{Q}'_{H_d} v_d Z_{h i1} + \mathcal{Q}'_{H_u} v_u Z_{h i2} + \mathcal{Q}'_S v_s Z_{h i3}),\\
\lambda_{\tilde{e}_R}^{h_i} = & \, \frac{g_Y^2}{2}\left(v_d Z_{h i1} - v_u Z_{h i2}\right)\nonumber\\
& -\gp2 \mathcal{Q}'_e (\mathcal{Q}'_{H_d} v_d Z_{h i1} + \mathcal{Q}'_{H_u} v_u Z_{h i2} + \mathcal{Q}'_S v_s Z_{h i3}).
\end{align}
Therefore we obtain for the third generations :
\begin{align}
\lambda_{\tilde{t}_j}^{h_i} = & \left[\lambda_{\tilde{u}_L}^{h_i} - h_t^2 v_u Z_{h i2}\right] Z^2_{\tilde{t} j1}
+ \left[\lambda_{\tilde{u}_R}^{h_i} - h_t^2 v_u Z_{h i2}\right] Z^2_{\tilde{t} j2}\nonumber\\
& + h_t\left[\sqrt{2}(\mu Z_{h i1} - A_t Z_{h i2}) + \lambda v_d Z_{h i3}\right]Z_{\tilde{t} j1}Z_{\tilde{t} j2},\label{eq:8.hstop}\\ 
\lambda_{\tilde{b}_j}^{h_i} = & \left[\lambda_{\tilde{d}_L}^{h_i} - h_b^2 v_d Z_{h i1}\right] Z^2_{\tilde{b} j1}
+ \left[\lambda_{\tilde{d}_R}^{h_i} - h_b^2 v_d Z_{h i1}\right] Z^2_{\tilde{b} j2}\nonumber\\
& + h_b\left[\sqrt{2}(\mu Z_{h i2} - A_b Z_{h i1}) + \lambda v_u Z_{h i3}\right]Z_{\tilde{b} j1}Z_{\tilde{b} j2},\label{eq:8.hsbottom}\\ 
\lambda_{\tilde{\tau}_j}^{h_i} = & \left[\lambda_{\tilde{e}_L}^{h_i} - h_\tau^2 v_d Z_{h i1}\right] Z^2_{\tilde{\tau} j1}
+ \left[\lambda_{\tilde{e}_R}^{h_i} - h_\tau^2 v_d Z_{h i1}\right] Z^2_{\tilde{\tau} j2}\nonumber\\
& + h_\tau\left[\sqrt{2}(\mu Z_{h i2} - A_\tau Z_{h i1}) + \lambda v_u Z_{h i3}\right]Z_{\tilde{\tau} j1}Z_{\tilde{\tau} j2},\label{eq:8.hstau}
\end{align}
where $h_f$ ($A_f$) with $f \in \{t,b,\tau\}$ are the Yukawa (trilinear) couplings and the $Z_{\tilde{f} jk}$ with $\tilde{f} \in \{\tilde{t},\tilde{b},\tilde{\tau}\}$ and $j,k \in \{1,2\}$ are the elements of the diagonalisation matrices in the sfermion sector which stem from the mixing between the chirality eigenstates $\tilde{F} = (\tilde{f}_L,\tilde{f}_R)$ with the mass eigenstates defined as $\tilde{f}_j=Z_{\tilde{f} jk} \tilde{F}_k$. As in the \NTools code we compute the running Yukawa couplings, $\l$ as well as the Higgs VEVs at the SUSY scale.\\

For the contributions in figure~\ref{fig:hloop}b, we need the couplings of the neutral Higgs bosons to the charginos. In the UMSSM they read :
\begin{align}
\lambda_{\chi^{\pm}_k}^{h_i} & = -\frac{1}{\sqrt{2}}\left[g_2 (Z_{u k2}Z_{v k1}Z_{h i1} +
Z_{u k1}Z_{v k2}Z_{h i2}) + \lambda Z_{u k2} Z_{v k2} Z_{h i3}\right],\label{eq:8.hevenchar}\\
\lambda_{\chi^{\pm}_k}^{A^0} & = -\frac{1}{\sqrt{2}}\left[g_2 (Z_{u k2}Z_{v k1}Z_{A 31} +
Z_{u k1}Z_{v k2}Z_{A 32}) - \lambda Z_{u k2} Z_{v k2} Z_{A 33}\right],\label{eq:8.hoddchar}
\end{align}
where we recall that the rotation matrices in the chargino sector $\mathbf{Z_u}$ and $\mathbf{Z_v}$ are defined as in the MSSM by $\chi^-_k=Z_{u kl}\psi^-_l$ and $\chi^+_k=Z_{v kl}\psi^+_l$, with $k,l \in \{1,2\}$ and the gauge eigenstates $\psi^-=(\widetilde{W}^-,\widetilde{H}_d^-)$ and  $\psi^+=(\widetilde{W}^+,\widetilde{H}_u^+)$.\\

Now that we saw the main contributions that enters in the calculation of the reduced couplings of Higgs bosons to $gg$, $\g\g$ and $\g Z_1$, we can compute an example of Higgs bosons signal strength. Here we look at the case of an Higgs boson decaying into photons when it is produced by gluon-gluon fusion. The signal strength reads 
\beq \mu_{H \ra \g\g}^{\rm ggF} =  {C_g^H}^2 \frac{\mathscr{B}(H\rightarrow \g\g)_{\mathrm{UMSSM}}}{\mathscr{B}(H\rightarrow \g\g)_{\mathrm{SM}}},\eeq
where $H \in \{h_1, h_2, h_3, A^0\}$. To compute the $H$ total width and then each decays we also use the {\tt NMHDECAY\;}routine with the couplings defined above. For some Higgs decays, like in other Higgs bosons, we use the Feynman rules derived with \lhep which are given to our modified {\tt NMHDECAY\;}routine.

\section{Flavour constraints on the UMSSM}

Indirect constraints coming from the flavour sector, especially those involving $B$-mesons, have been shown to play an important role in defining the allowed parameter space of BSM and especially supersymmetric models,\eg \cite{Bediaga:2012py,Bernal:2011pj,Arbey:2012ax,Domingo:2007dx}. Recent progress both on the experimental side (better determination of CKM matrix elements, improvement in the search for rare decays) and on the theoretical side (precise determination of several hadronic parameters using lattice QCD simulations, better estimation of SM contributions) force us to add more constraints coming from $B$-physics as compared to our previous study discussed in chapter~\ref{chapter:RHsneu}. Many of these constraints are generalized straightforwardly from the NMSSM case adapting the \NTools routines that calculate these observables to our model \cite{Domingo:2007dx}. By looking at MFV (namely flavour violation occurs only through the CKM matrix), we will show in the following how the Higgs sector and the supersymmetric particles of the UMSSM can impact the observables $\btau$, $\bsmu$, $\Delta M_s$, $\Delta M_d$, $\bsg$ and $\bXsmu$.

\subsection{$\btau$}

\begin{figure}[!htb]
\begin{center}
\includegraphics{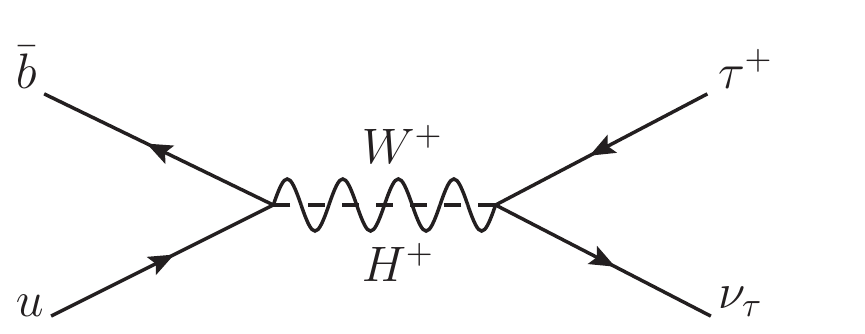} 
  \caption{$W$ boson (plain) and charged Higgs (dashed) contribution to $\mathscr{B}(B^+ \to \tau^+ \nu_\tau)$.}
\label{fig:btaunu}
\end{center}
\end{figure}

The branching ratio $\btau$ has the specificity to get a BSM contribution directly at tree level (figure~\ref{fig:btaunu}). This contribution is the same as in the MSSM, coming from a charged Higgs, so no peculiar pure UMSSM contribution appears here. A destructive interference is obtained between the $W^\pm$ and $H^\pm$ contributions and implies a modification in the expression for the branching ratio :
\beq
\mathscr{B}(B^\pm\ra\tau^\pm\nu_{\tau})=\frac{G_F^2M_{B^\pm}m_{\tau^\pm}^2}{8\pi}
\left(1-\frac{m_{\tau^\pm}^2}{M_{B^\pm}^2}\right)^2f_B^2
\Vub^2\tau_{B^\pm}\,r_H\,,
\eeq
with the parameterization of the deviation from the SM calculation given by $r_H$ whose expression, useful for large $\tan\beta$ values, was obtained in \cite{Akeroyd:2003zr} :
\beq
r_H=\left[1-\left(\frac{M_{B^\pm}}{m_{H^{\pm}}}\right)^2
\frac{\tan^2\beta}{1+\tilde{\varepsilon}_0\tan\beta}\right]^2,
\eeq
where the $\tilde{\varepsilon}_0$ parameter is computed using sparticle loops as defined in the MSSM \cite{Buras:2002vd} and the contribution coming from the neutralino sector is a straightforward generalization to the UMSSM of the expressions in the NMSSM \cite{Domingo:2007dx}.\\

The values of the different SM parameters are given in table~\ref{tab:btaunu}. In our preliminary scan these parameters will be fixed. However some of them could be considered as nuisance parameters as we did in the study described in chapter~\ref{chapter:ID} because of their large uncertainty, especially in the case of the puzzle in the inclusive/exclusive measurements of the CKM matrix element $\Vub$ (see\eg \cite{Bharucha:2013sn}).

\begin{table}[!htb]
\begin{center}
\begin{tabular}{cc}\hline \hline
\textbf{Parameter} & \textbf{Value} \\ \hline \hline
$G_F$ & 1.1663787 10$^{-5}$ GeV$^{-2}$\\ 
$\alpha_s(M_{Z_1})$ & 0.1184 $\pm$ 0.0007\\ 
$M_{B^\pm}$ & 5279.25 MeV\\ 
$f_B$ & 189 $\pm$ 4 MeV \cite{Na:2012kp}\\
$\Vub$ & (4.15 $\pm$ 0.49)$\times$10$^{-3}$\\ 
$\tau_{B^\pm}$ & (1.641 $\pm$ 0.008)$\times$10$^{-12}$ s\\ \hline \hline
\end{tabular}
\caption[SM parameters entering in the $\btau$ calculation and their uncertainty if non-negligible.]{SM parameters entering in the $\btau$ calculation and their uncertainty if non-negligible. From \cite{Beringer:1900zz} unless noted otherwise.}
\label{tab:btaunu}
\end{center}
\end{table}

Past discrepancy between the SM and experimental values of this branching ratio is now less obvious with the new result from the Belle Collaboration \cite{Adachi:2012mm}; a recent world average provided by the UTfit collaboration \cite{Tarantino:2012mq} reads
\beq
\mathscr{B}(B^\pm\ra\tau^\pm\nu_{\tau})_\textrm{exp} = (0.99 \pm 0.25)\times 10^{-4}.
\eeq
Meanwhile the CKMfitter global fit result provided at the ICHEP 2012 conference \cite{CKMfitICHEP} gives for the SM contribution
\beq
\mathscr{B}(B^\pm\ra\tau^\pm\nu_{\tau})_\textrm{SM} = (0.796^{\,+\,0.088}_{\,-\,0.087})\times 10^{-4}.
\eeq
Since we consider BSM effects on low energy observables, we will not use throughout this work the global SM fits of CKM matrix elements \cite{CKMfit,UTfit}. Then some discrepancies are expected between our \textit{SM-only} prediction of a low energy observable and the usual SM precise predictions. Here the discrepancy is mainly driven by $\Vub$; we get
\beq
\mathscr{B}(B^\pm\ra\tau^\pm\nu_{\tau})_{\textrm{SM-only}} = (1.09 \pm 0.26)\times 10^{-4}.
\eeq
We will then carefully analyse the effects of supersymmetric contributions by looking at $\tan\beta$ and $m_{H^\pm}$.

\subsection{$\bsmu$}
\label{sec:bsmu}

The branching ratio $\bsmu$ is known as one of the best observables to probe BSM effects on low energy physics, especially at the LHC. The LHCb collaboration released at the end of 2012 the first evidence for the decay $B^0_s \to \mu^+ \mu^-$ \cite{:2012ct}. They get
\beq
\bsmu_\textrm{exp} = (3.2^{\,+\,1.5}_{\,-\,1.2}) \times 10^{-9}.
\eeq
Comparing to recent SM predictions
\beq \begin{split}
\bsmu_\textrm{SM,1} = &  \, (3.23 \pm 0.27 ) \times 10^{-9} \, \textrm{\cite{Buras:2012ru}},\\
\bsmu_\textrm{SM,2} = &  \, (3.53 \pm 0.38 ) \times 10^{-9} \, \textrm{\cite{Mahmoudi:2012un}},\\
\bsmu_\textrm{SM,3} = &  \, (3.25 \pm 0.17 ) \times 10^{-9} \, \textrm{\cite{Buras:2013uqa}},
\end{split}\label{eq:SMbsmumu}\eeq
whose discrepancies come from the decay constant $f_{B^0_s}$ and the mean life $\tau_{B^0_s}$ choices, it leaves little room for BSM contributions in SUSY.\\

The calculation of this branching ratio, following the same assumptions as in \cite{Domingo:2007dx} reads 

\beq \begin{split}
& \mathscr{B}(B^0_s\ra\mu^+\mu^-)= \\
& \frac{G_F^2 \alpha_{_{em}}^2 M_{B^0_s}^5 f_{B^0_s}^2 \tau_{B^0_s}}
{64\pi^3\sin^4 \theta_W}\Vtbts^2
\sqrt{1-4\frac{m_{\mu^\pm}^2}{m_{B^0_s}^2}}
\left[\frac{1-4\frac{m_{\mu^\pm}^2}{M_{B^0_s}^2}}
{\left(1+\frac{m_s}{m_b}\right)^2}\left|c_S\right|^2
+\left|\frac{c_P}{1+\frac{m_s}{m_b}}+\frac{2m_{\mu^\pm}}{M_{B^0_s}^2}
c_A\right|^2\right].
\end{split}\eeq

\begin{figure}[!htb]
\begin{center}
\includegraphics[scale=0.8]{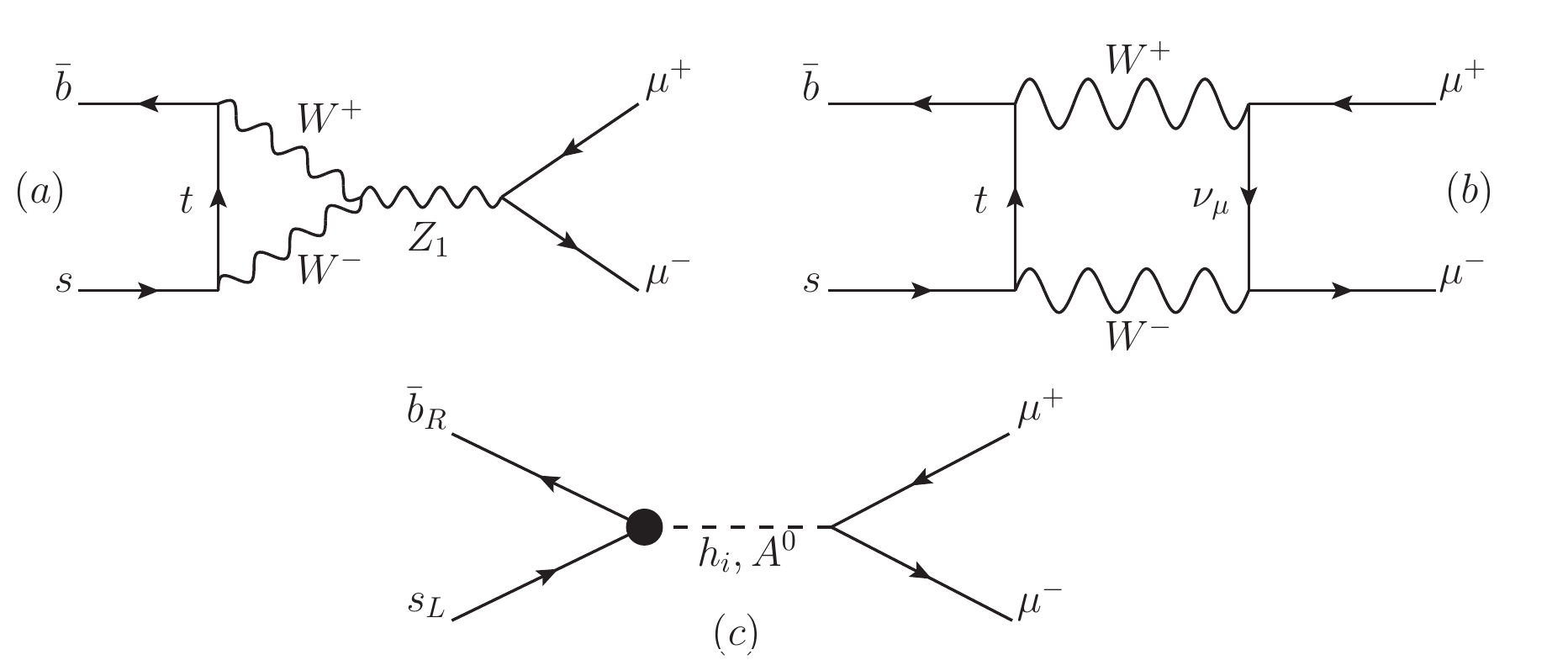} 
  \caption[Main SM (a, b) and UMSSM (c) contributions to $\bsmu$. The dot denote a flavour changing coupling.]{Main SM (a, b) and UMSSM (c) contributions to $\bsmu$ where $i \in \{1,2,3\}$. The dot denote a flavour changing coupling.}
\label{fig:bsmumu}
\end{center}
\end{figure}

In the SM, the main contributions arise from box and penguin diagrams as shown in figure~\ref{fig:bsmumu}a,b and are defined in the $c_A$ term while $c_S$ and $c_P$ include the neutral Higgs boson contributions. We assume the standard coupling of the $Z_1$ to $W$ bosons and muons and we disregard the $Z_2$ contribution, which makes sense given the upper bound we get on $\azz$ ($\azz \lesssim 10^{-3}$) and stringent lower bound on $M_{Z_2}$. Contributions from the new $U(1)$ symmetry are then safely neglected here.\\ 
The SM calculation is dependent on the uncertainty on the top quark mass since it includes the ratio $x_t = \left(\frac{m_t^{\overline{\textrm{MS}}}}{M_W}\right)^2$, where $m_t^{\overline{\textrm{MS}}}$ is the $\overline{\textrm{MS}}$ top quark mass.\\

Neutral Higgs boson contributions easily imply large deviation from the SM prediction through an effective flavour violating neutral Higgs vertex and a neutral Higgs propagator as shown in figure~\ref{fig:bsmumu}c. It happens if $\tan\beta$ is large ($\gg$1) and the heavy doublet-like Higgs is not so heavy (a few hundreds of GeV). These terms can be approximate in the UMSSM, given that we never get very light (some GeV's) neutral Higgs bosons, by
\beq \begin{split}
c_S \simeq & -\frac{\sqrt{2}g_2\pi^2m_{\mu^\pm}}{G_F^2M_W^3} \sum^3_{i=1} \frac{(Z_{h i2}-Z_{h i1}\tan\beta)Z_{h i1}\varepsilon_Y^{bs}}{m^2_{h_i} v\sin 2\beta (1+\tilde{\varepsilon}_0\tan\beta)(1+\tilde{\varepsilon}_3\tan\beta)},\\
c_P \simeq & +\frac{\sqrt{2}g_2\pi^2m_{\mu^\pm}}{G_F^2M_W^3} \frac{(Z_{A 32}-Z_{A 31}\tan\beta)Z_{A 31}\varepsilon_Y^{bs}}{m^2_{A^0} v\sin 2\beta (1+\tilde{\varepsilon}_0\tan\beta)(1+\tilde{\varepsilon}_3\tan\beta)},
\end{split}\label{eq:wilscoef}\eeq
where the $\varepsilon$'s are defined in \cite{Buras:2002vd}.\\
\begin{table}[!htb]
\begin{center}
\begin{tabular}{cc}\hline \hline
\textbf{Parameter} & \textbf{Value} \\ \hline \hline
$\alpha_{_{em}}$ & 1/137.035999074\\ 
$\sin^2 \theta_W (\overline{\textrm{MS}})$ & 0.23116\\ 
$M_{B^0_s}$ & 5366.77 MeV\\ 
$f_{B^0_s}$ & 225 $\pm$ 4 MeV \cite{McNeile:2011ng}\\
$\left|V_{cb}\right|$ & (40.9 $\pm$ 1.1)$\times$10$^{-3}$\\
$\left|V_{ts}\right|$ & (42.9 $\pm$ 2.6)$\times$10$^{-3}$\\ 
$\tau_{B^0_s}$ & (1.497 $\pm$ 0.015)$\times$10$^{-12}$ s\\ \hline \hline
\end{tabular}
\caption[A list of some of the parameters appearing in the determination of $\bsmu$ and their uncertainty if non-negligible. See table~\ref{tab:btaunu} for other relevant parameters.]{A list of some of the parameters appearing in the determination of $\bsmu$ and their uncertainty if non-negligible. From \cite{Beringer:1900zz} unless noted otherwise. See table~\ref{tab:btaunu} for other relevant parameters.}
\label{tab:bsmumu}
\end{center}
\end{table}

We consider throughout this work the $\overline{\textrm{MS}}$ value of the weak-mixing angle and the SM particle masses given in table~\ref{tab:SM}. We will also keep in our study the most precise determination of the $B^0_s$ leptonic decay constant. We summarize some of the parameters entering in the $\bsmu$ calculation in table~\ref{tab:bsmumu}. By assuming the unitarity of the CKM matrix we are able to determine $\left|V_{tb}\right|$ : 
\beq
\sum_{i \in \{u,c,t\}} \left|V_{ib}\right|^2 = 1.
\eeq
Using the experimental values of $\left|V_{ub}\right|$ and $\left|V_{cb}\right|$ given in table~\ref{tab:btaunu}~and~\ref{tab:bsmumu} we get
\beq
\left|V_{tb}\right| = 0.99915 \pm 0.00027.
\eeq
Note that this number is much more precise that those extracted from measurements of single top quark production cross section \cite{Beringer:1900zz}.\\
Taking into account the main uncertainties coming from $f_{B^0_s}$ and from the experimental determination of $\left|V_{ts}\right|$ and $m^{\textrm{pole}}_{t}$ leads to a SM prediction with a larger error than in eq.~\ref{eq:SMbsmumu} :
\beq
\bsmu_\textrm{SM-only} = (3.14 \pm 0.40) \times 10^{-9}.
\eeq

In our study we will also consider a theoretical uncertainty of 10\%.

\subsection{$\Delta M_s$ and $\Delta M_d$}

\begin{figure}[!htb]
\begin{center}
\includegraphics[scale=0.8]{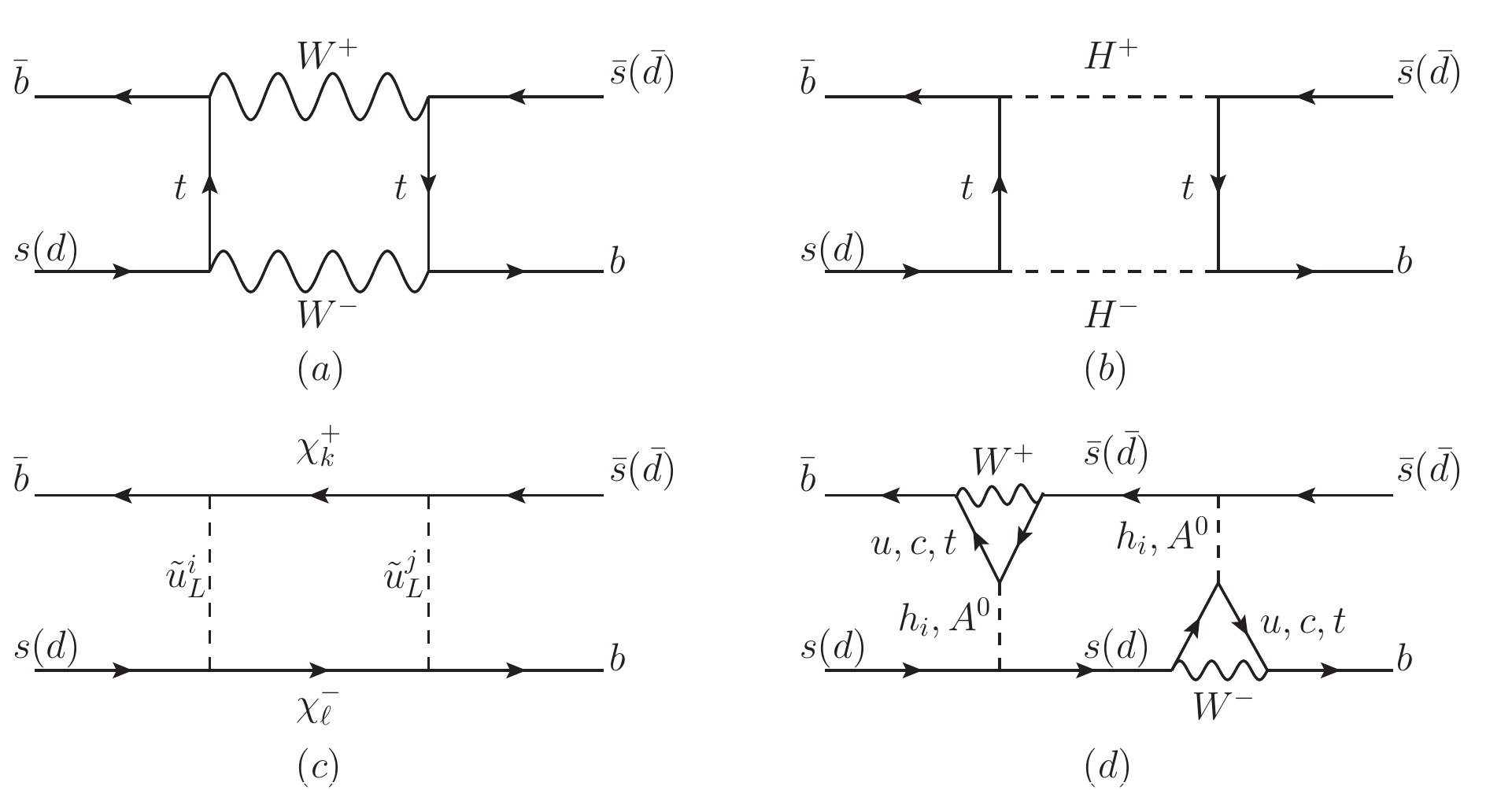} 
  \caption[Main SM (a) and UMSSM (b, c, d) diagrams contributing to $\Delta M_s$ and $\Delta M_d$.]{Main SM (a) and UMSSM (b, c, d) diagrams contributing to $\Delta M_s$ and $\Delta M_d$. $i,j \in \{1,2,3\}$ and $k,\ell \in \{1,2\}$ for the diagrams c and d. $\tilde{u}_L^{i,j}$ stands for LH up squarks.}
\label{fig:DeltaMds}
\end{center}
\end{figure}

Still in the $B$-mesons sector, we now look at the oscillations between the mesons $B^{0}_{q}$ and $\bar{B}^{0}_{q}$ where $q$ $\in$ $\{s,d\}$\footnote{Note that in the standard notation we have $B^{0}_{d} \equiv B^{0}$ ($\bar{B}^{0}_{d} \equiv \bar{B}^{0}$).}. The frequency of these oscillations is given by the mass difference relation
\beq
\Delta M_q = \frac{G_F^2M_W^2}{6\pi^2} M_{B^0_q} \eta_B
f_{B^0_q}^2\hat{B}_{B^0_q}\left|V_{tq}^{*}V_{tb}\right|^2\left|F^q_{tt}\right|,
\eeq
where $\left|F^q_{tt}\right|$ can be decomposed into four main parts, $F^q_{tt} = \Delta_\textrm{SM}+\Delta_{H^\pm}+\Delta^q_{\chi^\pm}+\Delta^q_\textrm{DP}$ with :
\begin{itemize}
\item The SM contribution with quark/$W^{\pm}$ box diagrams (see e.g figure~\ref{fig:DeltaMds}a). Below we show the main contribution from top quarks :
\beq
\Delta_\textrm{SM} \simeq \frac{1}{4}+\frac{9}{4(1-x_t)}-\frac{3}{2(1-x_t)^2}-\frac{3x_t^2\ln(x_t)}{2(1-x_t)^3},
\eeq
where $x_t =  \left(\frac{m_t^{\overline{\textrm{MS}}}}{M_W}\right)^2$;
\item The charged Higgs contribution, as in figure~\ref{fig:DeltaMds}b, is a powerful tool to exclude some regions of the UMSSM parameter space as it can easily exceed the experimental measurements, especially for $\Delta M_s$ and for very low $\tan\beta$ values ($\lesssim$ 1) as we saw in chapter~\ref{chapter:RHsneu} :
\beq
\Delta_{H^\pm} = f_1(m_{H^\pm})\cot^2\beta+f_2(m_{H^\pm})\cot^4\beta.
\eeq
Note that $f_1$ and $f_2$ decrease if the charged Higgs mass increases \cite{Bertolini:1990if};
\item Another SUSY contribution with squark/chargino box diagrams (see figure~\ref{fig:DeltaMds}c) was computed in \cite{Bertolini:1990if}. The main $\Delta^q_{\chi^\pm}$ dependence on CKM matrix element is on the ratios $\left|V_{cs}^{*}V_{cb}\right|/\left|V_{ts}^{*}V_{tb}\right|$ and $\left|V_{ub}^{*}V_{ud}\right|/\left|V_{td}^{*}V_{tb}\right|$ respectively for $\Delta M_s$ and $\Delta M_d$;
\item The last term concerns the Double Penguin diagrams contribution $\Delta^q_\textrm{DP}$ as in figure~\ref{fig:DeltaMds}d. In the UMSSM, the Wilson coefficients are modified in the same way as in eq.~\ref{eq:wilscoef}. This contribution is only interesting for large $\tan\beta$ values. Full expression is given in \cite{Bertolini:1990if};
\end{itemize}
A theoretical uncertainty of 10\% is also added here.
Some of the parameters entering in the $\Delta M_q$ calculation are listed in table~\ref{tab:DeltaMds}.\\

\begin{table}[!htb]
\begin{center}
\begin{tabular}{cc}\hline \hline
\textbf{Parameter} & \textbf{Value} \\ \hline \hline
$\eta_B$ & 0.551 \cite{Buras:2002vd}\\ 
$M_{B^0_d}$ & 5279.58 MeV \cite{Beringer:1900zz}\\ 
$\left|V_{ud}\right|$ & 0.97425 \cite{Beringer:1900zz}\\ 
$\left|V_{td}\right|$ & (8.4 $\pm$ 0.6)$\times$10$^{-3}$ \cite{Beringer:1900zz}\\
$f_{B^0_s}\sqrt{\hat{B}_{B^0_s}}$ & 258.7 $\pm$ 6.3 MeV \cite{McNeile:2011ng,CKMfitICHEP}\\
$f_{B^0_d}\sqrt{\hat{B}_{B^0_d}}$ & 214.8 $\pm$ 9.4 MeV \cite{Na:2012kp,CKMfitICHEP}\\ \hline \hline
\end{tabular}
\caption{Important input parameters in the $\Delta M_s$ and $\Delta M_d$ calculation and their uncertainty if non-negligible. See table~\ref{tab:btaunu} or table~\ref{tab:bsmumu} for other relevant parameters.}
\label{tab:DeltaMds}
\end{center}
\end{table}
The experimental measurements of the $\Delta M_q$'s given in \cite{Amhis:2012bh} are
\beq \begin{split}
(\Delta M_s)_\textrm{exp} = 17.719 \, \pm & \, 0.043 \, \textrm{ps}^{-1},\\
(\Delta M_d)_\textrm{exp} = 0.507 \, \pm & \, 0.004 \, \textrm{ps}^{-1}.
\end{split}\eeq
Assuming the unitarity of the CKM matrix ($\left|V_{cs}\right|$ = 0.973) our $(\Delta M_q)_{\textrm{SM-only}}$ results read
\beq \begin{split}
(\Delta M_s)_\textrm{SM-only} = 19.1 \, \pm & \, 2.5 \, \textrm{ps}^{-1},\\
(\Delta M_d)_\textrm{SM-only} = 0.497 \, \pm & \, 0.083 \, \textrm{ps}^{-1}.
\end{split}\eeq
Our choice of values for the CKM elements (respectively $\left|V_{td}\right|$ and $\left|V_{ts}\right|$) make a better (worse) agreement between $(\Delta M_d)_\textrm{SM-only}$ and $(\Delta M_d)_\textrm{exp}$ ($(\Delta M_s)_\textrm{SM-only}$ and $(\Delta M_s)_\textrm{exp}$) than for the fit from the CKMFitter group \cite{CKMfitICHEP} :
\beq \begin{split}
(\Delta M_s)_\textrm{SM} = 17.3&^{\,+\,2.4}_{\,-\,1.7} \, \textrm{ps}^{-1},\\
(\Delta M_d)_\textrm{SM} = 0.558&^{\,+\,0.045}_{\,-\,0.064} \, \textrm{ps}^{-1}.
\end{split}\eeq

\subsection{$\bsg$}
\label{sec:bsg}

\begin{figure}[!htb]
\begin{center}
\includegraphics[scale=0.8]{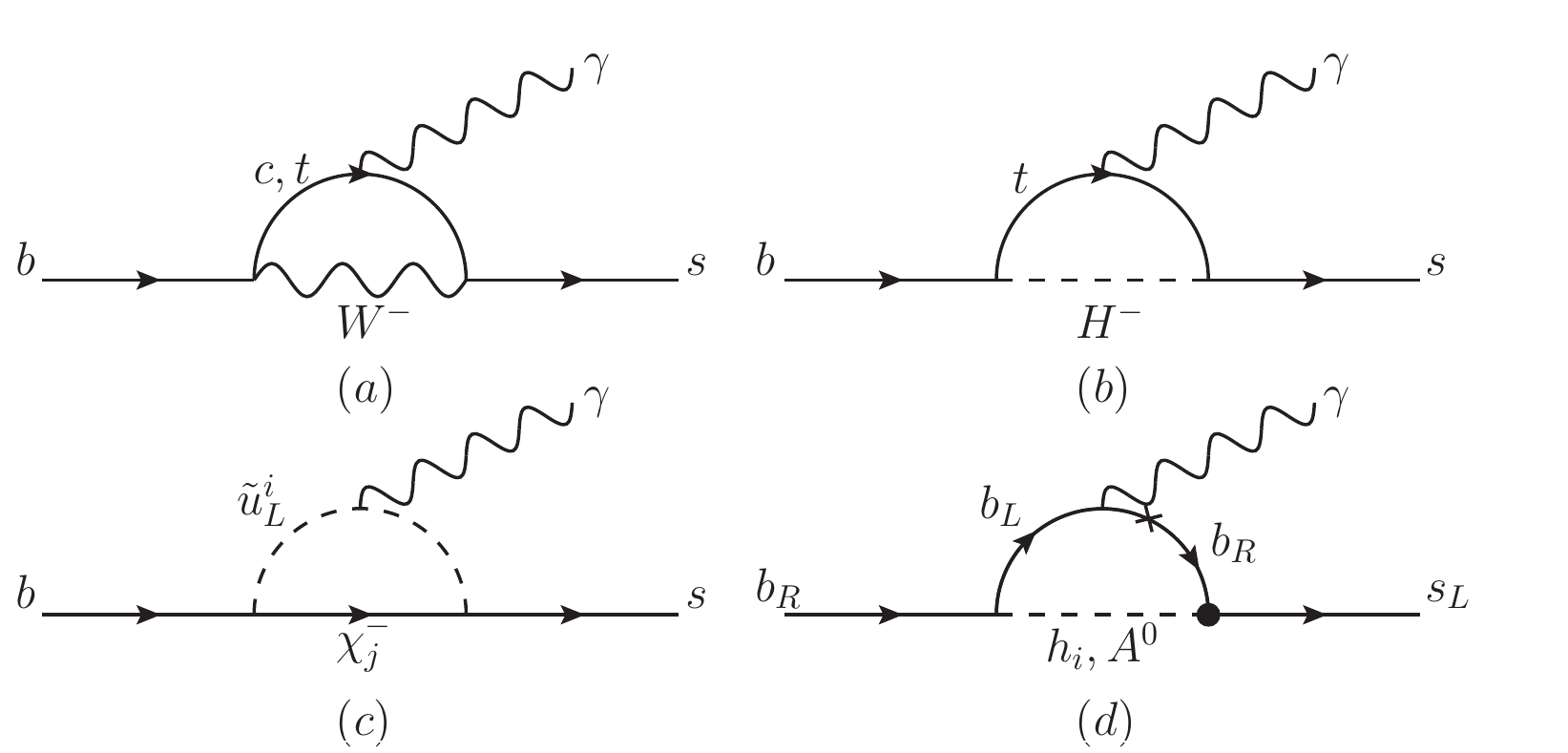} 
  \caption[Some contributions to the $K_c$ and $K_{t+\textrm{BSM}}$ terms for the $\bsg$ calculation. The dot and the cross in the lower right diagram denote respectively a flavour changing coupling and an helicity flip.]{Some contributions to the $K_c$ and $K_{t+\textrm{BSM}}$ terms for the $\bsg$ calculation. $i \in \{1,2,3\}$ and $j \in \{1,2\}$ for the diagram c and the specific UMSSM contribution in the diagram d. The dot and the cross in the lower right diagram denote respectively a flavour changing coupling and an helicity flip.}
\label{fig:bsg}
\end{center}
\end{figure}

Again following \cite{Domingo:2007dx}, we go now to the $\bsg$ calculation including BSM effects. The expression reads :

\beq \begin{split}
& \bsg_{E_\gamma > E_0} =\\
& \mathscr{B}(\bar{B}^0 \to X_c \ell^- \bar{\nu}_\ell)
\VtsbVcb^2 \frac{6\alpha_{_{em}}}{\pi C}
\left[\left|K_c+r(m_t^{\overline{\textrm{MS}}})K_{t+\textrm{BSM}}+\e_{ew}\right|^2+B(E_0)+N(E_0)\right]\,,
\end{split}\label{eq:bsg}\eeq
where $E_0$ = 1.6 GeV is a lower cutoff on the photon energy $E_\gamma$, $\e_{ew}$ denotes the EW radiative
corrections, $B(E_0)$ corresponds to bremsstrahlung contributions and $N(E_0)$ are nonperturbative corrections.
$K_{c}$ includes the charm quark contribution whereas $K_{t+\textrm{BSM}}$ includes the top quark and BSM ones, see figure~\ref{fig:bsg}. $r(m_t^{\overline{\textrm{MS}}})$ is the ratio $m_b^{\overline{\textrm{MS}}}(m_t^{\overline{\textrm{MS}}})/m^{1S}_{b}$. Again as some previous constraints, $\bsg$ can be a powerful way to exclude corners of the parameter space as large $\tan\b$ values through Higgs bosons contributions.\\

To calculate the $K_t$ terms and some bremsstrahlung contributions, we must determine the ratio of CKM matrix elements $\e_s = V^*_{us}V_{ub}/(V^*_{ts}V_{tb})$.
To do this we expand the CKM matrix elements using the standard parameterization in term of the three Cabibbo angles and the phase $\delta$ :
\beq \begin{split}
V_{us} & = \sin \theta_{12}\cos \theta_{13},\\
V_{ub} & = \sin \theta_{13}\, e^{-i\delta},\\
V_{ts} & = -\cos \theta_{12}\sin \theta_{23} -\sin \theta_{12}\sin \theta_{13}\cos \theta_{23}\, e^{i\delta},\\
V_{tb} & = \cos \theta_{13}\cos \theta_{23},
\end{split}\eeq
where
\beq \begin{split}
\sin \theta_{13} & = \left|V_{ub}\right|,\\
\cos \theta_{12} & = \frac{\left|V_{ud}\right|}{\cos \theta_{13}},\\
\sin \theta_{23} & = \frac{\left|V_{cb}\right|}{\cos \theta_{13}},\\
\delta & = 2 \arctan \left(\frac{1-\sqrt{1-(a^2-1)\tan^2 \gamma}}{(a-1)\tan \gamma}\right), \, \cos \gamma \geq 0,\\
a & = \frac{\cos \theta_{12}\sin \theta_{13}\sin \theta_{23}}{\sin \theta_{12}\cos \theta_{23}}.
\end{split}\eeq

Then, using the experimental measurement of the angle $\gamma$ and some CKM elements we find
\beq
\e_s = -0.0092^{\,+\,0.0038}_{\,-\,0.0042} + i\ 0.0214^{\,+\,0.0030}_{\,-\,0.0031},
\eeq 
where large uncertainties come from the experimental measurement of the angle $\gamma$ and from $\left|V_{ub}\right|$.

\begin{table}[!htb]
\begin{center}
\begin{tabular}{cc}\hline \hline
\textbf{Parameter} & \textbf{Value} \\ \hline \hline
$\mathscr{B}(\bar{B}^0 \to X_c \ell^- \bar{\nu}_\ell)$ & 0.1051 $\pm$ 0.0013 \cite{Amhis:2012bh}\\ 
$C$ & 0.546$^{\,+\,0.023}_{\,-\,0.033}$ \cite{Gambino:2008fj}\\ 
$\gamma$ & (68$^{\,+\,10}_{\,-\,11}$)$^\textrm{o}$\\ 
$m^{1S}_{b}$ & 4 650 $\pm$ 30 MeV\\ 
$\alpha^{-1}_{_{em,M_{Z_1}}}$ & 127.944 $\pm$ 0.014\\ \hline \hline
\end{tabular}
\caption[Some values for the calculation of $\bsg$ and the associated uncertainties if they are non-negligible. See table~\ref{tab:btaunu}, table~\ref{tab:bsmumu} or table~\ref{tab:DeltaMds} for other relevant parameters.]{Some values for the calculation of $\bsg$ and the associated uncertainties if they are non-negligible, from \cite{Beringer:1900zz} unless noted otherwise. See table~\ref{tab:btaunu}, table~\ref{tab:bsmumu} or table~\ref{tab:DeltaMds} for other relevant parameters. $\alpha_{_{em,M_{Z_1}}}$ corresponds to the running of $\alpha_{_{em}}$ at $M_{Z_1}$.}
\label{tab:bsg}
\end{center}
\end{table}
The relation eq.~\ref{eq:bsg} used to calculate $\bsg$ is not valid at NNLO where the contributions from charm quark and top/BSM cannot be separated. To reproduce the SM NNLO result obtained in \cite{Misiak:2006zs,Misiak:2006ab}
\beq
\bsg_{\textrm{NNLO}} = (3.15 \pm 0.23)\times 10^{-4},
\label{eq:bsgNNLO}\eeq
we must choose a convenient value of the ratio $m_c/m_b$ that enters in the calculation of $K_c$. We thus try to obtain this result using the input parameters listed in the Appendix A of \cite{Misiak:2006ab} and we find
\beq
\frac{m_c}{m_b} = 0.327,
\eeq
which is larger than in \cite{Domingo:2007dx}.\\

Our $\bsg_{\textrm{SM-only}}$, using the uncertainties on $\Vtbts$, $C$, $\e_s$ and $\mathscr{B}(\bar{B}^0 \to X_c \ell^- \bar{\nu}_\ell)$, is 
\beq
\bsg_\textrm{SM-only} = (3.69^{\,+\,0.24}_{\,-\,0.25}) \times 10^{-4}.
\eeq
The world average on the experimental measurements of the branching reads \cite{Amhis:2012bh}
\beq
\bsg_\textrm{exp} = (3.55 \pm 0.24 \pm 0.09) \times 10^{-4}.
\eeq
Note that we will consider a theoretical uncertainty of 10\% which includes the NNLO error presented in eq.~\ref{eq:bsgNNLO}.
\\

\subsection{$\bXsmu$}

The master formula for the $\bXsl$ branching ratio in the SM reads \cite{Huber:2005ig}
\beq
\frac{\textrm{d}\bXsl}{\textrm{d}\hat s} =
\mathscr{B}(\bar{B}^0 \to X_c \ell^- \bar{\nu}_\ell) \VtsbVcb^2  
\frac{4}{C} \frac{\Phi_{\ell\ell}(\hat{s})}{\Phi_u}, 
\eeq
where $\hat{s} = m_{\ell\ell}^2/m_{b,{\rm pole}}$, $m_{\ell\ell}$ being the dilepton invariant mass.\\
$\Phi_{\ell\ell}(\hat{s})$ and $\Phi_u$ are defined by
\beq \begin{split}
\frac{{\rm d}\Gamma(\bar{B}^0 \to X_s\ell^+ \ell^-)}{{\rm d}\hat{s}} & = \frac{G_F^2 m_{b,{\rm pole}}^5}{48\pi^3} \left| V_{ts}^{\ast}V_{tb}^{}\right|^2 \Phi_{\ell\ell}(\hat{s}),\\
\Gamma (\bar{B}^0 \to X_u \ell^- \bar{\nu}_\ell) & = \frac{G_F^2 m_{b,{\rm pole}}^5}{192 \pi^3} \left| V_{ub}^{}\right|^2 \Phi_u.
\end{split} \eeq 

The most precise experimental measurements of the integrated branching ratio $\bXsl$ have been obtained some years ago by BaBar \cite{Aubert:2004it} and Belle \cite{Iwasaki:2005sy} collaborations both in the low-dilepton invariant mass region $m_{\ell\ell}^2 \in [1,6]$ GeV$^2$ and the high $m_{\ell\ell}^2$ region $[14.4,25]$ GeV$^2$. An average given in \cite{Huber:2007vv} reads
\beq \begin{split}
\bXsl^\textrm{low}_\textrm{exp} & = (1.60 \pm 0.50) \times 10^{-6},\\
\bXsl^\textrm{high}_\textrm{exp} & = (0.44 \pm 0.12) \times 10^{-6}.
\end{split} \eeq 

Through flavour changing neutral Higgs couplings, neutral Higgs boson contributions to $\bXsmu$ are expected in the UMSSM as well as usual supersymmetric contribution shown in section \ref{sec:bsg}. Then following the \NTools implementation of the $\bXsmu$ we modified the Wilson coefficients, especially those linked to neutral Higgs bosons, to compute this observable within the UMSSM :
\beq \begin{split}
\bXsmu^{low} = \; & \bXsmu^\textrm{low}_{\mathrm{SM}} + \frac{4}{C} \mathscr{B}(\bar{B}^0 \to X_c \ell^- \bar{\nu}_\ell)
\left(\frac{\alpha_{_{em}}}{4\pi}\VtsbVcb \right)^2 \times\\
& \left\{ \left[8\ln6 - \frac{15}{(m^{1S}_{b})^2} + \frac{215}{3(m^{1S}_{b})^6}\right](C^2_{7 \textrm{all}}-C^2_{7t})
+ \frac32 (C_h+C_A) \right\},\\
\bXsmu^\textrm{high} = \; & \bXsmu^\textrm{high}_{\mathrm{SM}} + \frac{4}{C} \mathscr{B}(\bar{B}^0 \to X_c \ell^- \bar{\nu}_\ell)
\left(\frac{\alpha_{_{em}}}{4\pi}\VtsbVcb \right)^2 \times\\
& \left\{ \left[8\ln\left(\frac{(m^{1S}_{b})^2}{14.4}\right) - 4\left(1-\frac{14.4}{(m^{1S}_{b})^2}\right) \right. \right.\\
& \left. \left. \quad + \frac43 \left(1-\left(\frac{14.4}{(m^{1S}_{b})^2}\right)^3\right)\right](C^2_{7 \textrm{all}}-C^2_{7t})
+ \frac32 (C_h+C_A) \right\},
\end{split} \eeq
where $C_{7t}, C_{7 \textrm{all}}, C_h, C_A$ correspond to contributions from top quark, BSM and Higgs particles.
Here we directly use 
\beq \begin{split}
\bXsl^{low}_{\mathrm{SM}} & = 1.59 \times 10^{-6},\\
\bXsl^{high}_{\mathrm{SM}} & = 0.24 \times 10^{-6}.
\end{split} \eeq

\section{The anomalous magnetic moment of the muon in the UMSSM}

We saw in section \ref{sec:1.SMpbs} that the muon anomalous magnetic moment could reveal BSM physics. In this section we will describe carefully the new contributions in the context of the UMSSM. 

\subsection{Standard prediction}

First let us look at the SM contributions. Three standard types of contribution are expected here : QED, hadronic and EW.\\ 

Multi-loop QED corrections are now well calculated (see\eg \cite{Jegerlehner:2007xe}) and the complete tenth-order QED contribution was obtained last year \cite{Aoyama:2012wk} which leads to 
\beq
a^{\mathrm{QED}(1 \to 10 \, \mathrm{loop})}_\mu = (116~584~718~951 \pm 80) \times 10^{-14}.
\eeq
\begin{figure}[!htb]
\begin{center}
\includegraphics[scale=0.8]{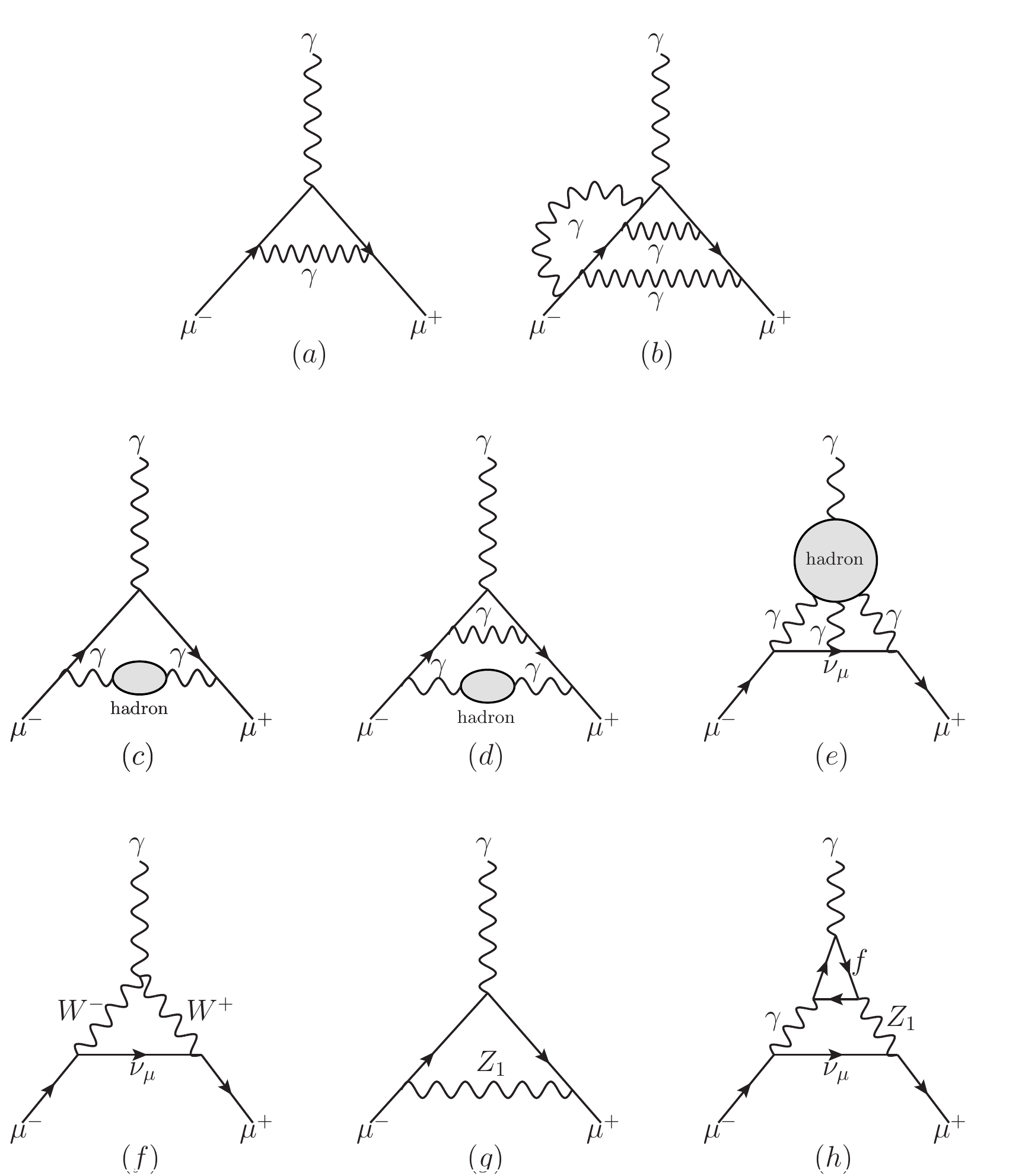} 
  \caption[Examples of SM contributions to $a_\mu$ : the one-loop (a) and a third order (b) QED contributions, hadronic part with the LO (c), an NLO (d) and light-by-light scattering (e) and finally the main leading (f, g) and a two-loop (h) EW corrections.]{Examples of SM contributions to $a_\mu$ : the one-loop (a) and a third order (b) QED contributions, hadronic part with the LO (c), an NLO (d) and light-by-light scattering (e) and finally the main leading (f, g) and a two-loop (h) EW corrections (where $f$ stands for fermions).}
\label{fig:gmusm}
\end{center}
\end{figure}
Hadronic contributions, which give the main theoretical uncertainty on the calculation of $a_\mu$, are devided into three parts :
\begin{itemize}
\item The leading hadronic input corresponds to vacuum polarization (v.p.) type corrections as in figure~\ref{fig:gmusm}c. Its evaluation and accuracy mostly depends on experimental information coming from $e^+e^-$ colliders. Several groups are working on this calculation \cite{Jegerlehner:2011ti,Bodenstein:2013flq} and we will use the highest contribution obtained \cite{Hagiwara:2011af}, to reduce the gap between the SM and the experimental values of $a_\mu$ :
\beq
a^{\mathrm{LO \, had. \, v.p.}}_\mu = (6949.1 \pm 37.2 \pm 21.0) \times 10^{-11}.
\eeq
\item Higher order hadronic vacuum-polarization contributions as in figure~\ref{fig:gmusm}d were also calculated in \cite{Hagiwara:2011af} :
\beq
a^{\mathrm{HO \, had. \, v.p.}}_\mu = (-98.4 \pm 0.6 \pm 0.4) \times 10^{-11}.
\eeq
\item Contributions from hadronic light-by-light scattering (see\eg figure~\ref{fig:gmusm}e) increase a lot the theoretical uncertainty on the $a_\mu$ calculation. We will consider the quantity obtained in \cite{Nyffeler:2009tw} :
\beq
a^{\mathrm{ll \, had. \, v.p.}}_\mu = (116 \pm 40) \times 10^{-11}.
\eeq
\end{itemize}

EW contributions to $a_\mu$ have been calculated up to two-loop order with\eg triangle fermionic-loops as in figure~\ref{fig:gmusm}h. We keep the result of \cite{Czarnecki:2002nt} : 
\beq
a^{\mathrm{EW}}_\mu = (154 \pm 2) \times 10^{-11}.
\eeq
Here we can safely neglect the contribution coming from the SM Higgs boson because $m_{h^0} \gg m_{\mu^\pm}$; with $m_{h^0} \sim 125$~GeV we have $a^{\mathrm{EW,}\, h^0}_\mu < 5 \times 10^{-14}$ \cite{Jegerlehner:2007xe}, which is far below current experimental sensitivity. As it was the case for the $\bsmu$ calculation in section \ref{sec:bsmu}, we consider as a good approximation to keep SM couplings for the $Z_1$ boson.

Adding all these contributions, the result of the $a_\mu$ calculation in the SM reads, \cite{Aoyama:2012wk}
\beq \begin{split}
a^{\mathrm{SM}}_\mu = \; & a^{\mathrm{QED}(1 \to 10 \, \mathrm{loop})}_\mu + a^{\mathrm{LO \, had. \, v.p.}}_\mu + a^{\mathrm{HO \, had. \, v.p.}}_\mu + a^{\mathrm{ll \, had. \, v.p.}}_\mu + a^{\mathrm{EW}}_\mu \\
= \; & (116~591~840 \pm 59) \times 10^{-11}.
\end{split} \eeq
On the experimental side the most precise determination of $a_\mu$ was obtained by the Brookhaven $(g-2)$ experiment E821. The current average given in \cite{Roberts:2010cj} is :
\beq
a^{\mathrm{exp}}_\mu = (116~592~089 \pm 63) \times 10^{-11}.
\eeq
This leads us to a discrepancy between the experimental and the SM value of
\beq
\amu = a^{\mathrm{exp}}_\mu - a^{\mathrm{SM}}_\mu = (249 \pm 86) \times 10^{-11} .
\eeq

\subsection{New contributions}

Now let us look at possible BSM contributions. Typically a new particle with mass $M_{\mathrm{new}}$ is expected to contribute to $a_\mu$ like
\beq
a^{\mathrm{new}}_\mu \propto \frac{m^2_{\mu^\pm}}{M^2_{\mathrm{new}}}.
\eeq
Before looking at supersymmetric particles, the contribution to $a_\mu$ of the new gauge boson $Z_2$ in the UMSSM should in principle be taken into account. This term has the opposite sign of $a^{\mathrm{SM}}_\mu$ \cite{Czarnecki:2001pv} so will increase the discrepancy with the measured value : 
\beq
a^{\mathrm{Z_2}}_\mu \simeq -\frac{M^2_{Z_1}}{M^2_{Z_2}} \times O(10^{-11}).
\eeq 
Nevertheless given the collider constraints on the mass of this new gauge boson ($\geq$ 2 TeV) it can be safely neglected, \cite{Barger:2004mr}.\\

\begin{figure}[!htb]
\begin{center}
\includegraphics[scale=0.8]{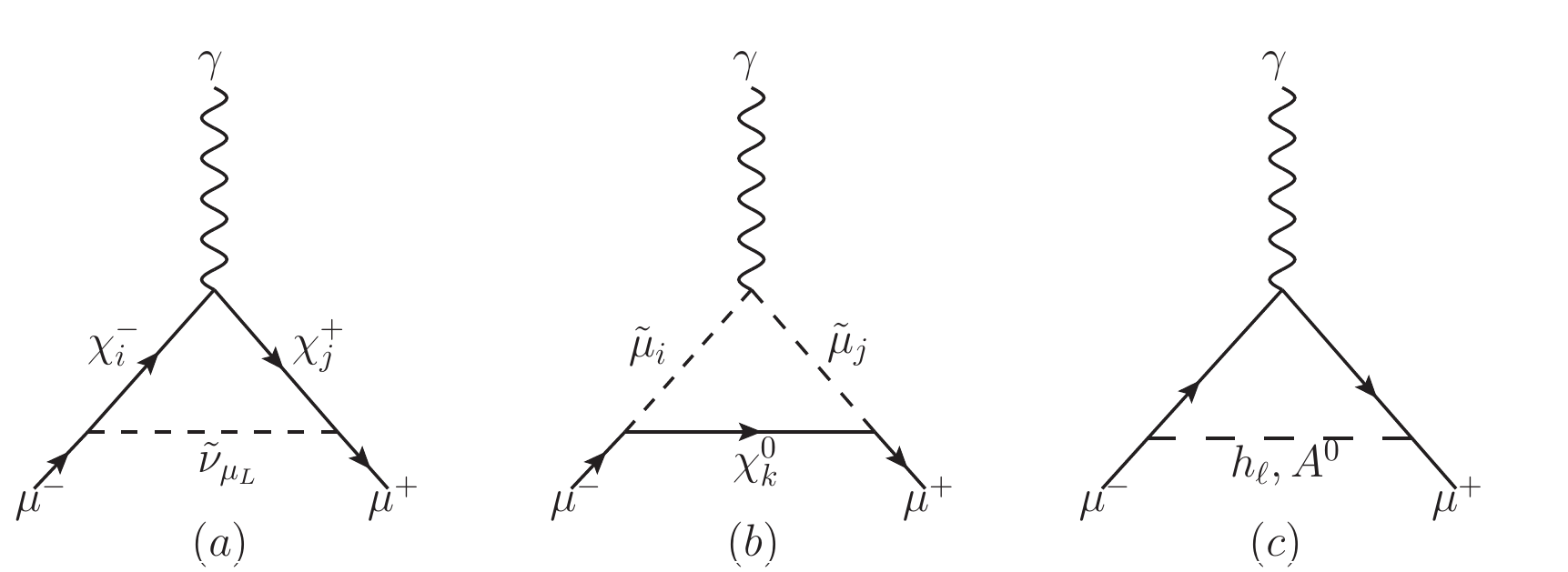} 
  \caption[Examples of one-loop SUSY contributions to $a_\mu$ : chargino/sneutrino (a), neutralino/smuon (b) and a diagram involving Higgs bosons (c).]{Examples of one-loop supersymmetric contributions to $a_\mu$ : chargino/sneutrino (a), neutralino/smuon (b) and a diagram involving Higgs bosons (c). Here $i,j \in \{1,2\}$, $k \in \{1,...,6\}$ and $\ell \in \{1,2,3\}$. We consider the smuon mass eigenstates $\tilde{\mu}_{i,j}$ as the result of the (weak) mixing between $\tilde{\mu}_{L}$ and $\tilde{\mu}_{R}$.}
\label{fig:gmubsm}
\end{center}
\end{figure}

Several supersymmetric particles contribute to $a_\mu$ \cite{Martin:2001st}. First we have one-loop contributions with chargino/sneutrino (figure~\ref{fig:gmubsm}a) and neutralino/smuon loops (figure~\ref{fig:gmubsm}b) approximated by
\beq
a^{\chi^{\pm, 0}, \tilde{\mu}, \tilde{\nu}_{\mu_L}}_\mu \propto \frac{m^2_{\mu^\pm}}{M^2_{\mathrm{SUSY}}} \tan\b \; \mathrm{sign}(\mu),
\eeq
where $M_{\mathrm{SUSY}}$ is a typical mass scale of the supersymmetric particles.
Here we just generalize the equations given in the NMSSM \cite{Domingo:2008bb} considering six neutralinos in the UMSSM.

We also add one-loop contributions from CP-even, CP-odd and charged Higgs bosons as in figure~\ref{fig:gmubsm}c. To obtain the relations for the neutral Higgs bosons in the UMSSM we must consider the reduced Higgs couplings $C^{h_i, A^0}_f$ defined in eq.~\ref{eq:9.redcoupfer}. Then, following \cite{Domingo:2008bb}, we adapt the one-loop contributions from Higgs bosons to the UMSSM and we get
\beq \begin{split}
a_{\mu}^{1 \, \mathrm{loop}\ h_i} & = \frac{G_F
m_{\mu^\pm}^2}{4\sqrt{2}\pi^2} \sum_{i=1}^3 
(C^{h_i}_{\mu^\pm})^2\int_0^1{\frac{x^2(2-x)\,dx}{x^2+
\left(\frac{m_{h_i}}{m_{\mu^\pm}}\right)^2(1-x)}},\\
a_{\mu}^{1 \, \mathrm{loop}\ A^0} & = -\frac{G_F
m_{\mu^\pm}^2}{4\sqrt{2}\pi^2}
(C^{A^0}_{\mu^\pm})^2\int_0^1{\frac{x^3\,dx}{x^2+
\left(\frac{m_{A^0}}{m_{\mu^\pm}}\right)^2(1-x)}}\label{amuhiggs1L},\\
a_{\mu}^{1 \, \mathrm{loop}\ H^\pm} & = \frac{G_F
m_{\mu^\pm}^2}{4\sqrt{2}\pi^2}\tan^2\beta 
\int_0^1{\frac{x(x-1)\,dx}{x-1+
\left(\frac{m_{H^{\pm}}}{m_{\mu^\pm}}\right)^2}} \; .
\end{split} \eeq

At two-loop we first add the leading contribution enhanced by large QED logarithms \cite{Degrassi:1998es}.
The bosonic EW two-loop contribution $a_{\mu}^{2 \, \mathrm{loop \; Bos}}$ is defined in \cite{Heinemeyer:2004yq} where the leading logarithmic contribution proportional to $c_L^{2 \, \mathrm{loop \; Bos}}$ is shown to be an excellent approximation of the full bosonic term. We then confine ourselves to the $c_L^{2 \, \mathrm{loop \; Bos}}$ term as in \cite{Domingo:2008bb}; the final expression reads 
\beq
a_{\mu}^{2 \, \mathrm{loop \; Bos}}=
\frac{5\, G_F\,m_{\mu^\pm}^2\,\alpha_{_{em,M_{Z_1}}}}{24\sqrt{2}\,\pi^3}
\ln \frac{m_{\mu^\pm}^2}{M_W^2} c_L^{2 \, \mathrm{loop \; Bos}},
\eeq
where $c_L^{2 \, \mathrm{loop \; Bos}}$ includes two-loop EW SM contributions :
\beq
c_L^{2 \, \mathrm{loop \; Bos}}=\frac{1}{30}
\left[98+9c_L^h+23\left(1-4\cos^2\theta_W\right)^2\right].
\eeq
We then took care as in \cite{Domingo:2008bb} to compute this SM contribution only once.
In the UMSSM, the expression of $c_L^h$ is defined as
\beq
c_L^h=\cos 2\beta M_{Z_1}^2
\sum_{i=1}^3\frac{Z_{h i1}\left(Z_{h i1}-\tan\beta Z_{h i2}\right)}
{m_{h_i}^2}.
\eeq
We also checked the rule $c_L^h$ = 1 \cite{Heinemeyer:2004yq}.

Starting from \cite{Domingo:2008bb,Cheung:2001hz}, the contributions from two-loop Higgs boson diagrams involving a closed SM fermion loop (confined to the lepton $\tau$ and the quarks t and b) in the UMSSM reads 
\beq \begin{split}
a_{\mu}^{2 \, \mathrm{loop}\ h_i} 
& = -\frac{G_F m_{\mu^\pm}^2 \alpha_{_{em}}}{2\sqrt{2}\pi^3}
\sum_{i=1}^3 \sum_{f=t,b,\tau} N_c^f C^{h_i}_{\mu^\pm} \mathcal{Q}^2_f C^{h_i}_{f}f_h\left(\frac{m_f^2}{m_{h_i}^2}\right)\\
a_{\mu}^{2 \, \mathrm{loop}\ A^0} & = \frac{G_F m_{\mu^\pm}^2 \alpha_{_{em}}}
{2\sqrt{2}\pi^3} \sum_{f=t,b,\tau} N_c^f C^{A^0}_{\mu^\pm} \mathcal{Q}^2_f C^{A^0}_{f} f_A
\left(\frac{m_f^2}{m_{A^0}^2}\right)
\end{split} \eeq
where $N_c^f$ represents the number of color degrees of freedom in $f$, $\mathcal{Q}_f$ its electric charge and the loop integral functions are given in \cite{Stockinger:2006zn,Domingo:2008bb} :
\beq \begin{split}\label{eq:gmu-fhA}
f_h(z) & = \frac{1}{2}z\int_0^1{\frac{1-2x(1-x)}{x(1-x)-z} \ln \frac{x(1-x)}{z}\,dx}\\
f_A(z) & = \frac{1}{2}z\int_0^1{\frac{1}{x(1-x)-z}\ln \frac{x(1-x)}{z}\,dx}.
\end{split} \eeq

Contributions coming from photonic Barr-Zee type diagrams involving closed sfermion and chargino loops were considered in the context of the MSSM \cite{Arhrib:2001xx} and the NMSSM \cite{Domingo:2008bb}.
The sfermionic contribution reads
\beq
a_\mu^{2 \, \mathrm{loop}\ \tilde{f}} = \frac{G_F m_{\mu^\pm}^2 M_W \alpha_{_{em,M_{Z_1}}}}{2 \sqrt{2}
\pi^3 g_2} \sum_{j=1}^2 \sum_{\tilde{f}=\tilde{t}_j,\tilde{b}_j,\tilde{\tau}_j} \left[ \sum_{i=1}^3
\frac{N_c^f C_{\mu^\pm}^{h_i} Q_f^2
\lambda_{\tilde{f}}^{h_i}}{m_{\tilde{f}}^2} f_{\tilde{f}}
\left(\frac{m^2_{\tilde{f}}}{m^2_{h_i}}\right) \right],
\eeq

where the $\lambda_{\tilde{f}}^{h_i}$ which are here the couplings of the CP-even Higgs bosons to stops, sbottoms and staus are given in eqs.~\ref{eq:8.hstop}, \ref{eq:8.hsbottom} and \ref{eq:8.hstau} and the loop integral function is
\beq
f_{\tilde{f}}(z) = \frac{1}{2}z\int_0^1{\frac{x(1-x)}{x(1-x)-z} \ln \frac{x(1-x)}{z}\,dx}.
\eeq

Finally we add the chargino contribution at two loops
\beq
a_\mu^{2 \, \mathrm{loop}\ \chi^{\pm}} =
\frac{G_F m_{\mu^\pm}^2 M_W \alpha_{_{em,M_{Z_1}}}}{4 \sqrt{2} \pi^3 g_2}
\sum_{k=1}^2\left[\sum_{i=1}^3\frac{C_{\mu^\pm}^{h_i}\lambda_{\chi_k^{\pm}}^{h_i}}
{m_{\chi_k^{\pm}}}f_h
\left(\frac{m^2_{\chi_k^{\pm}}}{m^2_{h_i}}\right)
-\frac{C_{\mu^\pm}^{A^0}\lambda_{\chi_k^{\pm}}^{A^0}}
{m_{\chi_k^{\pm}}}f_A
\left(\frac{m^2_{\chi_k^{\pm}}}{m^2_{A^0}}\right)\right],
\eeq
where the $f_{h,A}$ are given in eq.~\ref{eq:gmu-fhA} and the couplings of the neutral Higgs bosons to the charginos are given in eqs.~\ref{eq:8.hevenchar} and \ref{eq:8.hoddchar}.\\

As in \cite{Domingo:2008bb}, we follow \cite{Stockinger:2006zn} for the calculation of the theoretical errors : 2\% (one-loop contributions) + 30\% (two-loop contributions) + 2.8 $\times$ 10$^{-10}$ (additional corrections, no difference expected with respect to the same study in the MSSM).

\section{Scanning  the  $U(1)_\eta$ parameters} 

We are now able to examine how the Higgs sector and the low energy observables constrain the UMSSM parameter space. For simplicity we restrict ourselves to the $U(1)_\eta$ model. We consider scenarios where either one generation of RH sneutrino or the lightest neutralino is a DM candidate but we relax the lower bound on the $\lsp$ or $\chi^0_1$ relic density, thus keeping all scenarios with $\Omega_\chi h^2 \in [10\%\Omega^{-3\s}_\textrm{Pl} h^2, \Omega^{+3\s}_\textrm{Pl} h^2]$ with $\chi \in \{\lsp, \chi^{0}_{1}\}$ and $\Omega_\textrm{Pl} h^2$ is given in table~\ref{tab:cosmo_param}, namely $\Omega^{\pm 1\s}_\textrm{Pl} h^2 = 0.1187 \pm 0.0017$. Moreover we assume $M_1=M_2/2=M_3/6$ as in the previous chapter. We use the most recent ATLAS constraint on the $Z'_\eta$ gauge boson \cite{ATLAS:2013jma},\ie $\mzp >2.43$~TeV and we decide for simplicity to consider only $Z_2$ bosons above this limit. We also consider only points in the UMSSM parameter space that give a CP-even Higgs boson around the value derived in table~\ref{tab:SM}, namely 125.63~GeV, allowing a large theoretical uncertainty of $\pm 5$~GeV. Points which do not satisfy the LEP bounds on the mass of charged sparticles as the chargino are automatically rejected.

As in chapter~\ref{chapter:RHsneu} we fix the soft masses of sleptons and squarks to 2 TeV and then avoid LHC constraints on these particles. Nonetheless we keep the third generation of soft squark masses as free parameter, using a lower limit of 400 GeV to avoid most constraints from recent LHC searches as shown in figure~\ref{fig:atlassusy}. First we comment on constraints coming from the $\Delta \rho$ parameter. This observable, which measures the deviation of the $\rho$ parameter of the SM from unity, is computed using a \micro routine which contains third generation of MSSM sfermions and two-loop QCD contributions. This constraint is sensitive to the mass difference between sfermions. This could be an interesting constraint for several $U(1)'$ scenario since the new $D$-terms implied by the new symmetry and shown in eq.~\ref{eq:7.sferU} can give large mass splitting between the LH and RH sfermions. Because $Z_1$ is no longer purely the $Z$ boson in the UMSSM, the $\rho$ parameter also receives a specific UMSSM tree-level correction. In the limit where $M_{Z'}^2 \gg M_Z^2, \Delta_Z^2$ in the Abelian gauge boson mass matrix given in eq.~\ref{eq:7.matZZp}, which is safe for our TeV scale $Z'$, the correction which will be added to $\Delta \rho$ reads \cite{Babu:1996vt}
\beq \Delta \rho_Z = \frac{\Delta_Z^4 M_{Z_2}^2}{M_{Z'}^4 M_{Z_1}^2}.\eeq
Measurements of EW observables give the upper bound of $\Delta \rho < 2\times10^{-3}$.
 
The free parameters considered here are   
\begin{center}
$\{\mlsp, \mu, M_1, M_1', A_\l, A_t, A_b, M_{Z_2}, \azz, m_{\tilde{Q}_3}, m_{\tilde{u}_3}, m_{\tilde{d}_3}\}$
\end{center}
and the range associated to these parameters is given in table~\ref{tab:range_H_B}. 
\begin{table}[!htb]
\begin{center}
\begin{tabular}{cc}\hline \hline
\textbf{Parameter} & \textbf{Range} \\ \hline \hline
$\mlsp$ & [0.05, 2] TeV\\ 
$M_{Z_2}$ & [2.6, 4] TeV\\ 
$\azz$ & [-0.003, 0.003] rad \\ 
$A_\l$ & [0, 4] TeV\\
$A_t, A_b$ & [-4, 4] TeV\\
$m_{\tilde{Q}_3}, m_{\tilde{u}_3}, m_{\tilde{d}_3}$ & [0.4, 2] TeV\\ 
$\mu, M_1, M'_1$ & [0.1, 2] TeV\\ \hline \hline
\end{tabular}
\caption{Range of the free parameters in the $U(1)_\eta$ model.}
\label{tab:range_H_B}
\end{center}
\end{table}

\section{Numerical results}
 
\begin{figure}[!htb]
\begin{center}
\centering
\subfloat[]{\includegraphics[width=8.25cm,height=6.5cm]{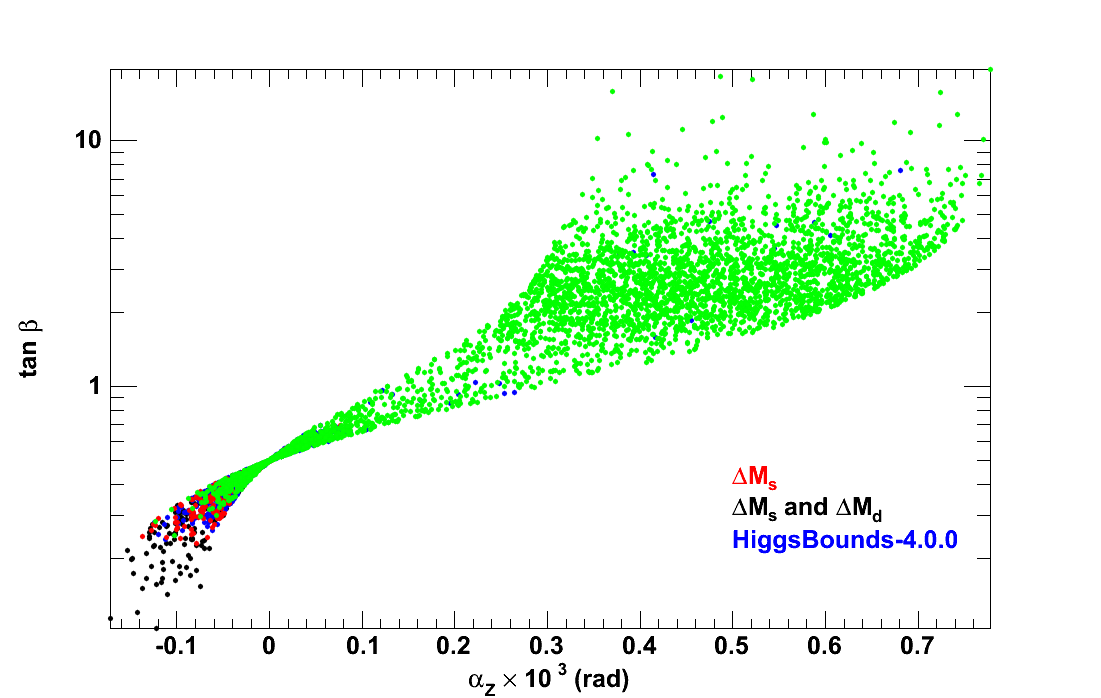}}
\subfloat[]{\includegraphics[width=8.25cm,height=6.5cm]{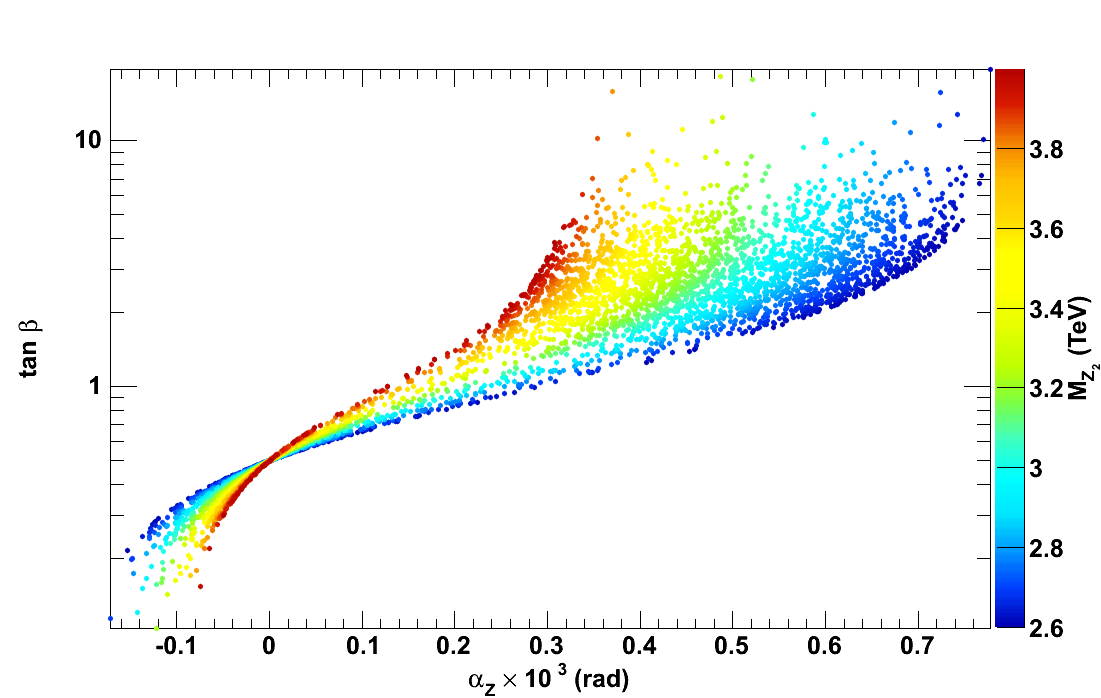}} 
\caption{UMSSM scenarios in the plane ($\tan\b, \azz$). In panel (a) the allowed scenarios are represented in green, scenarios excluded by $\Delta M_s$ (and $\Delta M_d$) in red (black) and the points excluded by {\tt HiggsBounds-4.0.0\;} in blue. Panel (b) shows the mass of the new gauge boson as indicated by the colour code.}
\label{fig:9.tbazz}
\end{center}
\end{figure}

Figure~\ref{fig:9.tbazz} illustrates the effect of low energy observables on the scenarios considered. Among these, only the mass differences of $B$-mesons are relevant in this scan. In red we plot the points excluded by $\Delta M_s$ and black points are excluded by both $\Delta M_s$ and $\Delta M_d$ observables. Note that none of these scenarios are able to explain the anomalous magnetic moment of the muon $\amu$ as observed in chapter~\ref{chapter:NUHM2} for the case of the MSSM. We also checked that the points are allowed by the latest version of {\tt HiggsBounds}, {\tt HiggsBounds-4.0.0\;}\cite{Bechtle:2013gu}. Note that we represent in blue the points excluded by {\tt HiggsBounds}\footnote{Recall that we allow the lightest Higgs boson mass to be within [120.63, 130.63]~GeV.}. As discussed in chapter~\ref{chapter:RHsneu}, low $\tan\b$ values are the most constrained regions as shown in figure~\ref{fig:9.tbazz}a. We also see that only scenarios with $|\azz| < 10^{-3}$ are obtained. Figure~\ref{fig:9.tbazz}b presents the reason of this observation : higher is the $Z_2$ mass lower is the mixing angle $|\azz|$. 

\begin{figure}[!htb]
\begin{center}
\centering
\subfloat[]{\includegraphics[width=8.25cm,height=6.5cm]{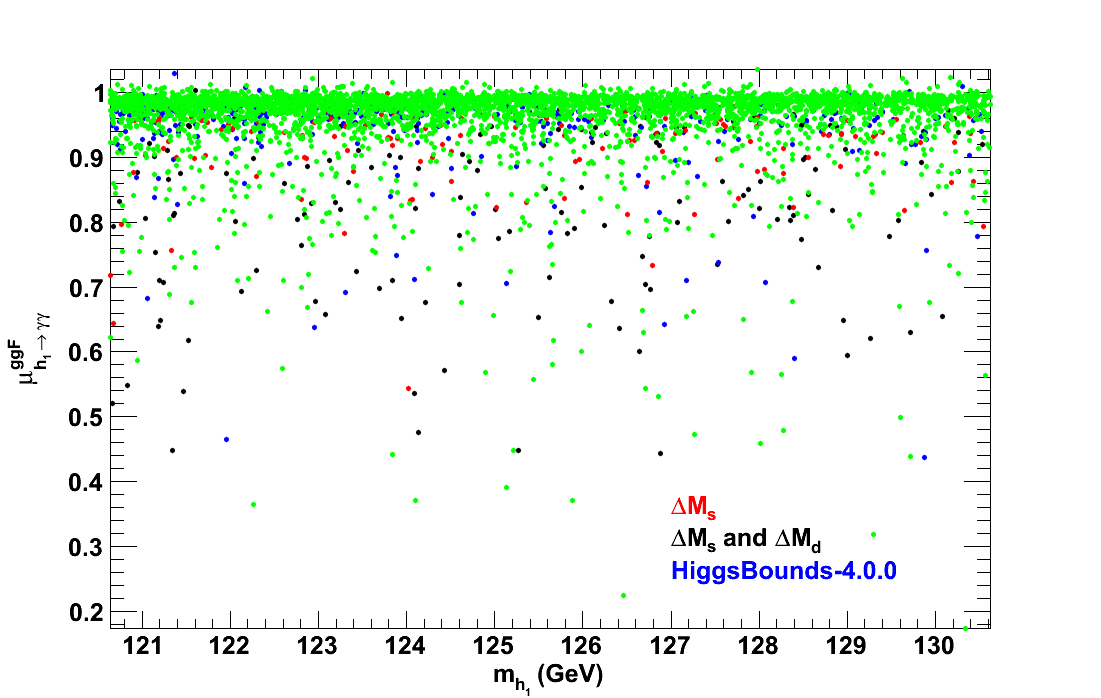}}
\subfloat[]{\includegraphics[width=8.25cm,height=6.5cm]{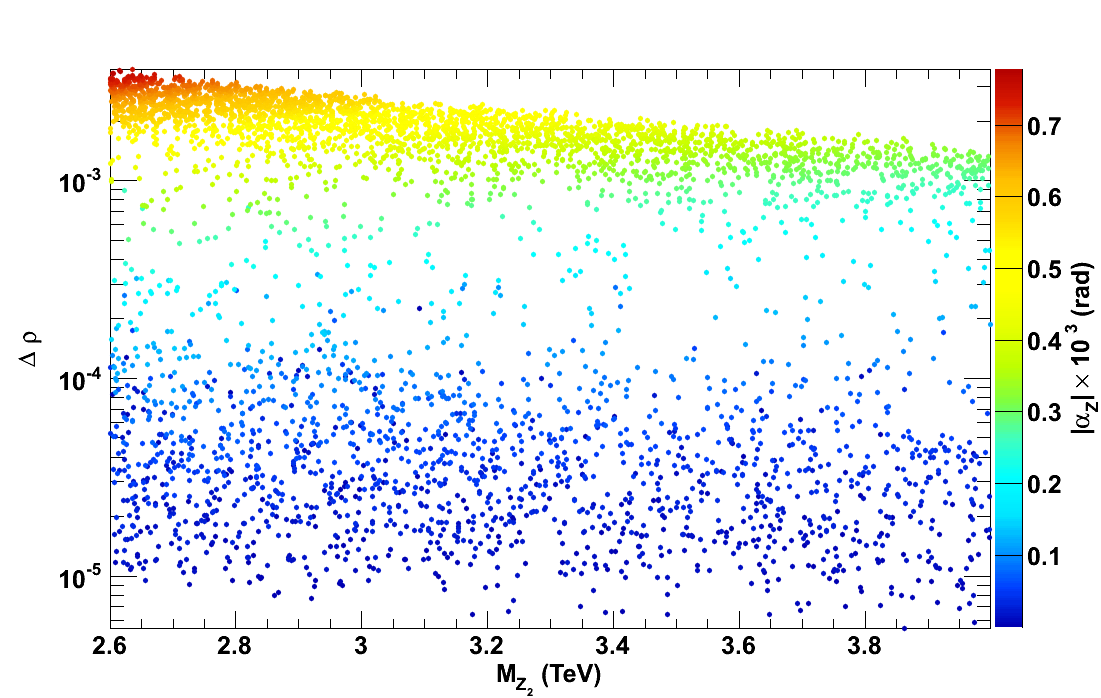}} 
\caption{(a) Signal strength $\mu_{h_1 \ra \g\g}^{\rm ggF}$ as a function of $m_{h_1}$, with the same colour code as in figure~\ref{fig:9.tbazz}a. (b) $\Delta \rho$\vs $M_{Z_2}$ with the absolute value of the mixing angle $\azz$ indicated by the colour code. }
\label{fig:collconstr}
\end{center}
\end{figure}

Figure~\ref{fig:collconstr}a shows that the signal strength $\mu_{h_1 \ra \g\g}^{\rm ggF}$ is not expected to be drastically modified as compared to the SM, only a few points are more than 10\% below 1. Panel (b) of figure~\ref{fig:collconstr} shows the predictions for the $\Delta \rho$ parameter as a function of the $Z'$ mass and the mixing angle $\azz$ : increasing the mass of the new gauge boson above 3 TeV and then decreasing $\azz$ allows to stay below the limit obtained with EW observables.

\begin{figure}[!htb]
\begin{center}
\centering
\subfloat[]{\includegraphics[width=8.25cm,height=6.5cm]{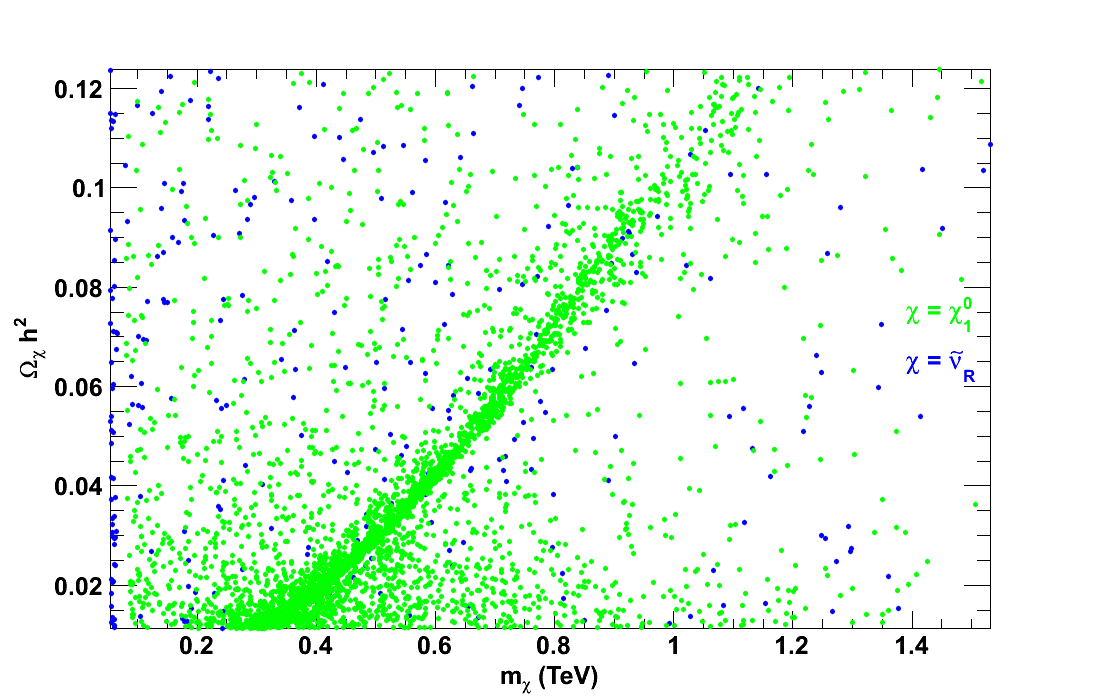}}
\subfloat[]{\includegraphics[width=8.25cm,height=6.5cm]{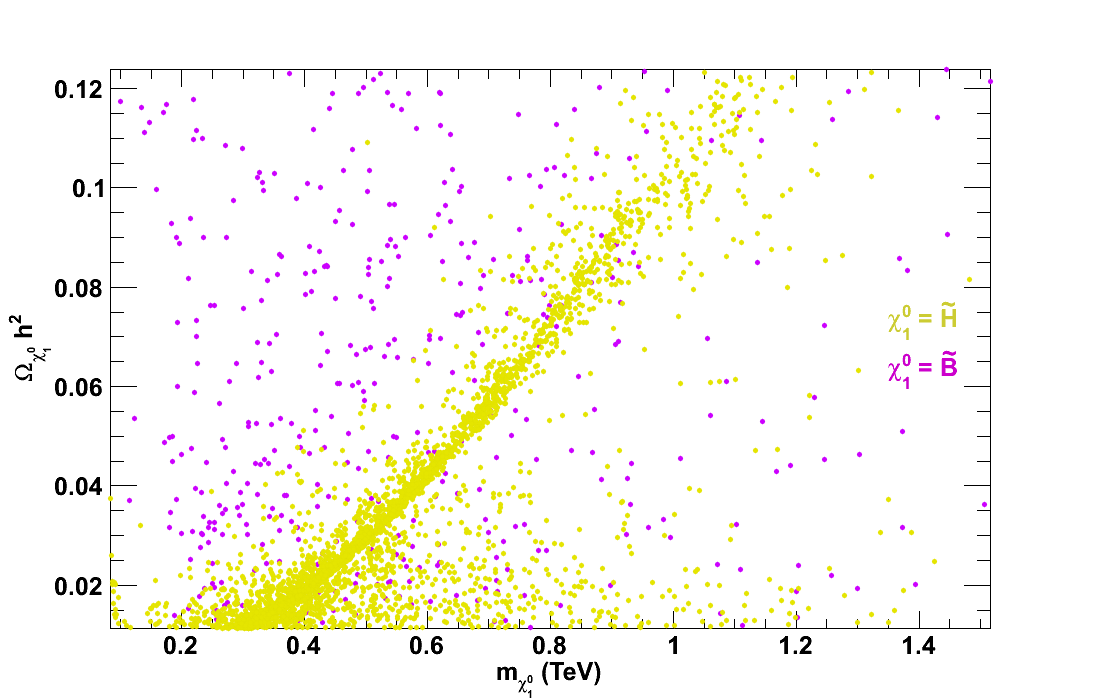}}\\ 
\subfloat[]{\includegraphics[width=8.25cm,height=6.5cm]{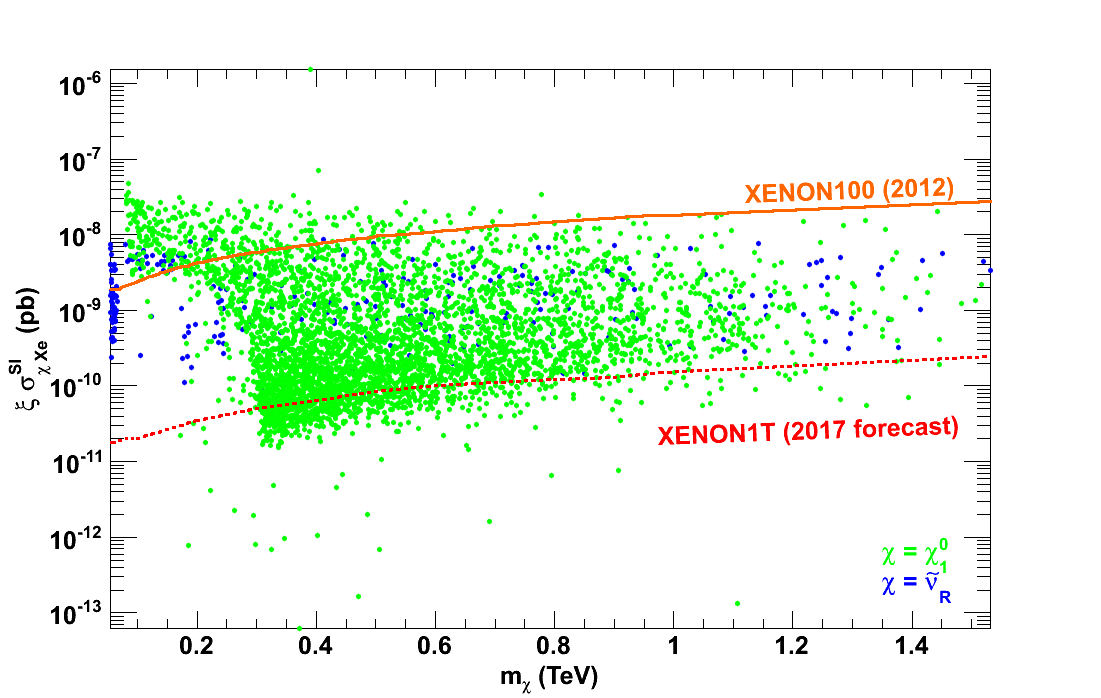}}
\caption{DM observables as functions of the DM mass. Panel (a) shows the relic density either for an $\lsp$ (blue) or an $\chi^0_1$ (green) LSP. Panel (b) represents the relic density either for an $\widetilde{B}$ (violet) or an $\widetilde{H}$ (yellow) LSP. Panel (c) compares the rescaled DD rate for an $\lsp$ (blue) or an $\chi^0_1$ (green) LSP compared to results or forecasts from Xenon-based experiments.}
\label{fig:astroconstr}
\end{center}
\end{figure}

Figure~\ref{fig:astroconstr} presents the characteristics of the DM sector in this scan. Panel (a) and (b) shows that most of the allowed points correspond to an higgsino LSP when we relax the relic density lower bound although we still find some bino LSP as well as sneutrino LSP. Furthermore since the range for the lightest Higgs boson mass is more restricted than in chapter~\ref{chapter:RHsneu}, the annihilation of a RH sneutrino near a light Higgs boson resonance requires even more fine-tuning. These scenarios are then difficult to obtain. As we saw in chapter~\ref{chapter:NUHM2} in the MSSM, many of the higgsino LSP scenarios can be probed by Xenon-based experiments.

\newpage

\section{Discussion}

These results indicate that even the more stringent Direct Detection and collider results of 2012 still allow scenarios with a viable Dark Matter candidate. We also found that the parameter space of the $U(1)_\eta$ model would be probe to a large extent by both future DD Dark Matter searches (Xenon) and by $Z'$ searches in the next run at the LHC. The impact of current experimental results on the properties of Dark Matter candidates in the generic UMSSM will be performed in the very near future.

% Conclusion in another part : 

\makeatletter
\newcommand{\ChapterOutsidePart}{%
   \def\toclevel@chapter{-1}\def\toclevel@section{0}\def\toclevel@subsection{1}}
\makeatother

\ChapterOutsidePart

\chapter*{Conclusion}
\addcontentsline{toc}{chapter}{Conclusion}
\markboth{Conclusion}{}

\section*{English version}

In this thesis we have seen how the theoretical and experimental developments in particle physics and cosmology, especially during the 20th century, have led to two main theories that describe our Universe in a very efficient way and with a lot of success. Nevertheless some issues, especially those concerning the Higgs sector and the Dark Matter problem, still need to be addressed. This thesis investigate these problems within the context of Supersymmetry. \\

We depicted how the Minimal Supersymmetric Standard Model can give meaningful solutions with the neutralino as a very interesting Dark Matter candidate. This model has also the nice feature to provide candidates to the cosmic inflation. Then a large range of collider, astroparticle and cosmological measurements can be used to point to the viable region of the parameter space of this supersymmetric model. To probe efficiently these viable regions we developed a code using the Markov Chain Monte Carlo approach and we found that, within the NUHM2 scenario which features non degenerate scalar masses at the Grand Unified Theories scale, most of our interesting points in the parameter space are characterized by a Lightest Supersymmetric Particle neutralino which is mostly higgsino and a Standard Model-like Higgs boson whose mass is in the preferred range set by the experiments. We also saw that the Direct Detection of Dark Matter is starting to probe this type of scenarios and that the expected sensitivity of forthcoming Direct Detection experiments could almost entirely span the favourite Dark Matter scenarios.\\

We also analysed how the Indirect Detection of Dark Matter can despite several drawbacks constrain the Dark Matter sector. Assuming that the neutralino Dark Matter candidate mostly annihilate into $W$ bosons and that there is a regeneration mechanism to raise the relic density of Dark Matter to the mesured value, we showed how lower bounds on the mass degeneracy between the neutralino and the NLSP chargino can be defined in a simplified version of the phenomenological Minimal Supersymmetric Standard Model, especially for a wino Lightest Supersymmetric Particle.\\

Despite the nice features of the Minimal Supersymmetric Standard Model, there are reasons to analyse some of its extensions. The NMSSM is the minimal extension which provides a very light Dark Matter candidate, the singlino, a Higgs with the appropriate mass and the possibility to fit possible non-standard signal strength in some Higgs decay channels determined at colliders. We saw by analysing some specific scenarios with light Dark Matter in this model that usual limits on coloured sparticles obtained at the LHC have to be taken with care in the context of such non-minimal scenarios.\\

We finally look at a gauge symmetry extension of the Minimal Supersymmetric Standard Model, the UMSSM. This model contains a new gauge boson which is now strongly constrained by current collider experiments. An interesting point of the UMSSM is that it provides a scalar Dark Matter candidate, the Right-Handed sneutrino. We found a variety of annihilation channels for this particle which allow to fit the Dark Matter constraints. Depending on the $U(1)$ scenario considered, the phenomenological predictions can be completely different. We finally made a first analysis of the Higgs sector of this model while considering a large number of low energy observables that can be affected by UMSSM contributions. In the near future we will pursue the analysis of this model to evaluate the impact of the recent collider results on Dark Matter properties. Our aim is also to include the UMSSM model in the public version of the \micro code.

\section*{Version fran\c{c}aise}

Cette th\`ese nous a permis de voir comment les d\'eveloppements th\'eoriques et exp\'erimentaux en physique des particules et en cosmologie, principalement pendant le 20e si\`ecle, nous ont conduit \`a se concentrer principalement sur deux th\'eories qui d\'ecrivent efficacement et avec succ\`es notre Univers. Cependant certaines questions, notamment celles qui concernent le secteur de Higgs et le probl\`eme de la Mati\`ere Noire, doivent encore \^{e}tre abord\'ees. Cette th\`ese traite de ces probl\`emes dans le contexte de la Supersym\'etrie. \\

Nous avons montr\'e \`a quel point le Mod\`ele Standard Supersym\'etrique Minimal peut donner d'int\'eressantes solutions avec un tr\`es attrayant candidat \`a la Mati\`ere Noire qui est le neutralino. Ce mod\`ele a \'egalement l'avantageuse caract\'eristique de fournir des candidats \`a l'inflation cosmique. Il s'ensuit qu'un large \'eventail de mesures en collisionneur, d'astroparticule et en cosmologie peut \^{e}tre utilis\'e pour pointer vers les r\'egions viables de l'espace des param\`etres de ce mod\`ele supersym\'etrique. Pour sonder de mani\`ere efficace ces int\'eressantes r\'egions un code utilisant l'approche Monte-Carlo \`a Cha\^{i}nes de Markov a \'et\'e d\'evelopp\'e et cela nous a permis de trouver que, dans le contexte du sc\'enario NUHM2 dont la caract\'eristique principale est de lever la d\'eg\'en\'erescence entre les scalaires \`a l'\'echelle de Grande Unification, la majeure partie de nos int\'eressantes configurations dans l'espace des param\`etres est caract\'eris\'e par une particule supersym\'etrique la plus l\'eg\`ere principalement higgsino et un boson de Higgs comparable \`a celui du Mod\`ele Standard dont la masse est dans l'intervalle fix\'e par les exp\'eriences. Nous avons en outre remarqu\'e que la D\'etection Directe de Mati\`ere Noire commence \`a sonder ce genre de sc\'enarios et la sensibilit\'e attendue des exp\'eriences \`a venir de D\'etection Directe peut presque enti\`erement couvrir l'ensemble des sc\'enarios pertinents de Mati\`ere Noire.\\

Nous avons aussi analys\'e comment la D\'etection Indirecte de Mati\`ere Noire peut malgr\'e plusieurs inconv\'enients contraindre le secteur de Mati\`ere Noire. En supposant que le candidat neutralino \`a la Mati\`ere Noire s'annihile principalement en bosons $W$ et qu'un m\'ecanisme de r\'eg\'en\'eration permet d'\'elever la densit\'e relique de Mati\`ere Noire \`a la valeur mesur\'ee, nous avons montr\'e comment des bornes inf\'erieures sur la d\'eg\'en\'erescence de masse entre le neutralino et la NLSP chargino peuvent \^{e}tre d\'efinies dans une version simplifi\'ee du Mod\`ele Standard Supersym\'etrique Minimal ph\'enom\'enologique, particuli\`erement pour une particule supersym\'etrique la plus l\'eg\`ere wino.\\

Nonobstant les int\'eressantes caract\'eristiques du Mod\`ele Standard Supersym\'etrique Minimal, certaines raisons nous poussent \`a analyser quelques-unes de ses extensions. Le NMSSM est l'extension minimale qui fournit un candidat \`a la Mati\`ere Noire l\'eg\`ere, le singlino, un boson de Higgs avec la masse appropri\'ee et la possibilit\'e d'\^{e}tre compatible avec de possibles signaux non standards dans certaines cha\^{i}nes de d\'esint\'egration du boson de Higgs obtenues en collisionneur. Nous avons vu en \'etudiant quelques sc\'enarios avec une Mati\`ere Noire l\'eg\`ere dans ce mod\`ele que les limites sur les sparticules color\'ees obtenues au LHC doivent \^{e}tre prises avec pr\'ecaution dans le contexte de ces sc\'enarios non minimaux.\\

Nous avons finalement consid\'er\'e une extension de la sym\'etrie de jauge du Mod\`ele Standard Supersym\'etrique Minimal, le UMSSM. Ce mod\`ele contient un nouveau boson de jauge qui est fortement contraint par les exp\'eriences actuelles en collisionneur. Un point int\'eressant de l'UMSSM est qu'il fournit un candidat scalaire \`a la Mati\`ere Noire, le sneutrino droit. Nous avons trouv\'e divers canaux d'annihilation pour cette particule qui permettent \`a des sc\'enarios UMSSM de satisfaire les contraintes de Mati\`ere Noire. Selon le sc\'enario $U(1)$ choisi, les pr\'edictions ph\'enom\'enologiques peuvent \^{e}tre compl\`etement diff\'erentes. Pour finir nous avons analys\'e le secteur de Higgs de ce mod\`ele tout en incluant un grand nombre d'observables de basses \'energies qui peuvent \^{e}tre impact\'ees par les contributions UMSSM. Dans un futur proche nous poursuivrons l'analyse de ce mod\`ele afin d'\'evaluer les cons\'equences des r\'esultats r\'ecents en collisionneur sur les propri\'et\'es de la Mati\`ere Noire. Notre objectif est aussi d'inclure le mod\`ele UMSSM dans la version publique du code {\tt micrOMEGAs}.

\begin{appendices}

\renewcommand{\thesection}{\Alph{section}}
\setcounter{section}{0}
\numberwithin{equation}{section}

\section{Cross section sneutrinos - nucleons : gauge bosons contribution}
\label{appen:sigmaZ1Z2}

Here the aim is to compute the cross section for the process below :
\begin{figure}[!ht]
\centering
\includegraphics[scale=0.8]{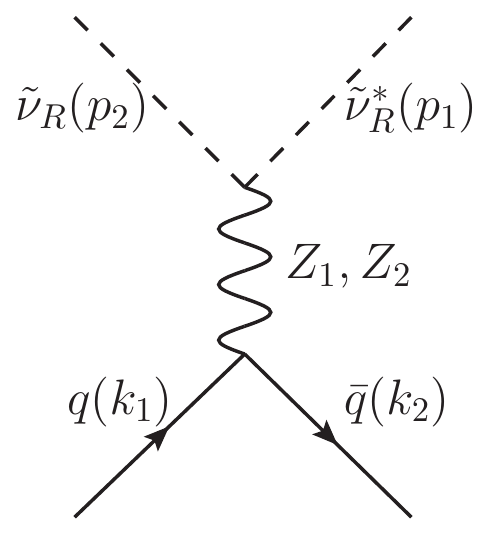}
\end{figure}

Since we consider DD and then low energy processes, we restrict ourselves to the vectorial part : 
\begin{align}
(f_{\bar{q} q Z_1})_{\mu} & = -(\frac{1}{2 s_W} \cos \alpha_{Z} g_Y Q^q_V + \frac{1}{2} \sin \alpha_{Z} {g'_1} Q^{'q}_V)\gamma_{\mu} = \xi_q^{Z_1}\gamma_{\mu},\\
(f_{\bar{q} q Z_2})_{\mu} & = (\frac{1}{2 s_W} \sin \alpha_{Z} g_Y Q^q_V - \frac{1}{2} \cos \alpha_{Z} {g'_1} Q^{'q}_V)\gamma_{\mu} = \xi_q^{Z_2}\gamma_{\mu},\\
(f_{Z_1 \tilde{\nu}^*_R \tilde{\nu}_R})_{\mu} & = {g'_1} \mathcal{Q}'_{\nu} \sin \alpha_{Z} (p_1 + p_2)_{\mu},\\
(f_{Z_2 \tilde{\nu}^*_R \tilde{\nu}_R})_{\mu} & = {g'_1} \mathcal{Q}'_{\nu} \cos \alpha_{Z} (p_1 + p_2)_{\mu}.
\end{align}

Then the matrix element for the interaction between the LSP and a nucleus $N$ with $Z$ protons and $A-Z$ neutrons is :

\begin{figure}[!ht]
\centering
\includegraphics[scale=0.5]{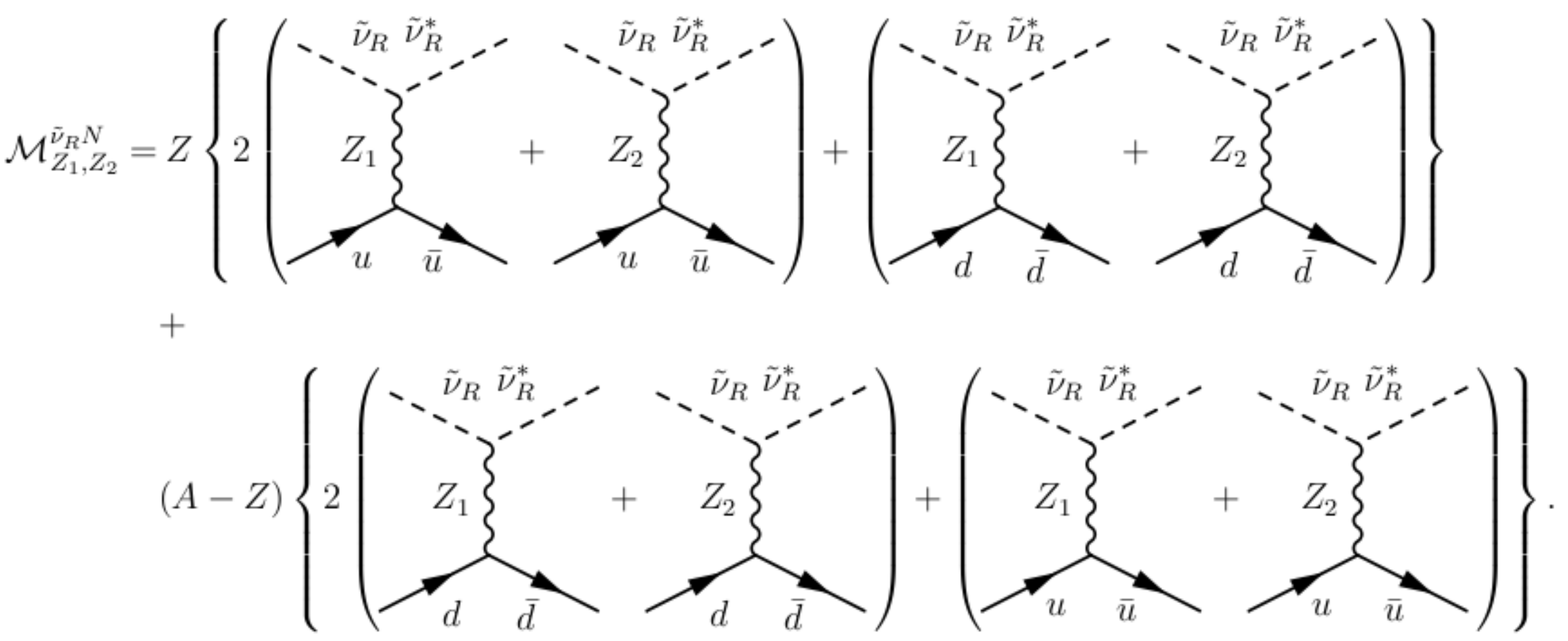}
\end{figure}

It follows that
 
\begin{align}
\mathcal{M}^{\tilde{\nu}_R N}_{Z_1, Z_2} = & \, -i {g'_1} \mathcal{Q}'_{\nu} (p_1 + p_2)_{\mu} \tilde{\nu}^*_R \tilde{\nu}_R \bar{N(k_2)} \gamma^{\mu} N(k_1)\nonumber\\
& \left\{ Z\left[ \frac{\sin \alpha_{Z}}{M^2_{Z_1}} (2 \xi_u^{Z_1} + \xi_d^{Z_1}) + \frac{\cos \alpha_{Z}}{M^2_{Z_2}} (2 \xi_u^{Z_2} + \xi_d^{Z_2}) \right] + \right. \nonumber \\ 
& \left. (A-Z)\left[ \frac{\sin \alpha_{Z}}{M^2_{Z_1}} (2 \xi_d^{Z_1} + \xi_u^{Z_1}) + \frac{\cos \alpha_{Z}}{M^2_{Z_2}} (2 \xi_d^{Z_2} + \xi_u^{Z_2}) \right] \right\}.
\end{align}
Given that the expressions
\begin{align}
Q^q_V & = I_3^q - 2 \mathcal{Q}^q s^2_W, \\
Q^{'q}_V & = \mathcal{Q}'_Q - \mathcal{Q}'_{\bar{Q}},
\end{align}
leads to
\begin{align}
Q^u_V = \frac{1}{6}(3 - 8 s^2_W),& \qquad Q^d_V = \frac{1}{6}(-3 + 4 s^2_W), \\
Q^{'u}_V& = 0,
\end{align}
we have for the $Z$ term : 
\begin{align}
& \: \frac{\sin \alpha_{Z}}{M^2_{Z_1}} (2 \xi_u^{Z_1} + \xi_d^{Z_1}) + \frac{\cos \alpha_{Z}}{M^2_{Z_2}} (2 \xi_u^{Z_2} + \xi_d^{Z_2}) \nonumber \\
= & \: \frac{g_Y \sin \alpha_{Z} \cos \alpha_{Z}}{4 s_W} (1 - 4 s^2_W) \left(\frac{1}{M^2_{Z_2}} - \frac{1}{M^2_{Z_1}}\right) - \frac{{g'_1}}{2} Q^{'d}_V \left(\frac{\sin^2 \alpha_{Z}}{M^2_{Z_1}} + \frac{\cos^2 \alpha_{Z}}{M^2_{Z_2}}\right),
\end{align}
while the $A-Z$ term reads :
\begin{align}
& \: \frac{\sin \alpha_{Z}}{M^2_{Z_1}} (2 \xi_d^{Z_1} + \xi_u^{Z_1}) + \frac{\cos \alpha_{Z}}{M^2_{Z_2}} (2 \xi_d^{Z_2} + \xi_u^{Z_2}) \nonumber \\
= & \: -\frac{g_Y \sin \alpha_{Z} \cos \alpha_{Z}}{4 s_W} \left(\frac{1}{M^2_{Z_2}} - \frac{1}{M^2_{Z_1}}\right) - g'_1 Q^{'d}_V \left(\frac{\sin^2 \alpha_{Z}}{M^2_{Z_1}} + \frac{\cos^2 \alpha_{Z}}{M^2_{Z_2}}\right).
\end{align}
Then in the non-relativistic approximation we have
\beq
| \bar{\mathcal{M}}_{Z_1, Z_2} |^2 = \frac{1}{2}\sum_{spins}| \mathcal{M}_{Z_1, Z_2} |^2 = 16C({g'_1} \mathcal{Q}'_{\nu})^2 m^2_{\tilde{\nu}_R} m^2_N.
\eeq
with
\begin{align}
C = & \:\: [ (y(1 - 4 \sin^2\theta_W) + y')Z + (-y + 2y')(A-Z) ]^2,\\
y = & \:\: \frac{g_Y \sin \alpha_{Z} \cos \alpha_{Z}}{4 \sin\theta_W} \left(\frac{1}{M^2_{Z_2}} - \frac{1}{M^2_{Z_1}}\right),\\
y' = & \:\: - \frac{{g'_1}}{2} Q^{'d}_V \left(\frac{\sin^2 \alpha_{Z}}{M^2_{Z_1}} + \frac{\cos^2 \alpha_{Z}}{M^2_{Z_2}}\right).
\end{align}
Finally the contribution of the gauge bosons to the cross section $\lsp$ - nucleus reads
\beq
\sigma^{Z_1, Z_2}_{\tilde{\nu}_R N} = \frac{\mu^2_{\tilde{\nu}_R N}}{\pi} ({g'_1} \mathcal{Q}'_{\nu})^2 [ (y(1 - 4\sin^2\theta_W) + y')Z + (-y + 2y')(A-Z) ]^2,
\eeq
with $\mu_{\tilde{\nu}_R N} = \frac{m_{\tilde{\nu}_R} m_N}{m_{\tilde{\nu}_R} + m_N}.$

\newpage

\section{Radiative corrections in the Higgs sector \textit{\`a la Coleman-Weinberg}}
\label{appen:radcor}

The dominant radiative corrections due to top and stops that are considered in chapter~\ref{chapter:RHsneu} come from \cite{Barger:2006dh}. The CP-even and CP-odd mass matrices are given in terms of the tree-level elements $\left({\mathcal{M}_{\pm}^0}\right)_{ij}$ given in eqs. \ref{treeUMSSM_even} and \ref{treeUMSSM_odd} where $i,j \in \{1,2,3\}$ plus the corrections $\left({\mathcal{M}_{\pm}^1}\right)_{ij}$ that are obtained through the Coleman-Weinberg correction to the scalar potential with 1-loop (s)top \cite{Coleman:1973jx}. They read :
\begin{align}
\left({\mathcal{M}_{+}^1}\right)_{11} & = k \left[ \left( \frac{({\widetilde m}^2_1)^2}{(m_{\widetilde t_1}^2 - m_{\widetilde t_2}^2)^2} {\mathcal G} \right) v_d^2 + \left( \frac{\lambda y_t^2 A_t}{2 \sqrt{2}} \mathcal{F} \right) \frac{v_s v_u}{v_d} \right],\nonumber \\
\left({\mathcal{M}_{+}^1}\right)_{12} & = k \left[ \left( \frac{{\widetilde m}^2_1 {\widetilde m}^2_2}{(m_{\widetilde t_1}^2 - m_{\widetilde t_2}^2)^2} {\mathcal G} + \frac{y_t^2 {\widetilde m}^2_1}{m^2_{\widetilde t_1} + m^2_{\widetilde t_2}} (2-{\mathcal G}) \right) v_u v_d -  \left( \frac{\lambda y_t^2 A_t}{2 \sqrt{2}} \mathcal{F} \right)v_s \right],\nonumber \\
\left({\mathcal{M}_{+}^1}\right)_{13} & = k \left[ \left(\frac{{\widetilde m}^2_1 {\widetilde m}^2_s}{(m_{\widetilde t_1}^2 - m_{\widetilde t_2}^2)^2} {\mathcal G} + \frac{\lambda^2 y_t^2}{2} {\mathcal F} \right) v_d v_s - \left( \frac{\lambda y_t^2 A_t}{2 \sqrt{2}} \mathcal{F} \right) v_u \right],\nonumber \\
\left({\mathcal{M}_{+}^1}\right)_{22} & = k \left( \frac{({\widetilde m}^2_2)^2}{(m_{\widetilde t_1}^2 - m_{\widetilde t_2}^2)^2} {\mathcal G} + \frac{2 y_t^2 {\widetilde m}^2_2}{m^2_{\widetilde t_1} + m^2_{\widetilde t_2}} (2-{\mathcal G}) + y_t^4 \ln \frac{m^2_{\widetilde t_1} m^2_{\widetilde t_2}}{m_t^4} \right) v_u^2\nonumber\\
& \quad + k \left( \frac{\lambda y_t^2 A_t}{2 \sqrt{2}} \mathcal{F} \right) \frac{v_s v_d}{v_u},\nonumber \\
\left({\mathcal{M}_{+}^1}\right)_{23} & = k \left[ \left(\frac{{\widetilde m}^2_2 {\widetilde m}^2_s}{(m_{\widetilde t_1}^2 - m_{\widetilde t_2}^2)^2} {\mathcal G} + \frac{y_t^2 {\widetilde m}^2_s}{m^2_{\widetilde t_1} + m^2_{\widetilde t_2}} (2-{\mathcal G}) \right) v_u v_s - \left( \frac{\lambda y_t^2 A_t}{2 \sqrt{2}} \mathcal{F} \right)v_d \right],\nonumber\\
\left({\mathcal{M}_{+}^1}\right)_{33} & = k \left[ \left(\frac{({\widetilde m}^2_s)^2}{(m_{\widetilde t_1}^2 - m_{\widetilde t_2}^2)^2} {\mathcal G} \right) v_s^2 + \left( \frac{\lambda y_t^2 A_t}{2 \sqrt{2}} \mathcal{F} \right) \frac{v_u v_d}{v_s}\right], \label{eq8}\end{align}

where $k ={3/(4\pi)^2}$, $y_t$ is the top Yukawa coupling and
\begin{align}
{\cal G} & = 2\left[1- \frac{m^2_{\tilde t_1}+ m^2_{\tilde t_2}}{m^2_{\tilde t_1}- m^2_{\tilde t_2}} \log \left( {m_{\tilde t_1} \over m_{\tilde t_2}} \right)\right], 
\nonumber\\
{\cal F} & = \log \left( \frac{m^2_{\tilde t_1} m^2_{\tilde t_2}}{\Lambda^4}\right) - {\cal G}(m^2_{\tilde t_1}, m^2_{\tilde t_2})
\end{align}
and
\begin{align}
\widetilde{m}_1^{2} & = y_t^2 \mu \left(\mu - \frac{A_t v_u}{v_d} \right),\nonumber\\
\widetilde{m}_2^{2} & = y_t^2 A_t \left(A_t - \frac{\mu v_d}{v_u} \right),\nonumber\\
\widetilde{m}_s^{2} & = \left(\frac{v_d y_t}{v_s} \right)^2 \mu (\mu - A_t t_\beta),\\
\end{align}
$\Lambda$ is defined as $\Lambda = \sqrt{m_{\widetilde t_1} m_{\widetilde t_2}}$. 
For the CP-odd sector we have
\beq
(\mathcal{M}^1_{-})_{ij}  = \frac{ \lambda v_u v_d v_s}{\sqrt 2 v_i v_j}\frac{k y_t^2 A_t}{2} {\cal F}, \quad i,j \in \{1,2,3\}, \label{eq9}
\eeq
where $v_1=v_d,v_2=v_u,v_3=v_s$.

For the charged Higgs mass we consider eq.~\ref{treeUMSSM_charged} plus the following corrections :
\beq
\Delta^2_{H^\pm} = {\lambda A_t v_s k y_t^2 {\cal F}\over \sqrt 2 \sin 2 \beta} + {3 {g_2}^2 \over 32 \pi^2 M_W^2}\left({2 m_t^2 m_b^2\over s^2_\beta c^2_\beta}-M_W^2\left({m_t^2\over s^2_\beta}+{m_b^2\over c^2_\beta}\right)+{2\over3}M_W^4\right) \log{ \Lambda^2\over m_t^2}. \label{eq10}
\eeq

\newpage

\section{Gauge invariance : Goldstones and ghosts}
\label{appen:gold_and_ghost}

As it was said in section~\ref{sec:8.DD} the implementation of the UMSSM model in a gauge invariant way needs a stringent definition of unphysical fields that are mandatory for the consistency of the gauge theory considered. These fields are the Goldstone bosons and the ghosts.
 
\subsection{Gauge fixing  : Goldstone of $Z$ and $Z'$}

To be able to only consider the physical degrees of freedom of the model, we must add a Lagrangian term that will allow to get rid of the Goldstone fields. Actually \lhep automatically defines the Godstones associated to the physical gauge bosons. Nevertheless we need to define the Goldstones associated to $Z$ and $Z'$ to be able to compute correctly the unitary matrix $\mathbf{Z_A}$ shown in section~\ref{subsec:7.higgs}. To do that we must introduce a mixing angle $\b_{Z}$ that parameterizes the transition between ($G_{Z_1}, G_{Z_2}$) and ($G_{Z}, G_{Z'}$). The expression of this link reads
\beq
\binom{G_Z}{G_{Z'}} = -\begin{pmatrix} \cos \b_{Z} & -\sin \b_{Z} \\ \sin \b_{Z} & \cos \b_{Z} \end{pmatrix} \binom{G_{Z_1}}{G_{Z_2}},
\eeq
with
\beq 
\tan \b_{Z} = \frac{\cos \alpha_{Z} \sqrt{g_Y^2 + g_2^2} (v_d Z_{A 21} - v_u Z_{A 22}) + \sin \azz g'_1 \sum_{i=1}^{3} \mathcal{Q}'_i v_i Z_{A 2i}}{\cos \alpha_{Z} \sqrt{g_Y^2 + g_2^2} (v_d Z_{A 11} - v_u Z_{A 12}) + \sin \azz g'_1 \sum_{i=1}^{3} \mathcal{Q}'_i v_i Z_{A 1i}}, \eeq
and with $(v_1,\mathcal{Q}'_1) \equiv (v_d, \mathcal{Q}'_{H_d})$, $(v_2,\mathcal{Q}'_2) \equiv (v_u,\mathcal{Q}'_{H_u})$ and $(v_3,\mathcal{Q}'_3) \equiv (v_s,\mathcal{Q}'_S)$. Then in the Feynman gauge the gauge fixing Lagrangian of the model can be expressed as
\beq \mathscr{L}_{\mathrm{UMSSM}}^{\mathrm{GF}}  = -\frac{1}{2} \left(|F^\g|^2 + |F^G|^2 + 2F^{W^+}F^{W^-} + |F^{Z_1}|^2 + |F^{Z_2}|^2 \right),\eeq
where we define
\beq \begin{split}
F^\g & = \p_\mu A^\mu,\\
F^G & = \p_\mu G^\mu,\\
F^{W^\pm} & = \p_\mu W^\pm \pm i M_W G_{W^\pm},\\
F^{Z_1} & = \p_\mu Z_1^\mu + M_{Z_1} G_{Z_1},\\
F^{Z_2} & = \p_\mu Z_2^\mu + M_{Z_2} G_{Z_2}.
\end{split} \eeq

\subsection{Fadeev-Popov ghosts}

The Fadeev-Popov ghosts $c$ \cite{Faddeev:1967fc}, anticommuting scalar fields which are needed to maintain the consistency of the gauge theory, have the following general Lagrangian term :
\beq \mathscr{L}^{\mathrm{FP}} = -\bar{c}_\alpha (D(c)[\Phi^\alpha]), \eeq
where $\Phi$ is a gauge boson and $D(c)$ is a gauge transformation. To do that we start from the Becchi-Rouet-Stora-Tyutin (BRST) tranformations \cite{Becchi:1974xu,Becchi:1974md,Becchi:1975nq,Tyutin:1975qk} given in \cite{Belanger:2003sd} and the relations given in \cite{Kuroda:1999ks}. For the contribution of the gauge bosons that we want to highlight for our model we have
\beq \begin{split}
D(c)[W^\alpha_\mu] & = g_2 \epsilon^{\alpha \beta \gamma} W^\beta_\mu c^\gamma + \partial_\mu c^\alpha \\
D(c)[B_\mu] & = \partial_\mu c_B \\
D(c)[B'_\mu] & = \partial_\mu c_{B'}.
\end{split} \eeq

\subsubsection{Ghosts in the MSSM}

In the MSSM, we have the following terms :
\beq \begin{split}
& -\bar{c}_{W^+} D(c)[\partial^\mu W^+_\mu + M_W G_{W^+}],\\ 
& -\bar{c}_{W^-} D(c)[\partial^\mu W^-_\mu + M_W G_{W^-}],\\ 
& -\bar{c}_Z D(c)[\partial^\mu Z_\mu + M_Z G_Z],\\ 
& -\bar{c}_\g D(c)[\partial^\mu A_\mu],
\end{split} \eeq
with
\beq \begin{split}
c_{W^\pm} & = \frac{1}{\sqrt{2}}(c_{W^1} \mp c_{W^2}),\\ 
c_\g & = s_W c_{W^3} + c_W c_B,\\
c_Z & = c_W c_{W^3} - s_W c_B.  
\end{split} \eeq
We can now write
\beq \begin{split}
D(c)[\partial^\mu W^1_\mu] & = g_2 W^2_\mu c_{W^3} -g_2 W^3_\mu c_{W^2} + \partial_\mu c_{W^1},\\
D(c)[\partial^\mu W^2_\mu] & = g_2 W^3_\mu c_{W^1} -g_2 W^1_\mu c_{W^3} + \partial_\mu c_{W^2}
\end{split} \eeq
which gives
\beq \begin{split} -\bar{c}_{W^\pm} D(c)[\partial^\mu W^\pm_\mu] & = - \frac{1}{\sqrt{2}} \bar{c}_{W^\pm} \partial^\mu(D(c)[W^1_\mu] \mp D(c)[W^2_\mu])\\
& = -\partial^\mu\partial_\mu \bar{c}_{W^\pm} c_{W^\pm} - i g_2 \bar{c}_{W^\pm} \partial^\mu (W^\pm_\mu c_{W^3} - c_{W^\pm} W^3_\mu).
\end{split} \eeq
Using the same method we have
\beq \begin{split}
-\bar{c}_Z D(c)[\partial^\mu Z_\mu] & = -\partial^\mu\partial_\mu \bar{c}_Z c_Z + i g_2 c_W \bar{c}_Z \partial^\mu (W^+_\mu c_{W^-} - c_{W^+} W^-_\mu),\\
-\bar{c}_\g D(c)[\partial^\mu A_\mu] & = -\partial^\mu\partial_\mu \bar{c}_\g c_\g + i g_2 s_W \bar{c}_\g \partial^\mu (W^+_\mu c_{W^-} - c_{W^+} W^-_\mu).
\end{split} \eeq
For the gauge transformations acting on Goldstones, we have to use the definition of the Higgs fields\footnote{The following definitions apply only for this appendix on ghosts in the (U)MSSM.} : 
\begin{align}
H_d = \left(\begin{array}{c} (v_d + \phi_d -i \varphi_d)/\sqrt{2}  \\ -\phi^-_d \end{array}\right), Y = -1 \qquad
H_u = \left(\begin{array}{c}  \phi^+_u \\ (v_u + \phi_u +i \varphi_u)/\sqrt{2} \end{array}\right), Y = 1.
\end{align}
With $Y$ the hypercharge. It follows that
\begin{align}
D(c) H_{d,u} & = i \frac{g_2}{2} c_{W^\a} \s_\a H_{d, u} + i \frac{g'}{2} Y c_B H_{d, u} \nonumber\\
& = i \frac{g_2}{2} (c_{W^\a} \s_\a + t_W Y c_B) H_{d, u} \nonumber\\
& = i \frac{g_2}{2} \left[ \sqrt{2} \begin{pmatrix} 0 & c_{W^+} \\ c_{W^-} & 0 \end{pmatrix} + c_{W^3} \begin{pmatrix} 1 & 0 \\ 0 & -1 \end{pmatrix} + c_B \begin{pmatrix} t_W Y & 0 \\ 0 & t_W Y \end{pmatrix}\right] H_{d, u} \nonumber\\
& = i \frac{g_2}{2} \left[ \sqrt{2} \begin{pmatrix} 0 & c_{W^+} \\ c_{W^-} & 0 \end{pmatrix} + s_W c_\gamma \begin{pmatrix} 1 + Y & 0 \\ 0 & Y -1 \end{pmatrix}\right. \nonumber\\
& \ \ \left. + c_W c_Z \begin{pmatrix} 1 - Y t^2_W & 0 \\ 0 & -(1 + Y t^2_W) \end{pmatrix}\right] H_{d, u}.
\end{align}
This leads to
\begin{align}
D(c) H_{d} & = i \frac{g_2}{2} \left[ \begin{pmatrix} \frac{c_Z}{c_W} & c_{W^+} \sqrt{2} \\ \sqrt{2} c_{W^-} & -2 s_W c_\gamma -c_W c_Z (1 - t^2_W) \end{pmatrix}\right] \left(\begin{array}{c} (v_d + \phi_d -i \varphi_d)/\sqrt{2}  \\ -\phi^-_d \end{array}\right)\nonumber \\
& = i \frac{g_2}{2} \left[ \begin{array}{c} \frac{c_Z}{c_W \sqrt{2}}(v_d + \phi_d -i \varphi_d) -\phi^-_d c_{W^+} \sqrt{2} \\ 
c_{W^-} (v_d + \phi_d -i \varphi_d) + \phi^-_d ( 2 s_W c_\gamma + c_W c_Z (1 - t^2_W)) \end{array}\right],\\
D(c) H_{u} & = i \frac{g_2}{2} \left[ \begin{array}{c} c_{W^+} (v_u + \phi_u +i \varphi_u) + \phi^+_d ( 2 s_W c_\gamma + c_W c_Z (1 - t^2_W)) \\ 
-\frac{c_Z}{c_W \sqrt{2}}(v_u + \phi_u +i \varphi_u) + \phi^+_u c_{W^-} \sqrt{2} \end{array}\right].
\end{align}
Defining the transformations
\begin{align}
\frac{D(c) H^0_{d} + i D(c) \varphi_{d}}{\sqrt{2}} & = \begin{pmatrix} 1 & 0 \end{pmatrix} D(c) H_{d},\\
\frac{D(c) H^0_{u} - i D(c) \varphi_{u}}{\sqrt{2}} & = \begin{pmatrix} 0 & 1 \end{pmatrix} D(c) H_{u},\\
i D(c) \phi^-_{d} & = \begin{pmatrix} 0 & 1 \end{pmatrix} D(c) H_{d},\\
-i D(c) \phi^-_{u} & = \left[ \begin{pmatrix} 1 & 0 \end{pmatrix} D(c) H_{u}\right]^\dag,\\
-i D(c) \phi^+_{d} & = \left[ \begin{pmatrix} 0 & 1 \end{pmatrix} D(c) H_{d}\right]^\dag,\\
i D(c) \phi^+_{u} & = \begin{pmatrix} 1 & 0 \end{pmatrix} D(c) H_{u},
\end{align}
we have
\begin{align}
D(c) H^0_{d, u} & = -i \frac{g_2}{2} ( c_{W^+} \phi^-_{d, u} -c_{W^-} \phi^+_{d, u} + \frac{c_Z}{c_W}i\varphi_{d, u}),\\
D(c) \varphi_{d, u} & = - \frac{g_2}{2} ( c_{W^+} \phi^-_{d, u} +c_{W^-} \phi^+_{d, u} - \frac{c_Z}{c_W}(v_{d, u} + \phi_{d, u})),\\
D(c) \phi^\pm_{d, u} & = \frac{g_2}{2} \left( c_{W^\pm} (v_{d, u} + \phi_{d, u} \pm i \varphi_{d, u}) + \phi^\pm_{d, u} ( 2 s_W c_\gamma + c_W c_Z (1 - t^2_W))\right).
\end{align}
With
\begin{align}
G_{W^\pm} & = c_\beta \phi^\pm_{d} + s_\beta \phi^\pm_{u},\\
G_{Z} & = c_\beta \varphi_{d} + s_\beta \varphi_{u},
\end{align}
the last Fadeev-Popov terms read
\begin{align}
-\bar{c}_{W^\pm} D(c)[M_W G_{W^\pm}] = & -M_W \bar{c}_{W^\pm} D(c)[c_\beta \phi^\pm_{d} + s_\beta \phi^\pm_{u}]\nonumber\\ 
= & -M_W c_\beta \frac{g_2}{2} \bar{c}_{W^\pm} \left( c_{W^\pm} (v_{d} + \phi_{d} \pm i \varphi_{d}) + \phi^\pm_{d} ( 2 s_W c_\gamma + c_W c_Z (1 - t^2_W))\right)\nonumber\\ 
& -M_W s_\beta \frac{g_2}{2} \bar{c}_{W^\pm} \left( c_{W^\pm} (v_{u} + \phi_{u} \pm i \varphi_{u}) + \phi^\pm_{u} ( 2 s_W c_\gamma + c_W c_Z (1 - t^2_W))\right)\nonumber\\ 
= & -M_W \frac{g_2}{2} \bar{c}_{W^\pm} ( 2 s_W c_\gamma + c_W c_Z (1 - t^2_W))G_{W^\pm} \nonumber\\ 
& -M_W \frac{g_2}{2} \bar{c}_{W^\pm} c_{W^\pm} ( c_\beta \phi_{d} + s_\beta \phi_{u} \pm i G_{Z}) + \textrm{quadratic terms} \nonumber\\ 
= & \mp i M_W \frac{g_2}{2} \bar{c}_{W^\pm} c_{W^\pm} G_{Z} -M_W \frac{g_2}{2} \bar{c}_{W^\pm} c_{W^\pm} \begin{pmatrix} c_\beta & s_\beta \end{pmatrix} \begin{pmatrix} \phi_{d} \\ \phi_{u} \end{pmatrix} \nonumber\\ 
& -e M_W \bar{c}_{W^\pm} c_\gamma G_{W^\pm} -M_W \frac{g_2}{2 c_W} (c^2_W - s^2_W)\bar{c}_{W^\pm} c_Z G_{W^\pm}\nonumber\\
& + \textrm{quadratic terms}, 
\end{align}

\begin{align}
-\bar{c}_Z D(c)[M_Z G_Z] = & -M_Z \bar{c}_Z D(c)[c_\beta \varphi_{d} + s_\beta \varphi_{u}]\nonumber\\ 
= & \: M_Z c_\beta \frac{g_2}{2} \bar{c}_Z ( c_{W^+} \phi^-_{d} +c_{W^-} \phi^+_{d} - \frac{c_Z}{c_W}(v_{d} + \phi_{d}))\nonumber\\ 
& + M_Z s_\beta \frac{g_2}{2} \bar{c}_Z ( c_{W^+} \phi^-_{u} +c_{W^-} \phi^+_{u} - \frac{c_Z}{c_W}(v_{u} + \phi_{u}))\nonumber\\ 
= & \: M_Z \frac{g_2}{2} \bar{c}_Z (c_{W^+} G_{W^-} + c_{W^-} G_{W^+}) + M_Z \frac{g_2}{2 c_W} \bar{c}_Z c_Z \begin{pmatrix} c_\beta & s_\beta \end{pmatrix} \begin{pmatrix} \phi_{d} \\ \phi_{u} \end{pmatrix} \nonumber\\
& + \textrm{quadratic terms}. 
\end{align}
It follows that the Fadeev-Popov Lagrangian in the MSSM is :
\beq \begin{split}
\mathscr{L}^{\mathrm{FP}}_{\mathrm{MSSM}} = & \: g_3 f^{\alpha \beta \gamma} \bar{c}^\alpha c^\beta \partial^\mu G^\gamma_\mu \\
& - i g_2 \bar{c}_{W^\pm} \partial^\mu (W^\pm_\mu c_{W^3} - c_{W^\pm} W^3_\mu) + i g_2 c_W \bar{c}_Z \partial^\mu (W^+_\mu c_{W^-} - c_{W^+} W^-_\mu)\\ 
& + i g_2 s_W \bar{c}_\gamma \partial^\mu (W^+_\mu c_{W^-} - c_{W^+} W^-_\mu) \mp i M_W \frac{g_2}{2} \bar{c}_{W^\pm} c_{W^\pm} G_{Z} -e M_W \bar{c}_{W^\pm} c_\gamma G_{W^\pm}\\
& -M_W \frac{g_2}{2} \bar{c}_{W^\pm} c_{W^\pm} \begin{pmatrix} c_\beta & s_\beta \end{pmatrix} \begin{pmatrix} \phi_{d} \\ \phi_{u} \end{pmatrix} -M_W \frac{g_2}{2 c_W} (c^2_W - s^2_W)\bar{c}_{W^\pm} c_Z G_{W^\pm} \\ 
& + M_Z \frac{g_2}{2 c_W} \bar{c}_Z c_Z \begin{pmatrix} c_\beta & s_\beta \end{pmatrix} \begin{pmatrix} \phi_{d} \\ \phi_{u} \end{pmatrix} + M_Z \frac{g_2}{2} \bar{c}_Z (c_{W^+} G_{W^-} + c_{W^-} G_{W^+}) \\ 
& + \textrm{quadratic terms}.
\end{split} \eeq

\subsubsection{Ghosts in the UMSSM}

Here we also have to consider $S = (v_s + \sigma + i \xi)/\sqrt{2}$, with $Y = 0$. Comparing to the MSSM we modify or add the following terms :
\begin{align}
& -\bar{c}_{W^\pm} D(c)[M_W G_{W^\pm}],\nonumber\\ 
& -\bar{c}_{Z_1} D(c)[\partial^\mu Z_{1 \mu} + M_{Z_1} G_{Z_1}],\nonumber\\ 
& -\bar{c}_{Z_2} D(c)[\partial^\mu Z_{2 \mu} + M_{Z_2} G_{Z_2}].\nonumber
\end{align}
The transformations read now :
\begin{align}
D(c) H_{d,u} & = i \frac{g_2}{2} \left[ \sqrt{2} \begin{pmatrix} 0 & c_{W^+} \\ c_{W^-} & 0 \end{pmatrix} + s_W c_\gamma \begin{pmatrix} 1 + Y & 0 \\ 0 & Y -1 \end{pmatrix} + c_W c_Z \begin{pmatrix} 1 - Y t^2_W & 0 \\ 0 & -(1 + Y t^2_W) \end{pmatrix}\right. \nonumber\\
& \qquad \quad \left. + 2 \frac{g'_1}{g_2}\mathcal{Q}'_{H_d,H_u} c_{Z'}\begin{pmatrix} 1 & 0 \\ 0 & 1\end{pmatrix}\right] H_{d, u} \nonumber \\
D(c) H_{d} & = i \frac{g_2}{2} \left[ \begin{array}{c} \left( \frac{c_Z}{c_W} + \frac{2 g'_1}{g_2}\mathcal{Q}'_{H_d} c_{Z'} \right)(v_d + \phi_d -i \varphi_d)/\sqrt{2} -\phi^-_d c_{W^+} \sqrt{2} \\ 
c_{W^-} (v_d + \phi_d -i \varphi_d) + \phi^-_d \left( 2 s_W c_\gamma + c_W c_Z (1 - t^2_W) - \frac{2 g'_1}{g_2}\mathcal{Q}'_{H_d} c_{Z'}\right) \end{array}\right] \nonumber\\
D(c) H_{u} & = i \frac{g_2}{2} \left[ \begin{array}{c} c_{W^+} (v_u + \phi_u +i \varphi_u) + \phi^+_d \left( 2 s_W c_\gamma + c_W c_Z (1 - t^2_W)+ \frac{2 g'_1}{g_2}\mathcal{Q}'_{H_u} c_{Z'}\right) \nonumber\\ 
\left( -\frac{c_Z}{c_W} + \frac{2 g'_1}{g_2}\mathcal{Q}'_{H_u} c_{Z'} \right)(v_u + \phi_u +i \varphi_u)/\sqrt{2} + \phi^+_u c_{W^-} \sqrt{2} \end{array}\right]\\
D(c) S & \: = i g'_1 \mathcal{Q}'_S c_{Z'}(v_s + \sigma + i \xi)/\sqrt{2}.\nonumber
\end{align}
Which gives
\begin{align}
D(c) H^0_{d, u} & = -i \frac{g_2}{2} ( c_{W^+} \phi^-_{d, u} -c_{W^-} \phi^+_{d, u} + \frac{c_Z}{c_W}i\varphi_{d, u}) \pm g'_1 \mathcal{Q}'_{H_d,H_u} c_{Z'}\varphi_{d, u}\\
D(c) \varphi_{d, u} & = - \frac{g_2}{2} ( c_{W^+} \phi^-_{d, u} +c_{W^-} \phi^+_{d, u} - \frac{c_Z}{c_W}(v_{d, u} + \phi_{d, u})) \pm g'_1 \mathcal{Q}'_{H_d,H_u} c_{Z'}(v_{d, u} + \phi_{d, u})\\
D(c) \phi^\pm_{d, u} & = \frac{g_2}{2} \left( c_{W^\pm} (v_{d, u} + \phi_{d, u} \pm i \varphi_{d, u}) + \phi^\pm_{d, u} ( 2 s_W c_\gamma + c_W c_Z (1 - t^2_W))\right)\nonumber\\
& \: \quad \mp g'_1 \mathcal{Q}'_{H_d,H_u} c_{Z'}\phi^\pm_{d, u}\\
D(c) S^0 & \: = -g'_1 \mathcal{Q}'_S c_{Z'}\xi\\
D(c) \xi & \: = -g'_1 \mathcal{Q}'_S c_{Z'}(v_s + \sigma).
\end{align}

$G_{W^\pm}$ has the same expression as in the MSSM. However in the UMSSM we have 
\begin{align}
G_{Z} = & -Z_{A 11}\varphi_{d} + Z_{A 12}\varphi_{u} + Z_{A 13}\xi \\
G_{Z'} = & -Z_{A 21}\varphi_{d} + Z_{A 22}\varphi_{u} + Z_{A 23}\xi.
\end{align}
It implies :
\begin{align}
-\bar{c}_{W^\pm} D(c)[M_W G_{W^\pm}] = & -M_W \frac{g_2}{2} \bar{c}_{W^\pm} c_{W^\pm} \begin{pmatrix} c_\beta & s_\beta \end{pmatrix} \begin{pmatrix} \phi_{d} \pm i\varphi_{d}\\ \phi_{u} \pm i\varphi_{u}\end{pmatrix} -e M_W \bar{c}_{W^\pm} c_\gamma G_{W^\pm} \nonumber\\ 
& -M_W \frac{g_2}{2 c_W} (c^2_W - s^2_W)\bar{c}_{W^\pm} c_Z G_{W^\pm}\nonumber\\ 
& + M_W g'_1 \bar{c}_{W^\pm} c_{Z'} \left((c^2_\beta \mathcal{Q}'_{H_d} - s^2_\beta \mathcal{Q}'_{H_u})G_{W^\pm} + c_\beta s_\beta \mathcal{Q}'_S H^\pm \right)\nonumber\\
& + \textrm{quadratic terms},
\end{align}

\begin{align}
-\bar{c}_{Z_1} D(c)[M_{Z_1} G_{Z_1}] = & M_{Z_1} \bar{c}_{Z_1} D(c)[\cos \beta_{Z} G_{Z} + \sin \beta_{Z} G_{Z'}] \nonumber\\
= & -\cos \beta_{Z} M_{Z_1}\frac{g_2}{2} \bar{c}_{Z_1} \left( -Z_{A 11} c_{W^\pm}(c_\beta G_{W^\pm} - s_\beta H^\pm) + Z_{A 12} c_{W^\pm}(s_\beta G_{W^\pm} + c_\beta H^\pm)\right) \nonumber\\
 & -\sin \beta_{Z} M_{Z_1}\frac{g_2}{2} \bar{c}_{Z_1} \left( -Z_{A 21} c_{W^\pm}(c_\beta G_{W^\pm} - s_\beta H^\pm) + Z_{A 22} c_{W^\pm}(s_\beta G_{W^\pm} + c_\beta H^\pm)\right) \nonumber\\
& +\cos \beta_{Z} M_{Z_1}\frac{g_2}{2 c_W} \bar{c}_{Z_1} c_Z (-Z_{A 11}\phi_{d} + Z_{A 12}\phi_{u}) \nonumber\\
& +\sin \beta_{Z} M_{Z_1}\frac{g_2}{2 c_W} \bar{c}_{Z_1} c_Z (-Z_{A 21}\phi_{d} + Z_{A 22}\phi_{u}) \nonumber\\
& -\cos \beta_{Z} M_{Z_1} g'_1 \bar{c}_{Z_1} c_{Z'} (Z_{A 11} \mathcal{Q}'_{H_d} \phi_{d} + Z_{A 12} \mathcal{Q}'_{H_u} \phi_{u} + Z_{A 13} \mathcal{Q}'_S \sigma)\nonumber\\
& -\sin \beta_{Z} M_{Z_1} g'_1 \bar{c}_{Z_1} c_{Z'} (Z_{A 21} \mathcal{Q}'_{H_d} \phi_{d} + Z_{A 22} \mathcal{Q}'_{H_u} \phi_{u} + Z_{A 23} \mathcal{Q}'_S \sigma)\nonumber\\
& + \textrm{quadratic terms},
\end{align}

\begin{align}
-\bar{c}_{Z_2} D(c)[M_{Z_2} G_{Z_2}] = & M_{Z_2} \bar{c}_{Z_2} D(c)[-\sin \beta_{Z} G_{Z} + \cos \beta_{Z} G_{Z'}] \nonumber\\
= & \sin \beta_{Z} M_{Z_2}\frac{g_2}{2} \bar{c}_{Z_2} \left( -Z_{A 11} c_{W^\pm}(c_\beta G_{W^\pm} - s_\beta H^\pm) + Z_{A 12} c_{W^\pm}(s_\beta G_{W^\pm} + c_\beta H^\pm)\right) \nonumber\\
& -\cos \beta_{Z} M_{Z_2}\frac{g_2}{2} \bar{c}_{Z_2} \left( -Z_{A 21} c_{W^\pm}(c_\beta G_{W^\pm} - s_\beta H^\pm) + Z_{A 22} c_{W^\pm}(s_\beta G_{W^\pm} + c_\beta H^\pm)\right) \nonumber\\
& -\sin \beta_{Z} M_{Z_2}\frac{g_2}{2 c_W} \bar{c}_{Z_2} c_Z (-Z_{A 11}\phi_{d} + Z_{A 12}\phi_{u}) \nonumber\\
& +\cos \beta_{Z} M_{Z_2}\frac{g_2}{2 c_W} \bar{c}_{Z_2} c_Z (-Z_{A 21}\phi_{d} + Z_{A 22}\phi_{u}) \nonumber\\
& +\sin \beta_{Z} M_{Z_2} g'_1 \bar{c}_{Z_2} c_{Z'} (Z_{A 11} \mathcal{Q}'_{H_d} \phi_{d} + Z_{A 12} \mathcal{Q}'_{H_u} \phi_{u} + Z_{A 13} \mathcal{Q}'_S \sigma)\nonumber\\
& -\cos \beta_{Z} M_{Z_2} g'_1 \bar{c}_{Z_2} c_{Z'} (Z_{A 21} \mathcal{Q}'_{H_d} \phi_{d} + Z_{A 22} \mathcal{Q}'_{H_u} \phi_{u} + Z_{A 23} \mathcal{Q}'_S \sigma)\nonumber\\
& + \textrm{quadratic terms}. 
\end{align}
Finally, the complete Fadeev-Popov Lagrangian in the UMSSM reads
\begin{align}
\mathscr{L}^{\mathrm{FP}}_{\mathrm{UMSSM}} = & \: g_3 f^{\alpha \beta \gamma} \bar{c}^\alpha c^\beta \partial^\mu G^\gamma_\mu - i g_2 \bar{c}_{W^\pm} \partial^\mu (W^\pm_\mu c_{W^3} - c_{W^\pm} W^3_\mu) + i g_2 c_W \bar{c}_Z \partial^\mu (W^+_\mu c_{W^-} - c_{W^+} W^-_\mu) \nonumber\\ 
& + i g_2 s_W \bar{c}_\gamma \partial^\mu (W^+_\mu c_{W^-} - c_{W^+} W^-_\mu) -M_W \frac{g_2}{2} \bar{c}_{W^\pm} c_{W^\pm} \begin{pmatrix} c_\beta & s_\beta \end{pmatrix} \begin{pmatrix} \phi_{d} \pm i\varphi_{d}\\ \phi_{u} \pm i\varphi_{u}\end{pmatrix} \nonumber\\ 
& -e M_W \bar{c}_{W^\pm} c_\gamma G_{W^\pm}  -M_W \frac{g_2}{2 c_W} (c^2_W - s^2_W)\bar{c}_{W^\pm} c_Z G_{W^\pm} \nonumber\\
& + M_W g'_1 \bar{c}_{W^\pm} c_{Z'} \left((c^2_\beta \mathcal{Q}'_{H_d} - s^2_\beta \mathcal{Q}'_{H_u})G_{W^\pm} + c_\beta s_\beta \mathcal{Q}'_S H^\pm \right) \nonumber\\
& -\cos \beta_{Z} M_{Z_1}\frac{g_2}{2} \bar{c}_{Z_1} \left( -Z_{A 11} c_{W^\pm}(c_\beta G_{W^\pm} - s_\beta H^\pm) + Z_{A 12} c_{W^\pm}(s_\beta G_{W^\pm} + c_\beta H^\pm)\right) \nonumber\\
& -\sin \beta_{Z} M_{Z_1}\frac{g_2}{2} \bar{c}_{Z_1} \left( -Z_{A 21} c_{W^\pm}(c_\beta G_{W^\pm} - s_\beta H^\pm) + Z_{A 22} c_{W^\pm}(s_\beta G_{W^\pm} + c_\beta H^\pm)\right) \nonumber\\
& +\cos \beta_{Z} M_{Z_1}\frac{g_2}{2 c_W} \bar{c}_{Z_1} c_Z (-Z_{A 11}\phi_{d} + Z_{A 12}\phi_{u}) \nonumber\\
& +\sin \beta_{Z} M_{Z_1}\frac{g_2}{2 c_W} \bar{c}_{Z_1} c_Z (-Z_{A 21}\phi_{d} + Z_{A 22}\phi_{u}) \nonumber\\
& -\cos \beta_{Z} M_{Z_1} g'_1 \bar{c}_{Z_1} c_{Z'} (Z_{A 11} \mathcal{Q}'_{H_d} \phi_{d} + Z_{A 12} \mathcal{Q}'_{H_u} \phi_{u} + Z_{A 13} \mathcal{Q}'_S \sigma)\nonumber\\
& -\sin \beta_{Z} M_{Z_1} g'_1 \bar{c}_{Z_1} c_{Z'} (Z_{A 21} \mathcal{Q}'_{H_d} \phi_{d} + Z_{A 22} \mathcal{Q}'_{H_u} \phi_{u} + Z_{A 23} \mathcal{Q}'_S \sigma)\nonumber\\
& +\sin \beta_{Z} M_{Z_2}\frac{g_2}{2} \bar{c}_{Z_2} \left( -Z_{A 11} c_{W^\pm}(c_\beta G_{W^\pm} - s_\beta H^\pm) + Z_{A 12} c_{W^\pm}(s_\beta G_{W^\pm} + c_\beta H^\pm)\right) \nonumber\\
& -\cos \beta_{Z} M_{Z_2}\frac{g_2}{2} \bar{c}_{Z_2} \left( -Z_{A 21} c_{W^\pm}(c_\beta G_{W^\pm} - s_\beta H^\pm) + Z_{A 22} c_{W^\pm}(s_\beta G_{W^\pm} + c_\beta H^\pm)\right)\nonumber\\
& -\sin \beta_{Z} M_{Z_2}\frac{g_2}{2 c_W} \bar{c}_{Z_2} c_Z (-Z_{A 11}\phi_{d} + Z_{A 12}\phi_{u}) \nonumber\\
& +\cos \beta_{Z} M_{Z_2}\frac{g_2}{2 c_W} \bar{c}_{Z_2} c_Z (-Z_{A 21}\phi_{d} + Z_{A 22}\phi_{u}) \nonumber\\
& +\sin \beta_{Z} M_{Z_2} g'_1 \bar{c}_{Z_2} c_{Z'} (Z_{A 11} \mathcal{Q}'_{H_d} \phi_{d} + Z_{A 12} \mathcal{Q}'_{H_u} \phi_{u} + Z_{A 13} \mathcal{Q}'_S \sigma)\nonumber\\
& -\cos \beta_{Z} M_{Z_2} g'_1 \bar{c}_{Z_2} c_{Z'} (Z_{A 21} \mathcal{Q}'_{H_d} \phi_{d} + Z_{A 22} \mathcal{Q}'_{H_u} \phi_{u} + Z_{A 23} \mathcal{Q}'_S \sigma)\nonumber\\
& + \textrm{quadratic terms}.
\end{align}

\end{appendices}

\newpage\null\newpage
\cleardoublepage
\phantomsection
\addcontentsline{toc}{chapter}{Bibliography}
\bibliography{biblio}

\newpage \pagestyle{empty} \null

\setlength{\topmargin}{-0.7in}
\setlength{\headheight}{0cm}
\setlength{\headsep}{0cm}
\setlength{\footskip}{-0.8in}
\setlength{\textheight}{19in}
\newpage
\begin{center}
\textbf{\Large{Supersymmetric Dark Matter candidates in light of constraints from collider and astroparticle observables}}
\end{center}

~~~\\
\footnotesize{\textbf{Abstract} : The Standard Model of particle physics has been strengthened by the recent discovery of the long-awaited Higgs boson. The standard cosmological model has met the challenge of the high precision observations in cosmology and astroparticle physics. However these two standard models face both several theoretical issues, such as the naturalness problem in the Higgs sector of the Standard Model, as well as observational issues, in particular the fact that an unknown kind of matter called Dark Matter accounts for the majority of the matter content in our Universe. Attempts to solve such problems have led to the development of New Physics models during the last decades. Supersymmetry is one such model which addresses the fine-tuning problem in the Higgs sector and provides viable Dark Matter candidates. Current high energy and high precision experiments give many new opportunities to probe the supersymmetric models. It is in this context that this thesis is written. Considering the Minimal Supersymmetric Standard Model (MSSM), the simplest supersymmetric extension of the Standard Model of particle physics, and its conventional Dark Matter candidate, the neutralino, it is shown that collider constraints could provide informations on the very early Universe at the inflation area. It is also demonstrated that the Indirect Detection of Dark Matter, despite several drawbacks, can be a powerful technique to probe supersymmetric Dark Matter models. Beyond the MSSM it is shown that unique characteristics of the Dark Matter candidate in the NMSSM could be probed at colliders. The study of a supersymmetric model with an extended gauge symmetry, the UMSSM, is also developed. The features of another Dark Matter candidate of this model, the Right-Handed sneutrino, are analysed. More general constraints such as those coming from low energy observables are finally considered in this model.}

~~~\\
\footnotesize{\textbf{Keywords} : Dark matter - Supersymmetry - Collider and astroparticles constraints - Neutralino - Right-Handed sneutrino - UMSSM}
~~~\\
\rule{\linewidth}{.5pt}
\begin{center}
\textbf{\Large{Les candidats supersym\'etriques \`a la mati\`ere noire \`a la lumi\`ere des contraintes provenant des observables en collisionneur et d'astroparticule}}
\end{center}

~~~\\
\footnotesize{\textbf{R\'esum\'e} : Le Mod\`ele Standard de la physique des particules a \'et\'e renforc\'e par la r\'ecente d\'ecouverte du tr\`es attendu boson de Higgs. Le mod\`ele standard cosmologique a lui relev\'e le d\'efi de la haute pr\'ecision des observations cosmologiques et des exp\'eriences d'astroparticules. Toutefois, ces deux mod\`eles standards sont encore confront\'es \`a plusieurs probl\`emes th\'eoriques, comme le probl\`eme de naturalit\'e dans le secteur de Higgs du Mod\`ele Standard, ainsi que des probl\`emes observationnels \`a l'image des nombreuses preuves de l'existence d'un genre inconnu de mati\`ere, appel\'e Mati\`ere Noire, qui repr\'esenterait la majeure partie du contenu en mati\`ere de l'Univers. Les tentatives visant \`a r\'esoudre ces probl\`emes ont conduit au d\'eveloppement de nouveaux mod\`eles physiques au cours des derni\`eres d\'ecennies. La Supersym\'etrie est un de ces mod\`eles qui traite du probl\`eme du r\'eglage fin dans le secteur de Higgs et fournit de bons candidats \`a la Mati\`ere Noire. Les exp\'eriences actuelles de physique des hautes \'energies et de haute pr\'ecision offrent de nombreuses possibilit\'es pour contraindre les mod\`eles supersym\'etriques. C'est dans ce contexte que cette th\`ese s'inscrit. En consid\'erant le Mod\`ele Standard Supersym\'etrique Minimal (MSSM), l'extension supersym\'etrique la plus simple du Mod\`ele Standard, et son candidat \`a la Mati\`ere Noire, le neutralino, il est montr\'e que les contraintes obtenues en collisionneur pourraient fournir des informations sur une p\'eriode de l'Univers jeune, l'\`ere inflationnaire. Il est \'egalement d\'emontr\'e que la D\'etection Indirecte de Mati\`ere Noire, en d\'epit de plusieurs inconv\'enients, peut se r\'ev\'eler \^{e}tre une technique efficace pour explorer les mod\`eles de Mati\`ere Noire supersym\'etrique. Au-del\`a du MSSM il est montr\'e que des caract\'eristiques uniques du candidat \`a la Mati\`ere Noire dans le NMSSM peuvent \^{e}tre explor\'ees aux collisionneurs. L'\'etude d'un mod\`ele supersym\'etrique avec une sym\'etrie de jauge \'etendue, le UMSSM, est \'egalement d\'evelopp\'ee. Les caract\'eristiques d'un autre candidat de la mati\`ere noire de ce mod\`ele, le sneutrino droit, sont analys\'ees. Des contraintes plus g\'en\'erales telles que celles provenant d'observables de basse \'energie sont finalement prise en compte.}

~~~\\
\footnotesize{\textbf{Mots-cl\'es} : Mati\`ere Noire - Supersym\'etrie - Contraintes en collisionneur et d'astroparticule - Neutralino - Sneutrino droit - UMSSM}
~~~\\
\rule{\linewidth}{.5pt}
\begin{center}
\begin{minipage}{2cm}
\href{http://lapth.cnrs.fr}{\includegraphics[width=2.2cm,height=1.5cm]{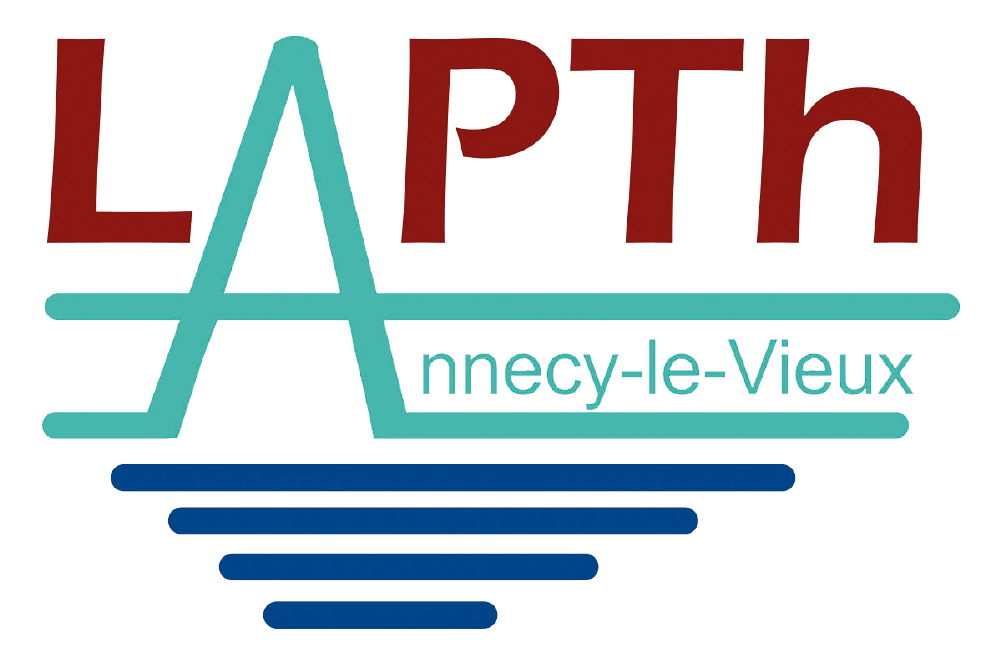}}
\end{minipage}
\begin{minipage}{9cm}
\begin{center}
~~\\
\scriptsize{
Laboratoire d’Annecy-le-Vieux de Physique Theorique (LAPTh)\\
UMR 5108, CNRS-INP/Universit\'e de Savoie\\
9, Chemin de Bellevue, BP 110\\
74 941, Annecy-le-Vieux Cedex\\
France}
\end{center}	
\end{minipage}
\begin{minipage}{3.1cm}
\href{http://www.cnrs.fr}{\includegraphics[width=1.5cm,height=1.5cm]{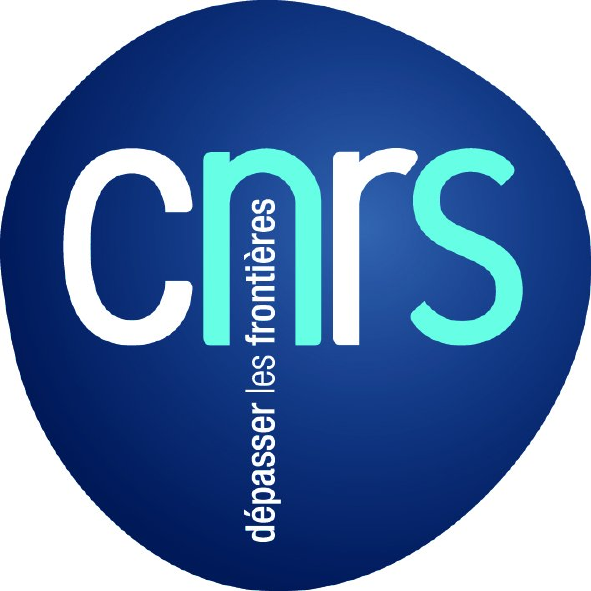}}
\href{http://www.univ-savoie.fr}{\includegraphics[width=1.5cm,height=1.5cm]{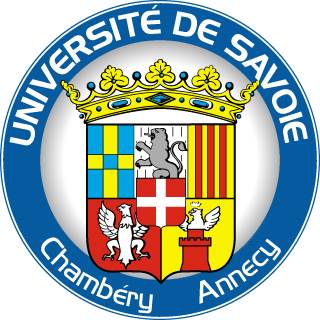}} 
\end{minipage}
\end{center}

\end{document}